\pdfoutput=1
\documentclass[11pt,twoside,a4paper,cmspaper,final,collab]{cms-tdr}
\def\svnVersion{ca0be6f}\def\svnDate{2024/05/23}\def\cmsCernNoTag{CERN-EP-2023-136}\def\cmsCernDate{\today}\def\cmsMessage{Published in the Journal of Instrumentation as \href{http://dx.doi.org/10.1088/1748-0221/19/05/P05064}{\doi{10.1088/1748-0221/19/05/P05064}.}}
\begin{document}\cmsNoteHeader{PRF-21-001}

\providecommand{\cmsTable}[1]{\resizebox{\textwidth}{!}{#1}}

\makeatletter
\newcommand{\footnotemarks}[2]{\textsuperscript{\@alph{#1},\@alph{#2}}}
\makeatother

\newcommand{\CentOSseven}{\mbox{CentOS 7}\xspace}
\newcommand{\DCDC}{\mbox{DC-DC}\xspace}
\newcommand{\ItwoC}{\ensuremath{\text{I}^2\text{C}}\xspace}
\newcommand{\LastAccessed}{July 28, 2023}
\newcommand{\layer}[1]{\mbox{layer~#1}\xspace}
\newcommand{\Linuxeight}{\mbox{Linux~8}\xspace}
\newcommand{\Phase}[1]{\mbox{Phase~#1}\xspace}
\newcommand{\proc}{\ensuremath{\text{PROC600}}\xspace}
\newcommand{\psiana}{\ensuremath{\text{PSI46}}\xspace}
\newcommand{\psidig}{\ensuremath{\text{PSI46dig}}\xspace}
\newcommand{\Run}[1]{\mbox{Run~#1}\xspace}
\newcommand{\SLINK}{\mbox{S-LINK64}\xspace}
\newcommand{\Tier}[1]{\mbox{Tier~#1}\xspace}
\newcommand{\uGMT}{$\mu$GMT\xspace}
\newcommand{\uGT}{$\mu$GT\xspace}
\newcommand{\uHTR}[1][]{$\mu$HTR#1\xspace}
\newcommand{\uROS}{$\mu$ROS\xspace}
\newcommand{\uTCA}{$\mu$TCA\xspace}

\newcommand{\egamma}{\ensuremath{\Pe/\PGg}\xspace}
\newcommand{\pp}{\ensuremath{\Pp\Pp}\xspace}
\newcommand{\Zee}{\ensuremath{\PZ\to\Pe\Pe}\xspace}

\newcommand{\abseta}{\ensuremath{\abs{\eta}}\xspace}
\newcommand{\avPU}{\ensuremath{\langle\text{PU}\rangle}\xspace}
\newcommand{\betastar}{\ensuremath{\beta^\ast}\xspace}
\newcommand{\Deltat}{\ensuremath{\Delta t}\xspace}
\newcommand{\Deta}{\ensuremath{\Delta\eta}\xspace}
\newcommand{\detajj}{\ensuremath{\Deta_{\mathrm{jj}}}\xspace}
\newcommand{\Dphi}{\ensuremath{\Delta\phi}\xspace}
\newcommand{\DTCL}{\ensuremath{D_{x,\text{TCL}}}\xspace}
\newcommand{\dTCL}{\ensuremath{d_{\text{TCL}}}\xspace}
\newcommand{\dTCT}{\ensuremath{d_{\text{TCT}}}\xspace}
\newcommand{\DXRP}{\ensuremath{D_{x,\text{XRP}}}\xspace}
\newcommand{\dXRP}{\ensuremath{d_{\text{XRP}}}\xspace}
\newcommand{\dxy}{\ensuremath{d_{xy}}\xspace}
\newcommand{\epsmax}{\ensuremath{\varepsilon_{\text{max}}}\xspace}
\newcommand{\mjj}{\ensuremath{m_{\mathrm{jj}}}\xspace}
\newcommand{\Mmin}{\ensuremath{M_{\text{min}}}\xspace}
\newcommand{\Mmax}{\ensuremath{M_{\text{max}}}\xspace}
\newcommand{\mominitial}{\ensuremath{\vec{p}_{\mathrm{i}}}\xspace}
\newcommand{\momfinal}{\ensuremath{\vec{p}_{\mathrm{f}}}\xspace}
\newcommand{\nTCT}{\ensuremath{n_{\text{TCT}}}\xspace}
\newcommand{\sigTCL}[1]{\ensuremath{\sigma_{x,\text{TCL#1}}}\xspace}
\newcommand{\sigTCT}{\ensuremath{\sigma_{x,\text{TCT}}}\xspace}
\newcommand{\sigXRP}{\ensuremath{\sigma_{x,\text{XRP}}}\xspace}
\newcommand{\tplus}{\ensuremath{t_+}\xspace}
\newcommand{\tminus}{\ensuremath{t_-}\xspace}
\newcommand{\Vapp}{\ensuremath{V_{\text{app}}}\xspace}
\newcommand{\Veff}{\ensuremath{V_{\text{eff}}}\xspace}
\newcommand{\Vfifty}{\ensuremath{V_{50\%}}\xspace}
\newcommand{\Vgas}{\ensuremath{V_{\text{gas}}}\xspace}

\newcommand{\AltwoOthree}{\ensuremath{\text{Al}_2\text{O}_3}\xspace}
\newcommand{\ArCOtwo}{\ensuremath{\text{Ar+CO}_2}\xspace}
\newcommand{\CFfour}{\ensuremath{\text{CF}_4}\xspace}
\newcommand{\COtwo}{\ensuremath{\text{CO}_2}\xspace}
\newcommand{\CsixFfourteen}{\ensuremath{\text{C}_6\text{F}_{14}}\xspace}
\newcommand{\CtwoHtwoFfour}{\ensuremath{\text{C}_2\text{H}_2\text{F}_4}\xspace}
\newcommand{\iCfourHten}{\ensuremath{\text{i-C}_4\text{H}_{10}}\xspace}
\newcommand{\HtwoO}{\ensuremath{\text{H}_2\text{O}}\xspace}
\newcommand{\Ntwo}{\ensuremath{\text{N}_2}\xspace}
\newcommand{\Otwo}{\ensuremath{\text{O}_2}\xspace}
\newcommand{\PbWOfour}{\ensuremath{\text{PbWO}_4}\xspace}
\newcommand{\SFsix}{\ensuremath{\text{SF}_6}\xspace}

\newcommand{\sixtyCo}{\ensuremath{{}^{60}\text{Co}}\xspace}
\newcommand{\onethreesevenCs}{\ensuremath{{}^{137}\text{Cs}}\xspace}

\newcommand{\Ccm}{\ensuremath{\,\text{C/cm}}\xspace}
\newcommand{\cmsq}{\ensuremath{\,\text{cm}^2}\xspace}
\newcommand{\deC}{\ensuremath{{}^\circ\text{C}}\xspace}
\newcommand{\eV}{\ensuremath{\,\text{e\hspace{-.08em}V}}\xspace}
\newcommand{\fC}{\ensuremath{\,\text{fC}}\xspace}
\newcommand{\GB}{\ensuremath{\,\text{GB}}\xspace}
\newcommand{\Gbs}{\ensuremath{\,\text{Gb/s}}\xspace}
\newcommand{\GBs}{\ensuremath{\,\text{GB/s}}\xspace}
\newcommand{\GHz}{\ensuremath{\,\text{GHz}}\xspace}
\newcommand{\Gy}{\ensuremath{\,\text{Gy}}\xspace}
\newcommand{\Hzcmsq}{\ensuremath{\,\text{Hz/cm}^2}\xspace}
\newcommand{\kA}{\ensuremath{\,\text{kA}}\xspace}
\newcommand{\kB}{\ensuremath{\,\text{kB}}\xspace}
\newcommand{\kHz}{\ensuremath{\,\text{kHz}}\xspace}
\newcommand{\kHzcmsq}{\ensuremath{\,\text{kHz/cm}^2}\xspace}
\newcommand{\kOhm}{\ensuremath{\,\text{k}\Omega}\xspace}
\newcommand{\kRad}{\ensuremath{\,\text{kRad}}\xspace}
\newcommand{\kV}{\ensuremath{\,\text{kV}}\xspace}
\newcommand{\kVcm}{\ensuremath{\,\text{kV/cm}}\xspace}
\newcommand{\kW}{\ensuremath{\,\text{kW}}\xspace}
\newcommand{\mA}{\ensuremath{\,\text{mA}}\xspace}
\newcommand{\mbar}{\ensuremath{\,\text{mbar}}\xspace}
\newcommand{\MB}{\ensuremath{\,\text{MB}}\xspace}
\newcommand{\Mbs}{\ensuremath{\,\text{Mb/s}}\xspace}
\newcommand{\MBs}{\ensuremath{\,\text{MB/s}}\xspace}
\newcommand{\mCcmsq}{\ensuremath{\,\text{mC/cm}^2}\xspace}
\newcommand{\mcub}{\ensuremath{\,\text{m}^3}\xspace}
\newcommand{\mcubh}{\ensuremath{\,\text{m}^3\!\text{/h}}\xspace}
\newcommand{\MHz}{\ensuremath{\,\text{MHz}}\xspace}
\newcommand{\MHzcmsq}{\ensuremath{\,\text{MHz/cm}^2}\xspace}
\newcommand{\MHzcmsqns}{\ensuremath{\text{MHz/cm}^2}\xspace}
\newcommand{\mmsq}{\ensuremath{\,\text{mm}^2}\xspace}
\newcommand{\mmcub}{\ensuremath{\,\text{mm}^3}\xspace}
\newcommand{\MOhm}{\ensuremath{\,\text{M}\Omega}\xspace}
\newcommand{\mrad}{\ensuremath{\,\text{mrad}}\xspace}
\newcommand{\Mrad}{\ensuremath{\,\text{Mrad}}\xspace}
\newcommand{\Mradns}{\ensuremath{\text{Mrad}}\xspace}
\newcommand{\ms}{\ensuremath{\,\text{ms}}\xspace}
\newcommand{\msq}{\ensuremath{\,\text{m}^2}\xspace}
\newcommand{\mTesla}{\ensuremath{\,\text{mT}}\xspace}
\newcommand{\muA}{\ensuremath{\,\mu\text{A}}\xspace}
\newcommand{\mumsq}{\ensuremath{\,\mu\text{m}^2}\xspace}
\newcommand{\mV}{\ensuremath{\,\text{mV}}\xspace}
\newcommand{\Neq}{\ensuremath{\,\Pn_{\mathrm{eq}}\text{/cm}^2}\xspace}
\newcommand{\nm}{\ensuremath{\,\text{nm}}\xspace}
\newcommand{\ns}{\ensuremath{\,\text{ns}}\xspace}
\newcommand{\Ohm}{\ensuremath{\,\Omega}\xspace}
\newcommand{\Ohmcm}{\ensuremath{\,\Omega\text{cm}}\xspace}
\newcommand{\pcmsq}{\ensuremath{\,\Pp\text{/cm}^2}\xspace}
\newcommand{\pcmsqs}{\ensuremath{\,\Pp\text{/(cm}^2\text{s)}}\xspace}
\newcommand{\percmsqns}{\ensuremath{\text{cm}^{-2}}\xspace}
\newcommand{\ppm}{\ensuremath{\,\text{ppm}}\xspace}
\newcommand{\ps}{\ensuremath{\,\text{ps}}\xspace}
\newcommand{\TB}{\ensuremath{\,\text{TB}}\xspace}
\newcommand{\TBq}{\ensuremath{\,\text{TBq}}\xspace}
\newcommand{\Tbs}{\ensuremath{\,\text{Tb/s}}\xspace}
\newcommand{\uHz}{\ensuremath{\,\text{Hz}}\xspace}
\newcommand{\uPB}{\ensuremath{\,\text{PB}}\xspace}

\newcommand{\nplus}{\ensuremath{\text{n}^+}\xspace}

\cmsNoteHeader{PRF-21-001}

\title{Development of the CMS detector for the CERN LHC \Run3}

\author*[cern]{CMS Collaboration}
\date{\today}

\abstract{Since the initial data taking of the CERN LHC, the CMS experiment has undergone substantial upgrades and improvements. This paper discusses the CMS detector as it is configured for the third data-taking period of the CERN LHC, \Run3, which started in 2022. The entire silicon pixel tracking detector was replaced. A new powering system for the superconducting solenoid was installed. The electronics of the hadron calorimeter was upgraded. All the muon electronic systems were upgraded, and new muon detector stations were added, including a gas electron multiplier detector. The precision proton spectrometer was upgraded. The dedicated luminosity detectors and the beam loss monitor were refurbished. Substantial improvements to the trigger, data acquisition, software, and computing systems were also implemented, including a new hybrid CPU/GPU farm for the high-level trigger.}

\hypersetup{%
pdfauthor={CMS Collaboration},%
pdftitle={Development of the CMS detector for the CERN LHC Run 3},%
pdfsubject={CMS},%
pdfkeywords={CMS, Run 3, physics, detector}}

\maketitle
\tableofcontents

\clearpage
\section{Introduction}

The CMS detector~\cite{CMS:Detector-2008} is a large, multipurpose apparatus located at the CERN LHC.
The detector was designed for the study of a variety of physics phenomena, including the search for the Higgs boson, which was discovered in 2012~\cite{CMS:HIG-12-028, CMS:HIG-12-036, ATLAS:2012yve}, and the measurement of its properties, the exploration of the electroweak sector and vector boson scattering, precision measurements of standard model (SM) particles and interactions, flavor physics, heavy-ion physics, and searches for new physics beyond the SM.

The LHC \Run1 started in 2009 and, until the end of 2012, proton-proton (\pp) collision data corresponding to a total integrated luminosity of about 30\fbinv were delivered at center-of-mass energies of 7 and 8\TeV.
In addition, CMS successfully recorded data from high-energy lead-lead collisions.
After a first long shutdown, referred to as long shutdown 1 (LS1), the second data-taking period, \Run2, followed in 2015--2018 at an energy of 13\TeV, during which an integrated luminosity of about 165\fbinv was delivered with peak instantaneous luminosities up to $2\times10^{34}\percms$, twice the original LHC design value.
During LS1 and \Run2, the first generation of detector upgrades, referred to as the \Phase1 upgrade program, was implemented.

After the second long shutdown (LS2, 2019--2021), the LHC \Run3 was started in 2022 and is expected to deliver about 250\fbinv of integrated luminosity.
In \Run3, the center-of-mass energy for \pp collisions is 13.6\TeV.
During the third long shutdown (LS3), scheduled to start in 2026, CMS will undergo a major upgrade program, referred to as the \Phase2 upgrade, in preparation for the data taking at the High-Luminosity LHC (HL-LHC), designed to deliver instantaneous luminosities up to $7.5\times10^{34}\percms$ at a \pp center-of-mass energy of 14\TeV.
At the end of the HL-LHC, the total integrated luminosity is expected to be 3000\fbinv.
The \Phase2 upgrade includes a new inner tracking system and a new endcap calorimeter, as well as substantial improvements for most other subsystems of CMS.
Upgrades are also in preparation in almost all other areas of CMS.
In this paper, we present the various upgrades of the CMS detector since \Run1 that are designed to optimize the detector for sustained or improved performance at increased luminosity and energy.

\begin{figure}[!t]
\centering
\includegraphics[width=\textwidth]{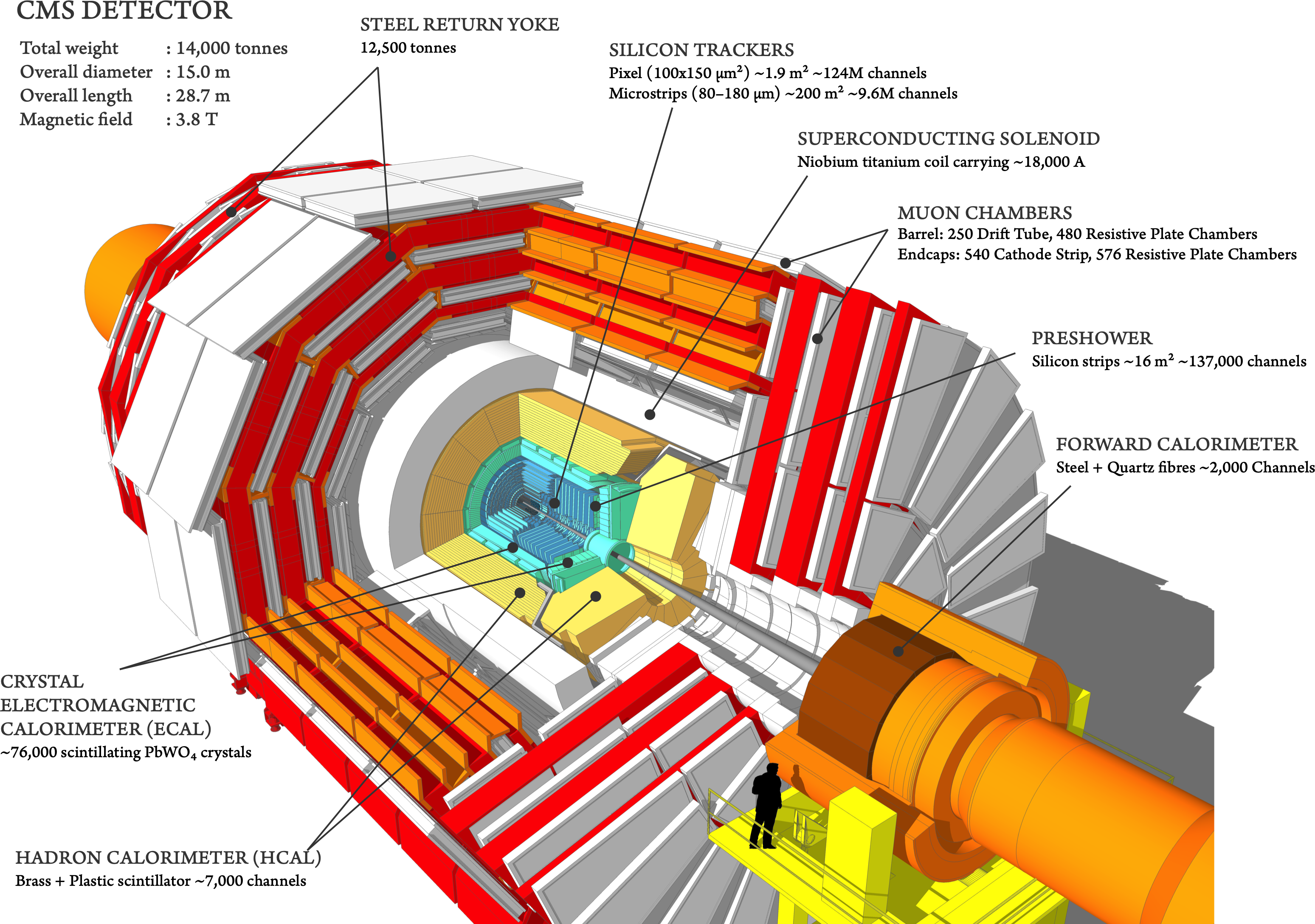}
\caption{%
    Schematic drawing of the CMS detector, from Ref.~\cite{CMS:OUTREACH-2019-001}.
}
\label{fig:introduction:cms}
\end{figure}

The CMS detector has an overall length of 22\unit{m}, a diameter of 15\unit{m}, and weighs 14\,000 tons.
A schematic view is shown in Fig.~\ref{fig:introduction:cms}.
The detector is nearly hermetic, designed to trigger on~\cite{CMS:TRG-17-001, CMS:TRG-12-001} and identify electrons, muons, photons, and (charged and neutral) hadrons~\cite{CMS:EGM-13-001, CMS:MUO-16-001, CMS:EGM-14-001, CMS:TRK-11-001}.
The central feature of the CMS experiment is a superconducting solenoid of 6\unit{m} internal diameter and 12.5\unit{m} length that provides a magnetic field of 3.8\unit{T} with a stored energy of 2.2\unit{GJ}.
Within the magnetic volume are a silicon pixel and strip tracker, a lead tungstate crystal electromagnetic calorimeter (ECAL), and a brass and scintillator hadron calorimeter (HCAL), each composed of a barrel and two endcap sections.
Forward calorimeters extend the pseudorapidity ($\eta$) coverage provided by the barrel and endcap detectors.
Muons are measured in gas-ionization detectors embedded in the steel flux-return yoke outside the solenoid.

Events of interest are selected using a two-tier trigger system.
The first-level (L1) trigger is composed of custom hardware processors and uses information from the calorimeters and muon detectors to select events at a rate of about 100\kHz within a latency of about 4\mus~\cite{CMS:TRG-17-001}.
The second level, known as the high-level trigger (HLT), consists of a cluster of commercial processors running a version of the full event reconstruction software optimized for fast processing.
It was originally designed to reduce the event rate to around 1\kHz before data storage~\cite{CMS:TRG-12-001}.
During \Run3, the L1 trigger and HLT operate at typical output rates of 110\kHz and 5\kHz, respectively.

A full description of the CMS detector, together with a definition of the coordinate system used and the relevant kinematic variables, is reported in Ref.~\cite{CMS:Detector-2008}.
In the remaining part of this section, the CMS detector components are briefly introduced.

In Section~\ref{sec:magnet}, the CMS solenoid magnet is described.

The inner tracking system (Section~\ref{sec:tracker}) is used to measure the trajectories of charged particles produced in the collisions at the LHC.
It is located in the innermost part of the CMS detector, closest to the interaction point.
Prior to the \Phase1 upgrade, the pixel detector had three barrel layers and two disks in each endcap.
In its current form, the pixel detector is composed of four barrel layers and three disks of silicon sensors on each side of the interaction point, with a total of 124 million readout channels.
During LS2, the innermost barrel layer was replaced to ensure optimal performance until the end of \Run3.
The strip tracker comprises ten layers of silicon strip sensors in the barrel, arranged in a cylindrical shape, and nine disks in the endcaps on each side of the detector.
The strip sensors are segmented into long, thin strips, which are used to measure the trajectories of the particles and provide a hit resolution of 20\mum for charged particles that cross the sensor perpendicularly.
The tracker is designed to have excellent momentum resolution and tracking efficiency.
It can detect and track particles with transverse momentum \pt as low as 50\MeV within the range $\abseta<2.5$.
Tracks with a momentum around 100\GeV in the central region of the detector have an impact parameter resolution of about 10\mum, and a transverse momentum resolution near 1\%.

The electromagnetic calorimeter (Section~\ref{sec:ecal}) is made of 75\,848 lead tungstate (\PbWOfour) crystals:\ 61\,200 crystals are located in the barrel (EB) and 7324 in each of the endcaps (EE) that provide a pseudorapidity coverage of $\abseta<3$.
The lead tungstate crystals have a depth of about 23\cm, corresponding to about 25 radiation lengths $X_0$.
Preshower detectors, consisting of two planes of silicon sensors interleaved with a total of $3X_0$ of lead, are located in front of each EE detector.
The ECAL is designed to identify electrons and photons, and measure their positions and energies, which are reconstructed from energy deposits using algorithms that constrain the clusters to the size and shape expected for electrons or photons.
The ECAL also contributes significantly to the reconstruction of jets and missing transverse momentum.
The electron momentum is estimated by combining the energy measurement in the ECAL with the momentum measurement in the tracker.
The momentum resolution for electrons with $\pt\approx45\GeV$ from $\PZ\to\EE$ decays ranges from 1.6\% to 5\%.
It is generally better in the barrel region than in the endcaps, and also depends on the bremsstrahlung energy emitted by the electron as it traverses the material in front of the ECAL.
The ECAL also provides information on the arrival time of the electrons and photons which can be used in physics analyses, such as searches for long-lived particles.

The hadron calorimeter (Section~\ref{sec:hcal}) is designed to measure the energy of charged and neutral hadrons.
It contributes to the identification of hadrons and the measurement of their properties.
It also aids in the reconstruction of jets and missing transverse momentum, and the identification of electrons and photons.
The HCAL comprises four subdetectors:\ the hadron barrel (HB), hadron endcap (HE), hadron outer (HO), and hadron forward (HF) calorimeters.
The HB and HE are located inside the solenoid magnet of the CMS detector and surround the ECAL.
They cover the pseudorapidity ranges $\abseta<1.392$ and $1.305<\abseta<3.0$, respectively, and are made of layers of brass plates interleaved with layers of scintillating tiles.
The HF, constructed from steel and quartz fibers, is located outside the solenoid at $\pm11.5\unit{m}$ from the collision point, and covers the $3.0<\abseta<5.2$ range.
Finally, the HO, made of plastic scintillator and the first layer of the barrel flux return covers the $\abseta<1.26$ range.
The HCAL is designed to have a good hermeticity, with the ability to detect hadrons in nearly the full $4\pi$ solid angle.
The \Phase1 upgrade of HCAL was installed in stages from 2016--2019.
In HB and HE, the hybrid photodiode detectors were replaced with silicon photomultipliers, which reduced anomalous signals, improved radiation tolerance, and allowed for finer longitudinal readout segmentation.
The HF photomultiplier tubes were also upgraded to reduce anomalous signals.
The readout electronics were upgraded to support the increased channel count, improve the precision, and add signal timing information.
When combining information from the entire CMS detector, the jet energy resolution typically amounts to 15--20\% at 30\GeV, 10\% at 100\GeV, and 5\% at 1\TeV.

The muon detectors (Section~\ref{sec:muon}) are used in the identification of muons and the measurement of their momenta.
The muon system comprises four subsystems:\ the drift tubes (DTs), the cathode strip chambers (CSCs), the resistive-plate chambers (RPCs), and the recently added gas electron multiplier (GEM) detector.
Altogether, the CMS muon detectors comprise almost one million electronic channels.

The DTs consist of chambers formed by multiple layers of long rectangular tubes that are filled with an Ar and \COtwo gas mixture.
An anode wire is located at the center of each tube, whereas cathode and field-shaping strips are positioned on its borders.
They create an electric field that induces an almost uniform drift of ionization electrons produced by charged particles traversing the gas.
The charged-particle trajectory is determined from the arrival time of the currents generated on the anode wires of the readout.

The CSCs are made of layers of proportional wire chambers with orthogonal cathode strips and are operated with a gas mixture of Ar, \COtwo, and \CFfour.
Signals are generated on both anode wires and cathode strips.
The finely segmented cathode strips and fast readout electronics provide good timing and spatial resolution to trigger on and identify muons.

The RPCs comprise two detecting layers of high-pressure laminate plates that are separated by a thin gap filled with a gas mixture of \CtwoHtwoFfour, \iCfourHten, and \SFsix.
The electronic readout strips are located between the two layers, and the high voltage is applied to high-conductivity electrodes coated on each plate.
The detectors are operated in avalanche mode to cope with the high background rates.
Due to their excellent time resolution, they ensure a precise bunch-crossing assignment for muons at the trigger level.

The key feature of the GEM is a foil consisting of a perforated insulating polymer surrounded on the top and bottom by conductors.
A voltage difference is applied on the foils producing a strong electric field in the holes.
The GEM is operated with a gas mixture of Ar and \COtwo.
When the gas volume is ionized, electrons are accelerated through the holes and read out on thinly separated strips.
This structure allows for high amplification factors with modest voltages that provide good timing and spatial resolution, and can be operated at high rates.

The precision proton spectrometer (PPS) (Section~\ref{sec:pps}) is designed to detect protons scattered at very small angles in interactions where the protons remain intact and only a small fraction of their initial energy goes into the production of particles at small rapidity.
In such events, the reconstruction of the kinematic properties of the protons uses their energy loss to determine the invariant mass of the system produced in the quasi-elastic collision.
The PPS detector includes tracking and timing stations, which are located inside the LHC tunnel on both ends of the CMS detector about 200\unit{m} from the CMS interaction point.
Precision tracking and timing is provided by silicon pixel and diamond detectors, respectively.
The detectors are enclosed in movable stations, referred to as ``roman pots'', within which the detectors can be positioned as close as a few millimeters from the proton beam.
The PPS first started in \Run2 as a joint project (CT-PPS) with the TOTEM Collaboration~\cite{CMS:PRO-21-001}.
The initial PPS system consisted of two tracking and one timing station on each side.
For \Run3, the PPS was upgraded for improved efficiency and precision with an additional timing station on each side.

The beam radiation instrumentation and luminosity (BRIL) system (Section~\ref{sec:bril}) comprises various detectors that measure the instantaneous luminosity and monitor in real time the beam-induced background, beam losses, and timing.
Three luminosity detector systems provide robust bunch-by-bunch luminosity measurements in real time.
They are:\ (i)~the fast beam condition monitor (BCM1F), which counts hits in silicon pad diodes; (ii)~the pixel luminosity telescope (PLT), which counts triple coincidences; and (iii)~the HF calorimeter.
The HF is instrumented with a dedicated readout for the real-time luminosity measurement and provides hit-tower counting (HFOC) and transverse-energy sums (HFET).
The beam condition monitor for losses (BCML), using diamond and sapphire sensors, provides protection against catastrophic beam loss and is part of the LHC beam-abort system.
The beam pickup timing device (BPTX) provides logical beam signals to the L1 trigger system.
The BRIL system includes the measurement of radiation in the experimental cavern.
The measurements are complemented by detailed simulations using the CMS radiation simulation applications.

The data acquisition (DAQ) system (Section~\ref{sec:daq}) is responsible for:\ the readout of all detector data for events accepted by the L1 trigger; the building of complete events from subdetector event fragments; the operation of the filter farm cluster running the HLT; and the transport of event data selected by it to the permanent storage in the \Tier0 computing center.
The DAQ consists of:\ custom-built electronics reading out event fragments; a data-concentrator network transporting the fragments to the surface; a cluster of readout and event-building servers interconnected via the event-building network; the filter-farm cluster of multicore servers connected by the data network running the HLT software; a distributed storage system where event data selected by HLT filtering are buffered; and a transfer system connected to the \Tier0 center via a high-speed network.
The DAQ also includes the trigger control and distribution system (TCDS), which distributes timing to the trigger and subdetector electronics, and implements trigger control logic as well as the trigger throttling system (TTS).

The L1 trigger (Section~\ref{sec:l1trigger}) consists of electronics responsible for making a fast selection of events based on the presence of high-energy particles in the detector.
The L1 trigger receives energy and position information, so-called trigger primitives (TPs), from the calorimeters and the muon detectors.
The TPs are evaluated by a trigger processor, which is composed of custom-built electronics and field programmable gate array (FPGA) devices that perform the trigger decision based on a set of predefined trigger algorithms.
The L1 trigger operates at trigger rates of about 110\kHz.
During LS2, the L1 trigger was upgraded to also process TPs that are designed to select long-lived particles.

The HLT (Section~\ref{sec:hlt}) is a software-based system in which the full event information is used to select events of interest based on their physics content.
The HLT is implemented as a parallel computing system that processes the event data in real time.
Since the start of \Run3, the HLT makes use of graphical processing units (GPUs) in the trigger filter farm.
The GPUs facilitate the offloading of specific parts of the reconstruction algorithms, \eg, tracking based on the pixel detector, as well as parts of the calorimeter reconstruction.
The use of GPUs has led to a substantial reduction of the overall event processing time.
With the performance improvements for \Run3, HLT-reconstructed analysis data, referred to as HLT scouting data, are recorded at a rate of about 30\kHz.
In parallel, the storage rate of normal triggers was increased to about 2\kHz.
Furthermore, the system also stores extra data samples, the so-called parking data sets, at a rate of around 3\kHz.
The parking events will only be reconstructed by the offline computing infrastructure at a later time, when the resources will not be needed for the core activities.
Other HLT reconstruction improvements in the areas of muon tracking, \PQb jet tagging, and tau lepton reconstruction were also implemented for \Run3.

The offline computing system (Section~\ref{sec:offline}) has three key roles:\ to process the recorded data; to generate sufficiently large Monte Carlo simulation samples based on theoretical models and detector response modeling; and to facilitate physics data analysis performed at the CMS institutes around the world.
The CMS data and simulation samples are stored and processed in a globally distributed network of centers, using an ever-growing array of heterogeneous resources.
Continuous improvements are made in software and computing performance.
Most notably, multithreaded processing and offloading to GPU resources have been introduced.

\clearpage
\section{Solenoid magnet}
\label{sec:magnet}

\begin{figure}[!h]
\centering
\includegraphics[width=0.7\textwidth]{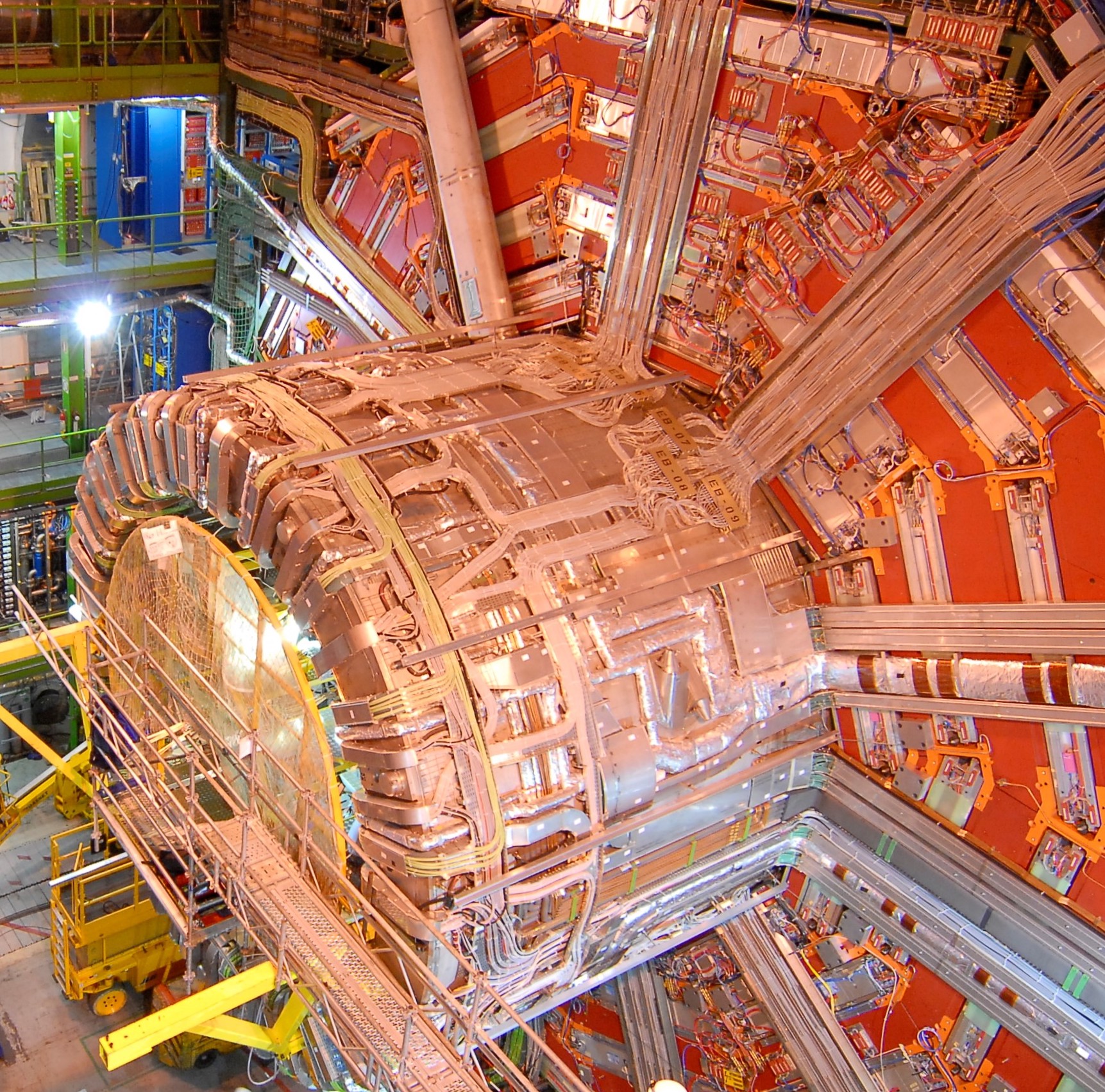}
\caption{%
    The solenoid magnet cryostat within the open CMS detector, from Ref.~\cite{CMS:PHO-GEN-2008-019}.
}
\label{fig:magnet:cryostat}
\end{figure}

The superconducting solenoid magnet provides a magnetic field of 3.8\unit{T}, and forms the center piece of the CMS experiment.
A picture of the open CMS detector with visible magnet cryostat is shown in Fig.~\ref{fig:magnet:cryostat}.

The original plan for the magnet during LS2 had been to turn it off, but to keep it cooled at 4\unit{K}.
The plan was changed substantially because of a water leak in the experimental cavern inside a diffusion pump of the magnet cryostat discovered during a routine check.
To intervene without putting the magnet at risk, it was decided to warm the magnet up to room temperature.

\begin{figure}[!htp]
\centering
\includegraphics[width=0.48\textwidth]{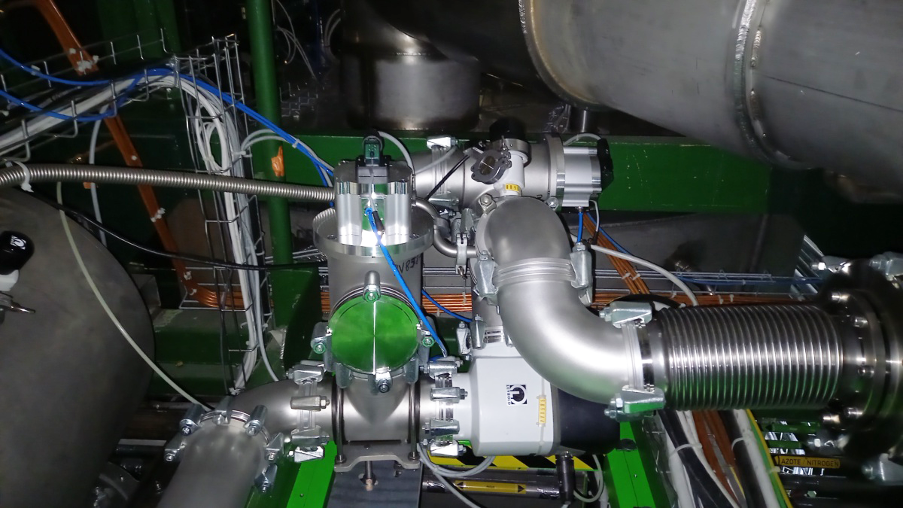}
\hfill
\includegraphics[width=0.48\textwidth]{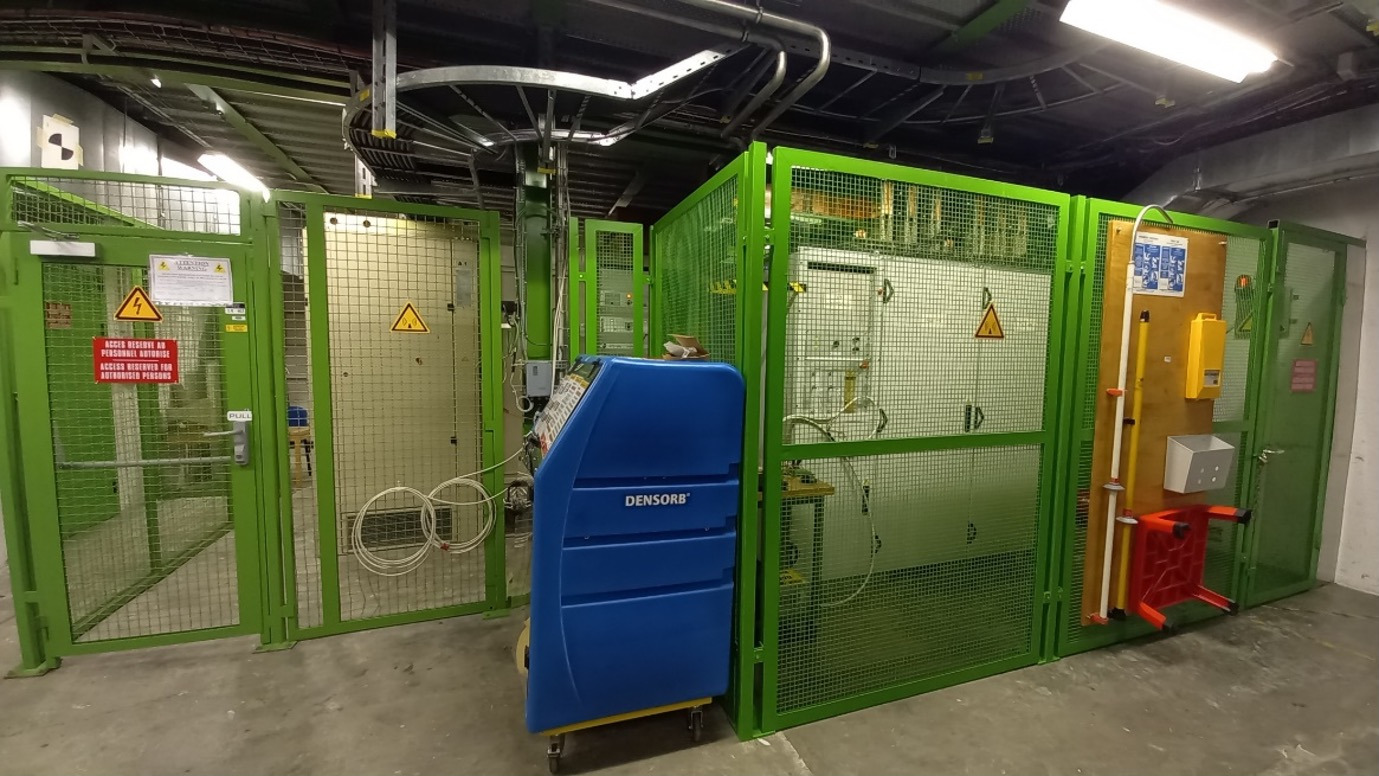}
\caption{%
    Photographs of a part of the new vacuum pumping circuit (left) and of the new CMS FWT system installed in the CMS service cavern (right).
}
\label{fig:magnet:updates}
\end{figure}

The procedure took place during the Covid-19 lockdown at CERN in April and May 2020.
After an outgassing period, the vacuum volume was brought back to atmospheric pressure, and the diffusion pumps were removed and replaced.
The vacuum circuit was also fully cleaned, and modifications were implemented for easier access and improved backup capabilities, with new valves and flanges allowing the connection of backup vacuum pumps if needed, as illustrated in the picture in Fig.~\ref{fig:magnet:updates} (left).

In parallel to this repair, the new free wheel thyristor (FWT) system~\cite{Yammine:2021edms} was installed on the powering circuit of the magnet, visible in Fig.~\ref{fig:magnet:updates} (right).
The FWT bypasses the power converter in a closed loop in case the converter is in a faulty state, \eg, in the event of a power failure or a lack of cooling, thus avoiding a slow discharge to zero current.
The FWT contributes to increasing the magnet's lifetime and the operational time at nominal field.

\begin{figure}[!htp]
\centering
\includegraphics[width=0.65\textwidth]{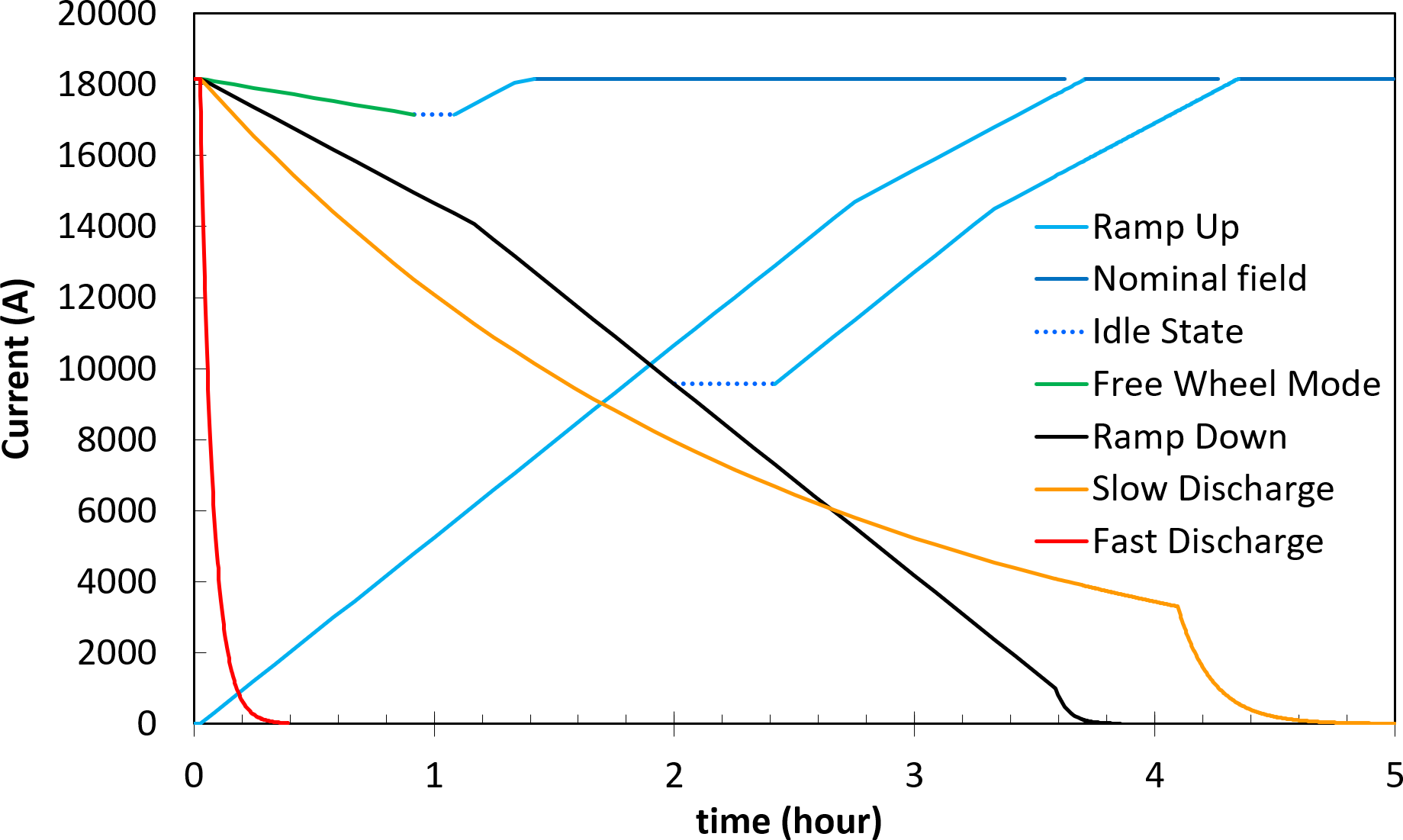}
\caption{%
    The CMS magnet current ramp and discharge modes representing the various magnet operation procedures and their duration.
    A current of 18164\unit{A} corresponds to a magnetic field of 3.8\unit{T}.
}
\label{fig:magnet:current}
\end{figure}

As shown in Fig.~\ref{fig:magnet:current}, a full discharge followed by a ramp up takes about eight hours.
A partial ramp down to 9.5\kA, corresponding to 2\unit{T} at the \pp interaction point, is implemented in case of a cryogenics stop.
The idle state of the power converter at constant current, indicated in Fig.~\ref{fig:magnet:current} as a dashed line, is devised to provide the time needed for the reconnection of the cold box and refill of the liquid helium dewar, this way avoiding the risk of a fast discharge caused by helium flow fluctuations which may trigger a quench on the power leads and superconducting busbars.
Also represented in Fig.~\ref{fig:magnet:current} is the ``free wheel'' mode, when the FWT is triggered, typically after a power glitch, followed by an idle state period  after the power converter restart and before the ramp up to nominal field can be resumed.

The control systems of the magnet and insulation vacuum were also fully upgraded during LS2.
New programmable logic controllers were installed, new control electronics for both the FWT and the renewed vacuum pumping circuit were integrated, and the magnetic measurement system, using Hall probes, flux-loops, and nuclear magnetic resonance devices, was consolidated.
The cryogenics system inside the cold box was improved by installation of a large filter with a reduced mesh to limit the recurrent clogging of the turbine filters as much as possible.

The magnet was successfully commissioned in September 2021, just ahead of the LHC pilot beam run, when it was operated at full magnetic field for two weeks with its upgraded powering system and repaired vacuum system.
In March 2022, the magnet was ramped up to its nominal field of 3.8\unit{T} and declared ready for \Run3.

\clearpage
\section{Inner tracking system}
\label{sec:tracker}

\subsection{Pixel detector}

The silicon pixel detector is the innermost part of the CMS inner tracking system.
It provides three-dimensional space points close to the LHC collision point, which allow for high precision tracking and vertex reconstruction.

\subsubsection{Detector design}
\label{sec:tracker:design}

The first CMS pixel detector~\cite{CMS:Detector-2008}, installed in 2008, consisted of three barrel layers at radii of 44, 73, and 102\mm and two endcap disks on each end at distances of 345 and 465\mm from the detector center.
It provided three-point tracking for charged particles and performed very well during \Run1.
However, already in \Run1 the instantaneous luminosity delivered by the LHC exceeded the design value of $1\times10^{34}\percms$, which resulted in a pixel detector readout inefficiency.
In order to maintain good tracking performance, this pixel detector was replaced with a more efficient and robust four-point tracking system.
In addition, the radius of the beam pipe was reduced in 2014 from 30 to 23\mm, which allowed the innermost pixel layer to be placed closer to the interaction point.
The improved pixel detector was installed at the beginning of 2017.

\begin{figure}[!htb]
\centering
\includegraphics[width=\textwidth]{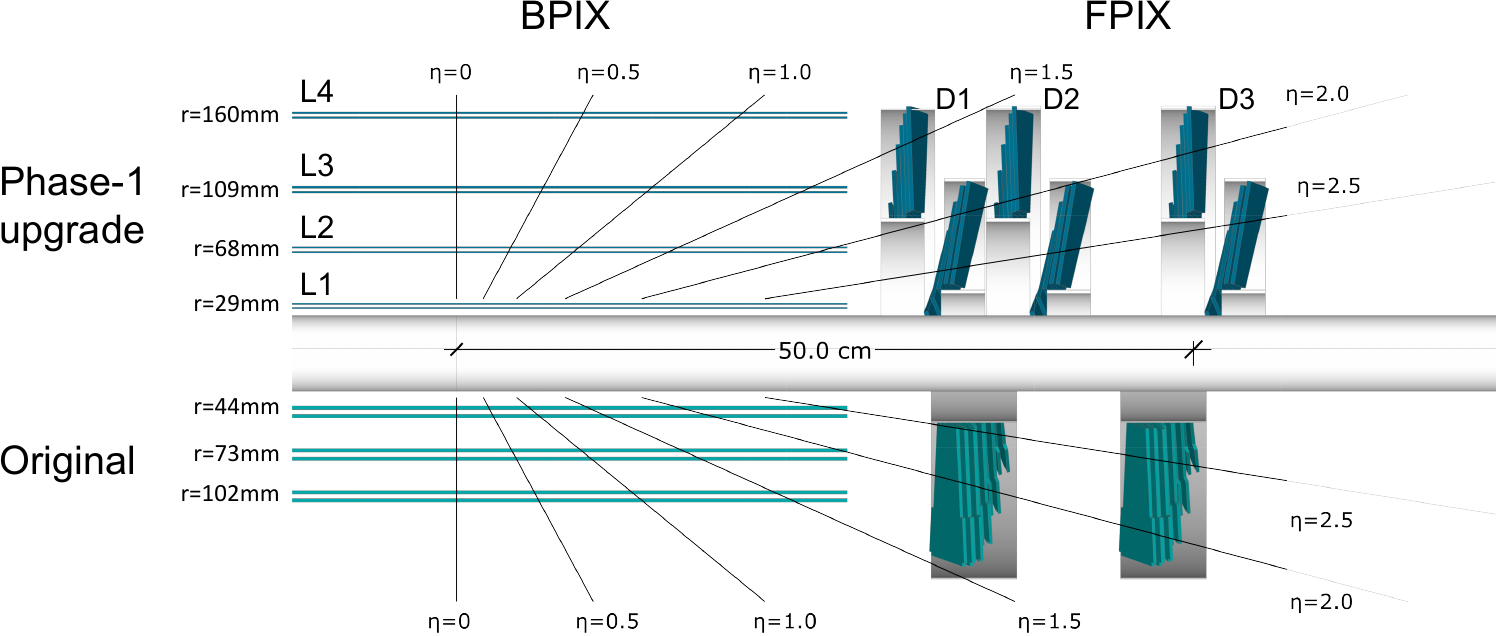}
\caption{%
    Longitudinal view of the \Phase1-upgraded pixel detector compared to the original detector layout, from Ref.~\cite{CMSTrackerGroup:2020edz}.
}
\label{fig:tracker:layout}
\end{figure}

The new detector, referred to as the \Phase1 pixel detector~\cite{CMS:TDR-011}, consists of four barrel layers (L1--L4) at radii of 29, 68, 109, and 160\mm, and three disks (D1--D3) on each end at distances of 291, 396, and 516\mm from the center of the detector.
The layouts of the two detectors, the original and the upgraded one, are compared in Fig.~\ref{fig:tracker:layout}.
The new layout provides four-hit coverage, instead of three, for tracks up to an absolute pseudorapidity of 3.0.
The details of the \Phase1 pixel detector design and construction have been published in Ref.~\cite{CMSTrackerGroup:2020edz}.

In addition to providing more tracking planes, the new detector contains several other improvements:\ \DCDC power converters are used to supply the necessary current while reusing the existing cables; a two-phase \COtwo cooling system is used; a 400\Mbs digital readout system is implemented instead of the 40\MHz analog one; and the readout chips are modified to handle higher data rates.

Table~\ref{tab:tracker:design} shows the main parameters of the \Phase1 pixel detector.
The detector consists of two parts, the pixel barrel (BPIX) and the pixel forward disks (FPIX), which are mechanically and electrically independent.
The BPIX detector consists of two 540\mm-long half-barrels divided in the $y$--$z$ plane.
The FPIX detector consists of twelve half-disks with radii ranging from 45 to 161\mm.
Each half-disk is further divided into two rings of modules.
The basic building block of both the BPIX and FPIX is a silicon sensor module comprised of a sensor with 160$\times$416 pixels and a size of $100{\times}150\mum^2$, connected to 16 readout chips (ROCs).
In total, 1184 and 672 modules are used in the BPIX and FPIX, respectively.
The entire \Phase1 pixel detector comprises a total of 124 million readout channels.

\begin{table}[!t]
\centering
\topcaption{%
    Summary of the average radius and $z$ position, as well as the number of modules for the four BPIX layers and six FPIX rings for the \Phase1 pixel detector.
}
\label{tab:tracker:design}
\renewcommand{\arraystretch}{1.1}
\begin{tabular}{lccc}
    \multicolumn{4}{c}{BPIX} \\
    Layer & Radius [mm] & $z$ position [mm] & Number of modules \\
    \hline
    L1 & \phantom{0}29 & $-270$ to $+270$ & \phantom{0}96 \\
    L2 & \phantom{0}68 & $-270$ to $+270$ & 224 \\
    L3 & 109 & $-270$ to $+270$ & 352 \\
    L4 & 160 & $-270$ to $+270$ & 512 \\
    \hline
    \multicolumn{4}{c}{} \\
    \multicolumn{4}{c}{FPIX} \\
    Disk & Radius [mm] & $z$ position [mm] & Number of modules \\
    \hline
    D1 inner ring & 45--110 & $\pm$338 & \phantom{0}88 \\
    D1 outer ring & 96--161 & $\pm$309 & 136 \\
    D2 inner ring & 45--110 & $\pm$413 & \phantom{0}88 \\
    D2 outer ring & 96--161 & $\pm$384 & 136 \\
    D3 inner ring & 45--110 & $\pm$508 & \phantom{0}88 \\
    D3 outer ring & 96--161 & $\pm$479 & 136 \\
\end{tabular}
\end{table}

{\tolerance=800
The sensor modules are mounted on light-weight mechanical structures, with thin-walled stainless steel tubes used for the \COtwo evaporative cooling (Section~\ref{sec:tracker:services}).
Carbon fiber and graphite materials with high thermal conductivity are incorporated into the detector mechanical structures (Section~\ref{sec:tracker:mechanics}).
Both the BPIX and FPIX are connected to four service half-cylinders.
They host the auxiliary electronics for readout and powering.
\par}

The installation of the \Phase1 pixel detector took place during the extended year-end technical stop of the LHC in 2016--2017.
The new detector performed successfully during the \Run2 data-taking period in 2017--2018.
However, since the detector installation in 2017 two major interventions took place.
Due to an unexpected failure (Section~\ref{sec:tracker:services}), all 1216 \DCDC converters had to be replaced in the LHC year-end technical stop 2017--2018.
The second intervention was done during LS2 and involved the replacement of the layer-1 modules and, again, all the \DCDC converters.
The new layer-1 modules, in addition to the planned replacement of the radiation-damaged silicon sensors, also included new versions of the readout ASICs, which fixed some of the shortcomings observed in the earlier versions, described in Section~\ref{sec:tracker:modules}.
The LS2 was also used to repair several faulty optical fibers and bad power connections, upgrade the electronic boards in the FPIX system, and fix broken FPIX cooling inlets.

With the innermost layer placed at a radius of 29\mm from the beam line, the modules in this region are exposed to very high radiation doses and hit rates, as shown in Table~\ref{tab:tracker:requirements}:\ the radiation fluence for L1 with 300\fbinv is $2.2\times10^{15}\Neq$, corresponding to the operational limit of the installed system~\cite{CMSTrackerGroup:2020edz}.
To ensure that the inner layer remains fully operational throughout all of \Run3, as planned from the beginning, the innermost BPIX layer was replaced during LS2 in 2019--2021.

\begin{table}[!ht]
\centering
\topcaption{%
    Expected hit rate, fluence, and radiation dose for the BPIX layers and FPIX rings.
    The hit rate corresponds to an instantaneous luminosity of $2.0\times10^{34}\percms$~\cite{CMSTrackerGroup:2020edz}.
    The fluence and radiation dose are shown for integrated luminosities of 300\fbinv for the BPIX L1 and 500\fbinv for the other BPIX layers and FPIX disks, well beyond the expected integrated luminosities for the detectors at the end of \Run3, of 250 and 370\fbinv, respectively.
}
\label{tab:tracker:requirements}
\renewcommand{\arraystretch}{1.1}
\begin{tabular}{lccc}
    & Pixel hit rate & Fluence & Dose \\
    & [\MHzcmsqns] & [$10^{15}\Neq$] & [\Mradns] \\
    \hline
    BPIX L1 & 580 & 2.2 & 100 \\
    BPIX L2 & 120 & 0.9 & \phantom{0}47 \\
    BPIX L3 & \phantom{0}58 & 0.4 & \phantom{0}22 \\
    BPIX L4 & \phantom{0}32 & 0.3 & \phantom{0}13 \\
    FPIX inner rings & 56--260 & 0.4--2.0 & 21--106\\
    FPIX outer rings & 30--\phantom{0}75 & 0.3--0.5 & 13--\phantom{0}28\\
\end{tabular}
\end{table}

\subsubsection{Silicon modules}
\label{sec:tracker:modules}

\begin{figure}[!ht]
\centering
\includegraphics[width=\textwidth]{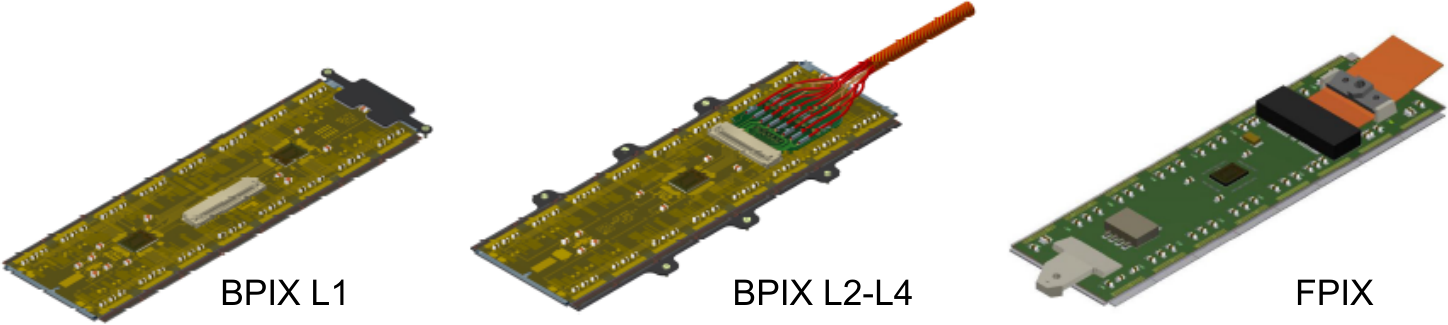}
\caption{%
    Drawings of the pixel detector modules for BPIX L1, BPIX L2--4, and the FPIX detector, from Ref.~\cite{CMSTrackerGroup:2020edz}.
}
\label{fig:tracker:modules}
\end{figure}

Schematic drawings of the \Phase1 pixel detector modules are shown in Fig.~\ref{fig:tracker:modules}.
A module consists of a $18.6{\times}66.6\mm^2$ silicon sensor that is bump-bonded to $2{\times}8$ ROCs.
Each ROC has $80{\times}52$ rectangular pixels with a size of $100{\times}150\mum^2$, the same as in the original pixel detector.
A high-density interconnect (HDI) flex printed circuit is glued to the sensor and wire-bonded to the 16 ROCs.
A token bit manager chip (TBM) is mounted on top of the HDI (two TBMs in the case of L1 modules).
The TBM controls the readout of a group of ROCs.

\paragraph{Sensors}

The silicon sensors are of the n-in-n type~\cite{CMSTrackerGroup:2020edz}, with strongly n-doped (\nplus) pixelated implants on an n-doped silicon bulk and a p-doped back side.
The \nplus implants collect electrons, which has the advantage of being
less affected by charge trapping caused by high irradiation~\cite{ROSE:2001avf, Rohe:2004hq}.
It is also advantageous as it allows sensors to be operated under-depleted after the detector is irradiated (when the so-called type inversion~\cite{CMSTrackerGroup:2020edz} is reached).

For the BPIX sensors, the n-side inter-pixel isolation was implemented through the moderated p-spray technique with a punch-through biasing grid.
More details can be found in the references of Ref.~\cite{CMSTrackerGroup:2020edz}.
The sensors are made from approximately 285\mum-thick phosphorus-doped
4-inch wafers produced in a float-zone (FZ) process (fabricated at CiS).
The FPIX sensors use open p-stops for n-side isolation, where each pixel is surrounded by an individual p-stop structure with an opening on one side.
They were produced on 300\mum-thick 6-inch FZ-wafers (produced at Sintef).

The radiation resistance of a module is, to a large extent, defined by the possibility of increasing the sensor bias voltage to obtain a sufficiently high signal charge.
As already mentioned in Section~\ref{sec:tracker:design}, BPIX L1 modules have the highest radiation exposure and therefore need to be exchanged after approximately 250\fbinv.

During operation in 2017--2018, L1 modules were run with high efficiency at a bias voltage of 450\unit{V}, up to an integrated luminosity of almost 120\fbinv.
After the replacement of the BPIX L1 during LS2, the new L1 modules must withstand a fluence expected to be about twice as high until the end of \Run3.
In order to maintain a high enough signal charge, the pixel detector power supplies were upgraded to deliver a maximum voltage of 800\unit{V} during LS2, as described in Section~\ref{sec:tracker:services}.

\paragraph{Readout chip}

The upgraded ROCs, \psidig and \proc~\cite{CMSTrackerGroup:2020edz}, are manufactured in the same 250\nm CMOS technology as the ROCs used in the original pixel detector (\psiana~\cite{Kastli:2013jkr}).
Their design requirements are summarized in Table~\ref{tab:tracker:roc}.

\begin{table}[!htb]
\centering
\topcaption{%
    Parameters and design requirements for the \psidig and \proc.
}
\label{tab:tracker:roc}
\renewcommand{\arraystretch}{1.1}
\begin{tabular}{lcc}
    & \psidig & \proc \\
    \hline
    Detector layer & BPIX L2--L4 and FPIX & BPIX L1 \\
    ROC size & $10.2{\times}7.9\mm^2$ & $10.6{\times}7.9\mm^2$ \\
    Pixel size  & $100{\times}150\mum^2$ & $100{\times}150\mum^2$ \\
    Number of pixels & 80$\times$52 & 80$\times$52 \\
    In-time threshold & $<$2000\,\Pem & $<$2000\,\Pem \\
    Pixel hit loss & $<$2\% at 150\MHzcmsq & $<$3\% at 580\MHzcmsq \\
    Readout speed & 160\Mbs & 160\Mbs \\
    Maximum trigger latency & 6.4\mus & 6.4\mus \\
    Radiation tolerance & 120\Mrad & 120\Mrad \\
\end{tabular}
\end{table}

The FPIX detector and layers 2--4 of the BPIX use the \psidig.
Its design follows very closely the original ROC with the readout architecture based on the column-drain mechanism~\cite{Horisberger:2000hi}.
The pixel cell remains essentially unchanged except for the implementation of an improved charge discriminator.
The improved discriminator reduces cross talk between pixels and the time walk of the signal~\cite{Kastli:2013jkr} and thus leads to lower threshold operation (below 2000\,\Pem).
The main modifications were made in the chip periphery in order to overcome the limitations of the \psiana at high rate.
They included a size increase of the data buffers (from 32 to 80 cells) and time-stamp buffers (from 12 to 24 cells) to store the hit information during the trigger latency, the implementation of an additional readout buffer stage to reduce dead time during the column readout, and the adoption of 160\Mbs digital readout.
In contrast to the previous ROC, the data readout is digital, using an 8-bit analog-to-digital converter (ADC) running at 80\MHz.
Digitized data are stored in a 64$\times$23 buffer, which is read out serially at 160\MHz.

The \proc is used for BPIX \layer1 and has to cope with hit rates of up to 600\MHzcmsq.
Therefore, the data transfer of pixel hits to the periphery must be much faster than the \psidig.
This was achieved by a complete redesign of the double column architecture.
The pixels within a double column are dynamically grouped into clusters of four and read out simultaneously, which significantly speeds up the readout process.

During operation in 2017 and 2018, the \proc delivered high-quality data.
However, two shortcomings of the \proc were identified, and are discussed in more detail in Ref.~\cite{CMSTrackerGroup:2020edz}.
The first was cross talk between pixels, which was higher than expected and generated noise at high hit rates.
The second was a lower efficiency caused by a rare loss of data synchronization in double columns.
Therefore, it was decided to develop for \Run3 a new, revised version of the \proc.
The main change in the new version addresses the rare cases of data synchronization loss.
The issue was tracked to a timing error in the time-stamp buffer of the double column, which leads to inefficiencies at low and high hit rates.
It was corrected in the buffer logic of the revised \proc.
The higher-than-expected noise was traced to inappropriate shielding of the circuitry for calibration pulse injection and has been fixed in the revised version of the \proc.
In addition, the routing and shielding of power and address lines were improved.
Both changes led to lower noise and lower cross talk between pixels.
The revised \proc was used to construct the modules for the new L1, which was installed during LS2 for use in \Run3.

\paragraph{Token bit manager chip}

The TBM is a custom, mixed-mode, radiation-hard integrated circuit that controls and reads out a group of 8 (L1) or 16 (L2--L4, FPIX) ROCs.
To increase the data output bandwidth from a module, two 160\Mbs ROC signal paths, with one path inverted, are multiplexed into a 320\Mbs signal, and then encoded into a 400\Mbs data stream that is optically transmitted to the downstream DAQ system.
The TBM has a single output (TBM08) version for L3, L4, and FPIX, and dual output versions, which are used in L2 (TBM09) and L1 (TBM10).
The TBM08 version has two independent 160\Mbs ROC readout paths, and the TBM09 and TBM10 versions have four separate, semi-independent, 160\Mbs ROC readout paths.

In addition to the increased output bandwidth, several critical features were added to the TBM for \Phase1.
As a result of adopting a faster digital readout, finer control over the timing of internal TBM operations and external TBM inputs was needed:
Delay adjustments were added for the ROC readouts, the token outputs, the data headers and the data trailers, and relative phase adjustments were added between the 40\MHz incoming clock, the 160\MHz clock, and the 400\MHz clock.
To prevent very long readouts from blocking the DAQ system, an adjustable token timeout was added that can reset the ROCs and drain buffered data.

Operation of the TBM during collision data taking revealed a vulnerability to a particular single event upset (SEU) that halts the TBM and requires a power cycle of the TBM to recover.
An additional iteration of the TBM chips was designed in the spring of 2018 to address this TBM SEU issue and to add an adjustable delay of up to 32\ns to the 40\MHz clock.
The delay was added to allow finer adjustment of the relative timing between modules.
The new chips are used in the new modules in BPIX L1, incorporated during the consolidation work in LS2.

\paragraph{BPIX module construction}

The BPIX detector contains 1184 modules, all having a similar design but coming in three different flavors depending on the requirements of the different layers.
The three module types are summarized in Table~\ref{tab:tracker:mod}.

\begin{table}[!ht]
\centering
\topcaption{%
    Overview of module types used in the \Phase1 pixel detector.
}
\label{tab:tracker:mod}
\renewcommand{\arraystretch}{1.1}
\begin{tabular}{lccccc}
    & \multirow{2}{*}{ROC} & Number & \multirow{2}{*}{TBM} & Number & Number of 400\Mbs \\[-1pt]
    & & of ROCs & & of TBMs & readout links/module \\
    \hline
    BPIX L1 & \proc & 16 & TBM10 & 2 & 4 \\
    BPIX L2 & \psidig & 16 & TBM09 & 1 & 2 \\
    BPIX L3, L4, \& FPIX & \psidig & 16 & TBM08 & 1 & 1 \\
\end{tabular}
\end{table}

\begin{figure}[ht!]
\centering
\includegraphics[width=0.8\textwidth]{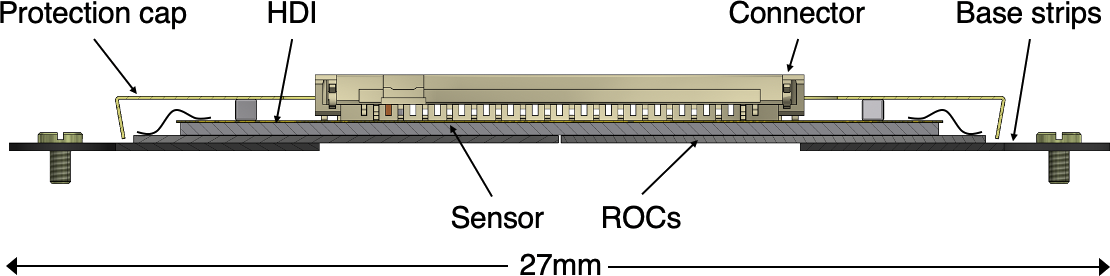}
\caption{%
    Cross sectional view of a pixel detector module for BPIX L2--4, cut along the short side of the module, from Ref.~\cite{CMSTrackerGroup:2020edz}.
}
\label{fig:tracker:modulexsec}
\end{figure}

A drawing of the detector module cross section for BPIX L2--4 is shown in Fig.~\ref{fig:tracker:modulexsec}.
During the module production process, bare modules are made by bump bonding of the 16 ROCs to the sensor.
In the next step, the thin four-layer HDI is glued onto the sensor side of the bare module and wire-bonded to the ROCs.
The TBMs are already mounted, wired-bonded, and tested on the HDI before joining the HDI to the bare module.
The BPIX L2--4 modules are mounted using silicon nitride base strips that are glued under the ROCs on the two long sides of the module (Fig.~\ref{fig:tracker:modules}, center).
The tight space requirements in the innermost BPIX layer required a different scheme with clamps between two modules (Fig.~\ref{fig:tracker:modules}, left).
A cap (not shown in the figure) made from a 75\mum-thick polyimide foil protects the wire bonds of the ROCs and TBMs against mechanical damage from the cables of other modules when mounted on the BPIX mechanics.
A single L2 module, excluding the cable, has a mass of about 2.4\unit{g} and represents a thickness corresponding to 0.8\% of a radiation length at normal incident angle.

The production of the BPIX modules was shared by five different module-assembly centers operated by groups from Germany, Switzerland, Italy, Finland, and Taiwan.
The assembly tools and procedures were mostly standardized, however, each center used different bump-bonding techniques, as described in more detail in Ref.~\cite{CMSTrackerGroup:2020edz}.
The overall yield of the module assembly ranged from 65 to 85\% in the different production centers, a bit lower than expected.
The main reason was low-yield phases during the production start-up.

\paragraph{FPIX module construction}

The 672 FPIX modules installed in the detector all have the same design and are instrumented with \psidig ROCs and a TBM08 with one readout link per module.
The FPIX modules (2.9\unit{g}/module without cable) use a different sensor, HDI, and aluminum/polyimide flat flex cable and are not interchangeable with BPIX modules.
The bump-bonding procedure was done at RTI and used an automated Datacon APM2200 bump bonder.
Bare modules were sent from RTI to two FPIX module assembly and testing sites in the USA.
Modules were assembled using an automated gantry and pick and placement equipment.
The wire bonds were encapsulated to protect against humidity and to mechanically support the wire bonds.
The average yield for FPIX production modules was 68\%, largely limited by bare module quality, especially in the early production.

\subsubsection{Mechanics}
\label{sec:tracker:mechanics}

\paragraph{BPIX mechanics}

{\tolerance=1200
The detector mechanics have been described in detail in Ref.~\cite{CMSTrackerGroup:2020edz}, and a brief outline will be given here.
The BPIX detector modules are mounted on ladders, with each ladder supporting eight modules.
The ladders are mounted on four concentric layers split into half-cylinders.
These are staggered by an alternating arrangement of ladders at smaller and larger radii.
Such a ladder placement provides between 0.5 and 1.0\mm of
sensor overlap in the $r$--$\phi$ plane.
Ladders are made from carbon-fiber reinforced polymer (CFRP) and have a length of 540\mm and a thickness of 500\mum.
The end rings, on which the ladders are suspended, consist of a CFRP/Airex/CFRP sandwich structure.
\par}

The 1.7\unit{m}-long BPIX service half-cylinders contain the \DCDC converters, the opto-electronic components, and the module connections.
The modules are connected to the service cylinders with micro-twisted-pair copper cables with a length of about 1\unit{m}.
These are contained in an additional structure placed in between the BPIX detector mechanics and the service half-cylinder.

The BPIX cooling is provided by complex looping tubes that cool the detector modules and the components on the service half-cylinders.
The tubes are made out of stainless steel with an inner diameter of 1.7\mm and a wall thickness of 50\mum.
The loops are between 957 and 1225\cm long and dissipate a power of up to 240\unit{W} each.
On the service half-cylinders the tubes have a wall thickness of 200\mum and an inner diameter of 1.8 and 2.6\mm for supply and return lines, respectively.

\paragraph{FPIX mechanics}

\begin{figure}[ht!]
\centering
\includegraphics[width=0.2\textwidth]{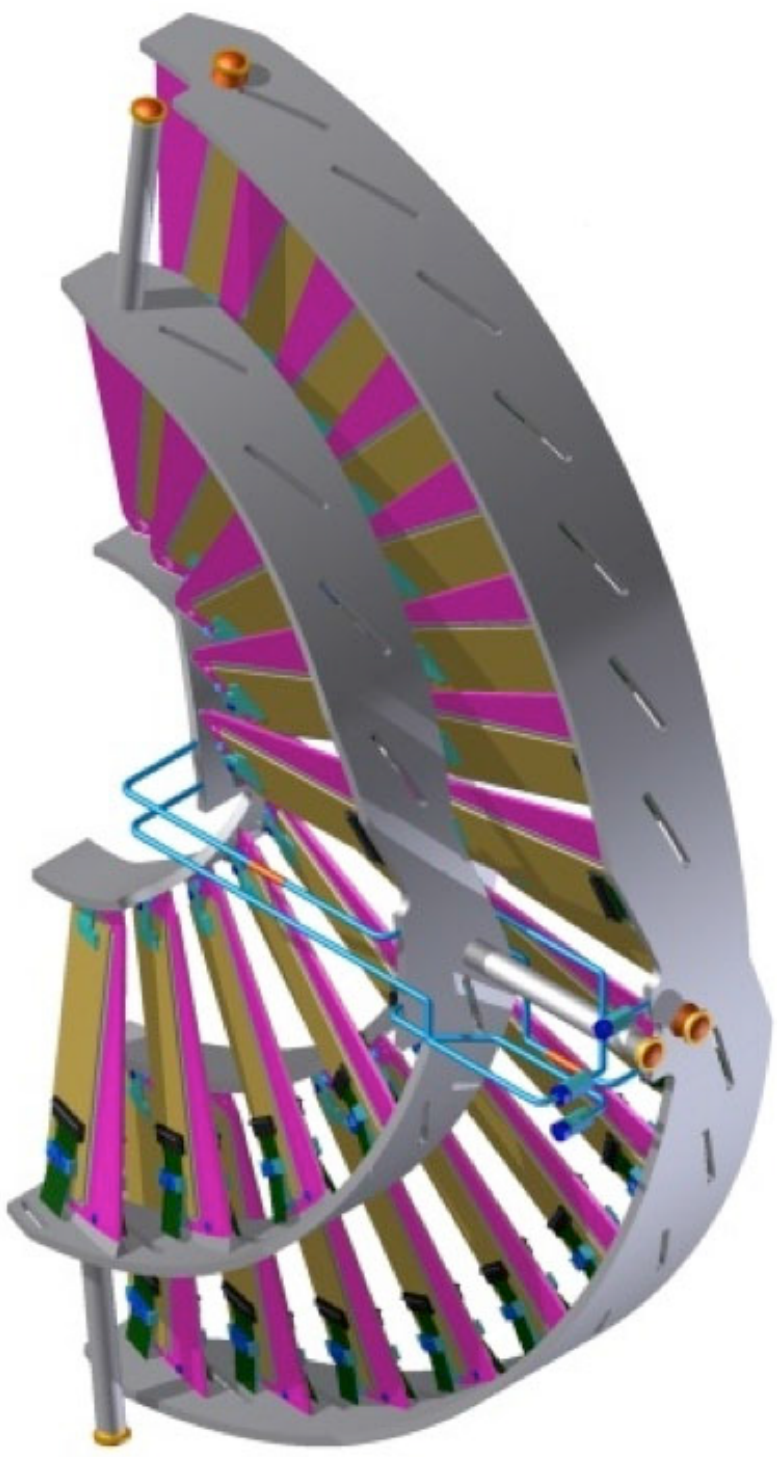}%
\hspace{0.1\textwidth}%
\includegraphics[width=0.4\textwidth]{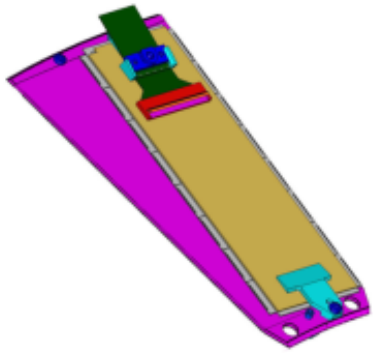}
\caption{%
    Drawing of an FPIX half-disk made from two half-rings of modules mounted on blades that are suspended between graphite rings (left), and close up view of a module mounted on a blade (right). Figures from Ref.~\cite{CMSTrackerGroup:2020edz}.
}
\label{fig:tracker:fpix}
\end{figure}

The three half-disks forming one FPIX quadrant are supported by a carbon-fiber composite service half-cylinder.
The FPIX half-disks consist of two turbine-like mechanical support structures with an inner assembly providing a sensor coverage from radii of 45 to 110\mm with 11 blades, while the outer assembly covers radii from 96 to 161\mm with 17 blades (Fig.~\ref{fig:tracker:fpix}).
The half-disks serve as the cooling isotherms for the sensor modules.
One module is mounted on each side of a blade, such that there is a small overlap in coverage at the outer edge of the blade.
The overlap is larger for adjacent modules closer to the beam.
The flat panel blades are made of 0.6\mm-thick sheets of thermal pyrolytic graphite (TPG) encapsulated between two 70\mum-thick single-ply carbon fiber face sheets.
The blades are suspended between an inner and an outer 2.4\mm thick graphite ring.
Each graphite ring houses the stainless steel cooling lines and is reinforced on the side facing away from the blades with a carbon fiber skin.

The FPIX service half-cylinders are 2220\mm long, with an outer radius of 175\mm and an inner radius of 165\mm.
The service half-cylinders consist of a double-walled carbon fiber rear section that is 1500\mm long for support elements, a corrugated single-wall carbon fiber front section that houses the half-disks, and an aluminum end flange for the cable and cooling tube pass through.
A ruby ball sliding in aluminum bushings on each of the three half-disk support stalks provides a kinematic mount.
A detailed description of the FPIX detector mechanics can be found in Ref.~\cite{CMSTrackerGroup:2020edz}.

\subsubsection{Services}
\label{sec:tracker:services}

\paragraph{Readout architecture and data acquisition system}

In order to be efficient, the CMS \Phase1 pixel detector readout architecture was copied from the original detector, with modifications made to handle the 400\Mbs readout and the increased number of channels.
The entirety of the CMS \Phase1 system readout is detailed in Ref.~\cite{CMS:2019kjw}, and a summary is provided below.

The auxiliary on-detector electronics were modified to include a crystal-driven phase-locked-loop chip (QPLL)~\cite{Moreira:2003cds} that reduced the jitter on the clock signal, a higher bandwidth chip (DLT)~\cite{Nageli:2013ozx} to translate the module readout into a level suitable for the laser driver chip, and a new optical transmitter (TOSA from Mitsubishi Electric Corp.) for the higher bandwidth digital signal.

The off-detector VME-based DAQ system used for the original detector was replaced by a \uTCA-based system using a common carrier board (FC7)~\cite{Pesaresi:2015qfa} with conventional 10\Gbs optical network links for the transfer of the data to the CMS central DAQ system.
The DAQ system, used to control and read out the full pixel detector, consists of 108 frontend driver modules (FEDs), which receive and decode the pixel hit information; three (2017--2018) to six ($\geq$2021) frontend controller modules (TkFECs), which serve the detector slow control; 16 pixel frontend controller modules (PxFECs) used for module programming and clock and trigger distribution; and 12 AMC13~\cite{Hazen:2013rma} cards providing the clock and trigger signals.
Application-specific mezzanine cards and firmware make an FC7 carrier board into a FED or FEC.

\paragraph{Power system}

The \Phase1 pixel detector requires low voltages to supply the readout chips on the pixel modules, a bias voltage to deplete the pixel sensors, and voltages to supply the control electronics components on the supply tubes and service cylinders.

The nominal control voltage of 2.5\unit{V} is supplied by A4602 CAEN power supply modules.
Power supplies by CAEN of type A4603DH provide the low and high voltages to the pixel modules.
The maximum bias voltage that can be delivered was raised from 600 to 800\unit{V} during LS2.
Each bias voltage channel can provide up to 16\mA, and serves several pixel modules.
While the auxiliary and bias voltage supply systems have conceptually not been changed for \Phase1, the low voltage supply system was changed from a direct parallel powering system to a \DCDC conversion powering system.
The number of ROCs, and thus the current consumption of the detector, has roughly doubled as compared to the original pixel detector.
Thanks to \DCDC conversion factors of 3--4, the currents on the typically 50\unit{m}-long supply lines between the detector and the power supplies are reduced with respect to a direct powering scheme.

The \DCDC converter modules are built around the CERN radiation-tolerant FEAST buck converter ASIC~\cite{Michelis:2012uin}, and were optimized in terms of dimensions, mass, and performance for the application in the pixel detector~\cite{Feld:2015zra, Feld:2016uwj}.
The \DCDC converter modules are installed on the BPIX supply tubes and in the FPIX service cylinders, at a distance of about 1\unit{m} from the pixel modules and outside of the sensitive pixel volume.
A total of 1216 \DCDC converter modules are used in the pixel detector.

Two types of \DCDC converter modules are needed to supply the pixel modules:\ one delivering 2.4\unit{V} to the analog circuitry of the ROC, and one delivering 3.3 or 3.5\unit{V} (depending on the position of the served pixel module in the detector) to the digital circuitry of the ROC and to the TBMs.
Each such pair of \DCDC converter modules serves between 1 and 4 pixel modules, depending on the layer and ring of the pixel modules.
The \DCDC converter output currents range from 0.4 to 1.7\unit{A} for the analog supply, and from 1.3 to 2.4\unit{A} for the digital supply.
The power efficiency is 80--84\%, depending on the output voltage and load.
The \DCDC converters can be disabled and enabled from an already existing chip (CCU)~\cite{Paillard:2002yn} used both in the original and in the \Phase1 pixel detector; when disabled the \DCDC converters do not provide an output voltage.
This feature was used during 2017 to power-cycle the TBM chips suffering from a SEU, which could not be recovered otherwise.

Up to seven \DCDC converters (of one variant) are connected to one low voltage power supply channel.
The low voltage part of the original A4603 power supplies has been adapted to the \DCDC conversion powering scheme.
The maximum output voltage was raised to 12.5\unit{V} and the fast remote sensing was abandoned, while a slow-control loop to compensate for voltage drops along the supply cables is still available.

In general, the \DCDC conversion powering system worked very well.
However, starting in October 2017, \DCDC converters started to fail, and at the end of the 2017 data-taking period, about 5\% of the \DCDC converters were defective.
This was traced back to a problem in the FEAST ASIC, namely a radiation-induced leakage current in a transistor~\cite{Faccio:2018web}.
When the chip is disabled, a voltage above the chip specification can build up on a certain node, damaging the chip.
For the 2018 data taking, all \DCDC converters were replaced with (almost) identical ones.
The input voltage was reduced from about 11 to 9\unit{V}, and disabling of the chip was replaced by power-cycling of the power supplies.
Due to these operational changes, no \DCDC converter failed in 2018.
During LS2, new \DCDC converters were again produced, featuring a new version of the \DCDC ASIC (FEAST2.3).
In this new version, the problem is fixed and operational changes are no longer needed.

The use of \DCDC converters meant that the LV and HV modularity of the detector no longer matched.
The LV could be switched off by disabling a \DCDC converter, however, the HV stayed on because of the limited number of HV wires.
The failure of individual \DCDC converters affected a small number of pixel modules that were kept under bias voltage in 2017 while their LV was off.
Having the sensor biased but the pixel readout amplifiers off damaged the amplifiers in the affected pixel modules.
Several of those damaged modules were replaced during LS2.
In addition, the LV/HV modularity was harmonized in FPIX; for BPIX this was not possible due to the limited number of HV wires.

\paragraph{Cooling}

To keep the silicon sensors below 0\deC and remove heat from the other detector elements, \COtwo evaporative cooling utilizing the two-phase accumulator controlled loop (2-PACL) approach~\cite{Tropea:2019xjr} is used in the \Phase1 pixel detector.
Evaporative \COtwo cooling provides low density, low viscosity, and high heat transfer capacity while allowing the use of an all passive, small-diameter, thin-walled stainless steel pipe network inside the detector volume.
This results in a lower contribution of cooling to the overall detector material budget.
During normal operation ($-22\deC$) the expected power~\cite{CMS:TDR-011} from the BPIX (6\kW) and FPIX detectors (3\kW) is removed using two dedicated 15\kW \COtwo plants, one for each detector, located in the CMS detector cavern.
There is enough flexibility and capacity in the system so that the eight cooling loops in each detector can be connected to either cooling plant with no loss of cooling capacity.
The typical temperature at the pixel module surface is about 12\deC higher than the coolant.
A detailed description of the CMS \Phase1 pixel detector cooling system can be found in Ref.~\cite{CMSTrackerGroup:2020edz}.

During the testing of the removed FPIX in 2019, one of the inlet cooling pipe connections at the detector end flange was broken.
It was therefore decided to replace all the FPIX inlet connectors with a more robust solution that required only a single wrench to make a connection.
This sped up the detector installation in 2021 and prevented another occurrence of a broken inlet connector.

\subsubsection{Detector operation}

\paragraph{Detector live fraction}

The detector live fraction, defined as the fraction of working ROCs, was 95.0 and 96.1\% for the BPIX and FPIX detectors, respectively, during the first collisions in 2017.
The main causes of failures in the BPIX detector were the loss of power due to faulty connectors and modules masked because of readout problems.
For the FPIX detector the main problem was an issue with the clock distribution in one sector.
Towards the end of 2017, the fraction of nonworking modules was dominated by the failure of the \DCDC converters described in Section~\ref{sec:tracker:services}.
The \DCDC problems lowered the working fraction to 90.9 and 85.0\% for the BPIX and FPIX detectors, respectively.

During the 2017/2018 LHC year-end technical stop, faulty components were repaired and all \DCDC converters were replaced.
A few broken BPIX modules, which were accessible without disassembling the detector layers, were also replaced.
These repairs improved the working detector fraction for the 2018 data-taking period.
For the BPIX it varied from initially 98 to 93.5\% at the year end, with the main drop due to faulty power connectors.
The FPIX detector working fraction was stable throughout the entire year at 96.7\%.

As already mentioned above, during LS2 a new L1 was installed in the BPIX.
Other repairs, in the BPIX and FPIX, involved faulty infrastructure components  like bad connectors and broken optical fibers.
In addition, eight modules were replaced in BPIX L2.
With these improvements, the working fractions for BPIX and FPIX at the start of \Run3 were 99.1 and 98.5\%, respectively.

\paragraph{Threshold adjustment}

Pixel charge thresholds are an important performance parameter since they directly influence the position resolution.
Lower thresholds increase the charge sharing between pixels, resulting in a better resolution.
However, too low thresholds result in noise saturating the readout.
Therefore, thresholds in all ROCs are adjusted to the lowest possible value, but well above the noise level itself.
More details about the threshold adjustment are given in Ref.~\cite{CMSTrackerGroup:2020edz}.

During the 2017-18 data-taking period, the BPIX L2--4 thresholds were about 1400\,\Pem and similarly for the FPIX detector at about 1500\,\Pem.
Because of the higher noise in BPIX L1, the thresholds had to be higher, about 2200\,\Pem.
With these pixel thresholds the number of noise hits was very low, below 10 pixels per bunch crossing per layer, resulting in a per pixel noise hit probability of less than $10^{-6}$.
Individual pixels that showed a hit probability exceeding 0.1\% were masked during operation.
The total fraction of masked pixels was less than 0.01\%.

The revised \proc used in the new L1 installed in 2021 has significantly lower cross-talk noise (as discussed in Section~\ref{sec:tracker:modules}).
The expected threshold, noise, and time-walk behavior of the revised \proc version is similar to that of the \psidig.
These improvements allow the BPIX L1 to be operated in \Run3 with significantly lower thresholds, similar to the ones for the other layers.

\subsubsection{Performance of the pixel tracker}

The two most important performance parameters for a pixel detector are hit efficiency and position resolution.
Both strongly affect the ability to perform pattern recognition and \PQb tagging, two main roles for a well-functioning pixel detector.

The hit efficiency has greatly improved with the \Phase1 upgrade, mostly due to the ability of the upgraded ROCs to read data at higher speed and operate at lower thresholds, as described in Section~\ref{sec:tracker:modules} and Ref.~\cite{CMSTrackerGroup:2020edz}.
The hit efficiencies measured at an instantaneous luminosity of $2\times10^{34}\percms$ are 97, 98, 99, and 99.5\% for BPIX L1--L4, respectively.
For the forward disks, the average efficiency is 99\%.
These numbers present a big improvement with respect to the \Phase0 detector, where the efficiencies would have been much smaller for the same data rates~\cite{CMSTrackerGroup:2020edz}.

The position resolution is measured using the ``triplet'' method~\cite{CMSTrackerGroup:2020edz}, where the expected hit position in a detector layer is interpolated from two other layers.
For the BPIX L3, this method yields a position resolution of 11.0\mum in the $r$--$\phi$ direction
and 24.3\mum in the $z$ direction.
For BPIX L1 and L2, the resolutions are somewhat worse due to the higher sensor radiation damage at lower radii.
For the forward disks, the resolution is 11.9\mum in the $r$ direction and 21.0\mum in the $z$ direction.
These numbers agree with the design expectations and are consistent with simulations.

Pixel sensors suffer from radiation damage induced by the high density of charge particle tracks.
In order to maintain good efficiency and position resolution, it is necessary to keep increasing the sensor's bias voltage.
The nominal voltage after installation was 150\unit{V} for the BPIX modules and 300\unit{V} for FPIX.  Especially for the inner BPIX layers, these voltages had to be increased in a few steps during data taking.
Presently, end of 2023, the voltages are 450, 350, 250, and 250\unit{V} for the BPIX L1--L4, respectively, and 350 and 300\unit{V} for the inner and outer rings of the FPIX disks.
The values are expected to increase further throughout \Run3, eventually reaching the maximum of 800\unit{V} for BPIX L1 and 600\unit{V} for all other layers.

\subsection{Strip detector}

\subsubsection{Detector description}

\begin{figure}[!ht]
\centering
\includegraphics[width=\textwidth]{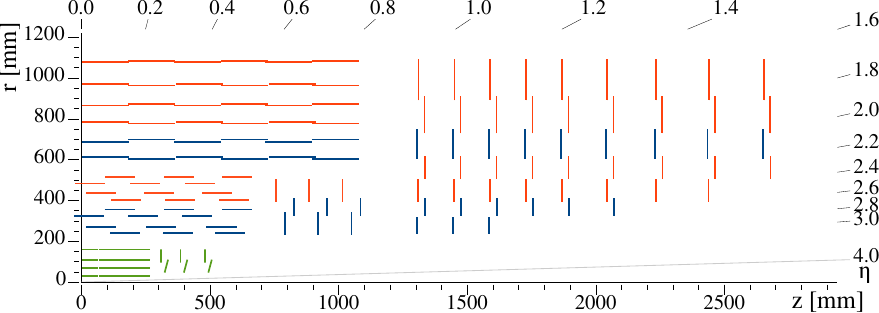}
\caption{%
    Schematic view of one quadrant in the $r$--$z$ view of the CMS tracker:\ single-sided and double-sided strip modules are depicted as red and blue segments, respectively.
    The pixel detector is shown in green. Figure from Ref.~\cite{CMS:TDR-014}.
}
\label{fig:tracker:sst}
\end{figure}

The silicon strip tracker (SST), together with the pixel detector, provides measurements of charged particle trajectories up to a pseudorapidity of $\abseta<2.5$.
The layout of the SST is shown in Fig.~\ref{fig:tracker:sst}.

The SST has 9.3 million silicon micro-strips and 198\msq of active silicon area distributed over 15\,148 modules.
Single-sided p-on-n micro-strip sensors are used.
The detector is 5\unit{m} long and has a diameter of 2.5\unit{m}.
It has ten layers in the barrel region with four layers in the tracker inner barrel (TIB) and six layers in the tracker outer barrel (TOB).
The TIB is supplemented with three tracker inner disks (TID) at each end.
In the forward regions, the detector consists of tracker endcaps (TEC).
Each TID is composed of three rings of modules and each TEC is composed of up to seven rings.
In TIB, TID, and in rings 1--4 of the TECs, sensors with a thickness of 320\mum are used, while in TOB and in rings 5--7 of the TECs, 500\mum thick sensors are used.
The modules in the barrel layers measure $r$ and $\phi$ coordinates, while the modules in the TECs and TIDs are oriented to measure the coordinates in $\phi$ and $z$.
In four layers in the barrel and three rings in the endcaps, stereo modules are used (Fig.~\ref{fig:tracker:sst}).
These modules have a second module mounted back-to-back with a stereo angle of 100\mrad.
The stereo modules provide coarse measurements of an additional coordinate ($z$ in the barrel and $r$ in the endcaps).

The analog signals from 128 strips are processed by one APV25 chip.
The chip has 128 readout channels, each consisting of a low-noise and charge-sensitive preamplifier, a 50\ns CR-RC type shaper, and a 192-element deep analog pipeline which samples the shaped signals at the LHC frequency of 40\MHz~\cite{French:2001xb}.
Signals from two APV25 chips are multiplexed, converted to optical signals by analog opto-hybrids (AOH)~\cite{CMS:Detector-2008}, and transmitted via optical fibers to front-end drivers (FED), located in the service cavern outside the radiation zone.
Pedestal and common mode subtraction, as well as cluster finding, are performed in the FEDs.
Clock, trigger information, and  control signal are trasmitted to the detector  by the frontend controllers (FEC), also located in the service cavern.
Configuration data for the modules is distributed via the I2C protocol to communication-and-control units (CCU)~\cite{Paillard:2002yn}, grouped in token ring networks (control rings).
The modules in the SST are grouped in power groups each of which shares one power supply channel.
There are 1944 power groups in total.
Each power group has two low-voltage channels with 2.5 and 1.25\unit{V} regulators and two high-voltage channels that can be regulated up to 600\unit{V}~\cite{CMS:Detector-2008}.
The detector is cooled with \CsixFfourteen monophase coolant by two cooling plants.

The SST has been operated stably and successfully since 2009, and operation is scheduled to continue until the end of \Run3.
During \Run1, the SST was operated with its primary cooling at $+4\deC$, significantly above the designed operating temperature, due to insufficient humidity control in the service channels and in the bulkhead region, \ie, the interface region between the detector volume and the outside seal.
In 2009, the detector suffered from an over-pressure incident.
Both inlet and outlet lines of 90 cooling loops of circulating \CsixFfourteen were closed on of the two cooling plants.
After that the detector warmed up.
As a result of the over-pressure, some of the cooling lines developed leaks or were detached from modules.
Due to this incident there are several regions in the detector that have closed cooling loops or degraded cooling contacts.

During LS1 in 2013--2014, a number of engineering changes were carried out on the detector infrastructure that allowed to lower the operating temperature of the SST below 0\deC.
Most prominently, a dedicated plant was installed to produce dry air or oxygen-depleted air.
The plant has a flow of about 250\mcubh, and is able to meet the dew point requirement of the SST (around $-60\deC$).
The plant is the primary source of dry gas injection to the detector and its services.
In LS1, the insulation of all service channels and the bulkhead of the detector was significantly improved  in order to control the humidity conditions in the closed volume of the detector.

All these modifications to the detector infrastructure facilitated the SST operation at $-15\deC$ since the beginning of \Run2.
However, with increasing irradiation, the leakage currents began to approach the power supply limits (12\mA) in the regions with no cooling or degraded cooling contacts.
Thermal runaway was observed in several power groups of the TIB.
As a consequence, since 2018, the SST was operated at $-20\deC$, which sufficiently reduced the leakage currents.
By the end of \Run3 it will be necessary to lower the operating temperature to $-25\deC$.
During LS2 a test at a temperature of $-25\deC$ confirmed that the detector can be operated at this temperature and that there is no degradation of the humidity conditions inside the detector and the service channels.

\subsubsection{Performance of the strip tracker}

The performance of the SST will be discussed in the following.
More details about the results in this section can be found in Ref.~\cite{CMS:TRK-20-002}.

\begin{figure}[!ht]
\centering
\includegraphics[width=0.6\textwidth]{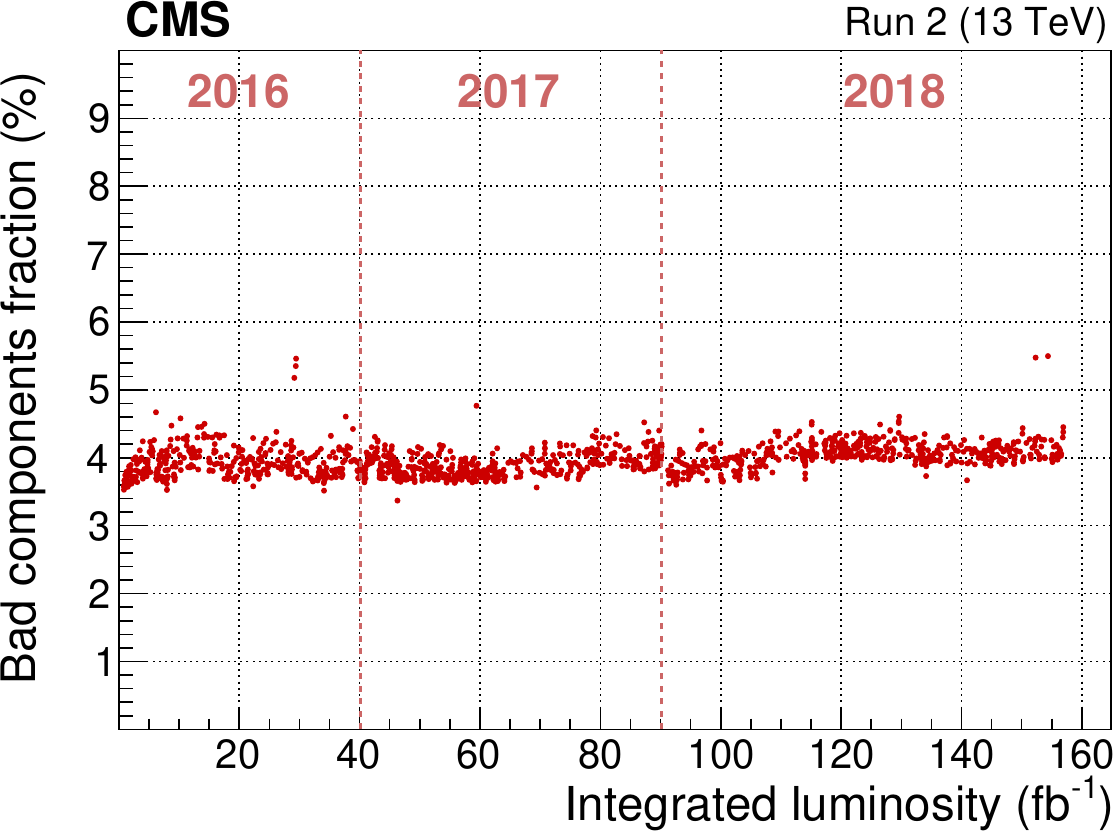}
\caption{%
    Fraction of bad components for the CMS silicon strip detector as a function of the delivered integrated luminosity~\cite{CMS:TRK-20-002}.
}
\label{fig:tracker:badcomponents}
\end{figure}

Throughout all the years of operation, no SST on-detector components were exchanged because the detector has been inaccessible.
The fraction of bad detector components has been largely stable during \Run1 and \Run2.
This includes the readout channels that are excluded from the data taking: failing control rings,  problems in LV or HV distribution, and individually switched-off modules, single APV25 chips or groups of strips.
As can be seen in Fig.~\ref{fig:tracker:badcomponents}, the fraction of bad components was stable throughout \Run2 and amounts to about 4\%.

\begin{figure}[!ht]
\centering
\includegraphics[width=0.48\textwidth]{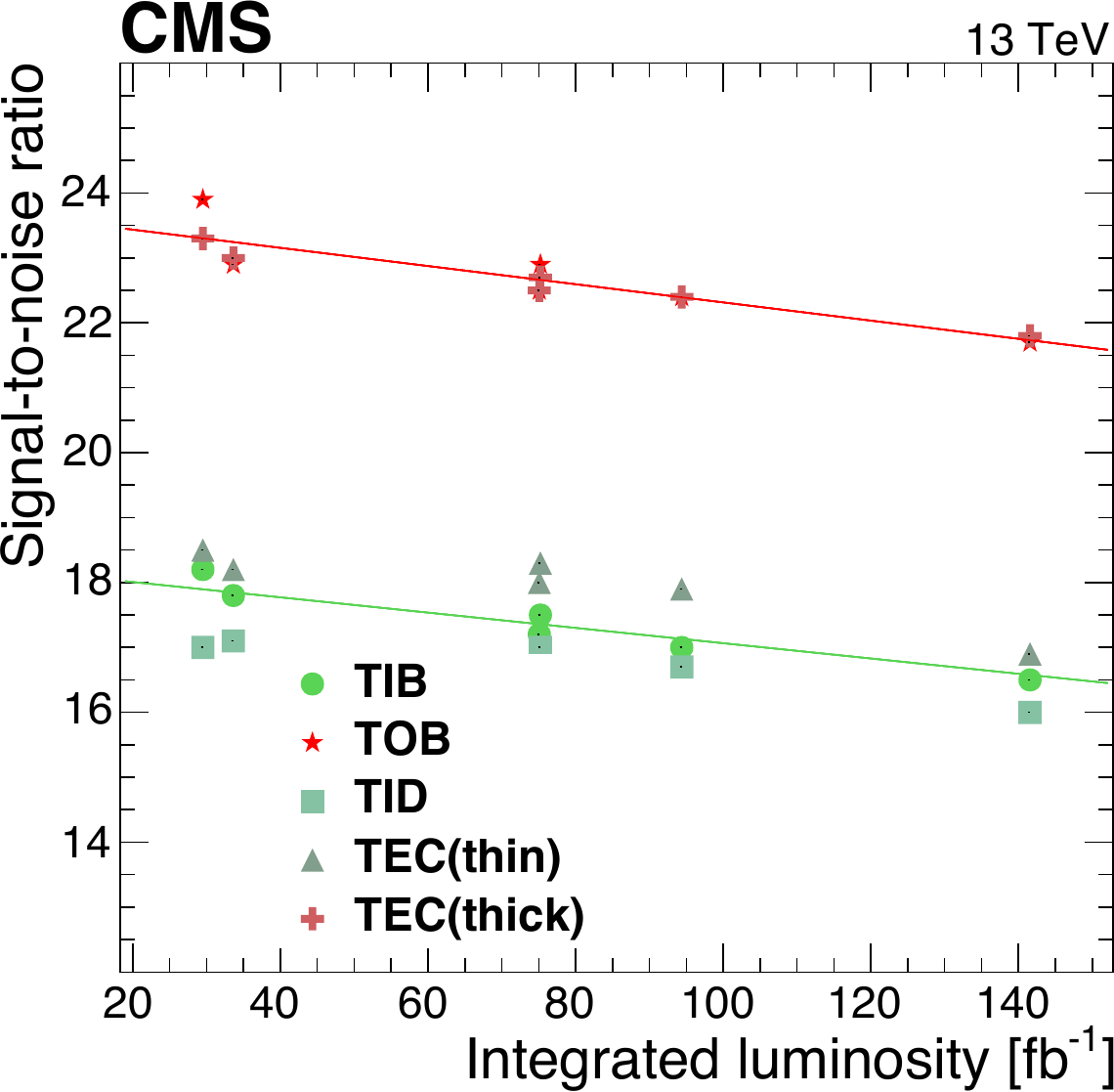}%
\hfill%
\includegraphics[width=0.48\textwidth]{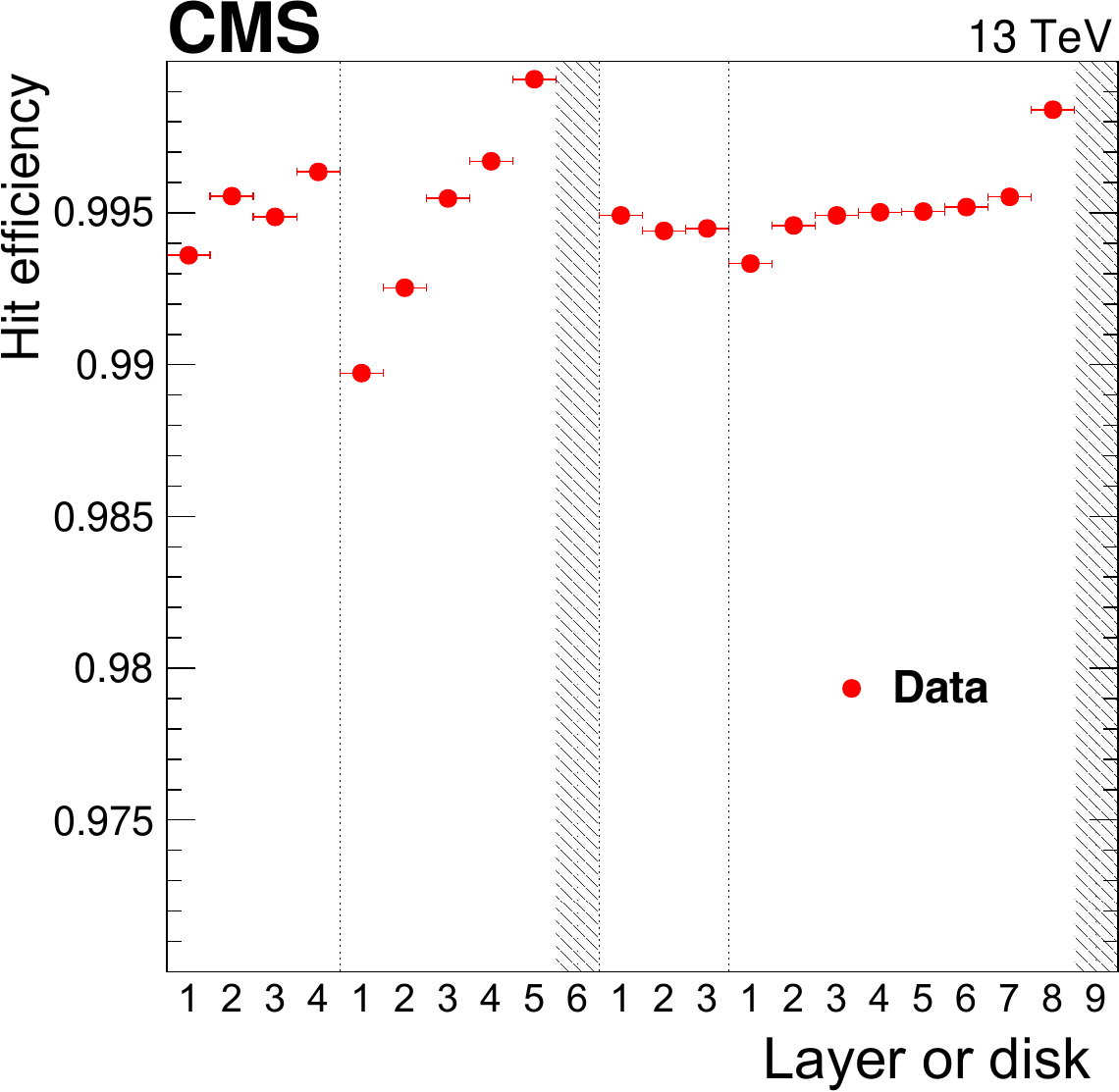}
\caption{%
    Left:\ signal-to-noise ratio as a function of integrated luminosity as accumulated in \pp collisions during \Run2, separately for the different detector partitions.
    Triangles and crosses indicate the results for sensors in the TEC of thickness 320 and 500\mum, respectively.
    Right:\ hit efficiency of the silicon strip detector taken from a representative run recorded in 2018~\cite{CMS:TRK-20-002} with an
    average hit efficiency under typical conditions at an instantaneous luminosity of $1.11\times10^{34}\percms$.
    The two gray bands represent regions where the hit efficiency is not measured due to the selection criteria of the analysis~\cite{CMS:TRK-20-002}.
}
\label{fig:tracker:sstperformance}
\end{figure}

One of the most important performance characteristics is the signal-to-noise ratio (S/N).
The evolution of S/N with accumulated integrated luminosity is shown in Fig.~\ref{fig:tracker:sstperformance} (left).
As expected from irradiation studies~\cite{CMS:Detector-2008}, the S/N degrades approximately linearly with the integrated luminosity~\cite{CMS:TRK-20-002}.
The decrease observed during \Run2 indicates that the SST will continue to provide high-quality data until its end of life, estimated to be at 500\fbinv, well beyond the expected end of \Run3.

Another important aspect of the SST is the hit efficiency, which is the detection efficiency for a particle traversing a sensor.
The measurement of the hit efficiency is performed using tracks that pass the quality criteria as defined in Ref.~\cite{CMS:DP-2018-052}.
In order to avoid inactive regions, trajectories that are close to sensor edges or their readout electronics in the studied layer are not considered.
The efficiency is determined from the fraction of traversing tracks with a hit in a module anywhere within a range of 15 strips from the expected position.
The measured hit efficiency under typical conditions during \Run2, at an average instantaneous luminosity of $1.11\times10^{34}\percms$, corresponding to about 31 \pp interactions per bunch crossing, is shown in Fig.~\ref{fig:tracker:sstperformance} (right).
The average hit efficiency is about 99.5\%, depending on the layer.
Since the inefficiency mainly depends on the particle flux, the inner layers have a somewhat lower efficiency than the outer ones.
Moreover, the inefficiency depends on the sensor thickness and on the pitch.

Radiation effects are also monitored during the operation of the detector, including the increase of the leakage currents in the sensors, the evolution of the full depletion voltage due to the change of the effective sensor doping concentration, and the evolution of the laser-driver performance in the optical readout chain.

During \Run3, due to increasing luminosity, the leakage currents will continue to rise.
It can therefore be expected that some modules in regions with closed loops or degraded cooling contact will experience thermal runaway, or that the corresponding HV power-supply channels will reach their limit of 12\mA.
Most of the modules are double-sided, and one way to reduce the self-heating effect is to switch off one side of the module.
A voltage reduction can also reduce the leakage currents significantly.
However, this is possible only if the applied voltage remains above the full depletion voltage.
Towards the end of \Run3, it is expected that a lowering of the detector temperature to $-25\deC$ will become necessary.
It is estimated that this measure will reduce the number of modules experiencing thermal runaway after 500\fbinv of integrated luminosity by roughly a factor of 2.

\begin{figure}[!ht]
\centering
\includegraphics[width=0.7\textwidth]{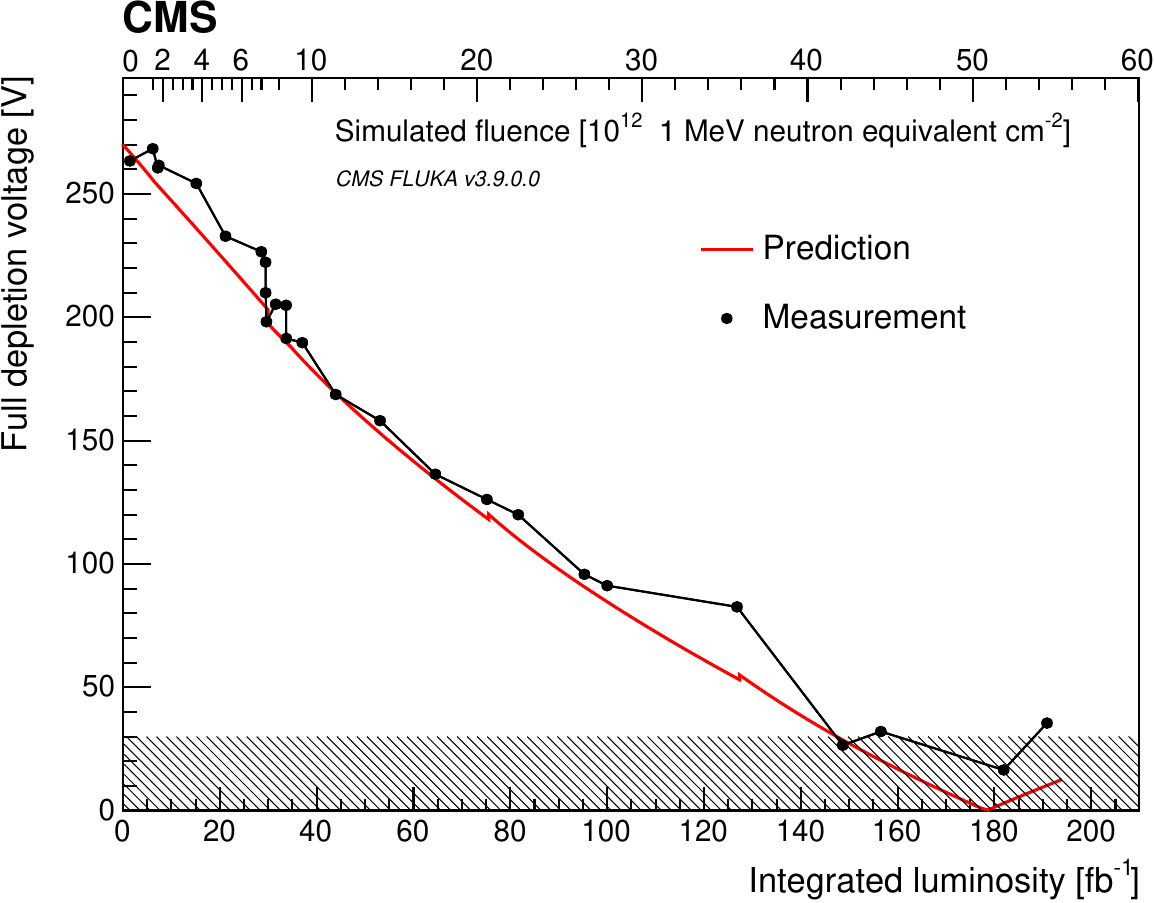}
\caption{%
    Evolution of the full depletion voltage for one TIB layer-1 sensor as a function of the integrated luminosity and fluence until the end of \Run2.
    The full depletion voltage is measured from the cluster-width variable, shown as black dots, and the predicted curve is based on a model that uses fluence and temperature history as inputs~\cite{Barth:2013zxa}.
    The hashed area highlights the region at low values of the full depletion voltage where the analysis loses sensitivity~\cite{CMS:TRK-20-002}.
}
\label{fig:tracker:voltagetib}
\end{figure}

The sensors of the SST are operated at an applied voltage of 300\unit{V} in over-depletion mode, because the sensors are p-on-n type and some of them have undergone type inversion of the bulk material.
The full depletion voltage is measured by performing bias voltage scans during \pp collisions.
A scan of the full detector is done usually twice per year during data taking and once per month on a selected set of modules.
The evolution of the full depletion voltage  with integrated luminosity of one module in TIB \layer1 is shown in Fig.~\ref{fig:tracker:voltagetib}.
The measurements of the full depletion voltage are also compared with simulations, which describe the change with integrated luminosity well.

\begin{figure}[!htp]
\centering
\includegraphics[width=0.7\textwidth]{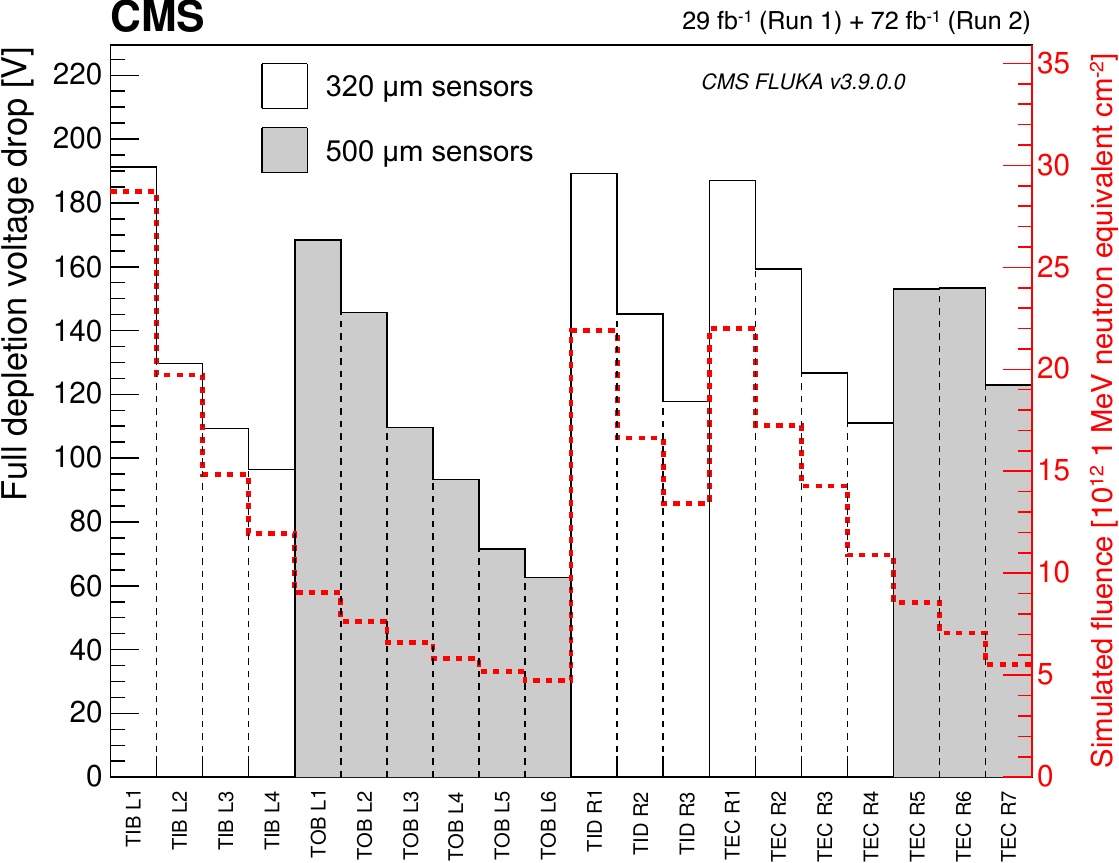}
\caption{%
    Decrease of the full depletion voltage for each layer, computed as the difference between the values measured at the time of the tracker construction and the values obtained by the analysis of a bias-voltage scan performed in September 2017 on all the tracker modules.
    The white (gray) histograms represent modules with 320 (500)\mum thick sensors.
    The average fluence for each layer is shown by the red line.
}
\label{fig:tracker:voltageall}
\end{figure}

The installation of the pixel detector and the cooling plant maintenance work caused extended periods of time when the silicon detector was not cooled as well as it would have been desirable from the point of view of radiation damage.
In Fig.~\ref{fig:tracker:voltagetib}, small increases due to annealing are visible in the simulation around integrated luminosities of 75 and 130\fbinv, corresponding to the winter shutdown periods.
As can be seen from measurements and simulation, at around 200\fbinv, the TIB layer-1 sensors are close to the inversion point.
The overall situation with the reduction of the full depletion voltage in the SST is shown in Fig.~\ref{fig:tracker:voltageall}.
For each subdetector a decrease of the full depletion voltage is observed that depends on the distance from the interaction point.
It is observed that the regions of the detector that are closest to the interaction point, namely TIB \layer1, TID ring 1, and TEC ring 1, are affected the most, as expected.

In summary, the SST has been delivering high quality data for the reconstruction of charged particle tracks since the start of the LHC operation.
The performance of the system continues to be excellent also after more than 200\fbinv of integrated luminosity.
Since the beginning of \Run3, the detector has been operated at $-20\deC$.
It is expected that the operation temperature will be lowered further to $-25\deC$ in order to reduce the leakage current.
While radiation effects are visible in all parts of the detector, the margins are large enough for the detector to be operated safely and efficiently, and to provide high-quality data until the end of \Run3.

\clearpage
\section{Electromagnetic calorimeter}
\label{sec:ecal}

The electromagnetic calorimeter (ECAL) is placed outside the inner tracking system of CMS.
It provides a measurement of the energy of electrons and photons, as well as their impact position and arrival time at the crystals.

\subsection{Experimental challenges}

The increase in the instantaneous and integrated luminosity, experienced during \Run1 and \Run2 of the LHC and expected to continue in the future, poses operational challenges for the ECAL.
The radiation dose deposited in the detector reduces the average light transmission of the \PbWOfour crystals, lowering the signal-to-noise ratio of the electronics readout.
The radiation also induces an increase in leakage currents in the barrel photodetectors, which are avalanche photodiodes (APDs), with a corresponding increase in the electronic noise~\cite{Deiters:2000ip, Addesa:2015loa}.
The instantaneous luminosity reached $2.1\times10^{34}\percms$ during \Run2, compared to $0.75\times10^{34}\percms$ achieved in 2012.
The increase in luminosity also poses challenges to the level-1 (L1) trigger system.
Specifically, signals from direct energy deposition by particles in the APDs, termed ``spikes'', must be rejected.
Such spikes occur at a rate that is proportional to the luminosity.
The higher radiation has also caused the silicon sensors of the preshower detector to have increased bulk currents, which require regular updates of the HV bias and correspondingly of the calibration of their response.

Additionally, the number of multiple \pp interactions in a single bunch crossing (BX), termed pileup, has increased on average from 21 (up to 40) during \Run1 to 34 (up to 80) during \Run2.
The bunch spacing in the machine reached its nominal value of 25\ns at the beginning of \Run2, half of what it was in \Run1.
Since the typical signals from the calorimeter, after shaping by the electronics, fall to 10\% of their peak value in about 250\ns, the changes in the LHC operation have resulted in an increased number of overlapping signals from neighboring BXs, referred to as out-of-time (OOT) pileup.

These effects will be discussed in more detail in the following sections, along with the improvements in the calibration of the calorimeter and the final performance achieved during \Run2.

\subsection{Response monitoring}

{\tolerance=800
The \PbWOfour crystals of the ECAL, when subjected to irradiation, undergo transparency changes.
This is discussed in greater detail in Ref.~\cite{CMS:NOTE-2009-016} and can be ascribed to the formation of color centers, which cause absorption bands in the crystal that reduce the light attenuation length.
The creation of color centers is a dynamic process depending on the dose rate absorbed by the crystals.
Its annealing process spontaneously takes place at room temperature and results in partial recovery of the transmittance.
Since the scintillation process remains unaltered, a reference light signal can be used to measure and monitor the transparency and response changes, and corrections can be applied to equalize the crystal-to-crystal response.
\par}

To monitor and correct the response of the ECAL, a dedicated laser monitoring system is used that operates primarily at a wavelength of 447\nm, near the peak of the scintillating light spectrum.
Additional monitoring wavelengths have been used, in particular a near-infrared one at 796\nm and a green one at 527\nm.
These probe the transparency in regions that are much less sensitive to radiation damage (infrared) and more sensitive to the permanent component of the radiation damage (green).

\begin{figure}[!ht]
\centering
\includegraphics[width=0.8\textwidth]{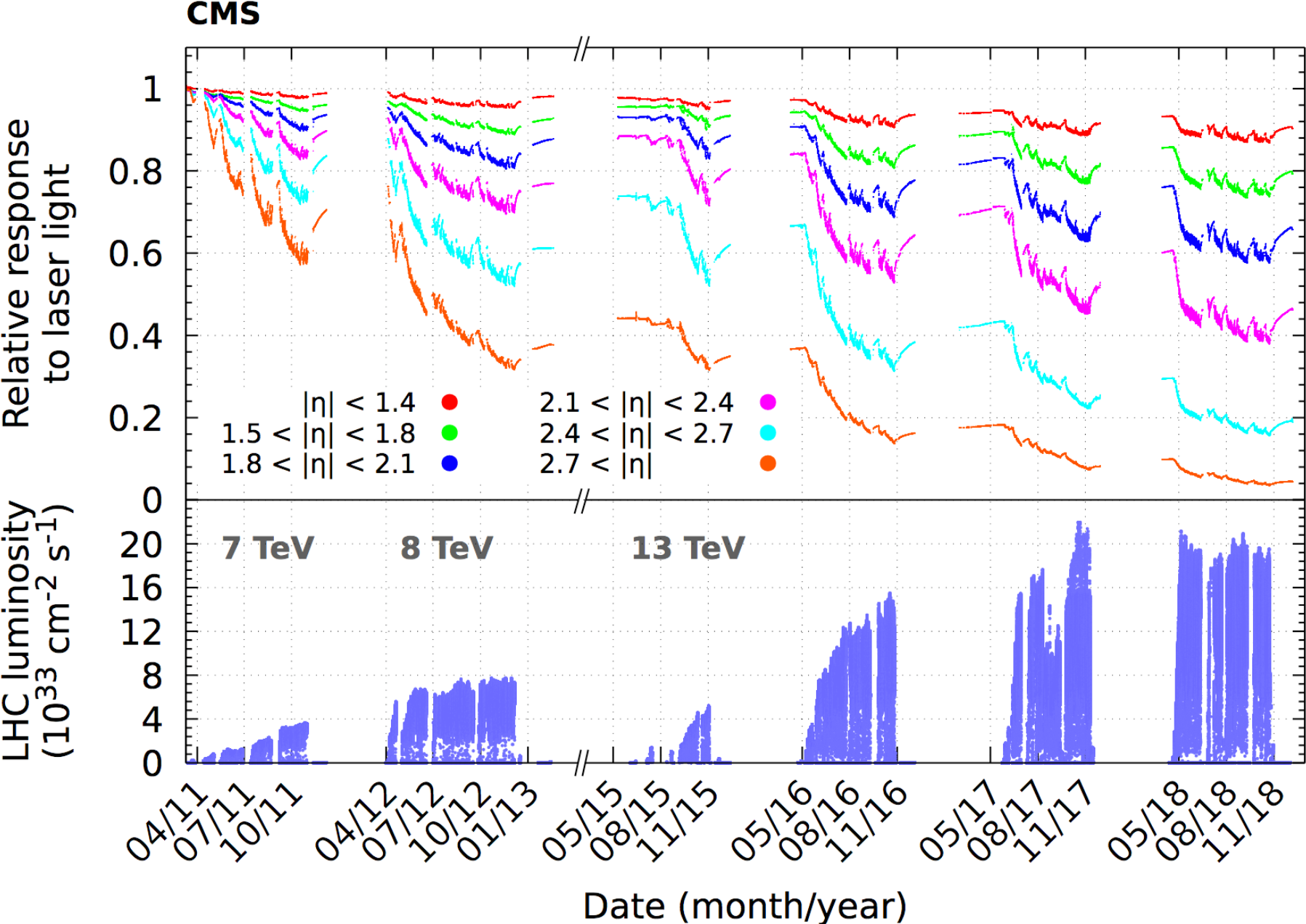}
\caption{%
    Relative response to laser light injected into the ECAL crystals, measured by the laser monitoring system, averaged over all crystals in bins of \abseta.
    The response change observed in the ECAL channels is up to 13\% in the barrel, $\abseta<1.5$, and reaches up to 62\% at $\abseta\approx2.5$, the limit of the CMS inner tracker acceptance.
    The response change is up to 96\% in the region closest to the beam pipe.
    The recovery of the crystal response during the periods without collisions is visible.
    These measurements, performed every 40 minutes, are used to correct the physics data.
    The lower panel shows the LHC instantaneous luminosity as a function of time.
}
\label{fig:ecal:transparency}
\end{figure}

The laser is operated at 100\uHz.
To avoid interference with signals from beam collisions, the light is injected into the crystals during the LHC abort gap where there are no bunches in either beam, in intervals of at least 3\mus.
The abort gap is necessary to accommodate the beam abort kicker rise time and is available in all LHC filling schemes.
The power of commercial lasers operating at a suitable repetition rate allows the injection of light into a few hundred crystals simultaneously.
This is achieved using a system of optical fibers and diffusing spheres acting as homogeneous splitters.
Light from a group of 200 fibers is measured by two p-n diodes.
The variation in response is obtained by comparing the signal acquired by the APDs with the reference p-n diode.
The time-dependent correction factor $LC_{i}(t)$, derived from the monitoring system for each crystal $i$, is defined as:\ $LC_{i}(t) = [{R_{i}(0)}/{R_{i}(t)}]^\alpha$, where $R_{i}(t)$ is the measured response to laser light at time $t$, and $\alpha$ is a parameter that takes into account the difference in path between the laser and scintillation light.
Figure~\ref{fig:ecal:transparency} summarizes the long-term evolution of the ECAL response to laser light during \Run1 and \Run2.

\subsection{Noise evolution}

\begin{figure}[!ht]
\centering
\includegraphics[width=0.47\textwidth]{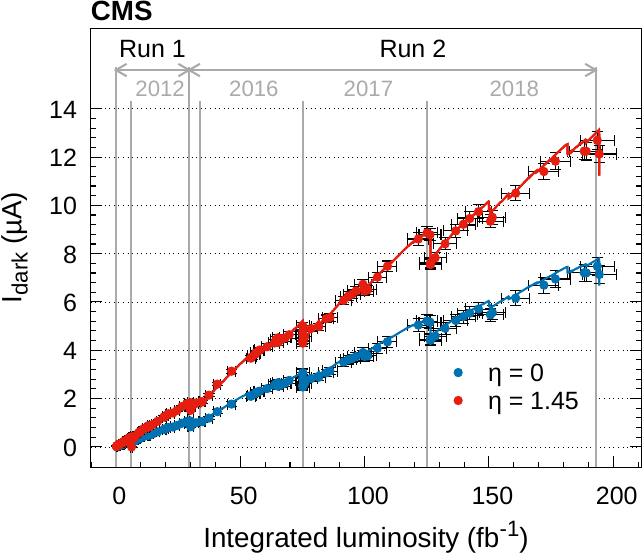}%
\hfill%
\includegraphics[width=0.48\textwidth]{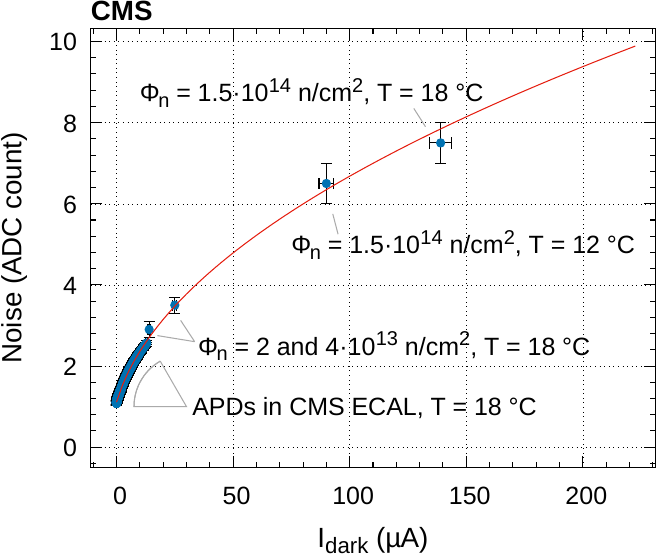}
\caption{%
    Left:\ evolution of the APD dark current as a function of the integrated luminosity since the beginning of \Run1.
    The gray vertical lines represent the ends of the \Run1 and \Run2 data-taking periods.
    Spontaneous annealing of the radiation-induced defects is visible as vertical steps in the measurements and corresponds to long stops in the data taking, \eg, year-end technical stops.
    Right:\ measurement of the electronic noise in the most sensitive amplification range as a function of the measured leakage current of an APD pair.
    The measurements are explained in the text.
    The red line is a fit to the data with a square root function.
    The maximum expected fluence for \Run3 is $4\times10^{13}\Neq$ at $\abseta=1.45$.
}
\label{fig:ecal:noiseapd}
\end{figure}

The electronic noise in the endcaps (EE) is approximately constant.
In the barrel (EB), radiation induces damage to the structure of the APD silicon lattice, causing an increase in the leakage current.
The evolution of the leakage current is shown in Fig.~\ref{fig:ecal:noiseapd} (left) as a function of the integrated luminosity for the central rapidity region and the most forward region of the barrel.
It is well in line with the expectation from irradiation studies shown in Fig.~\ref{fig:ecal:noiseapd} (right).
The studies were performed using a pair of APDs, equivalent to ones on each of the ECAL barrel crystals.
Measurements were done in CMS in situ for the points below 10\muA, while the points at higher currents are based on laboratory measurements of irradiation with neutrons at different fluences, as indicated in the figure.

\begin{figure}[!ht]
\centering
\includegraphics[width=0.48\textwidth]{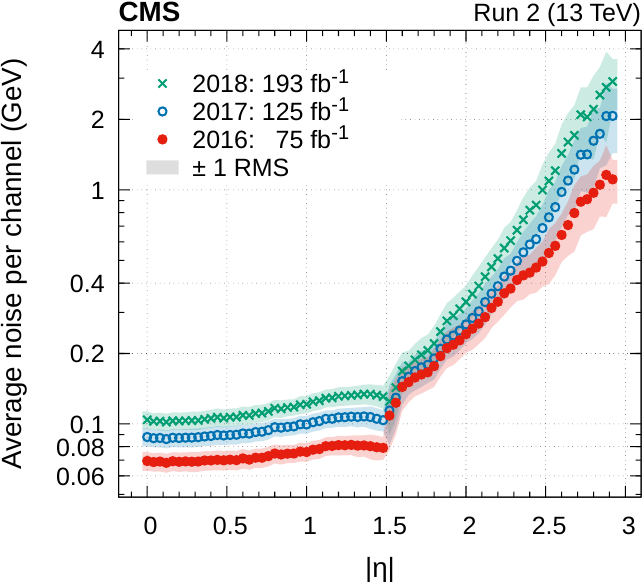}%
\hfill%
\includegraphics[width=0.48\textwidth]{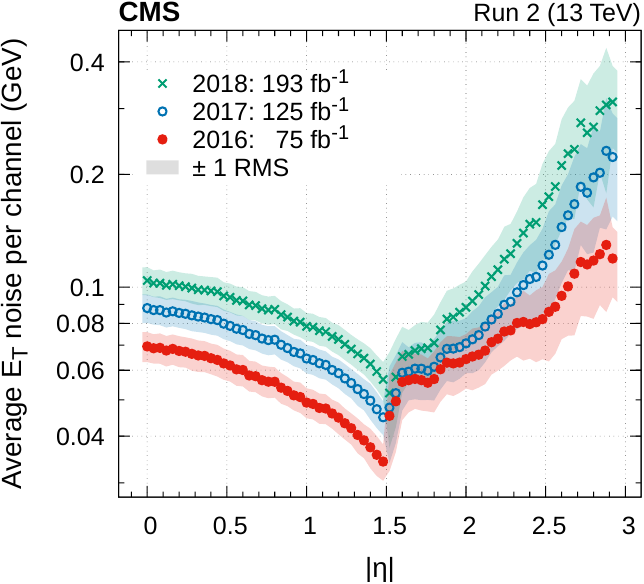}
\caption{%
    Average noise per channel as extracted from data using variations (RMS) of the signal baseline at the end of each year of \Run2 data taking.
    The amplitude RMS is converted into an energy equivalent (left) and a transverse-energy equivalent (right).
    The integrated luminosity accumulated since the start of \Run1 is indicated.
    The spread of the equivalent noise at a given pseudorapidity is indicated by the light-colored bands.
}
\label{fig:ecal:noisevseta}
\end{figure}

When the signal is corrected for the reduction in average light yield, the electronics noise is effectively amplified.
The effective noise, expressed in terms of equivalent energy and equivalent transverse energy, is shown in Fig.~\ref{fig:ecal:noisevseta}.
The measurements are extracted from the fluctuation of the signal baseline, and are shown as a function of the pseudorapidity, covering the barrel and endcap regions.
The three main data-taking periods of \Run2 are shown, along with the cumulative integrated luminosity since the beginning of \Run1.
The shape as a function of \abseta is the result of the noise increase as a function of rapidity, a consequence of the larger radiation dose received by the forward regions, and of the conversion to equivalent transverse energy, the relevant quantity for physics measurements.

\subsection{Signal reconstruction}

The signals, after analog processing, are digitized every 25\ns.
Upon a trigger, ten consecutive samples are transmitted to the backend electronics~\cite{Benetta:2004xh}.
In order to cope with the increased OOT pileup, a novel amplitude reconstruction algorithm was developed for \Run2, based on a template fit, called ``multifit''~\cite{CMS:EGM-18-001}.
The multifit algorithm replaced the \Run1 method based on a digital-filtering technique~\cite{CMS:NOTE-2006-037}, which estimated the energy by weighting five consecutive samples around the pulse maximum and subtracting the pedestals computed on the first three samples before the signal.

The multifit algorithm uses a template fit to extract the amplitude of the in-time pulse and the pulses coming from interactions occurring up to five BXs before and four BXs after, all within the 10-sample digitization window and potentially contributing to the total signal of one channel.
Because of the number of samples and free parameters in the fit, the baseline value of the signal is not computed dynamically but is instead obtained from regular measurements performed during data taking at least once per run.
Additional inputs to the fit are the template signal shapes, one per channel, and the noise covariance matrix.
Both inputs are regularly measured from data and updated when found to differ significantly from those in use.
This happens typically a few times per year, depending on the luminosity profile of the LHC.

The performance of the multifit algorithm has been measured using events from
$\PGpz\to\PGg\PGg$ and \Zee decays.
The energy resolution is excellent, as is the stability as a function of the OOT pileup.
The improvement with respect to a nonoptimized digital-filtering technique is more significant for low-amplitude pulses, where the relative contribution of OOT pileup pulses is larger.
The algorithm is sufficiently fast to be used in the HLT, and was adapted for execution on GPUs in the new processor farm used in \Run3.

The arrival time of the signal relative to the digitization window is measured by a digital-filtering technique based on the ratio of consecutive samples~\cite{CMS:CFT-09-006}.
The timing information is subsequently corrected for its dependency on the pulse amplitude, as derived from simulations.

\subsection{Trigger}
\label{sec:ecal:trigger}

The ECAL provides crystal energy sums, termed trigger primitives (TPs)~\cite{Paganini:2009zz}, to the CMS L1 trigger for every BX.
The trigger primitives are computed from energy sums of groups of 5$\times$1 crystals, referred to as ``strips''~\cite{Hansen:2003cds}.
Each strip is served by an individual FENIX chip that performs energy intercalibration, $E$-to-\ET conversion, amplitude estimation, and BX assignment functions.
In the EB, a sixth FENIX chip sums five strip-sums to compute the 5$\times$5 ``trigger tower'' transverse energy, calculates the ``fine-grain'' electromagnetic bit based on the compatibility of the deposits with those from an electromagnetic shower, and computes the strip fine-grain bit for the rejection of signals from direct energy deposition in the APDs (``spike killing'')~\cite{Petyt:2012sf}.
The strip fine-grain electromagnetic bit is configured to return a 0 for a spike-like energy deposit (a single channel above a configurable transverse energy threshold) or a 1 for a shower-like energy deposit (multiple channels above threshold).
In the EE, the five strip sums are transmitted to the off-detector trigger concentrator card (TCC) to complete the formation of the trigger towers.

The TCC is responsible for the transmission of the barrel and endcap TPs to the L1 calorimeter trigger every BX via the optical synchronization and link board (oSLB) mezzanine cards.
The TCC also performs the classification of each trigger tower, its transmission to the selective readout processor at each L1 trigger accept signal, and the storage of the trigger primitives for subsequent reading by the data concentrator card.

The ECAL L1 TPs are corrected for the effects of crystal and photodetector response changes due to LHC irradiation.
Correction factors are derived using measurements from the laser calibration system, and the same corrections are also applied in the HLT.
These corrections were first applied in 2012 only in the endcaps, for 22 individual rings of crystals of the same pseudorapidity, and were updated once per week.
During \Run2, because of the higher beam intensities and correspondingly larger response losses, the TP corrections were applied per crystal and extended to the EB.
From 2017 onwards, an automated validation procedure was developed to check the impact of the updated conditions on the L1 and HLT trigger rates, and the frequency of the updates was increased to twice per week to better track the response losses versus time.

Radiation-induced changes in the ECAL signal pulse shapes, in particular in the most forward regions of the EE, caused a continually growing probability for the BX to be misassigned in the ECAL TPs.
This effect, termed trigger primitive pre-firing, resulted in an inefficiency for recording potentially interesting events of about 0.1\% in any given primary data set, and about 1\% for events with two high-energy forward jets of invariant mass around 200\GeV~\cite{CMS:TRG-17-001}.
Following the discovery of this issue in early 2018, $\eta$-dependent timing offsets were applied to the ECAL frontend electronics, and throughout 2018, periodic updates to these offsets were made during every LHC technical stop, in order to minimize the level of pre-firing in both the EB and EE.

The ECAL spike-killer algorithm has been retuned for the more challenging beam and detector conditions of \Run2.
Spike-like energy deposits are rejected in the formation of the ECAL TPs by exploiting the additional functionality of the FENIX ASICs, the strip fine-grain electromagnetic bit.
If the deposit is considered spike-like, and the tower energy is above a second configurable threshold, the tower energy is set to zero and does not contribute to the triggering of the corresponding event.

The spike-killer parameters were updated in 2016 to account for the higher LHC luminosity and the larger single-channel noise observed in the EB during \Run2.
These new thresholds reduced the contamination of spikes in the ECAL TPs, corresponding to a transverse energy \ET of more than 30\GeV, by a factor of 2, with negligible impact on the triggering efficiency of electromagnetic signals with $\ET>20\GeV$.

The spike-killing efficiency is sensitive to drifts in the ECAL signal baseline.
Periodic updates in 2018 of up to twice per year of the baseline measurements used in the TP formation were therefore required in order to maintain a stable spike-killing efficiency.
By periodically updating the baseline values, the spike contamination for TPs with $\ET>30\GeV$ was maintained below 20\% during the 2018 run.
These improvements in TP calibration and spike rejection, together with improvements in the L1 trigger system itself, allowed the L1 electron/photon trigger to operate with high efficiency and at the lowest possible \ET thresholds throughout \Run2~\cite{CMS:TRG-17-001}.

\subsection{Channel calibration and synchronization}

While the principles and methods of the ECAL calibration have not changed and are described in Ref.~\cite{CMS:EGM-11-001}, a brief summary and update is given here to help discuss the results.

The calibration of the calorimeter proceeds in several steps:\ (i)~channels are corrected as a function of time for response changes as measured by the laser monitoring system;
(ii)~the response of channels at the same pseudorapidity, \ie, within the same $\phi$-ring, is intercalibrated using specific physics channels as reference;
(iii)~$\phi$-rings are intercalibrated with each other;
and (iv)~the absolute energy scale of the detector is fixed.
In a separate and independent procedure, channels are synchronized by using the average arrival time of particles in minimum-bias events.
Energy selections are applied to ensure an adequate signal-to-noise ratio, and additional criteria remove outliers in the timing distributions and ensure that the pulses have a good shape.

To complete the aforementioned steps (iii) and (iv), \Zee events are used.
For step (ii) the combination of a number of independent techniques is employed and is briefly summarized in the following.

The position of the two-photon invariant mass peak from \PGpz decays is a good physics standard candle for intercalibration, even if at low energy.
At the LHC, \PGpz are produced in abundance, and a dedicated trigger and data acquisition stream allow for an efficient collection of a large data set.
Events from this stream are saved in a reduced data format that contains only ECAL information in the proximity of the selected photon pair, to optimize the bandwidth at the HLT.
Starting from L1 electromagnetic candidates, the reconstruction applies a simplified clustering algorithm that collects energy in a 3$\times$3 crystal matrix centered around an energy deposit, called a ``seed'', greater than 0.5 (1.0)\GeV in the barrel (endcaps).
An offline analysis applies a correction derived from simulation to take into account effects of the readout, \eg, channel zero-suppression, energy lost in the vicinity of the detector boundaries, and dead channels.
An iterative fit to the invariant mass of the diphoton pair is performed, varying in each iteration the intercalibration coefficients and recomputing the clustered energy and candidate selection, until the variation from one step to the following is negligible.

The $E/p$ method exploits the distribution of the ratio between the reconstructed calorimeter energy $E$ and the momentum $p$ measured in the tracker of high-energy electrons from \PW and \PZ boson decays.
In order to obtain a pure electron sample, electron candidates are selected using kinematic, identification, and isolation requirements.
The algorithm evaluates the intercalibration in an iterative way.
In each iteration, the intercalibration coefficients are updated to constrain the peak of the $E/p$ distribution to equal unity, and the clustered energy is recalculated.
A correction is applied to take into account $\phi$-dependent biases in the momentum measurement due to the presence of inhomogeneous tracker support structures.
The correction is calculated from \Zee events using the tracker momentum measurement in a specific $\phi$ region for one of the electrons and the ECAL energy measurement in any $\phi$ region for the other.
The nonuniformity in $\phi$ is on the order of 1\%.

Electrons in \Zee events can also be used for the calibration of the detector.
Low-brems\-strah\-lung electrons are selected to minimize the influence of detector material upstream of the ECAL.
The definition of low-brems\-strah\-lung electrons is based on the narrowness of the electromagnetic shower detected in the calorimeter.
The well known invariant mass peak and distribution of the dielectron decay provide an almost background-free reference channel.
An unbinned likelihood is built for the distribution observed in data, assuming the invariant mass is well described by a classical Breit-Wigner function convolved with a Gaussian function that accounts for detector effects.
The resolution and scale of the Gaussian function peak value are the free parameters of the likelihood.
The granularity at which the parameters are allowed to vary permits the determination of the crystal intercalibration within $\phi$ rings, the relative calibration between rings, and the absolute energy scale of the detector.
Compared to the other two, this method of intercalibration is particularly effective at large pseudorapidities.
The calibration between rings ($\eta$-scale) is derived approximately every 5\fbinv, in order to correct for drifts in the detector response.

The azimuthal symmetry of the energy flow in minimum-bias events, which was successfully used during \Run1, became more challenging during \Run2 because of the increased effective noise.
While not competitive in precision for intercalibration purposes, it has been used to monitor the single-crystal response over time and provide useful insights into identifying and correcting residual imperfections in the light corrections.
An example of such imperfections are the slow regional drifts of approximately 1\% over one year of data taking, depending on the integrated luminosity and currently attributed to response changes in the p-n reference diode.

\subsubsection{Intercalibration precision}

Each of the intercalibration methods described above produces a set of constants with a corresponding statistical and systematic uncertainty.
The statistical precision can be evaluated by comparing sets of constants derived from disjoint event samples.
In the case of the \Zee method, the precision can be obtained from the fitting procedure.

The \PGpz method can be used to provide intercalibration constants for the barrel with a statistical precision that ranges from 0.1 to about 0.3\%, slowly increasing with pseudorapidity, for 10\fbinv of integrated luminosity.
In the endcaps the precision is around 1\%, due to the larger particle multiplicity, which requires a tighter event selection, and larger detector noise.

The $E/p$ method requires around 50\fbinv (about a calendar year in \Run2) to derive a set of intercalibration constants.
The corresponding statistical precision varies from 0.2\% for electrons in the inner barrel region to 0.4\% for electrons in the outer barrel.
The nonuniformity in $\phi$ of the material in front of the ECAL introduces a systematic uncertainty.

The \Zee method also requires about 50\fbinv to provide a set of constants of precision comparable to the $E/p$ ones in the barrel, while in the endcap it provides a precision far better than the other methods.
In \Run2, thanks to a sufficient integrated luminosity, it was possible to intercalibrate crystals in regions of $\abseta>2.5$ in the endcaps, where the other methods cannot be used either because of very high pileup contaminations or because the region is outside the tracker coverage.

The calibration constants determined using the three methods are combined using a weight that is proportional to the inverse square of the estimated precision.
For the \PGpz and $E/p$ methods, the systematic uncertainties are evaluated by studying the impact of the intercalibration constants on the \PZ boson lineshape, and found to be dominant for \PGpz and comparable to the statistical precision for $E/p$.
The intercalibration precision achieved in 2018 with the three methods combined is better than 0.5\% for the entire barrel, and between 0.5 and 1\% for the endcaps.

\subsection{\Run2 operations summary}

The ECAL DAQ operated during \Run2 with a luminosity-weighted efficiency larger than 99.6\% for the EB and EE, and larger than 99.2\% for the preshower detector.
The ECAL trigger system also operated with high efficiency and availability during \Run2.
The luminosity-weighted efficiency of the trigger system, accounting for trigger downtime and deadtime, \ie, automatic throttling of the readout decisions due to too high input rates, was larger than 99.9\%.
The fraction of ECAL channels that contributed to the DAQ was larger than 98.6\% at the end of \Run2, with a loss of less than 0.2\% over the course of the four years of \Run2 operations.
The fraction of channels that contributed to the trigger was larger than 99\%, and only a few problematic towers, strips, and individual channels were permanently masked.

A number of improvements to the firmware and software of the TCCs~\cite{Thiant:2019vlg} were implemented to achieve and maintain these high efficiencies in the more challenging beam conditions of \Run2.
These involved the automatic detection and masking of noisy or problematic signals from the frontend readout via configurable thresholds, without the need for manual intervention.
The algorithms allowed the setting of individual thresholds per strip in the EE, such that they could be adapted to changing LHC conditions, as well as to increased radiation-induced noise in the forward regions of the EE.
As a result of these improvements, which were fully implemented in both the EB and EE before the 2018 run, the number of incidents requiring manual intervention, as well as the deadtime and downtime associated with the ECAL trigger system, were significantly reduced in 2018 compared to 2017~\cite{Thiant:2019vlg}.

Additional improvements were made to the data acquisition boards, firmware, and software to be more resilient against and make automatic the recovery from single-event upsets.
Despite the luminosity increase, these remained a negligible source of downtime throughout \Run2.

To increase the reliability and ease maintenance, the crates of the high voltage system for the barrel and endcaps have been upgraded from the CAEN SY1525 to the CAEN SY4527, and the low voltage system has been made more redundant.

\subsection{\Run2 performance}

\begin{figure}[!ht]
\centering
\includegraphics[width=0.48\textwidth]{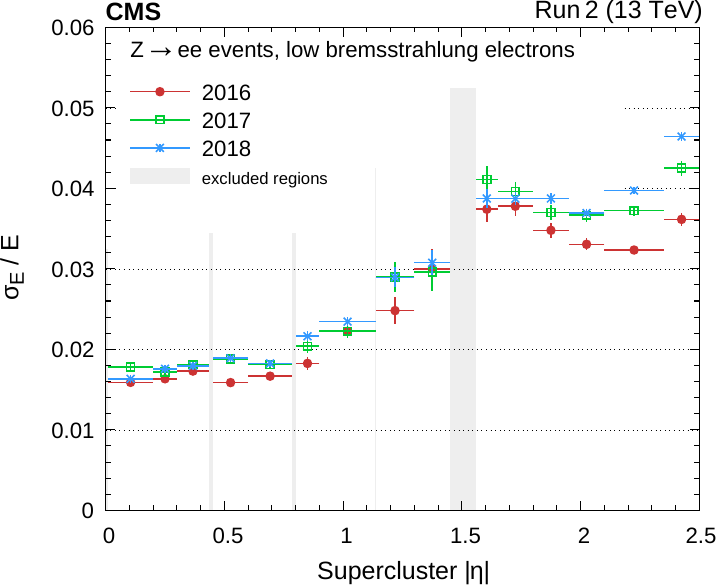}%
\hfill%
\includegraphics[width=0.48\textwidth]{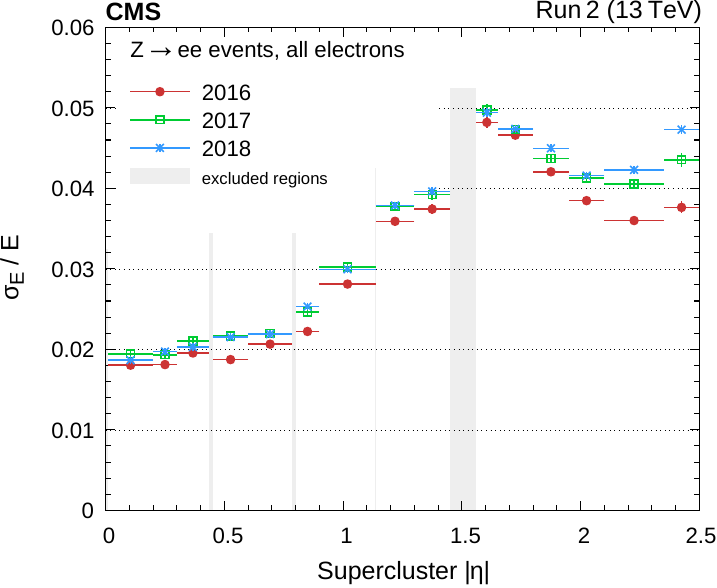}
\caption{%
    ECAL resolution for electrons having low bremsstrahlung emissions (left) and for an inclusive selection of electrons (right).
    The horizontal bars show the bin width.
}
\label{fig:ecal:resolution}
\end{figure}

The performance of the ECAL in terms of energy resolution and stability of the energy scale is evaluated using \Zee events reconstructed using the ECAL information alone.
The energy resolution is affected by pileup, noise, and accuracy of the calibration, in relative order of decreasing importance.
The resolution as a function of pseudorapidity is shown in Fig.~\ref{fig:ecal:resolution} for electrons with low bremsstrahlung emissions and for an inclusive electron sample.

The stability of the energy scale is monitored as a function of time by measuring the position of the peak in the dielectron invariant mass for \Zee events, and is found to be within 0.1\% in the barrel and a few times 0.1\% in the endcaps.
This figure is valid for electromagnetic deposits that are used in the reconstruction of jets and missing transverse momentum.
For precision physics studies involving electrons and photons, the energy scale is further corrected using \Zee events; after corrections, the stability is improved by a factor of 2 to 4, depending mostly on the pseudorapidity.

\begin{figure}[!ht]
\centering
\includegraphics[width=0.48\textwidth]{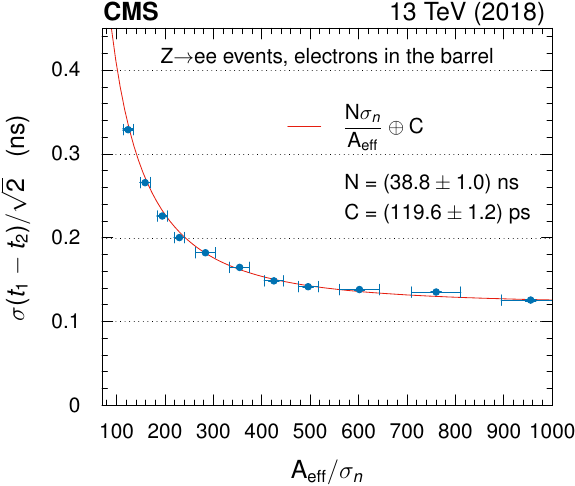}
\caption{%
    ECAL timing resolution as measured from \Zee events by comparing the arrival time of the two electrons.
    The performance has constantly improved over the years due to the frequency at which the synchronization constants have been updated, and is shown for 2018, the last year of \Run2.
    Updates in the constants are necessary to compensate for pulse shape changes induced by radiation.
    Vertical bars on the points showing the statistical uncertainties are too small to be seen in the plot.
    The red line correspond to a fit to the points with the parametrization of the resolution shown in the legend.
}
\label{fig:ecal:timing}
\end{figure}

The ECAL time resolution is measured using \Zee events by comparing the arrival times of the two electrons, defined as the time of the seed crystal in the supercluster.
This time is corrected for time-of-flight, determined from the two electron tracks from the primary vertex of the event.
Additional reconstruction quality and kinematic criteria ensure a pure sample of electrons.
The time resolution resulting from the analysis of 2018 data is displayed in Fig.~\ref{fig:ecal:timing}.
It is shown as a function of the effective amplitude normalized to the noise, defined as:
\begin{linenomath}\begin{equation}
    \frac{A_{\text{eff}}}{\sigma_n}=\sqrt{\frac{2}{(\sigma_1/A_1)^2+(\sigma_2/A_2)^2}},
\end{equation}\end{linenomath}
where $A_1$ and $A_2$ are the amplitudes of the two electron signals, and $\sigma_1$ and $\sigma_2$ the electronics noise of the corresponding channels.
Notable analyses profiting from the ECAL time resolution are those looking for long-lived particles, such as that described in Ref.~\cite{CMS:EXO-19-005}.

\subsection{Preparation for \Run3}

During LS2, the ECAL activities focused on improving the detector safety and control systems, on the algorithm to determine the trigger primitives, and on the development of a system to automatically compute, validate, and deliver updated calibrations.

\subsubsection{Safety and control system}

A significant upgrade of the ECAL safety and detector control systems, DSS and DCS, was performed during LS2.
The sensor readout system for temperature, humidity, voltage, and current levels, composed of custom readout units, was replaced by industrial analog input (AI) modules.
The 12 readout units based on RS-485 interfaces were exchanged by 45 AI modules, which are standard Siemens-certified peripherals (Simatic S7-300 analog input SM 331) connected through Profibus communication buses.
This type of connection provides access to extensive diagnostic information and enables the readout of a full sensor at sampling intervals as fast as 0.1\unit{s}.
This is about one order of magnitude faster than the previous system.
Following the update of the readout method, the programmable logic controller (PLC) of the safety system was reprogrammed with completely new software, and is now part of the pool of PLC framework applications in CMS.
The action matrix that defines the behavior of the PLC in terms of input and output signals and raises interlocks to protect the detector in case of alarm conditions, did not change, but benefits from the improved hardware.
The new system was extensively validated during the regular cosmic ray data-taking campaigns in 2020 and 2021.

\subsubsection{Trigger}

{\tolerance=800
Several improvements to the ECAL trigger-primitive formation and calibration were implemented for \Run3.
These include a further optimization of the spike-killer thresholds (Section~\ref{sec:ecal:trigger}) for the expected \Run3 pileup and noise levels, optimization of the digital-filter weights used to compute the trigger-primitive energies accounting for radiation-induced changes in pulse shapes, and the possible use of a second set of amplitude weights to further improve spike rejection and for the potential tagging of out-of-time signals.
\par}

More frequent corrections to account for crystal and photodetector response changes were also implemented for \Run3, and the frequency of the updates was increased from twice a week to once per LHC fill.
These corrections are important to maintain stable trigger rates and efficiencies, and improve the energy resolution of the related L1 and HLT objects, particularly electron/photon candidates.

\subsection{Calibration}

With the aim of reducing, as much as possible, the need for multiple reconstruction of CMS data sets with updated calibrations, a framework has been setup to provide automatic execution and bookkeeping of the workflows necessary to compute and validate updated and refined detector conditions as soon as enough data is available.
Not only does this allow to follow closely the prompt reconstruction of CMS data with the best foreseeable conditions, but it also permits to have the best conditions available for the data (re)reconstruction as soon as the data taking is finished.
While the techniques and physics standard candles used for calibrating the ECAL were well consolidated during \Run2, novelty and optimization have been introduced at the technical level of the data analysis needed to provide detector conditions.
Additionally, with a layer to validate conditions before deployment, key figures of detector performance can be controlled, such as stability, resolution, and projected rates in the HLT.
The workflows is synchronized with the online data taking and  orchestrated by a Jenkins instance, deployed through Red Hat OpenShift technologies, based on an InfluxDB backend and a Python server.

\clearpage
\section{Hadron calorimeter}
\label{sec:hcal}

\subsection{The hadron calorimeter in \Run1 and \Run2}

\begin{figure}[!ht]
\centering
\includegraphics[width=\textwidth]{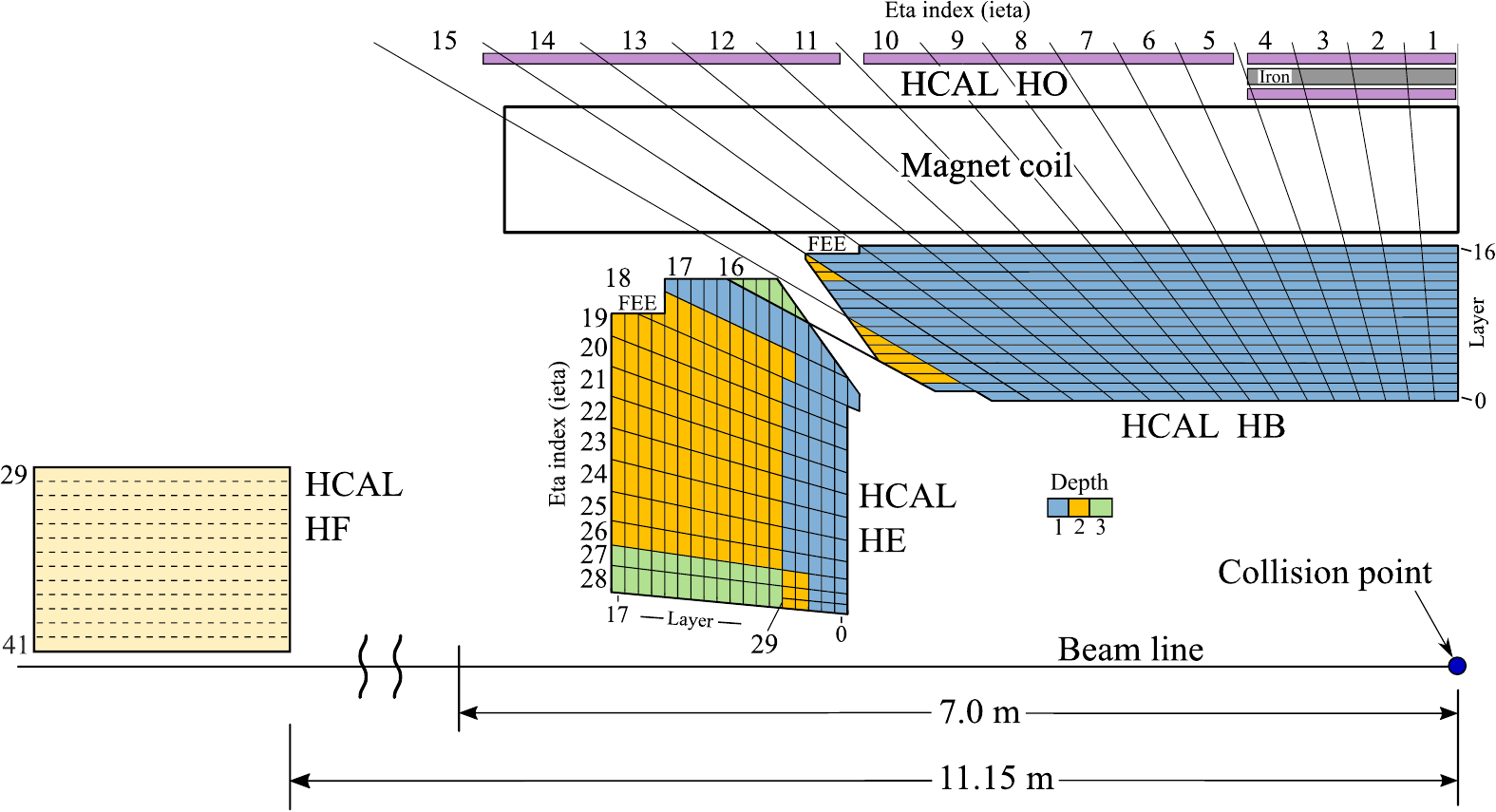}
\caption{%
    Schematic view of the HCAL as of 2016, showing the positions of its four major components:\ HB, HE, HO, and HF.
    The layers marked in blue are grouped together as ``depth~1,'' \ie, the signals from these layers of a given tower are optically summed and read out by a single photodetector.
    Similarly, the layers shown in yellow and green are combined as depths~2 and 3, respectively, and the layers shown in purple are combined for HO.
    The notation ``FEE'' denotes the locations of the HB and HE frontend electronics readout boxes.
    The solid black lines, roughly projective with the interaction point, denote the $\eta$ divisions in the tower $\eta$--$\phi$ segmentation, and the numbers at the edge of the tower denote the ieta index. Figure from Ref.~\cite{CMS:PRF-18-001}.
}
\label{fig:hcal:2016}
\end{figure}

{\tolerance=800
The CMS hadron calorimeter~\cite{CMS:TDR-2} (HCAL), shown schematically in Fig.~\ref{fig:hcal:2016}, is composed of four major subdetectors:\ the hadron barrel (HB)~\cite{CMS:NOTE-2006-138}, hadron endcap (HE)~\cite{CMS:TDR-2}, hadron forward (HF)~\cite{CMS:NOTE-2006-044}, and hadron outer (HO) calorimeters~\cite{CMS:NOTE-2008-020}.
The HB and HE cover the pseudorapidity regions $\abseta<1.392$ and $1.305<\abseta<3.0$, respectively.
The HO provides a measurement of the shower tails in the region $\abseta<1.26$, and the HF covers $3.0<\abseta<5.2$.
\par}

\begin{figure}[!ht]
\centering
\includegraphics[width=0.355\textwidth]{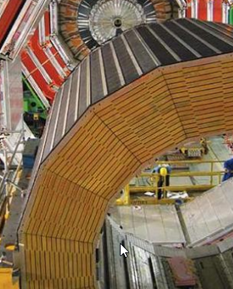}%
\hspace{0.04\textwidth}%
\includegraphics[width=0.35\textwidth]{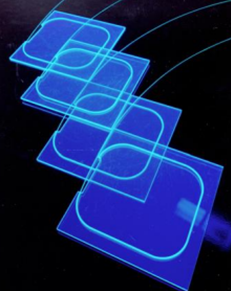}
\caption{%
    Left:\ brass absorber for the hadron barrel calorimeter HB.
    Right:\ scintillating tiles with wavelength shifting fibers used as the active media in the barrel, endcap, and outer hadron calorimeters.
}
\label{fig:hcal:photos}
\end{figure}

\begin{figure}[!ht]
\centering
\includegraphics[width=0.8\textwidth]{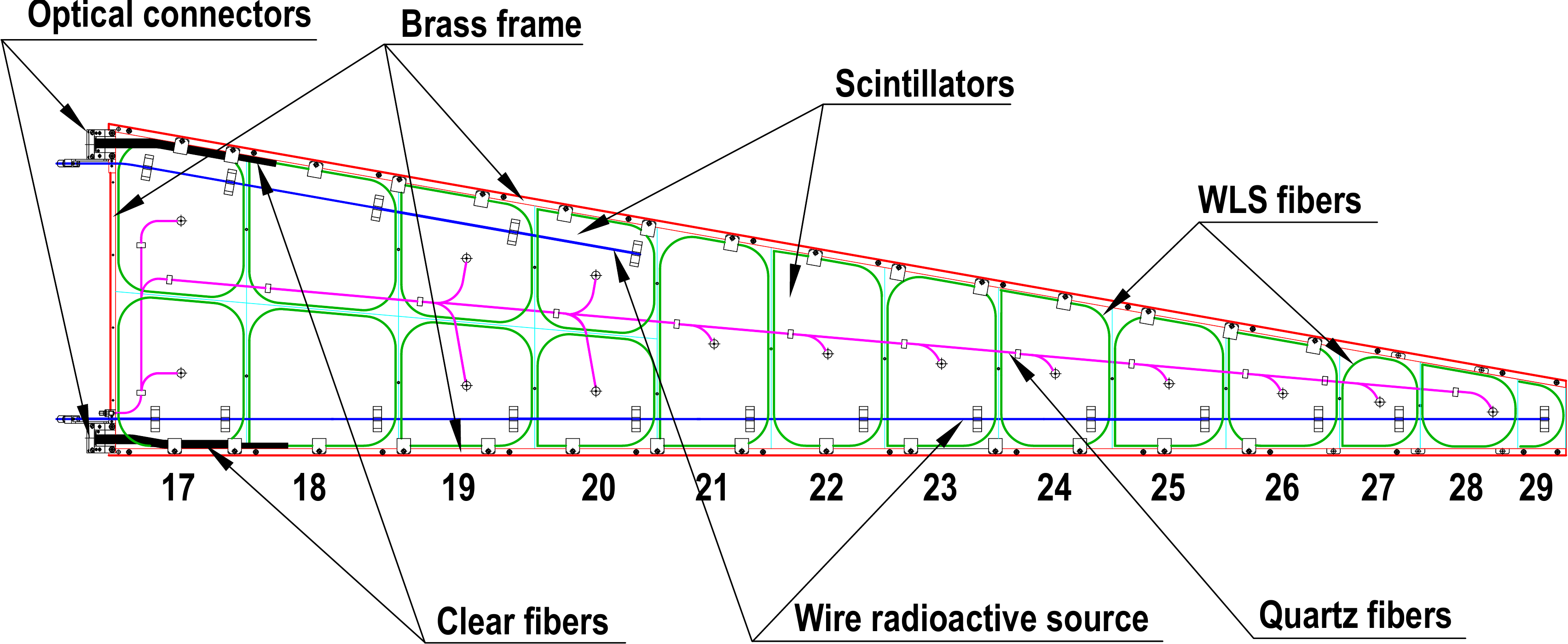}
\caption{%
    Physical arrangement of the scintillator and wavelength-shifting fibers into tiles for an HE megatile. Figure from Ref.~\cite{CMSHCAL:2016dvd}.
}
\label{fig:hcal:megatile}
\end{figure}

The HB and HE primarily use brass as the absorber, except for the inner and outer layers of HB, which are constructed from steel.
The HB absorber is shown in Fig.~\ref{fig:hcal:photos} (left).
The signals are produced in plastic scintillating tiles (Fig.~\ref{fig:hcal:photos}, right), and the resulting blue light is shifted to green via embedded wavelength-shifting fibers.
The towers in HB (HE) have up to 17 (18) scintillator layers, as shown in Fig.~\ref{fig:hcal:2016}.
Sequential layers are grouped into ``depth'' segments:\ the light from the layers in a given depth segment is optically summed and read out by a single photodetector.
Clear plastic fibers send the signal to the hybrid photodetectors (HPDs) in the original design or silicon photomultipliers (SiPMs) after the upgrades.
The segmentation is a tower structure in $\eta$--$\phi$ space.
The $\eta$ segmentation is indicated by the black solid lines in Fig.~\ref{fig:hcal:2016}.
The towers are referenced using integer indices ieta and iphi, where the ieta assignments are given in the figure and iphi runs from 0 to 71, corresponding to the 72 divisions in $\phi$.
Physically, the scintillators are arranged in ``megatiles'', which are trays that support an array of scintillator tiles, along with the fibers that route the light to the photodetectors.
An HE megatile is shown in Fig.~\ref{fig:hcal:megatile}.
All channels in a subdetector with the same iphi are defined as residing in the same wedge.

\begin{figure}[!p]
\centering
\includegraphics[width=0.7\textwidth]{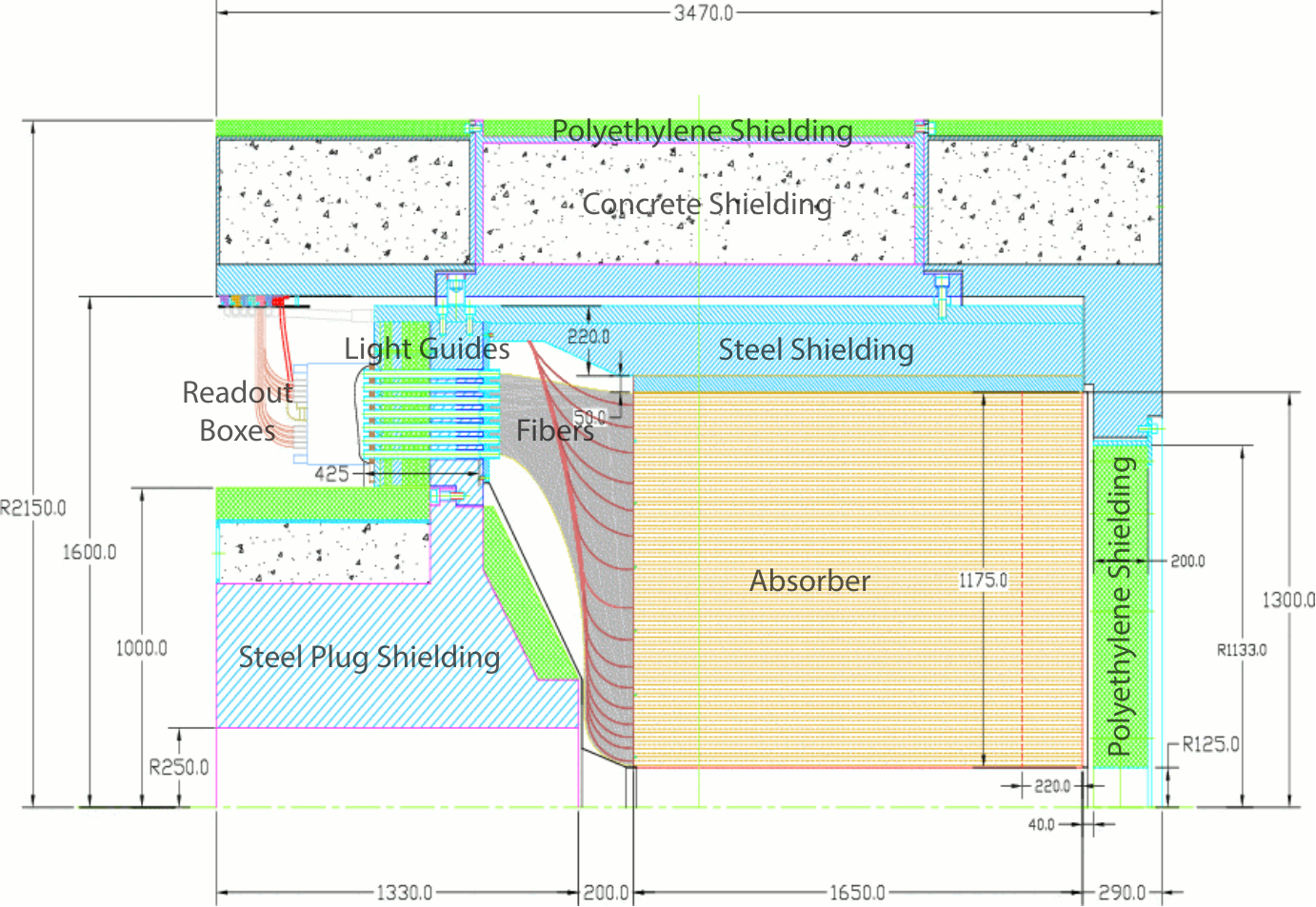}
\caption{%
    Schematic view of the CMS hadron forward calorimeter, HF.
    The yellow region represents the steel absorber with embedded quartz fibers; the grey shaded area to the left represents fibers which deliver the signals to light guides that penetrate the steel plug shielding; the white rectangle to the left of the light guides represents the frontend readout boxes, which house the photomultiplier tubes.
    The dimensions shown in the diagram are in millimeters.
}
\label{fig:hcal:hf2016}
\end{figure}

The HF, shown schematically in Fig.~\ref{fig:hcal:hf2016}, is a 1.65\unit{m}-long sampling calorimeter with steel absorber.
Plastic-clad quartz fibers with a diameter of 0.6\mm are the active elements, producing Cherenkov light.
The fibers are inserted into 1\mm grooves drilled in the steel absorber, 5\mm apart.
The fibers are parallel to the beam line.
To enable discrimination of electrons and photons from hadrons, the fibers have two lengths:\ 1.65\unit{m} ``long fibers'' and 1.40\unit{m} ``short fibers''.
The short fibers run from the outside of the HF to 25\cm from the front face of the HF.
The HF photodetectors are photomultiplier tubes (PMTs).

\begin{figure}[!p]
\centering
\includegraphics[width=0.7\textwidth]{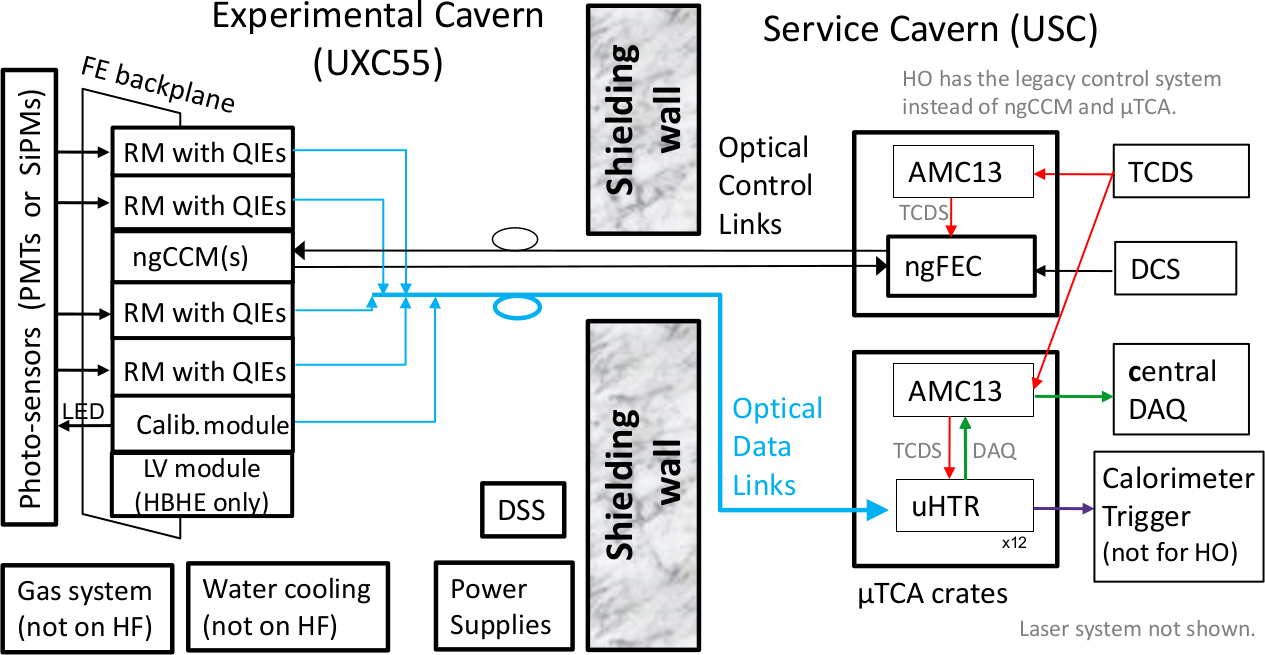}
\caption{%
    Schematic diagram of the HCAL readout.
    The frontend electronics chain begins with analog signals from the HPDs, SiPMs, or PMTs.
    The HB and HE used HPDs prior to the \Phase1 upgrade and SiPMs after, while the HF uses PMTs.
    These signals are digitized by the QIE chips in the readout modules (RMs). The ``next-generation clock and control module'' (ngCCM) is part of the frontend control system.
    The digitized data are sent to the backend \uHTR[s] (\uTCA HCAL trigger and readout cards).
    Data from multiple \uHTR[s] are concentrated into the AMC13 cards and forwarded to the CMS central data acquisition (DAQ) system.
    The \uTCA cards also send data to the trigger system.
    The AMC13 cards also distribute the fast commands arriving from the CMS timing and control distribution system (TCDS) within each \uTCA crate.
    The detector control system (DCS) software communicates with the ``next-generation frontend controller'' (ngFEC).
}
\label{fig:hcal:readoutchain}
\end{figure}

The HCAL readout control chain is shown schematically in Fig.~\ref{fig:hcal:readoutchain}.
In the frontend readout boxes (RBXs), mounted on the detector, the electronics chain begins with analog signals from the photodetectors.
The optical decoder unit (ODU) maps the incoming fibers from the scintillators to the photodetectors.
These signals are digitized by the QIE chips (QIE8, QIE10, and QIE11)~\cite{Zimmerman:2003gv, Baumbaugh:2014eya, Roy:2015bia, Hare:2016swd} in the readout modules (RMs), which provide both energy information via an analog-to-digital converter (ADC) and, for the QIE10 and QIE11 chips, rising-edge timing information via a time-to-digital converter (TDC).
The clock, control, and monitoring (CCM) module provides the LHC clock, synchronous fast commands, slow control, and monitoring for the detector frontend electronics.
Each RBX also contains a calibration module (CM), which provides calibration signals from two sources: an in situ LED, which sends light directly to the photosensors, and an off-detector laser, which can send light either to the photosensors or into the scintillator tiles for some of the detector layers.
The CM also provides a reference measurement of the LED and laser calibration signal intensities.

\subsection{Upgrades}

\subsubsection{Motivation and overview of the upgrade}

\begin{figure}[!ht]
\centering
\includegraphics[width=\textwidth]{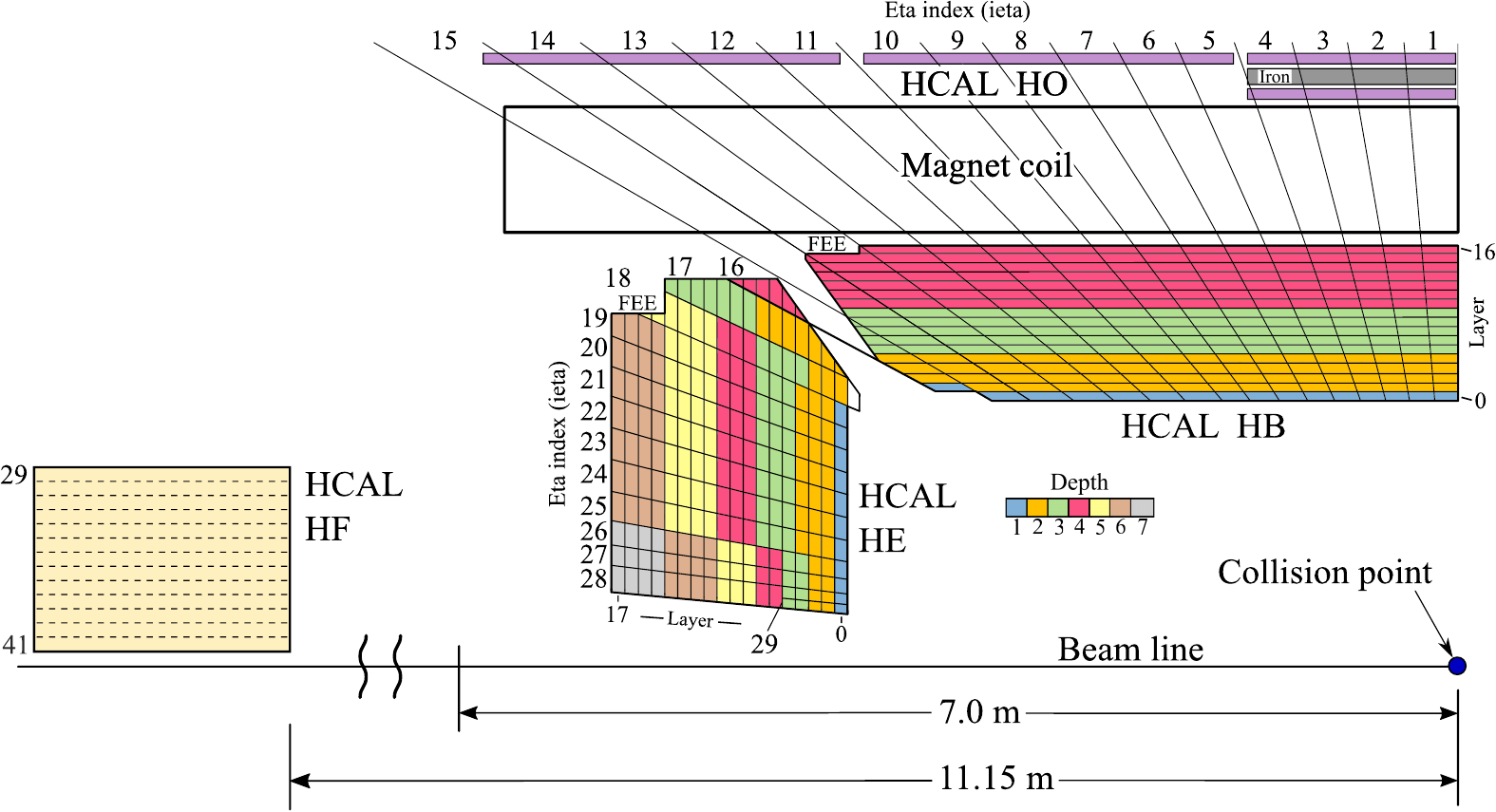}
\caption{%
    The longitudinal and transverse HCAL segmentation for \Run3.
    Within a tower, layers with the same color are routed to the same SiPM.
    The location of the frontend electronics is indicated by the letters FEE. Figure from Ref.~\cite{CMS:PRF-18-001}.
}
\label{fig:hcal:hbhesegmentation}
\end{figure}

The motivation for and the design of the CMS calorimeter system \Phase1 upgrades are given in Refs.~\cite{CMS:TDR-010, CMS:TDR-015}.
The main goals of the HB and HE upgrades were:\ to replace the HPD photodetectors, which produced anomalous signals~\cite{CMS:CFT-09-019} and showed signal degradation~\cite{Shukla:2019waq};
to increase the segmentation to that shown in Fig.~\ref{fig:hcal:hbhesegmentation}, to allow for both layer-dependent corrections for the observed radiation damage to the scintillating tiles~\cite{CMS:PRF-18-003} and better rejection of energy deposits from pileup interactions;
to increase the readout bandwidth to allow for the larger number of channels;
to add signal arrival time measurements in the HB, HE, and HF;
and to standardize the readout electronics across the different calorimeter systems.
For the HF, the PMTs were also replaced because they were a source of anomalous signals~\cite{CMS:CFT-09-019}.

The HCAL plastic scintillators are subject to radiation damage, which results in the reduction of the signal output.
The relevant primary characteristics of the LHC operation are the total delivered integrated luminosity, which determines the radiation dose received by the scintillator tiles, and the average instantaneous luminosity, which controls the dose rates.
Radiation effects were evaluated by studying the tile performance during the 2017 LHC operation, corresponding to a delivered integrated luminosity of about 50\fbinv.
Signal reduction was studied as a function of dose rate.
These measurements provide unique information on the radiation damage at dose rates significantly lower than previously studied.
The HE tile results were obtained using several complementary methods:\ a movable radioactive source that can access all the tiles to compare their signal output before and after the 2017 data-taking period; inclusive and isolated-muon energy deposits produced during \pp collisions; and a laser calibration system.

\begin{figure}[!ht]
\centering
\includegraphics[width=0.455\textwidth]{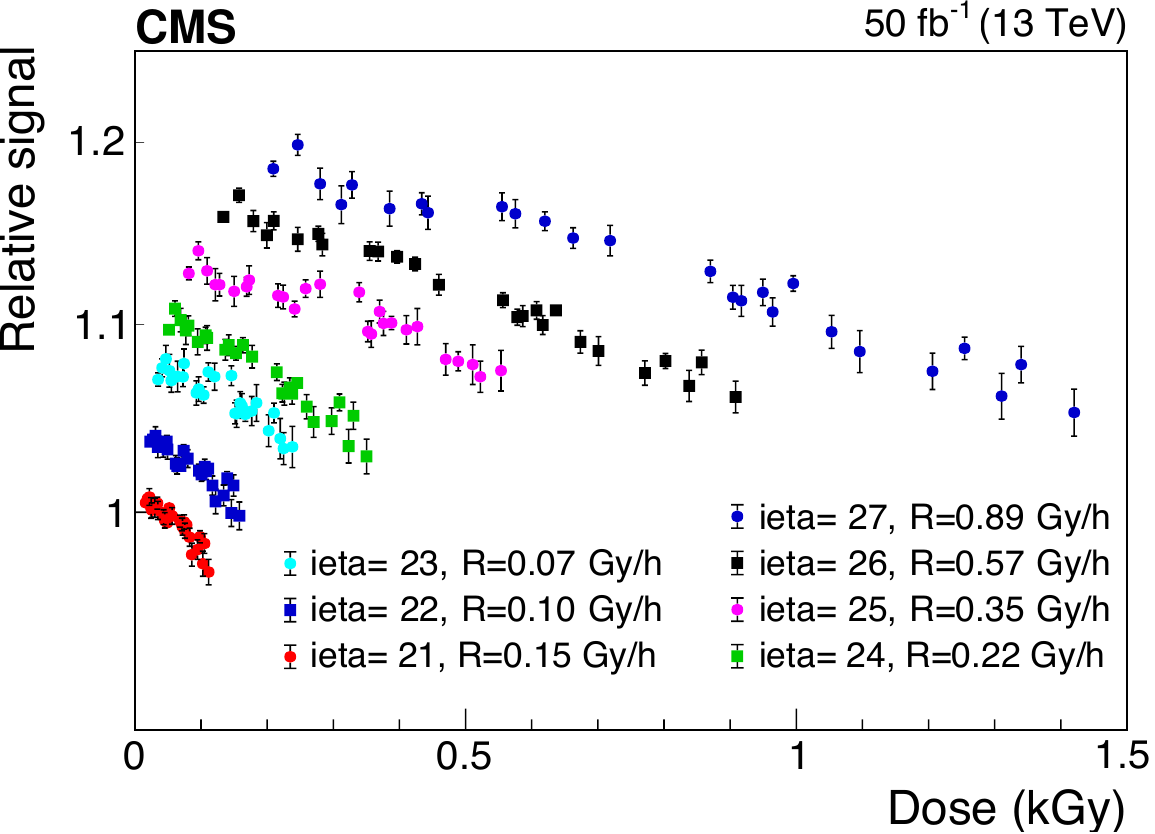}
\hfill
\includegraphics[width=0.505\textwidth]{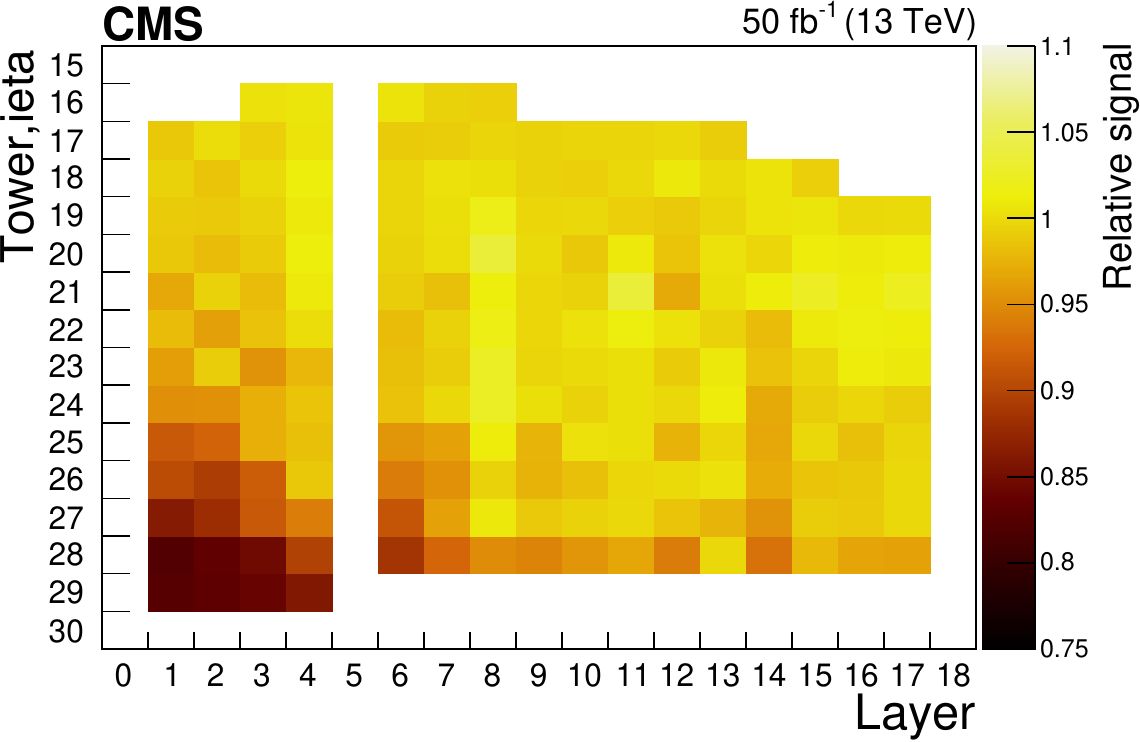}
\caption{%
    Left:\ relative laser light signal versus the accumulated dose for scintillator tiles in \layer1 and ieta range 21--27.
    The average dose rate $R$ for each set of points is given in the legend.
    The vertical scale is logarithmic and subsequent sets are shifted up by a factor of 1.03 relative to the previous set for better visibility.
    Each set starts at a dose corresponding to an integrated luminosity of 7\fbinv.
    The vertical bars give the scaled statistical uncertainties.
    Right:\ ratio of the signals from the \sixtyCo source observed before and after the 2017 data-taking period for scintillator tiles in the HE as a function of ieta and layer number.
    Tubes in layers 0 and 5 have obstructions and cannot be accessed.
}
\label{fig:hcal:raddam}
\end{figure}

The laser system consists of a triggerable excimer laser and light distribution system that delivers UV light (351\nm) to the scintillator tiles in layers 1 and 7 via quartz fibers, as well as directly to the photodetectors.
During the 2017 data-taking period, pulses of laser light were injected between LHC fills when there were no collisions.
Laser data were collected throughout the 2017 data-taking period.
Figure~\ref{fig:hcal:raddam} (left) presents the relative signals in \layer1 versus dose for tiles in the ieta range 21--27.
The signals show an approximately exponential decrease during periods of stable luminosity, with slopes that depend on the dose rate.
Tiles at smaller ieta show more damage per dose than those at larger ieta, implying that at a fixed dose the damage to the scintillators increases with decreasing dose rate, within the range of our measurements.

Each individual tile in the HCAL is designed to be serviced by a movable \sixtyCo radioactive source using small tubes that are integrated into the calorimeter.
The \sixtyCo source provides photons with energies of 1.17 and 1.33\unit{MeV}.
The source is attached to a wire that guides it through the tubes.
All tiles except those in layers 0 and 5, whose tubes have obstructions, can be accessed.
The data were collected during the periods when the LHC did not operate, before both the 2017 and 2018 data-taking periods.
Values of the ratio, averaged over iphi as a function of the scintillating tile layer number and tower index ieta, are shown in Fig.~\ref{fig:hcal:raddam} (right).
The signal loss is smaller for tiles at larger radial distance from the beam and for layers that are deeper in the calorimeter.

Because the damage depends on the layer number, increased segmentation allows for more accurate damage corrections to be applied as a function of integrated luminosity, reducing the degradation of the resolution.

The on-detector upgrade components must operate in a radiation field and withstand cumulative effects from both the total ionizing dose (TID) and nonionizing energy loss (NIEL), as well as single-event effects (SEE).
The required tolerance for NIELs is quoted as that corresponding to an equivalent fluence (\percmsqns) of 1\MeV neutrons.
Expected SEE rates are characterized by the fluence (\percmsqns) of hadrons with energy exceeding 20\unit{MeV}.
Requirements for the HB, HE, and HF are shown in Table~\ref{tab:hcal:radreq}.
Since the HE will be replaced after \Run3, the numbers given for the HE are the requirements for \Run3, while the HB and HF values correspond to the full HL-LHC duration.

\begin{table}[!ht]
\centering
\topcaption{%
    Radiation requirements for the \Phase1 upgrade.
    The HE numbers are for \Run3, while the HB and HF values correspond to the full HL-LHC duration.
}
\label{tab:hcal:radreq}
\renewcommand\arraystretch{1.2}
\begin{tabular}{cccc}
    Detector & TID & NIEL & SEE \\[-3pt]
    & [krad] & [$\percmsqns$] &  [$\percmsqns$] \\
    \hline
    HB & 3.1 & $1.1\times10^{12}$ & $2.0\times10^{11}$ \\
    HE & 0.9 & $9.0\times10^{10}$ & $1.6\times10^{10}$ \\
    HF & 4.1 & $7.0\times10^{11}$ & $1.8\times10^{11}$ \\
\end{tabular}
\end{table}

\subsubsection{HB/HE/HO photodetector upgrade}

A significant part of the HCAL \Phase1 upgrade was the replacement of the HPD photodetectors with SiPMs for HB, HE, and HO.
All three subdetectors used HPDs during \Run1~\cite{Cushman:2000iv}, and were converted to SiPMs at different times.
The HO SiPMs were installed prior to the start of \Run2.
For HE, a $\phi$ sector of 20\de\ was instrumented with SiPMs during the 2016/2017 year-end technical stop, and the rest of the detector was converted during the 2017/2018 technical stop.
The HB used HPDs for all of \Run2, with SiPMs installed prior to the start of \Run3.
The grouping of layers in a tower to individual photosensors, \ie, the longitudinal readout segmentation, is indicated by color in Fig.~\ref{fig:hcal:2016} prior to the upgrade and following the upgrade in Fig.~\ref{fig:hcal:hbhesegmentation}.
Prior to \Run3, most of the HB towers had a single readout segment, except for $\text{ieta}=15$ and $\text{ieta}=16$.
The HB readout segmentation increased from one to four channels starting in \Run3.
For HE, the readout segmentation increased from two to three channels per tower in \Run1 to six to seven channels during \Run2 as the SiPMs were installed.

The SiPM has many advantages over the HPD, including high photon detection efficiency (PDE), high gain, a large linear dynamic range, rapid recovery time, somewhat better radiation tolerance, and insensitivity to magnetic fields.
The active area of each SiPM is circular, with diameters of 2.8\mm (sensing up to four fibers) and 3.3\mm (sensing up to seven fibers).
In each HE readout unit, 19 HPD channels were replaced by 48 channels of SiPMs.
For the HB, each readout unit has 64 channels of SiPMs.

\begin{figure}[!ht]
\centering
\includegraphics[width=0.61\textwidth]{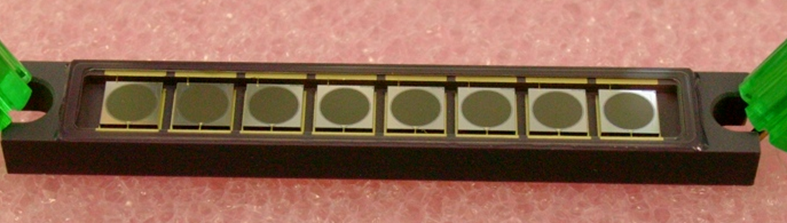}%
\hspace{0.04\textwidth}%
\includegraphics[width=0.3\textwidth]{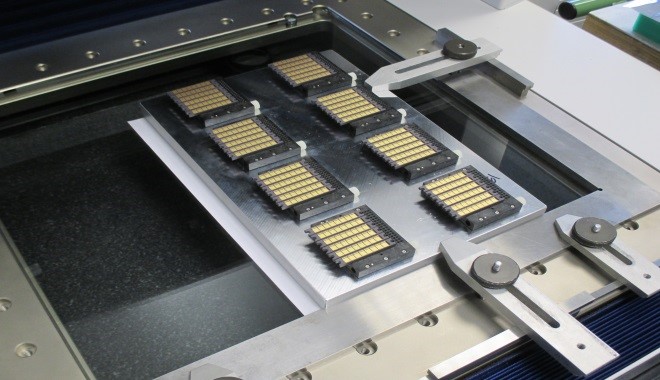}
\caption{%
    Left:\ top and side views of an eight-channel SiPM array in its ceramic package.
    Right:\ ceramic packages at the CERN Metrology Laboratory for characterization measurements.
}
\label{fig:hcal:sipmphotos}
\end{figure}

Arrays of eight individual SiPMs were placed into a ceramic carrier known as the ``SiPM package''.
The packages for the HE and HB are identical.
They were designed by CMS and fabricated at Kyocera.
Figure~\ref{fig:hcal:sipmphotos} (left) shows a top-and-side view of an eight-channel SiPM array in its package.
Each SiPM is connected electrically to the package via two wire bonds, one for the signal and one for the bias voltage.
The 16 pins at the bottom of the package connect the SiPMs to the downstream electronics.
The package precisely locates each SiPM on the ODU, so its mechanical dimensions are important.
Before sending them to the SiPM vendor, the mechanical tolerances of each package were measured at the CERN Metrology Laboratory, and more than 30 parameters were compared with the design specifications.
Figure~\ref{fig:hcal:sipmphotos} (right) shows 48 ceramic SiPM packages prepared for measurement.
The yield of good packages was very high with less than one percent rejected for being out of tolerance.

Before choosing a vendor and entering final SiPM production for the HE, and then later the HB, careful studies were carried out on preproduction samples from several vendors.
The preproduction samples were of order 1500 channels.
Key SiPM parameters included PDE, gain, recovery time, uniformity of breakdown voltage, and dark current~\cite{Heering:2016aqi}.
A small sample of SiPMs was also irradiated for radiation damage studies~\cite{Musienko:2015lia}.
Based on the results, the Hamamatsu SiPMs were chosen for final production.

\begin{figure}[!ht]
\centering
\includegraphics[width=0.48\textwidth]{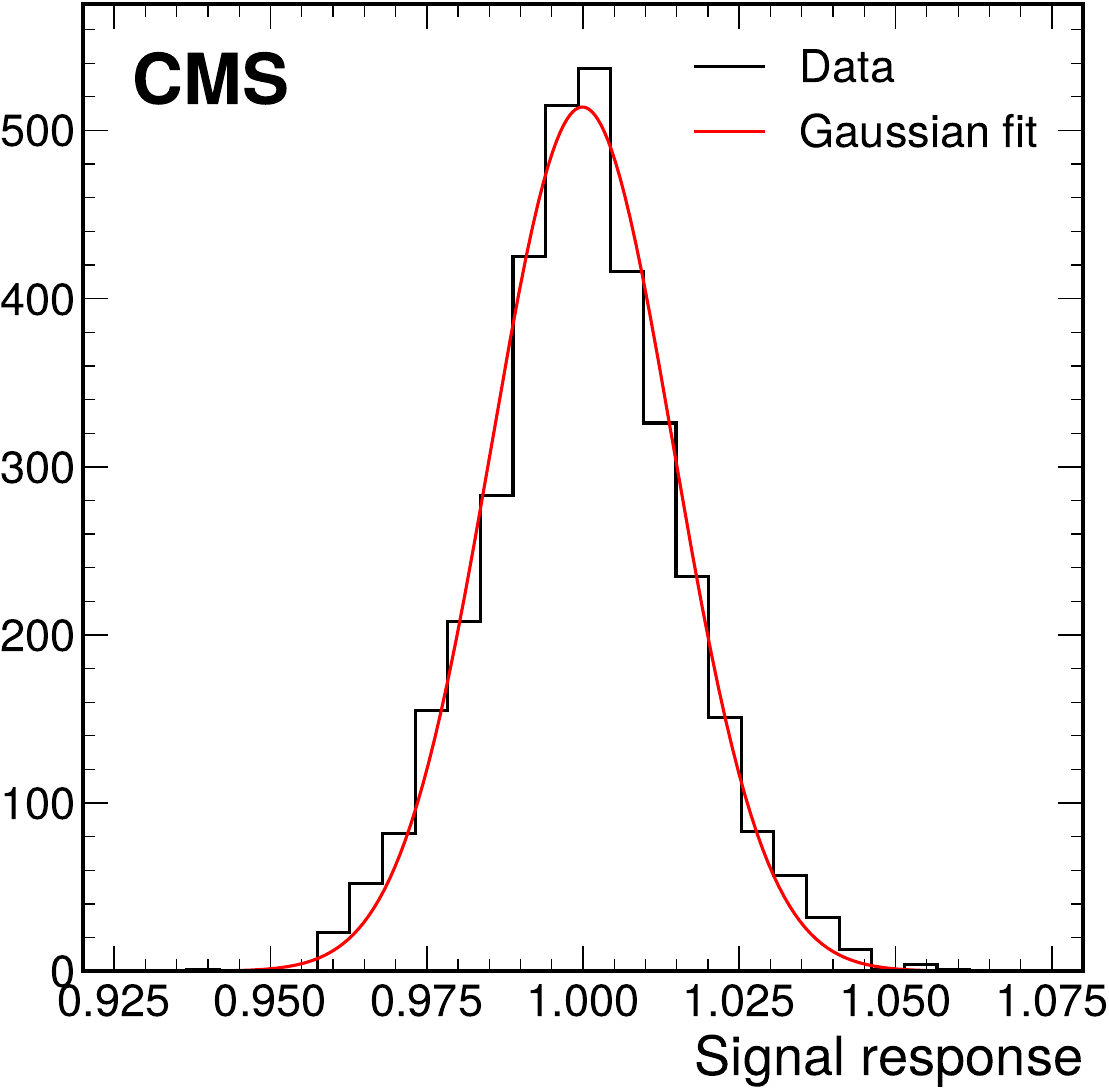}
\caption{%
    Distribution of the signal response (photon detection efficiency times gain) for 3600 HE SiPMs.
}
\label{fig:hcal:sipmresponse}
\end{figure}

The quality control program was virtually identical for the HE and HB.
We measured the signal response (PDE times gain), breakdown voltage, dark current, forward resistance, and capacitance for each channel.
We also looked for spurious noise pulses in every channel.
A total of 1400 production HE arrays were tested, and 104 arrays, or 7.43\%, were rejected.
Since there are eight SiPM channels per array, this corresponded to a yield of good SiPMs better than 99\%.
For the HB production, a total of 1680 production arrays went through the same quality control testing, with only 82 rejected, again giving a yield of good SiPMs better than 99\%.
The uniformity was also excellent throughout the two production runs.
As an example, Fig.~\ref{fig:hcal:sipmresponse} shows the signal response for 3600 HE SiPMs, with an RMS of about 1.5\%.
After quality control, a total of 864 (1152) SiPM arrays were installed into the HE (HB).

\subsubsection{Readout box}

\begin{figure}[!ht]
\centering
\includegraphics[width=.8\textwidth]{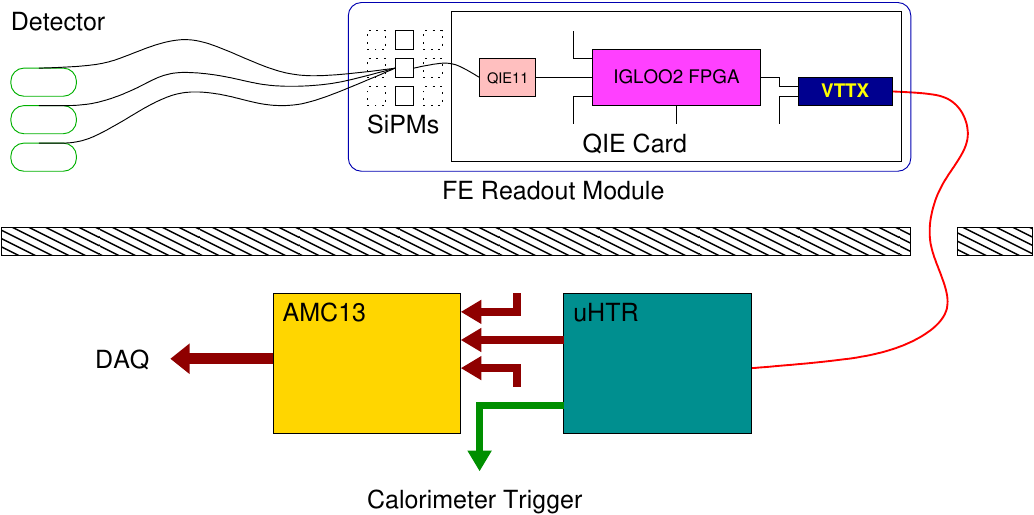}
\caption{%
    Upgraded HCAL endcap (HE) and barrel (HB) DAQ readout chain, including the SiPMs, frontend readout electronics, and backend system. Figure from Ref.~\cite{CMS:TDR-010}.
}
\label{fig:hcal:daqreadout}
\end{figure}

The RBX is an aluminum shell with snake-shaped grooves machined on the surface for insertion of copper cooling pipes.
The RBX houses the RMs, control electronics, and the QIE-based frontend electronics, which are connected to the backend in the control room via control and data fibers.
The HB and HE frontend electronics, shown schematically in Fig.~\ref{fig:hcal:daqreadout} are very similar, and primarily differ to reflect the mechanical, structural, and channel multiplicity differences between the two subsystems.

Each HE, HB, and HO RM has 48, 64, and 18 SiPMs, respectively.
Each RM houses a single SiPM control card and four frontend readout ADC cards (three cards for the HO).
Each RM also houses an ODU.
As shown in Fig.~\ref{fig:hcal:readoutchain}, the RBX consists of four RMs installed together with a ngCCM and a CU.
In HB, two ngCCMs are installed per RBX, each independently controlling two RMs.
All RBX components receive the power and communicate via a backplane PCB.
The RBX receives its fast and slow control signals through the ngCCM module.
This module receives the clock and fast reset signals via optical links from the backend, and distributes these signals to each card in the RBX.
Configuration information for the frontend is received over the optical link as well.
Finally, the ngCCM is responsible for aggregating and returning all status information about the RBX, which is transmitted to the backend via an optical link.

Each RBX is installed on top of a scintillator megatile in the locations indicated in Fig.~\ref{fig:hcal:hbhesegmentation}.
One RBX covers 20\de\ in $\phi$.
The total number of RBXs is 108, excluding those used for the HF.
The RBX is the part of the common CMS infrastructure; therefore, each RBX has individual power, water cooling, and dry gas lines connected to common water and gas manifolds in order to provide proper and stable cooling for the HCAL electronics and decrease the relative humidity around the SiPMs.
The maximum temperature of RBX components does not exceed 35\deC, and the relative humidity in the SiPM region stays constant within 5\%, except on rare occasions when there are cooling or dry gas issues.

The first upgraded RBX for the initial sector of the HE instrumented with SiPMs was installed in 2017.
One HE RBX was upgraded at the end of 2016, the rest of the HE in 2017/2018, and the HB RBXs in 2019 after \Run2.

\subsubsection{Photodetector control}

\begin{figure}[!ht]
\centering
\includegraphics[width=0.48\textwidth]{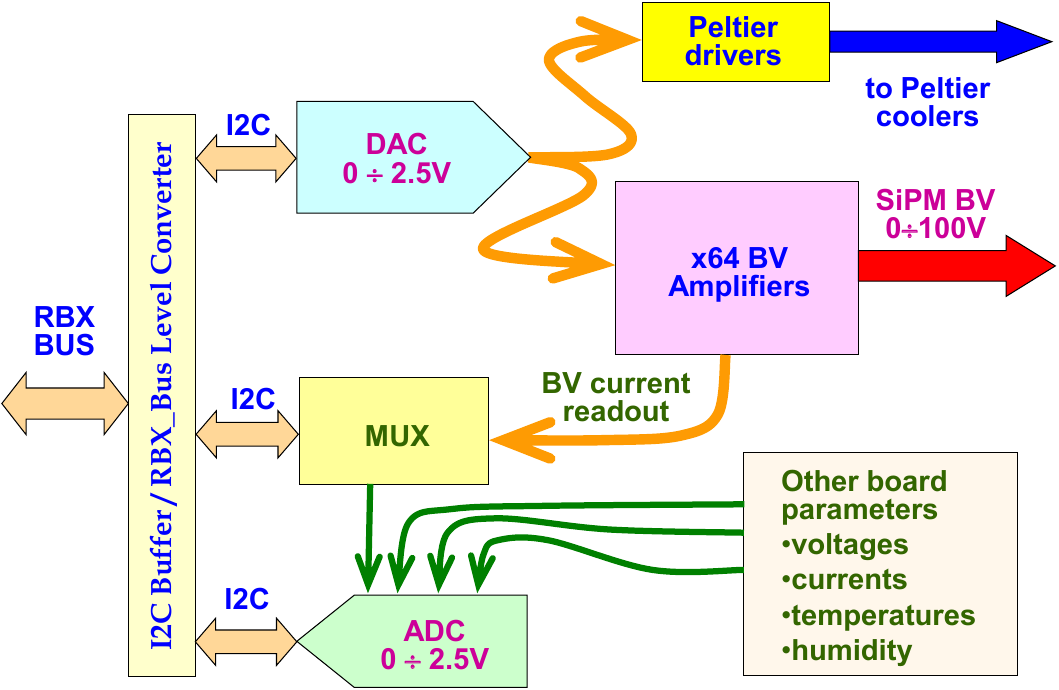}%
\hfill%
\includegraphics[width=0.48\textwidth]{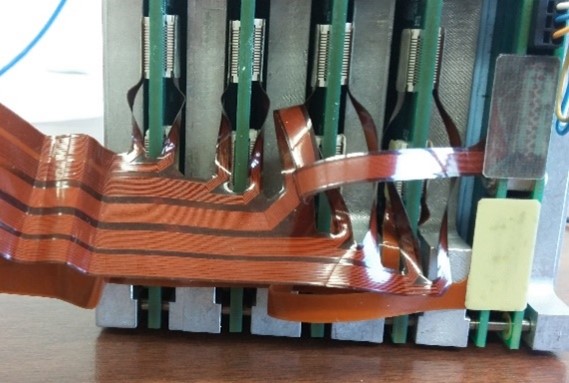}
\caption{%
    Left:\ control board block diagram.
    Right:\ HB card pack with one control and four frontend boards.
    The flex cables provide signal and bias voltage connections to 64 SiPMs.
}
\label{fig:hcal:sipmcontrol}
\end{figure}

The use of SiPMs required the design of completely new high-density photodetector control electronics.
The new design was done initially for the HO photodetector upgrade, and consisted of an additional board placed next to the QIE cards in the RM, shown schematically in Fig.~\ref{fig:hcal:sipmcontrol} (left) and as constructed in Fig.~\ref{fig:hcal:sipmcontrol} (right).
The HE and HB subdetectors reuse the same design topology to minimize design, testing, and validation time, although small changes were needed to accommodate the different SiPM types.
The control board uses a single $+8\unit{V}$ power voltage from the RBX backplane and is controlled via an I2C-compatible slow-control connection.
The board provides individual bias voltage regulation and leakage current measurements, limiting of bias voltage current for SiPM protection, a Peltier cooler control, and remote temperature/humidity readout.

Bias voltage and temperature regulation strongly affect the SiPM noise and stability.
In order to limit the variation in the gain of all the detector channels to 2\%, the bias voltage and temperature are strictly controlled, with several key parameters shown in Table~\ref{tab:hcal:sipmcontrol}.
The measured in situ temperature stability was 0.3\deC.

\begin{table}[!ht]
\centering
\topcaption{%
    HCAL SiPM requirements for the \Phase1 upgrade.
    The design (measured) parameter values are shown.
    For the  design (measured) value of the BV stability, the RMS (peak-to-peak) is quoted.
}
\label{tab:hcal:sipmcontrol}
\renewcommand\arraystretch{1.1}
\begin{tabular}{lcc}
    Parameter & HB & HE\\ \hline
    Channel number per RM & 48 (48) & 64 (64) \\
    Bias voltage [V] & 65--75 (0--80) & 65--70 (0--80) \\
    BV resolution [mV] & 40 (20) & 40 (20) \\
    BV stability [mV] & 40 (26) & 40 (26) \\
    BV ripple/noise (RMS above 10\kHz) [mV] & 7 (3--7) & 7 (3--7) \\
    Temperature monitoring [$\deC$] & 0.25 (0.1) & 0.25 (0.1) \\
    Max operating current [$\mu$A] & 500 (500) & 1000 (1000) \\
    BV current resolution [nA] & 122 (122) & 244 (244) \\
    Total ionizing dose [krad] & 3.1 (17) & 0.9 (17) \\
    1\MeV-equivalent neutron fluence [cm$^{-2}$] & $9\times10^{11}$ ($2\times10^{12}$) & $1.1\times10^{12}$ ($2\times10^{12}$) \\
    Power for Peltier cooler & ($6.5\unit{V}\times1\unit{A}$) & ($6.5\unit{V}\times2\unit{A}$) \\
\end{tabular}
\end{table}

\subsubsection{Optical decoder unit}

The optical decoder unit (ODU) is the interface between the clear optical fibers bringing light signals from the HCAL scintillator and the SiPMs that convert the light into electrical signals.
Each ODU contains a network of short (about 10\cm) clear plastic fibers that remap the planar geometry of the incoming light signals from the megatiles into a tower geometry that is more appropriate and useful for physics analysis.
While the size, shape, and some details are different for the HE and HB ODUs, the basic features are the same.
Each RM in HB and HE contains one ODU, for a total of 144 ODUs each in HB and HE.

\begin{figure}[!ht]
\centering
\includegraphics[width=0.533\textwidth]{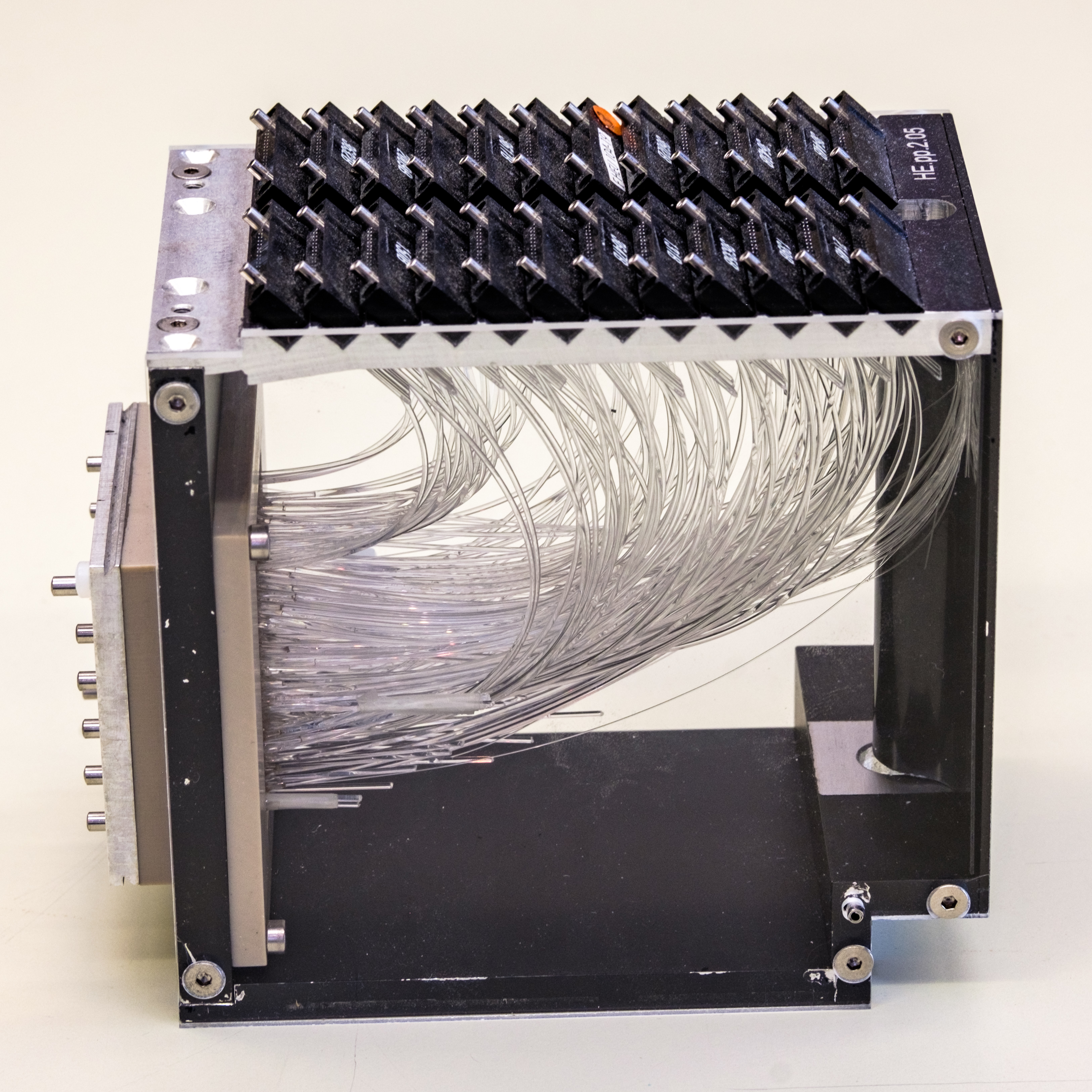}%
\hfill%
\includegraphics[width=0.4\textwidth]{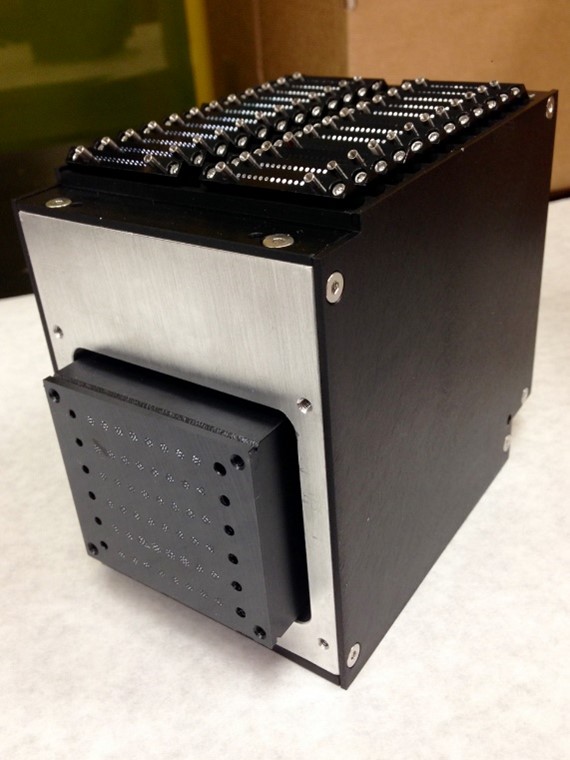}
\caption{%
    Left:\ view of a spare HE optical decoder unit (ODU), showing its light box and fiber mapping.
    The fibers route from the patch panel at the top to the ``cookie'' at the left.
    The side panels are clear rather than opaque for display purposes.
    Right:\ a production HE ODU.
    The clear fiber pigtail connectors attached to the patch panel are visible at the top.
    The plastic ``cookie'' is seen at the front of the ODU.
}
\label{fig:hcal:odu}
\end{figure}

The ODU is essentially a box containing the network of fibers described above.
The fibers are fabricated into cables with polished connectors at each end, which mate with the fiber connectors from the detector.
The cables are cut in half to create ``pigtails'', which form the basic units for assembly into the ODU.
Each HB (HE) pigtail has up to 18 (12) fibers.
The routing and connections of the fibers are shown in Fig.~\ref{fig:hcal:odu} (left).
At the top of the picture, the pigtail connectors are attached through open slots in a machined aluminum patch panel, where they connect one-to-one to the fibers coming from the HCAL detector layers.
At the left of the picture, the fibers are mapped to their designated holes in the so-called ``cookie,'' a precisely machined piece of plastic containing 64 (48) holes for the HB (HE) which directs the light to individual SiPMs.
The HE cookie is a solid piece of polyether ether ketone plastic, while the HB cookie has two plastic layers with an insulating foam layer in the middle to prevent heat from leaking to the SiPMs.
The number of fibers in each hole varies from one to seven, depending on the depth segment being read out.
The fibers are glued in place in the cookie, which is optically finished with a diamond flycutter.
All the precision-machined parts were fabricated in local industry, while the ODU assembly, inspection, and testing were done by CMS.
A completely assembled production HE ODU is shown in Fig.~\ref{fig:hcal:odu} (right).

Extensive quality control testing of both the individual fibers and the fully assembled ODUs was performed to ensure the integrity of the light transmission.
Each cable was tested by connecting it to a similar cable of wavelength-shifter fibers that were illuminated individually by LEDs.
The light throughput of each fiber in the cable was measured with a photodiode.
To make sure there were no problems with the connectors at either end, the cable was reversed, reconnected, and measured again.
To test the fully assembled ODUs, light was injected into the ODU through the connectors in the patch panel, fiber by fiber, and measured by a large photodiode at the face of the cookie.
A total of 144 fully qualified ODUs were installed into both the HB and HE.

\subsubsection{Frontend readout card}

The signals from the SiPMs are integrated and digitized by the QIE11 ASICs~\cite{Zimmerman:2003gv, Baumbaugh:2014eya, Roy:2015bia, Hare:2016swd}.
The QIE11 integrates negative input charge pulses in 25\ns buckets and digitizes the result with approximately constant resolution over a large dynamic range.
The QIE11 achieves an effective 17-bit dynamic range with approximately 1\% resolution with only 8 bits through the use of a custom floating point analog-to-digital converter (ADC) with pseudo-logarithmic response.
The ADC has a programmable gain via user-configurable input current shunts and a low input impedance ($<$15\unit{W}) that is suitable for use with SiPMs.
A 6-bit time-to-digital converter (TDC) is also included on the QIE11 chip, consisting of a programmable-threshold discriminator which detects arrival time of the input pulse in 0.5\ns bins.
The phasing of the charge integration window relative to the input clock can be adjusted by the user in 0.5\ns steps over the whole 25\ns window.
The nominal sensitivity of QIE11 is 3.1\fC per count at the low end (with no programmable input shunt selected).
The gain of any particular chip can vary from this nominal value due to process variations, so all chips are calibrated with better than 1\% precision.
The nominal maximum charge that can be digitized is approximately 350\unit{pC}, yielding the nearly 17-bit dynamic range.

Each readout card in the HE (HB) supports 12 (16) SiPM channels and contains a corresponding number of QIE11 ASICs.
In addition, each card contains one Microsemi ProASIC3L FPGA that acts as an I2C bridge between the slow-control unit (ngCCM) and the frontend chips, one or two Microsemi IGLOO2 FPGAs that serialize and format the data, and one VTTx~\cite{Vasey:2012xjw} module providing two 4.8\Gbs optical links.
The necessary power for operating the frontend readout card is supplied by the CERN FEASTMP \DCDC converters, which are radiation and magnetic field tolerant.

Digital data from the QIE chips are serialized and formatted by two (one) Microsemi IGLOO2 FPGAs in the HB (HE).
The flash-based IGLOO2 FPGA achieves sufficient radiation tolerance for the HCAL frontend, better than typical SRAM-based FPGA technologies.
In the HE, there is sufficient bandwidth to transmit all 8 bits of ADC and 6 bits of TDC data per channel per bunch crossing; however, in the HB, bandwidth constraints require the reduction of TDC information to two bits, which are used to encode four arrival time scenarios:\ prompt, slightly delayed, and significantly delayed times of arrival, plus the case where no valid pulse is present.
After formatting by the IGLOO2, data from each QIE card are transmitted to off-detector backend electronics via 5\Gbs VTTx optical links.

\subsubsection{Slow and fast-control systems}

The control systems for the HB, HE, and HF were updated to support the upgraded frontend electronics.
For the \Phase1 upgrade of the frontend, the fast-control system, synchronized with the LHC clock, and the slow-control system (also referred to as the DCS), which runs at a much slower frequency than the LHC clock, are handled together within the same hardware.
The fast-control system delivers the LHC clock with a maximum jitter of 1\ns, sends the orbit synchronization signal, delivers the so-called ``warning test enable'' (WTE) signal for the recording of calibration data, provides reset capabilities, and provides fast monitoring.

These requirements are similar for the HB, HE, and HF with a few exceptions:\ the HF has PMTs instead of SiPMs, and these do not need to receive configuration commands; the HB and HE have a secondary, redundant control link that can replace the primary control link; and each subsystem has a different number of channels and cards.
As a result, a similar architecture is used for the HB, HE, and HF control systems, although the physical realization of the hardware is different.
The redundant-link scheme has changed from what was described in the CMS HCAL \Phase1 Upgrade TDR~\cite{CMS:TDR-010}.
The redundant link was removed from the HF because it is relatively accessible.
In HB and HE, the hardware configuration has been simplified thanks to the installation of more optical fibers.
The redundant HE control link was successfully used during the last year of \Run2, when a primary control link failed due to a malfunction of its optical transmitter~\cite{Cummings:2022kgm}.
For \Run3, the control link with the highest optical power between primary and secondary is used.

\begin{figure}[!ht]
\centering
\includegraphics[width=0.7\textwidth]{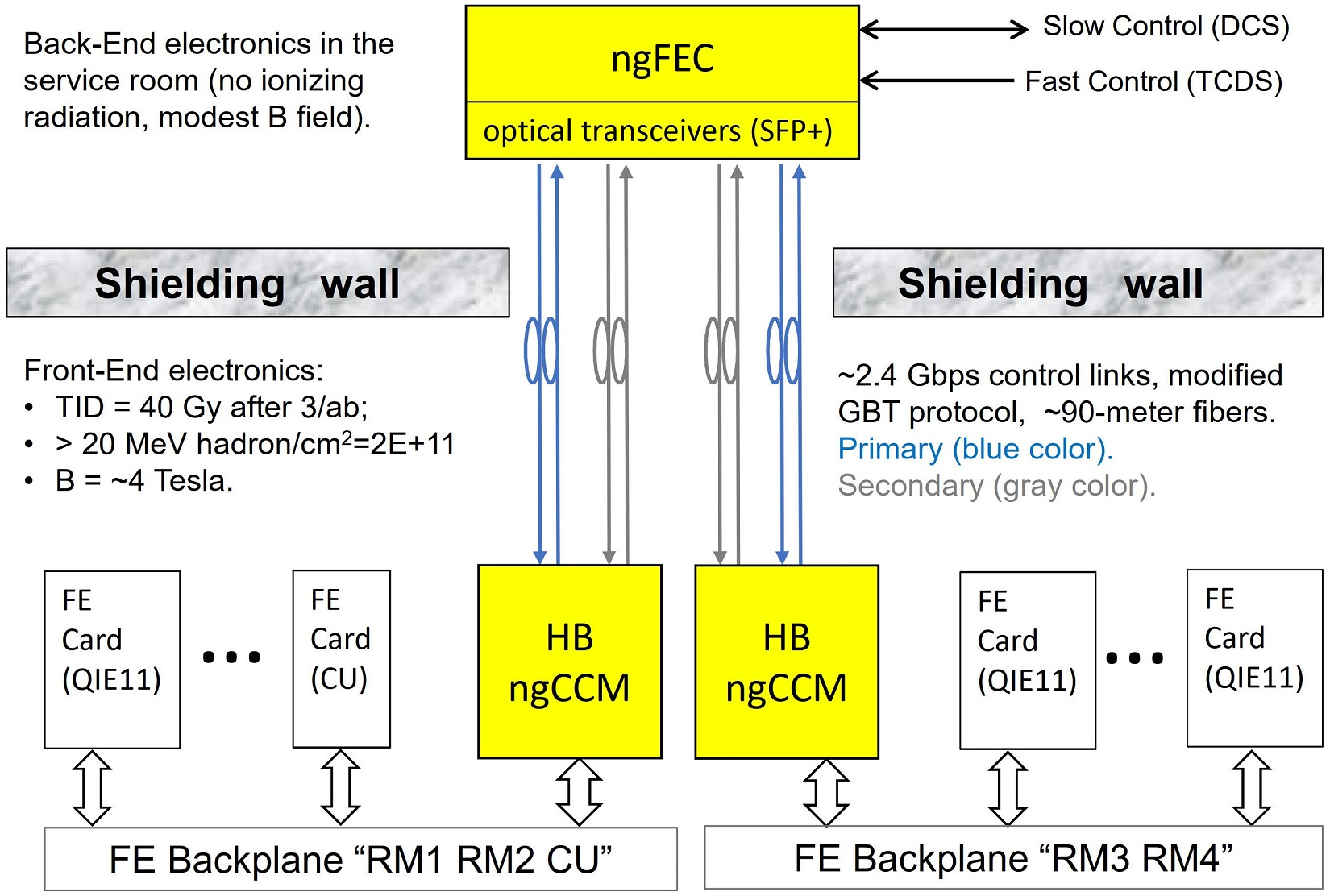}
\caption{%
    Block diagram of the HB controls.
    The ngFEC and ngCCM modules are needed to run and monitor the frontend electronics.
    All control links connecting the service rooms to the experimental cavern are optical.
    A secondary link connecting the ngFEC and the ngCCM is available in case of a primary link failure.
}
\label{fig:hcal:slowcontrols}
\end{figure}

The hardware includes the ngCCM modules in the frontend, the ngFEC modules in the backend, and pairs of optical fibers linking the modules (each pair supports bidirectional serial communication), as shown in Fig.~\ref{fig:hcal:slowcontrols}.
The ngFEC module is a \uTCA baseboard with mezzanine cards and pluggable optical transceivers (SFP+), and is based on the FC7 Kintex 7 FPGA AMC board~\cite{Pesaresi:2015qfa}.
Its main functions are to receive TCDS signals from the \uTCA backplane, provide an interface to the DCS computers, merge fast control and slow control over the same bidirectional link used to communicate with the ngCCM, maintain a fixed latency for the fast-control signals across power cycles and ngFEC optical ports, make use of the ngCCM redundant scheme, and support up to twelve bidirectional links.

The ngCCM is connected electrically to the rest of the frontend over backplanes.
The optical links run a modified scheme of the CERN-developed 4.8\Gbs GBT communication protocol~\cite{Vasey:2012xjw, Moreira:2010cjn}.
The GBT protocol was tailored for the GBTx frontend ASIC; however, the GBTx was not available in time for the original HCAL \Phase1 schedule.
In the ngCCM, the Microsemi IGLOO2 FPGA was used as a replacement for the GBTx.
The performance of the dynamic characteristics of the IGLOO2 FPGA are marginal at 4.8\Gbs; for CMS purposes, to allow an ample safety factor, it was decided to modify the protocol and run it at half speed.
The reduced bandwidth is still sufficient for all control requirements of the HCAL frontend.
The \Phase1 upgrade of the HCAL frontend and backend was successfully completed in October 2019.

\subsubsection{Backend electronics: readout}
\label{sec:hcal:backend}

During \Run2, the HCAL backend electronics were upgraded from VME-based to components based on the \uTCA architecture.
This replacement was the first step in the \Phase1 upgrade, and the \uTCA-based electronics have operated with a mixture of the original and upgraded frontend throughout \Run2.
The design of the backend is described in detail in Ref.~\cite{CMS:TDR-010}.

In \Run3, the backend and frontend host a much-increased channel count and data volume.
With the completion of the \Phase1 upgrade, the number of HB (HE) channels is increased by roughly a factor of 3.5 (2.6), and each channel has a larger number of bits from the upgraded QIE10/11 ADC and TDC.
This increase required efficient algorithmic development to calculate the trigger primitives within latency constraints and with the available resources of the \uHTR FPGA.
This firmware was achieved, along with the enhancements to the trigger primitive calculations discussed in Section~\ref{sec:hcal:trigger}.

The HCAL data acquisition system is managed by an online software suite that controls, configures, and monitors the HCAL electronics.
Development of the HCAL online software began in 2005, and it continues to evolve significantly; for example, new components were recently developed to support the upgraded frontend and backend electronics.
The software has two main layers that provide complementary functionality.
The first layer is composed of custom software written in C++ using the XDAQ~\cite{Gutleber:2002wc} framework for distributed data acquisition.
For communication with the \uTCA electronics, the IPBus architecture~\cite{Williams:2014cds} is used within XDAQ.
Configuration parameters are provided in human-readable format by a version-controlled system, and larger hardware parameter sets (\eg, SiPM bias voltages, phase delays, etc.) are read from databases.
The second software layer allows for coordination between different parts of the HCAL, and with other CMS subsystems.
It is written in Java, and based on the RCMS framework~\cite{Brigljevic:2003kv}.
The XDAQ and RCMS frameworks are described in detail in Section~\ref{sec:daq:software}.

The frontend electronics are controlled and configured by the ngCCM server, a separate program based on the C++ actor framework~\cite{Charousset:2016clss}.
It detects and corrects for single-event upsets automatically, and stabilizes the SiPM temperatures by controlling the Peltier voltages.
The ngCCM server communicates via IPBus with special command processors implemented in the ngFEC firmware for each I2C and JTAG channel in the frontend backplanes.
The XDAQ-based software communicates with the ngCCM server via JSONRPC over WebSocket~\cite{JSONRPC:2013web, Fette:2011rfc} to control, configure, and monitor the frontend electronics.
The HCAL DCS WinCC-OA communicates with the ngCCM server via a raw socket protocol to obtain voltage, current, temperature, and humidity readings and set the Peltier target temperature.
A command-line interface to the ngCCM server additionally allows updates to be uploaded to the frontend FPGAs.

\subsubsection{Voltage source upgrades}

The \Run1 HCAL frontend electronics power system was based on the CAEN Easy infrastructure.
The RBX electronics were fed with two power lines, nominally at 6.5 and 5\unit{V}, requiring a total power of around 90\unit{W} per RBX.
The HPD high voltage (HV) and bias voltage (BV) and the HF PMT HV were provided using power supplies custom made in Bulgaria.
The HB, HE, and HO RBXs were powered using CAEN A3016 power supplies, each module having six power channels capable of powering three RBXs.
The HF readout crates were powered using four CAEN A3100 modules for each HF module (positive and negative $\eta$).
Successive photodetector, frontend, and backend upgrades required some changes to the powering system.

The HO photodetector upgrade, with the replacement of HPD the sensors with SiPMs, required new power supplies, since the new frontend generates the BV internally from the low voltages (LV).
No change was required for the LV system, since the A3016 modules were able to provide the extra power demanded by the SiPMs.

The HF LV system for the new frontend electronics also required some changes.
Unlike the old electronics, the new electronics only use one supply voltage, in the range 8--10\unit{V}, but with more power.
During \Run2, the power was provided by eight A3100 units operating at 8\unit{V}, four each of the two HF detectors.
Each of the sets was in an CAEN EASY crate fed with an individual A3486 power converter.
However, a problem was observed during operation, correlated with instantaneous luminosity.
The power supplies, located in the HF racks near the detector, were subject to high levels of radiation, causing occasional single-event upsets (SEUs), \ie, incidents in which the control of the modules was lost, leading to the loss of power in the frontend electronics.
This resulted in the loss of about 150\pbinv of data in 2017.
For 2018, a mitigation protocol was developed, consisting of a fast software detection, reset, and recovery of the power supplies, followed by a reconfiguration of the frontend electronics.
The protocol reduced the time of data loss to typically 1--2 minutes, and reduced the total data loss to about 50\pbinv.
The rate of SEUs scaled linearly with instantaneous luminosity, with about 20 SEU events in 2018; further, the power cycling was expected to shorten the lifetime of the power supplies, and the high level of irradiation made maintenance prohibitively difficult.
Hence a more robust solution was implemented in LS2: the power supplies were moved to an area under CMS where radiation is minimal.
Long cables from this location lead the power to the frontends.
To compensate for the voltage drop on the cables, the A3100 power supplies have been replaced by A3100HBP units, described below, which can provide a high enough voltage to ensure the frontends receive the nominal operational voltage of 8\unit{V}.
Following these changes, no SEU incidents were observed during the first year of \Run3.

The HE and HB upgrades, with the increase in the number of SiPM readout channels, led to more changes to the powering system.
The HB and HE custom made power supplies were replaced by Keysight N6700C mainframes with N6736B modules.
Twenty N6700C units currently provide the BV for the HB and HE SiPMs.
The HE A3016 modules were replaced by A3100HBP units.
The CAEN A3100HBP model provides one channel with 8--14\unit{V}, 50\unit{A}, and 600\unit{W} output and can power up to two HE RBXs in parallel.
These units are also used in the HF, with ten units operating at 10\unit{V} used in each half of the HF.
A transient voltage spike in one of the HE power supplies, caused by a power cut in June 2018, damaged the inputs of two RBXs.
This affected HEM15 and HEM16, which cover 2\% of the total HCAL acceptance and 3\% of the HB+HE.
After identifying the cause of the problem, CAEN introduced a modification to the power supplies to prevent such transients from occurring again.
Independently, the HCAL Collaboration also developed an external circuit to suppress such transients.
Currently, all the HE RBXs are protected with both the CAEN modification and the external overvoltage protection circuit.

The extra readout channels of the HB SiPMs required more power than what was practical with the A3100HBP units, and Wiener Marathons (PL 508 with five channels 5--15\unit{V}/40\unit{A}) were chosen as replacements.
Eight Marathons are used to power the entire HB frontend electronics.
To prevent damage due to transients, overvoltage protection circuits similar to the ones for the HE are installed on each of the Marathon output channels.
New LV cables between the supplies and the RBXs were also installed to keep the voltage drop coming from the additional power within the safe operating range of the frontend electronics.

\subsubsection{Photodetector and system calibration instrumentation}

Changes in the light yield are expected due to aging of the scintillators, radiation damage, and variations in the photodetector response.
The HCAL calibration and monitoring systems are designed to determine the absolute energy scale, to monitor the calorimeter system for changes during the lifetime of the detector, and to derive the energy scale correction factors.
In particular, two complementary methods are used for monitoring.
The first method consists of a movable radioactive wire source, with source tubes installed in every megatile in such a way that the tubes cross all of its tiles.
Measurements for all tiles can be made during long LHC shutdowns to validate the readout-channel mapping.
This system was used during the \Phase1 upgrade to provide a relative calibration when the SiPMs replaced the HPDs.
The second method is based on light injection from two sources, a UV laser and an LED system.
The UV laser light is injected into two layers of each wedge to monitor radiation damage to the scintillator, and both the UV laser and LED light is injected into the optical decoder box that has the photodetectors, to monitor damage to the SiPMs.
A calibration module (CM) inside the frontend RBX is responsible for delivering the UV laser and LED light to the photodetectors.
The CM was redesigned as part of the \Phase1 upgrade.

Each RBX contains one CM, which includes an LED pulser for the SiPMs, as well as a laser light distribution system.
Fibers from the CM are connected to each of the interstitial microfibers in the light mixers, allowing for the LED or laser signals to be delivered to the SiPMs.
The CMs also contain pin diodes which directly measure the amplitude of the LED or laser signals, providing a reference value for the calibration.
The design of the LED pulser board is the same as that of the original HB/HE calibration module.
The controls for the pulser, however, are significantly enhanced.
In particular, the CMS global WTE signal is used by the CM to generate LED pulses during the LHC orbit gap.
The LED pulse can be positioned between 2 and 65\,535 triggering clock synchronization counts (MCLK) from a resetting WTE signal.
This pulse can be delayed relative to the MCLK signal from 0 to 25\ns in 0.5\ns increments.
The pulse amplitude and width are programmable from about 85\mV to 4.75\unit{V} and 0.5\ns to 25\ns, respectively.
The pulser board contains six TE Connectivity AMP 147323-1 connectors providing a total current of 0.75\unit{A} at 5.5\unit{V} DC and a ground for powering the PIN diodes.
The pulser board receives communication and MCLK from the ngCCM through the backplane header.

\subsubsection{HF upgrade}
\label{sec:hcal:hf}

\begin{figure}[!ht]
\centering
\includegraphics[width=0.55\textwidth]{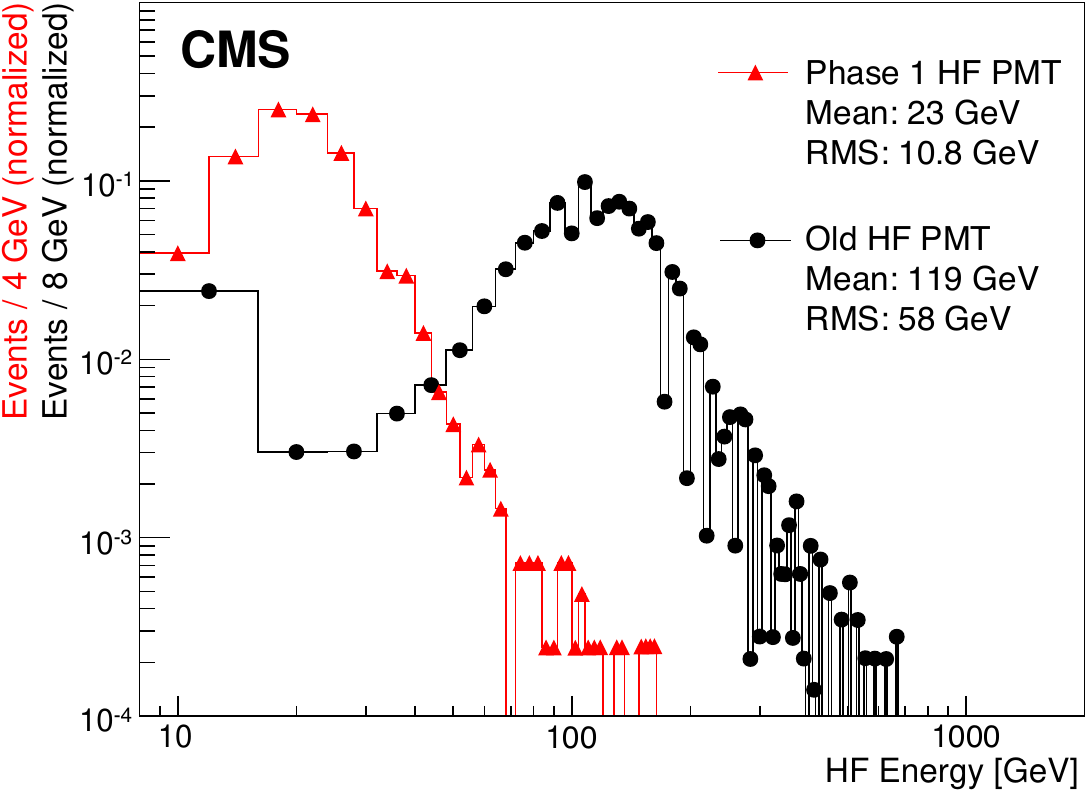}
\caption{%
    Cherenkov signals generated in the PMT windows from a muon test beam.
    Thin windows in the \Phase1 upgrade four-anode PMTs produce smaller signals (red) than those produced in the original PMTs with thick windows (black).
}
\label{fig:hcal:hfcherenkov}
\end{figure}

When shower particles pass through the HF quartz fibers, Cherenkov radiation is produced.
However, anomalous signals can also be produced by muons from \pp collisions or beam halo interactions whose trajectories pass in the vicinity of the RBXs.
Relativistic muons generate Cherenkov radiation when they pass through the glass window of the PMTs.
Since the sampling correction for the HF calorimeter is large, these ``window events'' result in very large signals.
Signals produced by muons can be distinguished from shower light through timing:\ the shower signal is delayed relative to the muon signal due to the longer path length and the lower speed of light in quartz.
The PMTs used in the original HF detector were Hamamatsu Corporation R7525s, with a glass window that is 1.2\mm thick at the center, increasing to 6\mm at the edge.
The HF PMT upgrade replaced these with Hamamatsu R7600-M4s.
The new PMTs have thinner windows (1\mm of UV glass) and four anodes arranged in a 2$\times$2 grid.
The light from the fibers of a given ieta-iphi tower is spread out over the face of the PMT and generates signals on all four anodes; the diagonal anodes in the 2$\times$2 grid are grouped together and read out by a single QIE10 chip.
The resulting dual-anode readout allows for the discrimination of signals from muons interacting directly with the PMT, which will typically produce a signal concentrated on a single anode closest to the impact point.
The amplitude of signals from muon window events between the old and new PMTs is compared in Fig.~\ref{fig:hcal:hfcherenkov}.

The new PMTs were extensively tested in several beam tests~\cite{Bilki:2011roa, CMS:NOTE-2010-003} and in CMS during data taking.
Algorithms utilizing the signal imbalance between the anodes were proposed and tested during the beam tests.
However, a new readout system was needed to take advantage of the additional channels.
To be cost-effective, frontend readout cards were redesigned for two-channel readout of the four-anode PMTs (where each readout channel sums two anodes).
The backend was also changed from the old VME-based to the new \uTCA-based system.

The new frontend cards included the upgraded QIE10 ASIC, which is similar to the QIE11, except for having a constant 20\Ohm input impedance and no programmable current shunting.
As described earlier, the QIE10 chips also have a TDC to measure the arrival time of the signals.
Each frontend readout card for HF contains 24 QIE10 ASICs, each digitizing a single dual-anode PMT channel, as well as two Microsemi IGLOO2 FPGAs and three VTTx modules providing six 4.8\Gbs optical links.
Similar to the upgraded electronics in HB and HE, the power is supplied by CERN FEASTMP \DCDC converters.
In addition to the replacement of the PMTs and the readout electronics, improvement in data transfer was planned to handle the increased load due to the TDC and two-channel readout.

The HF calorimeter is in a high-radiation area, leading to doses of 1000\Gy in fibers near the beam pipe.
To monitor radiation damage, some of the quartz fibers are equipped with a special laser system that allows both the incident laser light intensity and the intensity of the laser light after traversing the quartz fiber to be measured with the same PMT.
In \Run1, the light was provided by a laser system located in the underground service cavern, which is several meters away from the HF calorimeters.
The light was then split into several fibers before it reached the HF quartz fibers.

\begin{figure}[!ht]
\centering
\includegraphics[width=0.5\textwidth]{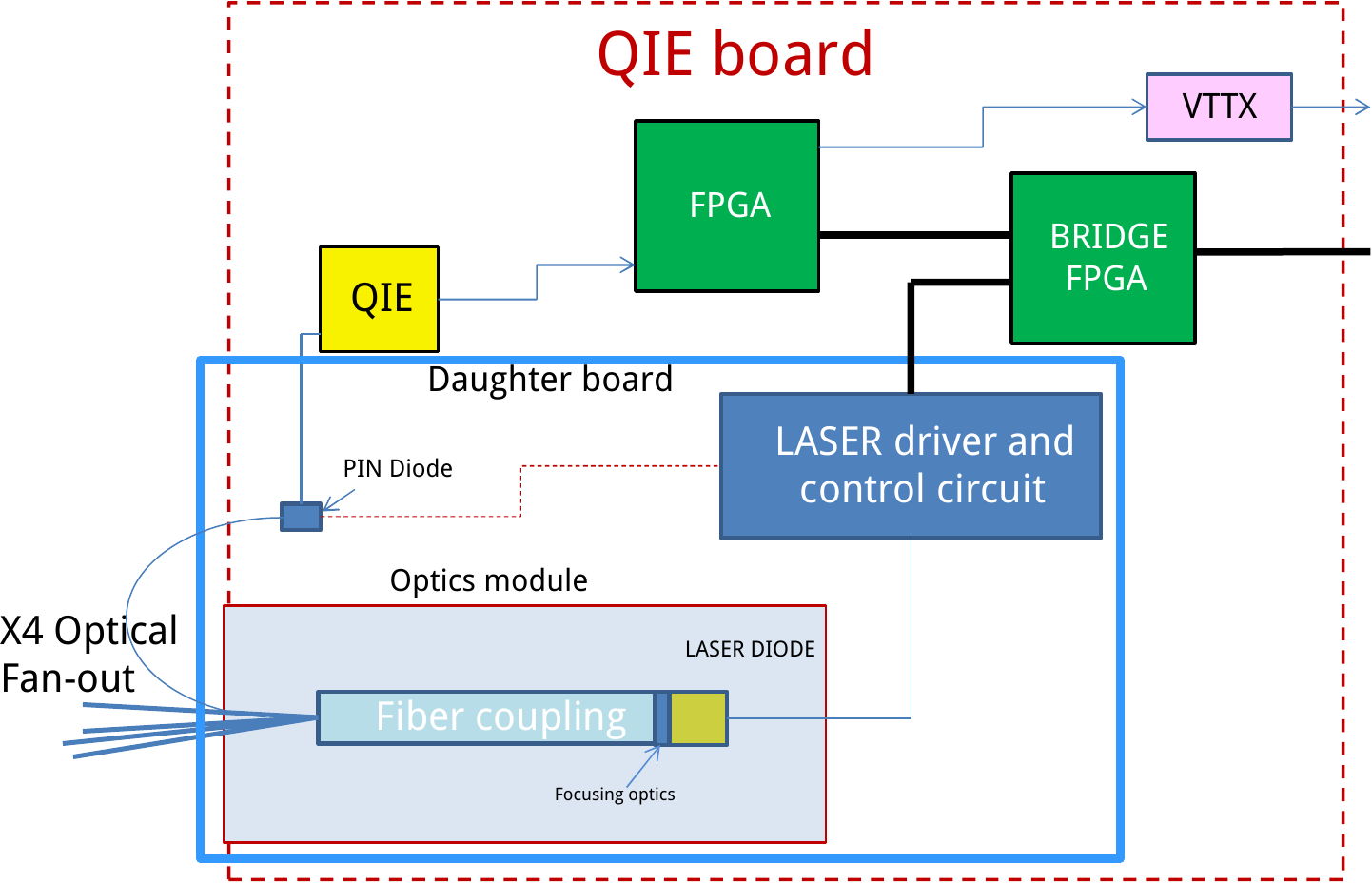}%
\hfill%
\includegraphics[width=0.46\textwidth]{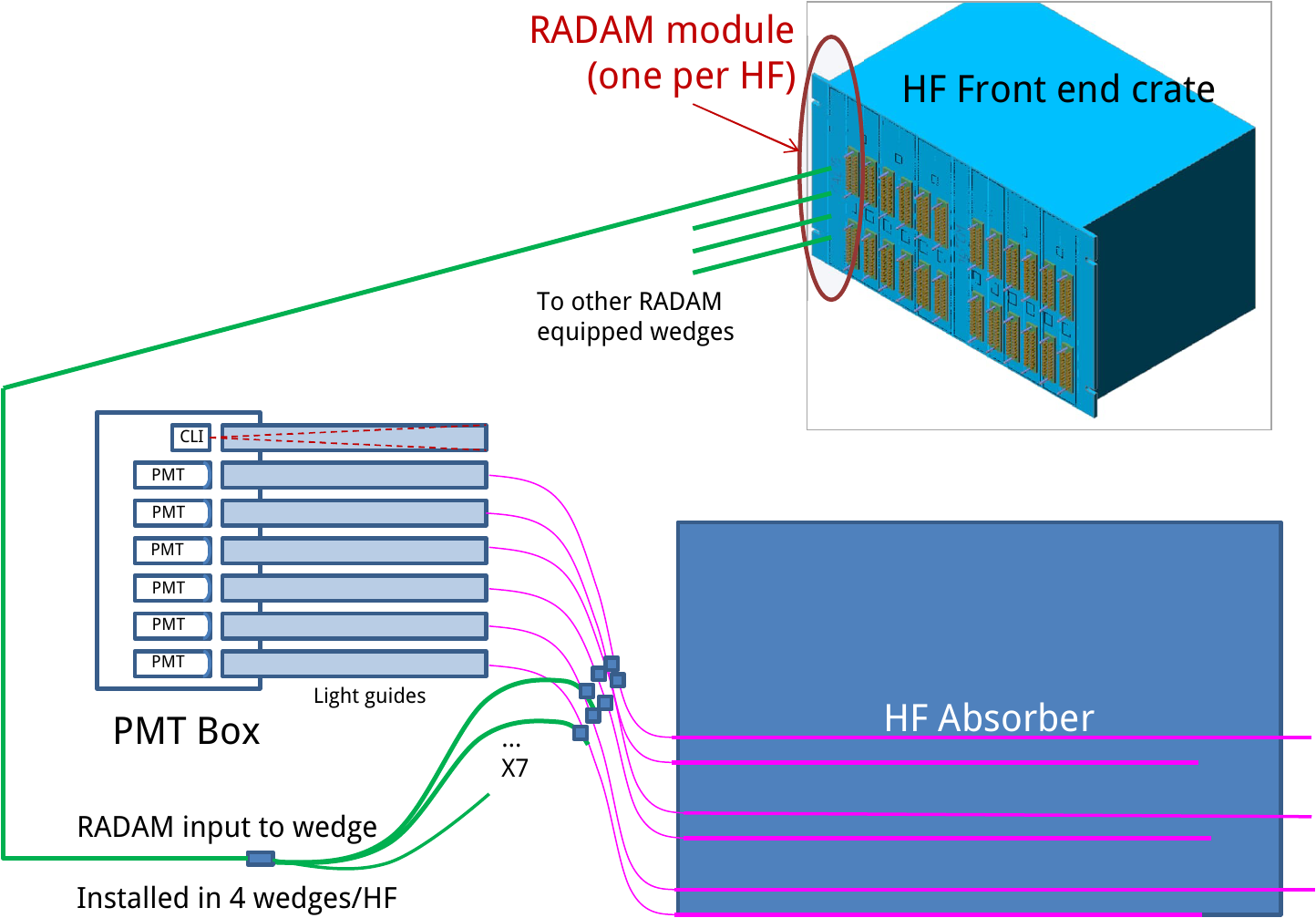}
\caption{%
    Left:\ schematic of the new HF QIE laser board, used in the HF radiation monitoring system.
    Right:\ sketch of the upgraded HF radiation damage monitoring system light distribution and electronics chain.
}
\label{fig:hcal:hfradmon}
\end{figure}

A new laser device was developed that utilized solid state laser diode technology with a wavelength of 450\nm (allowing for the removal of the wavelength shifters used in the old system), and is based on the HF \Phase1 LED calibration unit concept.
The laser diode driver and the control circuit reside on a mezzanine card mounted on the QIE board.
Optics mix the light and distribute it to four output fibers.
The module is installed in the HF frontend crates, which are attached to the HF calorimeters.
Figure~\ref{fig:hcal:hfradmon} (left) shows a schematic view of the laser daughter board as part of the QIE frontend card.
The upgraded HF light distribution system for monitoring radiation damage is shown in Fig.~\ref{fig:hcal:hfradmon} (right).
The settings of the new system are more flexible, since it is an integral part of the upgraded frontend, allowing for, \eg, a 32\ns delay range adjustable in 0.5\ns steps, and a pulse width adjustable from less than 3 to 30\ns full-width at half-maximum.
There is also an on-board PIN diode to independently monitor the performance of the light source.

\subsection{Trigger}
\label{sec:hcal:trigger}

The HCAL trigger is based on the HCAL calorimeter towers, which, in the HB and HE, are composed of projective calorimeter readout channels summed together in depth.
While most readout channels have $\Dphi=5\de$, the HE readout channels with $\abseta>1.74$ have $\Dphi=10\de$.
To form trigger towers with $\Dphi=5\de$, thereby matching the rest of the HCAL, these HE readout channels are split into two in $\phi$, with each unit assigned one-half of the measured energy.
In the HF, trigger towers are formed from six physical HF towers, with energies of the long and short fibers summed.

The HCAL trigger primitives (TPs) are formed for each trigger tower by combining the information from individual calorimeter towers.
Since the HCAL QIE10/11 chips use 8-bit nonlinear ADCs with several ranges, the energy reported by the QIE is first linearized as a 10-bit word via a dedicated look-up table.
The linearization depends on several parameters, including the gain, the conversion from the collected charge in fC to ADC counts, and the pedestal.
After linearization, the various depth samples are summed, and the amplitude of the pulse is estimated using a pulse filtering scheme that subtracts out-of-time pileup.
These sums define the various trigger channels, and a corresponding TP is produced by the \uHTR.
Finally, the trigger tower transverse energies are compressed to the range 0 to 128\GeV with a least significant bit (LSB) of 0.5\GeV, and transmitted to the L1 trigger.

The TP amplitude is reconstructed by taking the sample-of-interest (SOI) bunch crossing as the main contribution to the total signal amplitude.
The measurement from the SOI alone works well for HF, where the pulses are shorter than the 25\ns bunch spacing, but for HB and HE, a significant fraction of the pulse leaks into the subsequent bunch crossing.
In \Run2, the sum of energies in the SOI and SOI+1 was used to reconstruct the TP energy; this method accounts for the signal leakage into SOI+1, but also incorporates the energy from out-of-time pileup interactions in SOI+1.
For \Run3, a new algorithm was developed that instead uses the SOI and SOI-1, \ie, the preceding bunch crossing.
The scheme subtracts a weighted amount of the measurement from SOI-1, mitigating the inherent leakage of the out-of-time pileup from SOI-1 in the SOI.
To account for the signal leakage into SOI+1, a correction factor is derived from the known pulse shapes rather than using the measurement from SOI+1, thereby avoiding incorporating contributions from out-of-time pileup.

In addition to the TP generation in the HB and HE, six feature bits of information are also generated that can be transmitted to the L1 trigger.
These bits facilitate encoding information about:\ (i)~the longitudinal shower profile data for use in calibration, lepton isolation, and identification of minimum ionizing particles, and (ii)~the shower time data constructed from the TDC information available in each constituent channel of the trigger tower.
Configurable look-up tables determine which time windows within the bunch crossing of interest are represented by the available TDC codes.
These time window boundaries have a granularity of 0.5\ns.
Starting in \Run3, a subset of the feature bits is used to flag signals characteristic of exotic long-lived particle decays, using either the TDC timing to mark hits with late arrival times or the shower profile data to mark distinctive energy deposits in the various layers of the HCAL.
These bits are used by the L1 trigger to select hadronic signatures from long-lived particles with decay lengths of 1--2\unit{m} which decay prior to or within the HCAL.
The timing precision that can be achieved is estimated to be within 1--2\ns.
The resolution is dominated by interchannel synchronization uncertainties and shower-by-shower fluctuations, as determined from studies done using highly energetic hadronic showers in 2018 data.

\subsection{System and beam tests}

Beam tests were performed with the upgraded \Phase1 frontend at the H2 beam line of the SPS at CERN.
The upgrade electronics were installed in the 20\de-prototype HE wedge at H2, and were read out by four RMs split between two RBXs.
One RBX and two RMs were replaced with the upgraded electronics, while the remaining ones were left unchanged as a reference.
The system was tested with 150\GeV muons and with pions having energies ranging from 30 to 300\unit{GeV}.
Six time samples were recorded for each event, with the beam timed to arrive in the fourth time sample.
The total charge for an event was taken as the sum of the last three time samples centered around the beam arrival.
The pedestal for each event was estimated from the sum of the first three time samples before the beam trigger and was subtracted from the total charge.

\begin{figure}[!ht]
\centering
\includegraphics[width=0.5\textwidth]{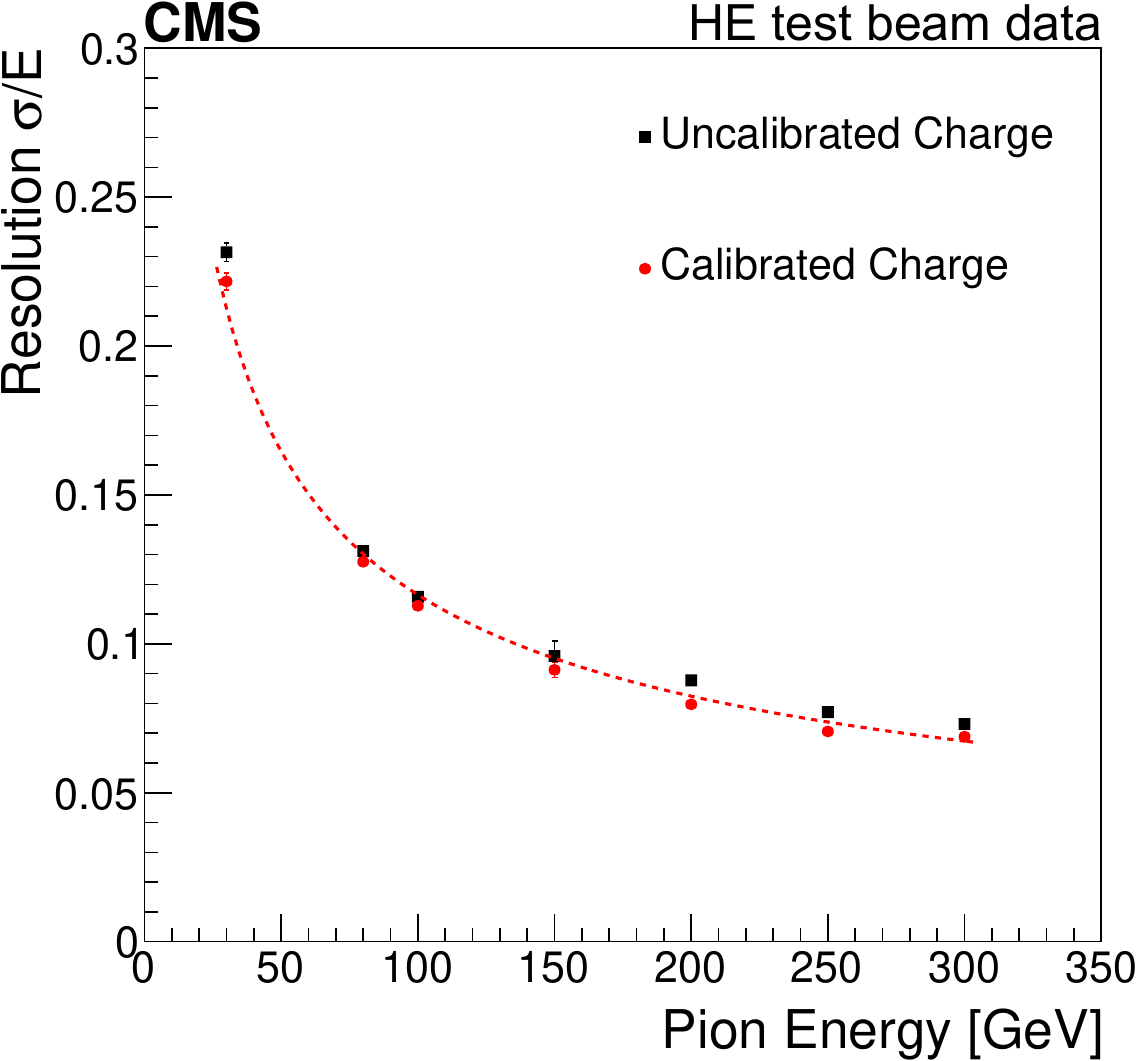}%
\hfill%
\includegraphics[width=0.45\textwidth]{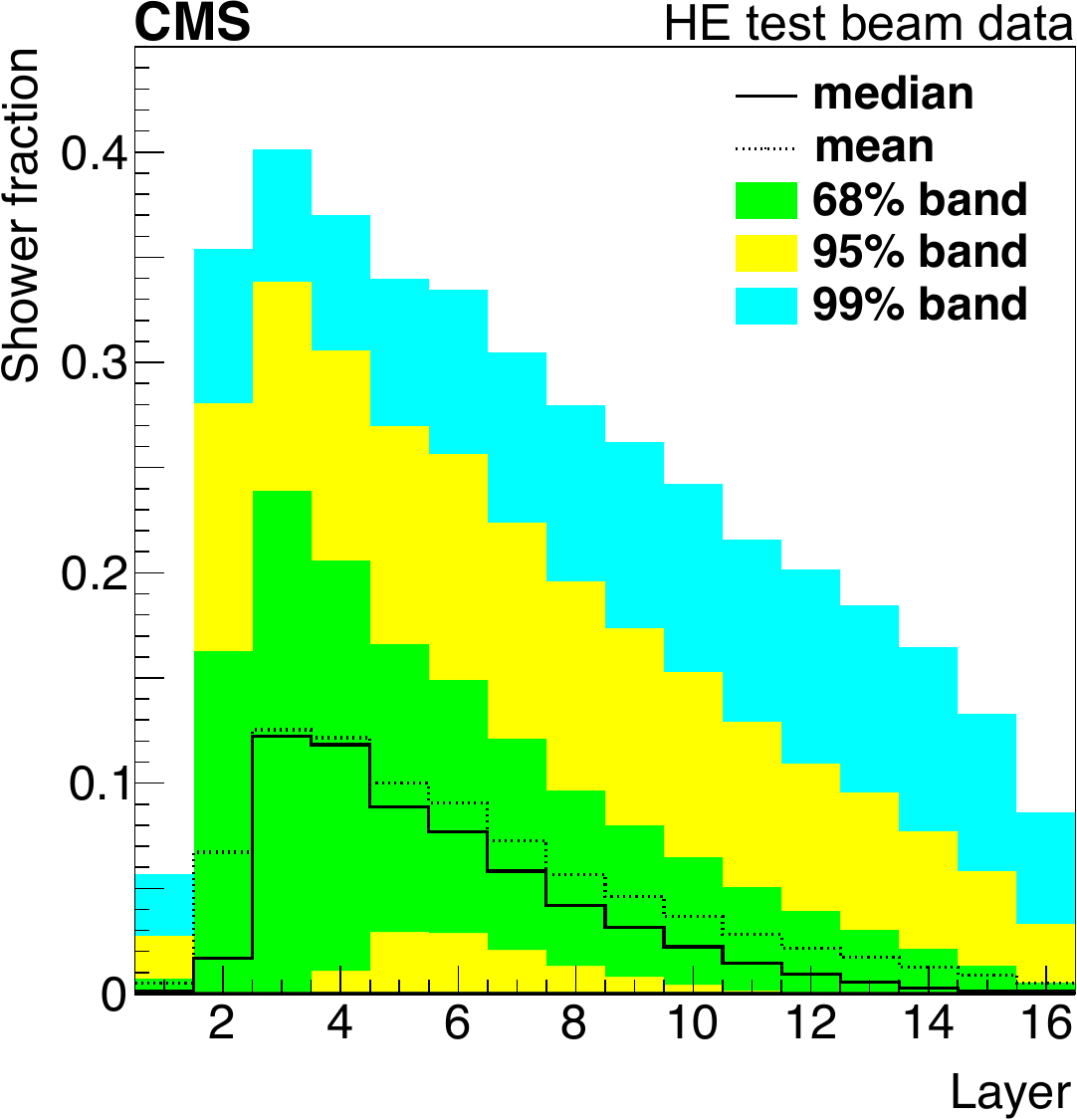}
\caption{%
    Left:\ energy resolution of the upgraded prototype detector as a function of the pion beam energy, shown with and without the channel calibrations derived from the response to muons.
    Right:\ longitudinal shower profile measured using the special ODU.
    Bands containing 68, 95, and 99\% of all events for each layer are shown.
\label{fig:hcal:hebeamtest}
}
\end{figure}

The 2015 beam tests served as the first full test of the entire upgraded readout system and proved the system was functional.
The 2017 test beam with production frontend electronics served to quantify the final performance of the upgraded detector.
Representative results from this data are shown in Fig.~\ref{fig:hcal:hebeamtest}.
Using the muon data, the response of the detector instrumented with SiPMs was measured to be 4.3 photoelectrons per layer.
The pion data were used to derive the energy response and resolution of the upgraded detector.
For pions, the number of photoelectrons per GeV was measured to be 32.1.
The shower energy was measured by taking the sum of all charge collected by each channel in a 3$\times$3 grid of towers around the beam direction.
The relative contribution of each detector channel to the energy sum was adjusted using its response to muons to account for differences in the detector response between channels.
The beam test was also used to make a unique measurement of the shower profile as a function of depth with a special ODU that allowed the individual readout of each layer in one tower.
These studies were particularly important because they measured the maximum energy that can be deposited in any particular layer.

Irradiation tests of the full suite of frontend electronics were performed at the CERN High Energy Accelerator Mixed-Field (CHARM) facility with the goal of determining the expected error rates from SEEs, as well as the expected lifetime performance related to the cumulative edamage from ionizing and nonionizing radiation.
The electronics were subjected to an ionizing dose of 202\Gy, a 1\MeV neutron equivalent fluence of $1.96\times10^{12}\cm^{-2}$, and a $>$20\MeV hadron equivalent fluence of $5.88\times10^{11}\cm^{-2}$.
At the end of the irradiation, the frontend electronics showed no indication of permanent damage, aside from the expected inability to program the FPGAs.
The SEE rates were determined based on bit error counting in the encoded data stream and the pseudo-random bit sequence in the control link.
The combined SEE rate was estimated to be less than 30 per \fbinv of data.
Other upsets requiring a reset or other interventions are estimated to occur at a rate of about one per day.

\subsection{Performance}

\subsubsection{Endcap photodetector performance in \Run2}

\begin{figure}[!p]
\centering
\includegraphics[width=0.46\textwidth]{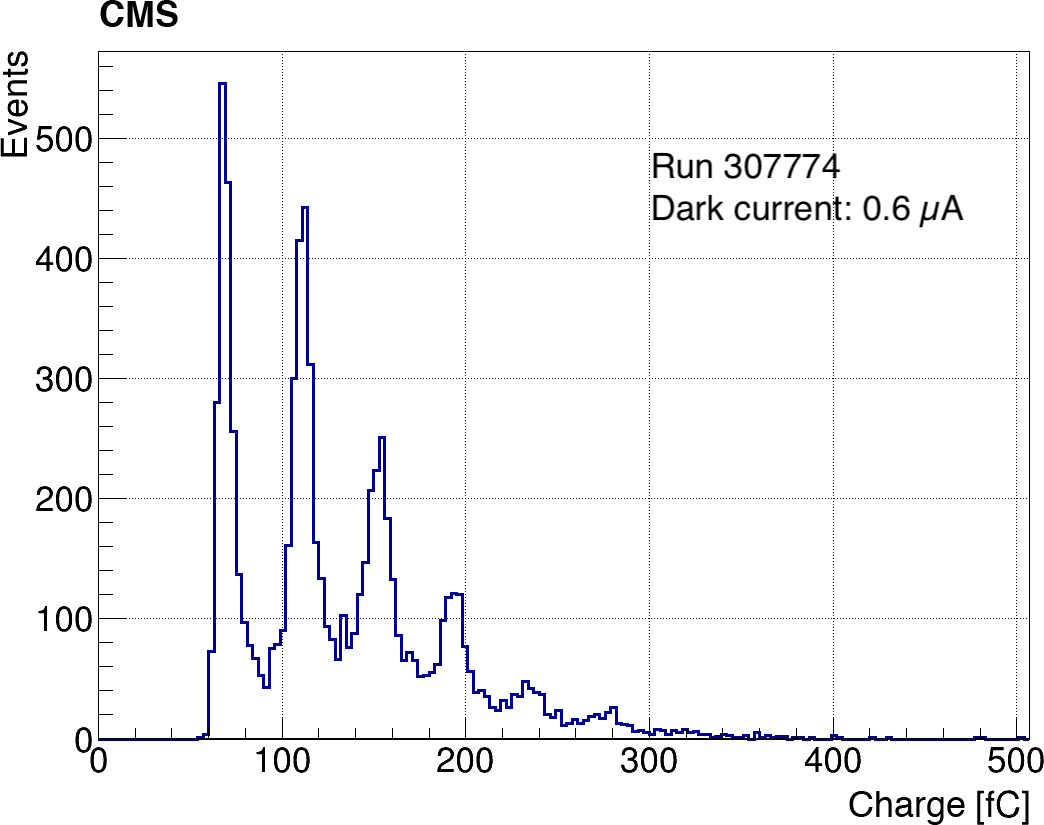}%
\hfill%
\includegraphics[width=0.5\textwidth]{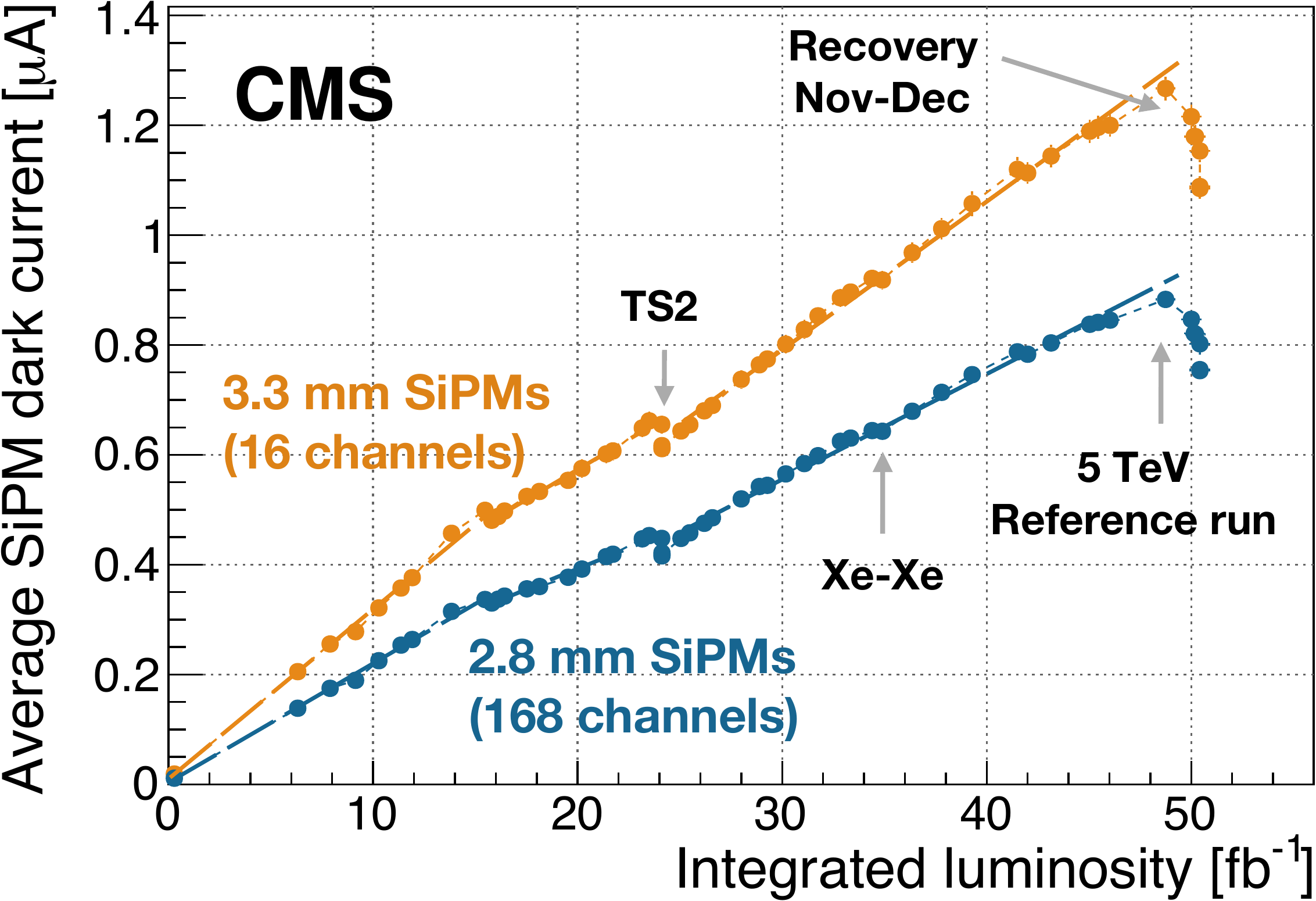}
\caption{%
    Left:\ pedestal distribution for a channel with QIE11 + SiPM readout.
    The charge is integrated in a time window of 100\ns.
    The QIE pedestal and photoelectron peaks are visible.
    Right:\ dark current increase with the integrated luminosity in 2017, where the slope of the fitted line is proportional to the SiPM area.
    The deviation from linear behavior is due to SiPM annealing in the absence of beam and variation in the instantaneous luminosity.
}
\label{fig:hcal:hesipm}
\end{figure}

The photoelectron peaks in the pedestal and low-intensity LED runs, shown in Fig.~\ref{fig:hcal:hesipm} (left), can be used to measure the SiPM gains using a single fit function to the observed multipeak charge spectrum.
The gain was found to be stable during the 2017 data taking at the 1\% level, corresponding to $42\pm1\fC$ per photoelectron, where the uncertainty is the RMS across the 184 channels equipped with SiPMs.
The SiPM dark current was also monitored during data taking, as shown in Fig.~\ref{fig:hcal:hesipm} (right).
The slope of the fitted line is proportional to the SiPM area.
The deviations from linearity are due to SiPM annealing when the beam is off and to variations in instantaneous luminosity.
With this SiPM dark current growth rate, we expect 110\MeV of noise at the end of \Run3 (500\fbinv).

\begin{figure}[!p]
\centering
\includegraphics[width=0.48\textwidth]{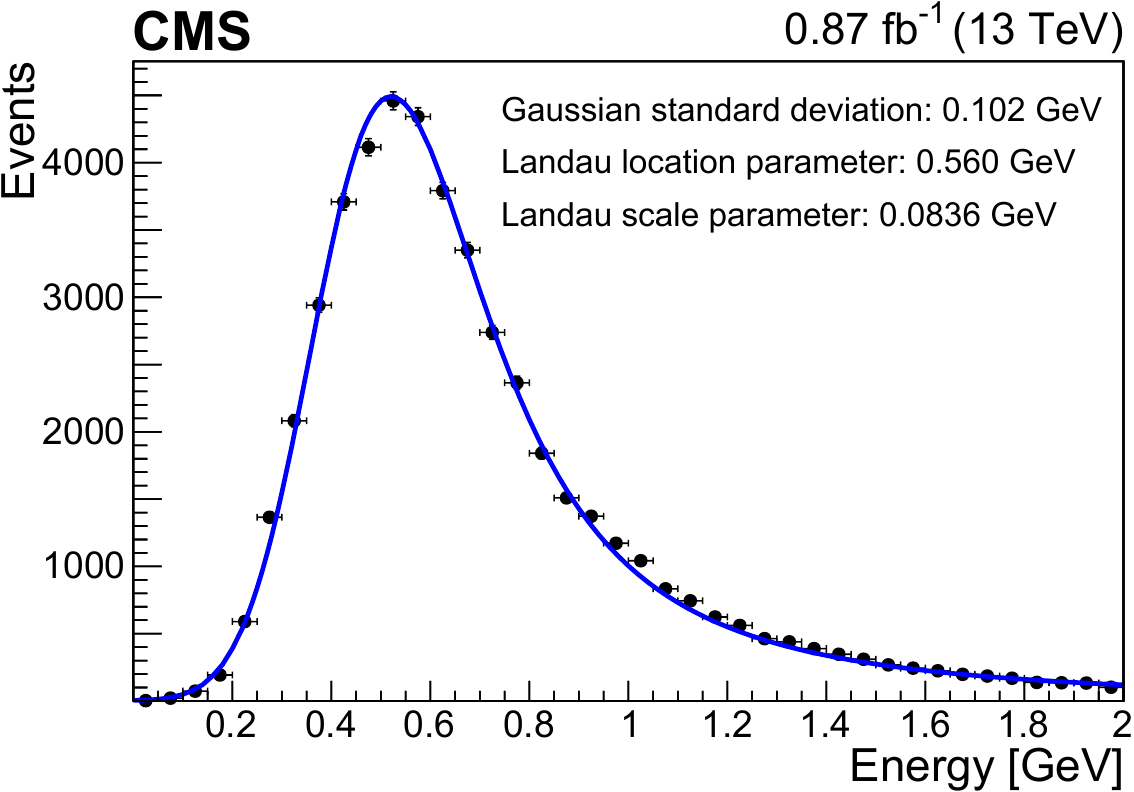}%
\hfill%
\includegraphics[width=0.48\textwidth]{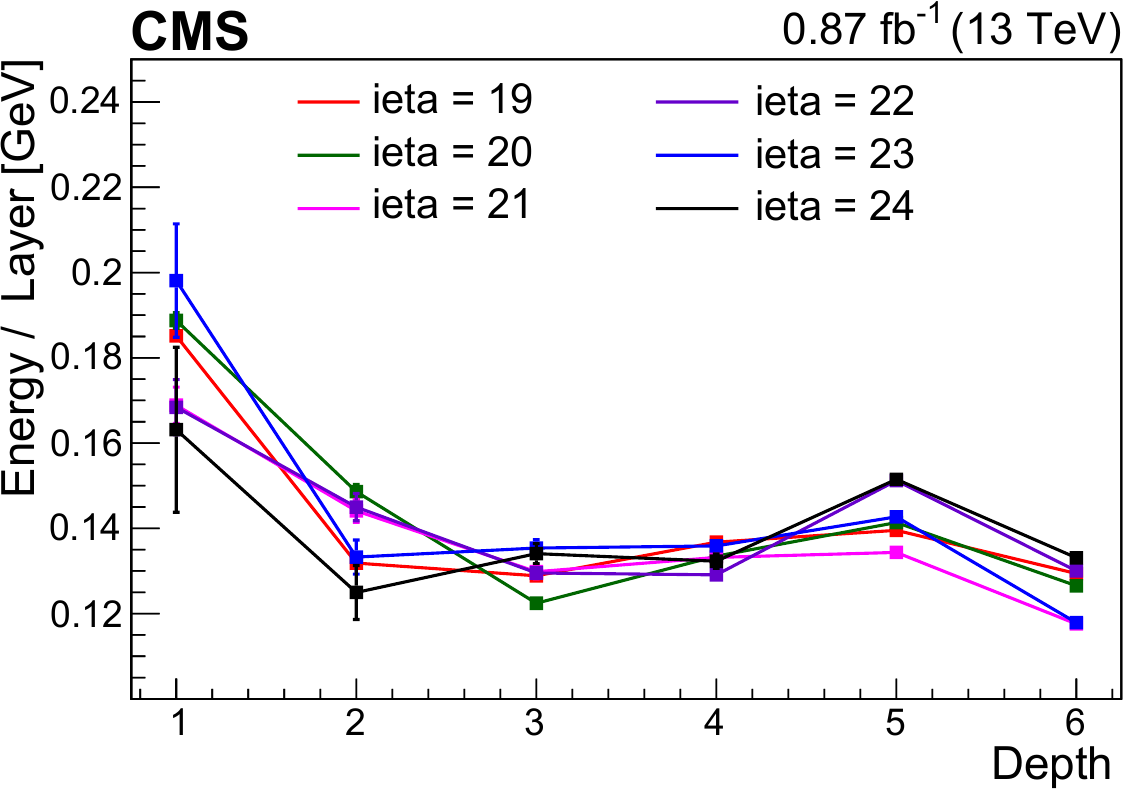}
\caption{%
    Left:\ energy deposit from muons in \pp collision events in the HE tower corresponding to $\text{ieta}=20$ and $\text{depth}=5$.
    The energy spectrum is fitted using the convolution of a Gaussian function with a mean of zero and a Landau distribution.
    The fitting function has three free parameters:\ the Landau location parameter, the Landau scale parameter, and the width of the Gaussian.
    Right:\ the most probable value of the muon energy deposit per layer in HE towers as a function of depth for different $\eta$ regions.
    The vertical bars represent the statistical uncertainty.
    Muons from collision events are considered when their trajectory is contained within a single HCAL tower.
    The muon signal peak is fitted with the convolution of a Gaussian and Landau functions.
    The Landau location parameter is divided by the number of scintillator layers in the considered depth.
}
\label{fig:hcal:hemip}
\end{figure}

In \Run1 and \Run2, the HCAL calibration used $E/p$ from isolated tracks for the energy scale measurement versus $\eta$ and $\phi$-symmetry for equalizing the response along $\phi$~\cite{CMS:PRF-18-001}.
After the \Phase1 upgrade, new calibration techniques are needed because of the increased depth segmentation.
The scintillator and SiPM aging depends on $\eta$ and depth, which necessitates depth intercalibration.
The signals from isolated, minimum-ionizing muons traversing the HCAL provide an excellent probe for interdepth calibration, corresponding to approximately five photoelectrons per layer.
As shown in Fig.~\ref{fig:hcal:hemip} (left), a clear minimum ionizing peak from muons can be observed in collision events
when the muon traverses one and only one HCAL tower.
The energy deposit per layer is shown in Fig.~\ref{fig:hcal:hemip} (right).
The HE detector is homogeneous in depths~2--6, while depth~1 has a thicker and brighter scintillator layer.
Because of this, only depths 2--6 are equalized.

\subsubsection{HCAL performance in \Run2}

For the combined ECAL and HCAL systems, including barrel and endcap, the relative charged pion energy resolution obtained from the test beam can be described as
\begin{linenomath}\begin{equation}
	\frac{\sigma}{E} = \frac{84.7\%}{\sqrt{E}}\oplus 7.6\%,
\end{equation}\end{linenomath}
where $E$ is in \GeVns.
Corrections for the nonlinearity of the calorimetry system due to its noncompensating response to hadronic and electromagnetic energy depositions have been made~\cite{CMS:NOTE-2008-034}.
The time resolution for energy deposits in the HB and HE, calculated by weighting the QIE digitization times by the associated energies, is 1.2\ns~\cite{CMS:CFT-09-018}.
The HF energy resolution from the test beam~\cite{CMS:NOTE-2006-044} is
\begin{linenomath}\begin{equation}
	\frac{\sigma}{E} = \frac{280\%}{\sqrt{E}}\oplus 11\%.
\end{equation}\end{linenomath}
The absolute calibration constants derived from the test beam modules were transferred to the full calorimeter system using an intercalibration from a radioactive \sixtyCo source~\cite{CMS:PRF-18-003}.
The calibration was improved using 13\TeV collision data from 2016, as described in Ref.~\cite{CMS:PRF-18-001}.
From the $\phi$-symmetry of energy flow in minimum-bias events, the HCAL was intercalibrated to within 3\%.
An absolute calibration uncertainty of 2\% was determined using isolated pions with track momenta between 40 and 60\GeV showering in the HCAL.
The energy resolution was found to be 19.4, 18.8, and 23.6\% in the HB, HE, and transition region, respectively.
Signal loss due to radiation damage is relevant in the HE at high \abseta.
It is monitored using a laser calibration system and the response to a \sixtyCo source, and is cross-checked with hadrons and muons from \pp collisions~\cite{CMS:PRF-18-003}.

During \Run2, new algorithms for reducing the effect of out-of-time pileup on reconstructed HCAL energies were developed.
The HCAL response to incoming particles rises to its maximum within 10\ns, followed by an exponential decay, with 90\% of the pulse contained within two 25\ns time samples (TSs).
In \Run1, the bunch spacing was 50\ns, and the energy of hits was reconstructed by a simple sum of charges in the SOI and SOI+1 after the contribution from leakage currents was subtracted, with a correction factor applied to account for the tail extending beyond two TSs.
In \Run2, the bunch spacing changed to 25\ns, and hence this method was expected to yield a poor energy resolution, since it incorporates contributions from pulses from preceding bunch crossings that overlap with the pulse from the SOI.
The new algorithms developed for \Run2 subtract the energy of out-of-time pileup using pulse template fits.

In 2016--2017, two separate pulse-template fitting algorithms, referred to as Method~2 and Method~3, were deployed for offline reconstruction and in the HLT, respectively.
The most recent algorithm, called ``minimization at HCAL, iteratively'' (MAHI), was developed and deployed for data taking in 2018~\cite{Lawhorn:2019yog}.
Notably, MAHI performs well enough to be executed within the HLT latency requirements, and, for \Run3, has also been ported to run on GPUs for further reduction in processing time.
The MAHI algorithm was used for the legacy reprocessing of the CMS \Run2 data.
Detailed results are presented in Ref.~\cite{CMS:PRF-22-001}.

\subsubsection{HF performance in \Run2}

\begin{figure}[!ht]
\centering
\includegraphics[width=0.48\textwidth]{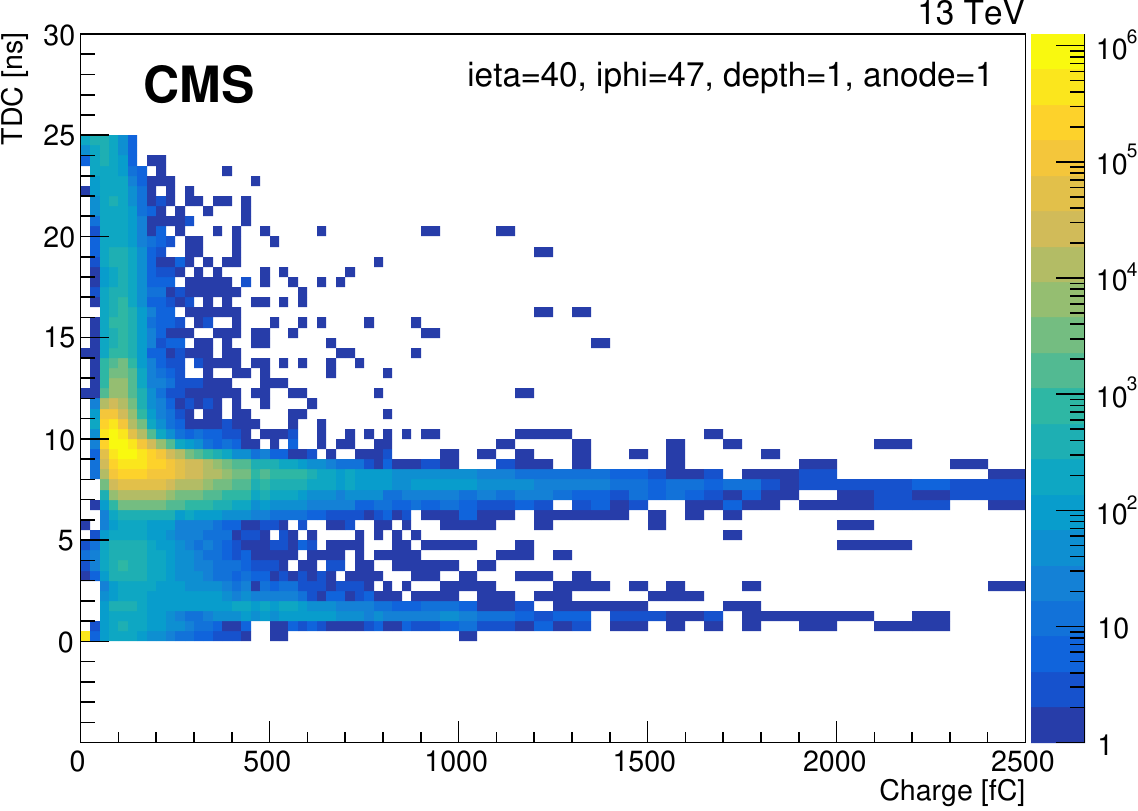}%
\hfill%
\includegraphics[width=0.48\textwidth]{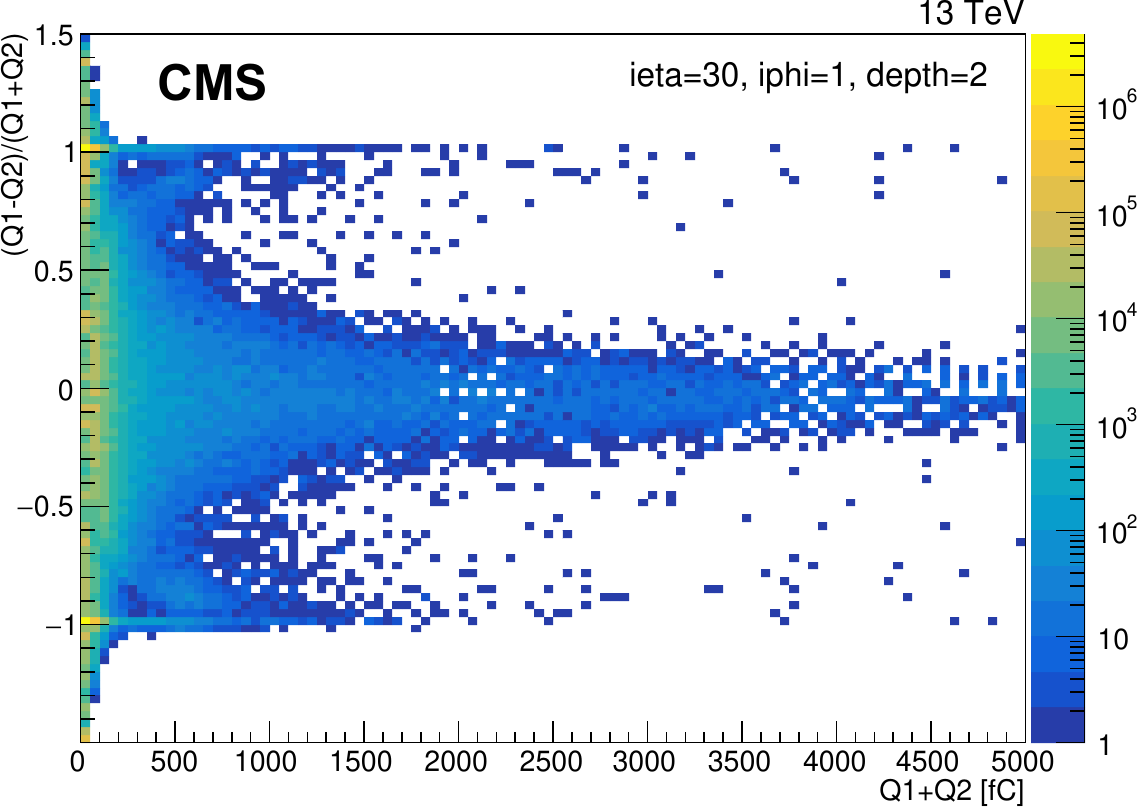}
\caption{%
    Left:\ HF signal arrival time, as measured in the TDC, versus the collected signal charge.
    All signals arriving within less than 5\ns are ``window events''.
    The color indicates the number of events using the scale to the right of each plot.
    The data were taken in early 2017.
    Right:\ charge asymmetry between the two channels of a PMT versus the total charge in both channels.
    The light from genuine collision events is well mixed in the light guides before falling on the PMT, hence similar signals are expected in all four anodes, which are grouped into two channels.
    The so-called ``window events'' due to Cherenkov radiation in the PMT window most likely fall on one or two anodes, producing asymmetric signals.
}
\label{fig:hcal:hfperf}
\end{figure}

The performance of the HF upgrade was evaluated during \Run2.
Figure~\ref{fig:hcal:hfperf} (left) shows the distribution of the signal arrival time as a function of the collected charge~\cite{CMS:DP-2017-034}.
Signals with a time around 7\ns are from showers in the calorimeter, while those at earlier times are due to Cherenkov radiation in the PMT window.
The arrival time can thus be used to identify window events.
While most signals due to PMT Cherenkov radiation come early, there is a tail to later times, due to imperfect synchronization.
Elimination of such background events can be improved by comparing the signals obtained in both channels of the PMT, as shown in Fig.~\ref{fig:hcal:hfperf} (right).
The charge asymmetry between the PMT channels is calculated as the difference divided by the sum of the signals in the two channels.
For genuine events, this value should be close to zero.
On the other hand, background events are observed well away from the central peak at zero since they are produced by stray muons hitting one quadrant or a side of the four-anode PMTs.
The combination of the arrival time and asymmetry methods improves the elimination of these background events.

\begin{figure}[!htp]
\centering
\includegraphics[width=0.48\textwidth]{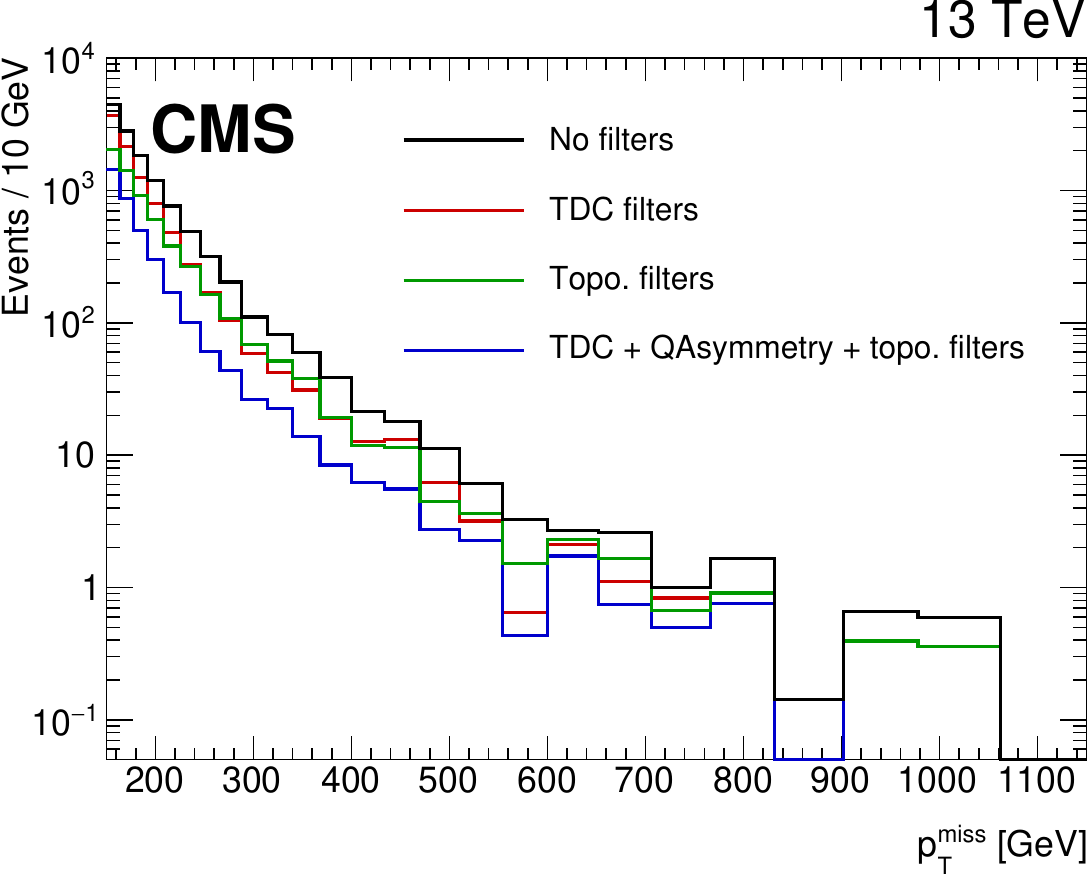}
\caption{%
    Effect of filters on the HF anomalous energy contributions to the missing transverse momentum measurement.
    The methods developed based on hardware improvements installed as part of the \Phase1 upgrade are as effective as the topological selections used previously.
    Including both the new and old filters further reduces the anomalous missing transverse momentum.
}
\label{fig:hcal:hfmet}
\end{figure}

Figure~\ref{fig:hcal:hfmet} shows the evolution of the missing transverse momentum with improvements to the HF anomalous signal identification based on the arrival time criteria (TDC filters), topological filters, and combined criteria (TDC, charge asymmetry, and topological filters).
Topological filters have been used since the beginning of \Run1 and are based on ratios of energies in the long and short fibers.
The new filters, based on the timing and ratios of the PMT channel energies, are as effective as this topological selection.
The combination of all the anomalous signal reduction techniques gives the best performance.
Additional topological filters based on the shape of jets versus $\eta$ and $\phi$ were developed to reject additional noise that escapes these filters~\cite{CMS:HIG-20-003}.

\clearpage
\section{Muon system}
\label{sec:muon}

A central feature of the CMS experiment is a powerful system for triggering on and detecting muons.
In the previous runs of the LHC, muons have been crucial to many of the physics results of CMS and have contributed to hundreds of published results, including the discovery of the Higgs boson.
The importance of muons in the CMS physics program continues to remain high in \Run3 and beyond.

The objectives of the CMS muon system are to identify muons, measure their momenta, and provide signals for triggering on them.
These goals are achieved with four complementary detector systems arranged in the steel flux-return yoke of the CMS solenoid.
These systems provide efficient detection of muons over a large range of pseudorapidity.
The location in the magnetized steel behind the calorimeters and solenoid ensures a low probability of penetration to the muon detectors by particles other than muons and neutrinos.

\begin{figure}[!ht]
\centering
\includegraphics[width=\textwidth]{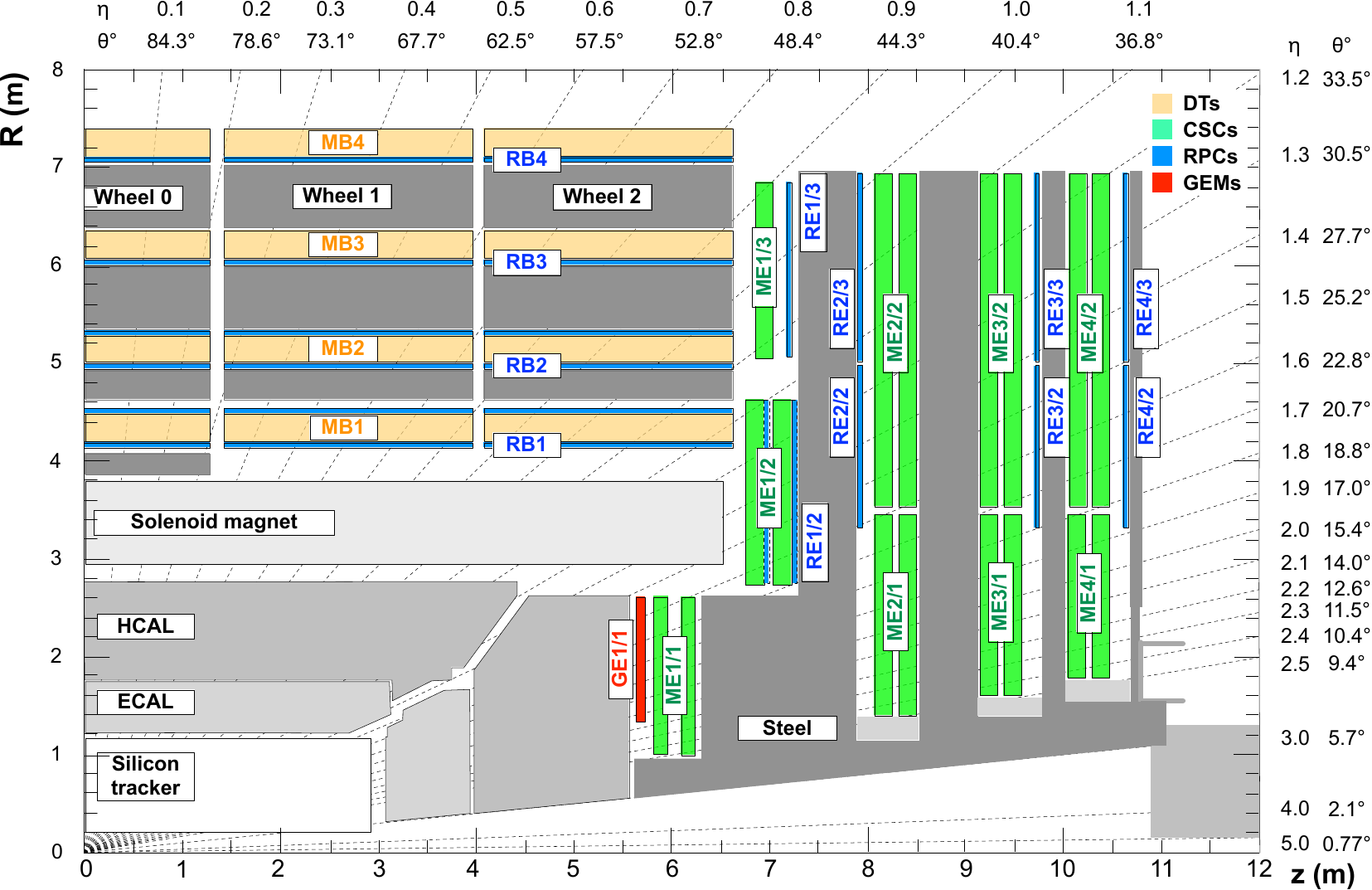}
\caption{%
    Schematic view in the $r$--$z$ plane of a CMS detector quadrant at the start of \Run3.
    The interaction point is in the lower left corner.
    The locations of the various muon stations are shown in color:\ drift tubes (DTs) with labels MB, cathode strip chambers (CSCs) with labels ME, resistive plate chambers (RPCs) with labels RB and RE, and gas electron multipliers (GEMs) with labels GE.
    The M denotes muon, B stands for barrel, and E for endcap.
    The magnet yoke is represented by the dark gray areas. Figure from Ref.~\cite{CMS:MUO-16-001}.
}
\label{fig:muon:quadrant}
\end{figure}

The physical arrangement of the muon detectors is shown in Fig.~\ref{fig:muon:quadrant}.
The central section is configured in a barrel geometry with four roughly cylindrical stations at different radii from the beam axis.
The endcap section is arranged in four planar stations in $z$ in each endcap.

The drift tube (DT) system in the barrel covers $\abseta<1.2$ and is composed of drift chambers with rectangular cells.
The DTs provide precise spatial measurements, as well as trigger information.
This system is described in more detail in Section~\ref{sec:dt}.

{\tolerance=800
The cathode strip chamber (CSC) system in the endcap comprises multiwire proportional chambers having cathode strips with an $R$--$\phi$ geometry and covering the region $0.9<\abseta<2.4$.
The CSC system provides both trigger and precision position information.
Because of the higher flux of particles in the endcap region, the CSCs are designed to have a faster response time than the DTs.
More details on the CSC system are found in Section~\ref{sec:csc}.
\par}

The resistive plate chambers (RPCs) are double-gap chambers operated in avalanche mode. The RPCs are located in both the barrel and endcap regions, and they complement the DTs and CSCs with a very fast response time that can be used to unambiguously identify the bunch crossing corresponding to a muon trigger candidate.
The RPCs are further described in Section~\ref{sec:rpc}.

Finally, GE1/1, a station of gas electron multiplier (GEM) chambers, is located in front of the inner ring of CSC chambers in the first endcap station.
The GEMs have both a fast response and good spatial resolution, and they augment the CSCs in a region of high particle flux.
More information about the GE1/1 station can be found in Section~\ref{sec:gem}.

The different groups of muon chambers are labeled with two letters:\ MB and RB for DTs and RPCs in the barrel region; ME, RE, and GE for CSCs, RPCs, and GEMs in the endcap regions.
For MB and RB, these letters are followed by a single number (1--4) indicating the barrel index.
Two indices follow ME, RE, and GE, where the first indicates the station in $\abs{z}$ and the second indicates the ring in $R$.
The value of each index increases with increasing distance from the center of the detector.

For each bunch crossing, fast trigger data (``trigger primitives'') are sent from each muon detector system (DT, CSC, RPC, and GEM) to the dedicated level-1 (L1) trigger muon track-finder hardware in the barrel (BMTF), endcap (EMTF), and overlap (OMTF) regions.
The L1 muon trigger is described in more detail in Section~\ref{sec:l1trigger:muon}.

The muon system is summarized in Table~\ref{tab:muon:summary}, which gives the number of chambers for each subsystem, the number of readout channels, and the spatial and time resolution.

\begin{table}[!ht]
\centering
\topcaption{%
    Properties of the CMS muon system at the beginning of \Run3.
    The resolutions are quoted for full chambers, and the range indicates the variation over specific chamber types and sizes.
    The spatial resolution corresponds the precision of the coordinate measurement in bending plane.
    The time resolution of the RPC of 1.5\ns is currently not fully exploited since the DAQ system records the hit time in steps of 25\ns.
}
\label{tab:muon:summary}
\renewcommand\arraystretch{1.2}
\cmsTable{\begin{tabular}{ccccc}
    \multirow{2}{*}{Muon subsystem} & Drift tube  & Cathode strip  & Resistive plate  & Gas electron \\[-3pt]
    & (DT) & chamber (CSC) & chamber (RPC) & multiplier (GEM) \\
    \hline
    \abseta range & 0.0--1.2 & 0.9--2.4 & 0.0--1.9 & 1.55--2.18 \\
    Number of & \multirow{2}{*}{250} & \multirow{2}{*}{540} & 480 (barrel) & \multirow{2}{*}{72} \\[-3pt]
    chambers & & & 576 (endcap) & \\
    Number of & 8 ($R$--$\phi$) & \multirow{2}{*}{6} & 1 & \multirow{2}{*}{2} \\[-3pt]
    layers/chamber & 4 ($z$, MB1--3) & & 2 (RB1, RB2) & \\
    Surface area & \multirow{2}{*}{18\,000\msq} & \multirow{2}{*}{7000\msq} & 2300\msq (barrel) & \multirow{2}{*}{60\msq} \\[-3pt]
    of all layers  & & & 900\msq (endcap) & \\
    Number of & \multirow{2}{*}{172\,000} & 266\,112 (strips) & 68\,136 (barrel) & \multirow{2}{*}{442\,368} \\[-3pt]
    channels & & 210\,816 (wire groups) & 55\,296 (endcap) & \\
    Spatial resolution & 100\mum & 50--140\mum & 0.8--1.3\cm & 100\mum \\
    Time resolution & 2\ns & 3\ns & 1.5\ns & $<$10\ns \\
\end{tabular}}
\end{table}

The performance of the muon system in \Run1 and the first part of \Run2 is documented in Refs.~\cite{CMS:MUO-11-001, CMS:MUO-16-001}.
Much of the muon system is unchanged from that used in \Run1 and described in 2008~\cite{CMS:Detector-2008}, but there have been additions and improvements.
Notably, there were three major additions to the endcap detector suite:\ the outer ring of CSC chambers in station four (``ME4/2''), the outer rings of RPC chambers in station four (``RE4/2'' and ``RE4/3'', collectively ``RE4''), and the GEM system in station one (``GE1/1'').
In addition, there were important upgrades to the electronics and trigger in many subsystems, which are described in the corresponding sections.

\subsection{Drift tubes}
\label{sec:dt}

\subsubsection{General description}
\label{sec:dt:description}

\begin{figure}[!ht]
\centering
\includegraphics[width=0.52\textwidth]{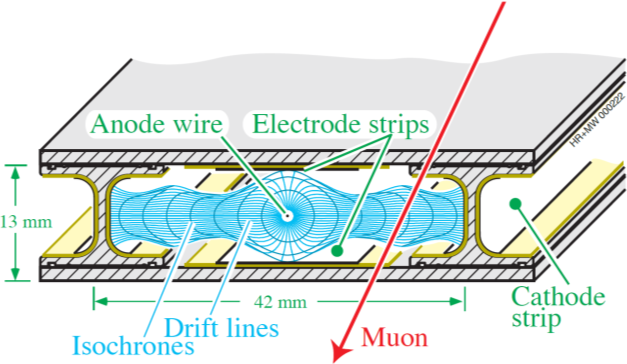}%
\hfill%
\includegraphics[width=0.44\textwidth]{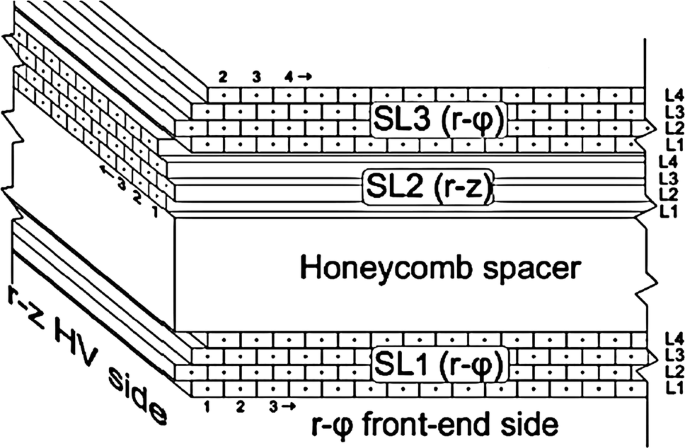}
\caption{%
    Left:\ layout of a CMS DT cell showing the drift lines and isochrones, from Ref.~\cite{CMS:Detector-2008}.
    Right:\ schematic view of a DT chamber, from Ref.~\cite{CMS:CFT-09-023}.
}
\label{fig:dt:detector}
\end{figure}

Drift tubes (DTs) equip the barrel part of the CMS muon detector, serving as offline tracking devices and providing standalone trigger capabilities.
The basic DT detector unit is a rectangular drift cell with a transverse size of $4.2{\times}1.3\cmsq$, whose layout is shown in Fig.~\ref{fig:dt:detector} (left).
A gold-plated stainless steel anode wire, with a diameter of 50\mum, is located at the center of the cell, and cathode strips are placed on its side walls.
Additionally, electrode strips are located at the top and the bottom of each cell to shape the drift field.
The cathodes and electrode strips are set at a voltage of $-1200$ and 1800\unit{V}, respectively, whereas the anode wires operate at applied voltages that vary between 3500 and 3600\unit{V}, depending on the individual chambers (more details are discussed in Section~\ref{sec:dt:longevity}).
Cells are filled with a gas mixture of 85\% Ar and 15\% \COtwo, which offers good quenching properties and, under the operational conditions described above, is characterized by a saturated drift velocity around 55\unit{$\mu$m/ns}.

A schematic layout of a DT chamber is shown in Fig.~\ref{fig:dt:detector} (right).
Within a chamber, cells are arranged parallel to each other to form layers (L), and groups of four layers, staggered by a half-cell, form superlayers (SL).
Each DT chamber is equipped with two SLs that measure the muon trajectory along the bending plane ($R$--$\phi$).
Additionally, chambers from the three innermost detector stations host an SL that measures the position along the longitudinal ($r$--$z$) plane as well.
A total of 250 DT chambers, covering a pseudorapidity range up to $\abseta<1.2$, are arranged in five wheels with an identical layout.
The wheels are placed parallel to each other along the CMS global $z$ axis, and are labelled W$-2$, $-1$, $0$, $+1$, and $+2$.
Within each wheel, chambers are organized in four concentric station rings, labelled from inside-out as MB1 to MB4, and segmented into 12 sectors (S) along the CMS global $\phi$ coordinate.

The performance of the DT system, measured over \Run1 and \Run2, was found to be remarkably stable and in line with the design expectations.
This is documented in Refs.~\cite{CMS:MUO-11-001, CMS:MUO-16-001}.
Track segments are typically reconstructed offline with an efficiency above 99\%, and they are characterized by spatial and time resolutions around 100\mum and 2\unit{ns}, respectively.
The efficiency to reconstruct a standalone DT segment in the trigger (also called a trigger primitive) and to correctly identify its bunch crossing (BX) of origin is above 95\%.
The position (direction) resolution of the DT trigger segments is approximately 1\mm (5\mrad).

Several upgrades, mostly concerning the off-detector electronics, occurred after the end of \Run1 to cope with the twofold increase in LHC instantaneous luminosity during \Run2 with respect to its original design.
They are part of the CMS \Phase1 upgrade~\cite{CMS:UG-TP-1} and are described in Section~\ref{sec:dt:ph1electronics}.

Even if a replacement of the DT on-board electronics will be required as part of the \Phase2 CMS muon system upgrade~\cite{CMS:TDR-021}, the DT chambers themselves will operate unchanged throughout the HL-LHC period.
For this reason, strategies to extend the detector longevity, and maximize the DT performance over a period longer than the one originally expected, were put in place in \Run2 and LS2.
They are documented in Section~\ref{sec:dt:longevity}.

\subsubsection{\Phase1 upgrades of the DT electronics}
\label{sec:dt:ph1electronics}

The increase in instantaneous luminosity during \Run2 required some changes to both the trigger and readout chains, particularly to the electronics hosted in the balconies surrounding the DT wheels in the experimental cavern, the so-called sector collector.
To accommodate the schedule of access opportunities in LS1 and the end-of-year technical stops, and to benefit also from the latest digital technologies, these upgrades took place in multiple stages:\ first a relocation of the sector collector and, subsequently, the trigger and readout upgrades.

On the other hand, with a single exception, no upgrade of the on-board DT detector electronics~\cite{CMS:Detector-2008} was performed or is foreseen until LS3.
This holds true for both the frontend (FE) electronics and the HV distribution, which are physically embedded in the chamber gas volume, and for the first level of the DT readout and local trigger electronics, which are hosted in aluminum structures attached to the DT chambers, called minicrates.

The only change that occurred in the minicrates was the replacement of 48 theta trigger boards (TRB), corresponding to the ones located in MB1 of W$\pm2$, which was performed over LS1.
The new TRBs host a Microsemi FPGA that offers the same functionality as the ASIC-based boards originally installed on the detector.
This replacement insures the availability of spares for the FE trigger components of the ASIC TRBs until the end of their operation.

\begin{figure}[!ht]
\centering
\includegraphics[width=\textwidth]{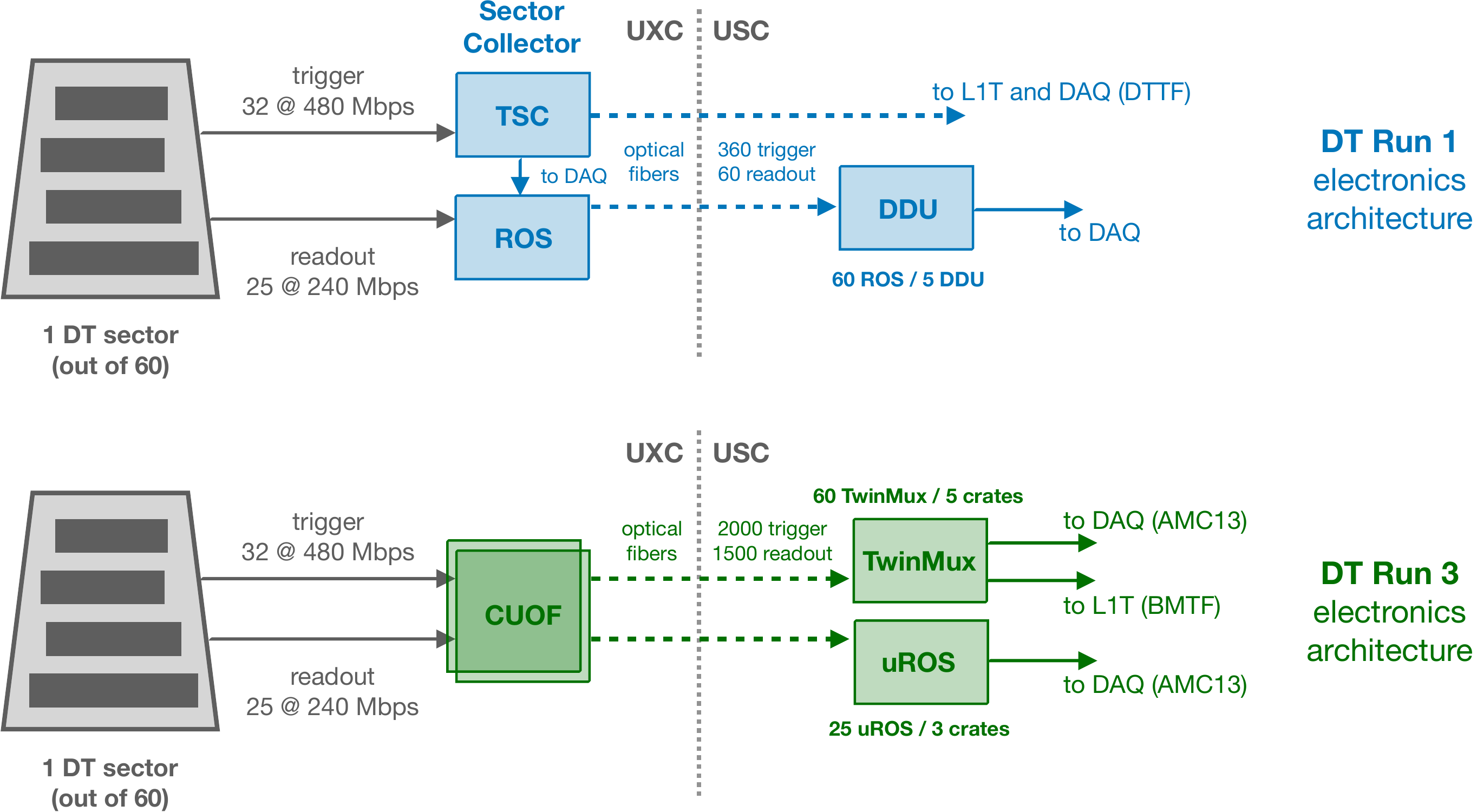}
\caption{%
    Schematic views of the \Run1 and \Run3 DT off-detector electronics architectures.
}
\label{fig:dt:phase1schema}
\end{figure}

A schematic view of the \Phase1 DT detector electronics architecture, as it was in \Run1 and is now in \Run3, is presented in Fig.~\ref{fig:dt:phase1schema}.
Details about this schema are described throughout the following sections.

\paragraph{Relocation of the DT sector collector}

The first stage of the DT electronics upgrade consisted of a relocation of the so-called sector collector (SC).
In its original layout, the SC was composed of ten 9U VME crates, hosting the second level of the DT trigger, the trigger sector collector (TSC) and the readout, the readout server (ROS) electronics, as well as different boards in charge of slow control, monitoring, timing signal distribution, and power distribution.
The SC was located on the balconies surrounding the detector in the experimental cavern (UXC).

The relocation project~\cite{Navarro-Tobar:2016hya} consisted of moving the TSC and ROS boards to the service cavern (USC).
As a consequence, the length of the readout and trigger links from the minicrates was increased by approximately 60\unit{m}.
To cope with the increased length, a dedicated system that converts electrical signals from the chambers into optical ones (CUOF) was installed in the racks previously occupied by the SC.
Data were then transmitted by means of optical fibers to the USC, where they were converted back to electrical signals by a second set of dedicated boards (OFCU) before being injected into the ROS and TSC.

The SC relocation brought an increase of operational reliability to the DT system.
Moving the TSC and ROS boards to the USC made them always accessible during LHC running periods, allowing the prompt solution of possible problems without the need to wait for technical stops when access to the UXC was possible.
It also paved the way for further stages of the \Phase1 DT upgrade, which led to an overall performance improvement.
Firstly, the region in which the SC was originally installed was characterized by high radiation levels (up to 0.2\Gy per year in nominal LHC conditions).
Without a relocation, these constraints would have led to a limitation in the choice of the electronics components that could be used for the upgrade to radiation-hard units.
Secondly, a nonnegligible residual magnetic field (up to 40\mTesla) is also present on the detector balconies, imposing further restrictions on the use of magnetic components such as inductors and ferrites.
The choice of cooling turbines used in the CMS balconies is also constrained by the need to operate in the presence of a magnetic field.
If the SC had remained in the UXC, this would have put stringent limits on the power consumption of the upgraded system, since the power dissipation of the original SC had already challenged the cooling capacity of the turbines operating on the balconies.
Thanks to the relocation, these constraints were relaxed, leading to more freedom in the design of the upgraded trigger and readout electronics.

The readout and trigger links from the minicrates are based on DS92LV1021 serializers from National Semiconductors, which have an embedded clock.
Serial words of 12 bits (10 bits for payload and overhead) are clocked at 20 (40)\MHz for the readout (trigger), resulting in a bit rate of 240 (480)\Mbs.
Unless an L1 accept trigger is issued, the readout transmits an idle payload that was designed to maximize the DC balancing, resulting in a 40\% duty cycle.
Instead, in the trigger payload, streams of zeros are preferentially transmitted unless trigger segments are built.
This results in a significant DC imbalance in the trigger links, which can be tolerated by operating the CUOF optical transmitters in a low-bias mode, among other improvements upstream, \eg, those to the TwinMux concentrator described below.

A CUOF board consists of a 9U motherboard where four mezzanine cards are typically plugged in.
Each mezzanine card carries out the conversion of information received from up to eight links, organized in two FTP cables from either the readout or trigger of one DT chamber.
The electrical signals enter the CUOF from RJ45 connectors located at the front of the crate.
They are then routed into a line equalizer, restoring levels and compensating for the distortions of the electrical transmission line, and finally are injected into a laser driver that controls the laser diodes.
Vertical cavity surface emitting laser (VCSEL) diodes are used.
There are eight VCSEL diodes for each mezzanine card, which are connected to a fiber fan-out using LC-type connectors.
Each CUOF motherboard also hosts two A3P600L ProASIC3L FPGAs from Microsemi, which control the configuration and the monitoring of the laser drivers.
Fine tuning of the drivers' bias and modulation settings is of prime importance to ensure correct transmission of the DC-unbalanced information from the trigger link.

\begin{figure}[!htp]
\centering
\includegraphics[width=0.3\textwidth]{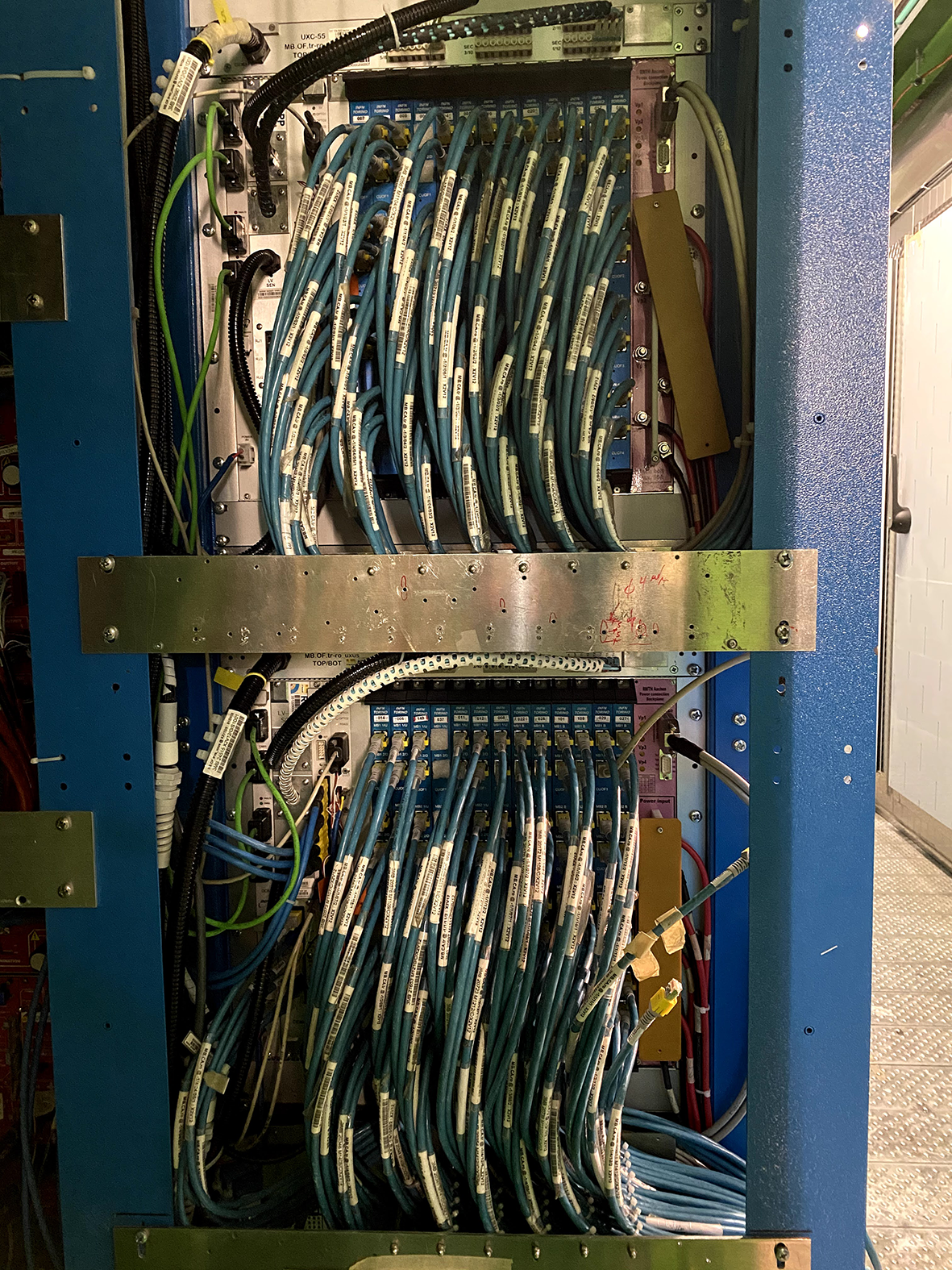}%
\hfill%
\includegraphics[width=0.3\textwidth]{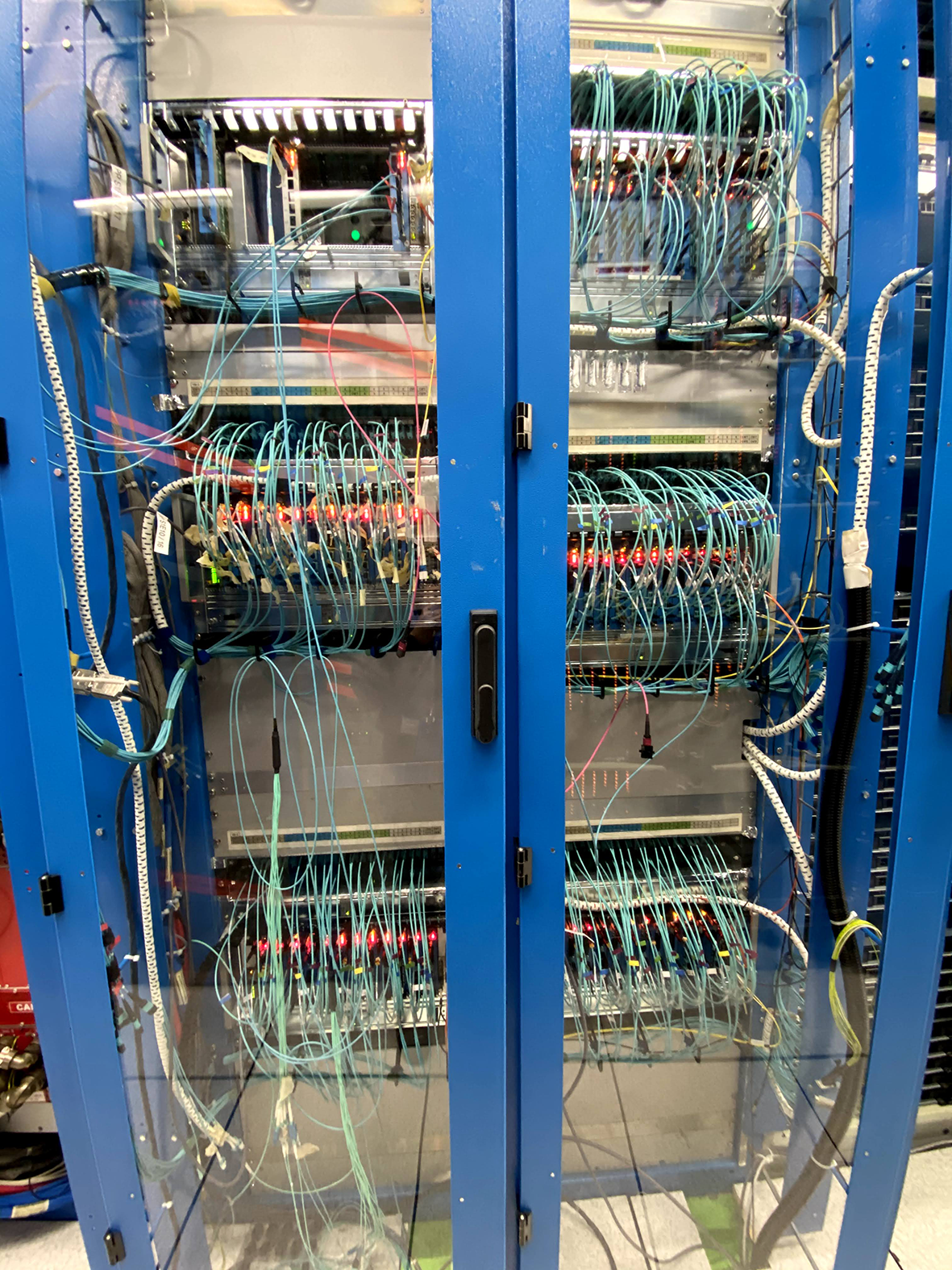}%
\hfill%
\includegraphics[width=0.3\textwidth]{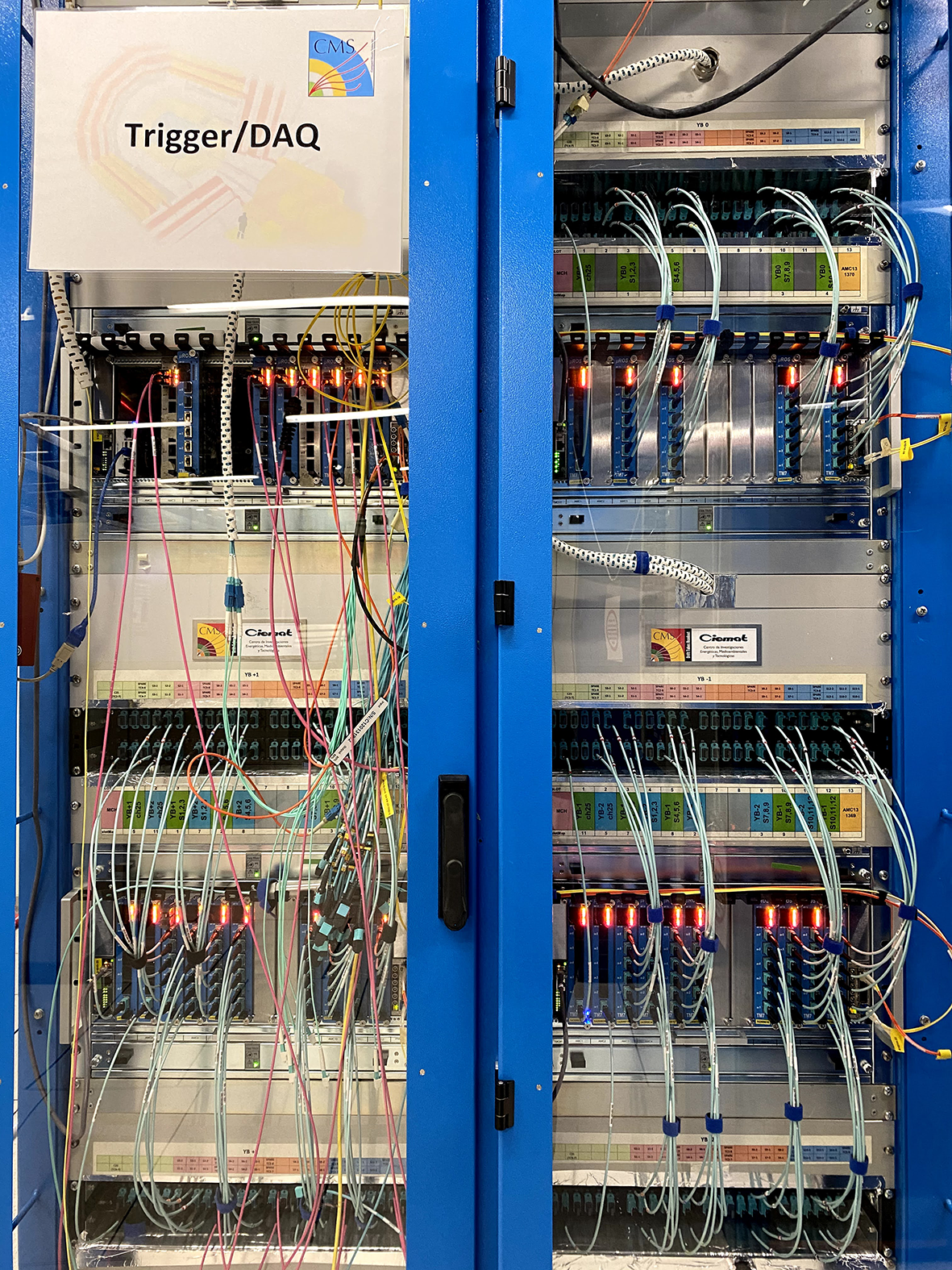}
\caption{%
    Left:\ front view of two out of the ten DT CUOF crates located in the UXC balconies surrounding the muon barrel (W$-1$).
    Center:\ front view of the five \uTCA crates of the TwinMux in the USC.
    Right:\ front view of the three \uTCA crates of the \uROS in the USC.
}
\label{fig:dt:phase1crates}
\end{figure}

A total of ten CUOF crates (corresponding to two crates per wheel) is installed in the UXC balconies. Each of them hosts thirteen CUOF boards, six (seven) of them dedicated to the transmission of readout (trigger) information, and covering a total of six DT sectors.
Two A3050 CAEN modules provide power to the system in each crate.
Each of them delivers two independent power supply channels, thus, for a given wheel, four power partitions exist.
The power consumption of the CUOF system corresponds roughly to half of that of the previous SCs.
A picture of the two CUOF crates instrumented in the balconies surrounding W$-1$, is shown in Fig.~\ref{fig:dt:phase1crates} (left).

Optical communication occurs as 850\nm transmission over OM3 multimode fibers with a 50 (125)\mum core (cladding) diameter.
Individual fibers are organized in MTP cords, each containing twelve fibers, which get further assembled in groups of eight to form trunk cables of 96 fibers.
Considering the arrangement of fibers into MTP cords, as well as the need for spares, a total of 60 trunk cables is used.
This corresponds to a total of 5760 fibers and thus easily covers the minimal need of the system of 3500 fibers.
The installation of the UXC-to-USC optical fibers through the trigger tunnels and cable chains to the wheels was labor- and access-intensive, but also leaves a legacy infrastructure that will be exploited by the DT \Phase2 upgrade in LS3.
A very important requirement, made at the time of purchasing from the vendor, was that the relative length of different fibers had to be carefully equalized.
This was needed to preserve the phase relationship between the signals coming from the different trigger links of a single chamber, which could not be compensated for at the input of the TSC.
The propagation delay along different fibers was measured, finding an excellent uniformity, with variations between fibers from the same MTP cord of under 1\ns~\cite{Navarro-Tobar:2013csa}.

The OFCU conversion was performed by dedicated boards that output LVDS signals into RJ45 connectors.
These, in turn, were used to transmit information for input to the ROS and TSC boards.
Due to different requirements for the readout and trigger electronics, different OFCU boards were developed, but they were both based on commercial parallel optics receivers from AVAGO (HFBR-782BEPZ).
At later stages of the \Phase1 DT upgrade, such receivers were re-used as components of the upgraded readout and trigger electronics.
The upgrade also included the design of various slow-control electronics for both the CUOF and OFCUs, and a link to maintain the injection of trigger data into the readout chain.

The entire relocation of the DT SCs took place during LS1, between February 2013 and August 2014.
Commissioning with cosmic rays and calibration runs were performed shortly after its completion.
It was confirmed that the performance of the upgraded system was consistent with the original one, and the DT system operated very successfully throughout \Run2.

\paragraph{Upgrade of the muon barrel local trigger: the TwinMux concentrator}

In the original CMS level-1 (L1) muon trigger architecture, tracks were reconstructed using three different track finders, each one mostly exploiting information from a single muon detector:\ DT, CSC, or RPC.
Candidates from the different track finders were only merged at the last stage of the muon trigger logic.
For the \Phase1 L1 trigger upgrade~\cite{CMS:TDR-012}, described in detail in Section~\ref{sec:l1trigger}, a different layout was chosen.
The overall muon trigger chain was designed to exploit information from all the detectors covering the area crossed by a given muon as early as possible in the online reconstruction.
This approach was adopted to maximize overall performance and to better control the data acquisition rates.
In the case of the muon barrel, the DTs provide excellent position resolution, whereas the RPCs are characterized by excellent time resolution.
Hence, in the \Phase1 L1 trigger, information from both detectors is combined by dedicated electronics that provide primitives of superior performance (called super-primitives) already at the input of the barrel muon track finder (BMTF).

To accomplish this, the DT TSC was replaced by a new component, called TwinMux~\cite{Triossi:2014oea}, which acts as a concentrator for the data coming from both the DT and barrel RPC chambers.
The TwinMux combines information from the two into superprimitives and transmits them to the BMTF and the overlap muon track finder (OMTF) using the 10\Gbs link protocol exploited by the \Phase1 L1 trigger system.

\begin{figure}[!ht]
\centering
\includegraphics[width=0.6\textwidth]{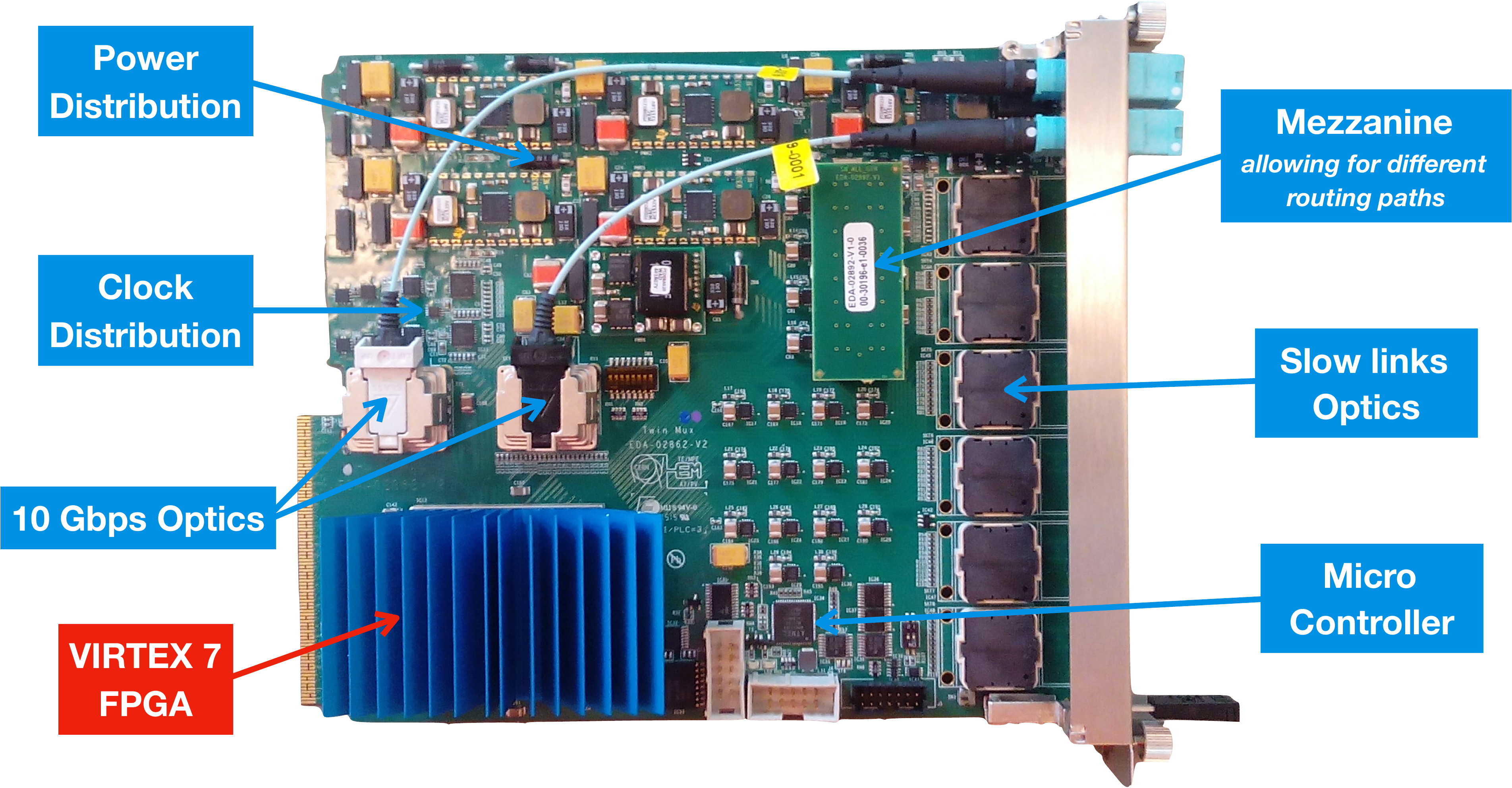}%
\hfill%
\includegraphics[width=0.35\textwidth]{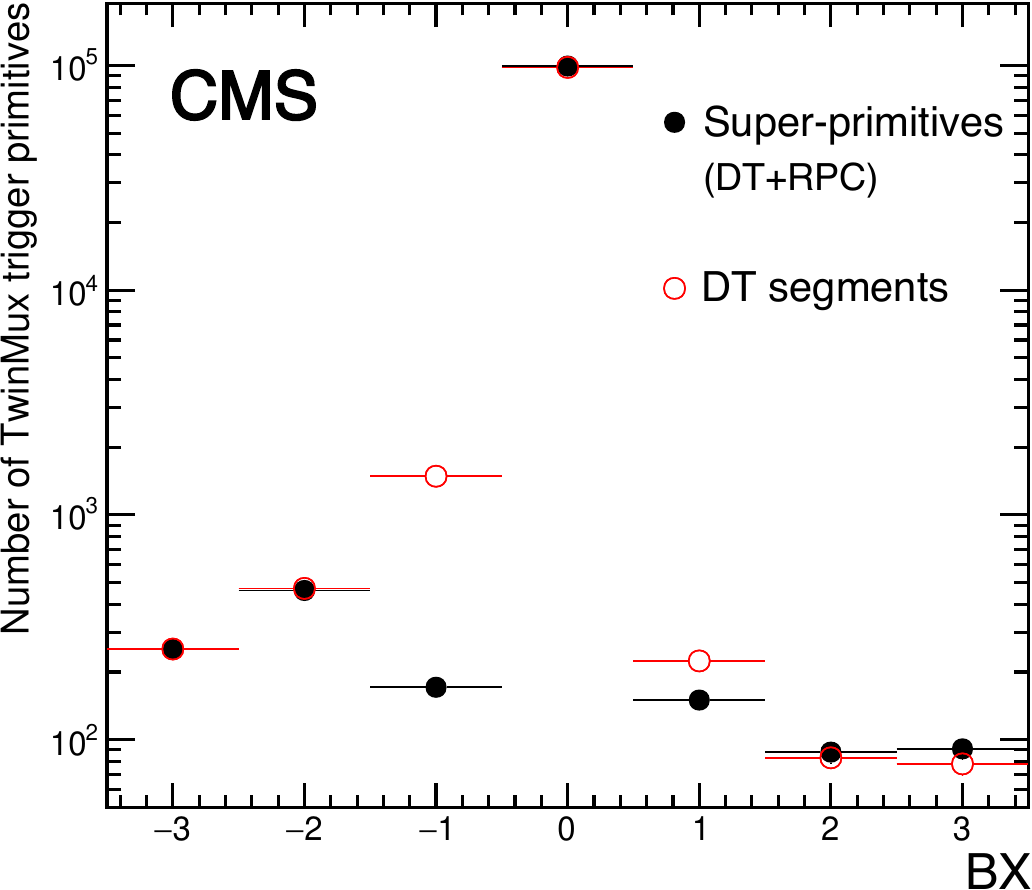}
\caption{%
    Left:\ picture of a TM7 board with the main modules highlighted.
    Right:\ BX distribution of L1 trigger primitives reconstructed in the muon barrel~\cite{CMS:DP-2016-074}.
    Open (red) circles show the performance of trigger primitives reconstructed using information from the DT detector only.
    Filled (black) circles show the same performance figure for super-primitives built combining information from the DT and RPC detectors.
}
\label{fig:dt:twinmux}
\end{figure}

For the TwinMux, a single slot double-width full-height \uTCA board, called TM7, is used.
A TM7 can reach a maximum of 96 optical connections thanks to six front panel Avago optical receivers (72 links limited to 2.7\Gbs) and two Minipods for high-speed data transmission and reception (up to 13\Gbs).
Figure~\ref{fig:dt:twinmux} (left) shows a picture of a TM7 board, where its main components are highlighted.
The TM7 board is based on a Xilinx Virtex-7 FPGA that, in the case of the trigger, achieves the merging of several 480\Mbs links to higher speed serial links and compensates delays to provide BX alignment of the trigger data coming from different inputs.
Twelve of the 72 inputs are optionally routable to GTH Gigabit Transceiver inputs~\cite{Xilinx:2018web} in order to handle the GOL-based~\cite{Moreira:2001cds} 1.6\Gbs links that receive the RPC links.
A small mezzanine PCB allows the desired path to be chosen for lines routed to one of the front optical transceiver.
Four additional GTHs are used to transfer data on the \uTCA backplane.

From each minicrate, DT trigger information is transmitted as described above.
For the RPC detector, five link board masters (LBMs) compress the trigger hit data relative to one muon barrel sector and serialize it through the GOL transmitters.
The TwinMux is in charge of forwarding this data to the BMTF and OMTF, by applying a scale-up in the transmission rate (and hence a reduction in the number of links).
It is also responsible for duplicating the data to be sent to different track finder processors (up to four times for the sectors of the outer wheels where DT data is shared between the barrel and overlap track finders).
Such redundancy is included to reduce the connections between the track finder processors and hence to increase the reliability of the system (which proved to be a weak point in the design of the legacy DT track finder).
The minimum bandwidth required for forwarding trigger data of one sector is 16 (8)\Gbs for the DT (RPC), implying the need for a total of three 10\Gbs links.
The clock distribution is based on two very-low-jitter PLLs that can broadcast two different clocks to all the FPGA transceivers for performing synchronous or asynchronous data transmission.
Finally, a microcontroller is responsible for managing the IPMI interface on the backplane.
It handles low-level operations like the hot swap of single boards without the need of switching off a full crate, or monitoring the temperature sensors.

To cover the full barrel, 60 TM7 TwinMux boards are hosted in five \uTCA crates.
Each of them is equipped with an AMC13 for clock and slow-control command distribution and for providing a connection to the CMS DAQ.
Each crate is also equipped with a commercial \uTCA carrier hub (MCH), a redundant power module, and a JTAG switch used for remotely programming the boards.
A picture of the five \uTCA crates hosting the TwinMux is presented in Fig.~\ref{fig:dt:phase1crates} (center).

The TwinMux firmware consists of six major blocks.
The DT and RPC receiver blocks handle input channels from the two detectors, which correspond to a given muon barrel sector, performing in both cases a link alignment to the same BX.
In the case of the DT receiver, a factor 3 oversampling, leading to a frequency of 1.44\Gbs, is applied to cope with potential data integrity errors caused by the aforementioned DC imbalance of trigger signals.
For the RPC, the GOL protocol is decoded.
Moreover, clusters are formed of nearby RPC hits and their coordinates are converted to the ones used by the local DT trigger.
A so-called super-primitive generation block combines information from the DT and RPC detectors according to algorithms described in more detail later in this section.
The output transmitter block formats and sends the trigger output data to the track finders using a 10\Gbs link protocol.
A readout implementation is also included that allows trigger-primitive readout through the CMS DAQ system.
Finally, an IPbus-based slow-control block also exists.
The TwinMux needs 34.5 BXs of latency for data to go through the FPGA from input to output.
Of these, only four BXs are actually used for super-primitive generation, whereas 26.5 are taken for handling of RPC data by the RPC receiver block.
Finally, four additional BXs are taken by the serialization of the data sent to the track finders.
The DT receiver block works in parallel with, and in the shadows of, the RPC one.

The TwinMux was deployed in production, as part of the \Phase1 L1 trigger upgrade, starting from 2016, after having been tested in a slice of the detector that was sending signals both to the TSC OFCU and TwinMux boards during the 2015 run.
The development of the algorithms to build super-primitives, which are used as input to the BMTF, occurred in stages.
In 2016, no combination was performed, hence only trigger primitives built by the DT on-board minicrate electronics were fed into the BMTF.

In 2017, an algorithm that combines information from the DT and RPC detectors was deployed to improve the super-primitive BX identification.
Within such algorithm, the compatibility of the DT trigger primitives and RPC clusters that are built in nearby chambers is checked by comparing the difference in azimuthal angle (\Dphi) between them.
If a match within a programmable window is found between a DT trigger primitive and (at least) one RPC cluster, and if the difference in terms of BXs between the DT trigger primitive and the RPC cluster is within $\pm$1 BX, a super-primitive is built using the DT trigger segment position and direction, but the RPC BX.
We note that no time correction is attempted if a DT trigger primitive is built exploiting hits from all layers of the $\phi$-SLs of a given DT chamber.
In any case, a dedicated quality flag, documenting the successful matching with the RPC, is set.
This combination better exploits the complementarity of the DT and RPC detectors, relying on the spatial resolution of the former and the time resolution of the latter.
The impact, in terms of BX identification performance for trigger primitives caused by muons from \pp collisions, is shown in Fig.~\ref{fig:dt:twinmux} (right).
The asymmetry in the BX distribution of DT primitives (red open dots), mostly due to the occasional presence of $\delta$ rays that can spoil the reconstruction of standalone DT trigger segments, is mitigated when the combination with the RPC is effective (black filled dots).
Because of the BX correction, the muon barrel trigger primitive BX assignment efficiency is also increased, on average, by 1.4\%.

In 2018, a further improvement was put into production.
In the MB1 and MB2 stations, where two RPC layers cover both surfaces of each DT chamber, as described in Section~\ref{sec:rpc:description}, the generation of RPC-only super-primitives is attempted if no DT trigger primitives are present in a given BX.
In that case, pseudosegments are built out of RPC cluster pairs that are reconstructed at the same BX in the two different RPC layers.
If more pseudosegments are generated at a given BX, only the one whose direction is closest to that of a straight track coming from the CMS interaction point ($\phi_b$) is retained.
Finally, only pseudosegments for which $\phi_b$ is below a programmable threshold are accepted.
Additionally, also in this case, the RPC-only primitive is marked with a dedicated quality flag.
Upon deployment of this algorithm in data taking, a further increase of the super-primitive efficiency around 3\% was observed in the MB1 and MB2 stations.
Due to the redundancy of the muon system, such local trigger efficiency improvement resulted in a marginal improvement of the BMTF efficiency.
Nevertheless, a rate reduction of a few percent was observed for the lowest unprescaled L1 muon trigger.
This is because, in a few cases, BMTF tracks now get built using a larger number of points along the muon trajectory, improving the measurement of the track transverse momentum.

Given the fact that it uses a different track-building algorithm, no super-primitive generation is attempted on primitives that are sent to the OMTF, since the OMTF receives a full list of DT trigger segments and combines them with RPC information directly at the track-finding step, as also described in Section~\ref{sec:rpc:upgrades}.
The super-primitive algorithm that operated throughout 2018 was also deployed at the start of \Run3.

\paragraph{Upgrade of the DT readout: the \uROS system}

Simulation studies have shown that the ROS was the most severe bottleneck in the DT readout chain.
The ROS combined information from a total of 25 readout boards (ROB) providing the minicrates with data for a full DT sector.
The time needed by a ROS to perform event building from all the input links depended on how hits were distributed within the ROBs.
A high noise rate from sporadic small groups of channels could be easily sustained.
However, high overall background hit rates, which come with a rather uniform distribution within each of the chambers of a sector, take much more time to be processed.
In addition, muons crossing the DT can produce up to 44 hits within a single ROS. When both the background and muon hit rates are considered, taking as a proxy an LHC instantaneous luminosity of $2\times10^{34}\percms$, the maximum acquisition rate that a ROS can sustain becomes close to the 100\kHz limit imposed by the DAQ and L1 trigger.

For this reason, throughout \Run2, the ROS, as well as the downstream components of the DT readout chain, called device-dependent units (DDU), were replaced with a new system, based on the \uTCA architecture, named \uROS.
For the \uROS~\cite{Navarro-Tobar:2017jbi}, the same TM7 boards designed for the TwinMux are used, but a different firmware is deployed to implement the functionalities needed by the DT readout.

Each TM7 features six 12-fiber MTP receivers, for a total of 72 input links.
A total of 25 ROBs provide the inputs from a given sector.
The data from each wheel are thus processed by five \uROS boards, four of them receiving three sectors each (24 channels per sector, 72 links), and the fifth receiving the 25th channel for each of the twelve sectors.
The production system comprises three \uTCA crates (central, positive, and negative wheels) and 25 \uROS boards.
Each crate is equipped with an AMC13 that provides clock and slow-control distribution, as well as a connection to the CMS DAQ.
A picture of the \uROS system is presented in Fig.~\ref{fig:dt:phase1crates} (right).
With this architecture, no further components of the DT readout system (DDU) are needed.
Under the conditions reached over \Run2, the maximum payload bandwidth varied, depending on the wheel, between 0.3\Gbs in W0 and 0.6\Gbs in W$+2$ and $-2$, remaining well within the AMC13 limits.

In terms of firmware, the components needed to handle the slow control and general board functionality are inherited from the TwinMux.
Special care was instead put into the design of the block in charge of data deserialization.
This firmware can recover input data with high quality and minimal data losses with respect to the input stream.
Carrying data at 240\Mbs, the receiver samples data at 1.2\Gbs (a factor of 5 oversampling).
Majority filtering is performed on the three central samples of each bit before reassembling the original input word.
Bits where a weak majority is found (two instead of three) are marked as transmission error candidates.
If the data frame parity error check fails and only one bit is marked as a transmission error candidate, the latter gets corrected.
The firmware implements a full verification of the ROB protocol and provides statistics for the different ROB failure cases, which are used for monitoring.
Finally, while the legacy ROS system masked channels in case of transmission errors until a resync was issued, the event builder from the \uROS is capable of recovering from all types of errors as soon as the condition disappears.

The transition to the \uROS occurred during the 2017/2018 year-end technical stop.
Prior to that, in 2017, signals from the ROBs for a slice of the detector were split and a \uROS slice-crate was instrumented and integrated into the CMS DAQ as a separate unit.
It was used to develop the \uROS prior to the deployment of the full system, which allowed a smooth transition.
In LS2, the FE signals of a sector in the external wheels (W$+2$ S12)
were split to allow the continuation of the same strategy for the \Phase2 upgrade, and another slice-test system will be operated in \Run3.
The data of such a \Phase2 slice-test are already read by the CMS DAQ.

\begin{figure}[!ht]
\centering
\includegraphics[width=0.48\textwidth]{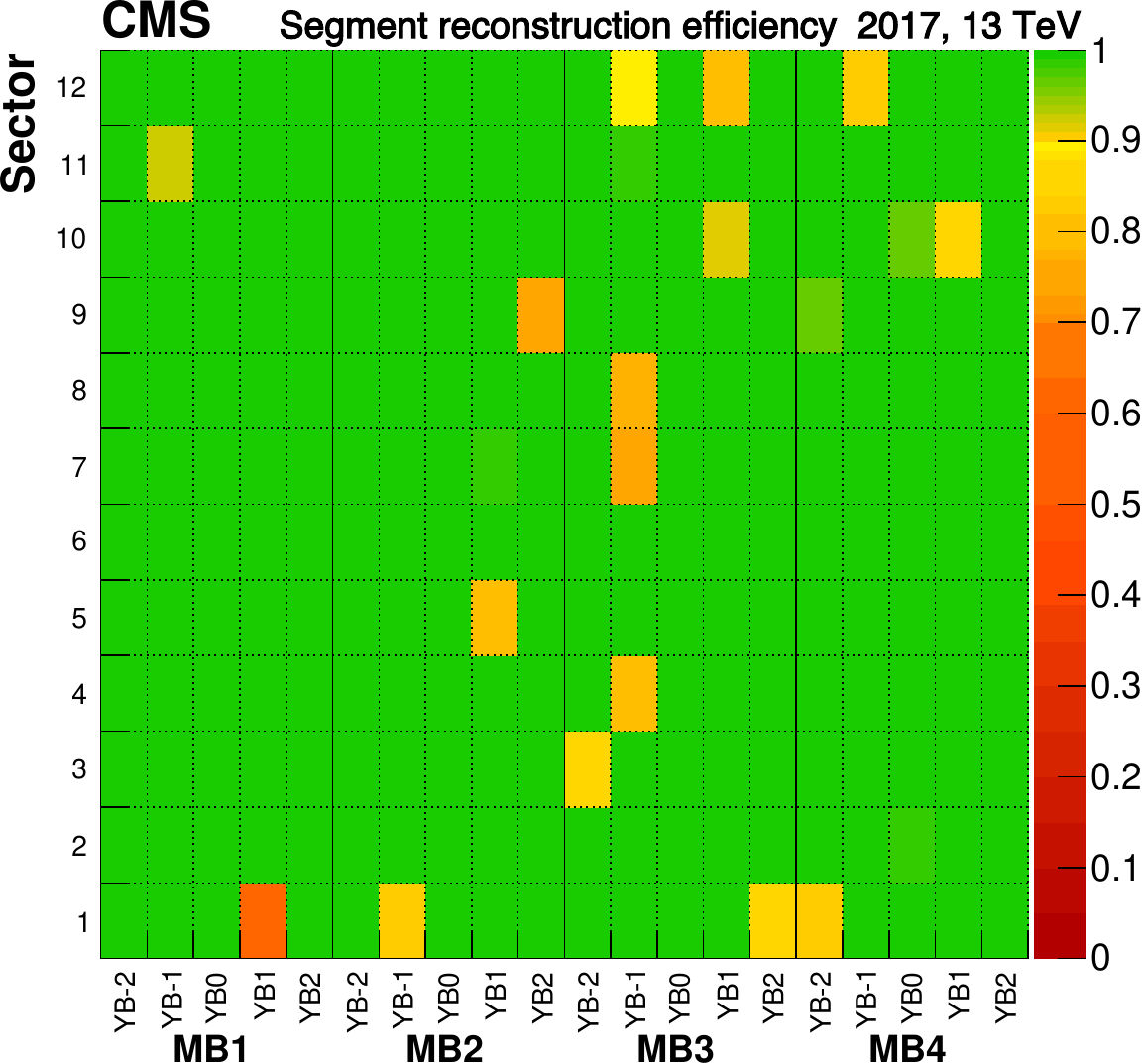}%
\hfill%
\includegraphics[width=0.48\textwidth]{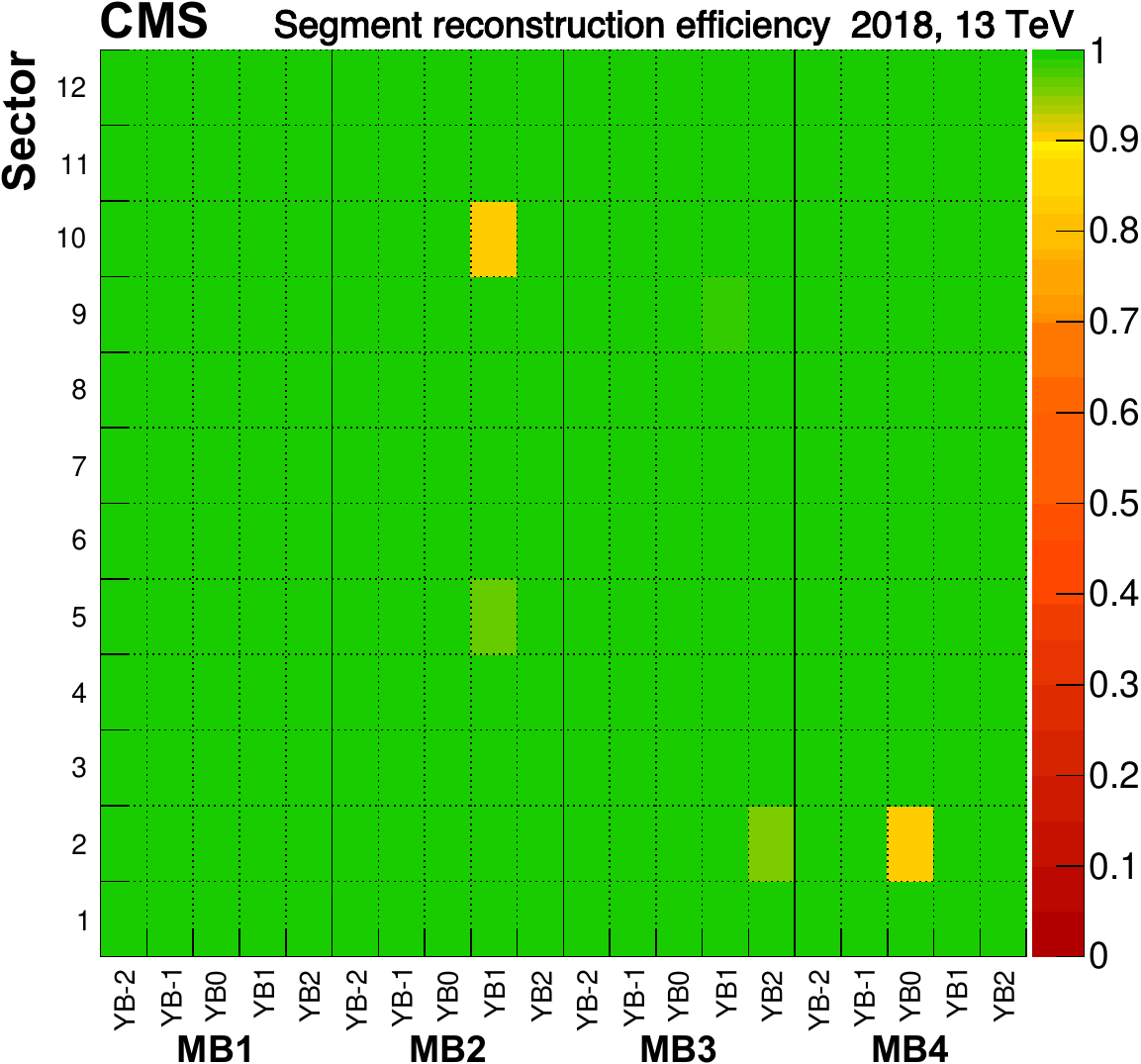}
\caption{%
    DT segment reconstruction efficiency measured with the tag-and-probe method using data collected by CMS in 2017 (left)~\cite{CMS:DP-2018-016} and 2018 (right)~\cite{CMS:DP-2019-008}, before and after the transition to the \uROS.
    The efficiency is usually above 99\% except for chambers affected by hardware problems, mostly coming from the DT readout.
    After the deployment of the \uROS, the overall efficiency improves.
}
\label{fig:dt:uros}
\end{figure}

The impact, in terms of DT performance, of the transition to the \uROS is presented in Fig.~\ref{fig:dt:uros}.
The two plots show a chamber-by-chamber map of the DT segment reconstruction efficiency, as measured with a tag-and-probe method.
In the measurement, no masking of chambers with hardware or readout issues is applied.
The left (right) side of the figure refers to results computed using 2017 (2018) data, collected before (after) the transition to the \uROS.
Bins where the efficiency is significantly lower than 99\% are due to possibly sporadic problems that affected chambers or their readout.
A better performance was observed after the upgrade to the \uROS, mostly thanks to a reduction of the number of chambers affected by readout problems.

\subsubsection{Detector longevity for \Run3 and beyond}
\label{sec:dt:longevity}

The DT system was designed and validated to sustain ten years of LHC operation in nominal conditions, corresponding to approximately 500\fbinv of integrated luminosity~\cite{Conti:2001kq}.
Accelerated aging studies, carried out prior to the installation of the chambers in CMS, indicated that no degradation in the performance of the present detector (and its electronics) is to be expected under those assumptions.
Furthermore, at the boundaries of long data-taking periods, data are regularly collected by varying the HV applied to the anode wires (HV scans) and measuring the detector efficiency at each HV point to assess its stability and exclude potential effects due to early aging.
Up to the end of \Run2, no efficiency degradation was observed during the HV scans.
Additionally, the original longevity studies were complemented with more stringent ones, performed using the CERN gamma irradiation facility (GIF++)~\cite{CMS:TDR-021}, which are summarized in the following paragraph.
Given the results of all the above studies, and the current projections in terms of integrated luminosity expected for \Run3, the DT performance is foreseen to remain almost constant over the coming run period.

{\tolerance=800
Nonetheless, though several electronics components will be replaced as part of the CMS \Phase2 upgrade program, existing muon detectors will operate throughout the HL-LHC era.
For this reason, further accelerated longevity studies are being performed at GIF++ on two spare DT chambers, as mentioned above.
Due to the complexity of the physics and chemistry phenomena driving the aging processes, accelerated studies have large uncertainties, typically taken into account by irradiating with a substantial excess with respect to the needed integrated values (safety factors).
The reference targets for instantaneous and integrated luminosities used in the aging studies to derive the DT performance are evaluated assuming safety factors of two, which double both the nominal HL-LHC luminosity ($5\times10^{34}\percms$) and the total integrated luminosity (3000\fbinv).
Considering these safety factors, by the end of the HL-LHC, the DT hit detection efficiency can possibly drop from its present value of around 96\% to approximately 70\% in the MB1 stations of W$\pm2$ (corresponding roughly to 10\% of the entire system).
Moreover, in a further 20\% of the detector (corresponding to the MB4 stations of the DT sectors covering the top half of CMS and to the MB1 stations of W$\pm1$), the efficiency is projected to range between 85 and 90\%.
Finally, in the rest of the system, efficiencies above 90\% are expected.
Given the redundancy in terms of the number of DT layers per chamber and of chamber stations in the muon system, the maximum inefficiency due to DT aging, localized in a narrow region around $\abseta=1.0$, is expected to be within 2 (5)\% for standalone offline reconstruction (trigger).
Though these expected losses are not very large, mitigation strategies, which are described in the following sections, have been put in place during \Run2 and LS2, to maximize the longevity of the DT detector and ensure the highest achievable performance in the long term.
\par}

\paragraph{Optimization of the operational working points}

For gaseous detectors, deterioration due to aging becomes more significant as the integrated charge released in the gas volume increases.
In turn, the integrated charge depends on the rate and type of particles crossing the different regions of the detector, the collected charge per particle, and the total integrated running time.

In the case of the DT detector, background dominates the hit rates and is largest in the MB1 stations of W$\pm2$ and the MB4 stations for the sectors covering the top half of CMS.
The collected charge per particle depends on both the particle type and the gas amplification factor, which is driven by the HV settings of the cell anode wires.
Therefore, for a fixed integrated luminosity, reducing the HV working points of the anode wires can result in a significant reduction of the integrated charge.
Of course, such an optimization can be performed only within the limits where the impact on detector performance is deemed acceptable.

During the 2017 and 2018 LHC runs, the HV of the DT anode wires was progressively lowered with respect to the default 3600\unit{V} used until 2016, in the chambers most exposed to background.
In 2017, the wires in MB1 of W$\pm2$, and the ones in MB4 of sectors 3, 4, and 5 for all wheels were operated at 3550\unit{V}.
In 2018, a further reduction was applied, reaching the HV values given in Table~\ref{tab:dt:hvsettings}.
The discrimination threshold values applied in the DT FE electronics were also lowered from 30 to 20\mV for the entire detector.

\begin{table}[!ht]
\centering
\topcaption{%
    HV settings for the anode wires of the different DT stations and wheels used for the 2018 LHC run.
}
\label{tab:dt:hvsettings}
\renewcommand{\arraystretch}{1.1}
\begin{tabular}{cccccc}
    & W$-2$ & W$-1$ & W0 & W$+1$ & W$+2$ \\
    \hline
    MB1 & 3500\unit{V} & 3550\unit{V} & 3550\unit{V} & 3550\unit{V} & 3500\unit{V} \\
    MB2 & 3550\unit{V} & 3550\unit{V} & 3600\unit{V} & 3550\unit{V} & 3550\unit{V} \\
    MB3 & 3550\unit{V} & 3600\unit{V} & 3600\unit{V} & 3600\unit{V} & 3550\unit{V} \\
    MB4 & 3550\unit{V} & 3550\unit{V} & 3550\unit{V} & 3550\unit{V} & 3550\unit{V} \\
\end{tabular}
\end{table}

For the MB1 of the external wheels, lowering the HV of the anode wires to 3500 (3550)\unit{V} resulted in a relative reduction of 58 (45)\% in the drained current with respect to the original 3600\unit{V} setting.

\begin{figure}[!ht]
\centering
\includegraphics[width=\textwidth]{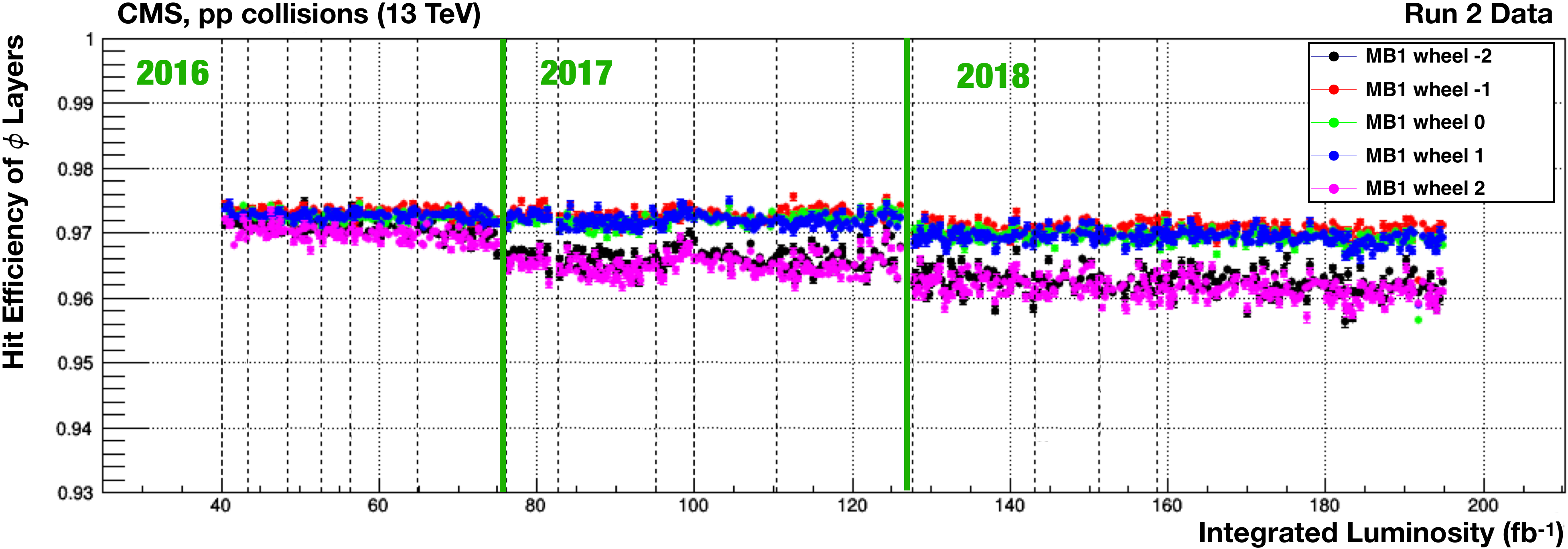}
\caption{%
    DT hit detection efficiency, computed as a function of the total CMS integrated luminosity, for the $\phi$-SLs of MB1 chambers~\cite{CMS:DP-2019-008}.
    Different colors refer to different DT wheels.
    The plot summarizes how the efficiency evolved during \Run2, mostly as a consequence of the different updates of the detector HV and FE threshold settings.
}
\label{fig:dt:hitefficiency}
\end{figure}

The overall performance of the system under the updated operational conditions was thoroughly studied using \pp collision data.
Firstly, the reduction of the FE thresholds to 20\mV resulted in a marginal increase of overall detector noise, which was handled by masking a few specific noisy wires.
Reducing the gain and the FE threshold resulted in shifts of a few ns of the effective drift time measured by the electronics caused by opposite-sign effects.
These partially cancel but were nevertheless carefully corrected for, by updating the trigger synchronization and offline calibration.
Fine-tuned corrections were needed to maintain the optimal performance of the system in terms of BX identification efficiency and time resolution of the offline reconstruction.
These remained well in line with what is reported in Section~\ref{sec:dt:description}.
The DT hit detection efficiency was then measured. Results for the $\phi$-SLs of the MB1 stations are given as function of the total integrated luminosity in Fig.~\ref{fig:dt:hitefficiency}.
Different colors in the plot refer to different DT wheels.
The efficiency is stable throughout the different run years, apart from step variations dominated by changes in the FE and HV, which amount at most to 1\%.
This was proven to be true for the rest of the detector as well and, given the redundancy of layers forming a DT chamber, a negligible impact on the DT local reconstruction efficiency was observed.
Finally, the effect of the updated settings on the DT hit resolution was also assessed.
In the case of $\phi$-SLs, where hit resolution impacts the muon transverse momentum measurement, only a slight worsening was observed.
In the chambers where the HV was lowered to 3550\unit{V}, the degradation is around 10\%.
In the MB1 of the external wheels, where the HV was set to 3500\unit{V}, the  effect is slightly larger, but the hit resolution remains within 250\mum, in line with the design expectations.

Given the reduction in terms of expected integrated charge, as well as the very mild impact on the overall performance, the operational working points used in 2018 were chosen as the default configuration for \Run3 as well.

\paragraph{Gas system upgrade and open-loop operation}

The DT gas system~\cite{CMS:Detector-2008} includes a facility on the surface that stores individual gas components (Ar and \COtwo) and combines them to obtain the gas mixture of 85\% Ar and 15\% \COtwo used by the DT.
A second facility, located in the USC, is dedicated to the predistribution of the final gas mixture to the five wheels, as well as to the gas analysis system.
Finally, five distribution racks located in the UXC provide the gas flows to the 250 DT chambers, with a flux around 40\unit{l/h} to each chamber.
The primary goal of the gas analysis system is ensuring the stability of the gas mixture.
It consists in the determination of the contamination of \Otwo and \HtwoO, and the drift velocity measurement.

The loss of efficiency caused by aging, described in Section~\ref{sec:dt:longevity}, is due to a decrease of gas gain which, in turn, is caused by deposits that form around the anode wires of the DT cells.
It is believed that this effect is mostly due to out-gassing of components inside a DT chamber.
For this reason, the re-injection and spread from hot detector regions of pollutants contaminating the DT gas should be avoided.
Therefore, starting from the 2018 run, the DT gas system operates in the, so-called, open-loop mode, where there is no re-circulation of used gas.
Prior to that, the system was operated in a closed-loop mode, where 85\% of the used gas was recirculated and only 15\% of fresh gas was injected.
The closed-loop mode can still safely be used during technical stops or longer periods of inactivity, where no aging due to out-gassing is expected.
In order to operate in open-loop mode, an upgrade of the gas system was performed.
It allows the intentional introduction of air from a bottle on the surface, to maintain the desired level of \Otwo, corresponding to 80\ppm.
Furthermore, a humidifier with a bypass that ensures the \HtwoO concentration is kept at 800\ppm was also installed.
The presence of small quantities of \Otwo and \HtwoO helps prevent effects that induce chamber aging (such as polymerization)~\cite{CMS:TDR-016}, and the target values for such contaminants were derived based on experience from the data taken in \Run1 and \Run2.
Because of this upgrade, prompt monitoring of the gas stability has become even more important than in the past.

\begin{figure}[!ht]
\centering
\includegraphics[width=0.6\textwidth]{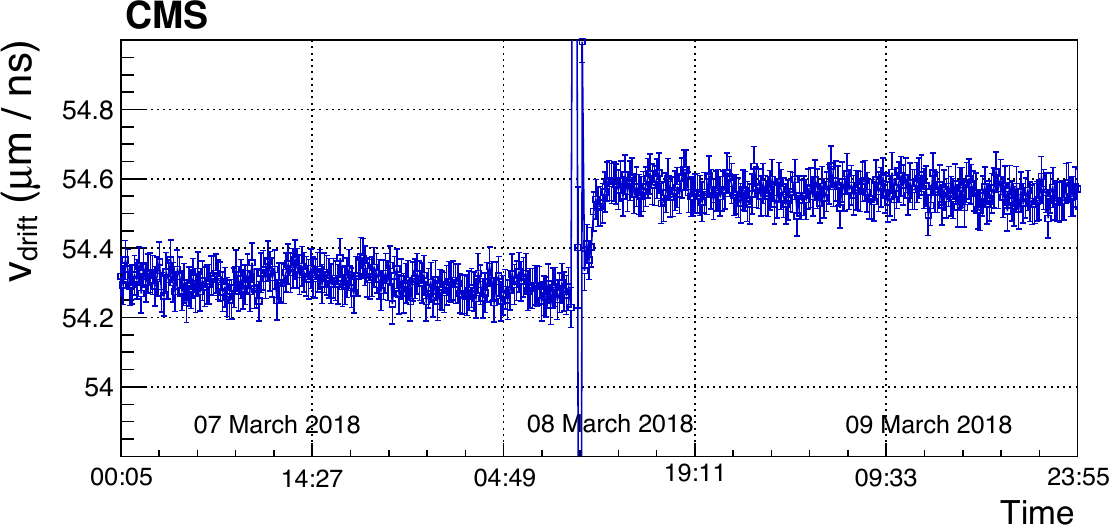}
\caption{%
    Drift velocity measurement using the fresh gas analyzed by the VDC system.
    The variation on 8th March 2018 corresponds to the transition between closed-loop and open-loop operation of the DT gas system.
}
\label{fig:dt:vdc}
\end{figure}

The \Otwo and \HtwoO measurements are performed using commercial sensors, which have to be recalibrated every year by injecting gas with known components of oxygen and humidity.
The drift velocity measurement is done by a more complex system, called VDC~\cite{Altenhofer:2017uux}, which aims to deliver, every ten minutes, a drift velocity measurement.
There is one \Otwo sensor, one \HtwoO sensor, and one VDC for each of the five DT wheels, and a separate sampling line for each wheel.
The gas sampling permits the selection of the output of any individual chamber, as well as the global output (or the global input) of an entire wheel.
The direct measurement of the drift velocity performed by the VDC system is sensitive to possible unknown components or contaminants in the gas.
The drift velocity depends on the Ar/\COtwo ratio, as well as on the level of contaminants, such as \Otwo and \Ntwo.
The drift velocity is a key parameter for the DT local reconstruction, so that any deviation spotted by the VDC system must be investigated and eventually requires immediate intervention on the operation of the DT gas system.
An example from the monitoring by the VDC system is presented in Fig.~\ref{fig:dt:vdc}.
The first transition from closed-loop to open-loop operation was correctly detected by the drift velocity measurement.
In this case, the gas analysis showed very quickly the impact of the injection of air and humidity into the mixture for the case of the open-loop mode, which resulted in a sub-percent variation of the drift velocity.

Other than direct monitoring of the gas mixture and drift velocity, the stability of the gas is also evaluated indirectly with event data, by looking at the stability of the performance of the DT local reconstruction.
If a track-segment fit is performed using a sufficient number of hits from a DT chamber, additional parameters, other than the segment position and direction, can be extracted from the fit itself.
In this way, the need for residual corrections on top of the calibration parameters used by the reconstruction, such as the drift velocity, can be evaluated on a segment-by-segment basis.
Overall biases in the distribution of such residual corrections are then  measured run-by-run for each DT chamber, and their stability across different runs is monitored.

Finally, the DT chambers are operated since \Run1 in a differential pressure mode.
The gas control system presently ensures that a positive differential pressure of $+3\mbar$ is applied at the bottom of the wheels.
This value, which is well within the mechanical maximum limit (50\mbar), protects against possible contamination.
A dedicated system of differential pressure sensors monitors continuously these values, which are transmitted to the DT online monitoring system.

\paragraph{Installation of shields over the outer MB}

As mentioned earlier, besides the MB1 stations of the external wheels (W$\pm2$), the parts of the MB characterized by the highest level of background are the MB4 stations in the sectors covering the top half of CMS (sectors 1 to 7).
Studies based on \pp collision data, as well as results from simulations, corroborate the hypothesis that background in this region is mostly generated from interactions of low-energy neutrons permeating the cavern.
Therefore, a strategy was put in place to effectively protect the top DT chambers by installing proper absorbing shields.

\begin{figure}[!ht]
\includegraphics[width=0.55\textwidth]{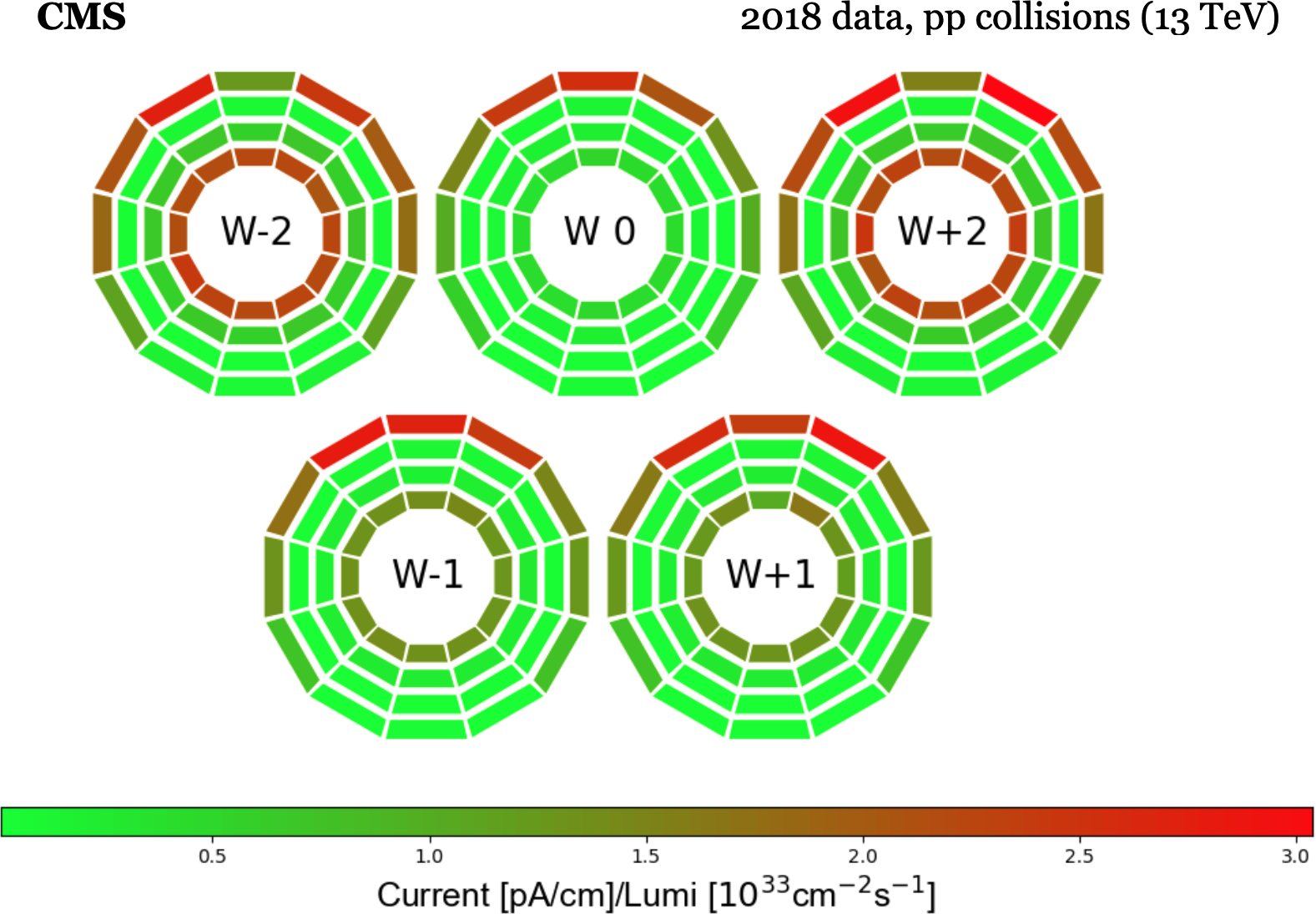}%
\hfill%
\includegraphics[width=0.40\textwidth]{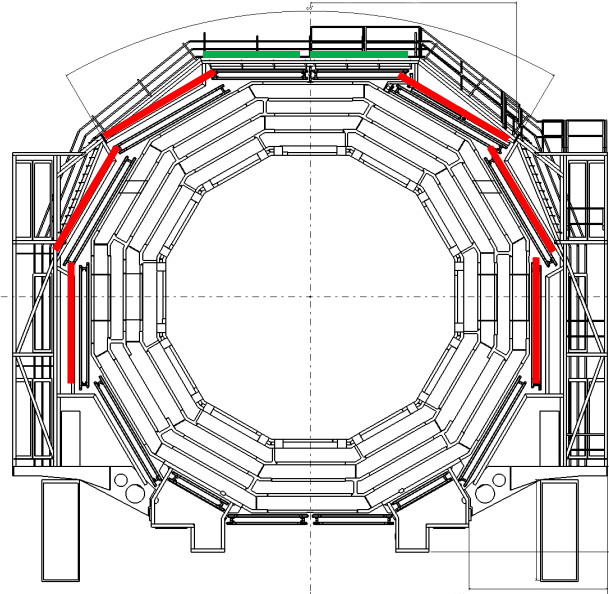}
\caption{%
    Left:\ magnitudes of the linear dependence between the currents from each DT chamber of the MB and the LHC instantaneous luminosity~\cite{CMS:DP-2020-011}.
    Results are computed after the optimization of the operational working points.
    Right:\ transverse view of an MB wheel highlighting the layout of the MB4 shield, as installed in W$-2$, $-1$, $+1$, and $+2$.
    Red (green) lines represent shield layers consisting of thin (thick) cassettes.
}
\label{fig:dt:shield}
\end{figure}

With the aim of designing such shields, absorbing layers were installed on top of the MB4 stations in sector 4 of W+2 and W$-2$ during \Run2.
Different configurations in terms of material and thickness of absorbers, suggested by simulations of the radiation field in the cavern, were tested each year from 2015 through 2018, and the impact in terms of background reduction was studied.
An example of these studies is shown in Fig.~\ref{fig:dt:shield} (left), where the magnitude of the linear dependence of the currents in each chamber with respect to the LHC instantaneous luminosity is presented for the 2018 run.
The impact of the test shielding configurations installed over the MB4 stations in sector 4 of W$\pm2$ is clearly visible.
Based on the results of these investigations and considering the outcome of simulations of the mechanical stress that different shield setups would induce on the supporting structures, a layout was proposed that is expected to reduce the background by about a factor 2.

The installation of the final DT MB4 shields was performed during LS2 and completed in October 2020.
A schematic view of the shielding layout is presented in Fig.~\ref{fig:dt:shield} (right).
The figure shows the shield installed over W$-2$, $-1$, $+1$ and $+2$.
In those wheels, shielding cassettes consisting of 9\mm of lead and 9\cm of borated polyethylene (thick cassettes) are put on top of sector 4 stations (green lines), whereas cassettes consisting of 9\mm of lead and 3\cm of borated polyethylene (thin cassettes), cover sectors 1, 2, 3, 5, 6, and 7.
Due to mechanical restrictions imposed by the presence of services from the vacuum tank and inner detectors, the shields deployed to cover the central wheel have a different layout.
In this case, no shielding is applied on top of sector 4, whereas, in the other six sectors, shields consisting of thin cassettes only cover the lowest part of half of each chamber.
The deformation on the supporting structures induced by the MB4 shields, which have a weight of approximately 40 tons in total, was measured accurately at the weakest points in the structure and found to be always smaller than 1\mm, in good agreement with finite-element calculations.

In summary, during \Run2, the DT off-detector electronics underwent a set of upgrades that improved the performance of both the readout and trigger.
Strategies to increase the detector longevity were also put in place to maximize the DT performance through the end of HL-LHC running.
Following a thorough recommissioning over LS2, the DT system successfully entered \Run3, showing offline and online tracking performances that are remarkably consistent with the ones achieved at the end of the previous LHC run.

\subsection{Cathode strip chambers}
\label{sec:csc}

\subsubsection{General description}

\begin{figure}[!ht]
\centering
\includegraphics[width=0.27\textwidth]{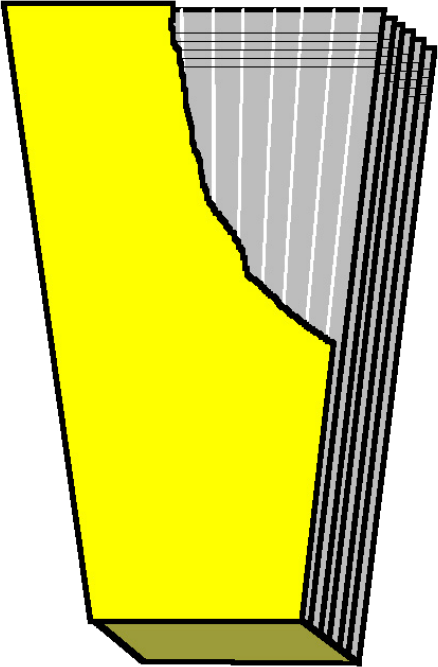}%
\hspace*{0.04\textwidth}%
\includegraphics[width=0.6\textwidth]{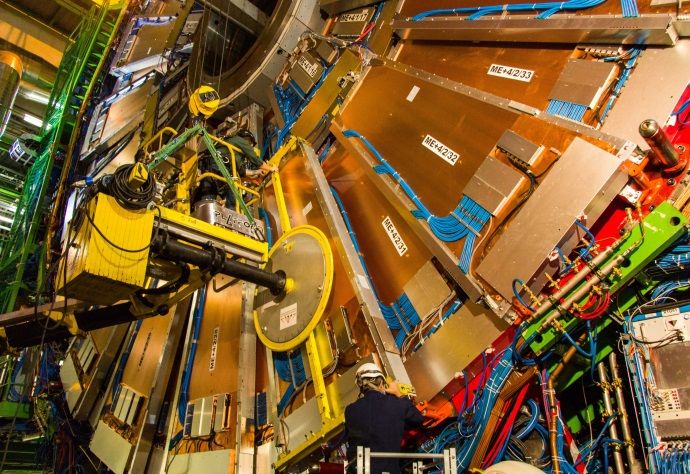}
\caption{%
    Left:\ layout of a CSC chamber, from Ref.~\cite{Paneva:2019baf}, with seven trapezoidal panels forming six gas gaps. Only a few wires (lines running from left to right) and strips (gray band running from top to bottom) on the upper right corner are shown for illustration.
    Right:\ installation of the outer CSC chambers (ME4/2) during LS1, from Ref.~\cite{CMS:PHO-MUON-2015-001}.
}
\label{fig:csc:layout}
\end{figure}

The cathode strip chambers (CSCs) are multiwire proportional chambers with a finely segmented cathode strip readout.
The strips run radially in order to measure the muon position in the bending plane, the plane perpendicular to the colliding beams axis, while the anode wires are oriented in azimuth and provide a coarse measurement in the radial direction.
A precise measurement of the muon coordinate in the azimuthal direction is obtained from charges induced on the cathode strips.
Each CSC module consists of six gas layers, each layer having a plane of radial cathode strips and a plane of anode wires running perpendicular to the strips.
Figure~\ref{fig:csc:layout} shows the CSC physical layout and a photograph of the installation of some CSCs into the CMS detector.

The gas mixture is 40\% Ar, 50\% \COtwo, and 10\% \CFfour.
The \COtwo component is a nonflammable quencher needed to achieve large gas gains, while the main function of the \CFfour is to prevent anode aging caused by polymerization processes on the wires.
Argon is the working gas that is ionized by traversing charged particles.
The primary design consideration for the CSC detectors is the ability to provide good spatial and temporal resolution for triggering and identification of a muon.
The typical position and time resolutions are 50--140\mum, depending on chamber type, and 3\ns per chamber, respectively.

Figure~\ref{fig:muon:quadrant} shows the geometry of the CSC system.
At the start of the LHC operation, a total of 468 trapezoidal CSC modules, placed between the steel magnetic flux-return plates, were arranged into four disks (stations).
The first station, closest to the interaction point, is further segmented into three rings (ME1/1, ME1/2, and ME1/3), while stations 2 through 4 are separated into just two rings.
The rings closest to the beam line (named ME1/1, ME2/1, ME3/1, and ME4/1, or collectively as ME1234/1) are subject to the highest particle rates.
Each of the chambers in the ME1/1 ring contains strips split at $\abseta=2.1$, which are read out separately and denoted ME1/1a ($2.1<\abseta<2.4$) and ME1/1b ($1.6<\abseta<2.1$).
Some key parameters of the CSCs are summarized in Table~\ref{tab:csc:summary}.

\begin{table}[!ht]
\centering
\topcaption{%
    Key parameters for different types of CSCs.
}
\label{tab:csc:summary}
\renewcommand{\arraystretch}{1.1}
\begin{tabular}{lcccc}
    & ME1/1 & ME1/23 & ME234/1 & ME234/2 \\
    \hline
    Wire diameter [$\mu$m] & 30 & 50 & 50 & 50 \\
    Wire spacing [mm] & 2.5 & 3.2 & 3.1 & 3.2 \\
    Strip width (narrow) [mm] & 3.2 & 6.6--11.1 & 6.8--8.6 & 8.5 \\
    Strip width (wide) [mm] & 7.6 & 10.4--14.9 & 15.6 & 16.0 \\
    Gap between strips [mm] & 0.35 & 0.5 & 0.5 & 0.5 \\
    Angle subtended by each strip [mrad] & 2.96--3.88 & 2.15--2.32 & 4.65 & 2.33 \\
    Gas gap [mm] & 7 & 9.5 & 9.5 & 9.5 \\
    Operating HV [V] & 2900 & 3600 & 3600 & 3600 \\
\end{tabular}
\end{table}

\begin{figure}[!htp]
\centering
\includegraphics[width=0.7\textwidth]{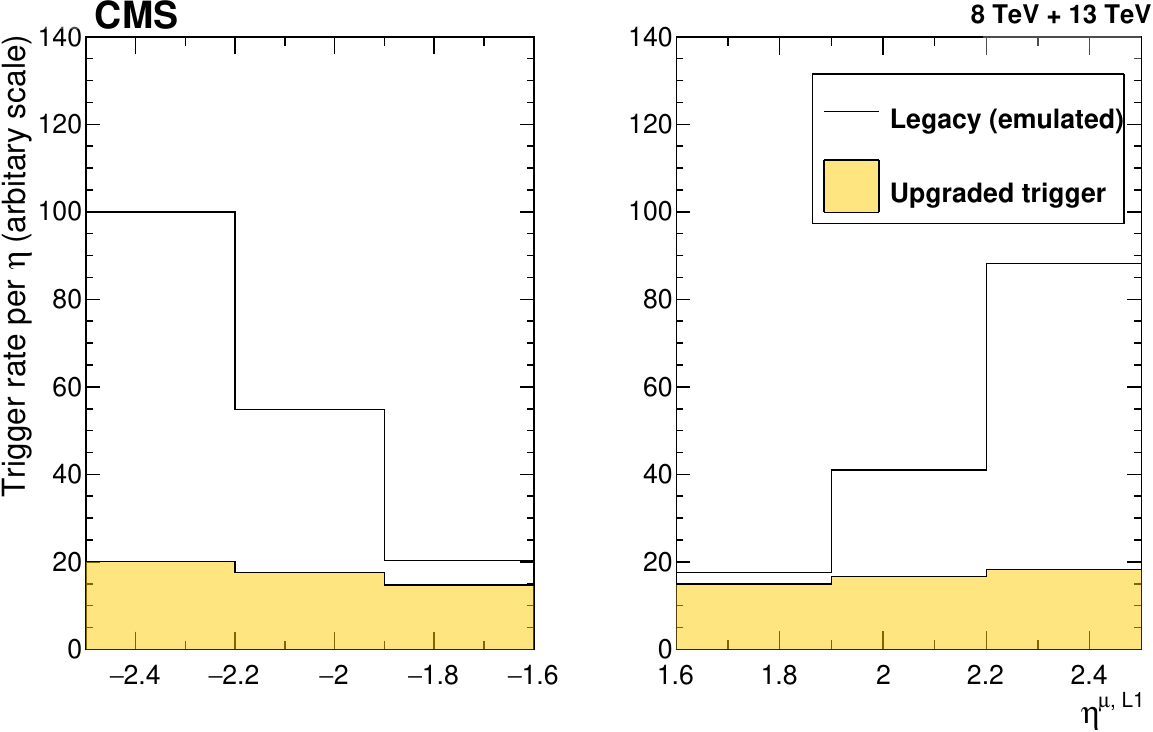}
\caption{%
    Emulation of the \Run1 algorithms compared to the upgraded \Run2 algorithms, as a function of the L1 muon $\eta$.
    The most common L1 single-muon trigger threshold used in 2017 was $\pt^{\mu\text{,L1}}\geq25\GeV$.
}
\label{fig:csc:triggerrate}
\end{figure}

The ME234/1 chambers each cover 20\de\ in $\phi$; all others cover 10\de.
All chambers, except for the ME1/3 ring, overlap by five strips at each edge and hence provide $\phi$ coverage without gaps.
The original design included 72 ME4/2 chambers that were descoped in the initial construction of the CSC system, and instead were built and installed during LS1 that occurred during 2013--2014.
The presence of ME4/2 chambers provides an additional track segment of six hits along the muon trajectory.
This allowed for a more robust muon \pt assignment in the L1 trigger described in Section~\ref{sec:l1trigger}, so that the rate with a given \pt threshold in the $1.2<\abseta<1.8$ region decreased during \Run2, as shown in Fig.~\ref{fig:csc:triggerrate}.

As illustrated in Fig.~\ref{fig:muon:quadrant}, in the $1.1<\abseta<2.4$ region, almost all possible muon paths cross at least three CSC chambers.
In the $0.9<\abseta<1.1$ overlap region, a combination of DT and CSC modules also typically provide at least three position measurements along muon paths.
Moreover, the four layers of RPCs between the CSCs provide additional muon hits at the trigger level in the $0.9<\abseta<1.6$ region.
The extension of the RPC layers and the addition of GEM detectors (GE1/1 in \Run3) further enhance the muon triggering and reconstruction in the $1.6<\abseta<2.2$ region.
The high redundancy of the muon system is a central feature of the design and is responsible for the robustness of the muon triggering and reconstruction in CMS.

\begin{figure}[!ht]
\centering
\includegraphics[width=\textwidth]{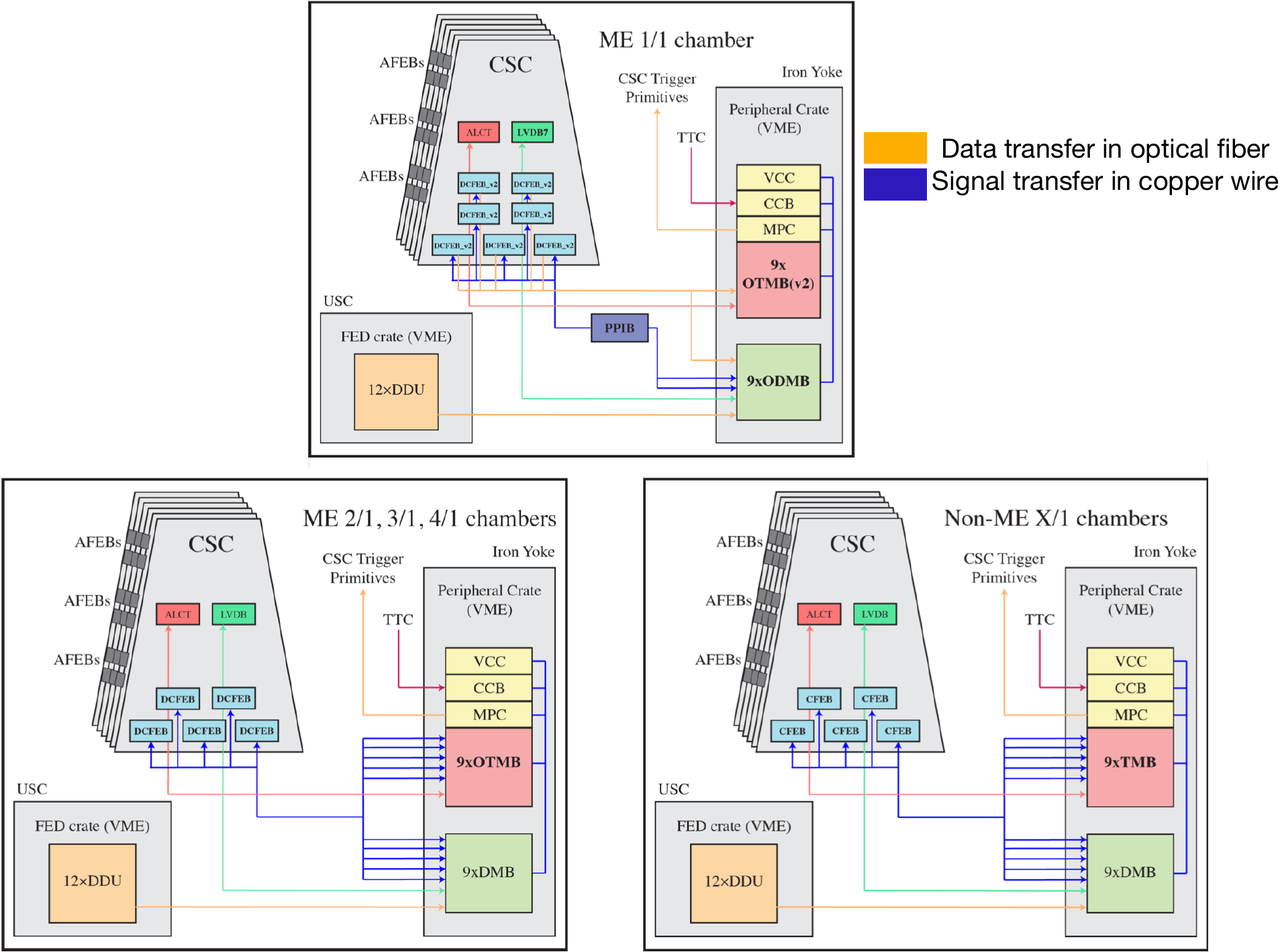}
\caption{%
    Schematic of the CSC electronics readout systems for \Run3:\ ME1/1 (upper), ME234/1 (lower left), all other chambers, ME1234/2 and ME1/3 (lower right). Figure from Ref.~\cite{Haddad:2022xfb}.
}
\label{fig:csc:electronics}
\end{figure}

The present electronics readout system of all CSC chambers, shown schematically in Fig.~\ref{fig:csc:electronics}, share the same architecture as in \Run1.
This is briefly summarized below, with a more detailed description available in Ref.~\cite{CMS:Detector-2008}.
On each CSC chamber, an anode local charged track (ALCT) board finds patterns among the six-layer anode hits (sampled by the ALCT board, based on the digitized pulses sent by the on-chamber anode frontend boards) that are consistent with muon tracks, and then assigns a quality rank to each of them.
These patterns are called ALCTs.
The ALCT board can then send up to two such tracks per each bunch crossing (BX) with the highest quality to the trigger motherboard (TMB).
Upon receiving a level-1 accept (L1A) signal from the trigger, coinciding within 3 BXs with an ALCT, the ALCT board sends all wire hits and ALCTs that have been found in a predefined time window to the data acquisition motherboard (DMB).
Both the TMB and DMB are located in VME crates mounted on the edge of the endcap steel yoke.

The cathode signals are processed by another set of on-chamber electronics, the cathode frontend boards (CFEB).
Each CFEB amplifies the signals from 16 strips on each of six layers of a CSC, and sends fast trigger information and charged particle hits localized within a half-strip width, as assessed by comparator ASICs (comparator data), to the TMB.
On receiving an L1A coinciding within 5 BXs with a cathode local charged track (CLCT), the CFEB digitizes the strip signal waveforms over a 400\ns long period, which is sampled every 50\ns by a 12-bit ADC, and sends the digitized samples to the DMB.
The transfer of CFEB data to the TMB and DMB comprises the trigger and data paths of the strip signals, respectively.
During \Run1, each CSC had five CFEBs, except for ME1/3, which had four per CSC (since the ME1/3 CSCs have only 64 strips).

The low-voltage power for the on-chamber electronics is distributed at the appropriate voltage levels by the on-chamber low-voltage distribution board (LVDB).

The TMB builds cathode hit patterns into CLCTs and finds coincidences with anode hit patterns that form ALCTs to make local charged tracks (LCTs), also called the CSC trigger primitives.
It sends the two LCTs with the highest quality per BX to the muon port card (MPC) and, on receiving an L1A, to the DMB.
The DMB controls the CFEBs on a chamber and collects anode, cathode, and trigger information to send to the detector-dependent unit (DDU), on the arrival of an L1A.
The DDU, situated in the frontend driver crates located in the underground service cavern, collects data from 15 DMBs in the CSC system and sends the information through the global DAQ path.

The MPC, situated in the VME crate, collects LCTs from each of up to nine TMBs in a trigger sector, and sends these trigger primitives to the muon track finders described in Section~\ref{sec:l1trigger:muon}.
There is one MPC per peripheral VME crate.
The operation of the VME crate is supported by the clock-control board (CCB) and custom VME crate controller (VCC).
The CCB serves as the interface between the CSC system and the trigger control and distribution system (TCDS) of CMS.
The VCC receives VME commands from the control room and distributes them to the other boards in the peripheral crate via the backplane.

\subsubsection{Upgrade of the CSC system since \Run1}

While the CSC system operated stably throughout \Run1 and \Run2, some CSC readout electronic boards needed to be upgraded in order to handle the expected longer latency and more stringent trigger requirements at the HL-LHC.
Specifically, if the electronics were not altered, the longer latency requirements would fill up and overflow the pipelines of the frontend boards in certain CSC stations, while the higher L1 trigger rates would overwhelm the output bandwidth of various on- and off-chamber electronics boards.
To reduce the installation load during LS3, the bulk of the CSC electronics upgrades that required chamber access (dismounting and re-installation) was already performed during LS2.

\begin{figure}[!ht]
\centering
\includegraphics[width=0.575\textwidth]{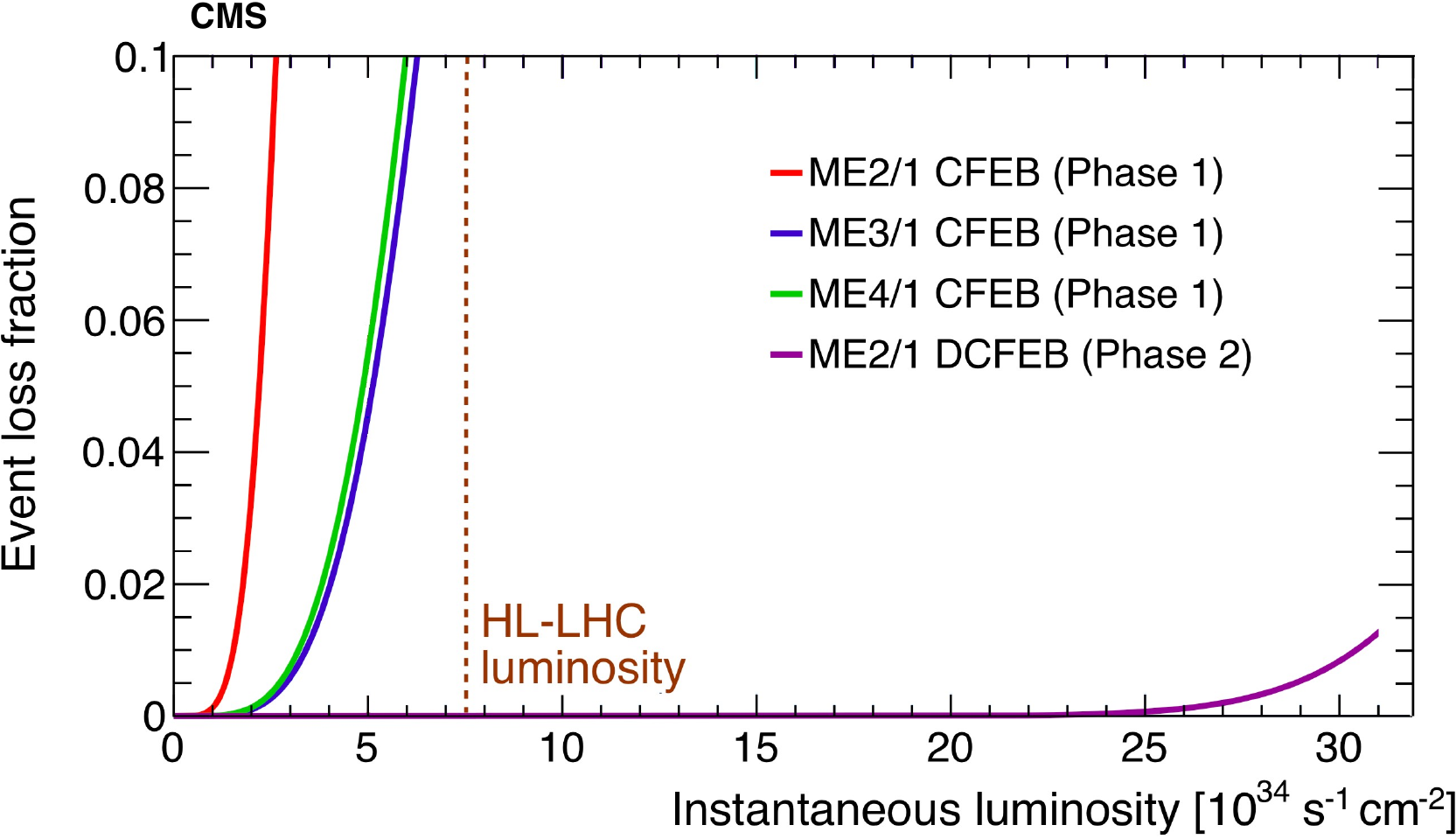}%
\hfill%
\includegraphics[width=0.385\textwidth]{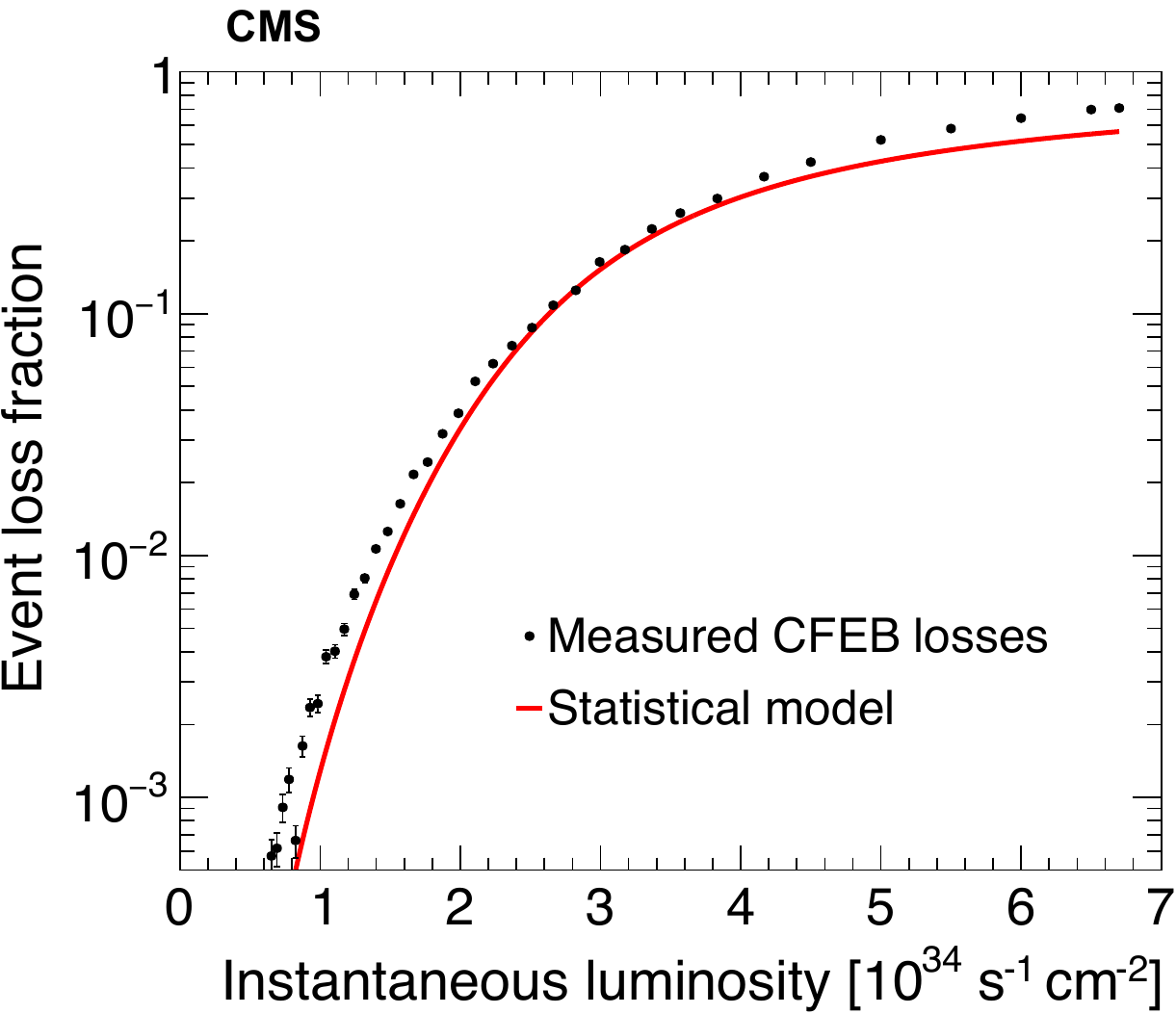}
\caption{%
    Left:\ event loss fraction as a function of instantaneous luminosity for different type chambers after different upgrades.
    The vertical dashed brown line indicates the design HL-LHC luminosity.
    Right:\ event loss rate measured in a CFEB under HL-LHC conditions for an ME2/1 chamber, compared to the statistical model~\cite{CMS:TDR-016}.
}
\label{fig:csc:cfebloss}
\end{figure}

The CFEBs use switch capacitor arrays to store the charge induced on the cathode strips.
These capacitor arrays are capable of storing 96 charge measurements (corresponding to six events worth of data) during the L1 trigger latency.
As mentioned above, the digitization of the analog signals and subsequent readout by the DMB only happens when a CFEB receives an L1A that is in coincidence with a CLCT.
As a consequence, for CSCs that are closest to the beam (ME1234/1) where the background rate is high, frequent memory overflows and large data losses are expected at the HL-LHC trigger latency.
Figure~\ref{fig:csc:cfebloss} shows the average event loss fraction for ME234/1 rings as a function of the instantaneous luminosity.
These curves are based on a statistical model that has been verified by measuring the loss rates in bench tests that emulate the expected background and L1 trigger rates.
These results show that data loss would be a severe issue at the HL-LHC with the original CFEBs.
The outer CSC chambers will not suffer from these problems because the trigger primitive rates in these rings are lower than those in the inner rings by factors of more than 3.

The CFEBs on ME1/1 have already been vulnerable to data loss at the highest instantaneous luminosities (up to $2\times10^{34}\percms$) in \Run2.
To mitigate such losses, the four CFEBs on the ME1/1b section of the chambers were replaced by four digital cathode frontend boards (DCFEBs) during LS1.
The DCFEBs use fast flash 12-bit ADCs that continuously digitize the cathode signals at 20\MHz, as well as having more powerful Virtex-6 FPGAs with large internal memory resources.
The resulting digital pipeline can hold up to 700 events, and thus there should be negligible dead time at the HL-LHC.
Two optical links running at 3.2\Gbs are employed per DCFEB to transmit the raw data for DAQ and the fast comparator data for building the L1 trigger to the new optical data acquisition motherboard (ODMB) and optical trigger motherboard (OTMB), respectively, which replaced the original DMB and TMB.

\begin{figure}[!ht]
\centering
\includegraphics[width=0.48\textwidth]{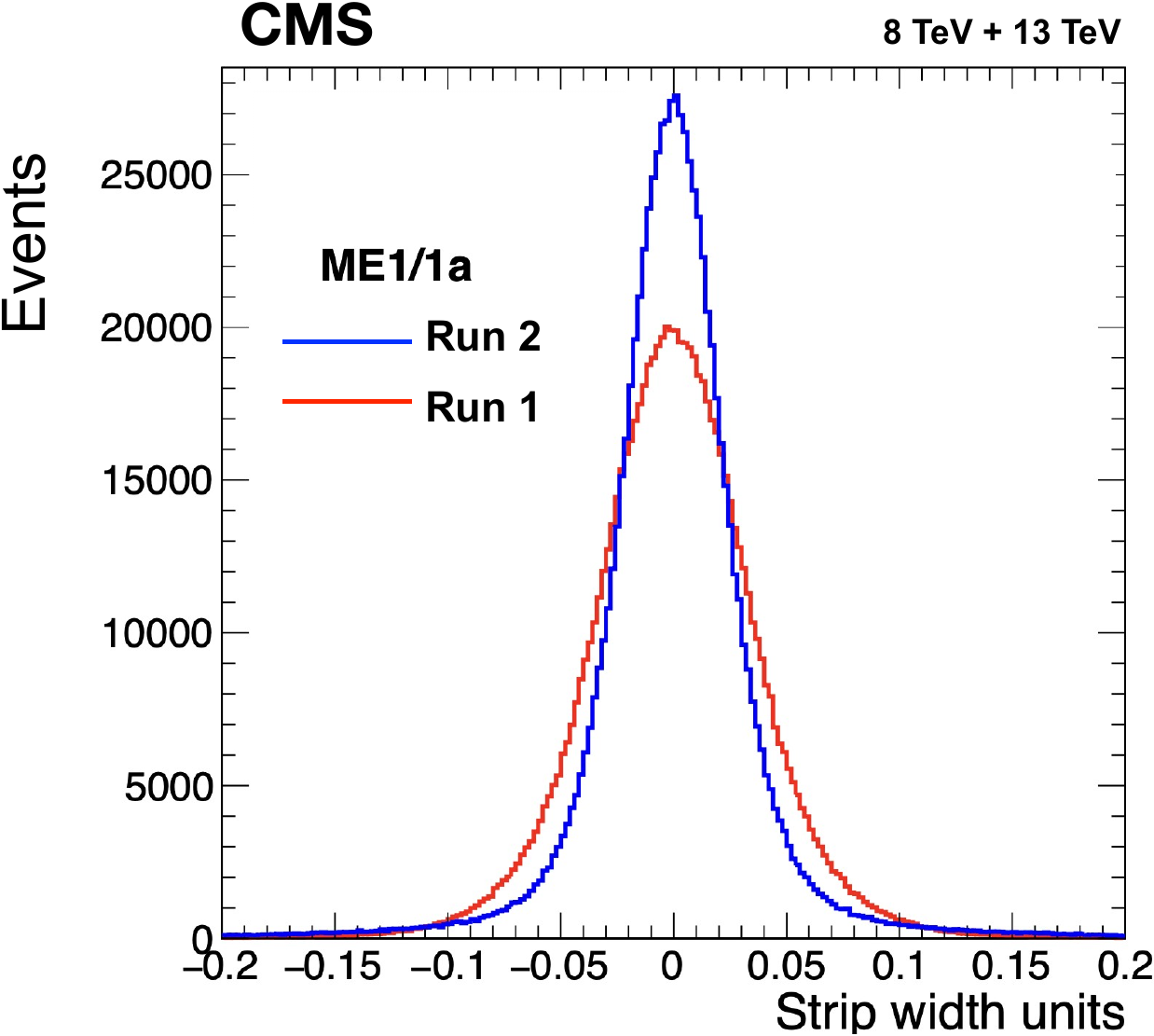}
\caption{%
    Difference between the position of a reconstructed hit in layer 3 of an ME1/1a chamber and the position obtained by fitting a segment with hits from the other five layers for \Run1 (red) and \Run2 (blue).
    The spatial resolution is improved by 27\% from $\sigma=64\mum$ in \Run1 to 46\mum in \Run2.
    This is due to the removal of the triple-grouping of strips in ME1/1a, which reduces the capacitance and hence the frontend noise.
}
\label{fig:csc:me11aresolution}
\end{figure}

In \Run1, due to cost constraints, the 48 strips in the ME1/1a section of the chamber were joined every 16 strips into groups of three, resulting in 16 readout channels served by one CFEB.
To remove the ambiguity in triggering and reconstruction, the grouping of the ME1/1a strip readout was removed during LS1, and the single CFEB was replaced by three DCFEBs for strip readout.
The ungrouping of the strips in ME1/1a leads to an improvement in spatial resolution of about 20\% (Fig.~\ref{fig:csc:me11aresolution}), as well as a decrease in the L1 trigger rate by at least a factor of 3 (Fig.~\ref{fig:csc:triggerrate}).

Operational experience during \Run2, and additional radiation tests, showed that the programmable read-only memory (PROM) employed in the current DCFEB may not withstand the radiation dose expected at the HL-LHC.
Thus, the new DCFEBs (DCFEBv2s) were designed with an option for remote programming of their FPGAs using the CERN-designed GBTx ASIC.
The performance of the two optical transceivers on the original DCFEBs has been occasionally unreliable, particularly due to single-event upsets (SEUs).
While mitigation measures in software and firmware have improved the reliability of these transceivers in the DCFEBv2s, they were replaced by the VTTx, a radiation-hard twin optical transmitter designed by CERN.
These new DCFEBv2s were installed in the ME1/1 chambers during LS2, thus mitigating the risks associated with the longevity of their PROMs.

The DCFEBs used in the ME1/1 ring during \Run2 were moved to the chambers in the ME2/1, ME3/1, and ME4/1 rings, where radiation levels are lower.
In \Run3, although the ME234/1 chambers are fitted with DCFEBs, the new ODMB is not scheduled to be installed before LS3, so these DCFEBs still send data to the DAQ via the original DMBs.

The ALCTs store raw wire-hit information in a pipeline within the FPGA while waiting for an L1A.
The ALCTs installed before \Run1 would also suffer significant data loss during HL-LHC operation.
This is because they do not have sufficient FPGA memory resources to hold all raw hit information before sending them to the DMB during the HL-LHC L1 trigger latency.
Moreover, the output bandwidth for the boards in the inner rings (ME1234/1) would not be capable of handling the expected HL-LHC data rates.
Both of these problems were solved by replacing the mezzanine cards in the affected chambers.

During LS1, the ME1/1 chambers were equipped with new ALCT mezzanine cards having more powerful FPGAs (Spartan 6), with 9--12 times more memory than those used previously (Virtex-E).
This, in turn, allows the pipeline to be deep enough to satisfy the HL-LHC latency requirements.
Similar ALCT mezzanine cards were also installed on the ME4/2 chambers during the same period.

During LS2, as a second phase of the ALCT upgrade, the ALCT mezzanine cards originally installed in ME1/1 during LS1 were moved to ME1/3, and all other mezzanine cards, except for those used in the ME4/2 rings, were replaced.
The new ALCT mezzanine cards are largely based on the ALCT design used for cards installed in ME1/1 during LS1, with a few exceptions discussed below.
The ALCT mezzanine cards servicing the ME234/1 rings use two new VTTx optical transmitters, each running at 3.2\Gbs, to increase the output bandwidth for the expected HL-LHC data rates.
The new ME1/1 ALCTs, in addition to a different FPGA from those used for the ME234/1 chambers, have a VTRx transceiver instead of two VTTx transmitters on each mezzanine card.
This allows for the same remote FPGA programming option (based on the GBTx ASIC) that the DCFEBv2s have, thus mitigating any risks associated with the aging of the PROMs in that ring.
From \Run4, the VTRx transceiver (as well as the VTTx transmitters from the ME234/1 ALCTs) will be connected to an updated ODMB board that transmits data to the CMS DAQ over an optical link at 4.8\GBs.
The new mezzanine cards serving the ME234/2 chambers have the same FPGA as those used for the ME1/1 ALCTs, but they will send data to the DMB through copper links even in the future runs due to the lower expected data rate.
Since the new design maintains backwards compatibility, the anode raw data sent to the DAQ from the ME1234/1 ALCTs can be transmitted to the DMBs via the copper path during \Run3, and to the new ODMBs via the optical link(s) from \Run4 onwards.

The DCFEBs that were installed on the ME1/1 chambers in LS1 transmitted data to the CMS DAQ and comparator data to the CMS trigger at 3.2\Gbs via optical fibers.
Since the previously installed TMBs and DMBs were designed to work with the CFEBs and have no optical receivers, the TMBs and DMBs for the ME1/1 chambers were upgraded to OTMBs and ODMBs during LS1 as well.
They receive optical data from the seven DCFEBs and use a more powerful FPGA.
During LS2, all TMBs used for ME234/1 chambers were upgraded to OTMBs to match the upgrade of their CFEBs to DCFEBs.
New OTMBv2s were produced for the ME1/1 and ME2/1 chambers.
These are almost identical to the OTMBs installed in LS1, except that obsolete components were replaced and the ability to build trigger primitives including GEM data was added, as described in Section~\ref{sec:gem:csc-gem-trig}.
The OTMBs serving the ME1/1 chambers during \Run2 were relocated to the ME34/1 rings.

The new electronics consume more power than the old so the low-voltage (LV) systems were upgraded to provide higher maximum currents at the appropriate voltage levels to ensure reliable operation of the CSCs at the HL-LHC.

For \Run1 and \Run2, the CSC high-voltage (HV) system comprised two independent subsystems:\ a custom one providing HV to the non-ME1/1 chambers (11\,016 channels), and a commercial subsystem supplying the ME1/1 chambers (432 channels)~\cite{Caen:2013web}.
In both systems, the voltage can be regulated on all channels individually, and the currents drawn are continuously monitored.

\begin{figure}[!ht]
\centering
\includegraphics[width=0.48\textwidth]{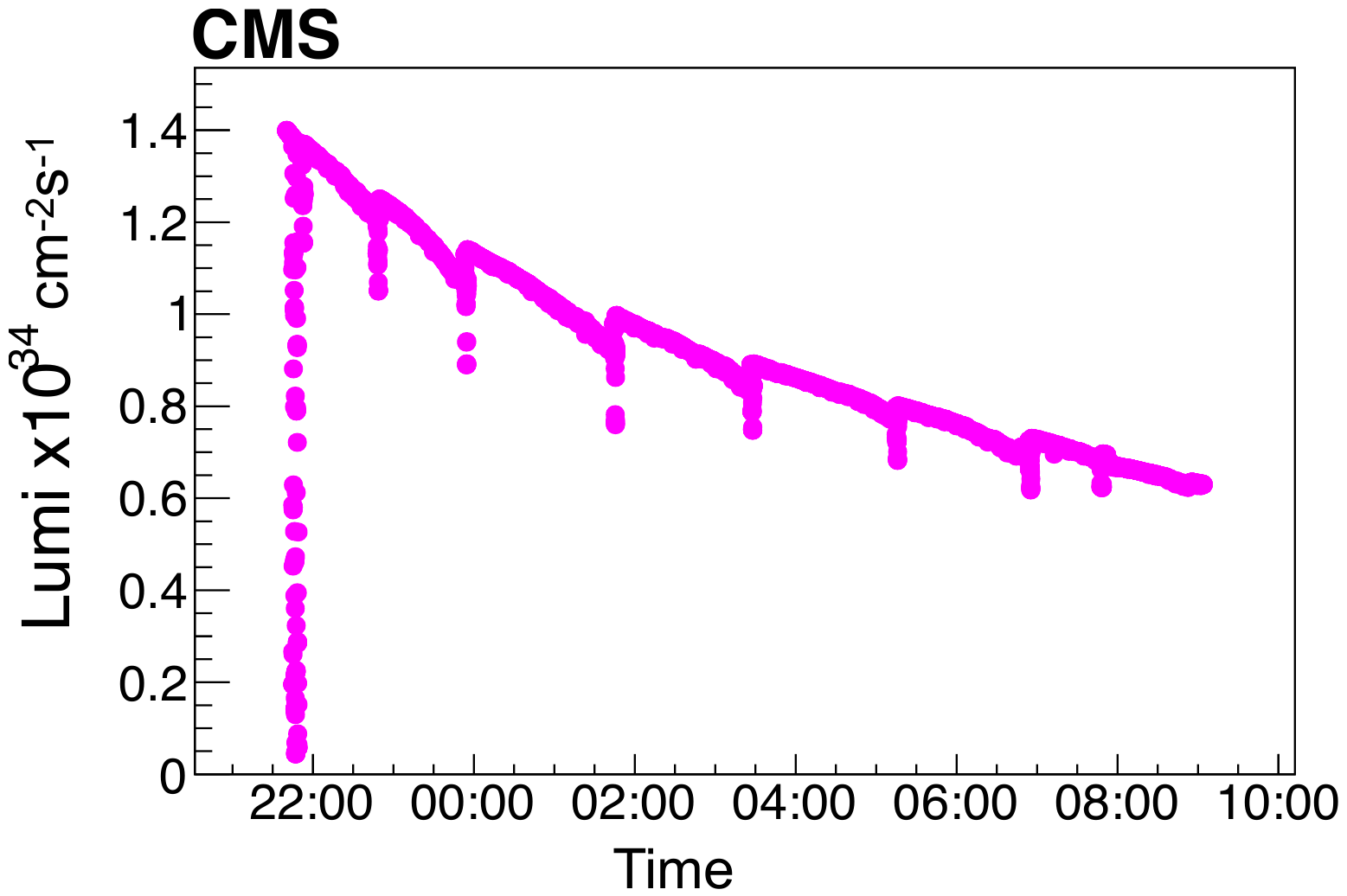}%
\hfill%
\includegraphics[width=0.48\textwidth]{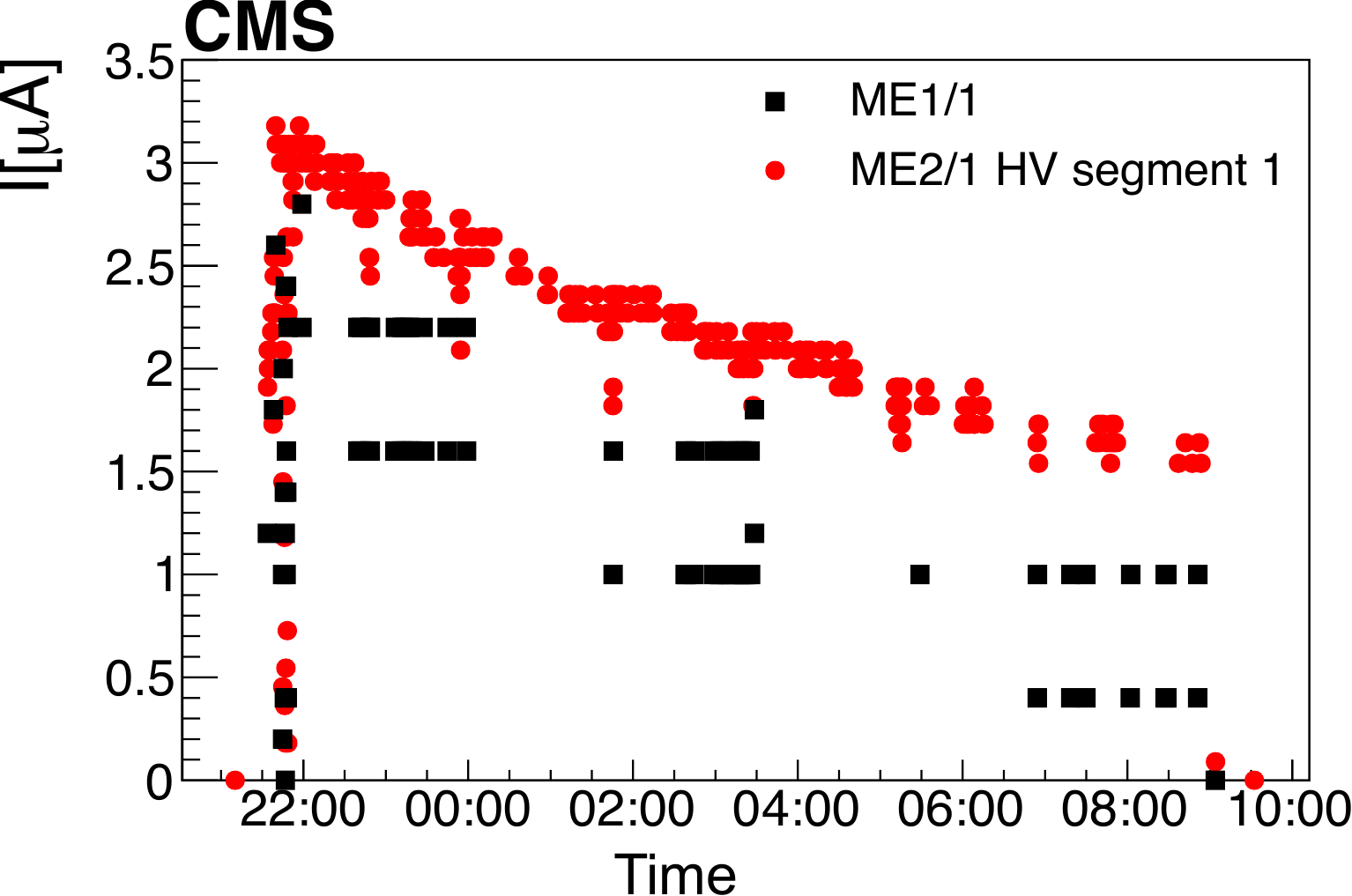}
\caption{%
    Left:\ instantaneous luminosity versus time for one of the LHC fills in 2016.
    Right:\ current (in nA) in an ME2/1 chamber (the HV segment closest to the beam) for the same fill, as measured with the custom-made HV subsystem used for non-ME1/1 chambers; current (in $\mu$A) in an ME1/1 chamber (one plane) for the same fill, as measured with the commercial HV subsystem.
}
\label{fig:csc:hvcurrent}
\end{figure}

The commercial system is no longer supported by the manufacturer and does not allow monitoring of the currents with as fine a precision as the HV system used for the non-ME1/1 chambers, as can be seen in Fig.~\ref{fig:csc:hvcurrent}.
For these reasons, during LS2, the commercial HV subsystem was replaced by a custom-made HV one, extending the system used for the non-ME1/1 CSCs.
Besides ensuring precise current measurements, such a replacement makes the entire CSC HV system homogeneous and simpler to operate and maintain.

\subsubsection{Longevity studies}

The CSC chambers will experience a much higher radiation dose than the muon chambers located in the barrel region.
The highest neutron fluence and total ionization dose, corresponding to 10 years of HL-LHC running, will reach a level of $1\times10^{14}\Neq$ and 10\Gy, respectively.
The total charge released in the gas volume per unit wire length will be about 0.3\Ccm for the CSCs closest to the beam line, corresponding to an integrated luminosity of 4000\fbinv, as maximally achievable by the end of the HL-LHC.

\begin{figure}[!ht]
\centering
\includegraphics[width=0.48\textwidth]{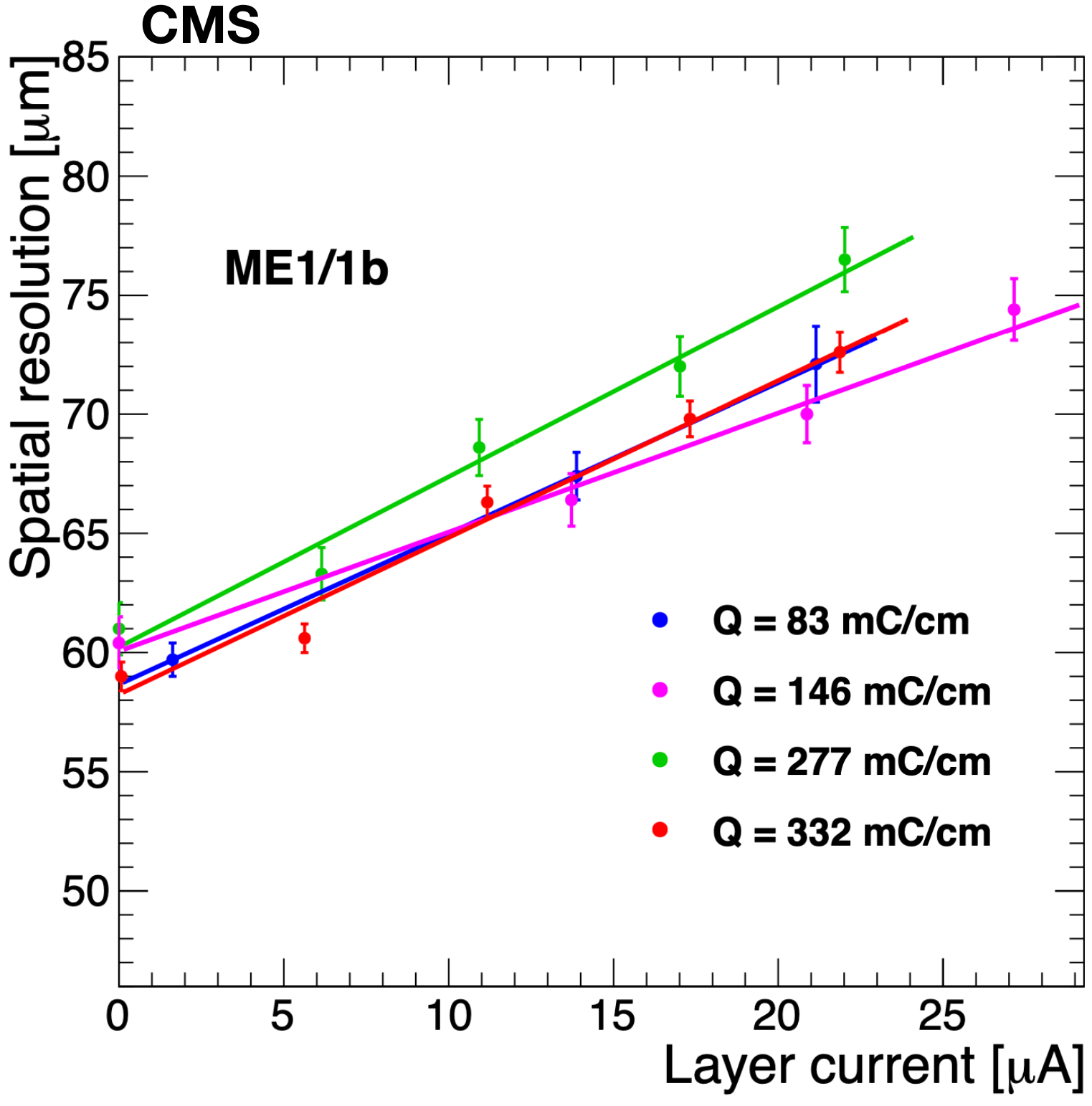}%
\hfill%
\includegraphics[width=0.49\textwidth]{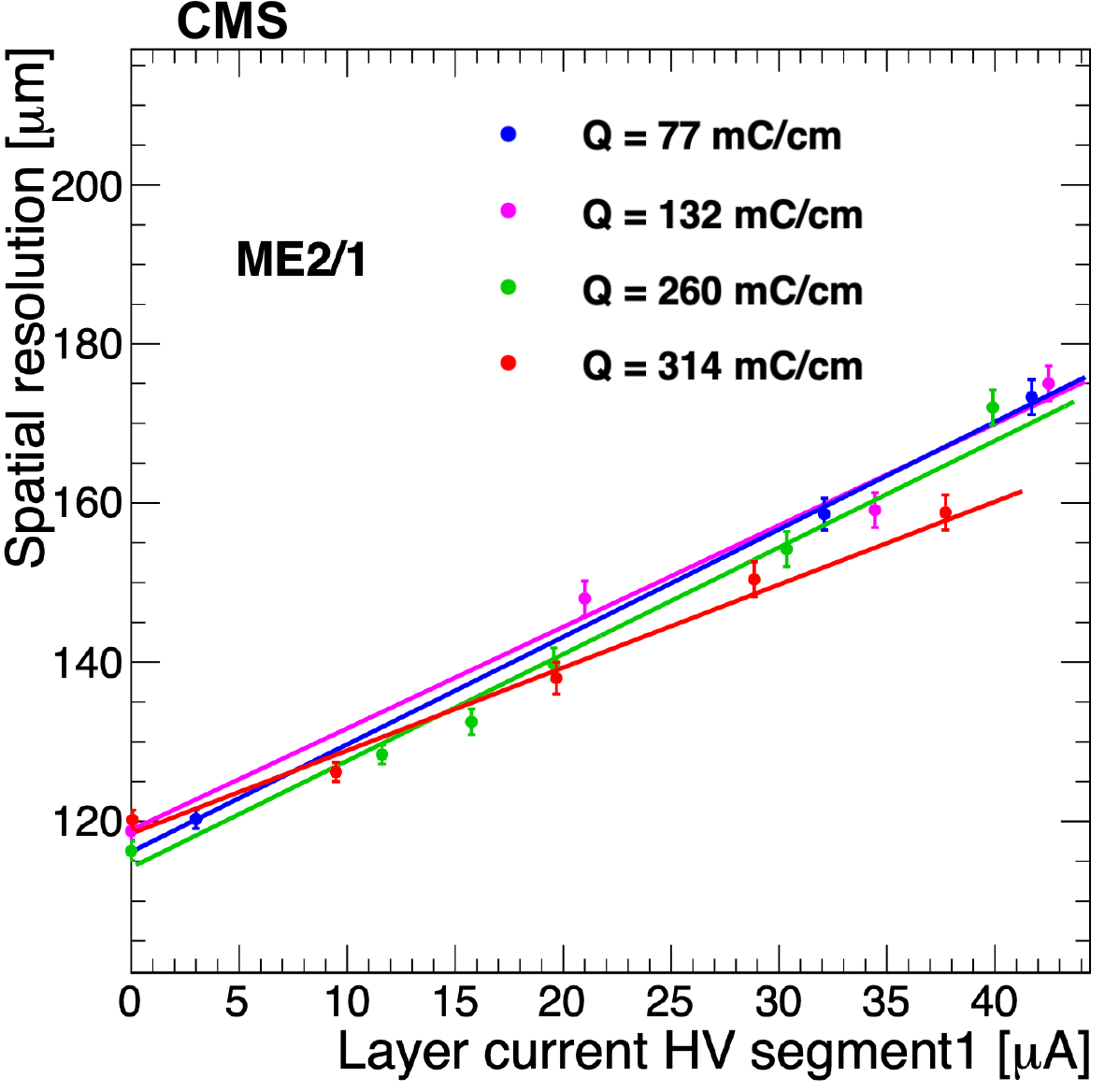}
\caption{%
    Spatial resolutions for an ME1/1b (left) and ME2/1 (right) chamber using a muon beam while being uniformly illuminated by a \onethreesevenCs photon source to simulate the background from high luminosity \pp collisions, as a function of the average current per layer in one HV segment.
    The results for four different accumulated charges per unit wire length are shown, along with linear fits to each set.
}
\label{fig:csc:gifstudies}
\end{figure}

Longevity tests of CSCs have been underway since 2016 at the CERN GIF++ facility~\cite{Pfeiffer:2016hnl}.
Typical ME1/1 and ME2/1 chambers have been used.
The ME1/1 chambers differ from those of the other rings, of which an ME2/1 chamber is typical, in having thinner wires,  smaller gas gaps, and being constructed of different materials.
The combination of a photon flux at high rates provided by the GIF++ gamma source and a muon beam periodically provided by the CERN SPS allows the study of not only the longevity of the CSCs but also their performance in an HL-LHC-like environment.
The CSCs installed in GIF++ have the same chain of trigger and data acquisition electronics as those installed in CMS during \Run2.
The working gas mixture of 40\% Ar, 50\% \COtwo, and 10\% \CFfour is supplied with a closed-loop gas system replicating the CSC gas system at CMS.
The gas flow and gas refreshing rate are kept at the same level as for nominal CSC operation.
While operating at the nominal gas gain under an intense irradiation for more than a year, the chambers accumulated a charge per wire length of 0.33\Ccm; this corresponds to about 1.1 and 1.7 times the amount of charge expected for ten years of HL-LHC running for the ME1/1 and ME2/1 chambers, respectively.
Figure~\ref{fig:csc:gifstudies} shows a modest deterioration of spatial resolution at higher background rates and no deterioration with an accumulated charge up to 0.33\Ccm, the measurements were made with the nominal CSC gas mixture.
The average current is proportional to the background intensity, which can be adjusted using a set of filters.
The expected typical currents for \Run3 (at $2\times10^{34}\percms$) and HL-LHC (at $7.5\times10^{34}\percms$) background conditions are about 4 and 30\muA for the ME1/1 chamber, and 4.4 and 22\muA for the ME2/1 chamber.
No significant deterioration of key chamber parameters such as gas gain, detection efficiency, spurious signal rates, strip-to-strip resistance, or dark currents has been observed.
Thus, it is expected that the CSCs themselves will function throughout the planned HL-LHC operation.

\begin{figure}[!ht]
\centering
\includegraphics[width=0.48\textwidth]{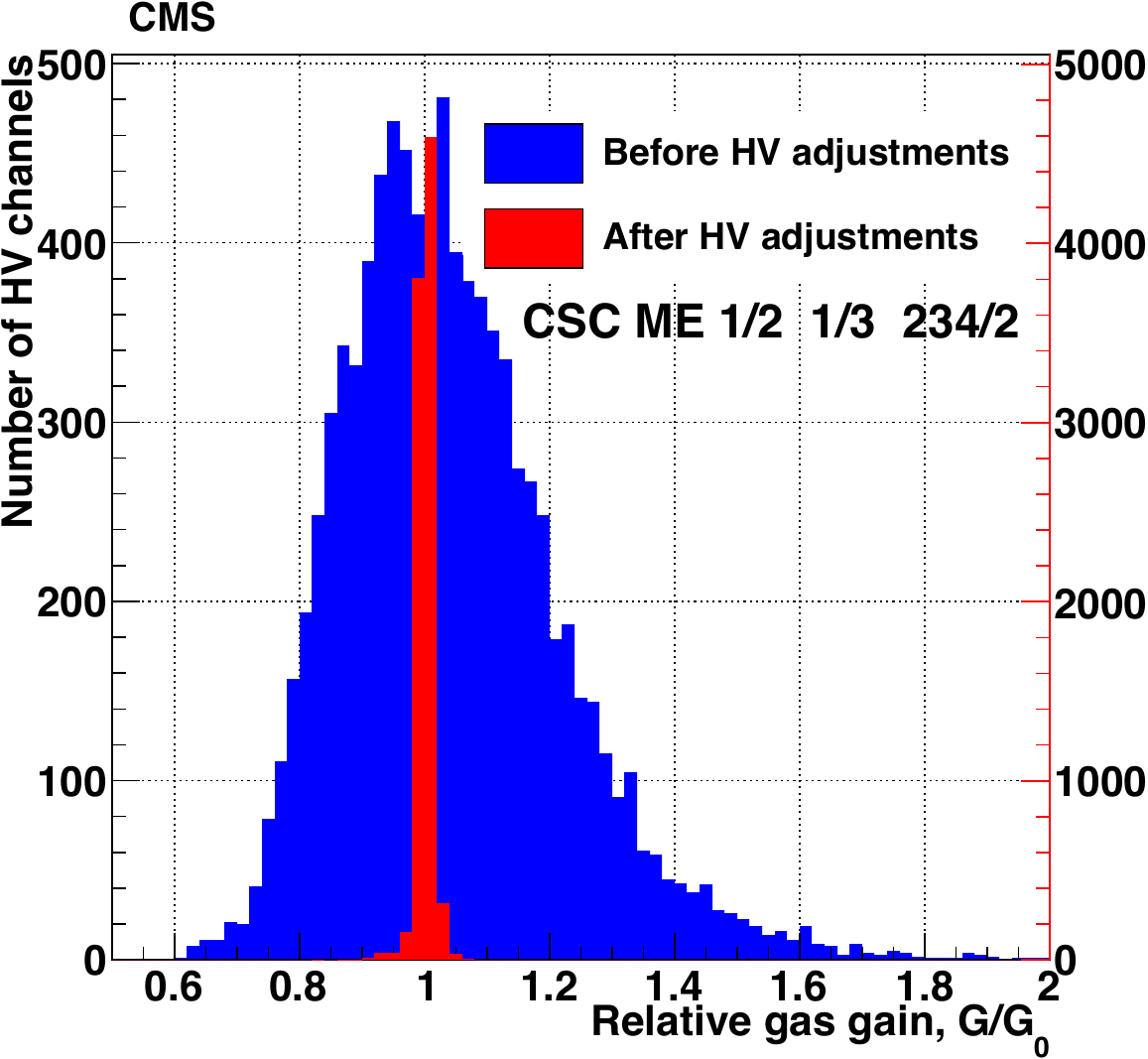}
\caption{%
    Relative gas gain distribution in CSCs before and after the gas gain equalization campaign in 2016~\cite{CMS:TDR-016}.
    Each entry in the histogram presents the mean value of gas gain in each HV channel.
    The scale of the blue histogram is on the left while the scale of the red histogram is on the right.
}
\label{fig:csc:gasgain}
\end{figure}

The chamber gas gain is set by the HV system, which can be individually adjusted in each of up to 30 segments per CSC.
Except for the ME1/1 chambers, each wire plane in other CSCs is divided into three or five independent HV segments, which allows the independent regulation or turning off the HV on any of the individual sections.
To maximize the lifetime of a CSC it should be operated at the lowest HV compatible with full efficiency.
Until mid-2016, all CSCs operated at the same HV, and the average gas gains in different HV segments varied by up to a factor of 2, as shown by the wide blue distribution in Fig.~\ref{fig:csc:gasgain}.
In 2016, the gas gains of individual CSCs were modified by tuning each HV channel to reduce the spread, as shown by the narrow red distribution in Fig.~\ref{fig:csc:gasgain}.
This optimized the CSC gas gains for good efficiency without having unnecessarily high HV on any chamber, thus maximizing chamber longevity.
Later, in 2018, the overall HV of all rings except ME234/1 was reduced by about 30\unit{V}, which decreased the gas gain by about 20\%.
The chambers now operate just above the knee of the efficiency plateau in gas gain versus HV and hence remain fully efficient, while the CSC system lifetime is expected to be extended by 20\%.

The CSC readout electronics can also degrade after exposure to large radiation doses.
A systematic program of irradiating the CSC electronic components was performed to identify any of those unable to operate reliably in the HL-LHC radiation environment.
All the components used in both the old and upgraded readout boards were found to withstand more than 3 times the expected HL-LHC doses, except for the PROMs used in the frontend boards installed in the ME1/1 chambers, which can only withstand 1--1.5 times the expected dose.

Before the production of the new electronics boards for the \Phase2 upgrade, radiation tests were carried out at the Texas A\&M cyclotron and nuclear reactor~\cite{Bylsma:2012mv}, and at the UC Davis cyclotron.
During the tests at the Texas A\&M cyclotron, the digital components on the test boards were operated with active data readout while being irradiated with 55\MeV protons.
The components tested for SEUs and single-event latch-ups (SELs) included the FPGAs, PROMs, level adapters, and optical transmitters and receivers.
In reactor tests, components were exposed to neutrons with energies up to a few \MeV, with exposures equivalent to a TID of 30\kRad.
This corresponds to a level of neutron radiation equivalent to about 50 years of that expected at the HL-LHC at the location where CSC electronics are exposed to the highest radiation flux (the inner portion of the ME1/1 chambers).
These tests targeted mostly nondigital components such as voltage regulators and power diodes.
The results showed that the components selected for the new electronics will operate reliably in the CMS radiation environment at the HL-LHC.

During \Run2, a new campaign of radiation testing was initiated for the upgraded electronics described above.
Particular attention was paid to the PROMs, which are known to have some failures after large radiation exposure~\cite{Ferrarese:2014upd}.
The PROMs in the DCFEBs and ALCTs installed during LS1 were tested at the CHARM II mixed radiation facility at CERN, the Texas A\&M reactor, and the UC Davis cyclotron.
The PROMs performed well up to an exposure of 10\kRad, but both types of PROMs showed some failures at exposures of 15--30\kRad.
To mitigate such failures, the option to perform promless programming of the frontend board FPGAs for the ME1/1 chambers, where the radiation is most severe, has been provided.
The optical receivers used on the OTMBs upgraded during LS2 were also tested at the UC Davis cyclotron and proved able to sustain the expected HL-LHC radiation dose.

The CSC system has been successfully operating during \Run3 after an enormous upgrade effort in LS1 and LS2.
Its longevity has been studied extensively, and it is expected to remain reliable throughout the future running until the end of the HL-LHC era.

\clearpage
\subsection{Resistive plate chambers}
\label{sec:rpc}

\subsubsection{General description}
\label{sec:rpc:description}

\begin{figure}[!ht]
\centering
\includegraphics[width=0.55\textwidth]{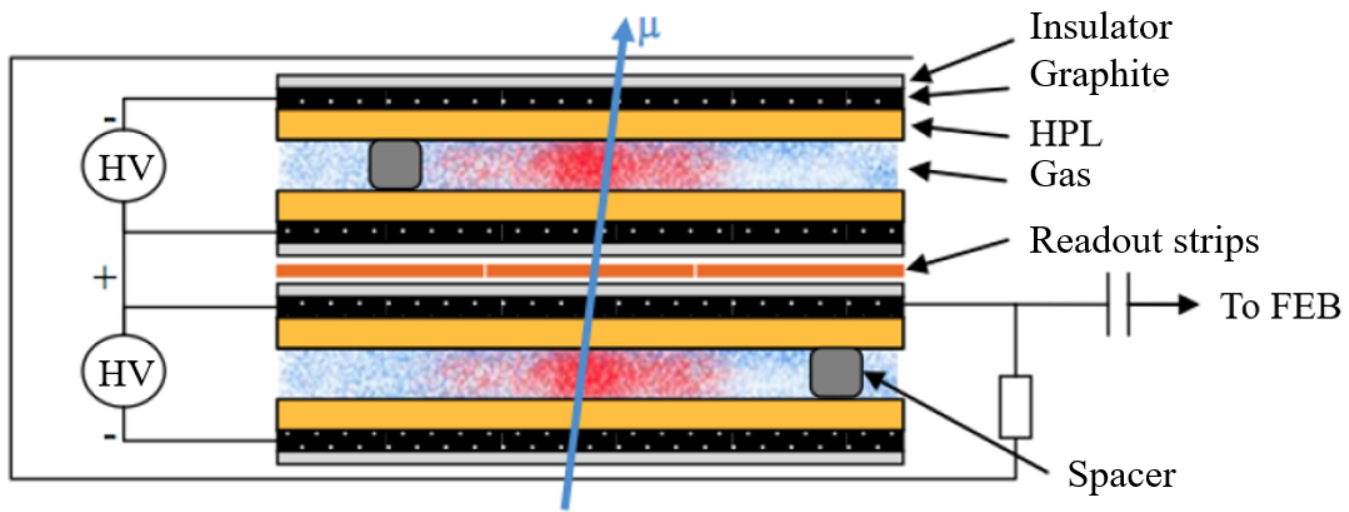}%
\hfill%
\includegraphics[width=0.42\textwidth]{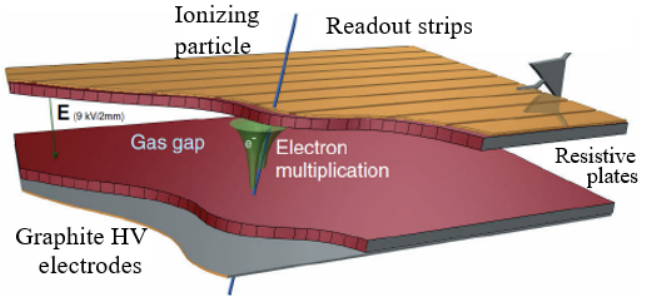}
\caption{%
    Left:\ schematic of the double-layer layout of the RPC chambers, from Ref.~\cite{Bunkowski:2009rsa}.
    Right:\ illustration of the RPC technology, from Ref.~\cite{Corrin:2002sgw}.
}
\label{fig:rpc:schematics}
\end{figure}

The CMS resistive plate chambers (RPCs) are gaseous detectors equipped with two gas gaps each having a width of 2\mm and a copper readout plane in between, as shown in Fig.~\ref{fig:rpc:schematics}.
High voltage is applied to the graphite electrodes, which are coated on the surface of high-pressure (HPL) laminate plates with bulk resistivity in the range of 2--$5\times10^{10}\Ohmcm$.
The chambers are operated in avalanche mode with a gas mixture of 95.2\% \CtwoHtwoFfour, 4.5\% \iCfourHten, and 0.3\% \SFsix.
This allows the chambers to cope with high background rates and ensures an excellent time resolution, as summarized in Table~\ref{tab:muon:summary}, facilitating a precise bunch-crossing assignment.

The RPC barrel detector is divided in the direction along the beam axis into five separate wheels labelled W$\pm$2, W$\pm$1, and W0.
The RPCs in each wheel consist of six layers.
The first four layers, called RB1in, RB1out, RB2in, and RB2out, are located on the inner and outer sides of the inner two stations of the DT chambers.
The other two layers, labelled RB3 and RB4, are located on the inner side of the third and fourth stations of the DTs.
There are four disks in each endcap, called RE$\pm$4, RE$\pm$3, RE$\pm$2, and RE$\pm$1.
Spanning the $\phi$ direction, each wheel is divided into twelve sectors and each disk into 36 sectors.
Due to requirements in the trigger logic, the chambers are divided into two or three pseudorapidity ($\eta$) partitions, called rolls.
In most of the barrel, there are two rolls:\ forward and backward.
Only RB2in in W$\pm$1 and W0 and RB2out in W$\pm$2 are divided into three rolls, called forward, middle, and backward.
The endcaps are divided into three rolls:\ A, B, and C.
The RPC system consists of 1056 chambers, covering an area of about 3950\msq, equipped with 123\,688 readout strips.
In the barrel, the strips are rectangular in shape with a pitch in the range between 2.28 and 4.10\cm, while in the endcaps they are trapezoidal with a pitch between 1.74 and 3.63\cm.

\subsubsection{RPC system upgrades since \Run1}
\label{sec:rpc:upgrades}

\paragraph{Endcap fourth station upgrade}

One of the major upgrades of the CMS experiment, performed during the first long shutdown (LS1) of the LHC in 2013--2014, was to add 144 new RPCs to the fourth stations of the detector endcaps.
Adding these stations increased the overall robustness of the muon spectrometer and improved the trigger efficiency in the endcap region over the range $1.2<\abseta<1.6$.

The new RE4 RPCs inherit their design from the previously installed endcap chambers~\cite{Colafranceschi:2014ija}.
They are grouped in 72 supermodules, 36 per disk, each consisting of two detectors and covering 10\de\ in azimuthal angle.
Prior to their installation in CMS, all RE4 chambers (both RE $\pm$4) passed a series of tests at all production stages in order to ensure their quality and performance.

\begin{figure} [!ht]
\centering
\includegraphics[width=0.48\textwidth]{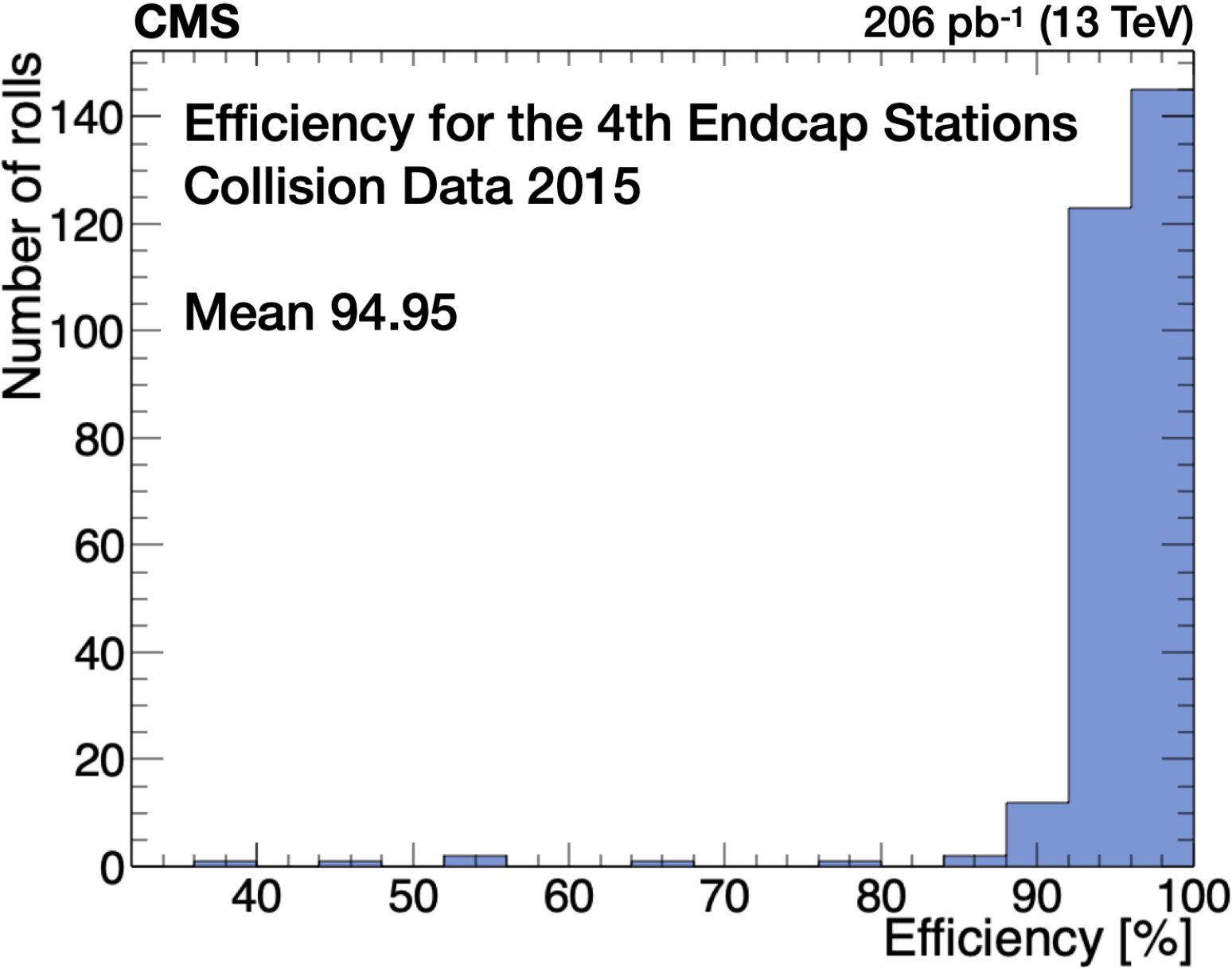}
\caption{%
    Efficiency distribution of the RE$\pm$4 stations in their first year of operation in 2015.
}
\label{fig:rpc:re4eff}
\end{figure}

This newest part of the RPC system was successfully commissioned in 2015 at the beginning of the \Run2 data taking.
The performance of RE4, shown in Fig.~\ref{fig:rpc:re4eff}, is in good agreement with expectations~\cite{CMS:TDR-8-2}.

\paragraph{Readout changes and \Phase1 RPC trigger contributions}

As mentioned in Section~\ref{sec:l1trigger:muon}, the \Phase1 level-1 (L1) trigger upgrade~\cite{CMS:TDR-012} moved from a muon detector-based scheme to a geometry-based system.
The three muon trigger systems were replaced by three track finders, each covering a specific pseudorapidity region.

The RPC pattern comparator trigger (PACT) has the function of reading out the information from the RPC frontend boards (FEBs).
These are installed at chamber level via the link board system, located in the CMS experimental cavern in the balconies next to the detector.
In the link boards, the low-voltage differential signaling (LVDS) signals from the FEBs are synchronized with the LHC clock, converted to optical signals, and sent to the RPC trigger boards in the service cavern.
Since the beginning of \Run2, the RPC signals coming from the link boards are split and sent to the muon processors TwinMux, as described in Section~\ref{sec:dt:ph1electronics}, and to the concentration pre-processing and fan-out (CPPF)~\cite{Cheng:2018app}.
Both TwinMux and CPPF do the hit clustering and cluster selection and forward the produced clusters to the track finders (TwinMux to the BMTF and CPPF to the EMTF).
The OMTF, described in Section~\ref{sec:l1trigger:muon}, reads the RPC data straight from the link boards, then combines them with the full list of DT trigger segments directly at the track-finding step and performs the clustering and selection.

\begin{figure}[!ht]
\centering
\includegraphics[width=0.48\textwidth]{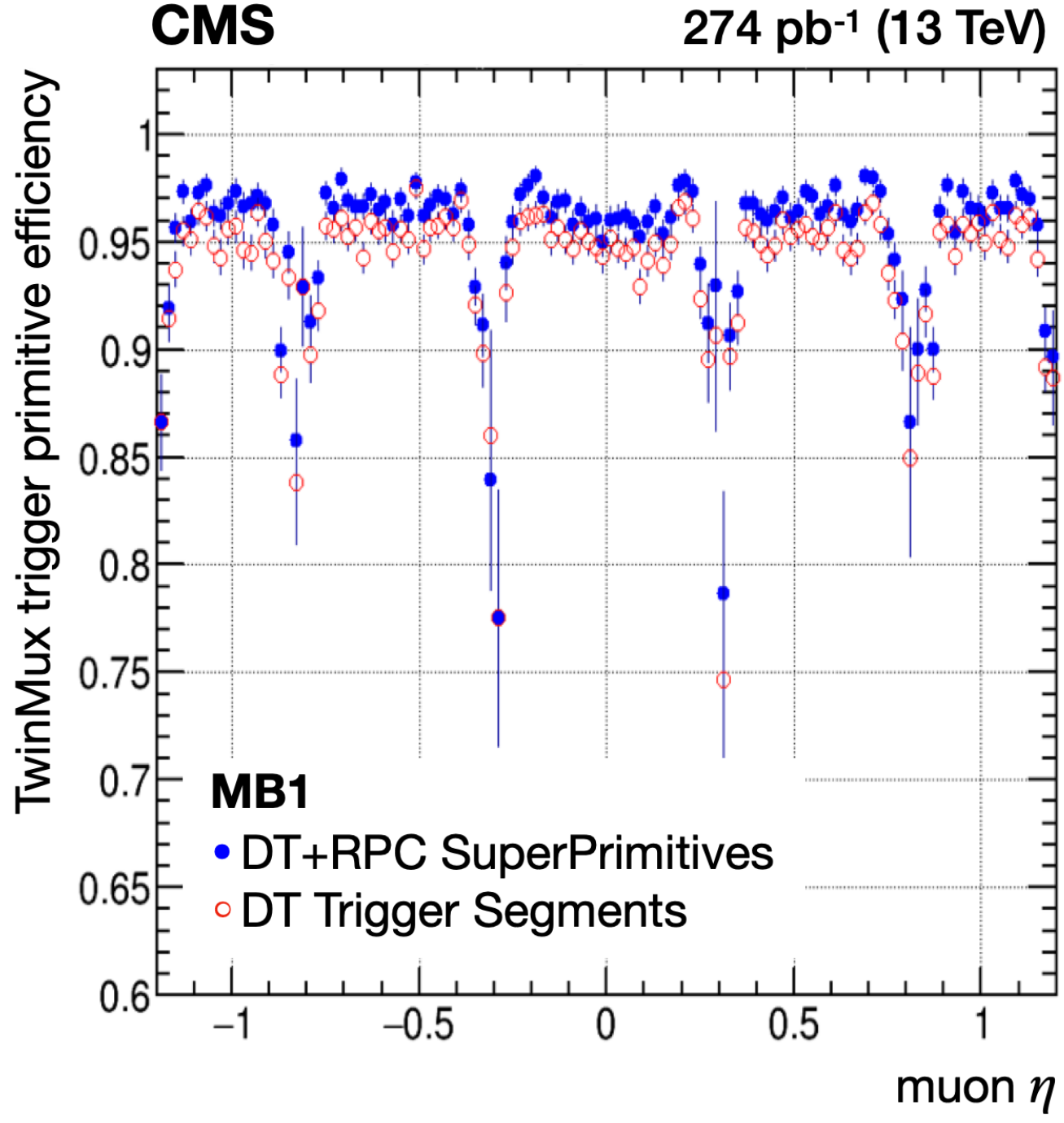}%
\hfill%
\includegraphics[width=0.48\textwidth]{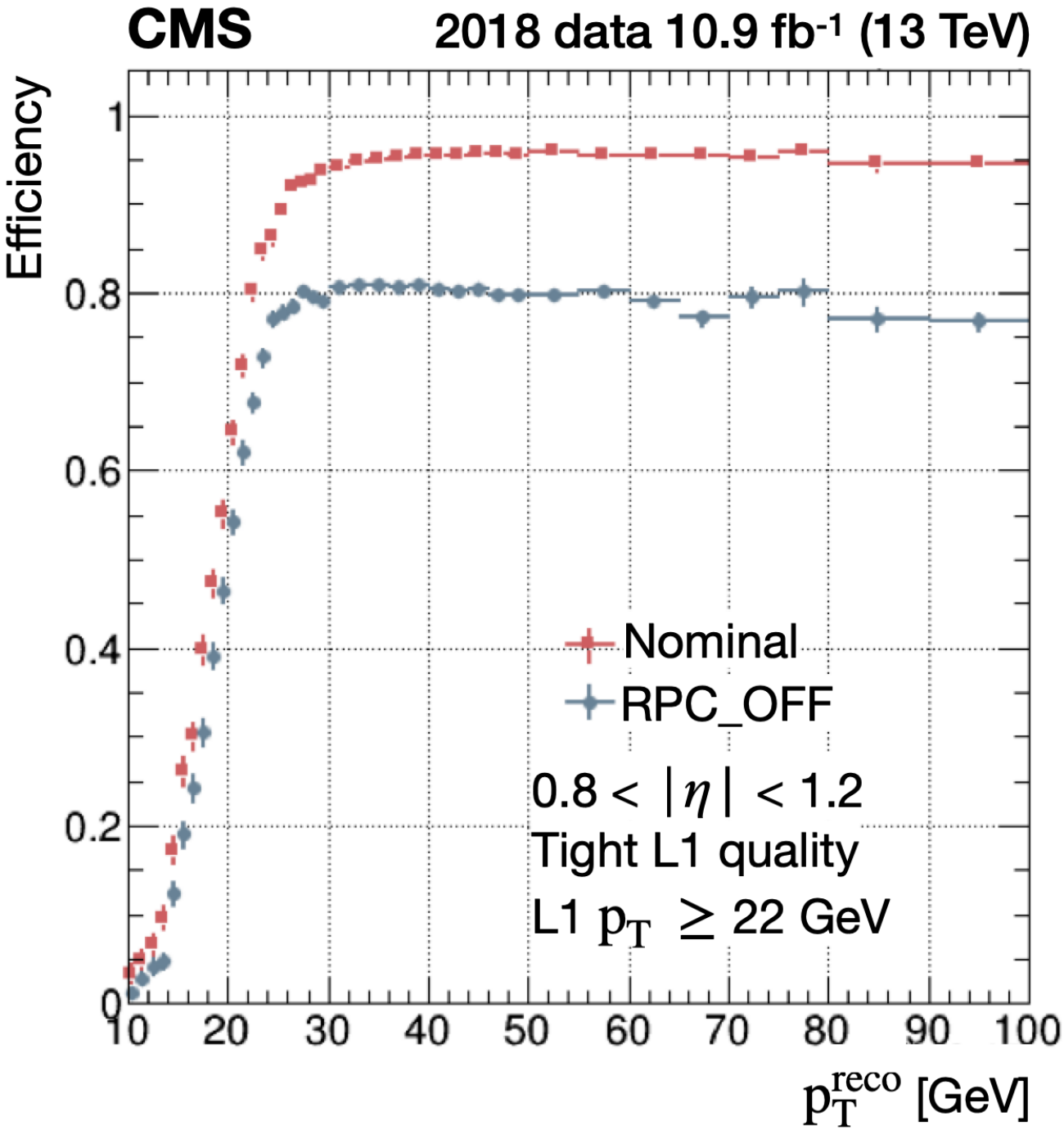}
\caption{%
    Left:\ station-1 barrel trigger-primitive efficiency as a function of muon pseudorapidity.
    Right:\ trigger efficiency as a function of muon \pt for the OMTF, derived from a trigger emulation applied to real data, using (red) and not using (blue) RPC information.
    Figures from Ref.~\cite{Samalan:2020zmg}.
}
\label{fig:rpc:dtwithrpc}
\end{figure}

The contribution of the RPC system to the refactored L1 muon trigger architecture is different for the three muon track finders and can be summarized as follows:
\begin{itemize}
\item BMTF: RPC timing information is used to improve the DT trigger primitives' bunch crossing assignment.
In the first two stations, where two RPC layers are present, a segment is built from the coincidence of hits in the inner and outer chambers.
The latter provides redundancy in case of DT inefficiencies.
\item OMTF: RPC hit position information is used standalone from the eight available chambers (five in the barrel and three in the endcap), per $\phi$ division (sector).
\item EMTF: RPC hit position information is used in case there is no corresponding CSC trigger primitive.
\end{itemize}
The combination of information from the barrel muon detectors at an early stage allows the exploitation of the system redundancy already at the step of building the trigger primitives.
As shown in the left plot of Fig.~\ref{fig:rpc:dtwithrpc}~\cite{CMS:DP-2016-074}, the combination of DT and RPC leads to an average increase of the station-1 barrel trigger-primitive efficiency of about 1.4\%, raising the plateau value from 95\% to about 96.5\%.
However, this is not the only gain from the new architecture.
The trigger also benefits from detector complementarity since the use of RPC timing information reduces the number of out-of-time DT trigger primitives, as shown in Fig.~\ref{fig:dt:twinmux}.

For the OMTF, the use of the RPC complementarity is even more prominent.
Figure~\ref{fig:rpc:dtwithrpc} (right) shows the efficiency as a function of the reconstructed muon \pt in the $\eta$ region of the OMTF.
If RPC data are not present, the efficiency decreases by about 15\%.

\subsubsection{RPC system longevity}

\paragraph{Detector stability studies}

{\tolerance=800
Continuous studies are performed throughout the data-taking periods to ensure correct operation and performance stability of the RPC system.
\par}

\subparagraph{RPC working point calibration}
\label{sec:rpc:wp}

To ensure the most stable performance possible, the operational high voltage \Vapp of the RPCs is controlled such that the effective high voltage \Veff is constant even when environmental conditions change during the data taking~\cite{Abbrescia:1995uj}.
The relation between \Vapp and \Veff is as follows~\cite{Colafranceschi:2012qf}:
\begin{linenomath}\begin{equation}
    \Vapp=\Veff \left[1 -\alpha +\alpha\left(\frac{P}{P_0}\right)\left(\frac{T_0}{T}\right)\right],
\end{equation}\end{linenomath}
where $P$ and $T$ are the actual pressure and temperature in the CMS cavern, $\alpha$ is a special pressure correction that equals 0.8, and $P_0$ and $T_0$ correspond to values of 965\mbar and 273\unit{K}, respectively.

To determine the optimal operating voltage for every chamber, a series of high voltage scans are regularly performed, typically in low and high luminosity conditions.
Data are collected for particular detector configurations:\ the \Veff values are equidistantly chosen within a range of 8.8 and 9.8\kV.
The data are selected using triggers from the DT and CSC and then reconstructed using the standard CMS muon reconstruction~\cite{CMS:MUO-16-001}.
During HV scan periods, no correction due to pressure or temperature is applied to the voltage.
The segment extrapolation method~\cite{CMS:MUO-11-001} is used to measure the RPC hit efficiency for each HV point.
Segments from the nearest DT and CSC chambers are extrapolated to the RPC planes, and the reconstructed RPC clusters (rechits) are matched to the extrapolated points.
The efficiency is then calculated as a ratio of the numbers of matched RPC clusters to the extrapolated segments.
The resulting efficiency distribution ($\varepsilon$) is fitted for every chamber using a sigmoid function defined as
\begin{linenomath}\begin{equation}\label{eq:rpc:sigmoid}
    \varepsilon(\Veff)=\frac{\epsmax}{1+\exp[-\lambda(\Veff - \Vfifty)]},
\end{equation}\end{linenomath}
where $\lambda$ characterizes the slope of the sigmoid at the inflection point, \epsmax represents the asymptotic efficiency for $V\to\infty$, and \Vfifty is the inflection point for which \epsmax reaches an efficiency of 50\%.

The optimal working point (WP) for each chamber is determined by interpolating the efficiency distribution $\varepsilon$ using the fitting parameters in Eq.~\eqref{eq:rpc:sigmoid}, and calculating the voltage corresponding to 95\% of \epsmax.
Then, the WP for each chamber is defined as $V_{95\%}+100\unit{V}$ (120\unit{V}) for the barrel (endcap) chambers, respectively.
The application of individually different offsets leads to similar overall efficiencies in the barrel and endcaps~\cite{Abbrescia:1995uj, Colafranceschi:2012qf}.

Since each HV channel supplies two chambers in the endcaps (with a few exceptions), the WP in the endcaps is the average of the WP of the two corresponding chambers for each HV channel.
All chambers that operate in single-gap mode or have HV problems are excluded from this method.

\begin{figure}[!p]
\centering
\includegraphics[width=0.48\textwidth]{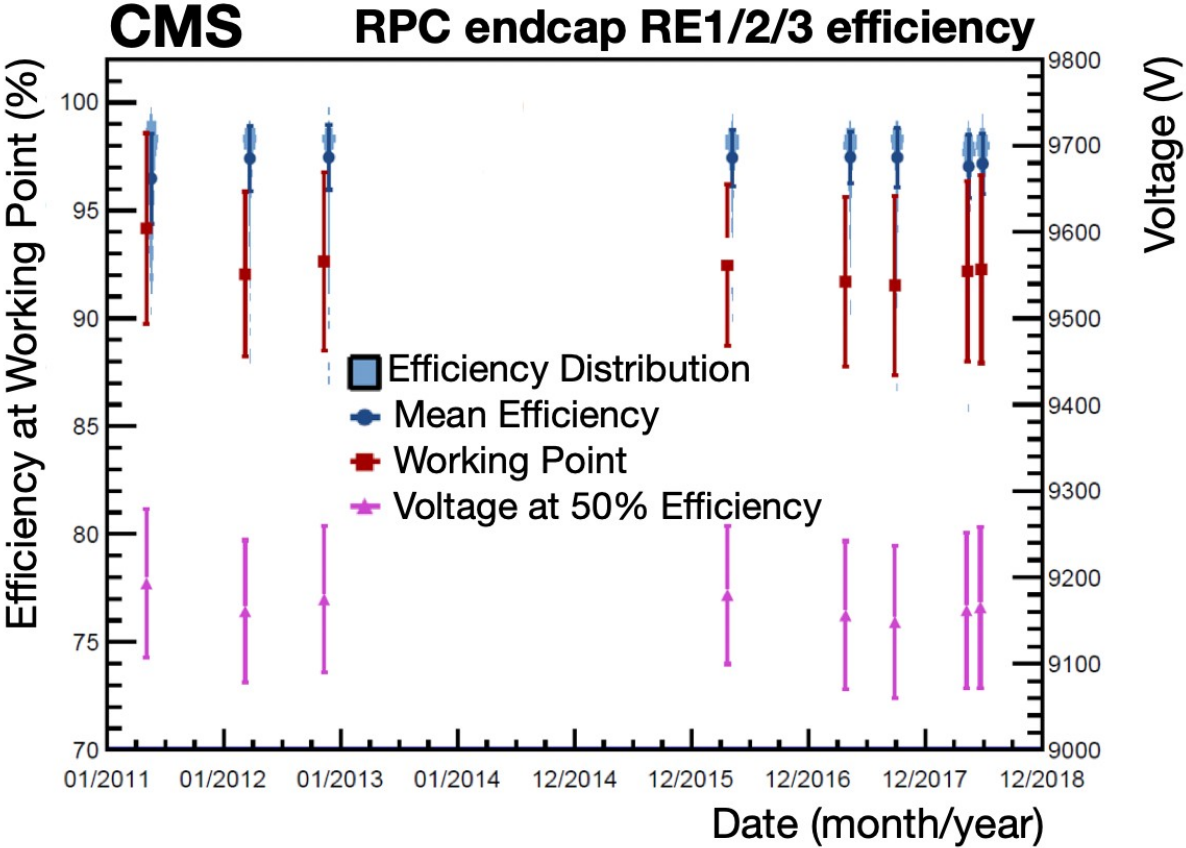}%
\hfill%
\includegraphics[width=0.48\textwidth]{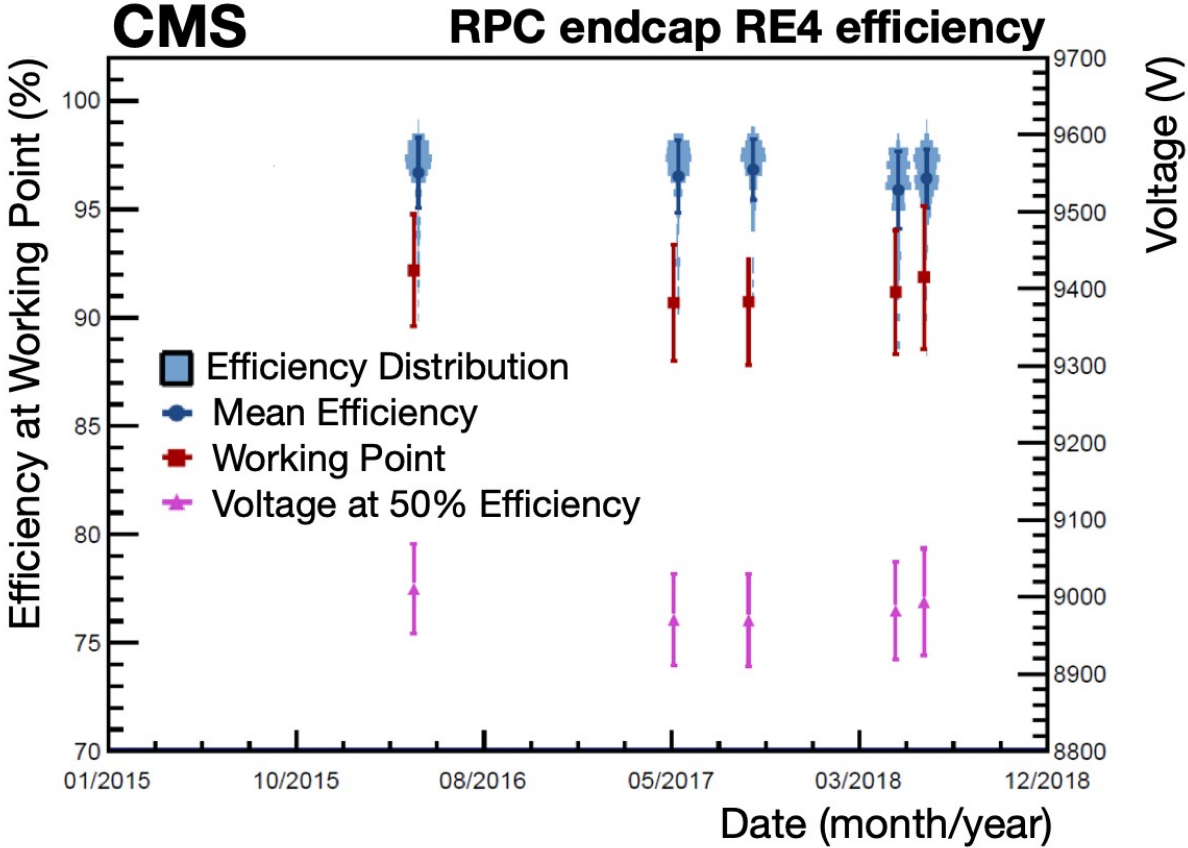} \\[1ex]
\includegraphics[width=0.48\textwidth]{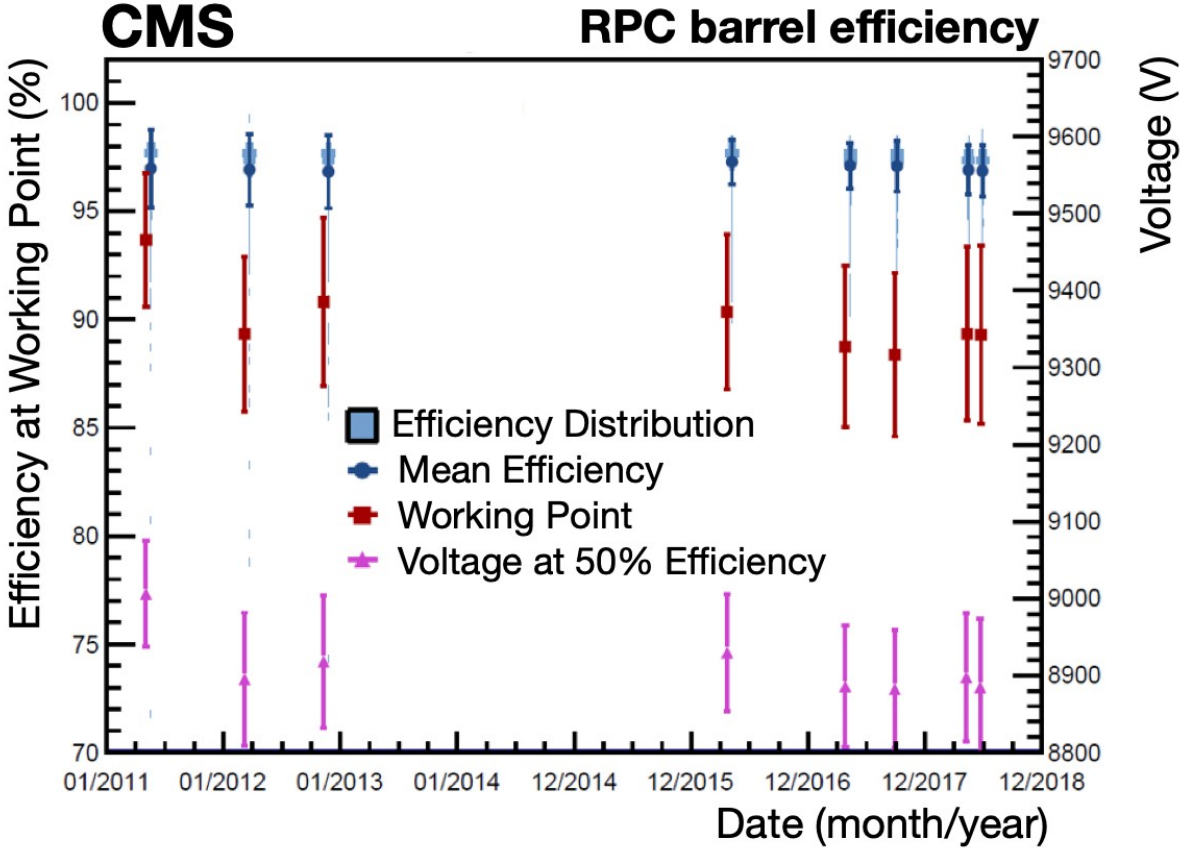}
\caption{%
    Temporal evolution of efficiencies determined from HV scan data at the WP and at \Vfifty, for RE1/2/3 (upper left), RE4 (upper right), and the barrel (lower).
    The light blue bands show the histograms of the distributions, where the width of the band represents the population of channels having the corresponding efficiency value.
}
\label{fig:rpc:history}
\end{figure}

Figure~\ref{fig:rpc:history} shows the evolution of the efficiency as a function of time, as determined from the HV scan data, evaluated at the WP and at \Vfifty during the LHC \Run1 and \Run2.
Despite the changes in the environmental and luminosity conditions, the different calibrations implemented in the detector allowed us to keep the efficiency stable.

\subparagraph{RPC efficiency and cluster size stability}
\label{sec:rpc:efficiency}

One of the most important measures of the RPC system performance is the efficiency of the rolls, described in Section~\ref{sec:rpc:description}.
The efficiency of a roll can be measured as the ratio of the number of reconstructed hits to the number of muons passing through the roll.
To obtain the position of the muon trajectory in the RPC layers, the muon track is extrapolated using either the track parameters or the direction of the track segment from the closest other subdetector in the muon system.
During the \Run2 data-taking periods, the analysis used both methods:\ the track~\cite{CMS:2014gzw} and segment~\cite{CMS:MUO-11-001} extrapolation.

\begin{figure}[!p]
\centering
\includegraphics[width=0.48\textwidth]{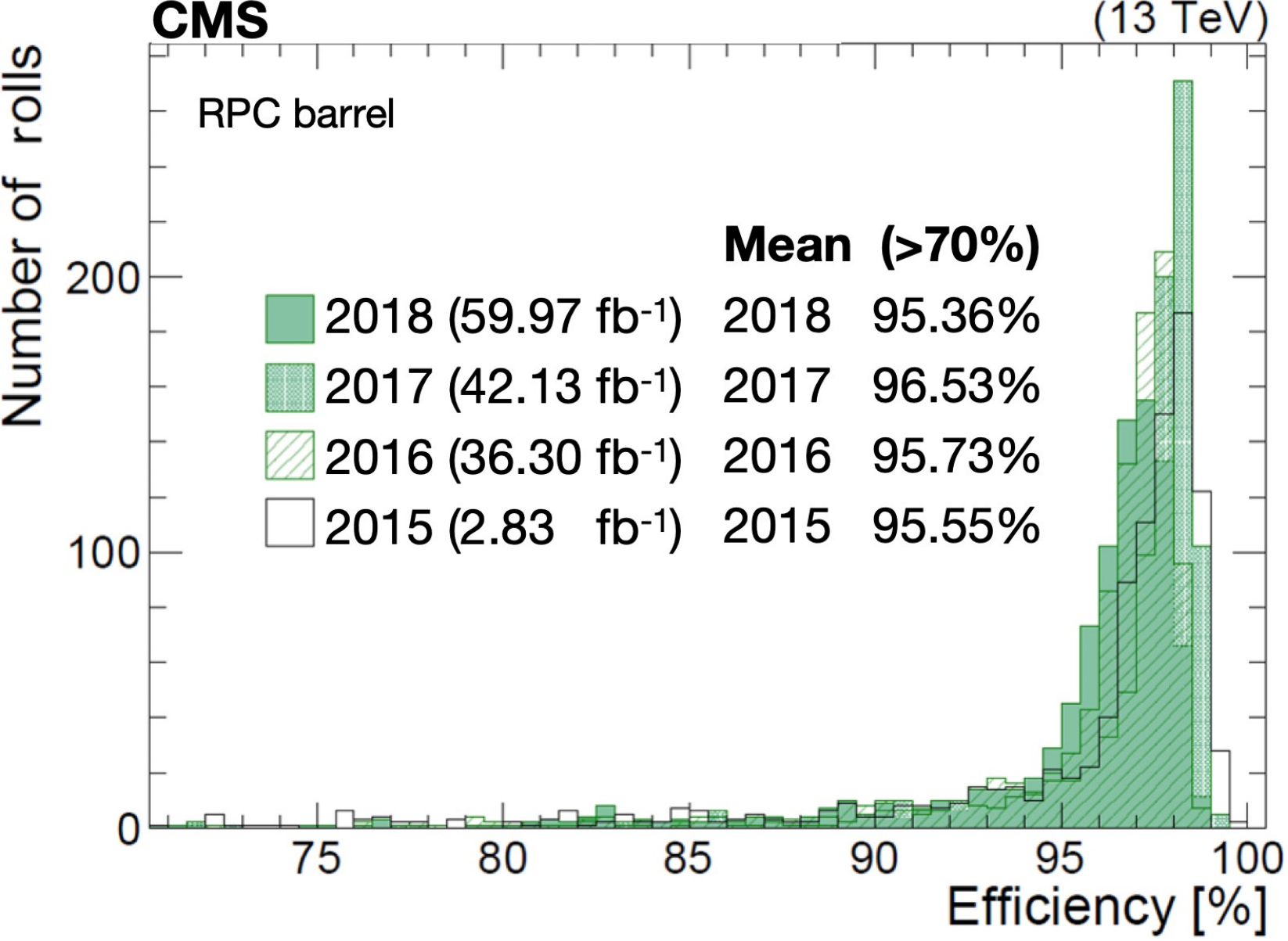}%
\hfill%
\includegraphics[width=0.48\textwidth]{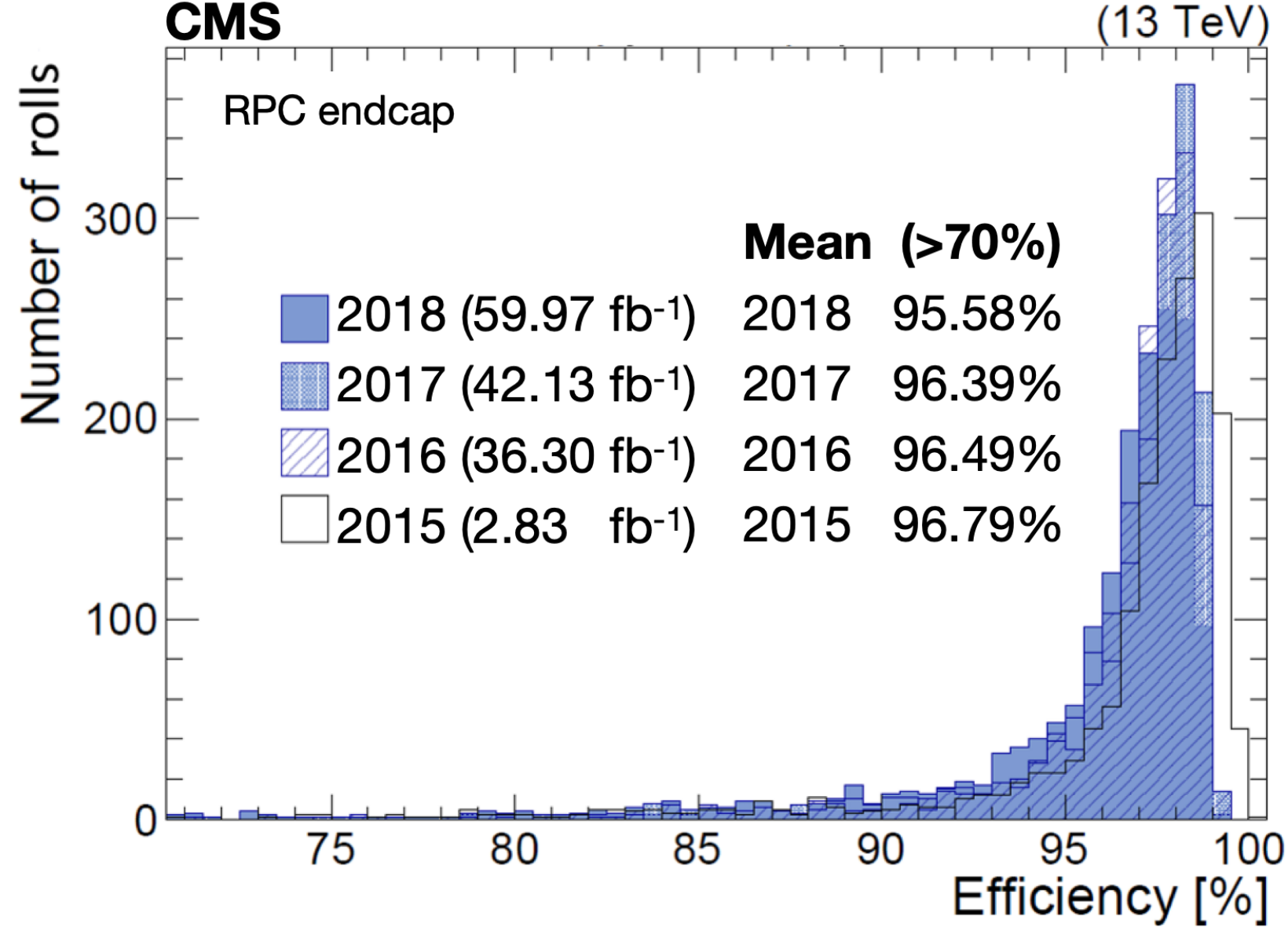}
\caption{%
    Distributions of the overall RPC efficiencies in the barrel (left) and endcaps (right) during \pp data taking in \Run2~\cite{Pedraza-Morales:2016ysr}. Figures from Ref.~\cite{CMS:2020mpl}.
}
\label{fig:rpc:efficiencydistr}
\end{figure}

Figure~\ref{fig:rpc:efficiencydistr} shows the overall efficiency distribution of the RPC rolls in \Run2.
The numbers are for all well-performing RPC rolls.
Rolls with an efficiency lower than 70\% are excluded if the efficiency drop is caused by a known hardware problems, such as a gas leak.

The cluster size (CLS) is another important quantity since it affects the RPC spatial resolution.
It is defined as the number of adjacent strips firing when a discharge is produced in the RPC.
The RPC system has an average cluster size of less than three strips, constant over the years, in agreement with expectations~\cite{CMS:Detector-2008}.
Keeping the cluster size stable over time is one of the most important prerequisites in the correct operation of the RPC system.
During \Run2, the chamber cluster size was monitored run-by-run to guarantee the stability of the system.

\begin{figure}[!ht]
\centering
\includegraphics[width=\textwidth]{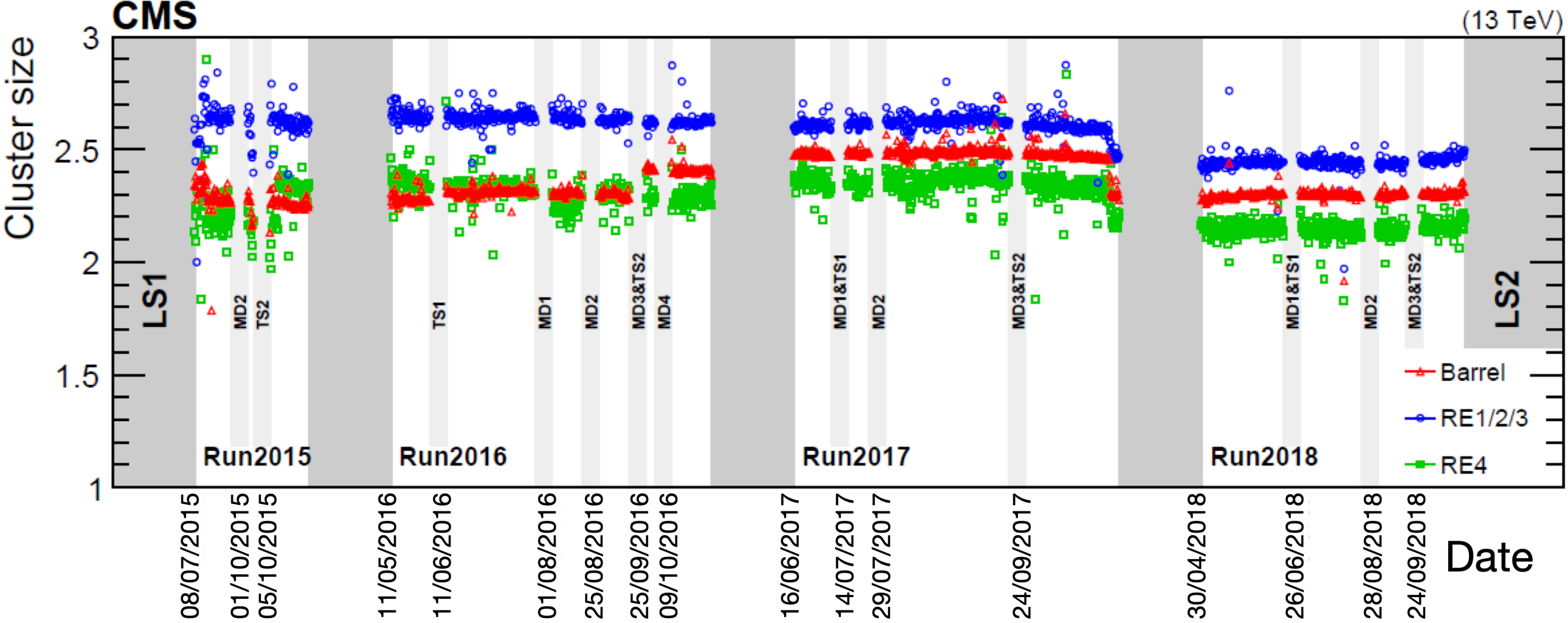} \\[1ex]
\includegraphics[width=\textwidth]{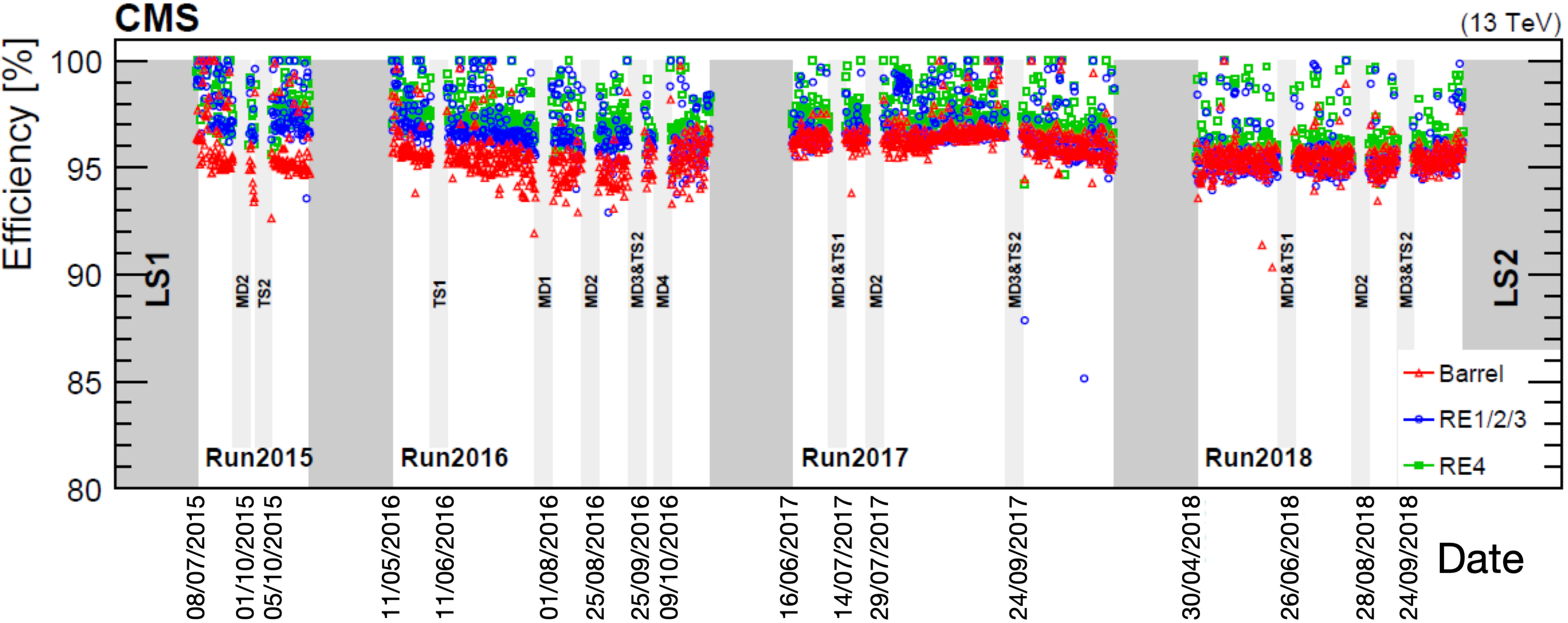}
\caption{History of the RPC efficiency (upper) and cluster size (lower) during \Run2. Gray areas correspond to the scheduled technical stops.}
\label{fig:rpc:performancehistory}
\end{figure}

The efficiency and cluster-size history for the \Run2 data taking are shown in Fig.~\ref{fig:rpc:performancehistory}.
The history follows the changes of the applied HV WPs and changes of the isobutane concentration in the gas mixture.
The spread in the cluster size distribution in 2015 was caused by a threshold control problem, which was resolved in October 2015, while the small drop of the RE$+4$ cluster size in 2016 was due to a temporary LV problem, solved at the end of August 2016.
The drop of efficiency in the period 01--19 August 2018 came from a configuration problem.
Detailed \Run2 performance results can be found in Refs.~\cite{Pedraza-Morales:2016ysr, CMS:2018gys}.

\begin{figure}[!ht]
\centering
\includegraphics[width=0.48\textwidth]{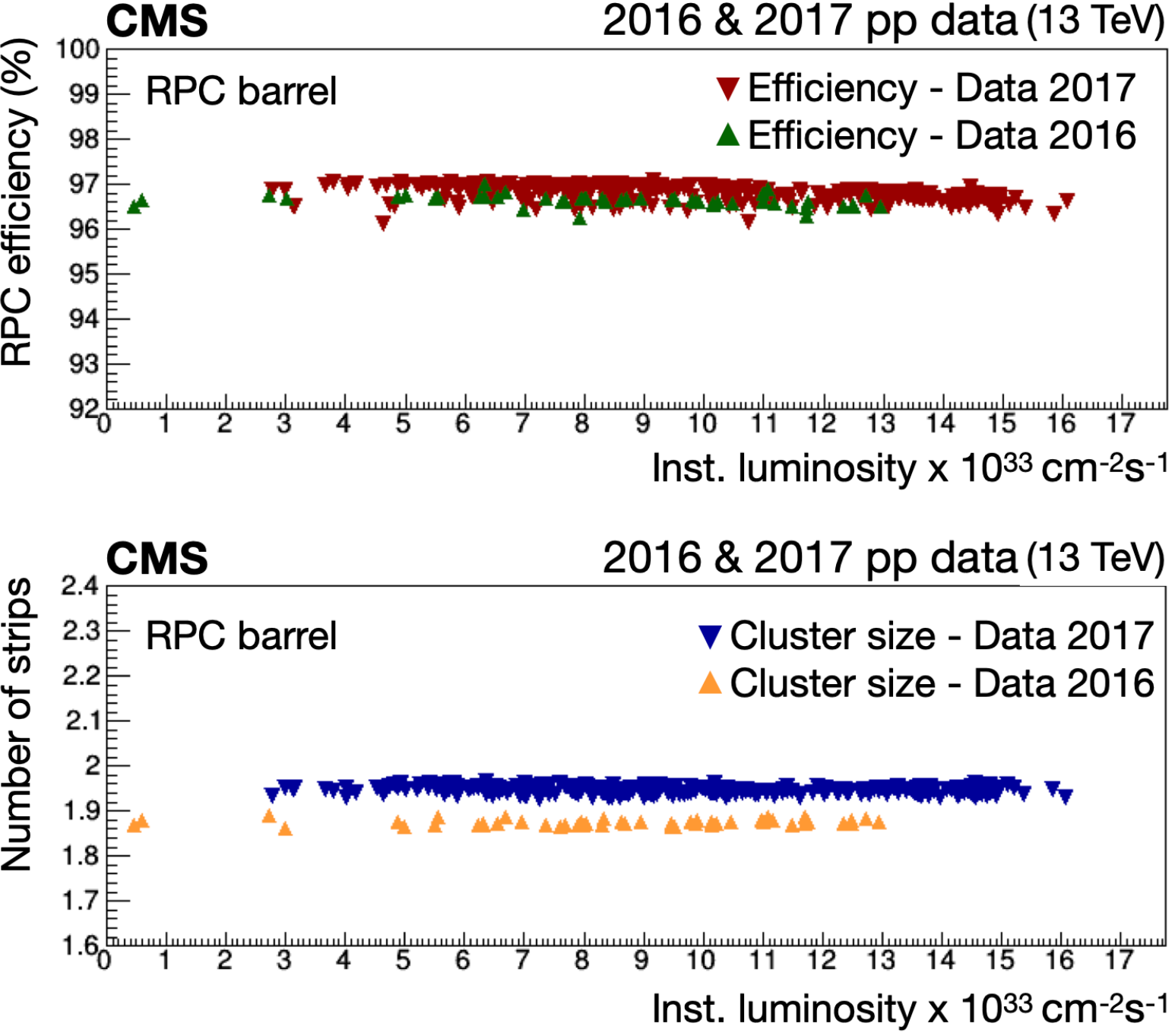}%
\hfill%
\includegraphics[width=0.48\textwidth]{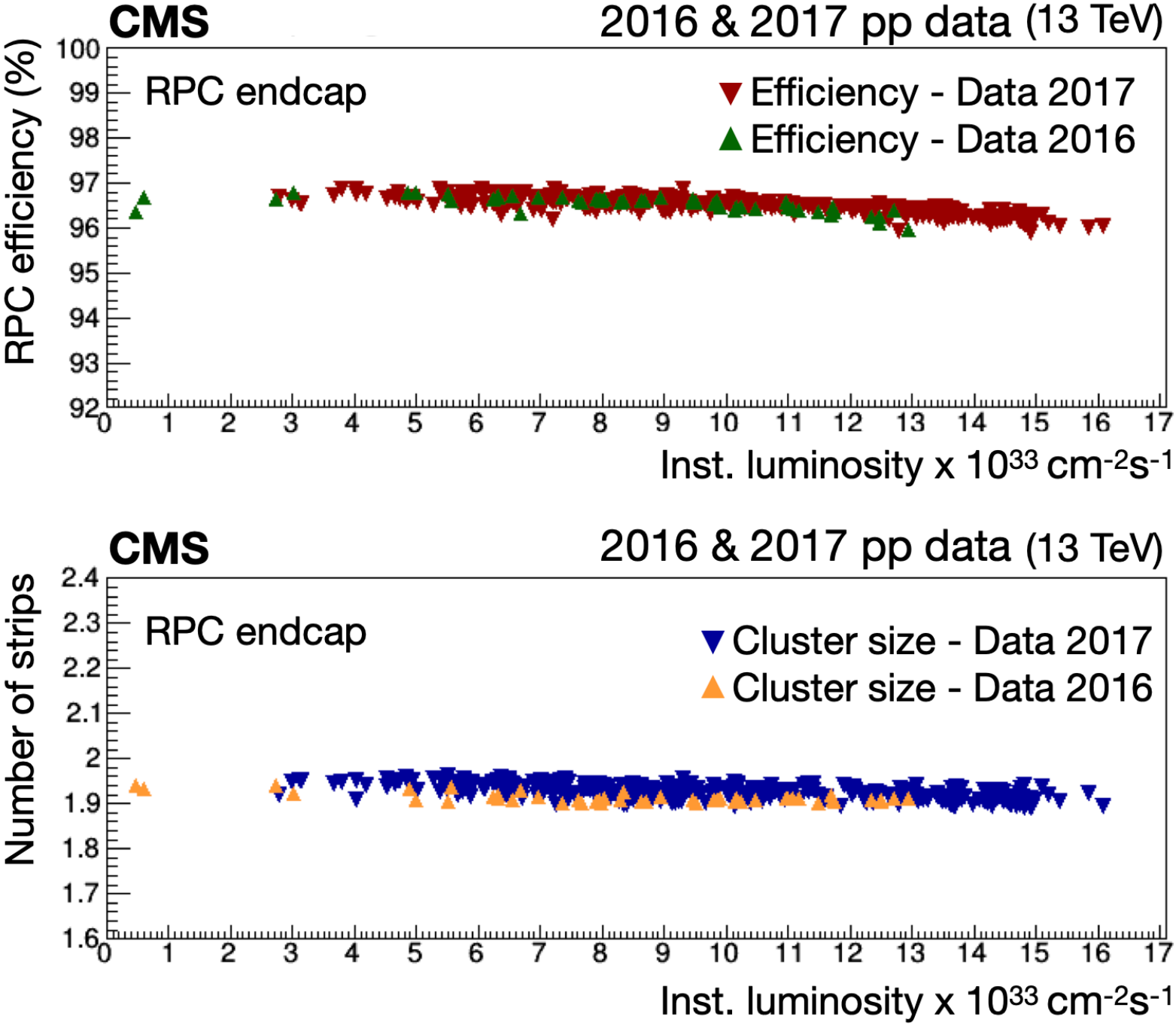}
\caption{%
    RPC barrel and endcap efficiency (upper) and cluster size (lower) as a function of the LHC instantaneous luminosity for \pp collisions in 2016 and 2017.
    The linear extrapolation to the instantaneous luminosity expected at the HL-LHC of $7.5\times10^{34}\percms$ shows a 1.35\% reduction in efficiency for the barrel and 3.5\% for the endcap. Figures from Ref.~\cite{CMS:2018gys}.
}
\label{fig:rpc:segmenteff}
\end{figure}

One of the main objectives of the RPC analysis is to monitor the system performance with respect to the luminosity.
Figure~\ref{fig:rpc:segmenteff} displays the barrel and negative endcap efficiency and cluster size as a function of the instantaneous luminosity during \pp data taking in 2016 and 2017.
For this comparison, data recorded with the same WP are used.
The lower efficiency and cluster size for the barrel in 2016 are caused by the higher isobutane concentration during that year.

The comparison between the 2016 and 2017 results shows a stable efficiency and cluster size.
The obtained results can be linearly extrapolated to the expected luminosity at the HL-LHC of $7.5\times10^{34}\percms$.
A reduction in efficiency of 1.35\% is predicted for the barrel and 3.5\% for the endcaps.
No change is expected for the cluster size under the HL-LHC conditions.

\subparagraph{Monitoring of RPC currents}

The current drawn by the RPC detectors is one of the base parameters to be considered, both as a measure of the working condition and as an indicator of possible problems.
The current should be kept as low as possible to guarantee long-term stability of the RPC gas gaps, and its time evolution is one of the most important input data for longevity studies.

The ohmic current of the RPC system is defined as the current with no beam in the accelerator, at a HV around 7000\unit{V}, where the current follows an ohmic law and there is no contribution from gas amplification.
In practice, the ohmic current values are monitored at 6500\unit{V}.
The ``cosmic current'' is defined as the current without beam, at the WP voltage, in the region of gas amplification.

For \Run3, the monitoring capability for the RPC detector currents was improved.
An especially designed tool was implemented that automatically stores and classifies the currents in a database as a function of different parameters, such as the CMS magnetic field, the instantaneous luminosity, and other parameters.
The tool can identify blocks of data of special importance, such as the periods of HV scans, described in Section~\ref{sec:rpc:wp}.

\begin{figure}[!b]
\centering
\includegraphics[width=\textwidth]{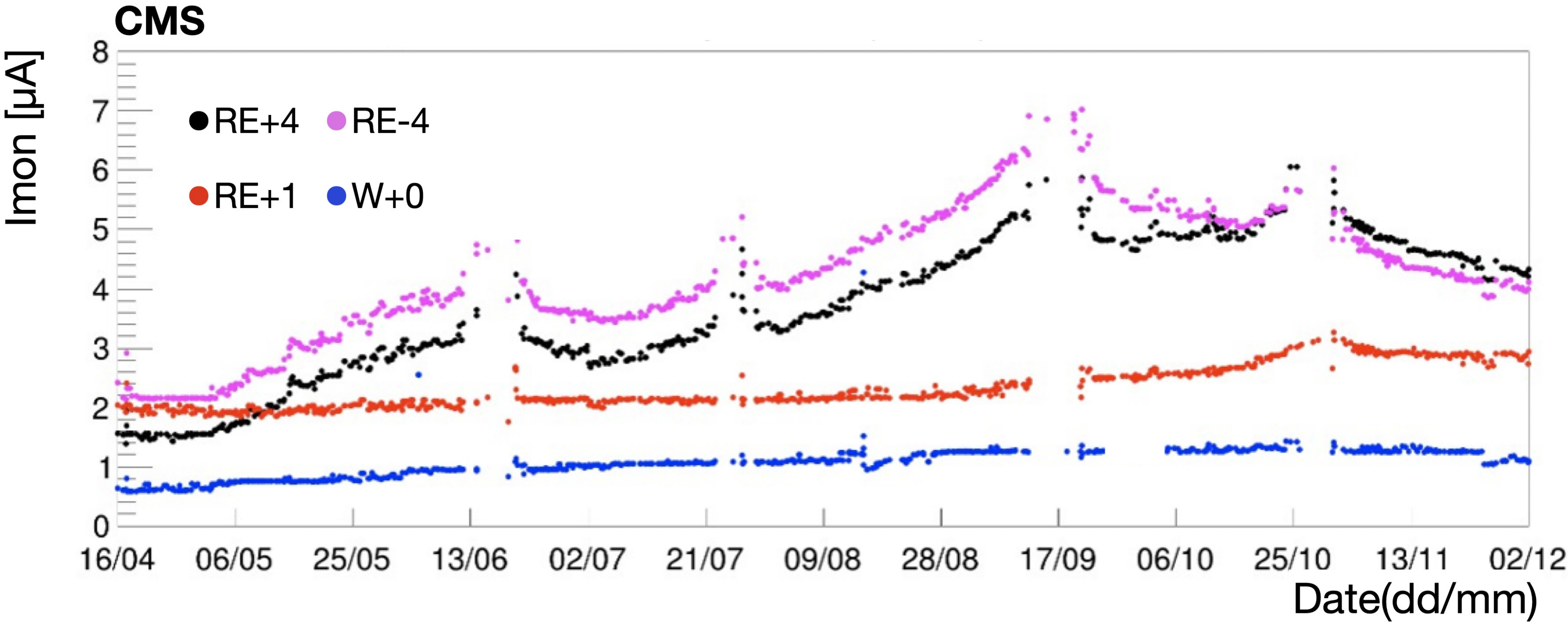}
\caption{%
    Ohmic current history in W$+0$, RE$+1$, RE$+4$, and RE$-4$. Figure from Ref.~\cite{CMS:2020mpl}.
}
\label{fig:rpc:ohmiccurrent}
\end{figure}

Figure~\ref{fig:rpc:ohmiccurrent} shows the currents measured in four RPC stations, W$+0$ in the barrel, and RE$+1$, RE$+4$, and RE$-4$ in the endcaps.
The measured currents are shown as a function of time.
An increase of the ohmic currents is observed, in particular for chambers that are more exposed to background, and a linear dependence of the rate on the instantaneous luminosity is seen.
In low-background regions such as W$+0$, the ohmic current increases very slowly.
In RE$+1$ and W$+0$, the background rate is less than 10\Hzcmsq.
Both have similar gas flows, corresponding to a volume exchange per hour (v/h) of 0.7 and 0.6, respectively.
In contrast, in RE$4$, the background rate is about 40\Hzcmsq and the gas flow is 0.35\unit{v/h}, which is lower than the rest of the system, due to a wrong calibration of the flow cells.
These rates are the maximum values measured in the upper sectors, \ie, the detector sectors above the beam pipe, at instantaneous luminosities of 1.5--$2\times10^{34}\percms$.

\begin{figure}[!ht]
\centering
\includegraphics[width=\textwidth]{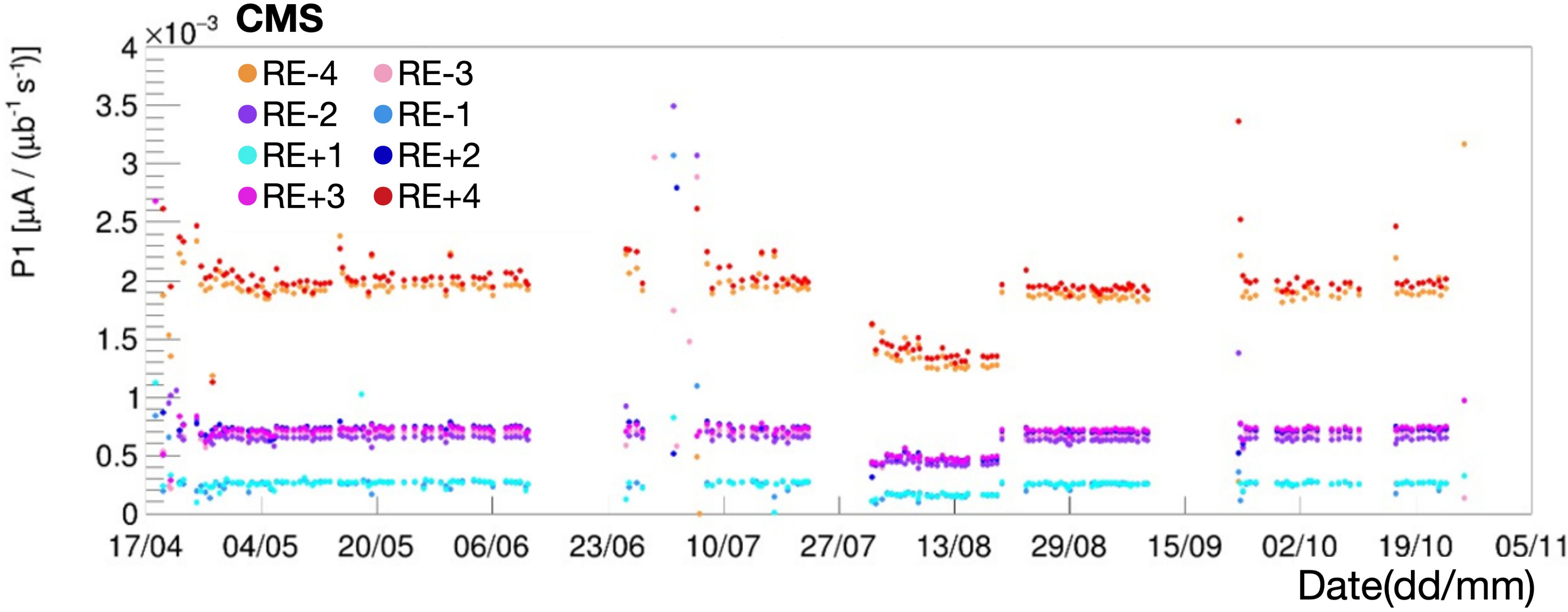}
\caption{%
    RPC physics current history in RE$\pm1$, RE$\pm2$, RE$\pm3$ and RE$\pm4$. Figure from Ref.~\cite{CMS:2020mpl}.
}
\label{fig:rpc:cosmiccurrent}
\end{figure}

The RPC currents depend linearly on the instantaneous luminosity~\cite{Gelmi:2021woo}.
For each LHC fill, the distributions were fit to a linear function to obtain the slope $P_1$, also known as the physics current ($i=P_1L$).
Due to the nature of the linear fit, an offset $P_0$ absorbs the cosmic current (offset + ohmic + gas gain).
The slopes $P_1$ as a function of time for the endcap stations are shown in Fig.~\ref{fig:rpc:cosmiccurrent}.
The slope of the RPC current distribution is stable in time.
The changes in the middle of August 2018 are due to different HV WPs.
Endcap stations that are located at equal distances from the interaction point along the beam pipe have similar slopes.
They also have similar rates~\cite{Gelmi:2021woo}.
No increase due to the integrated luminosity is observed for the slopes over the entire year.

Hydrogen fluoride (HF) is produced in the gas under high electrical discharge.
The HF has a high chemical reactivity and electrical conductivity~\cite{Abbrescia:2006hk}, and it is therefore expected to be a source of inner-detector surface damage and relative ohmic current increase that accelerates detector aging.
In summer 2018, measurements of the HF concentration were performed at the gas exhausts of three regions:\ W$+0$ in the barrel, and RE$+1$ and RE$+4$ in the positive endcap.

\begin{figure}[!ht]
\centering
\includegraphics[width=0.6\textwidth]{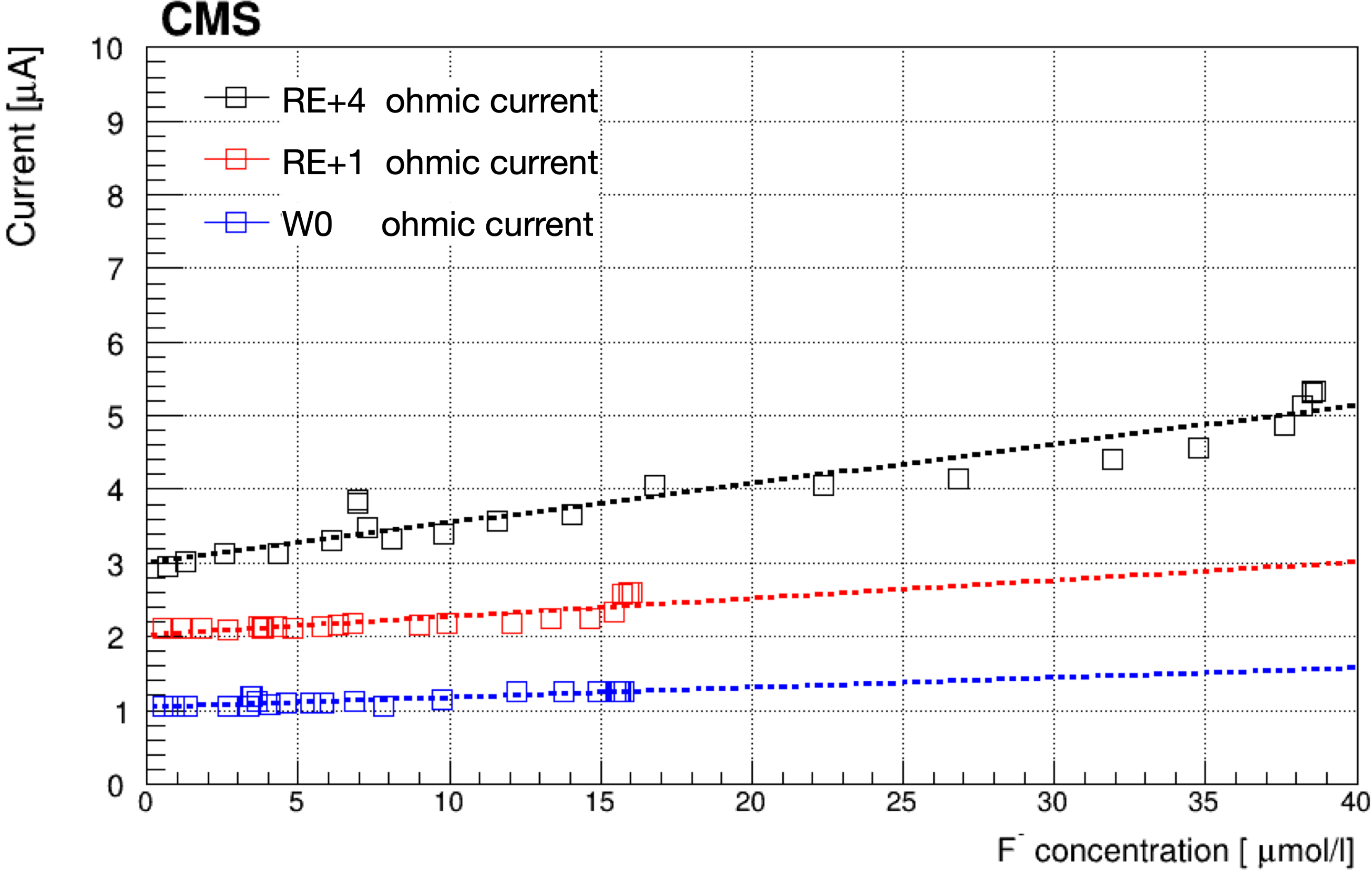}
\caption{%
    Ohmic current as a function of HF concentration. Figure from Ref.~\cite{CMS:2020mpl}.
}
\label{fig:rpc:ohmichf}
\end{figure}

The ohmic currents as a function of HF concentration are shown in Fig.~\ref{fig:rpc:ohmichf}.
The RE$+1$ and W$+0$ chambers have a similar HF concentration and gas flow (0.7 and 0.6\unit{v/h}), and a background of less than 10\Hzcmsq.
In RE$+4$, the amount of HF accumulated is higher by a factor 2 at a background of 40\Hzcmsq, and the gas flow is lower (0.35\unit{v/h}) than in W$+0$ and RE$+1$.
There is a clear linear dependence between the ohmic current and the HF concentration, which implies that HF trapped in the gas gap may form a thin conductive layer.
The HF can be efficiently removed by increasing the gas flow in the chambers, depending on the background rate.

\paragraph{GIF++ longevity studies}

The present RPC system was originally certified for ten years of LHC operation at a maximum background rate of 300\Hzcmsq and a total integrated charge of 50\mCcmsq~\cite{CMS:TDR-3, CMS:2019jwc}.
Based on the data collected in \Run2, assuming a linear dependence of the background rates as a function of the instantaneous luminosity, and including a safety factor of three, the expected background rates and integrated charge at the HL-LHC will be about 600\Hzcmsq and 840\mCcmsq, respectively~\cite{CMS:TDR-016}.
In such operating conditions, irrecoverable aging effects can appear, as the higher collision rates affect the detector properties and performance.

Therefore, since July 2016, a long-term irradiation test has been carried out at the CERN gamma irradiation facility (GIF++)~\cite{Guida:2016ivc} to study whether the present RPC detectors can survive the difficult background conditions during the HL-LHC running period~\cite{CMSMuonGroup:2020arl}.

Four spare RPC chambers have been irradiated, two each of type RE$2/2$ and RE$4/2$~\cite{CMS:Detector-2008, CMS:TDR-016}.
These are from the endcaps where the backgrounds are expected to be maximal~\cite{CMS:2014mpn}.
Two different RPC production types have been tested, reflecting the fact that the RPC endcap production was done in two different periods, 2005 for the RE2 detectors (both RE$\pm$2) and 2012--2013 for the RE4 detectors (both RE$\pm$4).
Two chambers, one from each period, are continuously being irradiated while the other two of the same type are kept as reference and are switched on only from time to time.
The detector parameters, such as dark currents, noise rates, currents, and count rates for various background conditions are monitored continuously and compared with the measurements from the reference chambers.

The integrated charge is calculated as the average density current accumulated in time in the three gaps that constitute the detector, since the gamma flux, provided by the 14\TBq \onethreesevenCs source at the GIF++~\cite{Guida:2016ivc} is uniformly distributed over the detector surface.
The collected integrated charge from the beginning of irradiation until September 2022 is about 813 and 478\mCcmsq for the RE2 and RE4 chambers, respectively, which corresponds to approximately 97 and 57\% of the expected integrated charge at the HL-LHC.

\subparagraph{Dark current and noise rate studies}

\begin{figure}[!ht]
\centering
\includegraphics[width=0.48\textwidth]{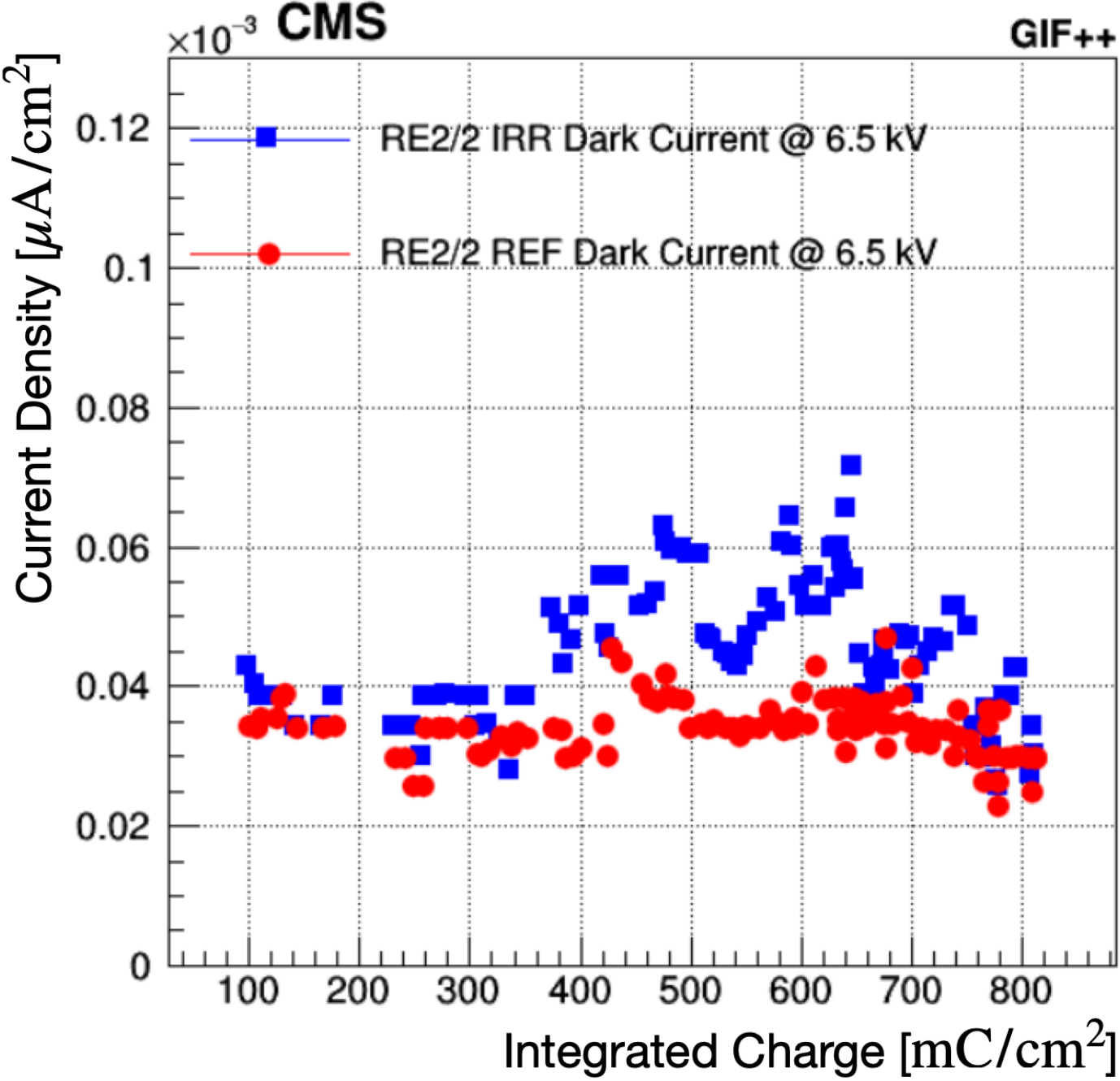}%
\hfill%
\includegraphics[width=0.48\textwidth]{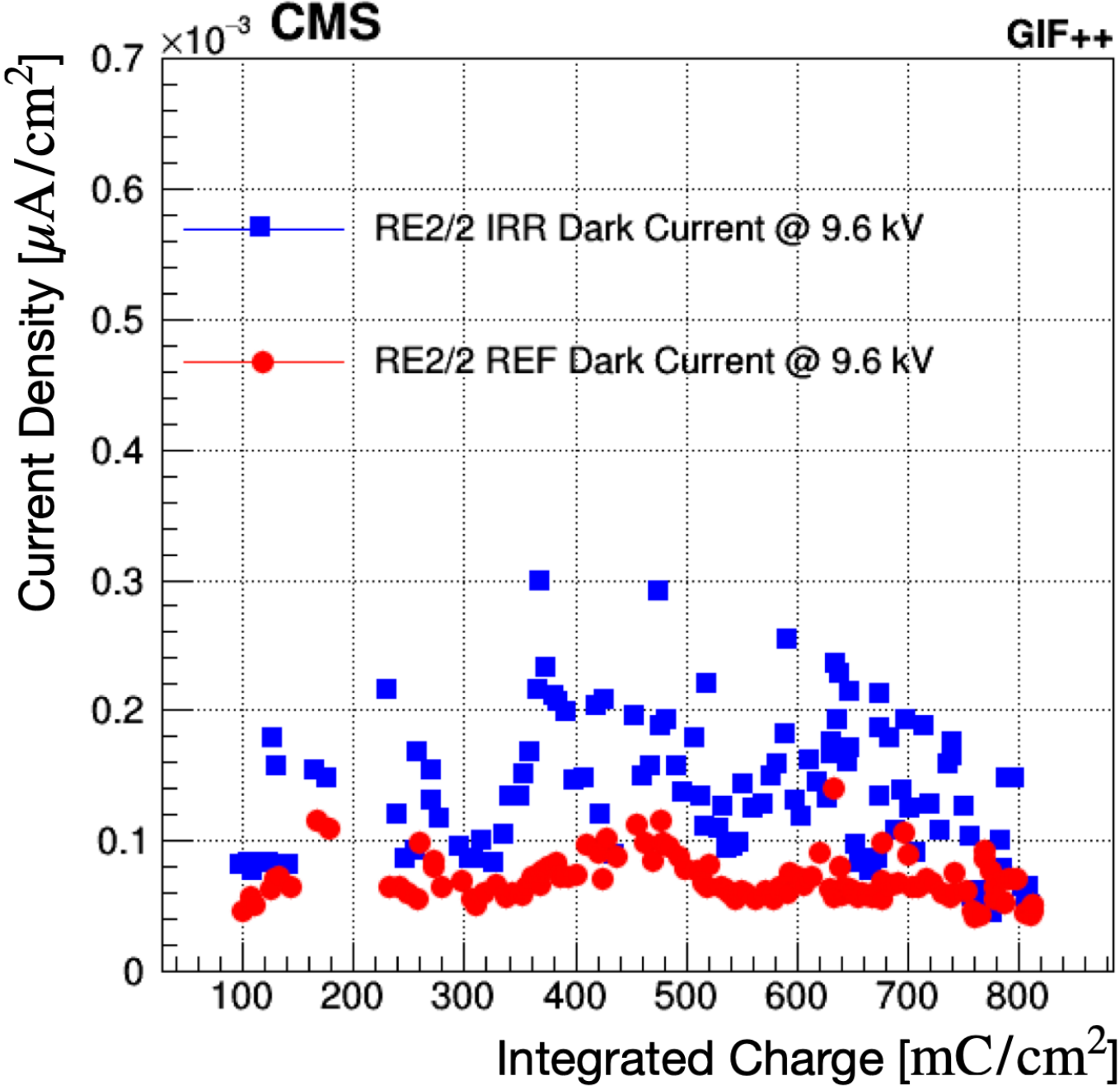}
\caption{%
    Dark-current density for the irradiated (blue squares) and reference (red circles) RE2 chambers as a function of the collected integrated charge at 6.5 (left) and 9.6\kV (right). Figures from Ref.~\cite{CMSMuon:2023skt}.
}
\label{fig:rpc:darkcurrent}
\end{figure}

Dark currents and noise rates are monitored periodically in order to spot aging effects due to irradiation.
The dark-current density, \ie, the currents normalized to the surface area, for both the irradiated and reference RE2 chambers are shown in Fig.~\ref{fig:rpc:darkcurrent} as a function of the collected integrated charge.
The dark currents were measured at 6.5\kV to determine the ohmic contribution, and at 9.6\kV to include the contribution from gas amplification.

\begin{figure}[!t]
\centering
\includegraphics[width=0.48\textwidth]{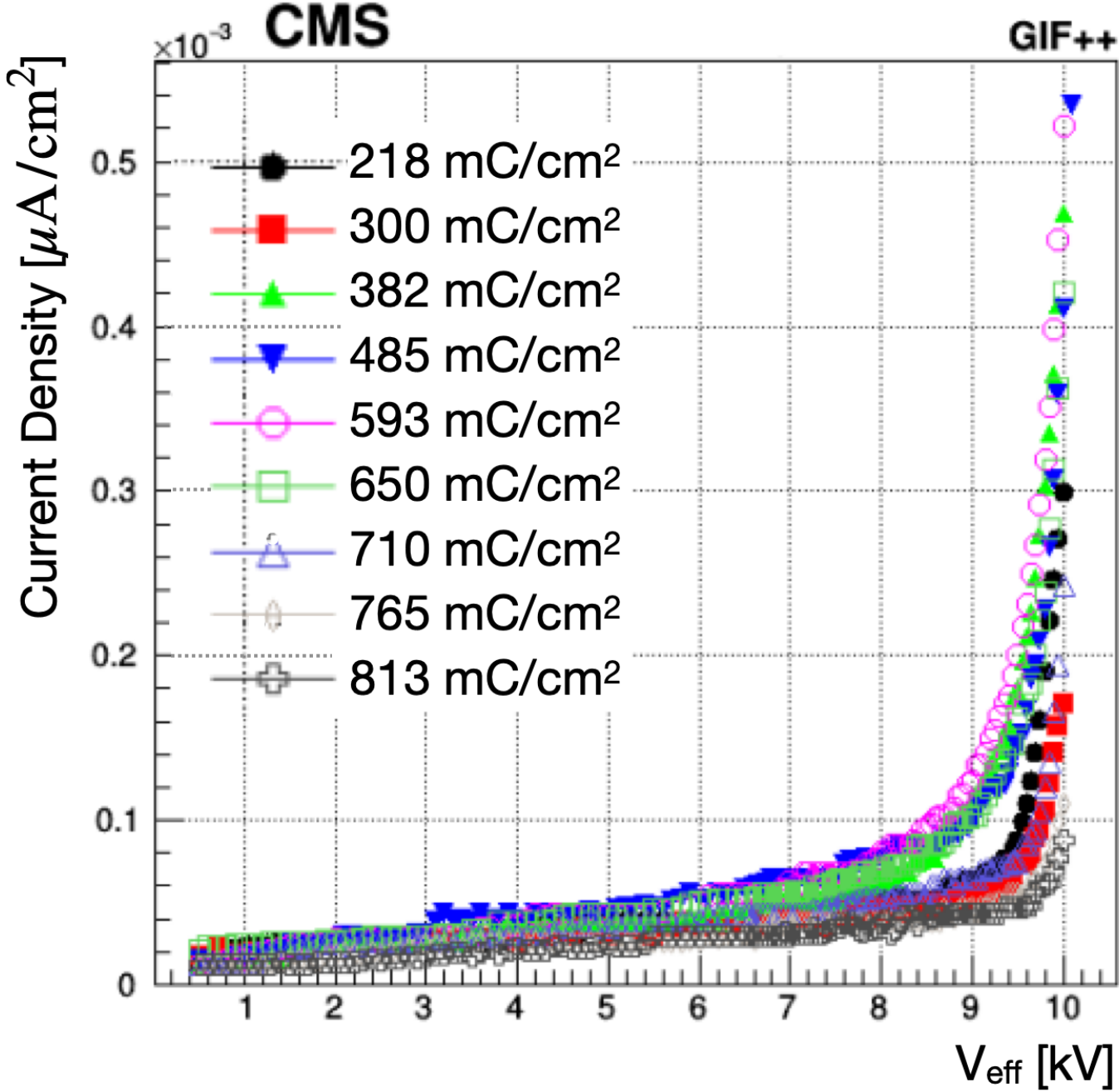}%
\hfill%
\includegraphics[width=0.48\textwidth]{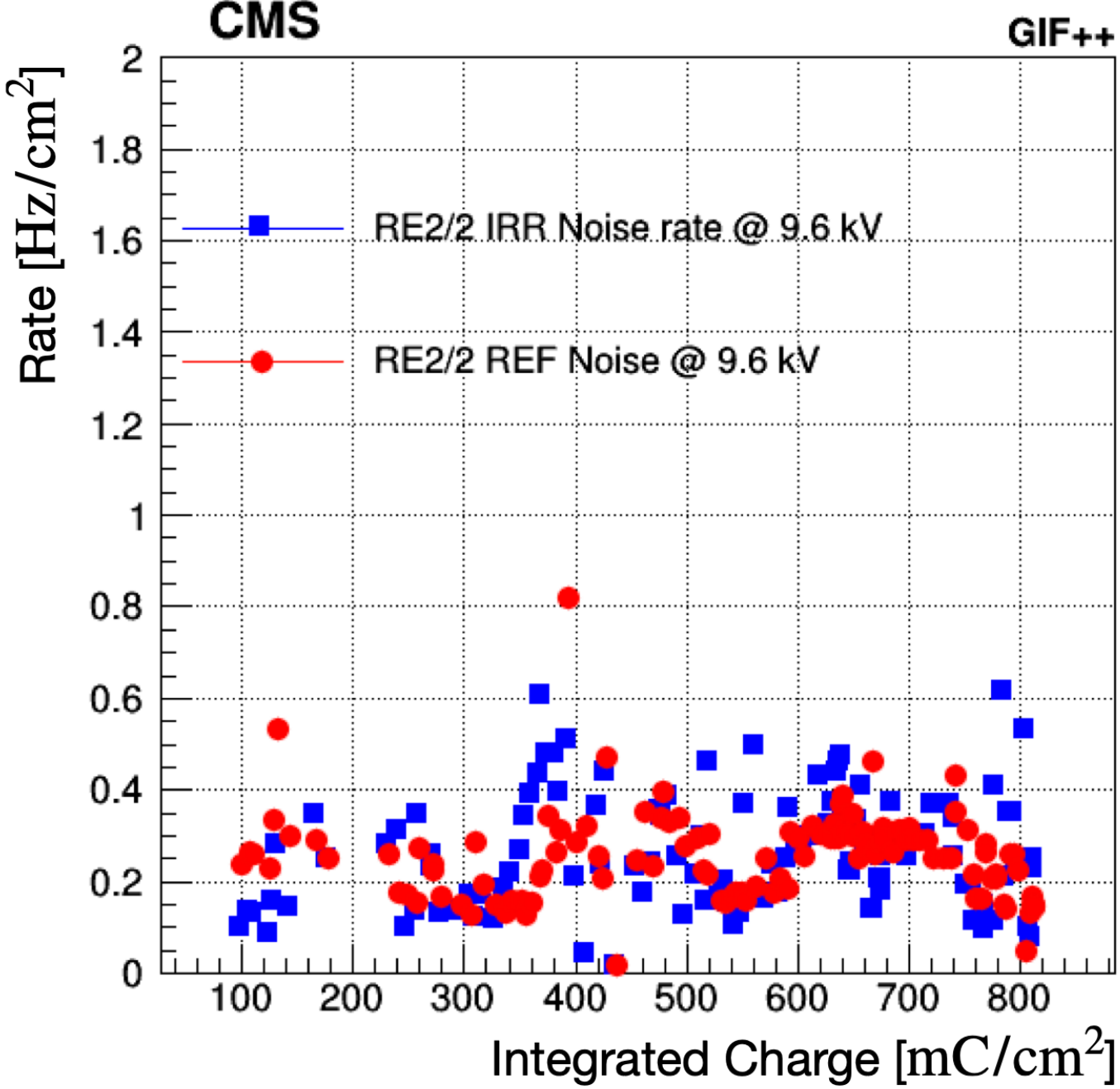}
\caption{%
    Left:\ dark-current density monitored as a function of the effective high voltage at different values of the collected integrated charge for the irradiated RE2 chamber.
    Right:\ average noise rate as a function of the collected integrated charge for the irradiated (blue squares) and reference (red circles) RE2 chambers.
    Figures from Ref.~\cite{CMSMuon:2023skt}.
    }
\label{fig:rpc:hvscan}
\end{figure}

{\tolerance=800
Figure~\ref{fig:rpc:hvscan} (left) shows the dark-current density monitored as a function of the effective high voltage, \ie, the voltage normalized to the standard temperature of 20\deC and pressure of 990\mbar~\cite{Colafranceschi:2014ija} at different values of collected integrated charge.
Since the beginning of irradiation, the dark currents have been stable in time, with only small acceptable variations.
Figure~\ref{fig:rpc:hvscan} (right) displays the average noise rate for the irradiated and reference RE2 chambers as a function of the collected integrated charge.
The average noise rate is stable with time and less than 1\Hzcmsq.
\par}

\subparagraph{Resistivity and current studies}

Other important parameters that are measured periodically are the current in the presence of background radiation and the resistivity of the electrodes.
The latter is measured several times per year, since it is a crucial performance parameter.
The resistivity is measured by filling the detector with pure argon and operating it in a self-sustaining streamer mode, which occurs when the gas-quenching components such as isobutane are removed.
The streamers propagate over the entire detector area, and by measuring the current at a given high voltage, the resistance, and hence the resistivity, can be calculated.
The measured resistivity values are normalized to 20\deC to allow comparison of the values for different temperatures~\cite{Carboni:2004gb}.

\begin{figure}[!ht]
\centering
\includegraphics[width=0.48\textwidth]{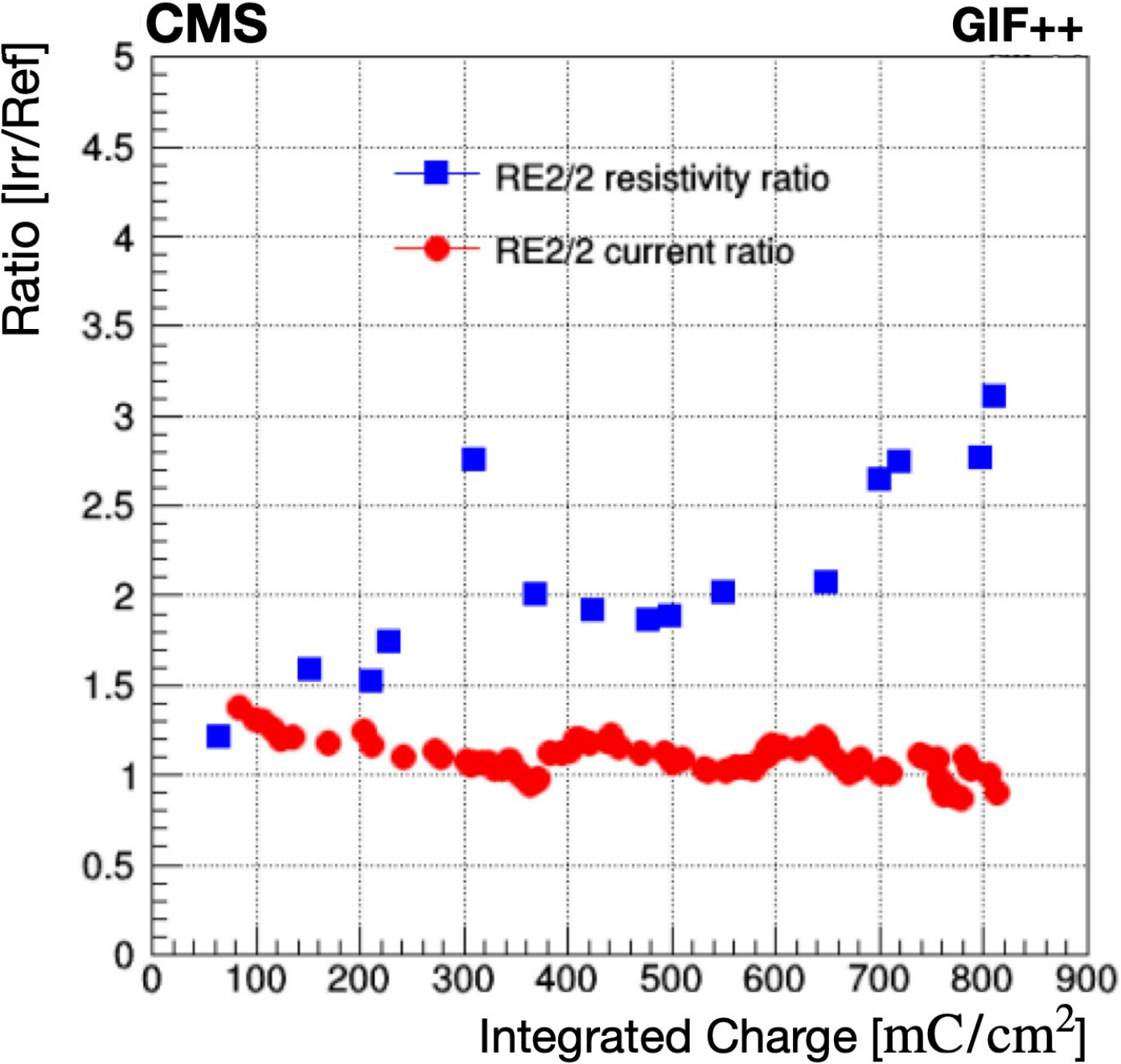}
\caption{%
    Resistivity ratio (blue squares) and current ratio (red circles) between the irradiated and reference RE2 chambers as a function of the collected integrated charge.
    Figure from Ref.~\cite{CMSMuon:2023skt}.
}
\label{fig:rpc:resistivity}
\end{figure}

To exclude effects from external parameters, the ratios of resistivity and current between irradiated and reference chambers are taken for a gamma background rate of about 600\Hzcmsq, as shown in Fig.~\ref{fig:rpc:resistivity}.
An increase in the resistivity is observed in the irradiated chamber during the first irradiation period, up to $\approx$300\mCcmsq, when the detectors are operated in conditions similar to those in CMS:\ one gas volume exchange per hour and a relative gas humidity of 35--45\%.
While the RPC system operates in these conditions of humidity and gas volume exchange, they are not the optimal ones for operation at GIF++, where the high background gamma rate is about 600\Hzcmsq, causing the HPL plates to dry and their resistivity to increase.
During the longevity test, after an integrated charge of $\approx$300\mCcmsq was reached, the relative gas humidity was increased and maintained at about 60\%, and the gas flow was increased to three gas volume exchanges per hour.
With these changes, the HPL resistivity decreased and the variations were reduced.
After $\approx$650\mCcmsq, an increase in resistivity is observed, which is related to the return to a low gas humidity at about 40\% due to sharing the same gas system with other RPCs.
The increase of resistivity is confirmed by the decrease of the measured current.

\subparagraph{Detector performance monitoring}

The detector performance has been measured during test beams before irradiation and at different periods of irradiation.
Comparisons between the irradiated RE2 chamber efficiency measured as a function of the effective HV without background radiation and in the presence of a background of 600\Hzcmsq, at different values of collected integrated charge, shows that the efficiency is stable over time in the absence of background radiation, and no shift in the WP is observed~\cite{Thyssen:2012zz, CMSMuonGroup:2020arl}.
In the presence of background, the efficiency is stable at the WP, but a WP shift of 100\unit{V} after collecting 260\mCcmsq of integrated charge is introduced.
This shift in the detector WP is related to the increase in the resistance ($R$) of the electrodes, which causes an increase of the voltage drop ($RI$) across them.
The latter leads to a difference between the effective voltage \Veff applied to the electrodes and the effective voltage across the gas gap \Vgas, which is compensated by introducing the shift~\cite{Pugliese:2001ab}.
A similar increase in resistivity is seen in the irradiated RE4 chamber.
The quantity \Vgas is defined as:
\begin{linenomath}\begin{equation}
    \Vgas = \Veff - RI,
\end{equation}\end{linenomath}
where $R$ is the HPL resistance and $I$ is the total current.

\begin{figure}[!ht]
\centering
\includegraphics[width=0.48\textwidth]{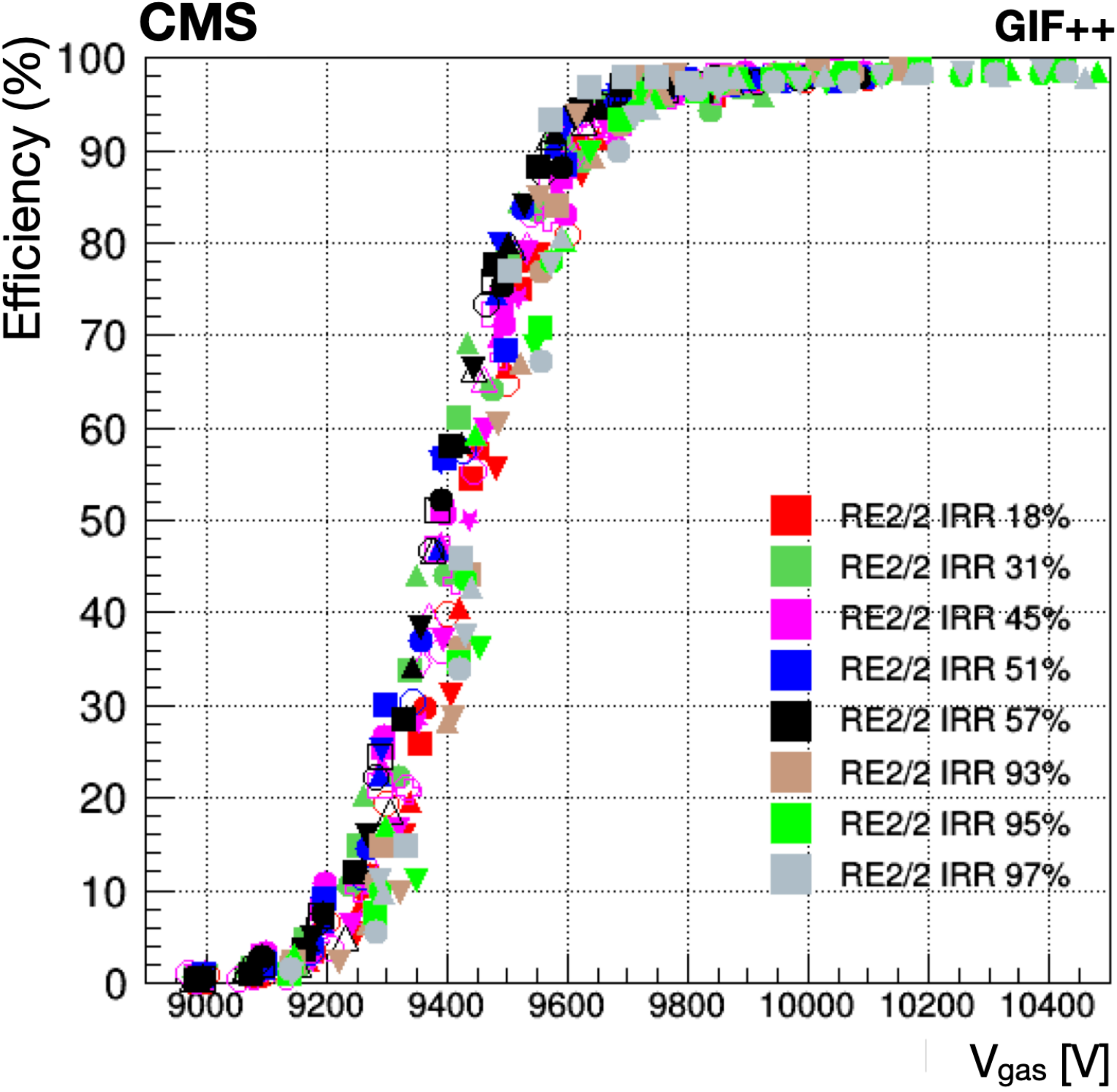}%
\hfill%
\includegraphics[width=0.48\textwidth]{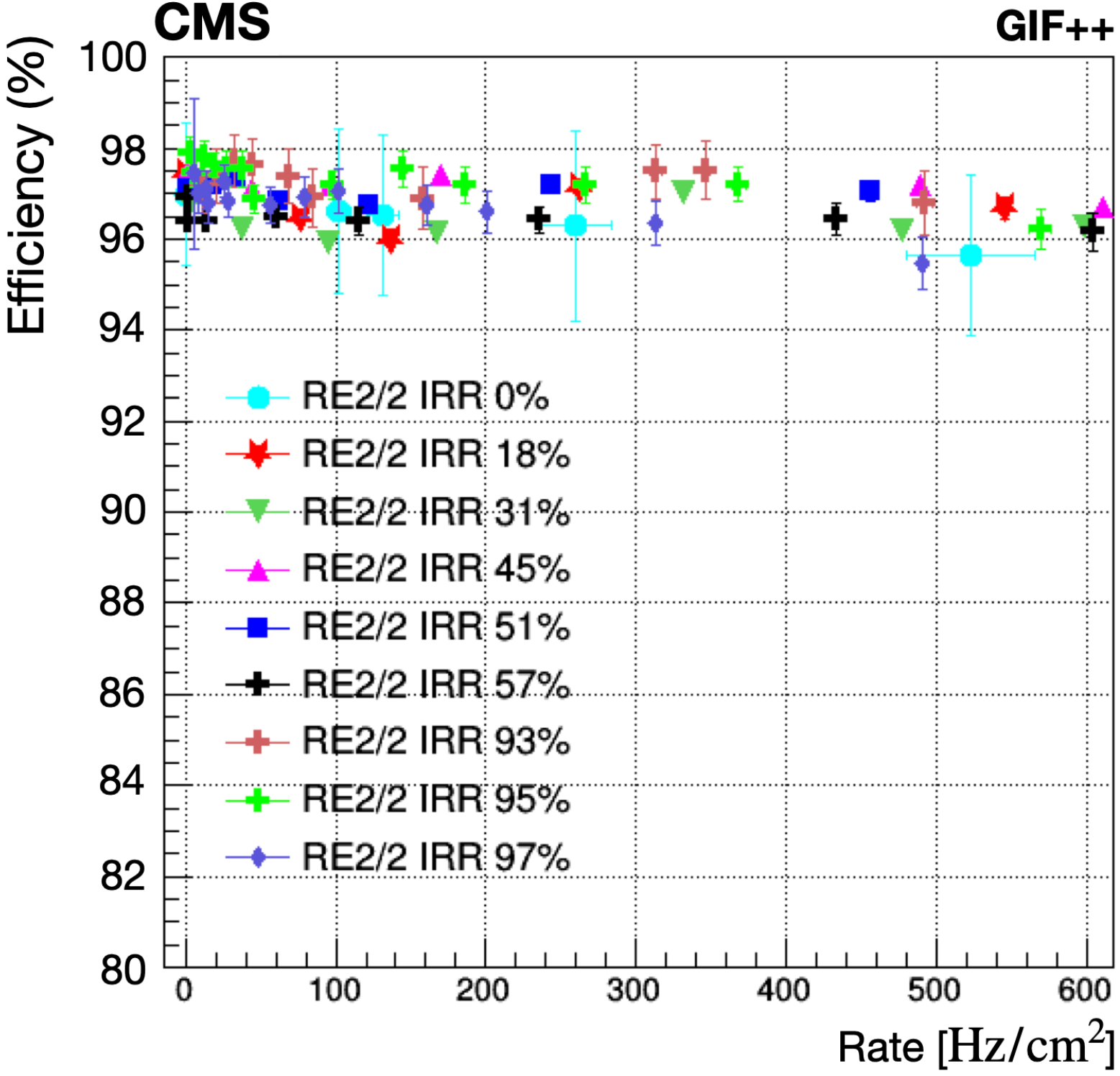}
\caption{%
    Left:\ irradiated RE$2/2$ chamber efficiency as a function of \Vgas for different background irradiation rates, up to 600\Hzcmsq, and different integrated charge values.
    Different marker shapes of the same color represent different background rates at the same integrated charge values.
    Right:\ irradiated RE2 chamber efficiency at the WP as a function of the background rate at different values of the collected integrated charge.
    Figures from Ref.~\cite{CMSMuon:2023skt}.
    }
\label{fig:rpc:hvgas}
\end{figure}

Since the detector operation regime is invariant with respect to \Vgas, the efficiency as a function of \Vgas does not depend on the HPL resistance, as shown in Fig.~\ref{fig:rpc:hvgas} (left).
After removing the $R$-increase effect on the electrodes by introducing \Vgas instead of \Veff, all the efficiency curves overlap, and no shift in the WP is observed.

The irradiated RE2 chamber efficiency at the WP is measured at different background rates and at different integrated charge values, as shown in Fig.~\ref{fig:rpc:hvgas} (right).
The efficiency is stable over time up to the highest background rate expected at the HL-LHC.

Longevity studies on spare RPCs are ongoing at GIF++ under controlled conditions.
Preliminary results show no evidence of aging effects. The main detector parameters and performance are stable.

\subsubsection{Changes to the RPC system in LS2}

\paragraph{Gas leak repairs and green-house gas emission strategy for \Run3}

The highest priority during LS2 was the consolidation of the RPC gas system.
The aim was to minimize the gas emissions and thus the environmental impact, since the standard gas mixture is composed mainly of F-gases with high global-warming potential.
To accomplish this, actions were taken to minimize the number of leaks and to implement a newly developed Freon recuperation system prototype~\cite{Guida:2020yqs}.

The RPC gas system is a 13\mcub closed-loop volume with re-circulation of 7.3\mcubh nominal mixture flow:\ 5\mcubh for the barrel and 2.3\mcubh for the endcaps.
A combination of factors during production of the components and operation of the system can lead to an increase of the leak rate.
Environmental conditions in the experimental cavern, such as humidity and temperature, as well as switching from Freon to \Ntwo and back to Freon, can accelerate the degradation process of the different components of the RPC gas system.

The gas leaks in the RPC system are mainly caused by the T-shaped or L-shaped polycarbonate gas connectors that are broken due to stress applied through the gas pipes, and the low-density polyethylene pipes that are brittle, deteriorated, or cut.

\begin{figure}[!ht]
\centering
\includegraphics[width=0.9\textwidth]{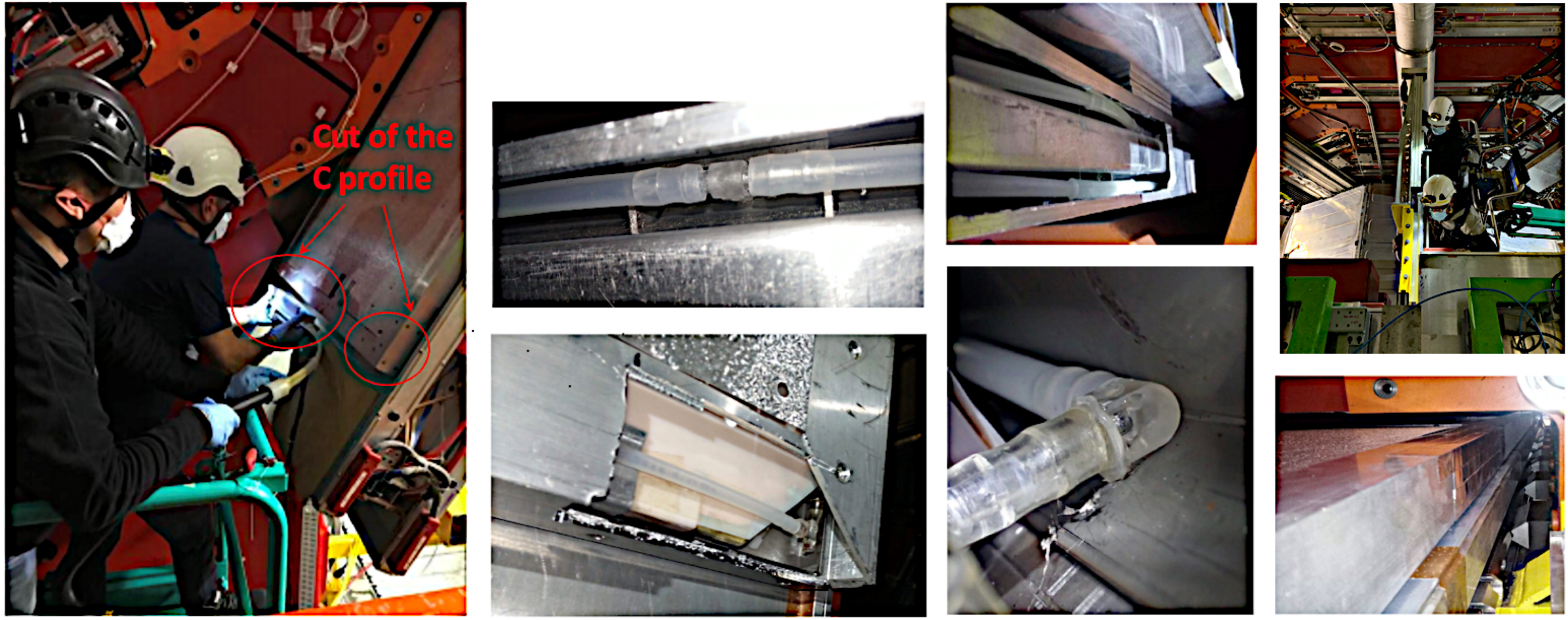}
\caption{%
    Photographs illustrating the gas-leak repair procedures.
    Left:\ scientists working on the repair procedure.
    Middle left:\ access to broken component.
    Middle right:\ repairing of components.
    Right:\ closing and validation.
}
\label{fig:rpc:leakrepairs}
\end{figure}

A special gas-leak repair procedure was developed to correctly identify the leaks using an endoscope, which allowed us to determine the exact location and the components that are sources of leaks.
The RPC barrel chambers are coupled with the DT chambers and inserted into the iron yoke.
Therefore, in order to reach the leak location, a partial extraction of the entire muon station (RPC and DT) by a distance of 80\cm from the back or front side is required.
A surgical cut of the C-aluminum profile is then performed to gain access to the broken gas pipe or T/L connector.
In Fig.~\ref{fig:rpc:leakrepairs}, an example of the newly developed gas-leak repair procedure is shown.
In one scenario, the broken pipe connecting the two chambers is removed, changing the splitting of the gas flow from internal to external (Fig.~\ref{fig:rpc:leakrepairs}, upper center) and preserving the parallel gas distribution of the chambers.
In another scenario, the repair is accomplished by gluing the
T/L connector (Fig.~\ref{fig:rpc:leakrepairs}, lower center).

By November 2020, when all possible repairs were finalized, a total number of 88 gas leaks due to cracked or broken pipes were identified in the barrel chambers.
Of those, 50 chambers were successfully repaired and returned to normal operation.
This included 17 chambers that did not work during \Run2.
The remaining 38 leaking chambers are either not accessible or the partial extraction technique is not applicable, either because the source of the leak is inaccessible or the source of the leak is not identified.
In the endcaps, a total of 11 chambers were replaced.

\begin{figure}[!ht]
\centering
\includegraphics[width=0.48\textwidth]{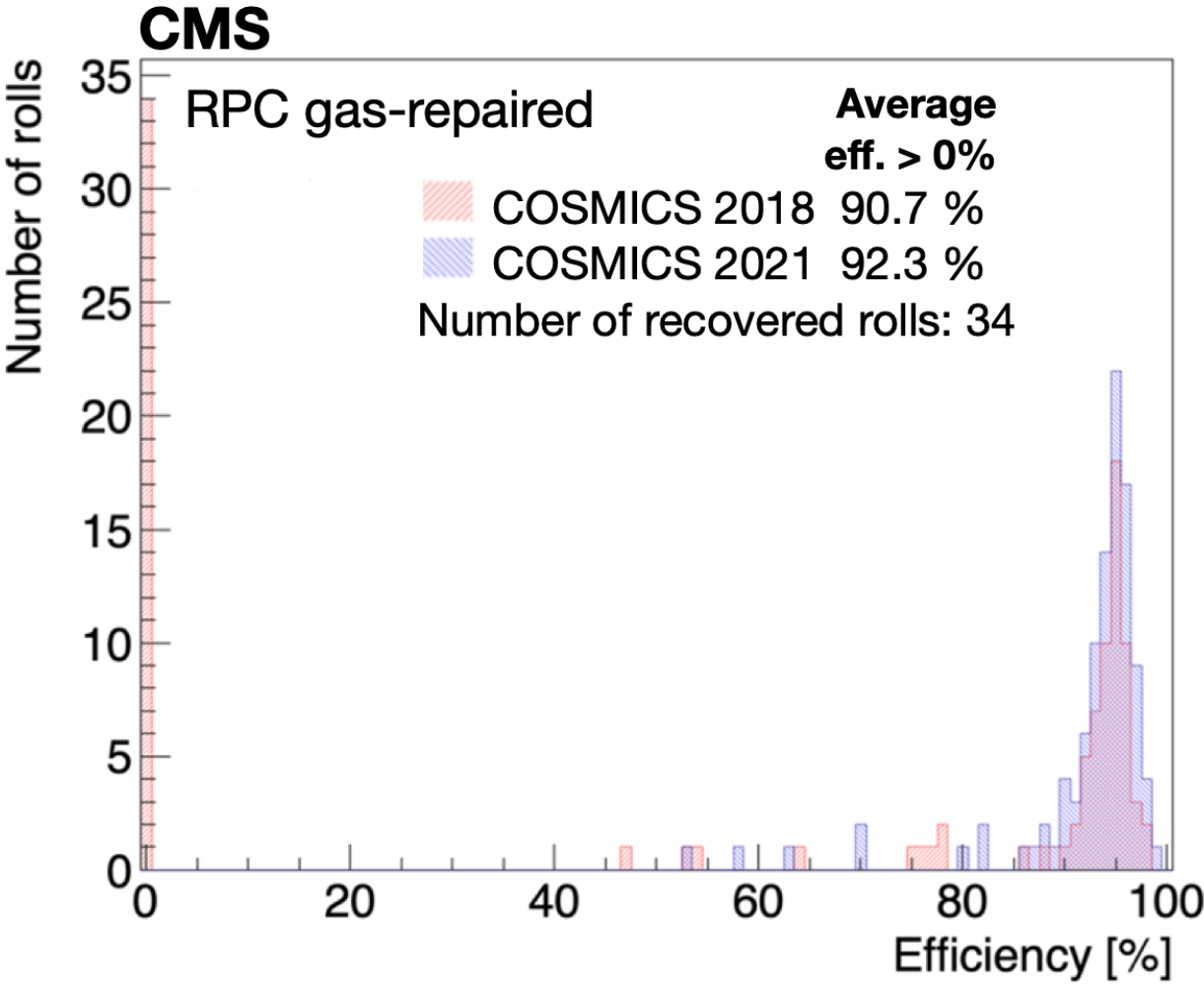}
\caption{%
    Efficiency comparison between 2018 (red) and 2021 (blue) cosmic ray data for chambers with repaired gas leaks.
}
\label{fig:rpc:gasleakeff}
\end{figure}

The gas-leak repairs had a large impact on the chamber performance.
Figure~\ref{fig:rpc:gasleakeff} shows a comparison of the efficiency of repaired chambers using cosmic ray data between the end of \Run2 before the reparation campaign and after all the repairs.
This led to an overall gain in efficiency of 1.5\%.

To minimize chamber pressure variations, which are considered to be a cause of leaks, automatic regulation valves on pre-distribution racks in the underground service cavern were installed in November 2021.
This operation has substantially reduced the development of new leaks in the system.

In the beginning of \Run3 in June 2022, the total number of disconnected barrel chambers was 108, corresponding to about 10\% of the entire RPC system.
This number includes 14 dysfunctional chambers with gas leaks, 63 operational chambers with gas leaks, and 31 gas-tight chambers that share gas channels with leaking chambers.

Another major activity was testing a Freon-recuperation prototype connected to the RPC gas system exhaust.
The aim is to recuperate the \CtwoHtwoFfour from the RPC gas mixture and re-use it~\cite{Guida:2020yqs}.
The \CtwoHtwoFfour and \iCfourHten gases form an azeotrope, \ie, a mixture of liquids whose proportions can not be altered or changed by simple distillation, because the intramolecular force of same species is much higher than the reciprocal attraction.
The separation of the two gases is done by slowly heating the liquefied azeotrope, allowing the enrichment of the liquid R134a and the \iCfourHten vapor.
The CERN EP-DT gas team is finalizing the R\&D on the first custom-built \CtwoHtwoFfour-recuperation system with an expected efficiency of 80\%~\cite{Guida:2022pyg}.
To maximize the amount of gas in the exhaust line in order for efficient recuperation system operation, the procedure is to turn off and disconnect all leaking chambers.
The recuperation system is expected to be ready and operational by summer 2023.

\paragraph{RPC commissioning during LS2}

\begin{figure}[!ht]
\centering
\includegraphics[width=0.48\textwidth]{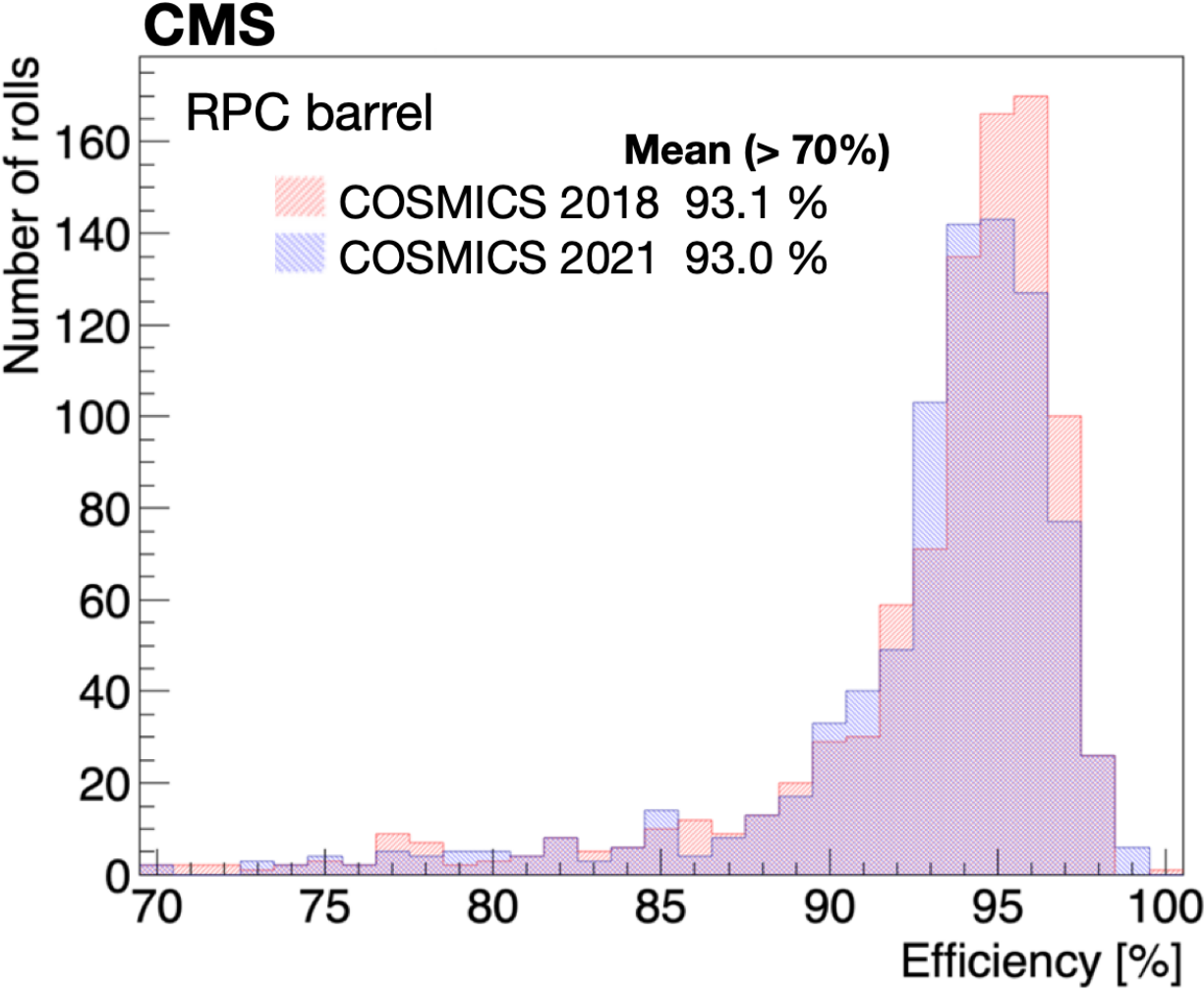}%
\hfill%
\includegraphics[width=0.48\textwidth]{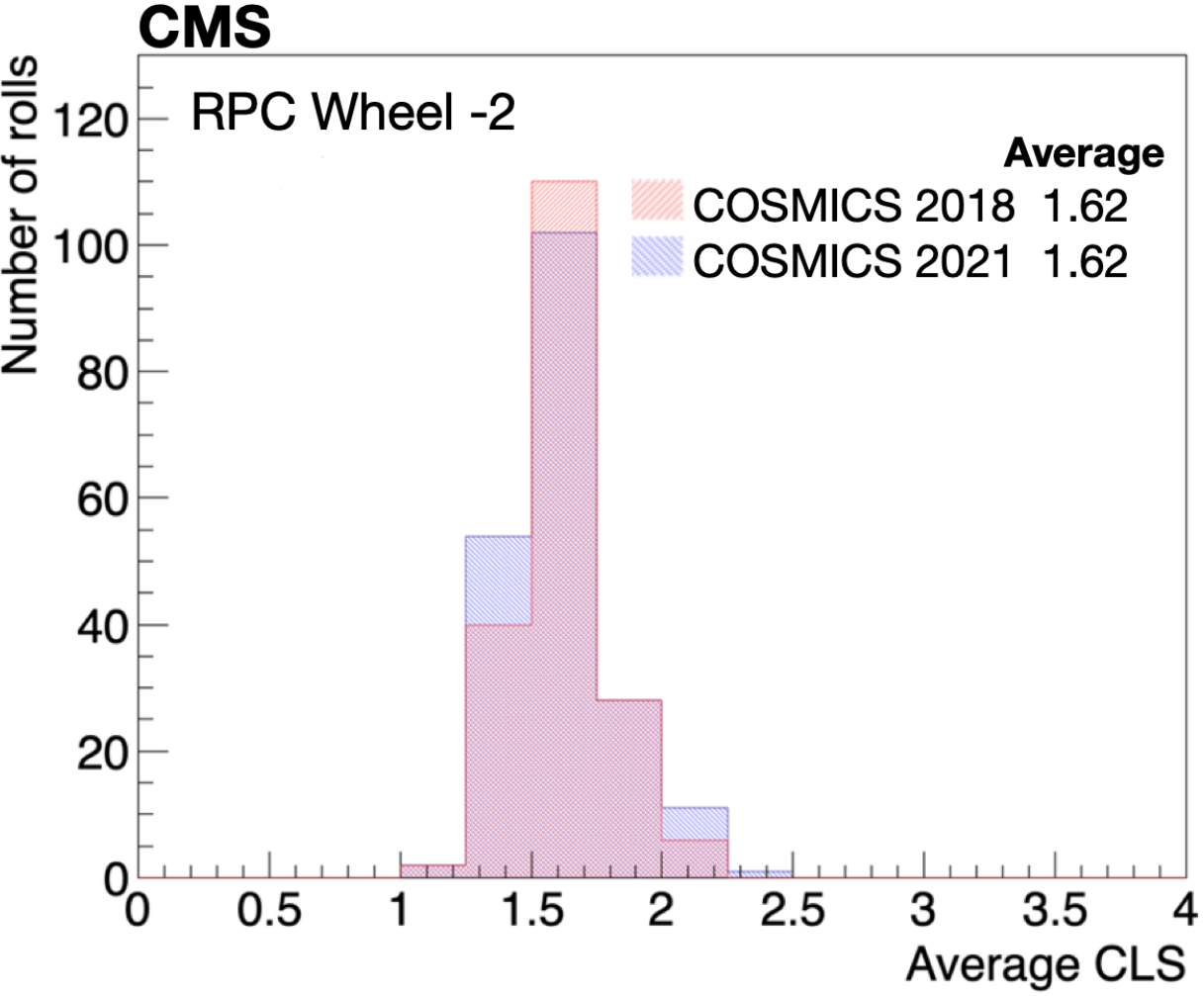}
\caption{%
    Left:\ distribution of the efficiency per roll in the RPC barrel chambers.
    Only chambers with rolls of efficiency greater than 70\% are considered.
    Right:\ average cluster size for the RPC barrel chambers in wheel W$-2$.
    In both figures, the cosmic ray muon data from 2018 (2021) are indicated in red (blue).
}
\label{fig:rpc:fulleffs}
\end{figure}

Figure~\ref{fig:rpc:fulleffs} (left) shows the efficiencies per roll, calculated using the segment extrapolation method described in Section~\ref{sec:rpc:efficiency}.
Comparing the average efficiencies for 2018 and 2021, an increase of 0.6\% is observed.
This increase can mainly be attributed to the recovery of chambers from single-gap to double-gap operation mode.
The fraction of chambers with more than 70\% efficiency rises by 6.1\%, considerably raising the redundancy in the measurement of muons.
These improvements are a result of the extensive maintenance and repairs during LS2.

In Fig.~\ref{fig:rpc:fulleffs} (right), the cluster size distribution for barrel wheel W$-2$ is displayed.
The average cluster size is kept around two strips, far below the possible maximum of three, in accordance with the trigger requirements.
The comparison between the 2018 and 2021 cosmic ray muon data shows a stable cluster size.

The efficiency measurement using cosmic ray muon data in the endcaps can have large systematic uncertainties, related to the vertical geometry of the CMS muon system in the endcap region.
Even so, the current and occupancy values at the WP during the cosmic ray muon data taking show good agreement with the expectation.

After extensive maintenance and repairs, as well as commissioning and longevity studies during LS2, the RPC detector successfully entered the data-taking period of \Run3.

\subsection{Gas electron multiplier chambers}
\label{sec:gem}

\subsubsection{Motivation and general description}

During the future operation of the High-Luminosity LHC (HL-LHC), the maximum hit rate in the forward region of the muon system is expected to reach 5\kHzcmsq in the first muon layer, with an integrated charge per unit area of 100\mCcmsq~\cite{CMS:TDR-016}.
This increase compared to the present values represents a challenge to the forward muon system, which must be resistant to radiation, have a high rate capability, and maintain an adequate pattern recognition for efficient muon reconstruction, while minimizing the number of misidentified tracks and keeping the level-1 (L1) trigger rate at acceptable levels.

\begin{figure}[!ht]
\centering
\includegraphics[width=0.4\textwidth]{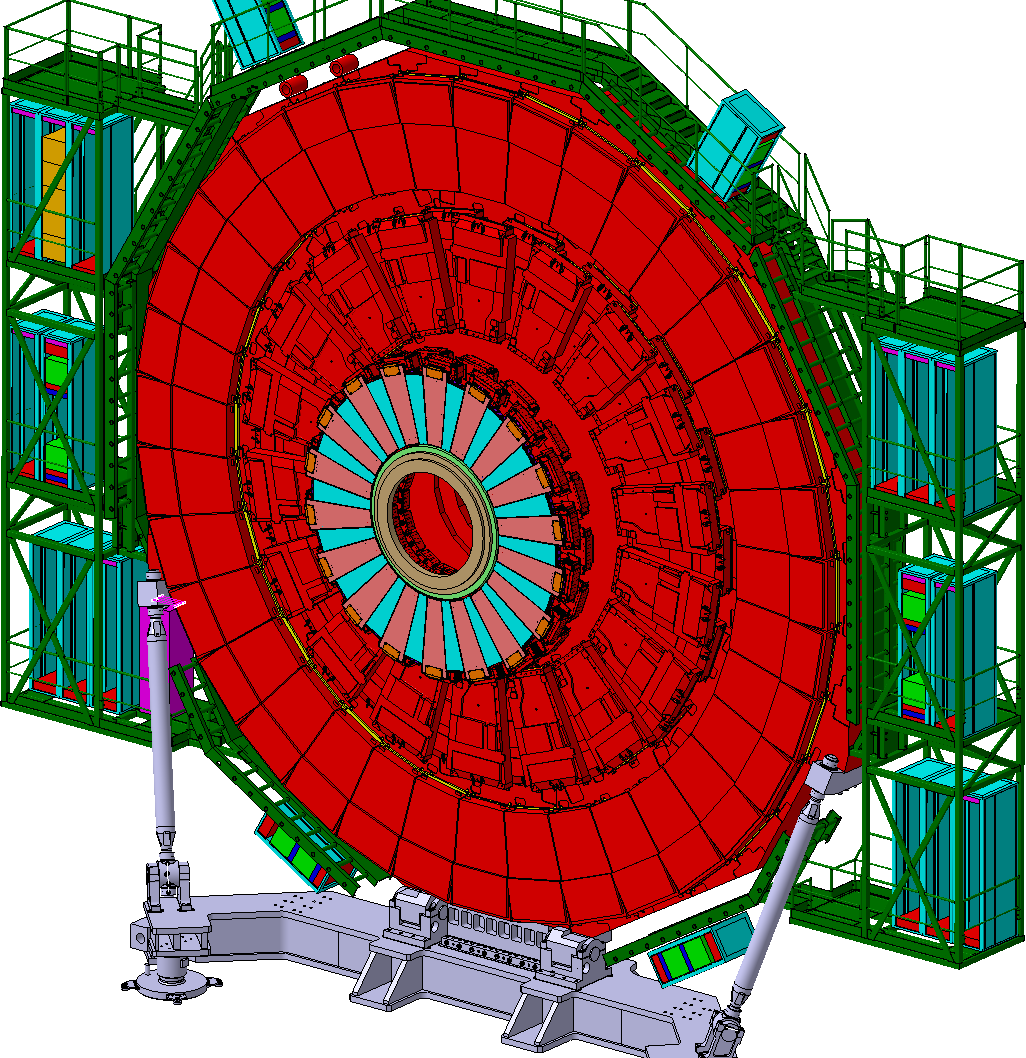}
\caption{%
    Sketch of GE1/1 system of one endcap indicating its location relative to the full endcap muon system, the endcap calorimeter, and the shielding elements, from Ref.~\cite{CMSMuon:2018wlc}.
}
\label{fig:gem:fullGE11}
\end{figure}

To enhance the track reconstruction and trigger capabilities of the endcap muon spectrometer, large-area triple-layer gas electron multiplier (GEM) detectors~\cite{Sauli:1997qp} were installed in the region covering $1.55<\abseta<2.18$ of the CMS detector for the start of \Run3.
This station, denoted GE1/1, is the first of three GEM rings that will be installed for the HL-LHC.
Figure~\ref{fig:muon:quadrant} shows a quadrant of the CMS detector with the location of the GE1/1 highlighted and outlined in red in the $r$--$z$ plane.
A detailed drawing is shown in Fig.~\ref{fig:gem:fullGE11}.

\begin{figure}[!htp]
\centering
\includegraphics[width=0.48\textwidth]{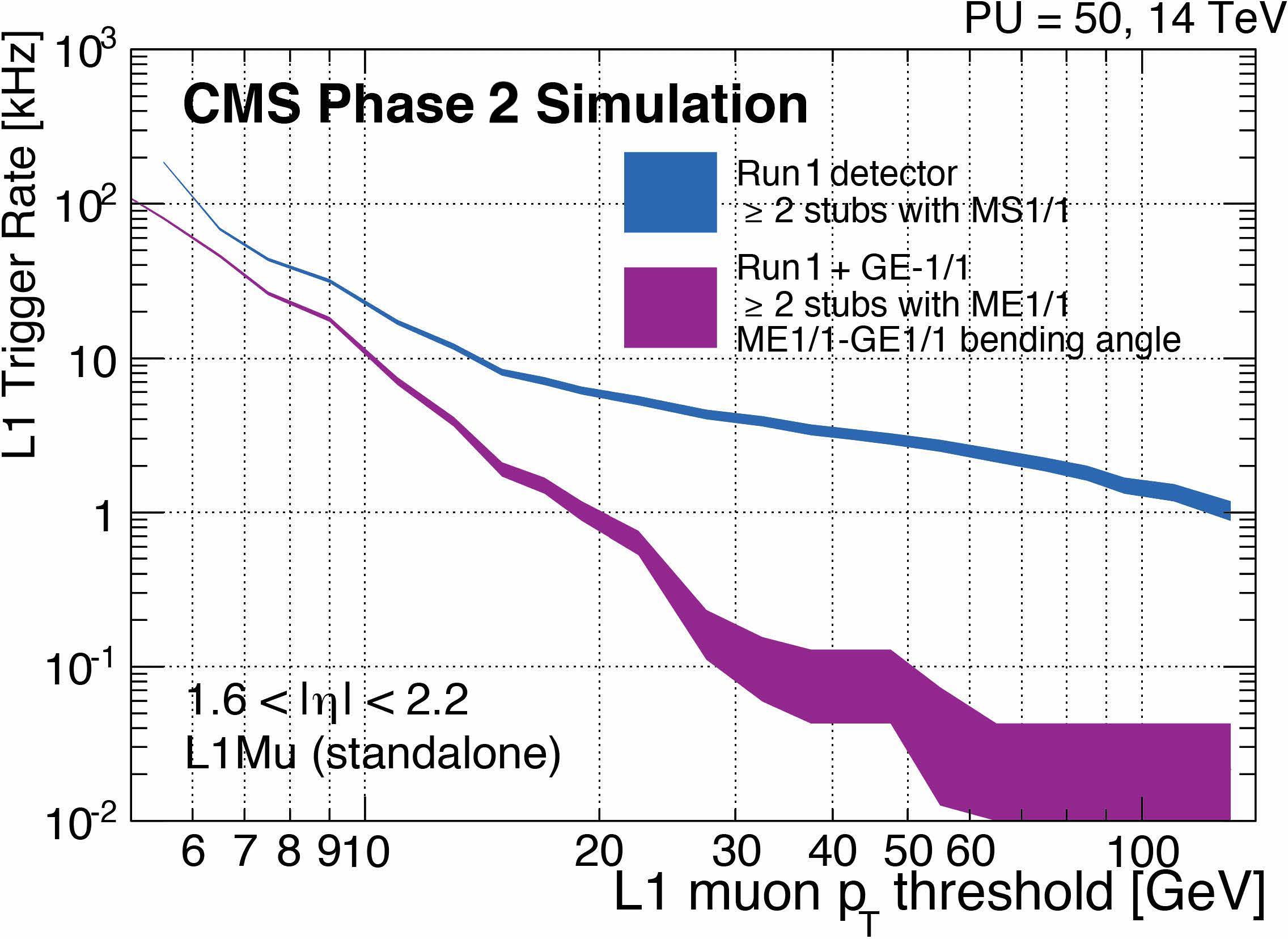}
\caption{%
    L1 muon trigger rate with and without the GE1/1 upgrade, assuming an instantaneous luminosity of $2\times10^{34}\percms$, where MS1/1 indicates the first muon station~\cite{CMS:TDR-016}.
}
\label{fig:gem:gemcsctrig}
\end{figure}

Muons experience the largest bending in the muon spectrometer at the position of the first muon station as the magnetic field lines bend around the endcap flux return.
Due to the increasing background rates at large $\eta$ and the reduction in the magnetic field in the first muon station, the trigger rate in this region is large and difficult to mitigate.
The insertion of the GE1/1 chambers increases the lever arm traversed by muons by a factor of 2.4--3.5, relative to ME1/1 alone, leading to a significant improvement in the muon trigger momentum resolution and a large reduction in the L1 trigger rate.
Figure~\ref{fig:gem:gemcsctrig} shows the expected trigger rates in the forward muon spectrometer with and without the GE1/1 upgrade.
This indicates that the upgrade will lower the trigger rates by a factor of 3--10 depending on the \pt threshold.
The muon trigger is described in detail in Section~\ref{sec:l1trigger:muon}.

\begin{figure}[!ht]
\centering
\includegraphics[width=0.43\textwidth]{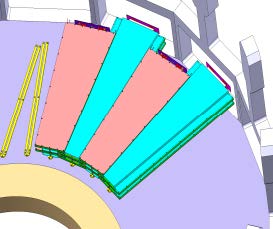}%
\hfill%
\includegraphics[width=0.53\textwidth]{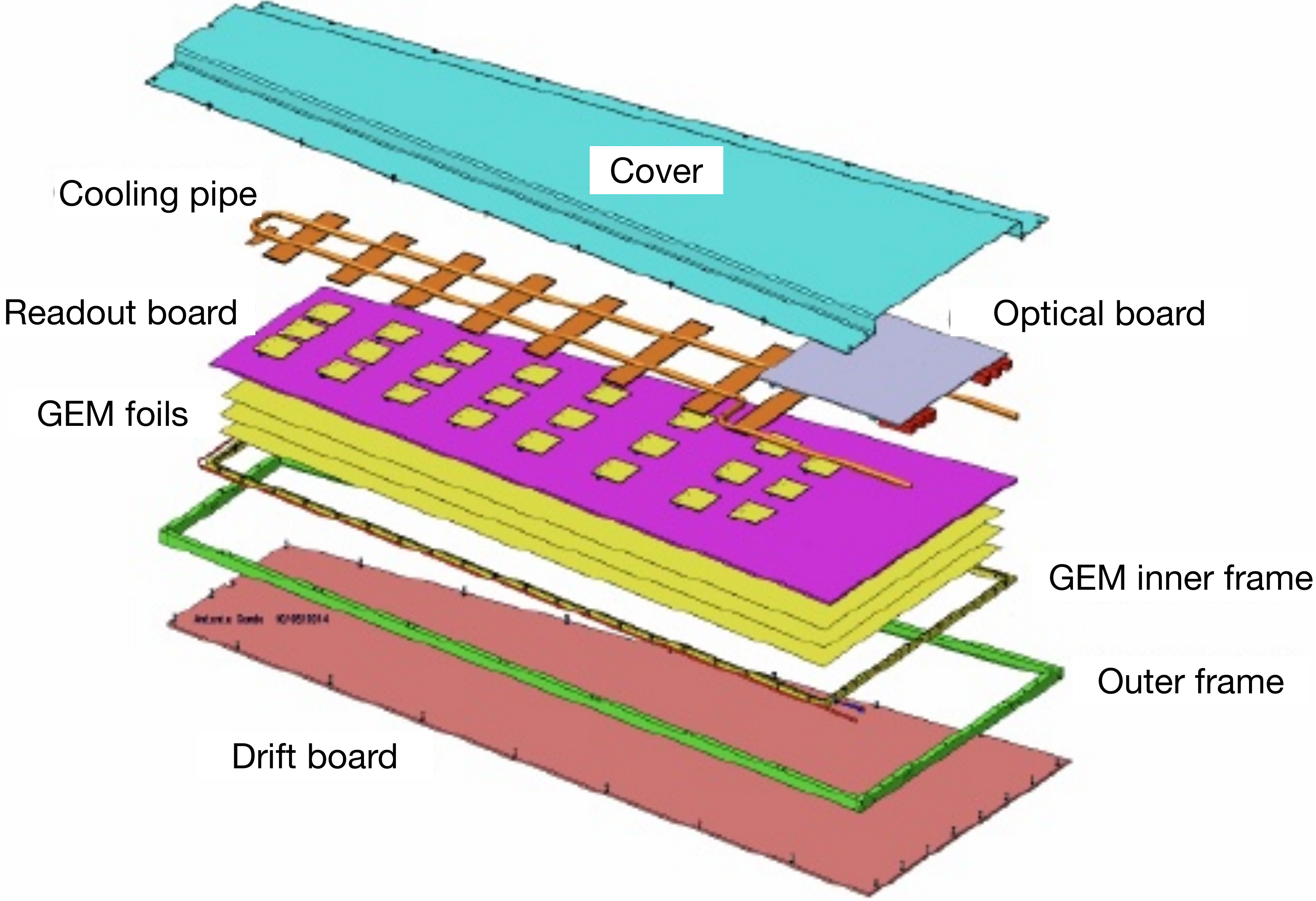}
\caption{%
    Left:\ layout of the GE1/1 chambers along the endcap ring, indicating how the short and long chambers fit in the existing volume. Figure from Ref.~\cite{Calabria:2016lez}.
    Right:\ blowup of the trapezoidal detector, GEM foils, and readout planes, indicating the geometry and main elements of the GEM detectors, from Ref.~\cite{CMS:2021bhf}.
}
\label{fig:gem:layout}
\end{figure}

Each endcap GE1/1 detector consists of 36 double-layered triple GEM chambers located just in front of the first CSC station, labeled ME1/1, each covering a 10\de\ sector in azimuth.
The chambers provide full coverage in $\phi$ and were constructed in two sizes, the odd-numbered GE1/1 are slightly longer in order to maximize the pseudorapidity coverage while fitting in the available space constrained by the support structure, as shown in Fig.~\ref{fig:gem:layout}.
The GEM detector technology can withstand rates up to 1\MHzcmsq and has excellent spatial and timing resolution of approximately 250--500\mum and $<$10\ns per layer, respectively.
The combined GE1/1 station spatial resolution is on the order of 100\mum.
The GE1/1 detector is placed, in global CMS coordinates, in $z$ between 566 and 574\cm, and at a radius between 145 and 230\cm.

\subsubsection{Technical design}

\begin{figure}[!t]
\centering
\includegraphics[width=0.4\textwidth]{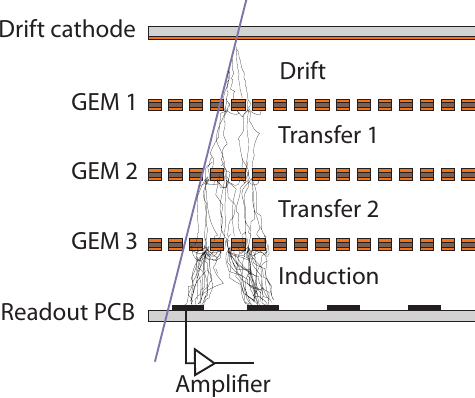}
\caption{%
    Sketch of a triple GEM detector showing the three foils, cathode, readout PCB and amplification, from Ref.~\cite{CMS:TDR-016}.
}
\label{fig:gem:triplegemsketch}
\end{figure}

The CMS triple GEM detector is a micro-pattern gas detector that comprises four gas gaps separated by three GEM foils, as shown in Fig.~\ref{fig:gem:triplegemsketch}.
It has an active area of 990$\times$(220--455)\mmsq with a 3/1/2/1\mm wide drift/transfer-1/transfer-2/induction field gap configuration~\cite{CMSMuon:2018wlc}.
The bottom of the GEM assembly is a printed circuit board that holds the drift electrode and voltage divider, while the top of the assembly is the readout board consisting of radially oriented readout strips along the long side of the chamber.
The strip pitch ranges from 0.6 to 1.2\mm, and the readout board is segmented in up to 10$\times$3 $\eta$--$\phi$ partitions, each with 128 strips.
The triple GEM arrangement allows for a charge amplification factor of up to a factor of several $10^5$, while limiting the probability of electrical breakdown or discharge.
The amplified charge induces a measurable signal on the readout electrode, which is segmented to provide positional information.
The gas mixture was chosen to be \ArCOtwo in a 70:30 proportion.

In the GE1/1 station shown in Fig.~\ref{fig:gem:fullGE11}, pairs of triple GEM detectors are matched to form a ``super-chamber'', providing two measurement planes and maximizing the detection efficiency for the station.
Thus, if each chamber operates at 98.0\% efficiency, the logical OR of the two signals from the super-chamber provides a 99.9\% efficiency.
The structure of a GE1/1 chamber is shown in Figs.~\ref{fig:gem:layout} (right) and~\ref{fig:gem:triplegemsketch}, while the dimensions and specifications of the two chamber types, ``short'' and ``long'', are given in Table~\ref{tab:gem:dim}.

\begin{table}[!htp]
\centering
\topcaption{%
    Dimensions and specifications of the GE1/1 short and long chambers, from Ref.~\cite{CMSMuon:2018wlc}.
}
\label{tab:gem:dim}
\renewcommand{\arraystretch}{1.1}
\begin{tabular}{lcc}
    Specification & Short & Long \\
    \hline
    Chamber length [cm] & 113.5 & 128.5  \\
    Chamber width [cm] & 28--48.4 & 26.6--51.2 \\
    Chamber thickness [cm] & 1.42 & 1.42 \\
    Active readout area [cm$^2$] & 3787 & 4550 \\
    Active chamber volume [liters] & 2.6 & 3 \\
    Geometric acceptance in $\eta$ & 1.61--2.18 & 1.55--2.18 \\
\end{tabular}
\end{table}

The main elements of an individual chamber are:\ the drift board holding the drift electrode, the GEM foils that amplify the ionization signal, the readout board where induced charge is read out on segmented strips, the internal and external frames, and a gas distribution system.
A more complete technical description of the layout and assembly of the GEM chambers can be found in Ref.~\cite{CMSMuon:2018wlc}.

\begin{figure}[!ht]
\centering
\includegraphics[width=0.44\textwidth]{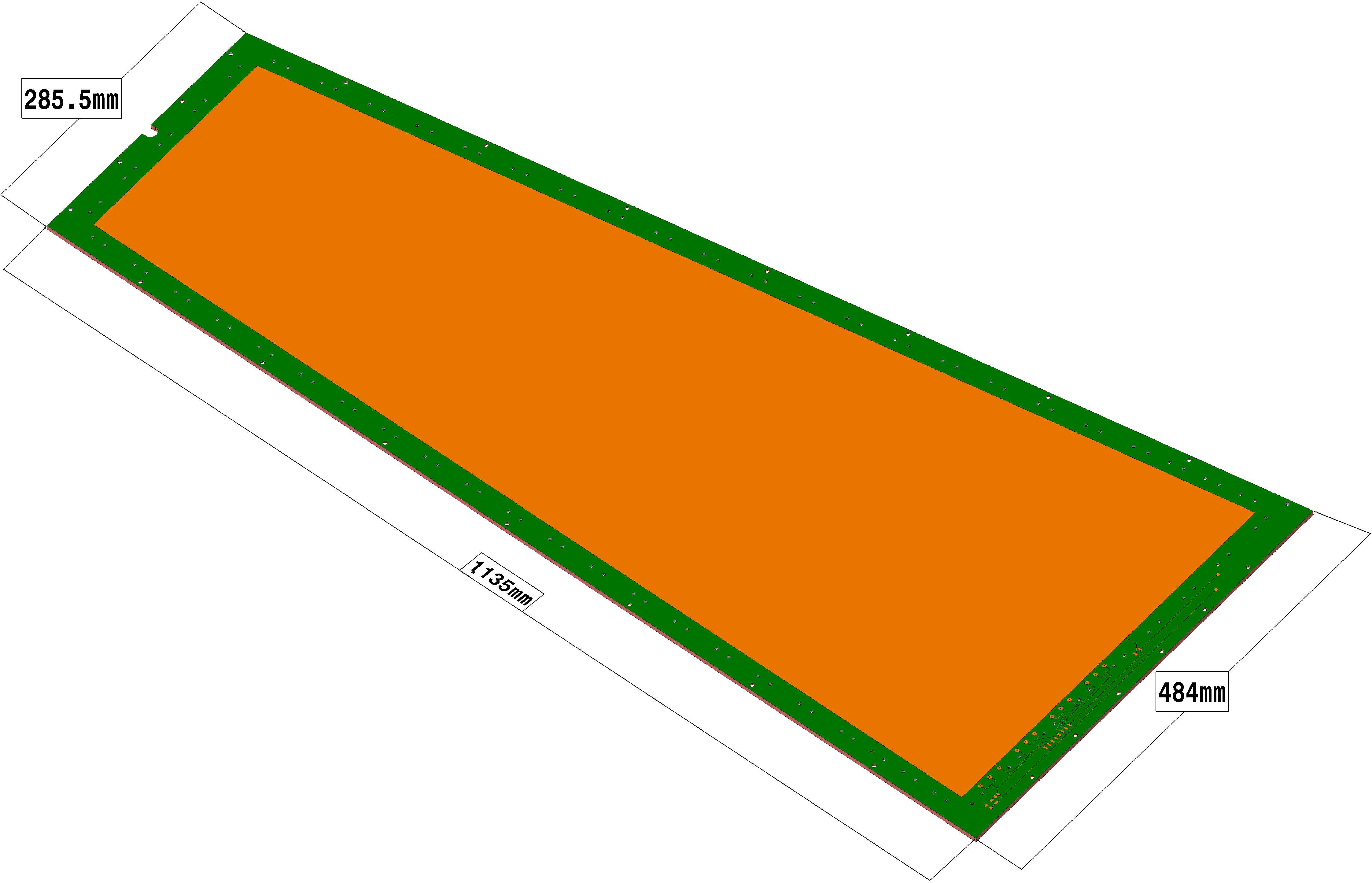}%
\hfill%
\includegraphics[width=0.52\textwidth]{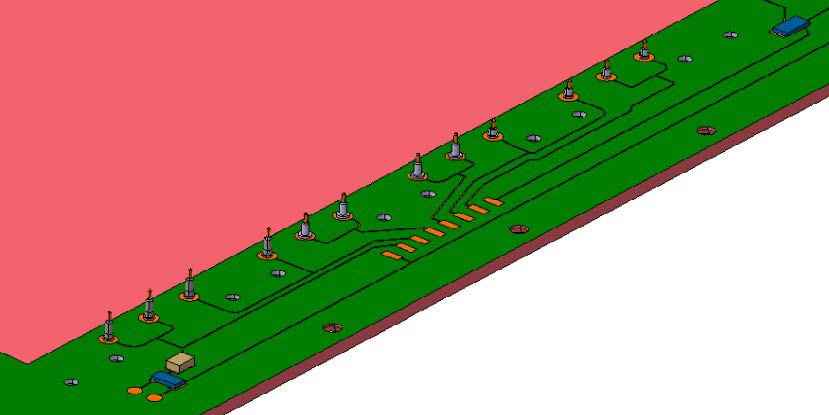}
\caption{%
    GE1/1 drift board (left) and a magnified view of the drift board (right) showing the HV pins and the resistor and capacitor network connecting to the chamber HV supply.
    Figures from Ref.~\cite{CMSMuon:2018wlc}.
}
\label{fig:gem:drift}
\end{figure}

The drift board is a trapezoidally shaped printed circuit board (PCB) inside the active gas volume and coated with a copper layer that serves as the chamber's drift electrode.
Outside the active gas volume, a 100\kOhm resistor and 330\unit{pF} capacitor are installed on pads on the PCB to limit the current from the high-voltage (HV) power supply and decouple the signal from the HV.
Twelve pins that carry the HV to the GEM foils are mounted as shown in Fig.~\ref{fig:gem:drift}.

Four internal frames composed of halogen-free glass epoxy with thicknesses of 3/1/2/1\mm are coated with polyurethane varnish and are used to define the spacing between the drift board, the three GEM foils, and the readout board.
The GEM foils are attached to the frames by screws that pass through the entire chamber assembly to hold it in place.

\begin{figure}[!ht]
\centering
\includegraphics[width=0.8\textwidth]{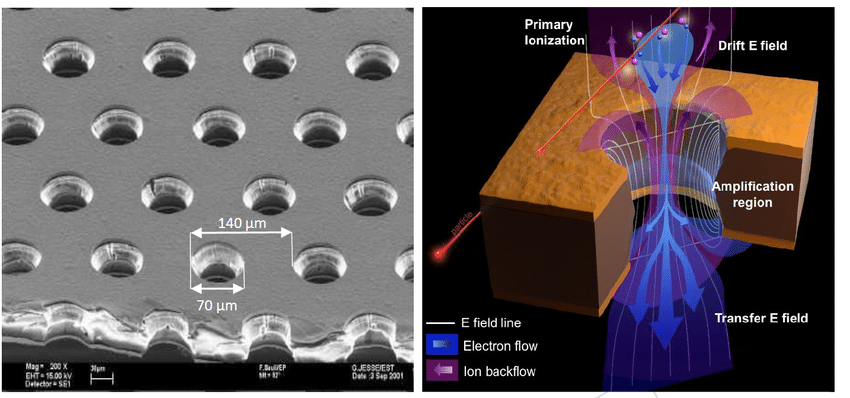}
\caption{%
    Electron scanning microscope image of a GE1/1 foil (left) and a diagram of the multiplication principle (right). Figures from Ref.~\cite{CMS:TDR-016}.
}
\label{fig:gem:hole}
\end{figure}

The GE1/1 detector uses three identical GEM foils shown in Fig.~\ref{fig:gem:hole}, which are thin polyimide foil clad with a thin copper layer containing micro-pattern holes etched in a periodic grid.
A voltage up to 400\unit{V} is applied across the copper-clad surface producing a strong electric field of between 60 and 100\kVcm.
The triplet structure allows for an amplification of around $10^{5}$ when moderately high voltages are applied.
The side of the foil facing the readout boards is a continuous conductor, while the strips facing the drift board are segmented into sectors of approximately equal area of 100\cmsq.
The segmentation ensures that, in the extreme case of a large discharge creating a short circuit between the GEM electrodes of a single sector, the affected dead area is limited to the 100\cmsq of a single sector instead of deactivating the entire detector.
Each sector is connected separately to the HV supply via a 10\MOhm resistor to limit currents from the HV supply and to quench any possible discharge.

The readout board is a printed circuit board with 3072 radially oriented readout strips on the side facing the interior of the chamber.
The strips are connected by metalized vias to the outer side of the board, which are routed to readout pads on the exterior-facing side of the board.
The board is segmented into 8$\times$3 $\eta$--$\phi$ partitions with three sets of 128 strips in $\phi$.
Each set of 128 strips is read out by the first stage of the frontend electronics, as described below.
The strip pitch varies between 0.6 and 1.2\mm on the shorter and longer ends of the trapezoidally shaped board.

\subsubsection{Gas system}

\begin{figure}[!htp]
\centering
\includegraphics[width=0.32\textwidth]{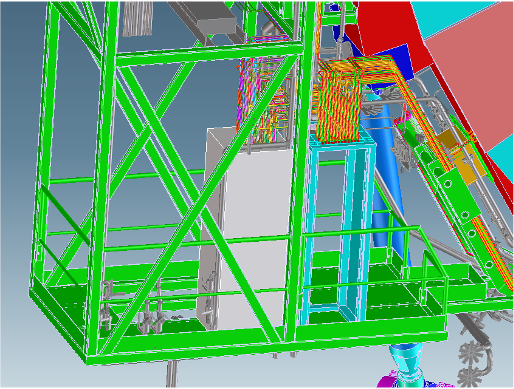}%
\hfill%
\includegraphics[width=0.27\textwidth]{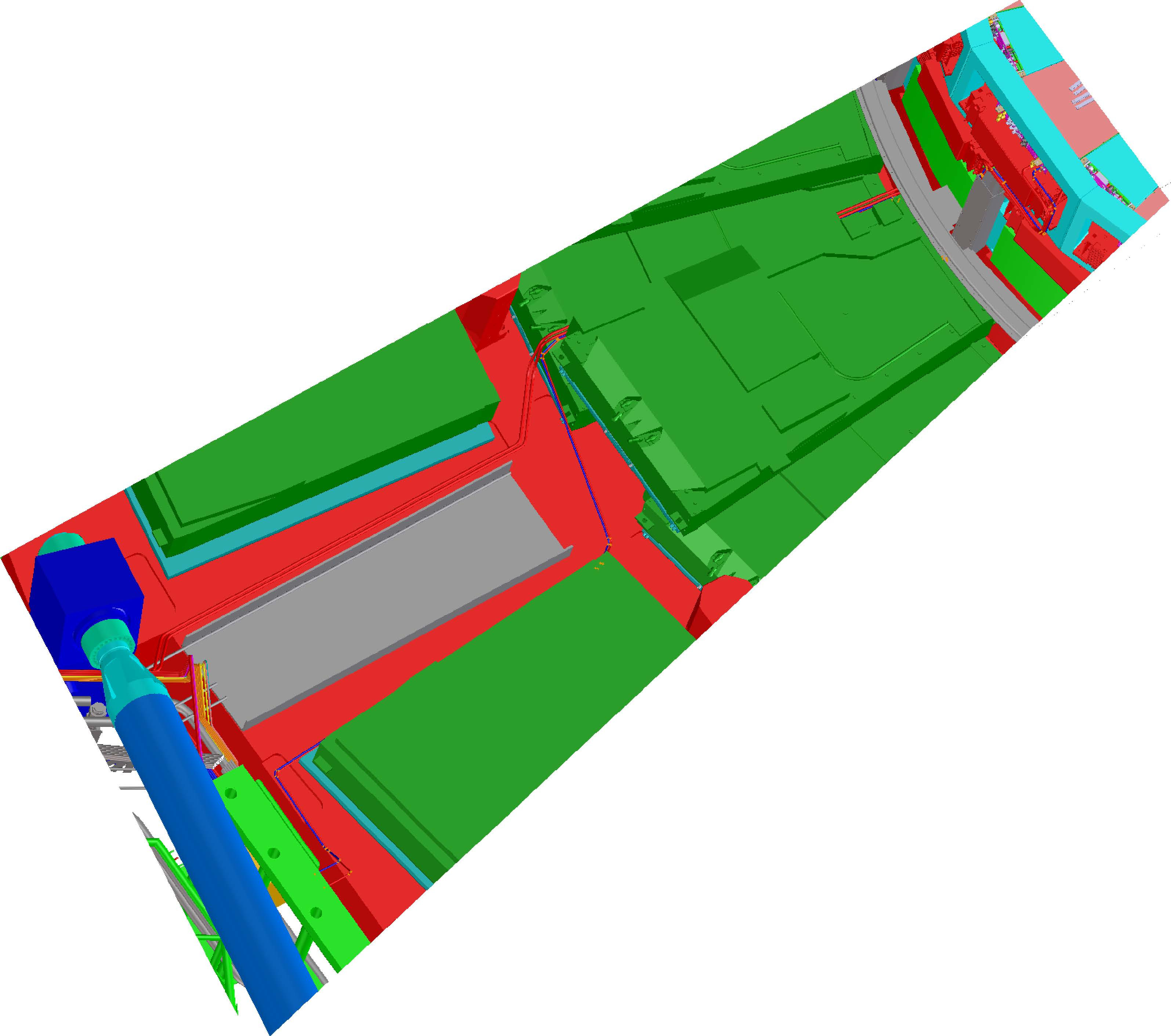}%
\hfill%
\includegraphics[width=0.31\textwidth]{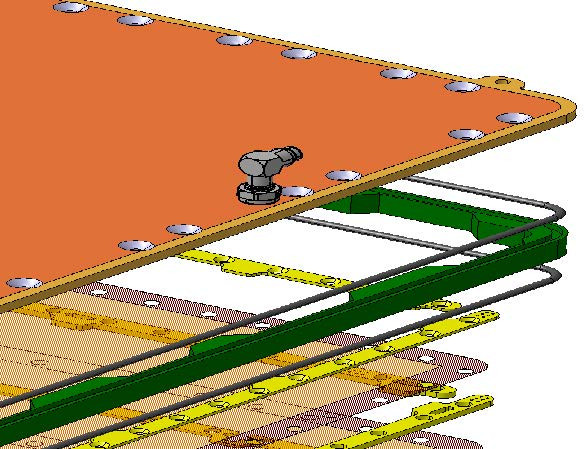}
\caption{%
    Left:\ periphery GE1/1 gas rack.
    Center:\ location of the gas lines feeding the GE1/1 chambers.
    Right:\ exploded view of a GE1/1 chamber showing the gas plug attached to the readout board.
    Figures from Ref.~\cite{CMSMuon:2018wlc}.
}
\label{fig:gem:gasrack}
\end{figure}

During operation, the chambers are run using an Ar:\COtwo gas mixture with a ratio of 70/30.
A gas mixer is installed in the surface gas barrack (SGX) where the argon and carbon dioxide gases are mixed to the required percentages.
The mixer has input and output lines that go to the underground gas room (UGX).
Four racks in the UGX are used to pump and control the gas from the surface using input and output lines to a crate on the periphery of the first muon ring on each endcaps.
The rack on the periphery uses twelve gas lines, each of which provides the gas mixture to six consecutive single chambers.
Each chamber in turn has a single inlet and outlet through which the gas enters and exits the chamber.
In Fig.~\ref{fig:gem:gasrack} a diagram of the periphery crate (left) and the location of the gas lines (center) are shown.
The entire system is monitored at different stages (mixer, pre-distribution, and distribution), and the pressure and flow-rate values are supervised by the gas control system (GCS).

Each chamber is connected to the gas system via a single inlet and single outlet attached to the readout board, as shown in Fig.~\ref{fig:gem:gasrack} (right).
The gas flows from an inlet placed at one corner of the readout board across to the outlet placed at the diagonally opposite corner of the trapezoid and through holes in the internal frames.

\subsubsection{Electronics}

\begin{figure}[!ht]
\centering
\includegraphics[width=0.6\textwidth]{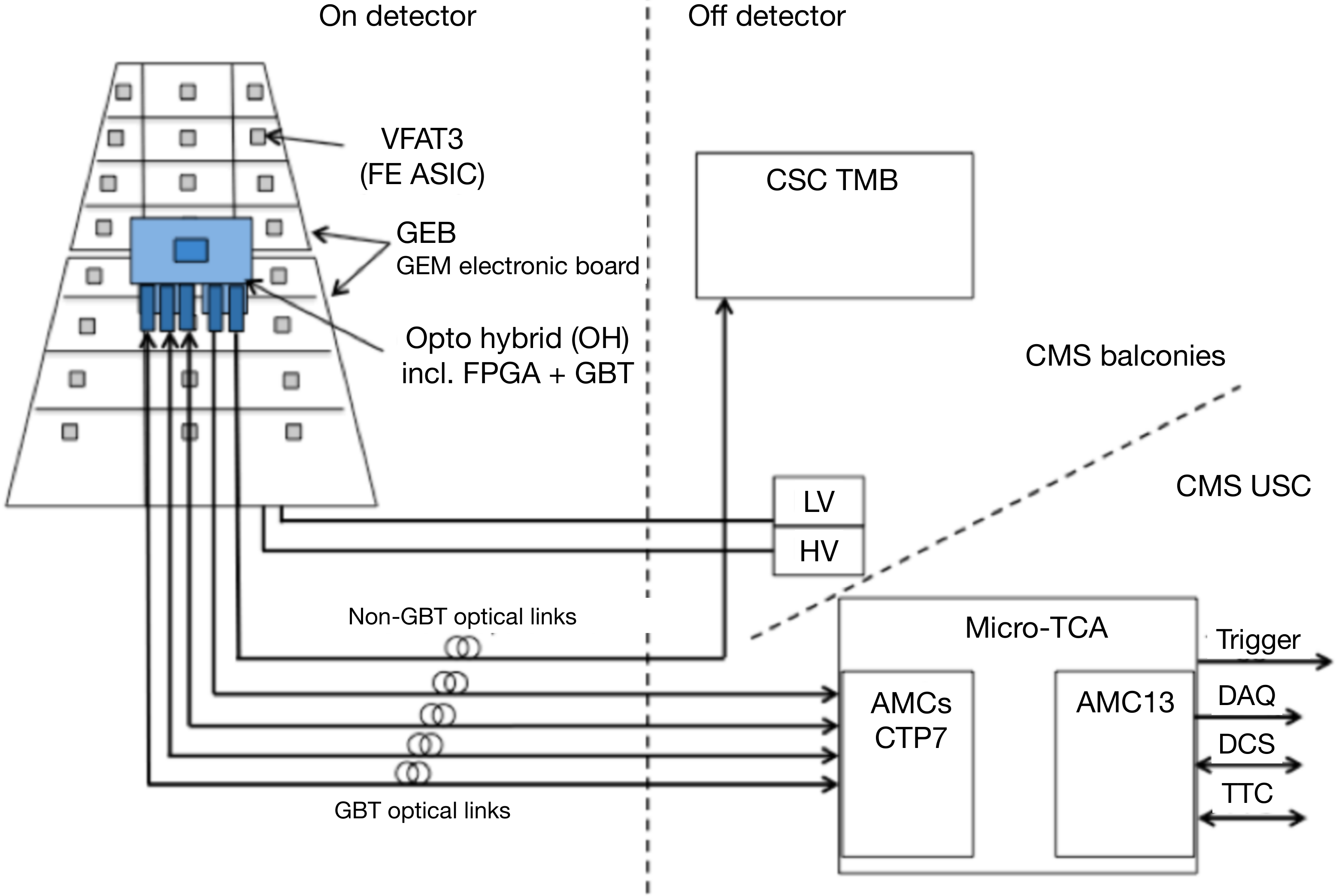}
\caption{%
    GE1/1 electronics overview showing the frontend electronics, VFAT, GEB, and opto-hybrid boards, as well as the optical links and backend readout with connections to the L1 trigger and DAQ.
}
\label{fig:gem:electronicsoverview}
\end{figure}

The GEM readout electronics are divided into frontend components that are mounted on the detector itself, and backend electronics that are outside the experimental cavern.
A schematic overview of the different components is shown in Fig.~\ref{fig:gem:electronicsoverview}.
The readout is organized according to 128 readout strips by the VFAT3 chip~\cite{Aspell:2018fmg}, which is the ASIC on the readout board responsible for the digitization of the induced signals.
The signals from individual VFAT3 chips are then sent via traces on a large chamber-sized PCB called the GEM electronics board (GEB)~\cite{Aspell:2014zma}.
The VFAT3 and opto-hybrid~\cite{Abbaneo:2016zsf} boards are attached to the GEB, which also routes high and low voltages to the chambers from external off-chamber power supplies.
The opto-hybrid board transmits the data via optical links to the off-chamber backend readout and also sends data to the CSC trigger board for processing at L1.
The optical links are routed from the opto-hybrid boards on the chambers to patch panels on the periphery crates and then to off-detector electronics in a \uTCA crate~\cite{Lenzi:2017ufj}.
The backend boards are CTP7 \uTCA boards, which are described in Ref.~\cite{Svetek:2016xvm}.
They contain bidirectional links to communicate with the frontend chamber electronics and are connected via AMC13~\cite{Hazen:2013rma} cards to the CMS L1 trigger and DAQ systems~\cite{Mommsen:2018csk}, as discussed in Section~\ref{sec:daq}.

\begin{figure}[!htp]
\centering
\includegraphics[width=0.45\textwidth]{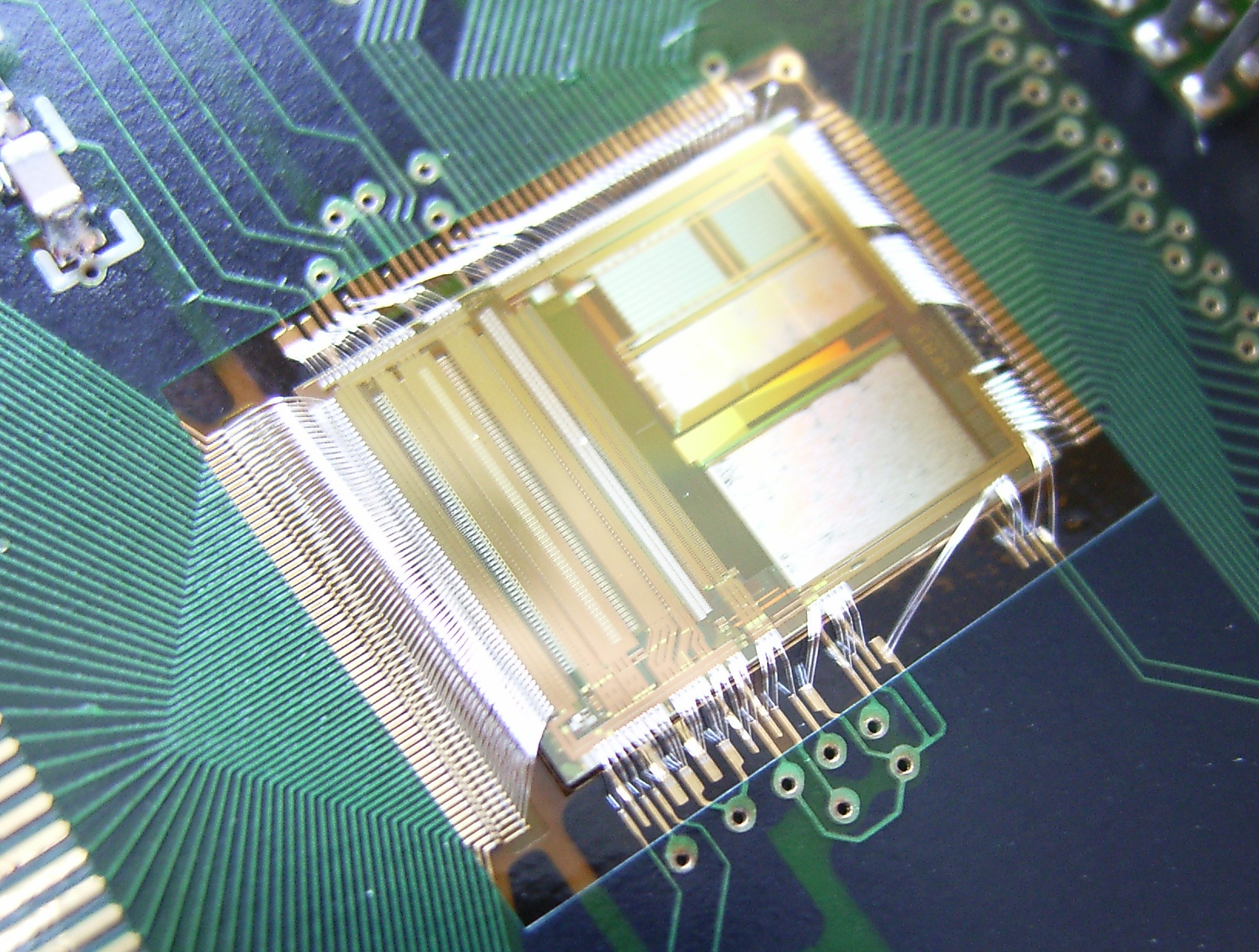}%
\hfill%
\includegraphics[width=0.51\textwidth]{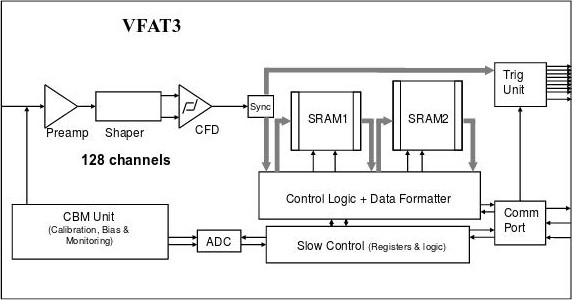}
\caption{%
    Picture of the VFAT3 ASIC (left) and a high-level schematic (right). Figures from Ref.~\cite{Abbaneo:2016zwv}.
}
\label{fig:gem:vfat3}
\end{figure}

The VFAT3 chip was developed from an existing frontend readout chip, the VFAT2~\cite{Aspell:2008vfa} (very forward ATLAS and TOTEM) ASIC, developed for the ATLAS and TOTEM readout.
The VFAT3 design was modified for application in the CMS GEM to provide readout in a high-rate environment, collect the total charge produced by particles crossing the GEM volume, and be radiation hard.
A picture and a high-level schematic of a VFAT3 chip are shown in Fig.~\ref{fig:gem:vfat3}.

The VFAT3 must collect, amplify, shape, digitize, and buffer the signal produced on each of the 128 channels connected to the readout strips on the GEM readout board.
The preamplifier and shaper are charge sensitive and are followed by a constant-fraction discriminator that digitizes the incoming charge pulses.
Following the discriminator, the binary comparator results are synchronized with the LHC clock in the synchronization unit.
A programmable threshold allows a channel-by-channel calibration to optimize the signal-to-noise ratio, and is set to obtain approximately 98\% efficiency per single hit so that the logical OR of two layers has a 99.9\% efficiency.

The data are then split into two paths.
A trigger signal, with fixed latency, is sent to the trigger unit, while an asynchronous tracking signal is sent to a buffer for full granularity readout via the communications port.
The signal is accompanied by a time stamp, the bunch crossing, for further processing.

The signal can last up to approximately 60\ns depending on the gas mixture, and the VFAT3 chip allows the shaping time to be adjusted to fully integrate the charge and maximize the signal-to-noise.
The timing resolution is optimized by the use of a constant-fraction discriminator and a gain with a programmable dynamic range.
The buffer for the tracking data is implemented by a SRAM memory that operates as a circular buffer 128 channels wide and 1024 bunch crossings deep.
It continuously samples all channels in every cycle of the LHC clock.
After 1024 bunch crossings, the buffer overwrites the first entry.

Since the L1 trigger operates with fixed latency, there is a fixed time between the level-1-accept (L1A) trigger signal and the data corresponding to the bunch crossing of interest.
Upon receiving an L1A, the appropriate data are transferred from the SRAM1 buffer to the SRAM2 buffer, which is 512 bunch crossings deep and stores the data until they are transmitted off the chip.
The chip supports a latency of up to 12.5\mus with an L1A rate of up to 1\MHz.
The data can either be sent in a lossless mode or zero suppressed.

The fixed latency, or trigger path, is used to provide fast hit information synchronous with the LHC clock, which can be put in coincidence with other detectors to decide if the event will generate an L1 trigger.
The logical OR of two adjacent channels is sent to the trigger unit.

The slow-control signals are used to send calibration, bias, monitoring, and control information via the communication port and to control and monitor the settings and status of the VFAT3 chip.

The output of each VFAT3 chip is routed along the GEM electronics board (GEB), which is mounted directly on top of the readout board.
The GEB is divided into two sections that cover the entire chamber.
The GEB routes the signals between the 24 VFAT3s on each chamber and the low voltage from the external power supplies to the chamber and frontend electronics.
CAEN power supplies provide voltage to the FEAST boards mounted on the GEB.
The FEAST boards perform the \DCDC conversion to provide the proper low voltages for the readout electronics, as well as the HV for the operation of the chamber.

\begin{figure}[!ht]
\centering
\includegraphics[width=0.48\textwidth]{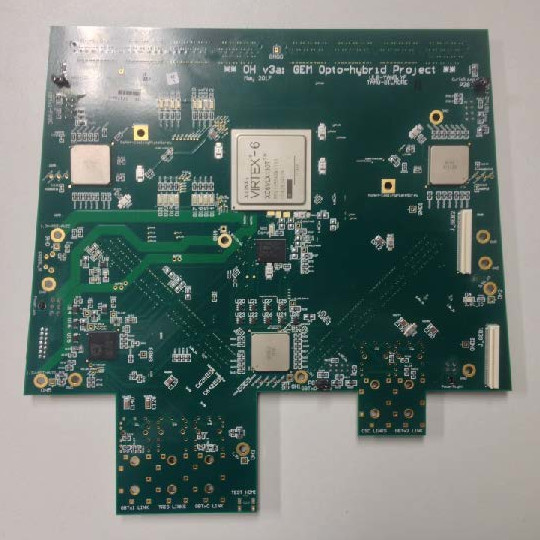}%
\caption{%
    Photo of the GE1/1 opto-hybrid board with the Xilinx Virtex-6 FPGA at the center.
}
\label{fig:gem:ohphoto}
\end{figure}

\begin{figure}[!ht]
\centering
\includegraphics[width=0.48\textwidth]{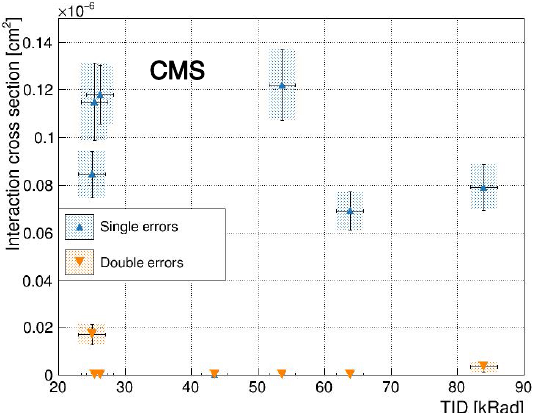}
\caption{%
    Single- and double-error cross sections measured in irradiation tests as a function of the TID.
}
\label{fig:gem:ohxsec}
\end{figure}

The opto-hybrid board serves as the chamber hub for data communication, transfer, and control.
It receives data from all the VFAT3 ASICs, formats it for the entire chamber, and sends the data using the GBT protocol and optical links to the backend electronics for final readout and use in the L1 muon trigger.
Additionally, each opto-hybrid board contains three GBTx chipsets~\cite{Moreira:2009pem}, one Virtex-6 FPGA~\cite{Alfke:2009ieee}, three versatile link transceivers (VTRx), and two versatile link transmitters (VTTx)~\cite{Vasey:2012xjw}.
Trigger data are sent over the VTTx via optical fibers to both the backend readout and the CSC trigger motherboard, while the three VTRx transmit tracking data to the backend readout (CTP7 card).
Each GBT can handle up to ten frontend chips at a transfer rate of 320\MHz, and the tracking data are transferred at 4.8\Gbs through the VTX.
The trigger data are transferred at 3.2\Gbs using 10b/8b encoding.
In operation, the trigger data are formatted within the Virtex-6 FPGA, while the tracking data flow directly from the frontend by the GBT VTRx links to the backend.
Calibration, and trigger, timing, and control (TTC) signals are sent to the opto-hybrid board and distributed to the frontend via the GEB.
Irradiation tests have shown that the single- and double-error rates from single-event upsets (SEUs) are well within acceptable margins for the total ionizing dose (TID) expected over the lifetime of the HL-LHC project.
A photo of the opto-hybrid board and the measured error rates are shown in Figs.~\ref{fig:gem:ohphoto} and~\ref{fig:gem:ohxsec} .

The backend electronics are \uTCA cards that were originally developed for a calorimeter trigger \Phase1 upgrade project~\cite{Svetek:2016xvm}.
The CTP7 has 67 optical receivers, 48 optical transmitters, a Xilinx Virtex-7 FPGA, and a Zync processor.
The firmware was adapted for the GEM detector with 36 GBT cores servicing 12 triple GEM detectors with one link to a CMS \uTCA AMC-13 card operating the standard DAQ and trigger links.
Using this hardware, all of the GE1/1 can be read out with one \uTCA crate hosting six CTP7 cards and one AMC-13 card.

\subsubsection{CSC/GEM trigger for \Run3}
\label{sec:gem:csc-gem-trig}

\begin{figure}[!htp]
\centering
\includegraphics[width=0.38\textwidth]{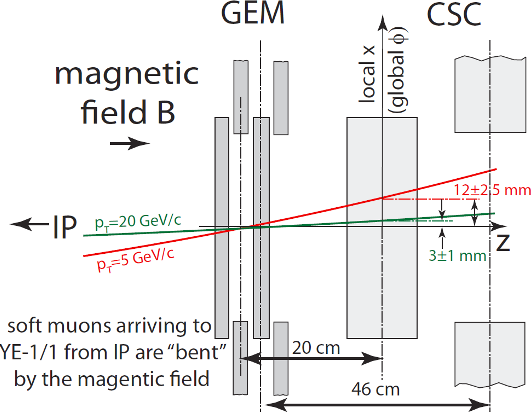}%
\hfill%
\includegraphics[width=0.55\textwidth]{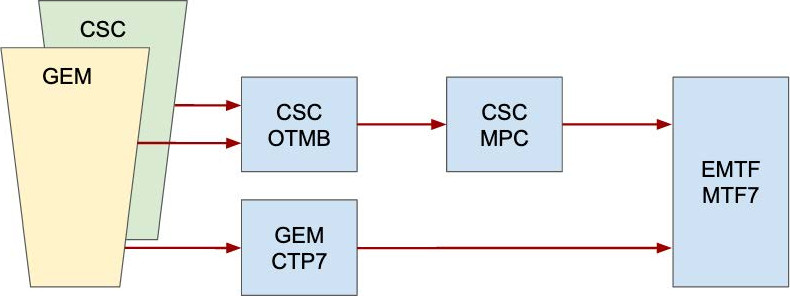}
\caption{%
    Side view of the GEM-CSC trigger coincidence (left) and schematic overview of the data flow in the GEM and CSC trigger processor (right).
    The addition of the GE1/1 station significantly increases (by a factor of 3 to 5) the lever arm of the distance traveled in $z$ by a muon originating from the interaction point.
    The bending angle between the entrance of the muon to the GE1/1 station and the exit from the ME1/1 CSC station can be used to estimate the momentum of the muon.
    Figures from Ref.~\cite{Tytgat:2014xxa}.
}
\label{fig:gem:gemcsctrigdialog}
\end{figure}

Figure~\ref{fig:gem:gemcsctrigdialog} shows a view in the global CMS transverse ($\phi$--$z$) plane indicating the location of the GE1/1 and the first CSC station, ME1/1.
The diagram indicates the relative position of GE1/1 and ME1/1 in the global $z$ coordinate and shows two muon trajectories as they traverse the stations' volume.
The deflection of muons with a \pt of 5 and 20\GeV is approximately 3 and 12\mm, respectively, in the local $x$ (global $\phi$) direction.
By reconstructing the $\phi$ direction in both the entrance to the GE1/1 station and the exit from the ME1/1 station, an estimate of the track \pt can be made and used for the trigger decision in the L1 muon trigger system.
The GE1/1 chambers add two new independent position measurements, which improves the redundancy, lowers the muon misidentification rate, increases the robustness of segments found in the muon spectrometer, and raises the efficiency of the stub-finding algorithm from 90 to 96\%.
Finally, the joint segments are utilized in the endcap muon track finder (EMTF) to optimize the final track reconstruction and resolution of the L1 trigger, as described more fully in Section~\ref{sec:l1trigger:muon}.

\begin{figure}[!ht]
\centering
\includegraphics[width=0.495\textwidth]{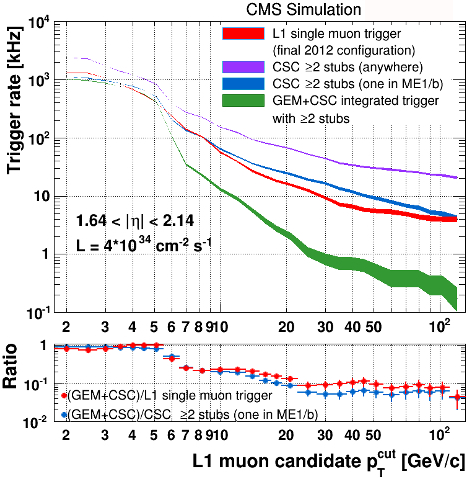}%
\hfill%
\includegraphics[width=0.465\textwidth]{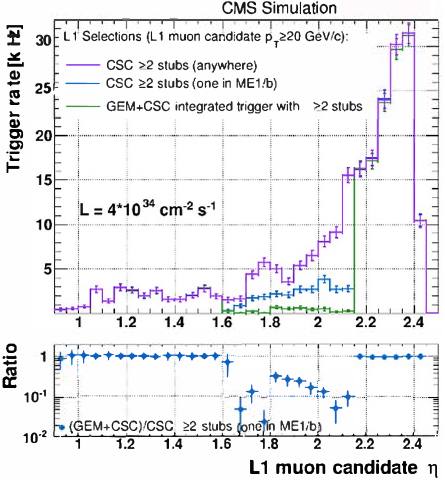}
\caption{%
    Muon rate from simulation as a function of \pt (left) and $\eta$ (right) with and without the integration of the GEM chambers under various assumptions. Figures from Ref.~\cite{CMS:TDR-016}.
}
\label{fig:gem:gemcscmuonrate}
\end{figure}

The data flow from the GEM chambers to the EMTF via two paths is shown in Fig.~\ref{fig:gem:gemcsctrigdialog}.
First, trigger data are sent through the opto-hybrid board to the CSC optical trigger motherboard (OTMB)~\cite{Ecklund:2012wk} that uses hits in the GE1/1 detector and the CSC ME1/1 to form stubs from the combined stations.
These stubs are in turn sent to the CSC muon port card that sorts the stubs from multiple chambers by sector and sends them to the appropriate EMTF card where the muon momentum is estimated from hits in the CSC, RPC, and GEM using a neural network algorithm~\cite{CMS:2018wav}, as described in Section~\ref{sec:l1trigger:muon}.
In Fig.~\ref{fig:gem:gemcscmuonrate}, simulated distributions of the muon rate in the endcap as a function of \pt and $\eta$ are shown~\cite{CMS:TDR-016}.

\subsubsection{Detector control system}

The CMS detector control system (DCS)~\cite{Brigljevic:2003kv} is based on the Siemens Simatic WinCC Open Architecture (WinCC OA), previously known as Prozess-Visualisierungs- und Steuerungssystem (PVSS), and is described in Ref.~\cite{Golonka:2014lxa}.
The  DCS~\cite{Abbas:2020era} was designed to control and monitor the high and low voltage and to monitor the gas conditions.

Even though the gas system is controlled by the central CERN gas group, the DCS maintains the function of monitoring the mixture composition (fraction of Ar and \COtwo), as well as the pressure and flow rate to the chambers.
In particular, the DCS reflects the structure of the gas system, which in CMS is divided into three main elements:\ mixer, rack, and flowcells.

The DCS system also controls both the high voltage used to operate the GEM foils and the low voltage used to power the frontend electronics on each chamber.
The system allows turning on and off either the high or low voltage for every component, as well as monitoring the voltages and currents in the system.

\subsubsection{Chamber assembly and installation}

To ensure high-quality performance of all the components, a detailed quality control process was put into place to assure the quality of every element.
The ten-step process, described in greater detail in Ref.~\cite{Venditti:2020pjq}, can be summarized as follows:
\begin{itemize}
\item quality control of the components for production (QC1--2),
\item assembly and commissioning of the GE1/1 chambers at the production sites (QC3--5),
\item assembly and commissioning at CERN before installation in CMS (QC6--10).
\end{itemize}
First, all components were cleaned using ultrasonic baths, baked, sand-blasted, and visually inspected for faults.
The components were then optically and electrically inspected including verification measurements, specifically measuring the $I$--$V$ curves and checking for shorts between readout strips on the drift board.
The foils were optically inspected, and checked that the leakage currents were less than 30\unit{nA} when 500\unit{V} was applied between the two sides of the GEM foils.
Each chamber was pressurized by dry nitrogen at 30\mbar and checked for gas leakage.
The chambers were then flushed for several hours with \ArCOtwo and monitored for output at moderate HV.
The uniformity of the gas gain was verified by placing an X-ray source 1\unit{m} away from a chamber enclosed in a copper protective box, which simultaneously illuminated the full chamber.

Once shipped to CERN, the gas leakage and HV tests were repeated for every chamber, and the frontend electronics and chambers were mounted into super-chambers with cooling plates.
Electrical conductivity tests, gas leak tests, and electronic noise measurements were repeated with the cooling on.
Finally, the super-chambers were mounted on a cosmic ray test stand where efficiency, noise, and tracking studies were done to confirm the correct and uniform operation of each chamber.
While in storage and before installation into CMS, the gas-leak and HV stability were tested again and monitored over one month before each GEM was installed and declared ready for commissioning in the CMS detector.

\subsubsection{Preliminary commissioning results}

Before installation in the CMS cavern, a GE1/1 chamber was tested at the CERN H4 beam, which was extracted from the SPS~\cite{CMSMuon:2019pzw}.
A secondary beam of pions was produced when the proton beam struck a beryllium target.
Finally, that secondary beam was filtered by collimators to produce a beam of 150\GeV muons from the decay of the pions in the secondary beam.

\begin{figure}[!ht]
\centering
\includegraphics[width=0.48\textwidth]{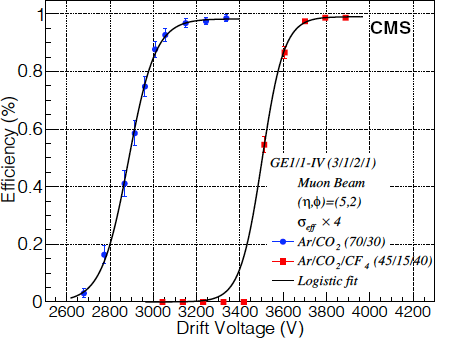}%
\hfill%
\includegraphics[width=0.48\textwidth]{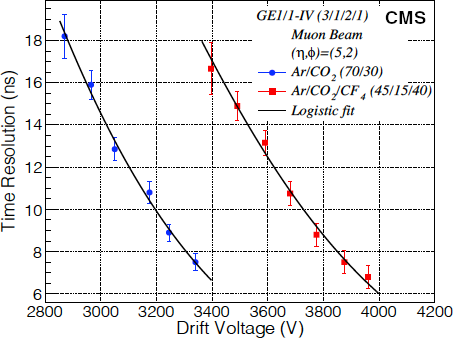}
\caption{%
    GEM test beam results showing the efficiency (left) and time resolution (right) as a function of the drift voltage placed on the GEM foils.
    The chosen gas mixture Ar/\COtwo (70/30) has very similar properties to Ar/\COtwo/\CFfour (45/15/40) and was selected as \CFfour, a greenhouse gas, is being phased out of industry. Figures from Ref.~\cite{CMSMuon:2019pzw}.
}
\label{fig:gem:testbeamresults}
\end{figure}

Several key performance parameters were measured in the test beam.
Two are the single-hit efficiency and the time resolution of the GEM chambers.
A set of scintillators read out by photomultiplier tubes (PMTs) were placed upstream of the chamber under test.
Events were selected requiring a triple coincidence in each of the three layers and the efficiency was estimated by counting the number of hits in the GE1/1 tracking chamber in that region.
Time resolution information was obtained by measuring the standard deviation of the distribution of the measured time between the trigger and GE1/1 detector signal.
As can be seen in Fig.~\ref{fig:gem:testbeamresults}, an efficiency greater than 98\% and a time resolution less than 10\ns were obtained.
Other results, such as discharge rate and rate capacity as a function of voltages and gas mixture are given in Ref.~\cite{CMSMuon:2019pzw}.

During LS2, the GE1/1 chambers were installed in CMS, connected to all services (LV/HV, gas, and optical fiber connection for readout), and the chambers were commissioned in situ.
Latency scans were done for all GE1/1 chambers on both endcaps to account for the variable cable and optical fiber lengths, and HV values were optimized for signal-to-noise ratio using cosmic ray muons.
It should be noted that, unlike in LHC collisions, cosmic ray muons do not arrive at a fixed and known time.
Thus, the latency and other operational parameters cannot be fully optimized for LHC operations until all chambers are exposed to both muons originating from the interaction point and the LHC background conditions.

\begin{figure}[!ht]
\centering
\includegraphics[width=\textwidth]{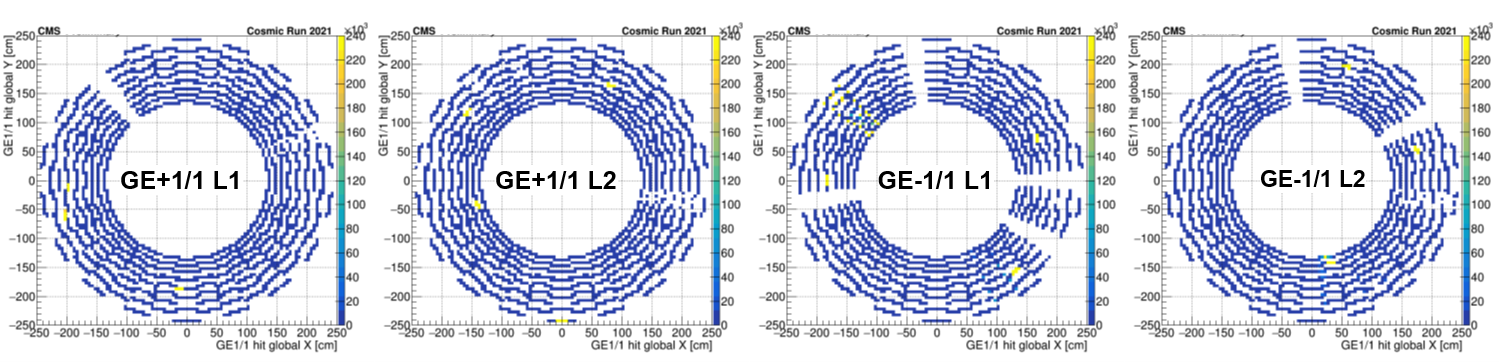}
\caption{%
    GE1/1 occupancy in cosmic ray muon data recorded in 2021.
}
\label{fig:gem:cosmicoccupancy}
\end{figure}

Figure~\ref{fig:gem:cosmicoccupancy} shows the occupancy in both GE1/1 endcap rings.
Not all chambers were included in the data taking with cosmic rays due to minor temporary commissioning issues.
This explains the missing sectors in the $-z$ endcap.
Furthermore, since electronic noise was being addressed, some chambers show higher-than-average occupancy.
The GE1/1 system was fully commissioned with cosmic rays and has been successfully operated since the beginning of LHC \Run3 in summer 2022.

\clearpage
\section{Precision proton spectrometer}
\label{sec:pps}

The precision proton spectrometer (PPS)~\cite{TOTEM:TDR-003} is a set of near-beam detectors, located in the LHC tunnel at a distance of about 200\unit{m} from the CMS IP, on both sides.
Originally conceived and developed as a joint project of the CMS and TOTEM collaborations (CT-PPS), it then evolved into a standard CMS subdetector following a Memorandum of Understanding between CERN, CMS, and TOTEM in 2018.
For the sake of clarity, only the PPS acronym is used throughout this section.
During \Run2, PPS took part in the CMS data taking, with various configurations between the years 2016 and 2018, and recorded data corresponding to an overall integrated luminosity of about 110\fbinv.
Following approval of the extension of the PPS experimental program to the LHC \Run3, a broad upgrade plan has been launched, involving replacement of all PPS detectors.

\subsection{Forward protons and the roman pot system}

The measurement of very forward (``leading'') protons~\cite{TOTEM:TDR-003} implies the detection of such protons at large distance from the IP and in proximity of the LHC beam.
The main variable characterizing the proton kinematics is its fractional momentum loss, defined as $\xi = (\abs{\mominitial} - \abs{\momfinal})/\abs{\mominitial}$, where \mominitial and \momfinal are the initial and final proton momentum vectors, respectively.
The coverage in $\xi$ at the PPS location is limited by the LHC magnet and collimator lattice in \Run2 and \Run3 to a range below 0.2.
In order to extend it as much as possible in the low-$\xi$ region, scattered protons must be detected as close as few (2--3) mm from the beam.
This can be accomplished by means of special movable beam pockets, the so-called roman pots (RPs), which host the particle detectors such that they can be moved close to the beam during stable beam operations.
The PPS experimental setup includes both tracking detectors, to measure $\xi$ from reconstructed proton tracks, and timing detectors, to suppress the contribution of protons originating from different primary interactions in the same LHC bunch crossing (pileup).
Tracking and timing detectors are hosted in separate RPs.

\begin{figure}[!b]
\centering
\includegraphics[width=\textwidth]{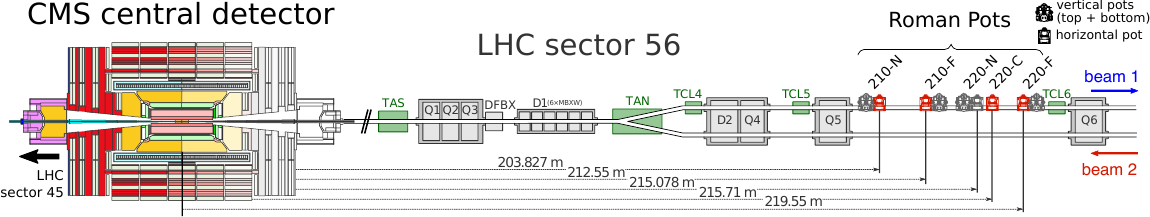}
\caption{%
    Schematic layout of the beam line between the interaction point and the RP locations in LHC sector 56, corresponding to the negative $z$ direction in the CMS coordinate system.
    The accelerator magnets are indicated in gray; the collimator system elements in green.
    The RP units marked in red are those used by PPS during \Run2; the dark gray ones are part of the TOTEM experiment.
    In \Run3 the ``220 near'' horizontal unit is used in place of the ``210 near''.
    Figure from Ref.~\cite{CMS:PRO-21-001}.
}
\label{fig:pps:rplayout}
\end{figure}

The TOTEM experiment~\cite{TOTEM:TDR-001} has been using an extensive system of box-shaped roman pots to conduct its physics program since the start of LHC operations in dedicated, low-luminosity runs.
The system has been conceived as a two-arm spectrometer, composed of two pairs of stations on each side of the CMS IP, at about 147 and 220\unit{m} along the beam line, each station comprising two RPs approaching the beam vertically, and one approaching it horizontally.
Within each pair, the station installed closer to the IP is called ``near'', and the other ``far''.
For \Run2, the layout of the spectrometer has been modified as sketched in Fig.~\ref{fig:pps:rplayout} for one of the two arms.
The stations at 147\unit{m} have been relocated in proximity of those at 220\unit{m} and identified as ``210\unit{m}''.
Following these conventions, the four stations are usually named 210-N, 210-F, 220-N, 220-F, where N stands for near, and F for far.
In order to allow operations at standard LHC luminosity with PPS, some of the horizontal RPs have been equipped with cylindrical ferrite shields to reduce their radio-frequency impedance~\cite{TOTEM:TDR-003} and have been reserved for the PPS tracking detectors (Section~\ref{sec:pps:tracking}).
Moreover, an additional horizontal RP with modified geometry (cylindrical RP) has been installed in each arm between the near and far stations at 220\unit{m} to host timing detectors; it is referred to as 220-C (Section~\ref{sec:pps:timing}).
All vertical RPs remain reserved for the TOTEM physics program.
However, their role is crucial for PPS operations at the beginning of each data-taking period, when the tracking detectors they host are used in special runs, in conjunction with those in the horizontal RPs, to determine the global alignment parameters~\cite{CMS:PRO-21-001}.

\begin{figure}[!ht]
\centering
\includegraphics[width=0.48\textwidth]{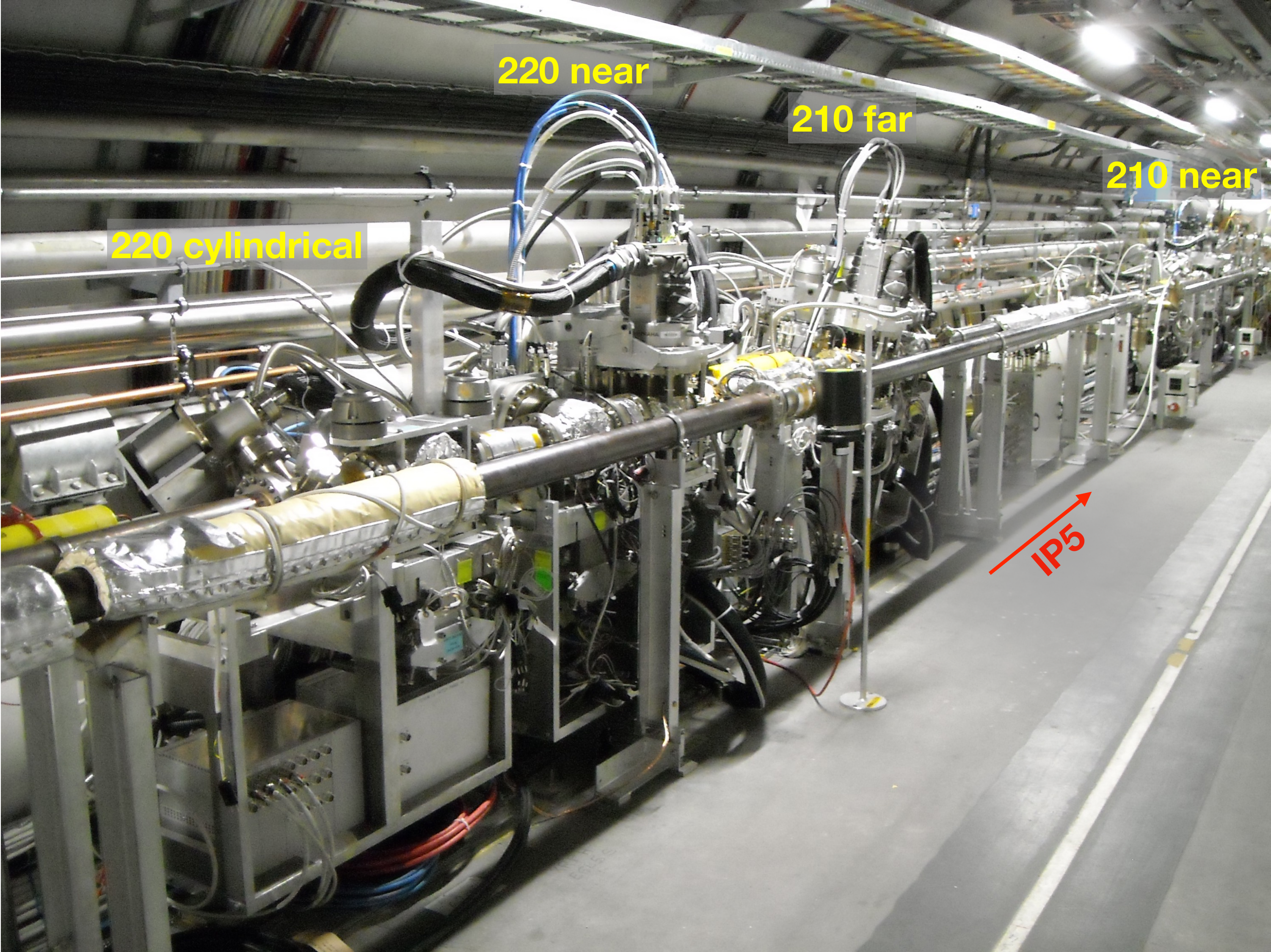}
\caption{%
    View of a section of the LHC tunnel in sector 45, with part of the PPS and TOTEM RP stations along the beam line.
}
\label{fig:pps:rptunnelview}
\end{figure}

\begin{figure}[!ht]
\centering
\includegraphics[width=0.25\textwidth]{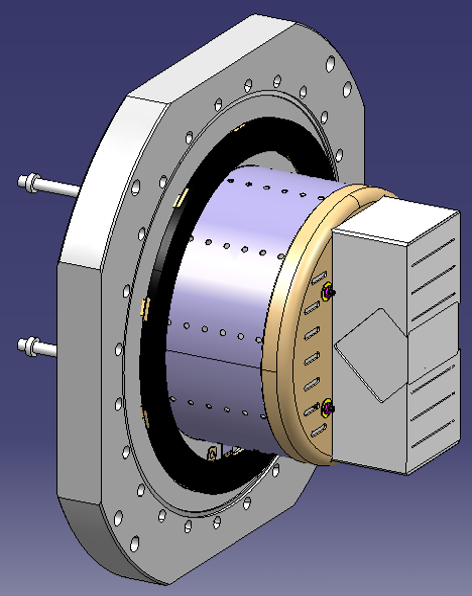}%
\hfill%
\includegraphics[width=0.25\textwidth]{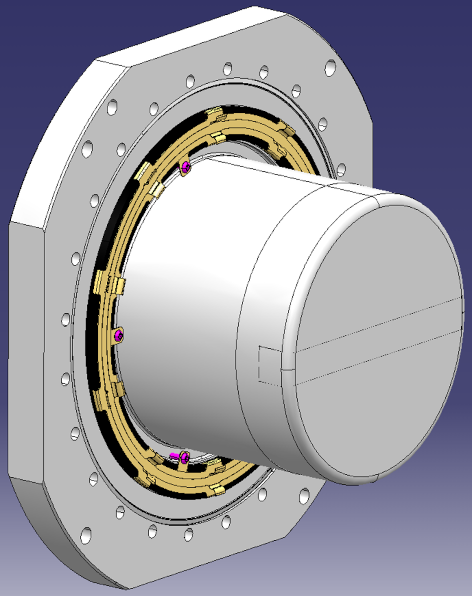}%
\hfill%
\includegraphics[width=0.46\textwidth]{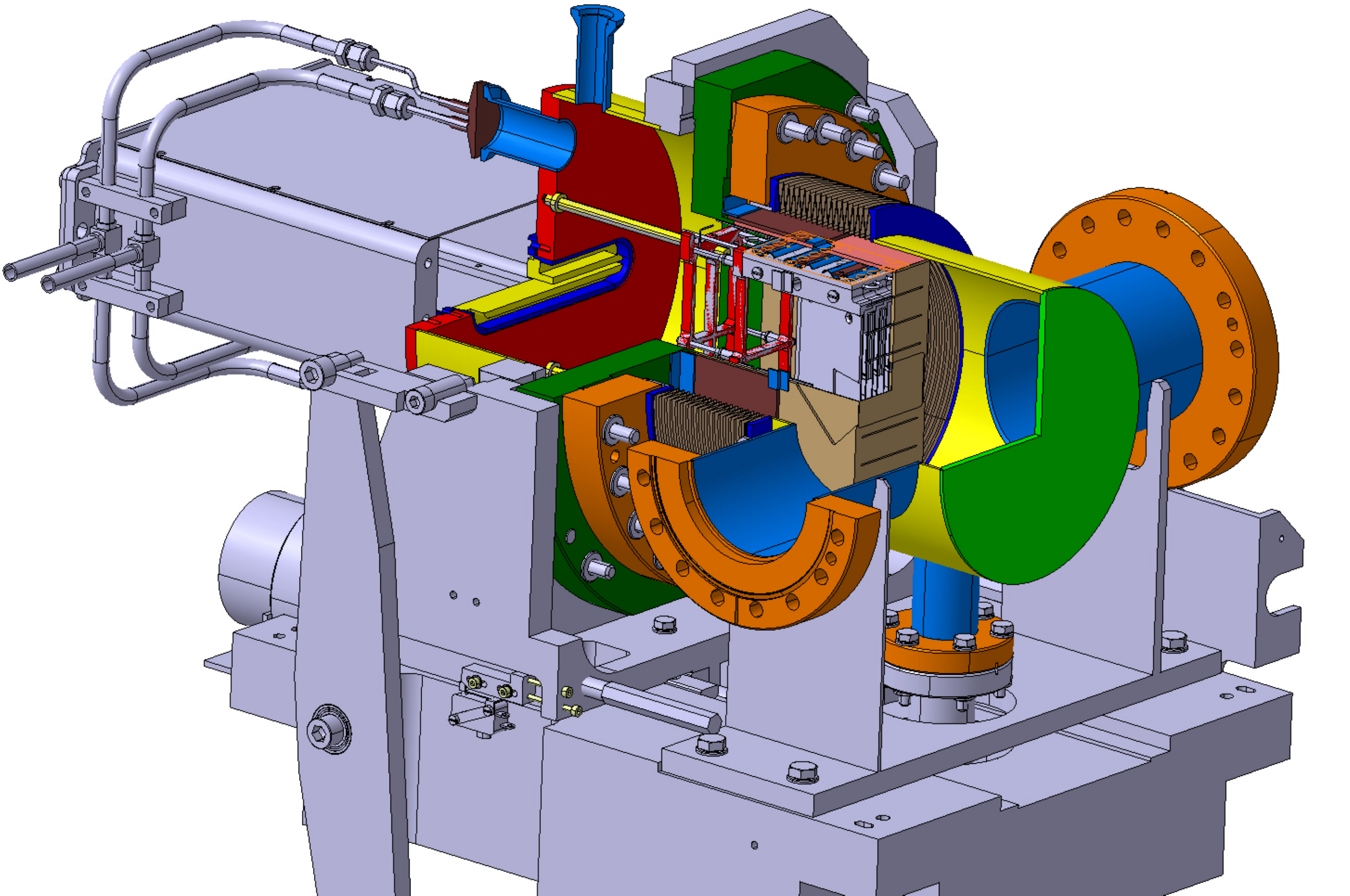}
\caption{%
    Sketches of the horizontal roman pots, from Ref.~\cite{Albrow:2015ois}.
    Left:\ box-shaped pot, with the ferrite RF-shield in place.
    Center:\ cylindrical pot where the ferrite RF-shield is integrated as a ring in the flange.
    The thin window shapes, different for the two cases, are visible in the rightmost, vertically centered part of the pots.
    Right:\ overview of the insertion system of a box-shaped pot in a section of the LHC beam pipe.
}
\label{fig:pps:rpdetails}
\end{figure}

Figure~\ref{fig:pps:rptunnelview} shows a view of the LHC tunnel, in the region where the RPs are installed in one of the two sectors.
The detector housings are made of 2\mm thick stainless steel, with a useful depth to contain the detectors of about 110\mm; the box-shaped pots have a rectangular cross section of 124$\times$50\mmsq, while the cylindrical pots have a diameter of 141\mm.
In order for detectors to approach the LHC beam as closely as possible, and to minimize the amount of material traversed by scattered protons, the thickness of the walls is reduced to 200 (300)\mum around the active sensor region in the box-shaped (cylindrical) RPs:\ this section is called ``thin window''.
Figure~\ref{fig:pps:rpdetails} shows the structure of a box-shaped and a cylindrical RP, as well as the insertion system installed on a section of the LHC beam pipe.

Moderate vacuum conditions have to be kept inside the detector housings, in order to minimize the pressure gradient between the detector volume and the ultra-high vacuum of the LHC beam pipe.
The exact value of the pressure depends on the detector technology, but does not exceed 100\mbar.
Several RPs are serviced by the same vacuum system.

The solid-state detectors equipping PPS need to operate at low temperatures (--20 to 5\deC, depending on the detector) to mitigate the effects of radiation damage.
This is accomplished by means of a fluorocarbon evaporative system~\cite{Vacek:1999qoa}:\ the cooling fluid is transported to the RPs, where suitable evaporation circuits, thermally coupled to the detectors, supply the needed cooling power~\cite{TOTEM:TDR-001, TOTEM:TDR-003}.

A more detailed description of the TOTEM and PPS roman pot system and its operations can be found in Refs.~\cite{TOTEM:TDR-001, TOTEM:TDR-003} and references therein.

\subsection{Tracking detectors}
\label{sec:pps:tracking}

The energy loss of protons can be measured, through detailed knowledge of the LHC optics parameters and dedicated calibrations~\cite{CMS:PRO-21-001}, from the parameters of their reconstructed tracks.
In order to obtain the needed accuracy in $\xi$, a resolution of few tens of $\mu$m is required on the coordinates of the track impact point.
The hit rate distribution is highly nonuniform in the region covered by the tracking detectors, with peak particle flux values of about 3\ten{9}\pcmsqs and a range spanning over three orders of magnitude.
This reflects in the high radiation dose the detectors have to withstand, with reference fluence values in the most irradiated areas of 1--3\ten{15}\Neq for an integrated luminosity of 100\fbinv.

\subsubsection{Detector units}

Proton tracks are reconstructed in PPS by two tracking stations per arm.
During \Run2, the roman pots hosting the trackers and the technology chosen for the detectors have changed over time.

In 2016, in order to advance data collection by one year with respect to the original plan, two of the TOTEM tracking stations, equipped with silicon strip detectors~\cite{Ruggiero:2009zz, TOTEM:2013vay}, were installed in the horizontal RPs at the 210-N and 210-F locations.
The TOTEM strip detectors were originally designed to operate in special LHC runs at very low luminosity, up to about $10^{30}\percms$.
Within PPS, they have been operated successfully well beyond design specifications, providing excellent position measurements.
However, they suffered from severe limitations related to the high hit rate.
Firstly, since the hit position in the detector planes is determined by the coincidence of pairs of perpendicular strips, the frequent presence of multiple proton tracks in the same bunch crossing leads to ambiguities that cannot be resolved.
Secondly, the radiation dose in the regions with the highest hit rate induced damages that caused an early loss of efficiency:
For those regions, the efficiency dropped to zero after $\mathcal{O}(10\fbinv)$ of integrated luminosity.

In 2017, new silicon pixel detectors, based on the 3D technology~\cite{Ravera:2016odg}, were installed in the RPs at the 220-F locations, replacing the strip trackers at 210-N.
These detectors were specifically developed for PPS, with much improved radiation tolerance and rate capability with respect to the strip detectors, allowing for the reconstruction of multiple tracks per bunch crossing.
Finally, in 2018, both stations at 210-F and 220-F were equipped with pixel detectors.

Despite the improved resistance to radiation, the dose accumulated by the pixel sensors and readout chips during \Run2 has been such to degrade significantly the detector performance, as briefly described in the following.
This, and the lack of enough replacement parts, called for the construction of new detector modules; with the aim of mitigating the adverse effects of radiation damage, a new design for the support mechanics has been developed for \Run3, which in turn implied a redesign of the frontend electronics.
Here, the main differences of the \Run3 design from that used in \Run2 are outlined.
A more complete description of the \Run2 setup can be found in Ref.~\cite{Ravera:2017cds}.

\begin{figure}[!ht]
\centering
\includegraphics[height=0.3\textwidth]{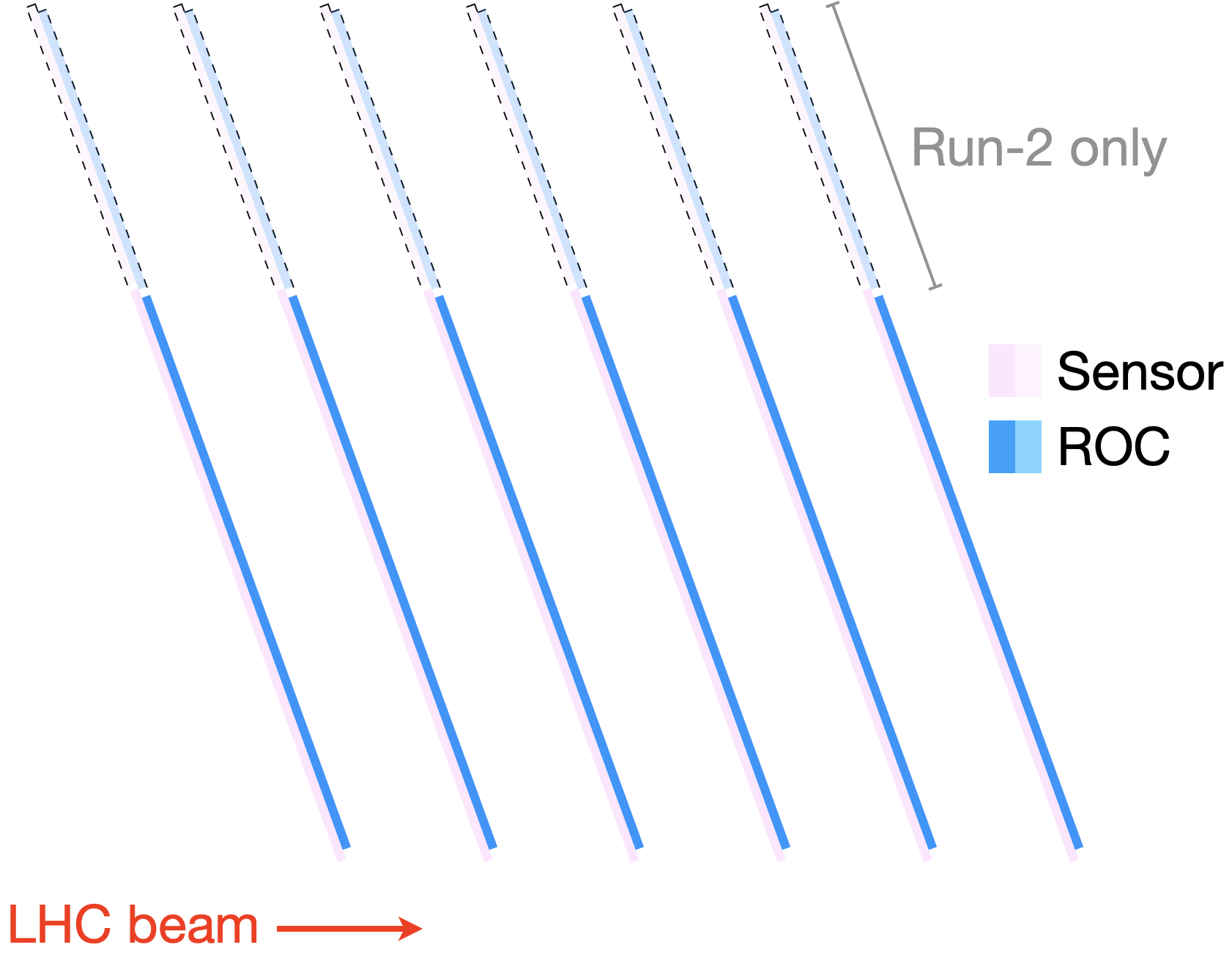}%
\hfill%
\includegraphics[height=0.3\textwidth]{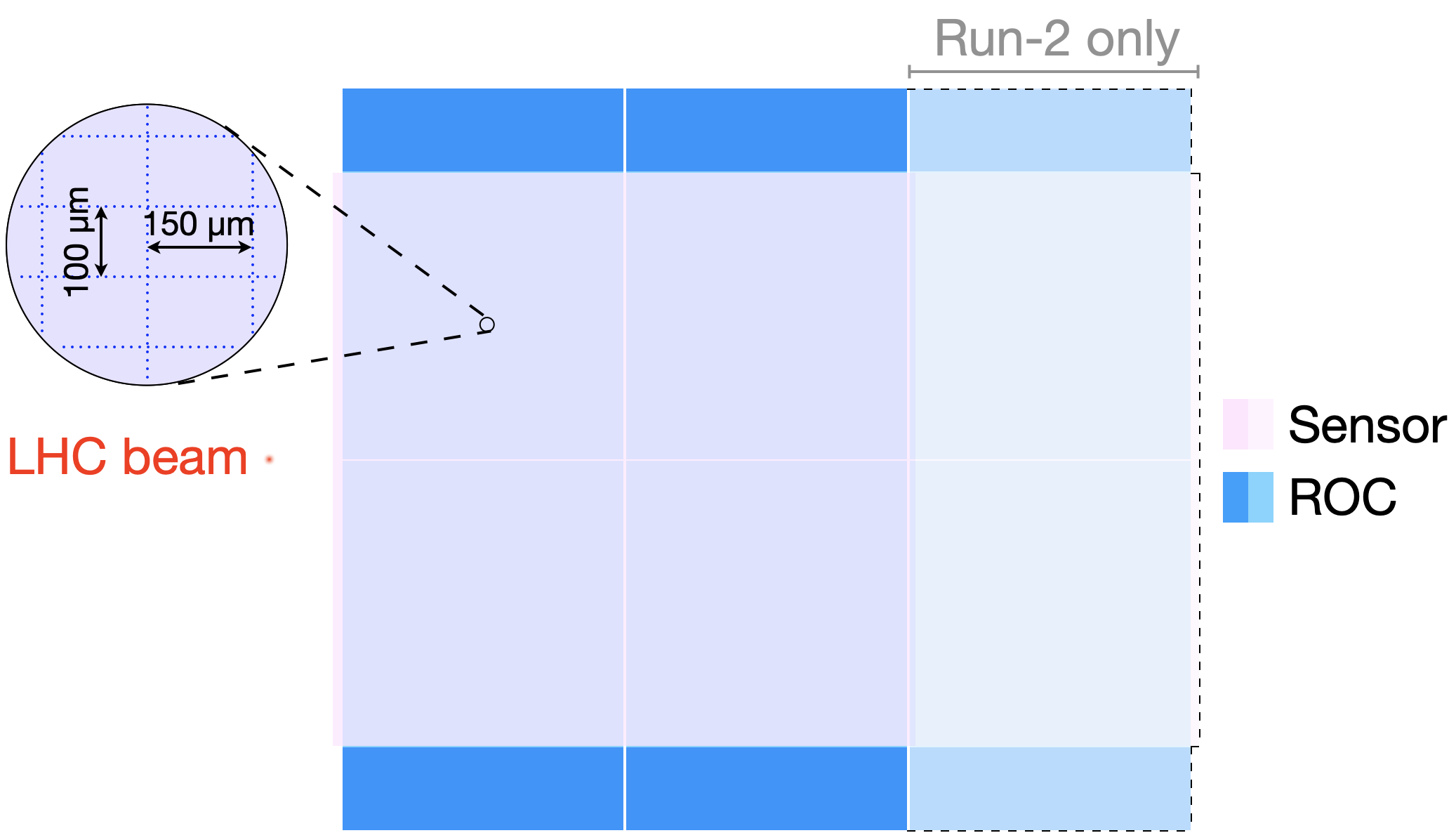} \\[1ex]
\includegraphics[height=0.3\textwidth]{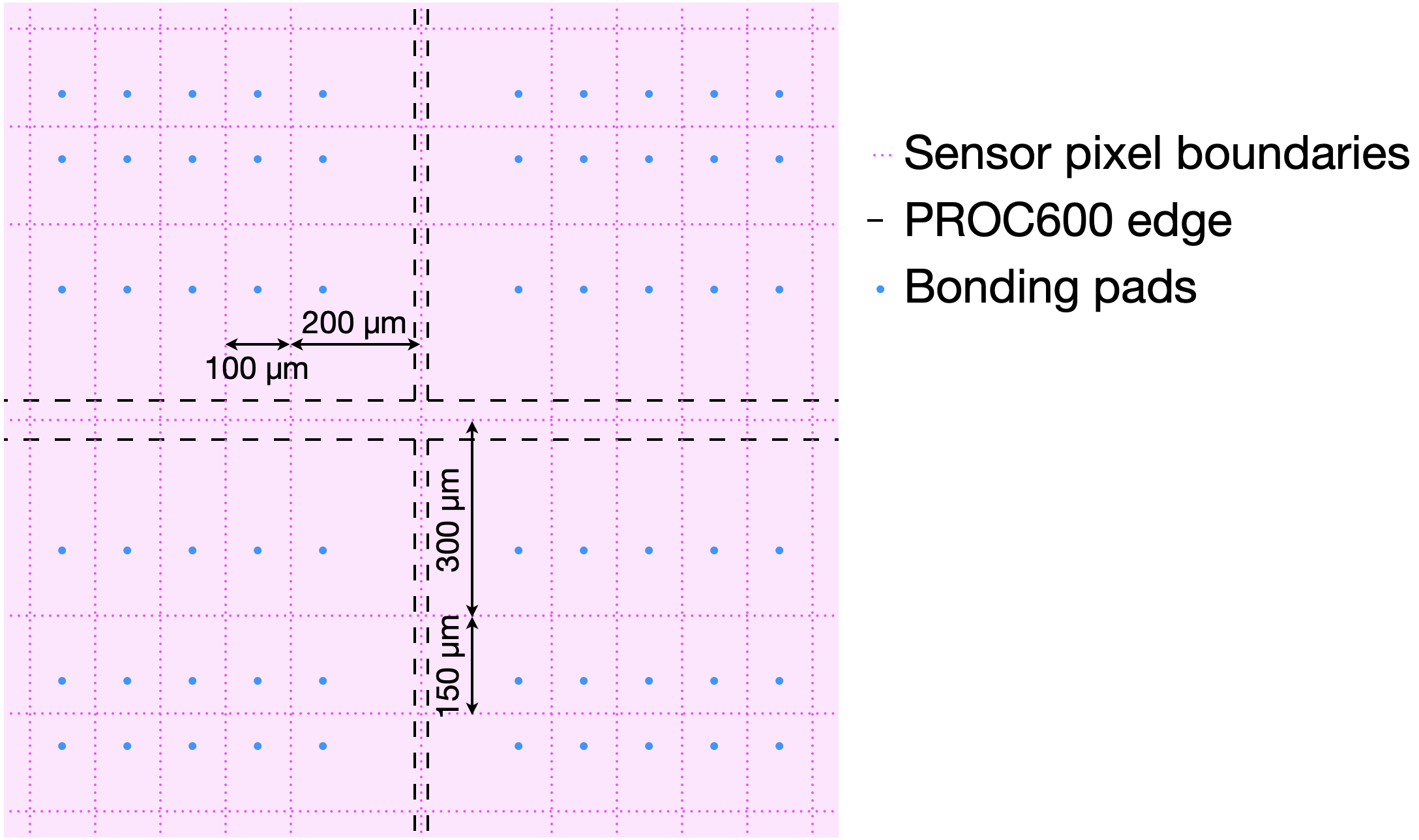}
\caption{%
    Geometry and arrangement of the pixel sensor modules.
    Upper left:\ arrangement of the sensor modules in a tracking station, relative to the LHC beam.
    Upper right:\ sensor, ROCs, and pixel geometry.
    The shaded areas with dashed contours refer to the larger, 3$\times$2 modules used in \Run2.
    Lower:\ detail of pixels at internal ROC edges.
}
\label{fig:pps:pixelmodulegeom}
\end{figure}

Each pixel tracking station consists of six detector planes, oriented at an angle of 70\de\ with respect to the thin window, \ie, 20\de\ with respect to the plane perpendicular to the beam axis (it was 18.4\de\ in \Run2).
This orientation removes the geometrical inefficiency associated to the junction and ohmic columns of the sensor, and increases the probability of charge sharing between nearby pixels for particle signals.
The sensor size is driven by that of the readout chip (ROC):\ for \Run3 it is 16.20$\times$16.65\mmsq, corresponding to a 2$\times$2 ROC matrix; in \Run2 most sensors were longer, 20.40$\times$16.65\mmsq, and read out by a 3$\times$2 ROC matrix.
The pixel size is 150$\times$100\mumsq, also adapted to that of ROC cells; the pixels located at the horizontal or vertical edge of a ROC have double size in the corresponding coordinate, except for those at the sensor edge.
The details of the sensor module geometry and the arrangement in the tracking station are shown in Fig.~\ref{fig:pps:pixelmodulegeom}.

\begin{figure}[!ht]
\centering
\includegraphics[width=0.9\textwidth]{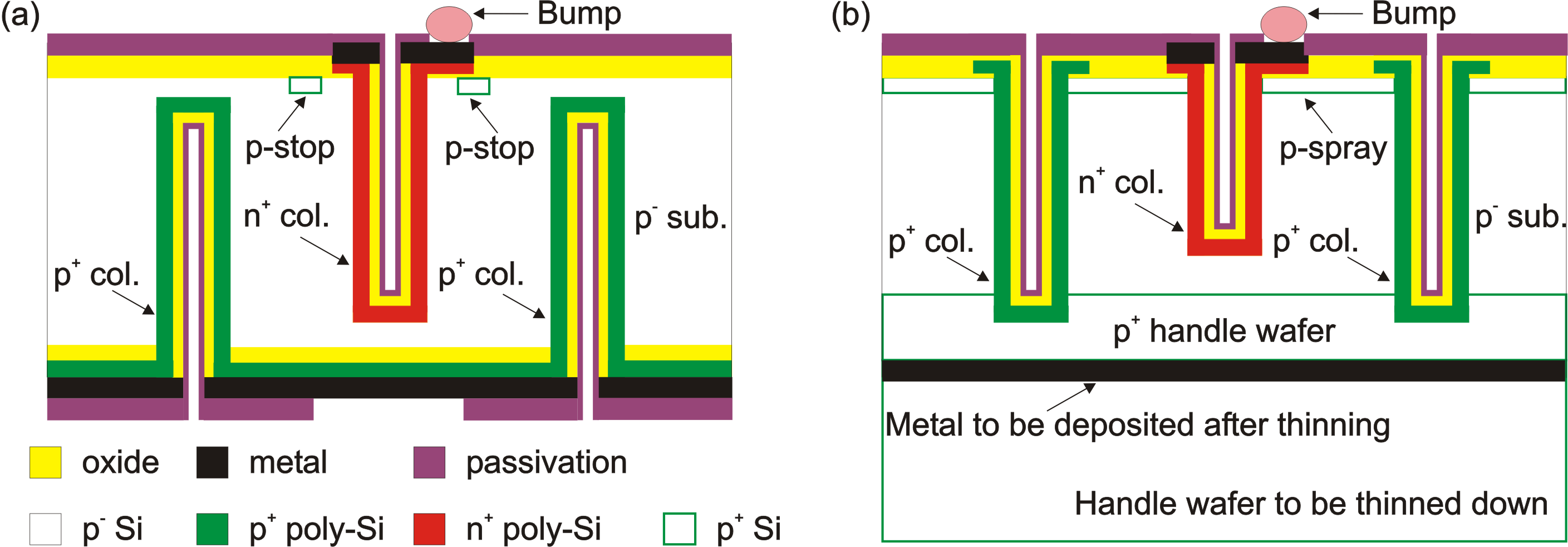}
\caption{%
    Schematic layout of the columns of 3D pixels.
    Left:\ double-sided technology, used in \Run2.
    Right:\ single-sided technology, used in \Run3.
    Metal bumps on top are used for interconnection with the readout chips; p-type surface implants, such as p-stop or p-spray, provide isolation between contiguous pixels.
    Reproduced from Ref.~\cite{DallaBetta:2017pjr}.
}
\label{fig:pps:pixel3dconf}
\end{figure}

{\tolerance=800
The 3D sensors have been produced by the Foundation Bruno Kessler (FBK) in Trento, Italy~\cite{Sultan:2016vzg}.
They are realized with ``single-side'' technology, where both the junction and ohmic columns are etched from the same side of the silicon wafer; the thickness of the active volume is 150\mum.
A so-called 2E configuration is used, where each pixel has two readout (junction) columns, thus improving the radiation resistance characteristics.
In \Run2, the sensors, produced by Centro Nacional de Microelectr\'onica (CNM) in Barcelona, Spain~\cite{Pellegrini:2013nq}, were realized with ``double-side'' technology, with the two types of columns etched on opposite sides of the wafer, with 230\mum active thickness, in both 1E and 2E configurations.
Figure~\ref{fig:pps:pixel3dconf} shows schematically a cross section of the column layout for the two productions.
Due to the 3D configuration, bulk depletion occurs at inverse polarization values of a few Volts.
Currents  below 1\muA are required at the working point, but are generally much lower (a looser requirement was set on \Run2 sensors); the breakdown tension must exceed 60\unit{V}.
\par}

\subsubsection{Readout electronics}

\begin{figure}[!p]
\centering
\includegraphics[width=0.9\textwidth]{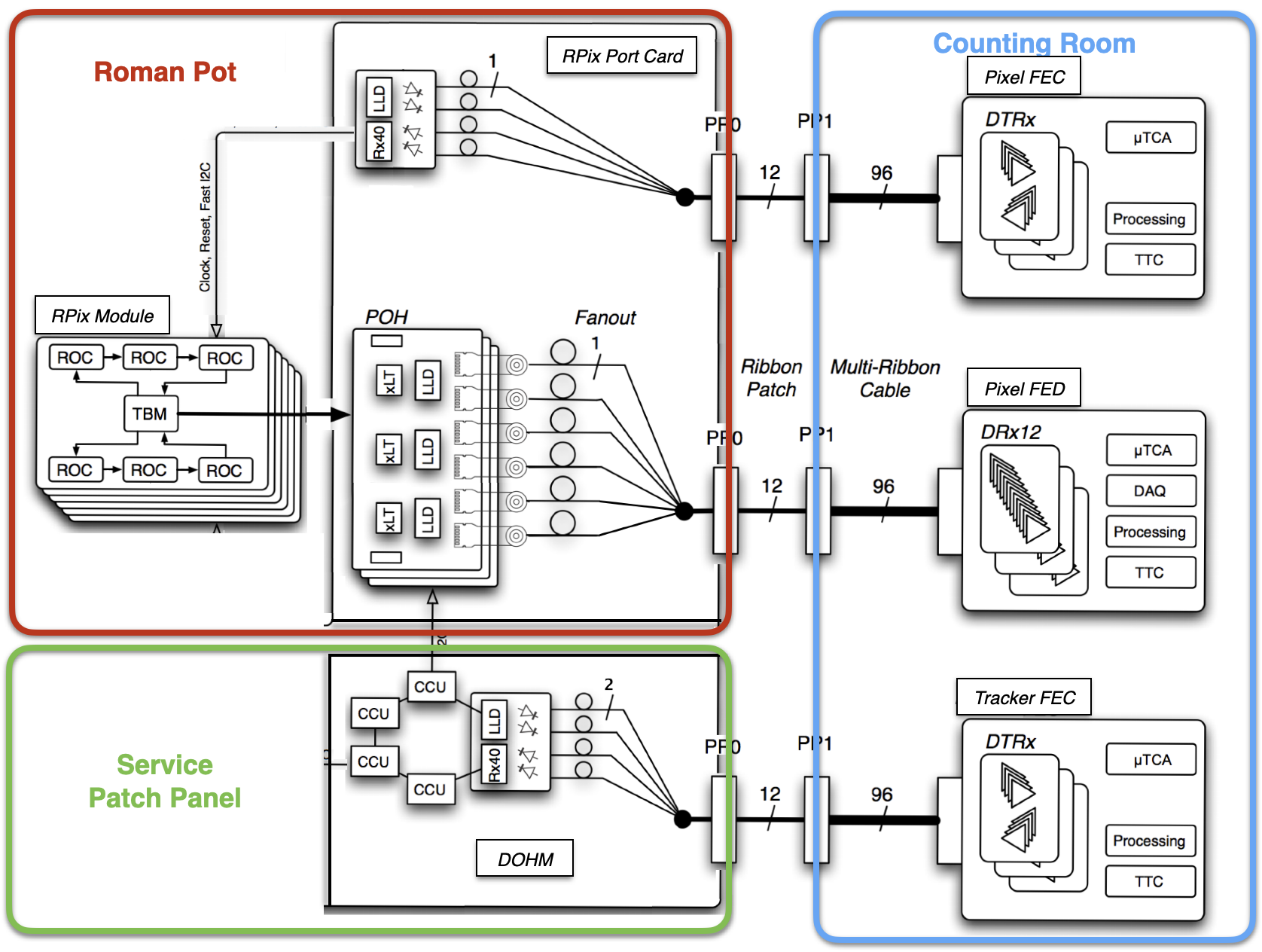}
\caption{%
    Schematic diagram of the PPS pixel tracker readout chain.
    The various components shown are described in Ref.~\cite{CMS:TDR-011}.
}
\label{fig:pps:pixelreadout}
\end{figure}

The readout of pixel detectors is based on the frontend electronics developed for the \Phase1 upgrade of the CMS pixel tracker~\cite{CMS:TDR-011}.
A schematic view of the readout chain is shown in Fig.~\ref{fig:pps:pixelreadout}.

\begin{figure}[!p]
\centering
\includegraphics[width=0.8\textwidth]{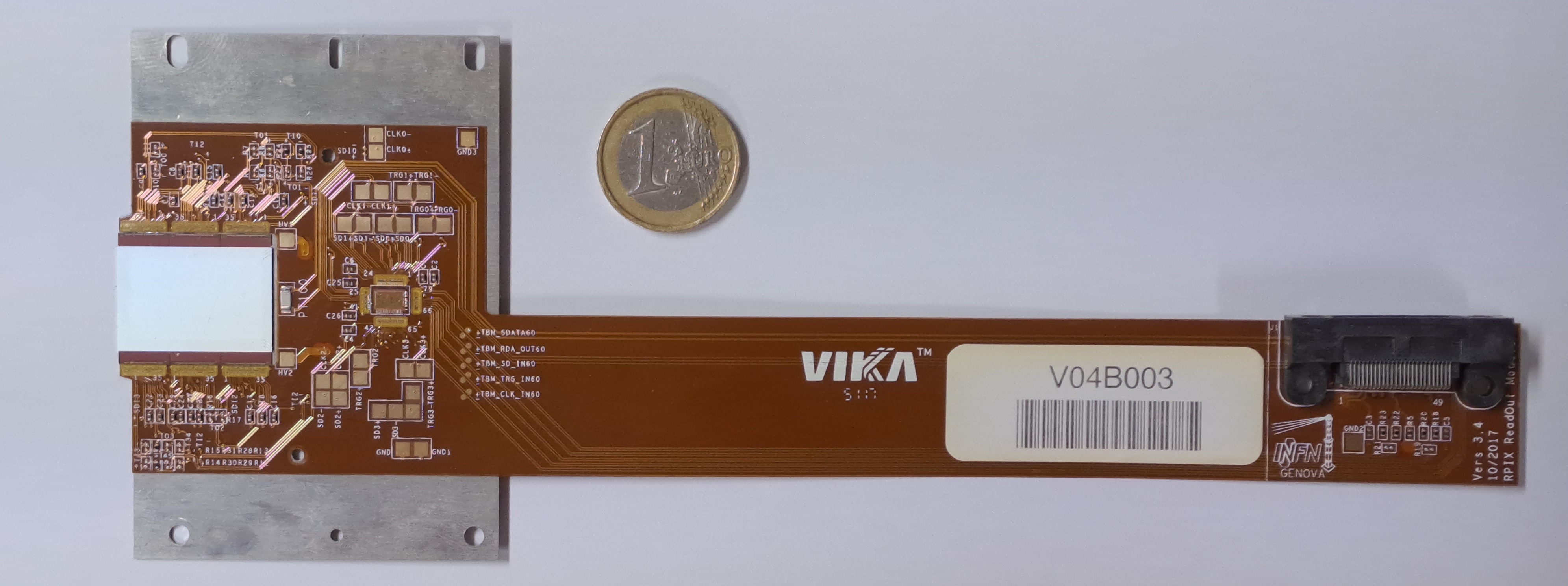} \\[1ex]
\includegraphics[width=0.8\textwidth]{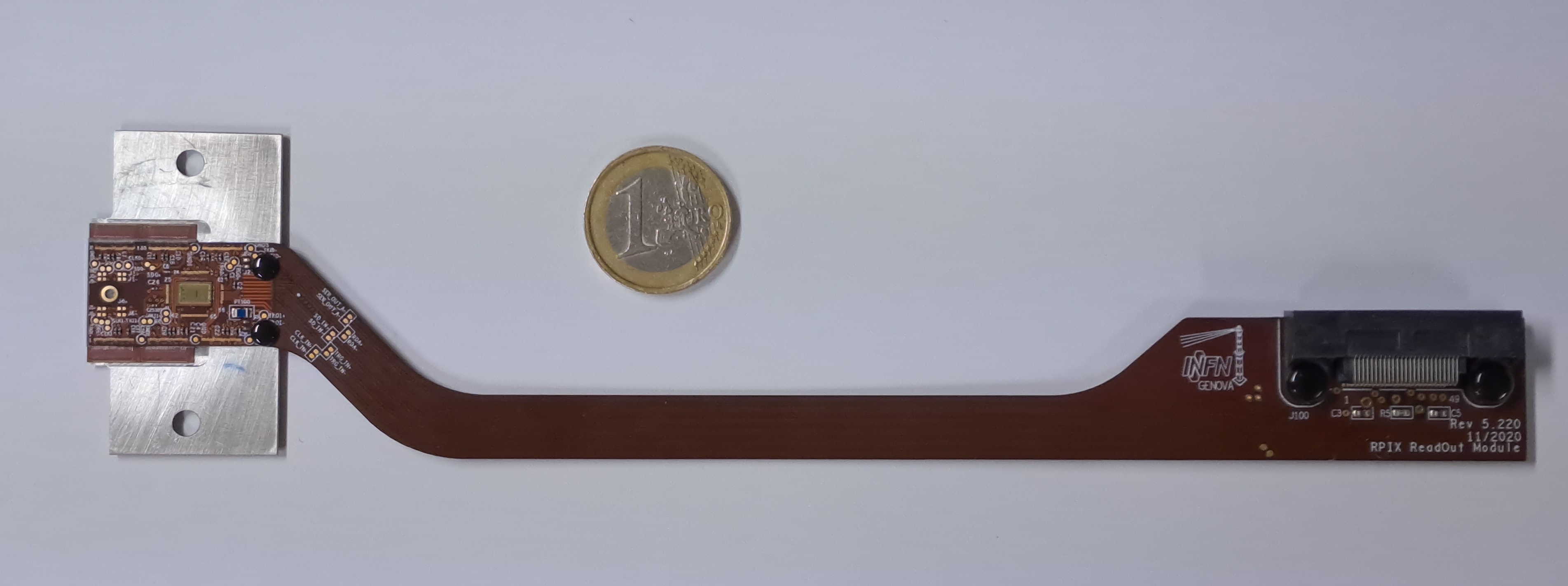}
\caption{%
    The detector module of the silicon pixel tracker, in its \Run2 (upper) and \Run3 (lower) versions.
}
\label{fig:pps:pixelmodules}
\end{figure}

Signals collected from the sensor electrodes are processed by the \proc chips developed for the \layer1 of the barrel pixel detector, described in Section~\ref{sec:tracker:services}:\ four chips are connected to each sensor via bump bonds.
In \Run2, the \psidig chips~\cite{Hits:2015jsa} were used.
A flexible printed circuit board is glued on top of the sensor; the ROCs are connected to the board via wire bonds.
A TBM10d chip (token bit manager), mounted on the board, distributes clock, fast commands and configuration instructions to the ROCs, and collects data through a token ring architecture; only one of the two TBM cores is used, resulting in a single output data stream.
In \Run2, the flexible board was mounted around the sensor module, and carried one TBM08c chip.
The two versions of the flexible board are shown in Fig.~\ref{fig:pps:pixelmodules}.

\begin{figure}[!t]
\centering
\includegraphics[width=0.48\textwidth]{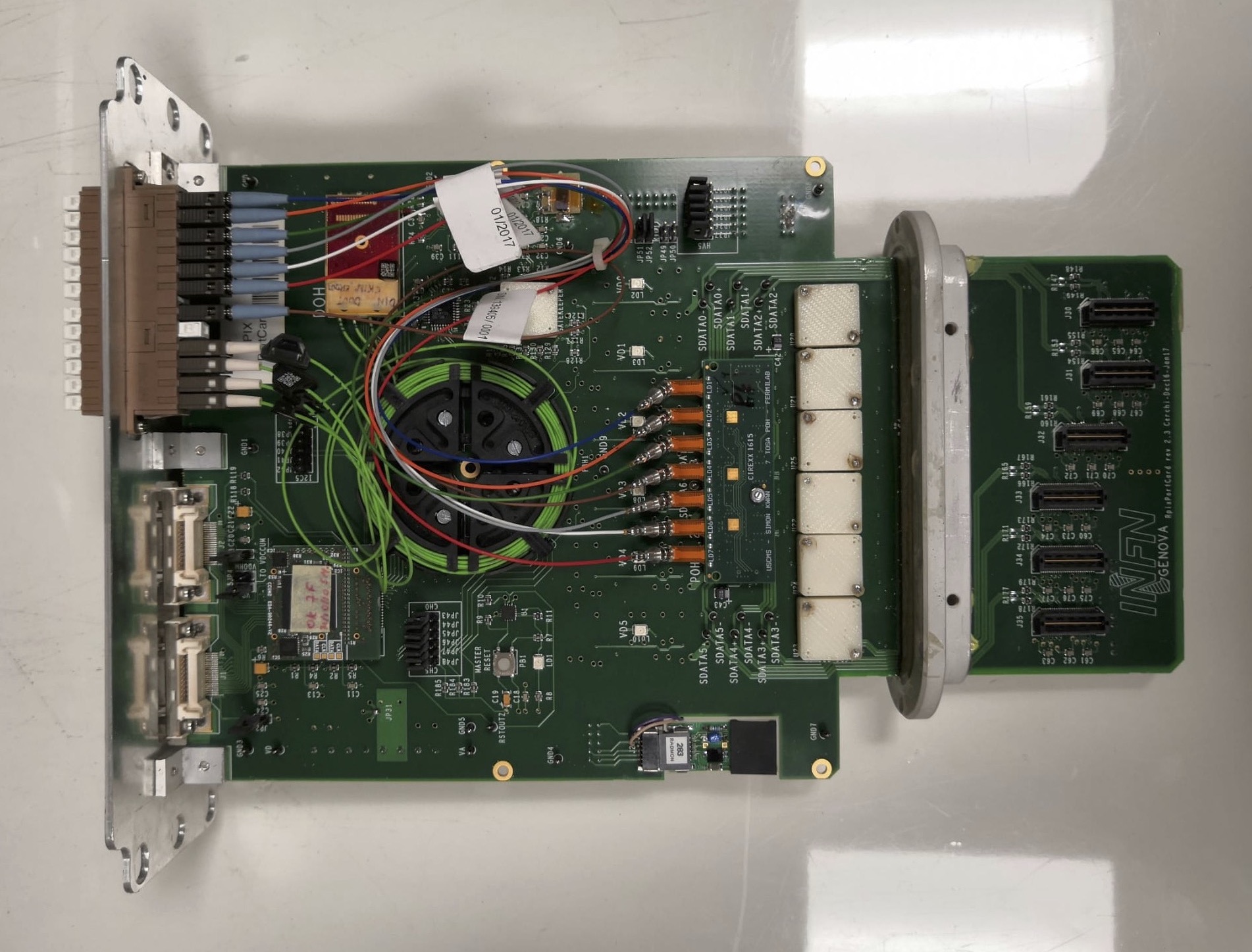}
\caption{%
    The pixel PortCard for \Run3; the small flange near the bottom acts as a feed-through for the board. The section below is housed inside the secondary vacuum volume of the pot.
}
\label{fig:pps:pixelportcard}
\end{figure}

The six detector modules of a tracking station are connected, through connectors on the flexible boards, to a concentrator card, called PortCard (Fig.~\ref{fig:pps:pixelportcard}).
The connections are made on a section of the PortCard that enters the secondary vacuum of the RP through a vacuum-sealed feed-through.
The rest of the board, operating at atmospheric pressure, contains all the active electronics needed to perform the board tasks:
\begin{itemize}
\item manage the configuration of the integrated circuits equipping the board itself via \ItwoC commands sent on the slow-control ring;
\item send the clock signal, fast commands, and configuration instructions to the frontend readout (via the TBM) through an optical link connection;
\item collect and send, through individual optical links, the data received from the detector modules;
\item route lines needed to read out environmental parameters in the detector housing and control the internal movement system (described later).
\end{itemize}
The PortCard can also control, through a parallel port register, the selective powering of the detector planes.

\subsubsection{Support structure and internal motion system}

The sensor modules and the flexible boards are glued, using a flowable silicon-based sealant, to aluminum support plates, ensuring thermal connection to the cooling circuit on the sides of the detector package (Fig.~\ref{fig:pps:pixelportcard}).
Precision holes in the plates are used to control the position of the sensors with respect to the thin window of the detector housing.
In \Run2, larger support plates were employed, realized in thermal pyrolytic graphite (TPG) enclosed in thin aluminum sheets.

\begin{figure}[!ht]
\centering
\includegraphics[width=0.3\textwidth]{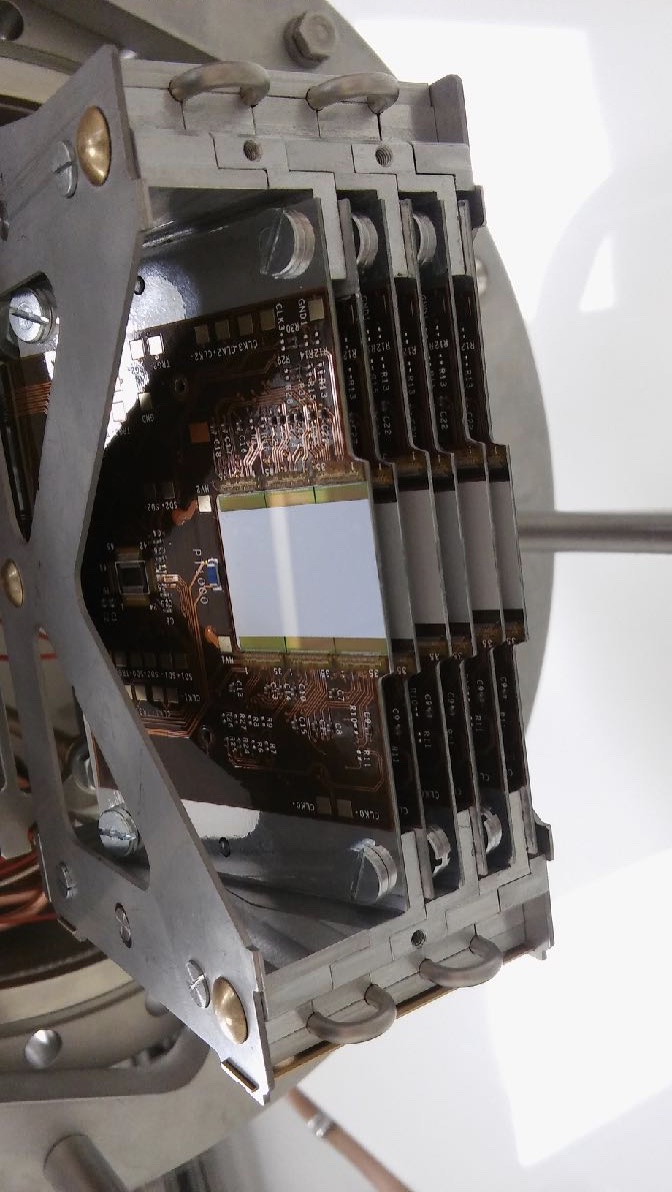}%
\hfill%
\includegraphics[width=0.3\textwidth]{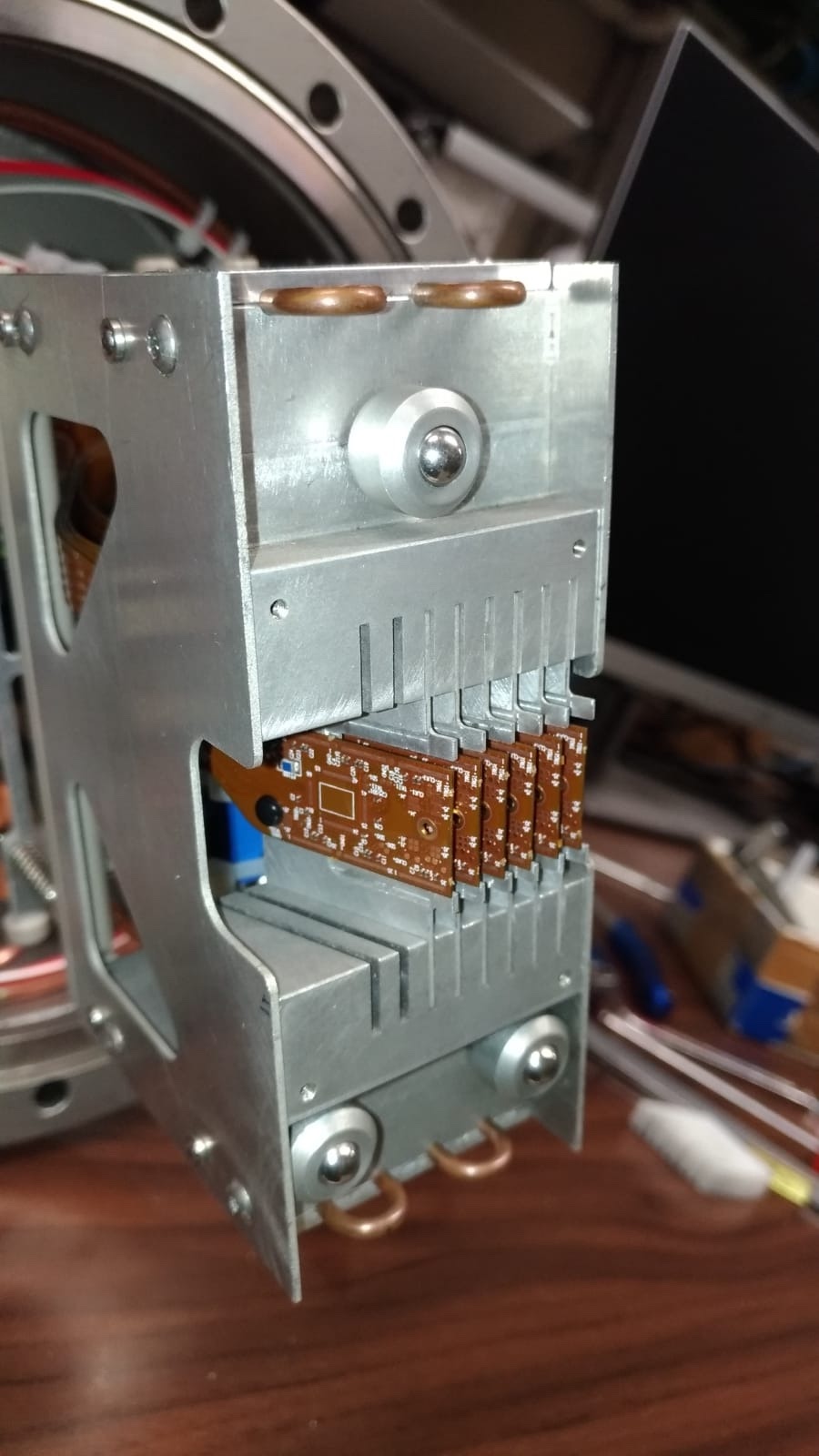}%
\hfill%
\includegraphics[width=0.3\textwidth]{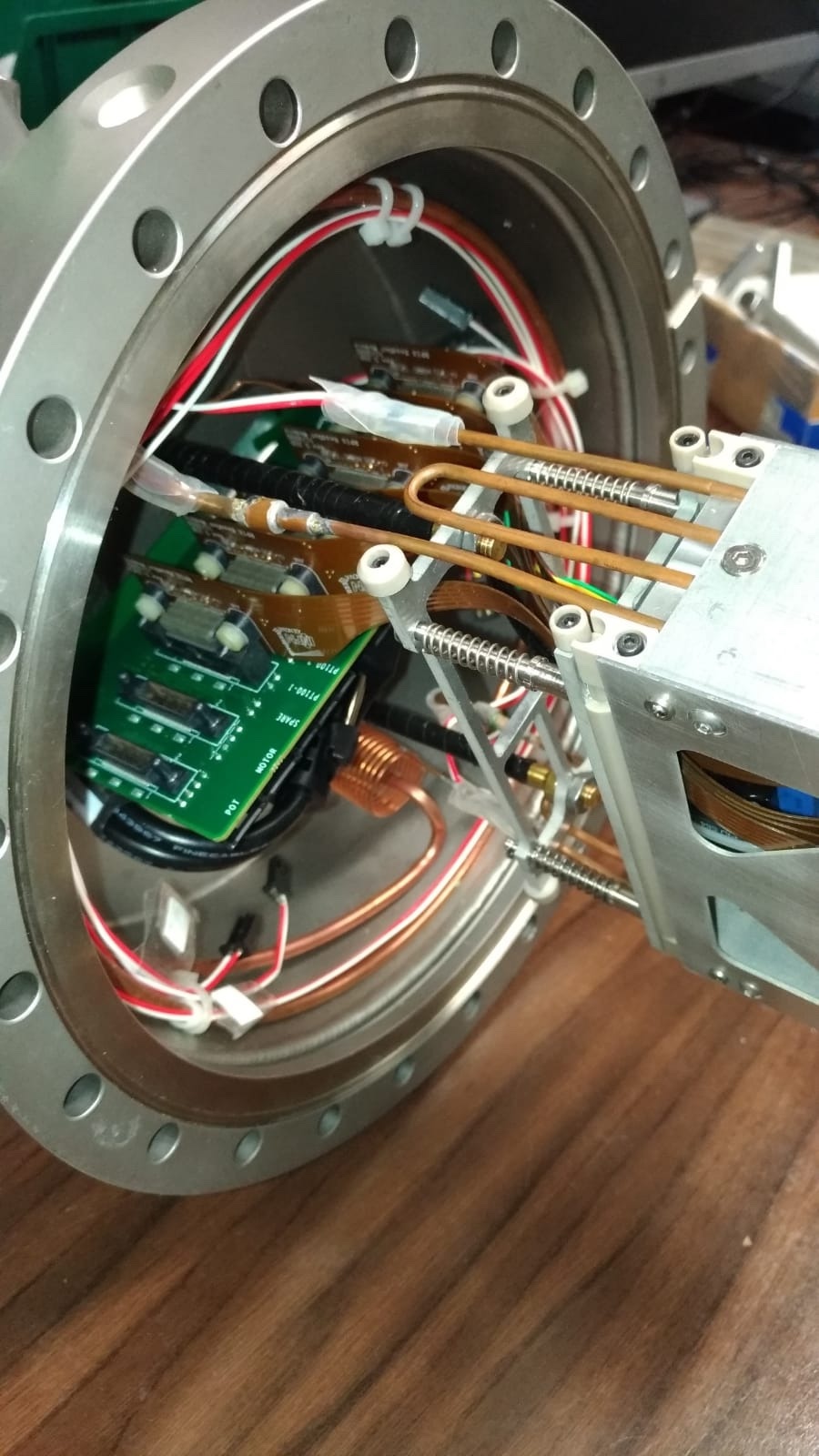}
\caption{%
    Details of the pixel detector package.
    Left:\ \Run2 version.
    Center:\ \Run3 version.
    Right:\ mechanical, cooling, and electrical connections to the upper part of the RP in \Run3:\ the vacuum section of the PortCard is visible, as well as the cooling circuit and the connections to the environmental sensors.
}
\label{fig:pps:pixeldetpackage}
\end{figure}

An aluminum support structure carries the six detector modules, ensuring their precise positioning and coupling them to the evaporators of the cooling system.
Six slits on the sides of the support structure host the aluminum plates of the modules, and pins on both sides keep them in place.
Three adjustable transfer-ball feet provide contact with the floor of the detector housing; on the opposite side, two spring-loaded frames allow the correct sealing of the upper flange of the RP and the transverse movement of the detector package, as described in the following.
In \Run2, planes were arranged in pairs in separate sections of the support structure, eventually assembled together and to the rest of the package mechanics.
Feet were static and no internal movement was allowed. Figure~\ref{fig:pps:pixeldetpackage} shows the assembled detector packages, as well as the connections to the readout and services outside the secondary vacuum volume.

\begin{figure}[!ht]
\centering
\includegraphics[width=0.48\textwidth]{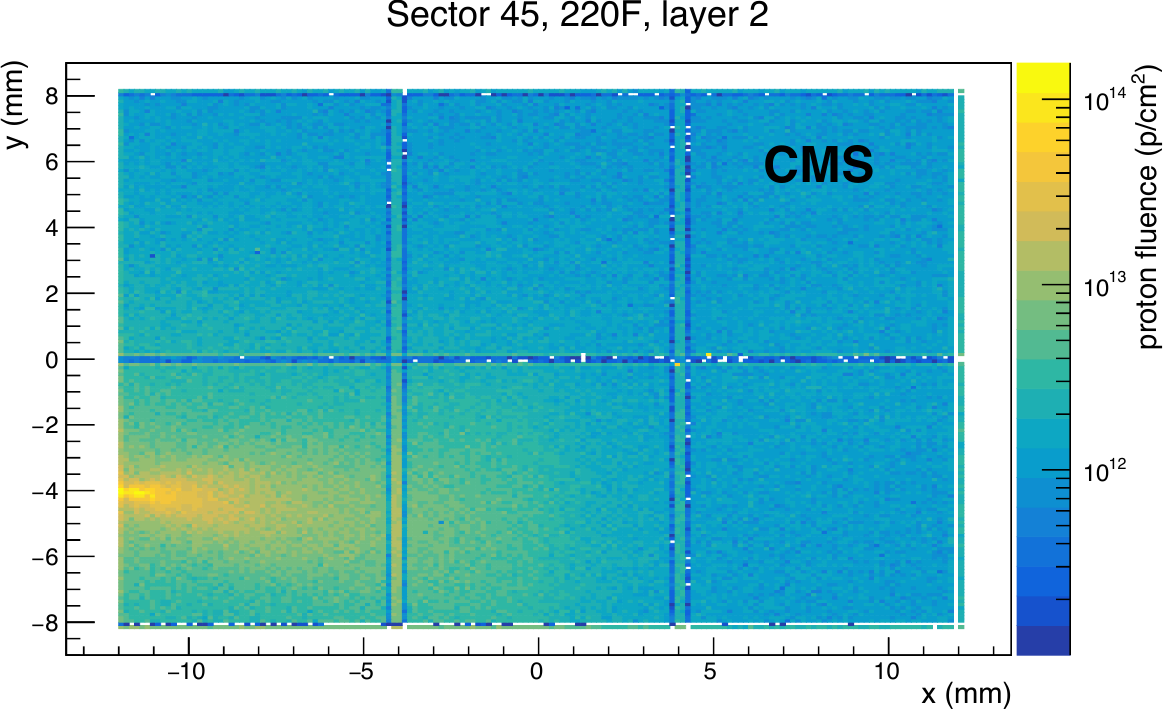}%
\hfill%
\includegraphics[width=0.48\textwidth]{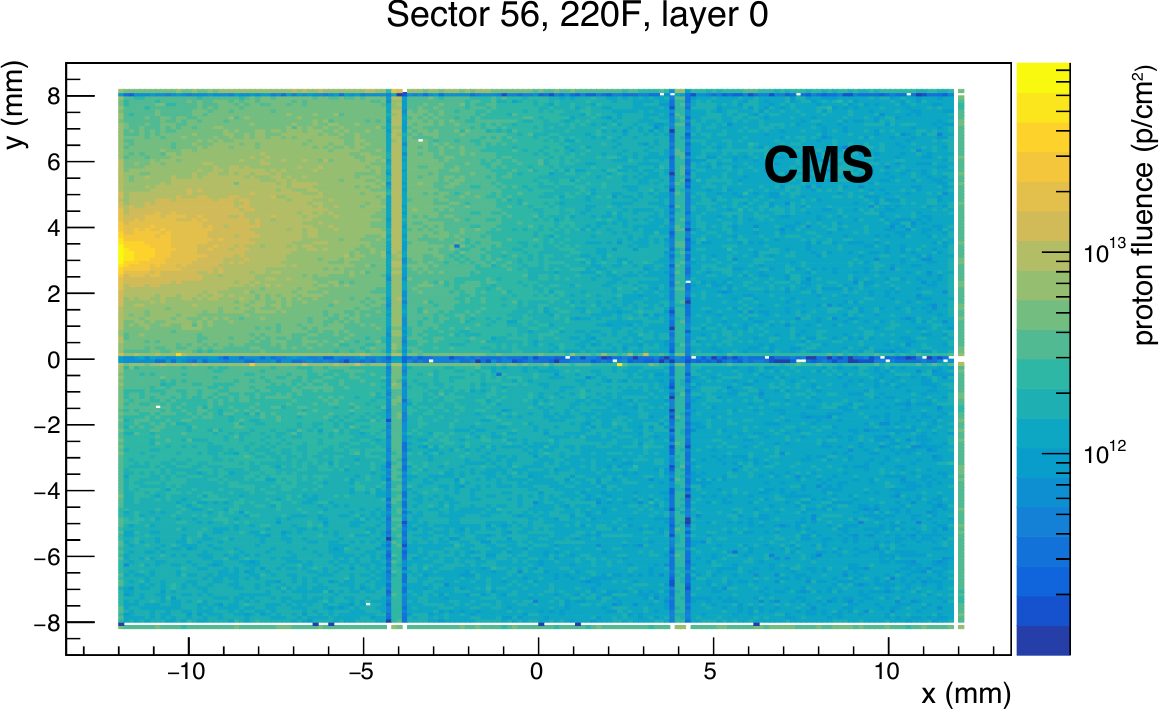}
\caption{%
    Proton fluence in two planes (one per arm) of the PPS pixel tracker in a sample run in 2017, normalized to an integrated luminosity of 1\fbinv.
    The horizontal and vertical lines in the distributions are an artifact due to the different pixel size of the sensors in those regions.
}
\label{fig:pps:pfluence}
\end{figure}

Both the sensors and the readout chips used in \Run2 were designed to withstand fluence values up to 2--3\ten{15}\Neq~\cite{Ravera:2016odg, Hoss:2016yrh}.
However, the highly nonuniform distribution of the hit rate (Fig.~\ref{fig:pps:pfluence}) implied localized radiation damage in the readout chip structures that could not be compensated by changing global configuration registers of the chip itself.
In particular, the position of the time window to accept hits (WBC register) could not be optimized for all pixels simultaneously:\ after a data-taking period corresponding to about 20\fbinv of integrated luminosity, the regions with the highest rate could not be read out.

With the aim of extending the lifetime of the detectors, in \Run2 the support of the corresponding RP was raised with respect to the floor of the LHC tunnel, in two steps of 0.5\mm, during technical stops.
Because the inefficient area was so localized, these small shifts were sufficient to move it to a region with much smaller hit rate, and thus almost restore the overall initial efficiency.
However, the range for this kind of movements cannot be further extended; moreover, the operation is onerous and not without risk.

In order to perform the same kind of movement in a more practical and safe fashion, an internal (vertical) movement system for the pixel detector package has been developed.
The very design of the new detector modules has been driven by this requirement.
Their width has been minimized (by placing the flexible board on top of the sensor) so as to exploit all the available space in front of the RP thin window.
The width of the thin window is 29\mm, while that of the sensor module is just below 22\mm.
A movement range of 5\mm for the detectors has been foreseen, including some safety margin.
This will allow to shift vertically the sensor modules, during the data-taking period, in 11, 500\mum-wide steps, thus distributing the radiation damage.

\begin{figure}[!ht]
\centering
\includegraphics[width=0.425\textwidth]{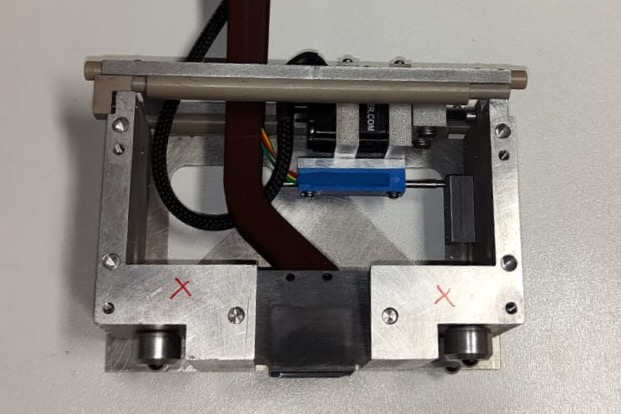}%
\hfill%
\includegraphics[width=0.515\textwidth]{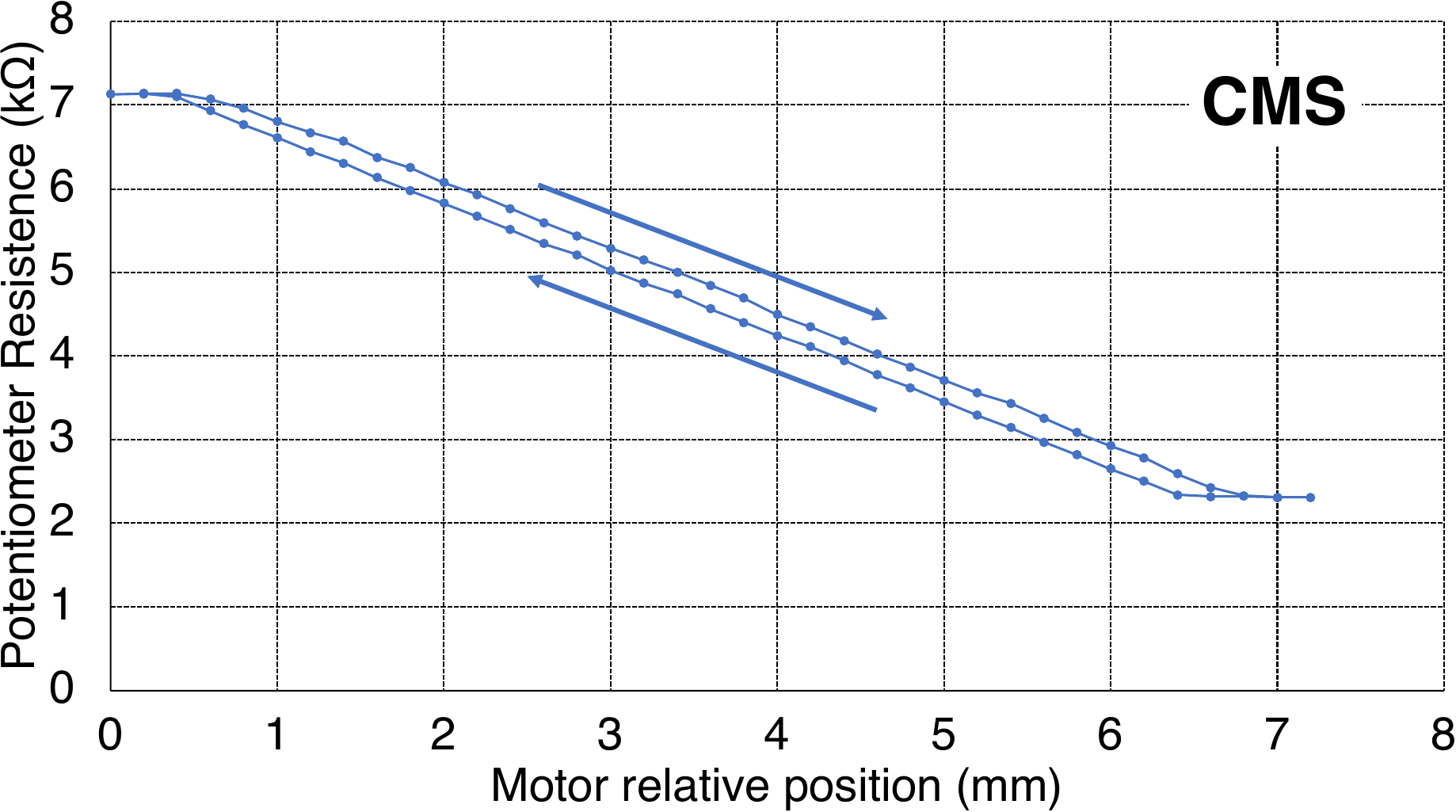}
\caption{%
    The system for internal motion of the pixel detector package.
    Left:\ detail of a partially assembled detector package:\ the stepper motor is the black object on top; the blue object below is the potentiometer used to monitor the position; both of them have their body tied up to the sliding slit on top, while their mobile tip is screwed to the support structure for the modules.
    Right:\ results of a motion test inside a RP at standard working conditions ($-20\deC$ and about 5\mbar), with measured versus nominal position.
    Two sets of points can be identified for forward and backward movements, revealing a hysteresis effect.
}
\label{fig:pps:pixelinternalmotion}
\end{figure}

The task is accomplished by means of a miniaturized stepping-motor linear actuator (Zaber LAC10A-T4A), mounted inside the detector package (Fig.~\ref{fig:pps:pixelinternalmotion}).
The actuator has an excursion of 10\mm and a step size of about 23.8\nm.
Its body is fastened to the lower frame of the spring-loaded structure; the head of the actuator is screwed into one of the side support parts.
These two parts can slide on each other thanks to low-friction elements realized in polyether ether ketone (PEEK).
Their relative position is measured by a simple linear motion potentiometer (Bourns 3048L-5-103), mounted in a similar fashion as the actuator, whose resistance can be read out externally.
Extensive tests have been performed on the fully assembled system, in particular performing repeated movements inside an exact reproduction of the RP and in conditions similar to the working ones ($T\approx-20\deC$, $P\approx15\mbar$).
Some results are shown in Fig.~\ref{fig:pps:pixelinternalmotion}.
The tests have demonstrated that the measured variation in resistance is consistent with the expected shift, showing the adequacy of the position measurement, where only a moderate precision is required.
Some hysteresis effect is observed, most likely due to the mechanics of the potentiometer, when the shift direction is inverted; however, this effect is reproducible over several movement cycles.

The linear actuators are controlled by dedicated two-channel controllers (Zaber X-MCC2).
One controller per arm is installed inside the RR53 and RR57 shielded areas (``alcoves'') along the tunnel.
They are connected to Raspberry PI microcomputers via USB link, which also reads out the detector position through an external ADC board; the microcomputers can be accessed from outside via the 4G network in the tunnel.
This kind of setup is adequate in view of the noncontinuous operation of the motors. In fact, only a limited number of detector shifts is performed over the whole data-taking time, in interfill periods.

\subsection{Timing detectors}
\label{sec:pps:timing}

In the presence of multiple proton-proton interactions in the same bunch crossing (pileup), the separation of overlapping events in CMS relies on the reconstruction of multiple primary interaction vertices.
However, because the proton tracks reconstructed by PPS have scattering angles very close to zero, the tracking system described in the previous section cannot associate them to CMS primary vertices.
For events with two protons detected on opposite sides, this limitation can be overcome if a precise measurement of the arrival time of protons is available:\ from the difference in time $\Deltat=\tplus-\tminus$, where \tplus and \tminus are the time measurements in the positive and negative arm of PPS, respectively, the position in $z$ of the \pp vertex can be inferred from $z_{\pp} = c\Deltat/2$.
Dedicated studies~\cite{TOTEM:TDR-003} have shown that a time resolution of $\mathcal{O}(10\ps)$ is needed to achieve the correct association of forward protons to the events reconstructed centrally by CMS.
For pileup conditions corresponding to $\mu=50$, a time measurement with 10 (30)\ps resolution could reject the combinatorial background from random combination of uncorrelated protons by about a factor 30 (20) while keeping about 60 (50)\% of the signal.
However, larger time resolution values, within 100\ps, can still help in reducing the background.
As in the case of the tracking system, timing detectors must be able to operate with large and highly nonuniform particle fluxes, to tolerate high radiation doses and to work in a vacuum.
Moreover, the material thickness must be limited.

During \Run2, one roman pot on each arm was used for timing, with four planes per RP.
For \Run3, instrumenting an additional station at the 220-N location will allow a second timing RP on each arm, giving a total of eight planes per arm.

\subsubsection{Detector modules}

The PPS timing detectors~\cite{TOTEM:2017mgc, Bossini:2020ycc, Bossini:2020pme} are based on synthetic single crystal chemical vapor deposit (scCVD) diamonds.
The detectors combine good time resolution, extreme hardness against large and nonuniform radiation, low material budget, and fine segmentation near the beam.

\begin{figure}[!b]
\centering
\includegraphics[width=\textwidth]{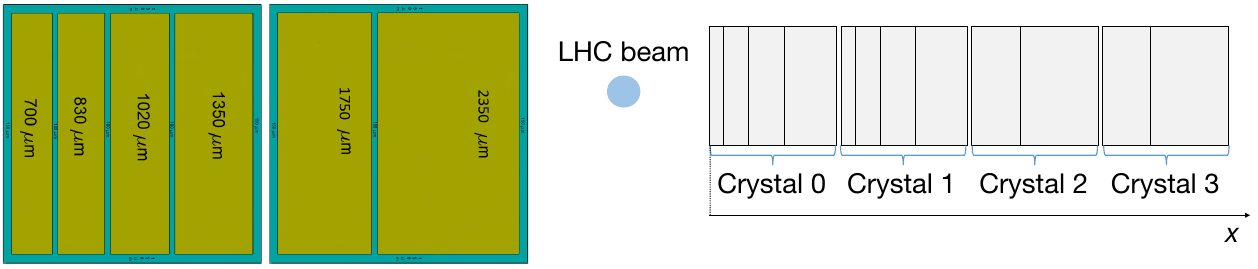}
\caption{%
    Left:\ details of the four-pad and two-pad segmentation of the diamond sensors used in the \Run3 modules.
    Right:\ arrangement of the four crystals in a \Run3 module, adapted from Ref.~\cite{CMS:DP-2019-034}, where the position of the beam is indicated by the spot on the left.
}
\label{fig:pps:diamondsegmentation}
\end{figure}

At the conclusion of \Run2, half of the planes were based on an
improved double-diamond architecture, in which two diamonds are connected to the same amplification channel.
This nearly doubles the signal, while keeping a similar noise level, resulting in a significant improvement of the timing resolution compared to a single-diamond design.
In \Run3, all timing planes are of the double-diamond design.
The crystals have dimensions of 4.5$\times$4.5\mmsq, with a total active surface area of approximately 20$\times$4.5\mmsq per plane.
The final segmentation is achieved during the metalization process, resulting in a total of 12 channels per plane. Due to the highly nonuniform flux of particles, the segmentation varies with the distance from the beam in the $x$ direction.
Several configurations of pad dimensions and sensor layout have been used in \Run2; in \Run3, sensors with two-pad and four-pad segmentation are used, with dimensions as detailed in Fig.~\ref{fig:pps:diamondsegmentation}, where the layout of four of these sensors in a detector module is also shown.
Pads close to the beam position have smaller size (as small as 0.55\mm), while a coarser segmentation is used farther from the beam.
This results in a more uniform occupancy, with low inefficiency due to multiple hits in the same channel.
Because of the horizontal crossing of the LHC beams at the IP, hits are more widely distributed along the $x$ axis.
For this reason, the detectors are only segmented horizontally.

The diamonds for \Run3 include newly produced crystals, and crystals that were previously used in \Run2, cleaned and remetalized.
Samples of the latter were first studied in beam tests after dismounting, and found to maintain high efficiency and achieve single plane time resolutions of $\sim$80--95\ps, after being exposed to fluences as large as 5\ten{15}\pcmsq~\cite{CMS:NOTE-2020-007}.
Compared to the nominal resolution of about 50\ps per plane for a new double-diamond detector, the decrease in resolution was found to be largely consistent with radiation damage to the preamplification electronics.

\begin{figure}[!ht]
\centering
\includegraphics[width=0.343\textwidth]{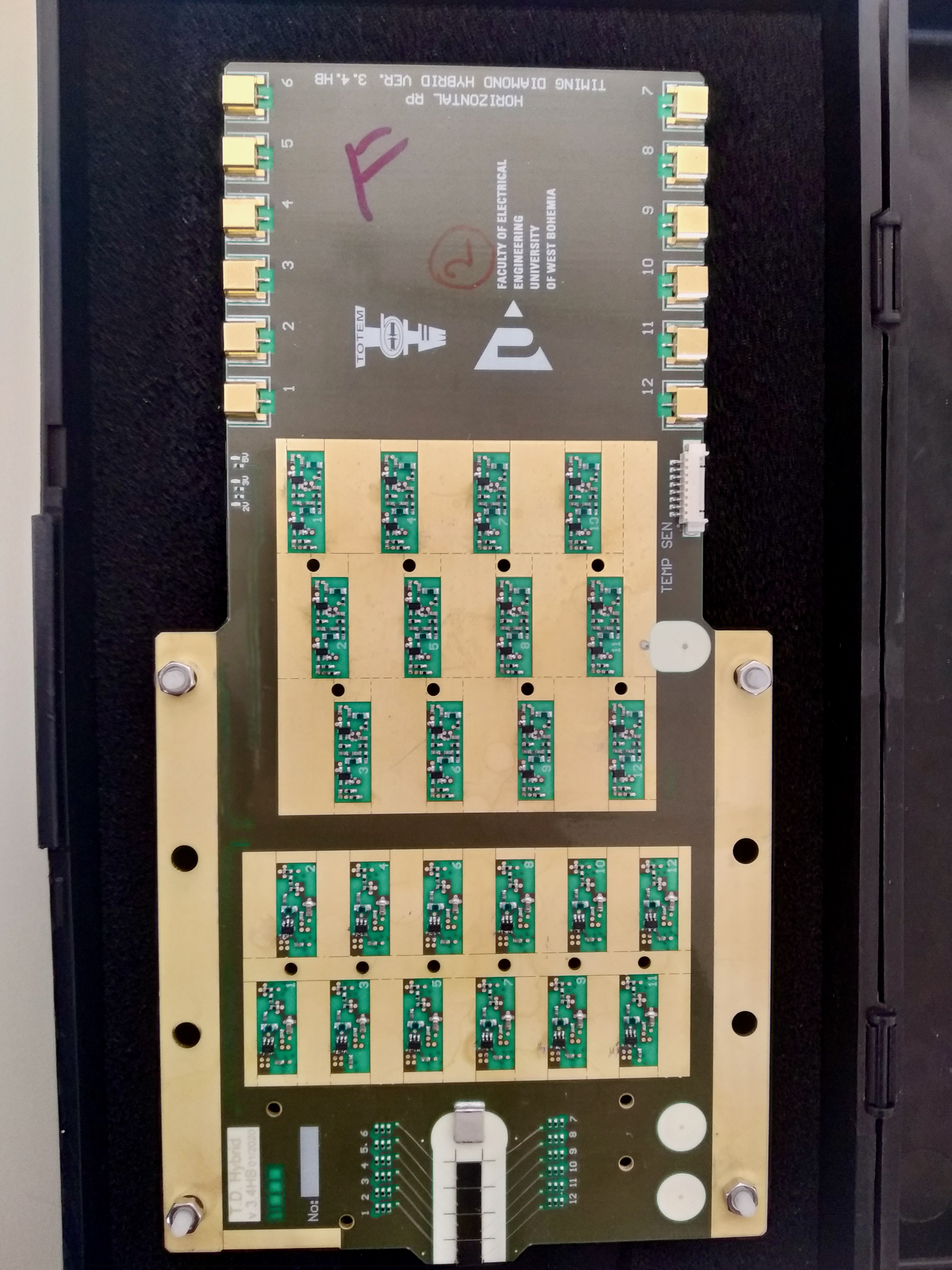}%
\hfill%
\includegraphics[width=0.609\textwidth]{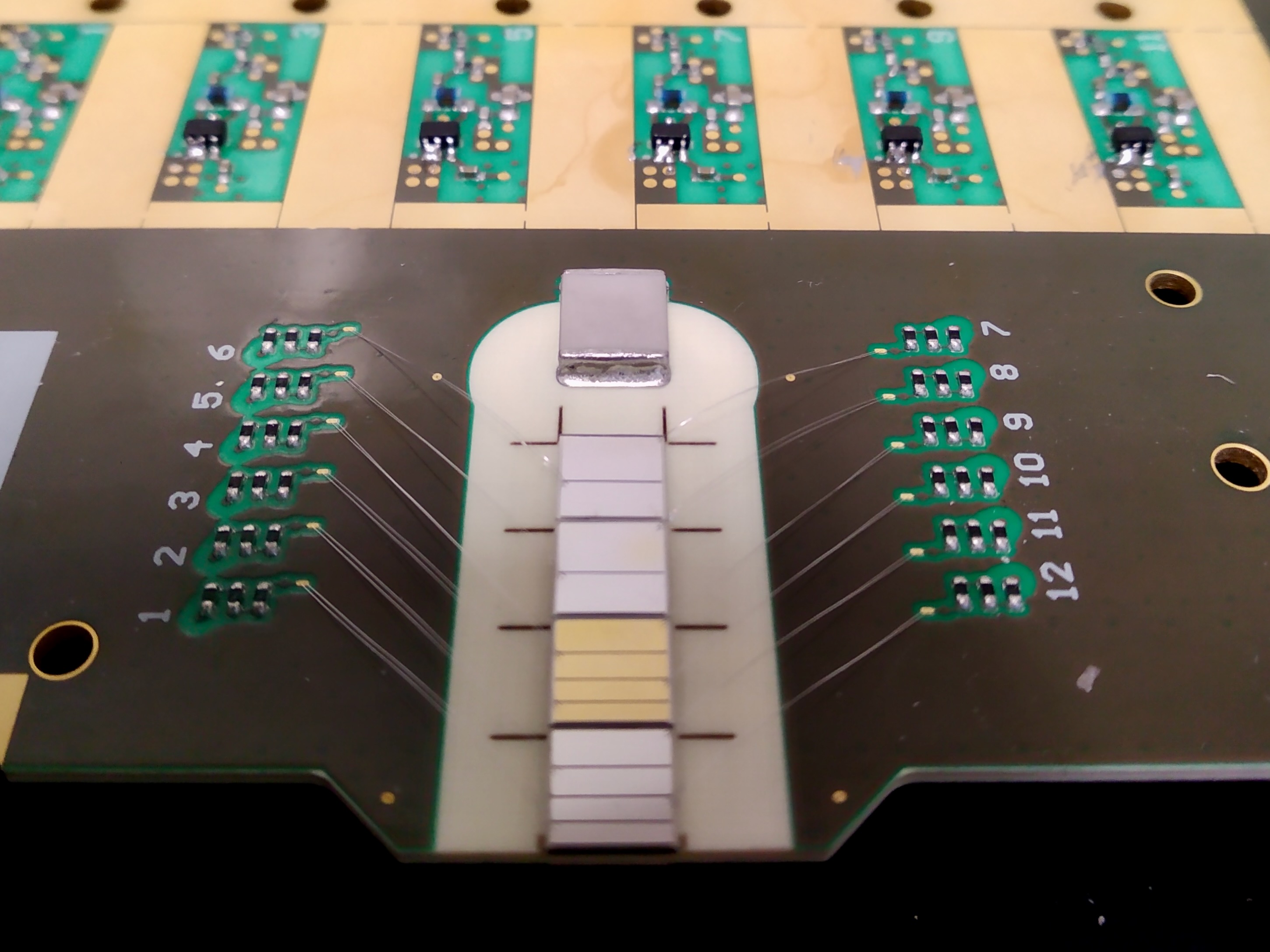}
\caption{%
    Left:\ the hybrid board for the \Run3 timing detectors' readout; the lower, wider section hosts the sensors and is housed inside the secondary vacuum volume of the pot.
    Right:\ detail of the diamond sensors on one side, connected via wire bonds to the frontend electronics.
}
\label{fig:pps:diamondhybrid}
\end{figure}

The diamonds are glued to a hybrid board (Fig.~\ref{fig:pps:diamondhybrid}), containing both the sensors and a multistage amplification chain for each of the 12 channels.
As mentioned above, radiation damage to the amplification chain was identified as a limiting factor for the timing performance in \Run2.
Therefore, a revision of the hybrid board was designed for \Run3, with a modified layout to mitigate the exposure of the preamplifiers to radiation.
Additional modifications to the design were made to improve the high-voltage isolation and the stability against RF noise pickup.
Finally, remote control of the amplifier gains was implemented, giving the opportunity to better fine tune the settings and compensate for any degradation during data taking.

\subsubsection{Readout electronics}

\begin{figure}[!ht]
\centering
\includegraphics[width=0.46\textwidth]{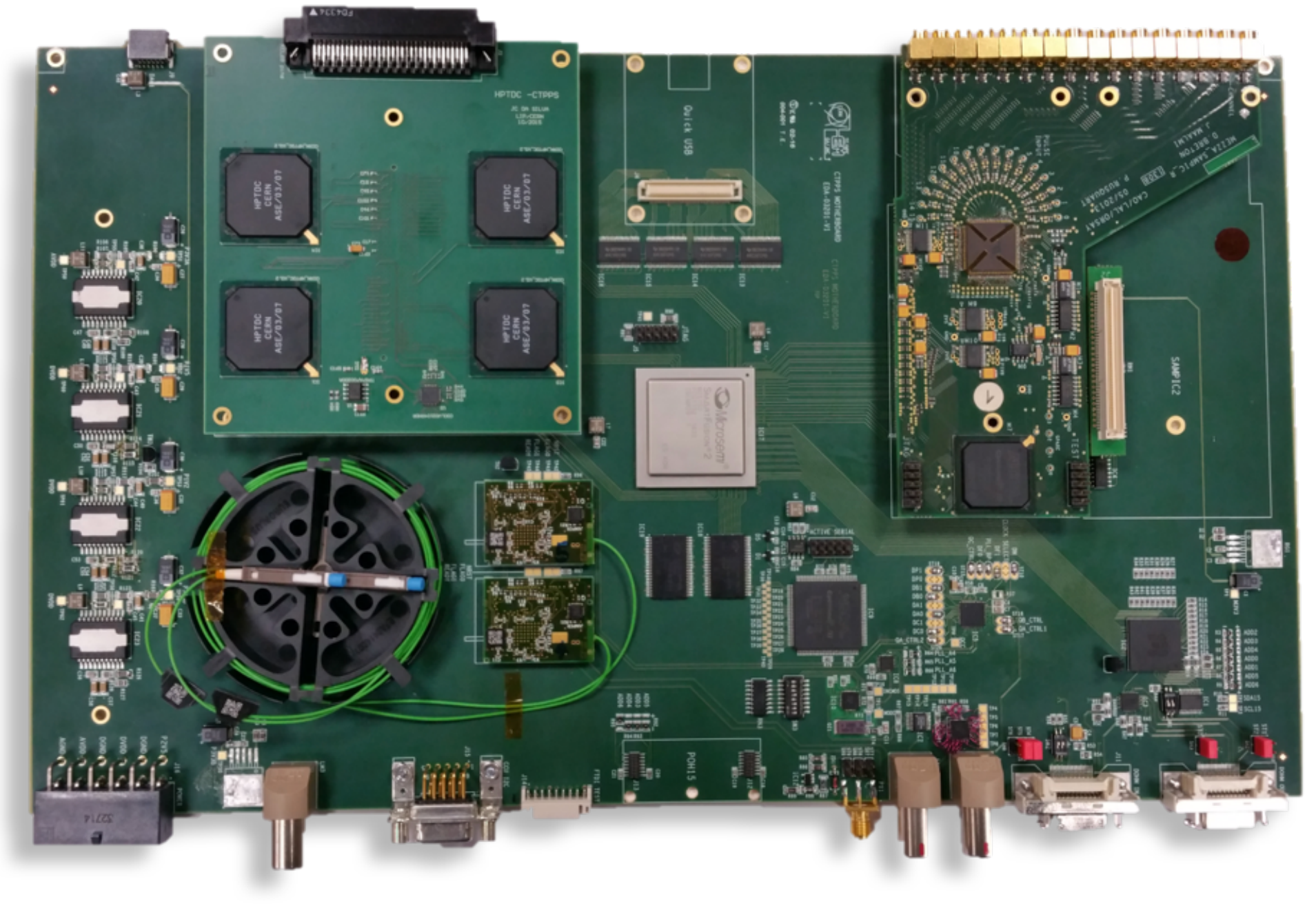}%
\hfill%
\includegraphics[width=0.5\textwidth]{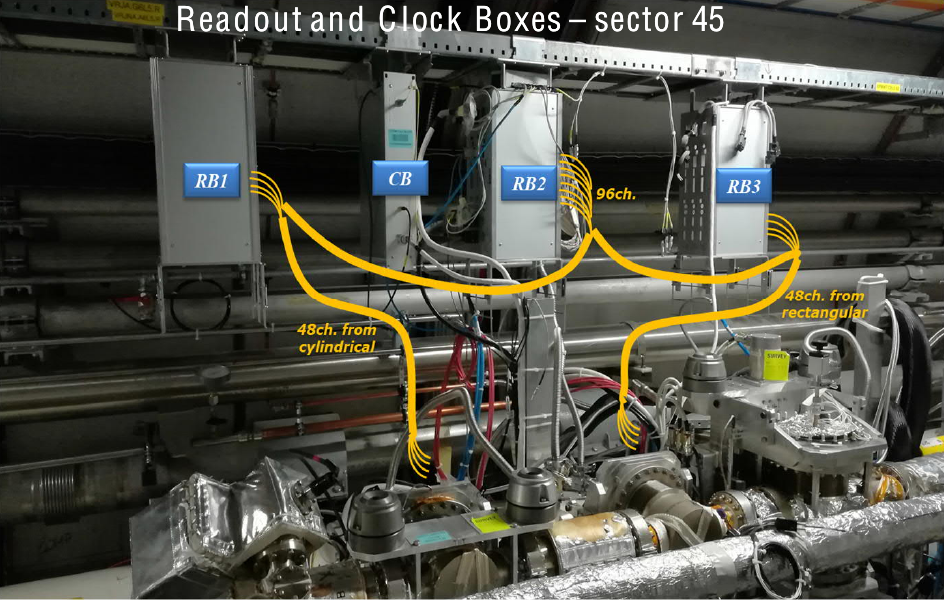}
\caption{%
    Left:\ the digital readout unit.
    In this example both an HPTDC mezzanine (upper left) and a SAMPIC mezzanine (upper right) are mounted on the motherboard.
    Right:\ location of PPS mini-crates installed above the LHC beam pipe, for both the cylindrical timing RP already used in \Run2, and the rectangular 220-N RP newly equipped for timing in \Run3.
    The readout boxes (labeled ``RB'') contain the NINO boards and DRUs as described in the text.
    The box labeled ``CB'' holds components for the precise reference clock.
}
\label{fig:pps:digitizerboard}
\end{figure}

The signals from the hybrid boards are transmitted by individual coaxial cables to custom ``NINO boards'', mounted in a mini-crate about 1\unit{m} above the LHC beam pipe (Fig.~\ref{fig:pps:digitizerboard}).
The main data path for reading all channels at the full trigger rate is based on the fast, low-power NINO ASIC~\cite{Anghinolfi:2004gg}, with four chips per board.
The NINO performs discrimination of the input signals above an adjustable threshold.
The width of the output signal is proportional to the input charge, and is then stretched by a constant value for compatibility with the next piece in the readout chain.
This serves as a proxy for the signal amplitude, allowing for time-walk corrections to be derived from the data.
In \Run3, a revision of the NINO board also allows for signals from a subset of channels to be split off to record the full waveform information, at reduced rate.

{\tolerance=800
From the NINO boards, the signals are transmitted to the digital readout units (DRUs), mounted in the same mini-crate above the beam pipe.
Each DRU consists of a digitizer motherboard supporting two types of mezzanine for timing measurements.
The first, used in \Run2 and for all channels in \Run3, includes four high-performance time-to-digital converter (HPTDC) chips~\cite{Christiansen:2004cds}.
The HPDTC for PPS is used in very high resolution mode, where an on-chip four-point sampling and interpolation of a delay-locked loop is performed to improve the time resolution.
This results in a TDC with a binning of about 25\ps, at the cost of limiting the number of output channels per chip to eight.
The HPTDC records the time of both the leading and trailing edges of the signal, allowing for time-walk corrections to be performed as a function of the reconstructed input charge value.
In \Run3, additional DRUs are also equipped with a second type of mezzanine, based on the SAMPIC~\cite{Royon:2015sfa} fast waveform sampler.
This allows the readout of the full analog signal shape, for calibration purposes, on a fraction of the data, for trigger rates up to 100\kHz.
Each digitizer motherboard includes a radiation-hard MicroSemi SmartFusion2 FPGA, which formats the data from the HPTDC (or SAMPIC) mezzanine into the final data frames.
The data from each board are then transmitted over optical fibers to the backend using two GOH opto hybrid mezzanines.
\par}

In addition to its role in the data acquisition, the DRU is also responsible for receiving and distributing slow-control commands, clock, and fast commands.
It receives both the standard clock and the dedicated precise reference clock~\cite{TOTEM:2017mgc} that is used for the timing measurement.
Along with the configuring of components on the DRU itself, it can propagate commands via \ItwoC interface to set the thresholds on the NINO board, and, for
\Run3, the low voltage settings on the diamond hybrid.
The ability to remotely set the low voltage will allow for better fine tuning of the operational settings, and provide the ability to compensate for possible radiation damage effects.

\begin{figure}[!ht]
\centering
\includegraphics[width=\textwidth]{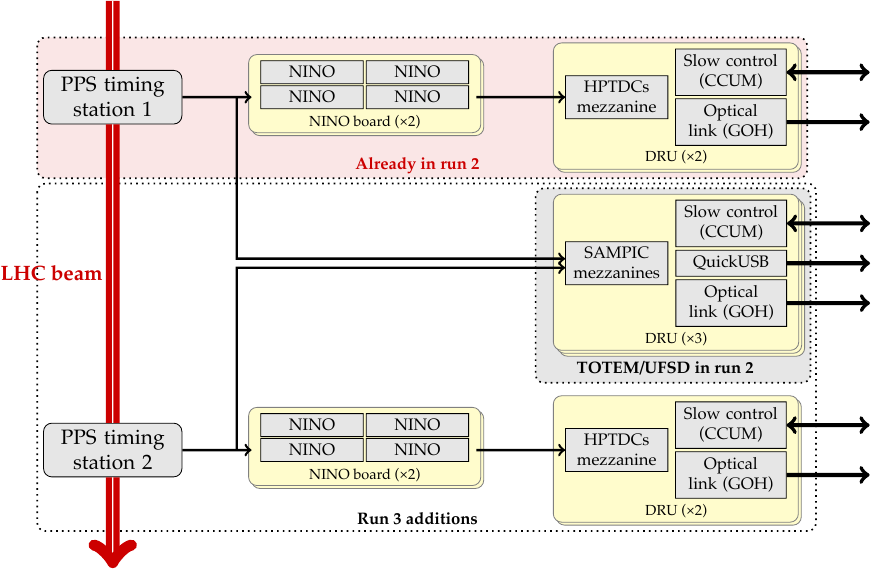}
\caption{%
    Schematic diagram of the full readout chain for the PPS timing detectors in \Run3, also showing the configuration used in \Run2.
    The arrows on the right represent electrical connections to the slow-control ring and optical connections to components in the underground service cavern of CMS, described in Section~\ref{sec:pps:daqdcs}.
    The central set of DRUs (3 units) was already present in \Run2, but was employed for the diamond detectors equipping the TOTEM experiment at that time, and for one layer of UFSD detectors used in 2017 for the PPS timing.
}
\label{fig:pps:timingreadoutchain}
\end{figure}

A schematic view of the timing control and readout chain components installed in the LHC tunnel is shown in Fig.~\ref{fig:pps:timingreadoutchain}.

\subsubsection{Reference clock distribution}

For precise timing measurements, a clock distribution system that
provides time information at points separated by large distances, with a picosecond precision, is needed.
This requires a system capable of the highest precision and of the utmost time stability.

PPS employs the clock distribution system developed by the TOTEM Collaboration~\cite{TOTEM:TDR-002}, an adaptation of the universal picosecond timing system~\cite{Bousonville:2009zz, Moritz:2011zz}, developed for FAIR (Facility for Antiproton and Ion Research) at GSI, Darmstadt.
The system has been installed and used during the TOTEM and PPS runs since 2017.

\begin{figure}[!p]
\centering
\includegraphics[width=\textwidth]{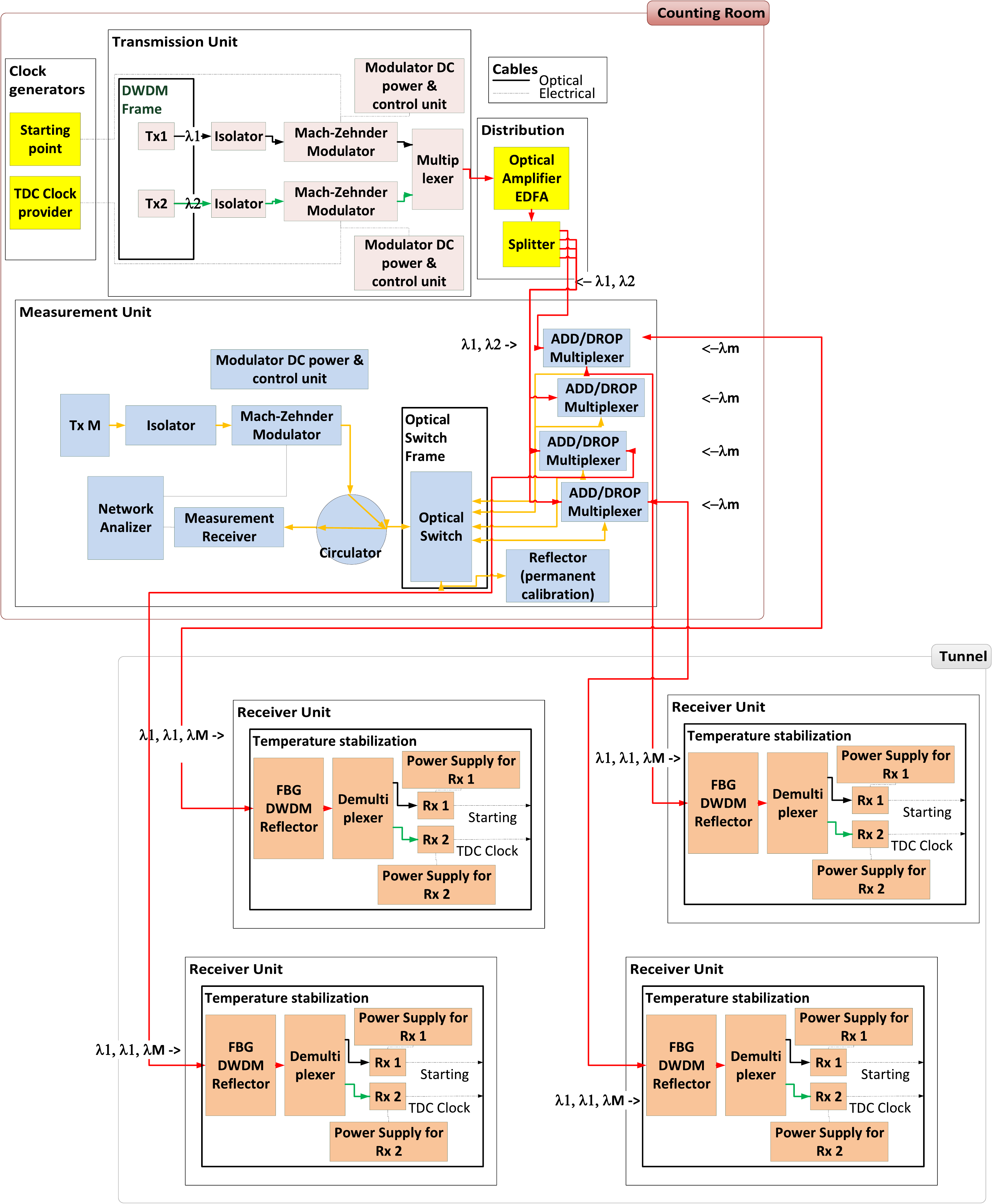}
\caption{%
    A block scheme of the clock distribution system for the PPS timing detectors.
    The four receiver units correspond each to a timing station in the tunnel; the remaining elements of the system are all located in the counting room.
    Different colors of the fiber lines represent the different wavelength carriers used by the system, $\lambda_1$, $\lambda_2$, $\lambda_\text{M}$ (black, green, and yellow, respectively), and the multiplexed signals (red).
}
\label{fig:pps:clockdistribution}
\end{figure}

The system can be logically subdivided into four major blocks:\ the
transmission unit, the distribution unit, the measurement unit, and the
receiving unit.
Receiving units are installed in the LHC tunnel, as close as possible to each timing detector, while the transmission, distribution, and measurement units are located in the IP5 counting room.
A block diagram of the entire system is shown in Fig.~\ref{fig:pps:clockdistribution}.
The system leverages the use of a dense wavelength division multiplex (DWDM) technique, exploiting the transmission of multiple signals of different wavelengths over a common single-mode fiber.
This allows the use of standard telecommunication modules compliant with ITU (International Telecommunications Union) standards.

{\tolerance=800
The transmission unit optically modulates two reference clock signals
using two different DWDM wavelength carriers:\ $\lambda_1$ and
$\lambda_2$.
These optical signals are multiplexed into a single fiber and transmitted to the distribution unit, where they are split to be distributed to all receiving units located in the tunnel on both sides.
The multiplexed signal is then optically amplified via an erbium-doped fiber amplifier (EDFA) to compensate for the attenuation due to the multiplexing and splitting steps.
\par}

These signals are further multiplexed with a third one, of wavelength
$\lambda_\text{M}$, with the goal of measuring the transmission delay over each fiber.
This is done in the measurement unit, where a network analyzer drives the optical modulation of this third DWDM signal, which is then sent to and reflected back by the receiving unit.
In this way, the signal delays can be determined and possible drifts can be monitored.

In the tunnel, receiving units separate and convert the multiplexed optical signals generated in the transmission unit back to electrical signals, and reflect, via a fiber Bragg grating reflector, the one generated in the measurement unit.
The electric signals are then routed to frontend and readout electronics.

Measurements performed with the installed system have shown an additional contribution to jitter of slightly less than 1\ps, mainly due to the inherent jitter of the clock source signal, the noise added by the optical components, and the bandwidth of the transmission system itself.

\subsection{Data acquisition and detector control}
\label{sec:pps:daqdcs}

The backend data acquisition systems for the various detectors used in PPS have been developed independently and are based on different hardware.
The strip trackers have maintained the original structure developed for TOTEM, and are integrated into the CMS data acquisition (DAQ) system, described in Section~\ref{sec:daq}.
A similar architecture is used for the timing detectors.
The pixel tracker system is based on the DAQ scheme employed by the \Phase1 upgrade of the CMS pixel detector, described in Section~\ref{sec:tracker:services}.

\subsubsection{Pixel detectors}

\begin{figure}[!ht]
\centering
\includegraphics[width=0.53\textwidth]{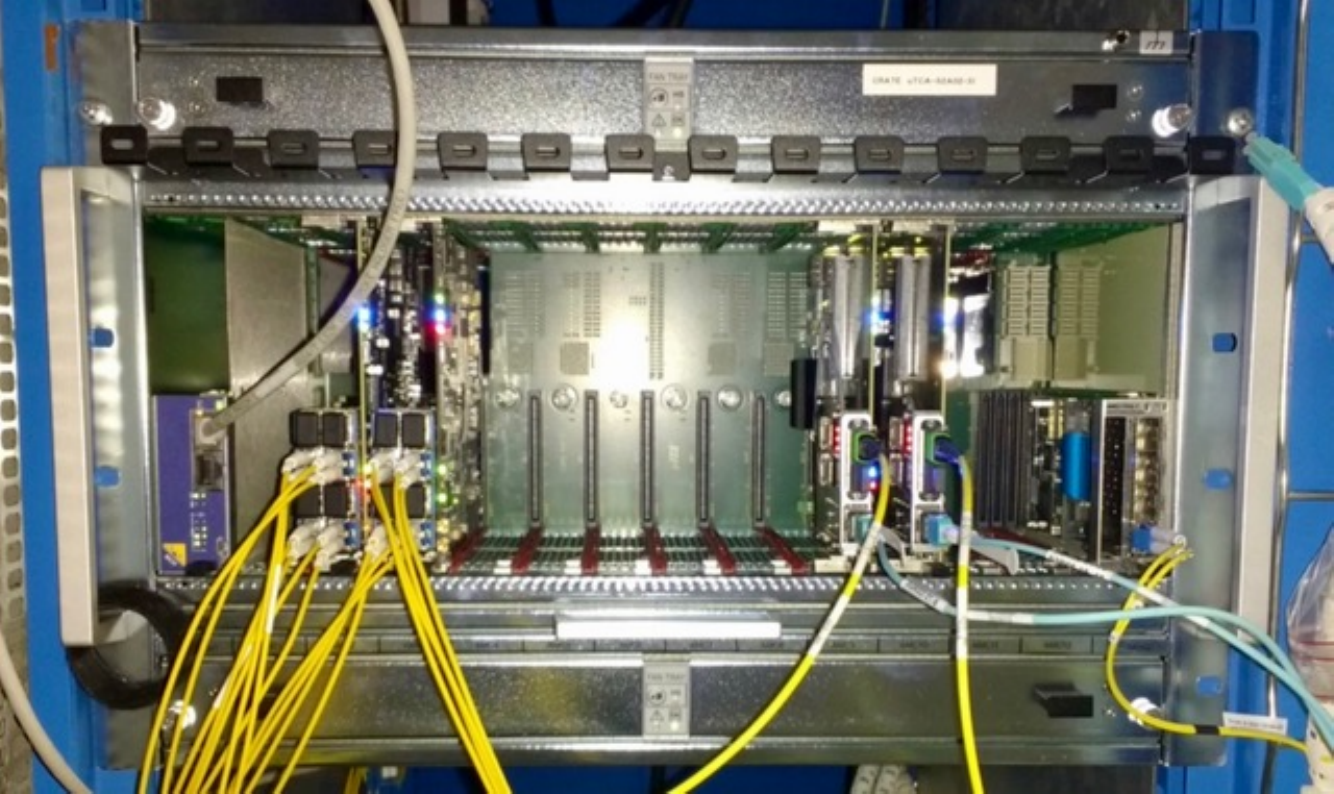}%
\hfill%
\includegraphics[width=0.42\textwidth]{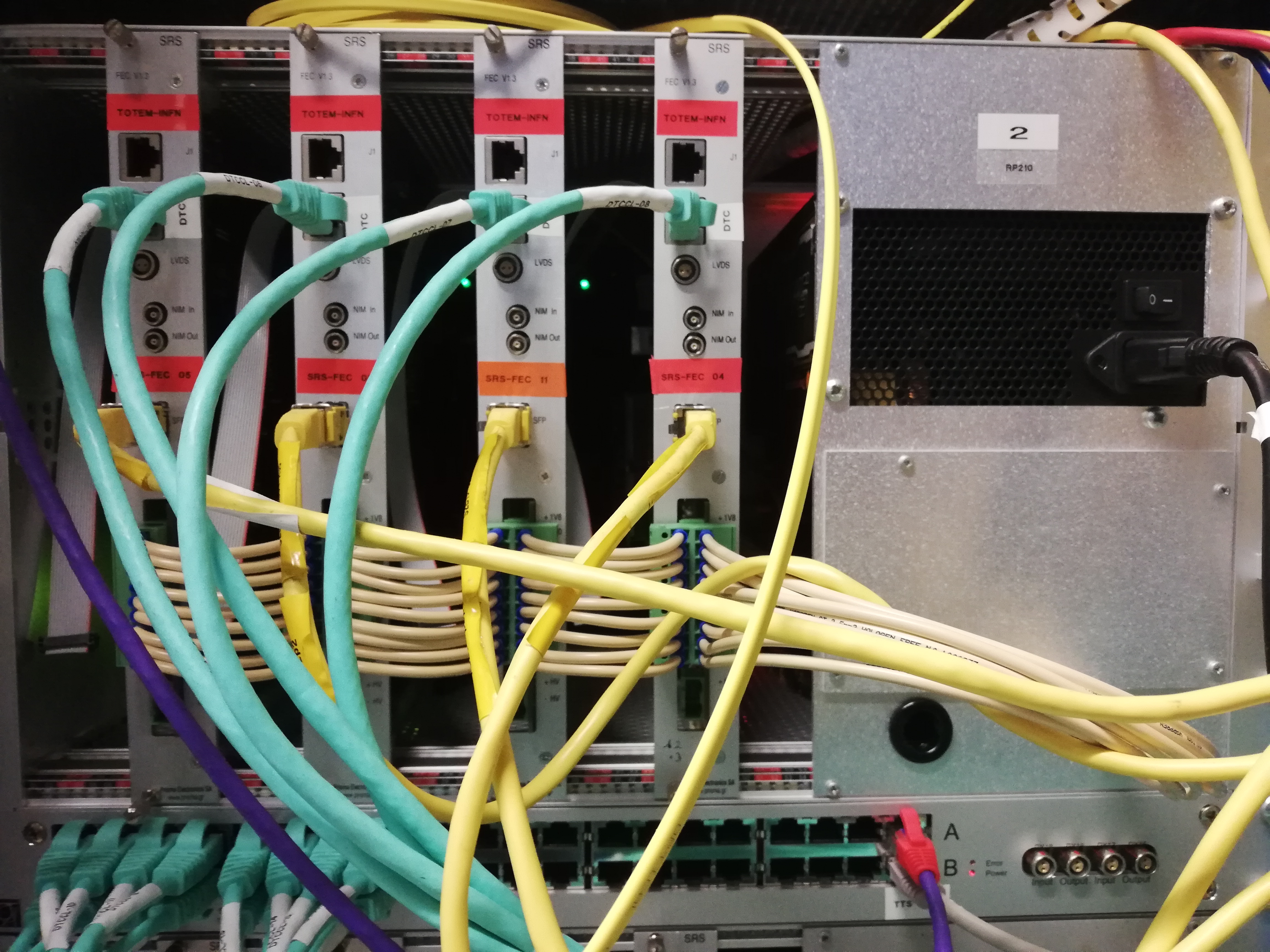}
\caption{%
    Left:\ the PPS pixel DAQ crate.
    On the left are two FC7 FECs, sending clock, trigger and slow-control commands; in the center-right are two FC7 FEDs, receiving data from the two arms of the PPS spectrometer; on the far left is the MCH crate controller, on the far right is the AMC13.
    Right:\ the timing and strips DAQ crate.
    The SLinks are placed in the backplane of the OptoRx boards delivering data to upstream CMS DAQ.
}
\label{fig:pps:daqcrates}
\end{figure}

The pixel DAQ is based on \uTCA FC7 carrier boards, following the design illustrated in Fig.~\ref{fig:pps:pixelreadout}, and described in more detail in Ref.~\cite{CMS:2019kjw}.
The boards are employed as either FEDs or FECs, depending on the type of attached mezzanine card and the firmware deployed.
Two FC7 boards are equipped as FEDs, and receive data transmitted by POHs over optical fibers from the pixel detectors on the two arms of the PPS spectrometer.
The typical data volume in \Run2 was about 0.5\unit{kB/event} or less, depending on the instantaneous luminosity and other LHC conditions, and is similar in \Run3.
Additional FC7s are equipped with FEC mezzanines, transmitting signals over fibers to the frontend boards in the LHC tunnel.
Depending on the firmware, the FECs are used as either PxFECs, sending the clock and trigger, or TkFECs, responsible for sending slow-control commands.
The entire system, shown in Fig.~\ref{fig:pps:daqcrates} (left), is housed in a single \uTCA crate, containing one AMC13 for clock and trigger distribution, and one MCH crate controller.

\subsubsection{Strip and timing detectors}

The DAQ for the PPS timing detectors, and for the strip detectors used in special alignment fills, was developed by the TOTEM Collaboration, based on a scalable readout system (SRS)~\cite{Quinto:2013zha}.
Optical receiver mezzanines (OptoRx) are connected to a custom designed board, which further connects to a frontend concentrator board.
The OptoRx receives data transmitted over fibers from GOHs on the timing digital readout units.
A scalable readout unit (SRU) is responsible for receiving and distributing the clock, trigger, and fast commands within the SRS crate (Fig.~\ref{fig:pps:daqcrates} right).
The system was designed for full compatibility with the CMS DAQ.
Data are transmitted via SLink to the FRLs of the central CMS DAQ, where they are treated in the same way as those from other FEDs.
During \Run2, a total of two VME-based FEDs were used for the timing detectors in normal data taking, while four additional such FEDs were used to read out the strip detectors in the vertical RPs for special alignment runs.
In \Run3, a total of six timing FEDs are used in standard runs, to support the addition of a second timing detector station, and the readout of a limited amount of data with a SAMPIC waveform digitizer.

The SRS firmware was originally designed to read data from the VFAT chip~\cite{Aspell:2008vfa} used by the TOTEM strip detectors.
In order to use this system with minimal changes, the timing data from the HPTDCs are packed into a VFAT-like frame, with no zero-suppression.
In \Run2, the data volume from the full timing system was around 2\unit{kB/event}.
In \Run3, this will more than double with the addition of a second timing station and the SAMPIC readout option.
The system is capable of sustaining the CMS trigger rate of more than 100\kHz.

The distribution of (non-reference) clock, trigger, and slow-control commands for the timing electronics is handled by a VME FEC, of the same type and configuration as used in the electromagnetic calorimeter preshower (Section~\ref{sec:ecal}).
The commands are transmitted over fibers to the frontend boards of the timing system.
One control loop is dedicated to the strip detectors, while another is dedicated to the timing detectors.
In addition, firmware for the FPGA of the timing digitizer motherboard can be deployed over the slow-control loop, using a new software development that takes advantage of the JTAG master of the CCU25 communication and control unit~\cite{Paillard:2002yn}.
This allows firmware updates to be made remotely, without requiring access to the LHC tunnel.

\subsubsection{Detector control system}

The PPS detector control system (DCS) is a legacy from the TOTEM experiment built with the industrial WinCC OA software (PVSS) used to control the low and high voltage of detector packages, to monitor the radiation dose, pressure, and temperature sensors, as well as the roman pots movements and the status of interlocks of the detector safety system (DSS).
The TOTEM DCS system includes the automatic matrix actions to manage the power-supply state of the roman pots according to the LHC beam status.
The system  for \Run3 is integrated within the CMS DCS framework, and runs with WinCC OA version 3.16.

\subsection{Roman pot insertion and running scenarios}
\label{sec:pps:rpinsertion}

The position of the roman pots with respect to the LHC beam is controlled by the LHC operators through standardized procedures.
During the injection, acceleration and luminosity tuning stages, the RPs are kept in a retracted (``garage'') position, at about 40\mm from the proton beam.
When ``Stable Beams'' are declared, the RPs are moved closer to the protons, at a distance depending on the beam width at that location, \sigXRP.
This distance, \dXRP, was given in \Run2 by the following rule from machine protection:
\begin{linenomath}\begin{equation}\label{eq:pps:insertionrule}
    \dXRP = \max[(\nTCT + 3) \sigXRP + 0.3\mm, 1.5\mm],
\end{equation}\end{linenomath}
where \nTCT is the distance of the tertiary collimator (TCT) from the beam center in units of the beam width \sigTCT at its location.
The $3\sigma$ retraction ensures that the RPs stay in the shadow of the TCT, while the additional 0.3\mm margin protects against accidental beam orbit deviations.
From arguments of mechanical rigidity of the RP thin window, which could bulge towards the beam in case of a secondary vacuum loss, an absolute lower limit of 1.5\mm was imposed.
In \Run2, with $\nTCT=8.5$, the resulting \dXRP ranged between 2.2\mm (210-F) and 1.5\mm (220-F).

\begin{figure}[!ht]
\centering
\includegraphics[width=0.48\textwidth]{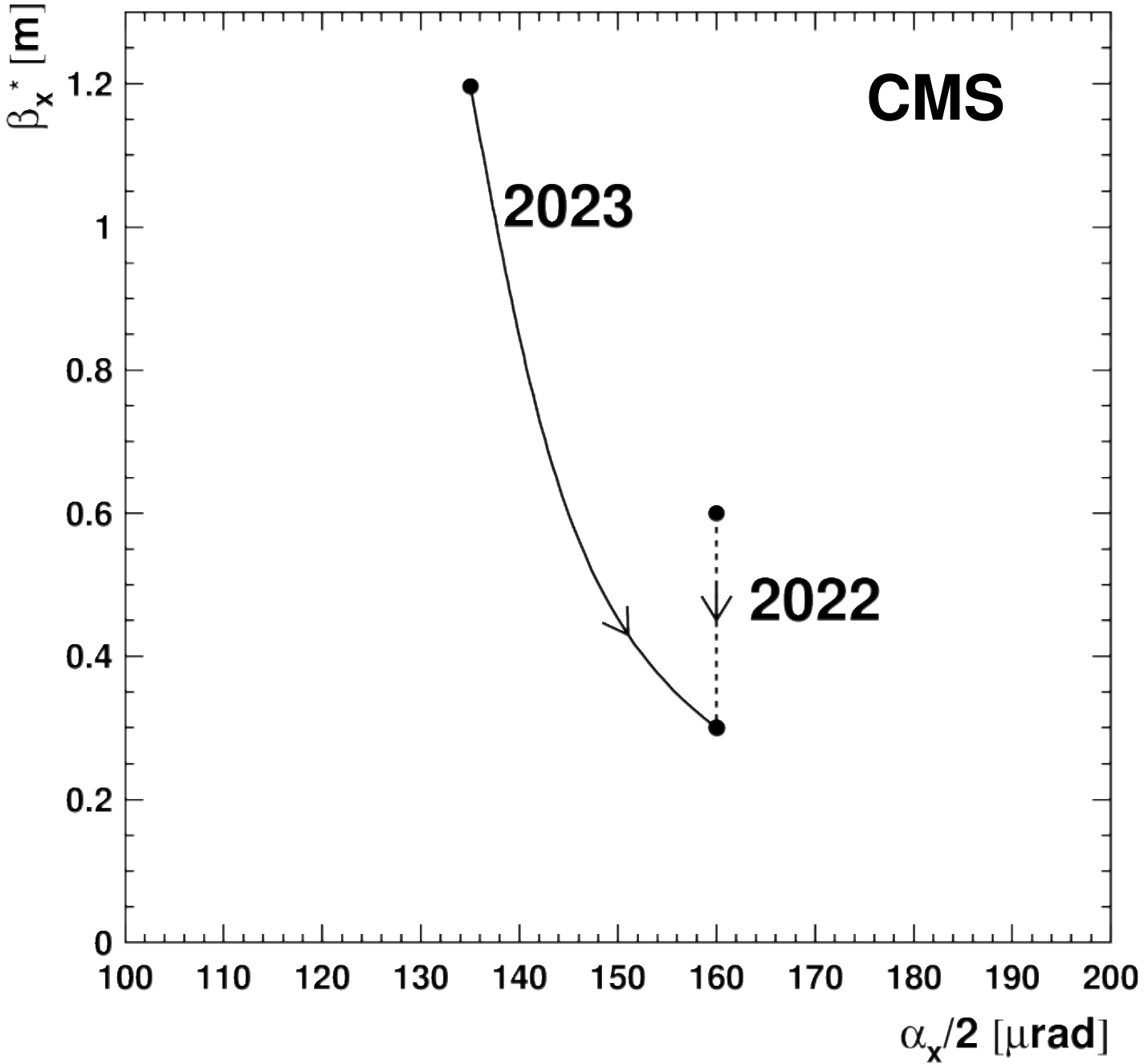}
\caption{%
    Luminosity-leveling trajectories for typical LHC fills in the $(\alpha/2, \betastar)$ plane for 2022 (dashed line) and 2023 (continuous line).
}
\label{fig:pps:levellingtrajectory}
\end{figure}

Operations in \Run3 are characterized by a far more complex luminosity-leveling scheme with concurrent changes in the crossing angle, called $\alpha$ in this section, and the beta function at the IP, \betastar.
The most recent concept of the leveling scheme can be represented by the trajectories shown in Fig.~\ref{fig:pps:levellingtrajectory} for the years 2022 and 2023.

Both $\alpha$ and \betastar, as well as the collimation scheme, have a decisive impact on operation and performance of the RP spectrometer.

\subsubsection{TCT collimator and roman pot insertion scheme}

The RP approach distance \dXRP, as given in Eq.~\eqref{eq:pps:insertionrule}, depends on \betastar through the beam width, \sigXRP, and through the TCT distance, $\nTCT=\dTCT/\sigTCT$.
The function $\dTCT(\betastar)$ characterizes the TCT collimation scheme.

\begin{figure}[!ht]
\centering
\includegraphics[width=0.48\textwidth]{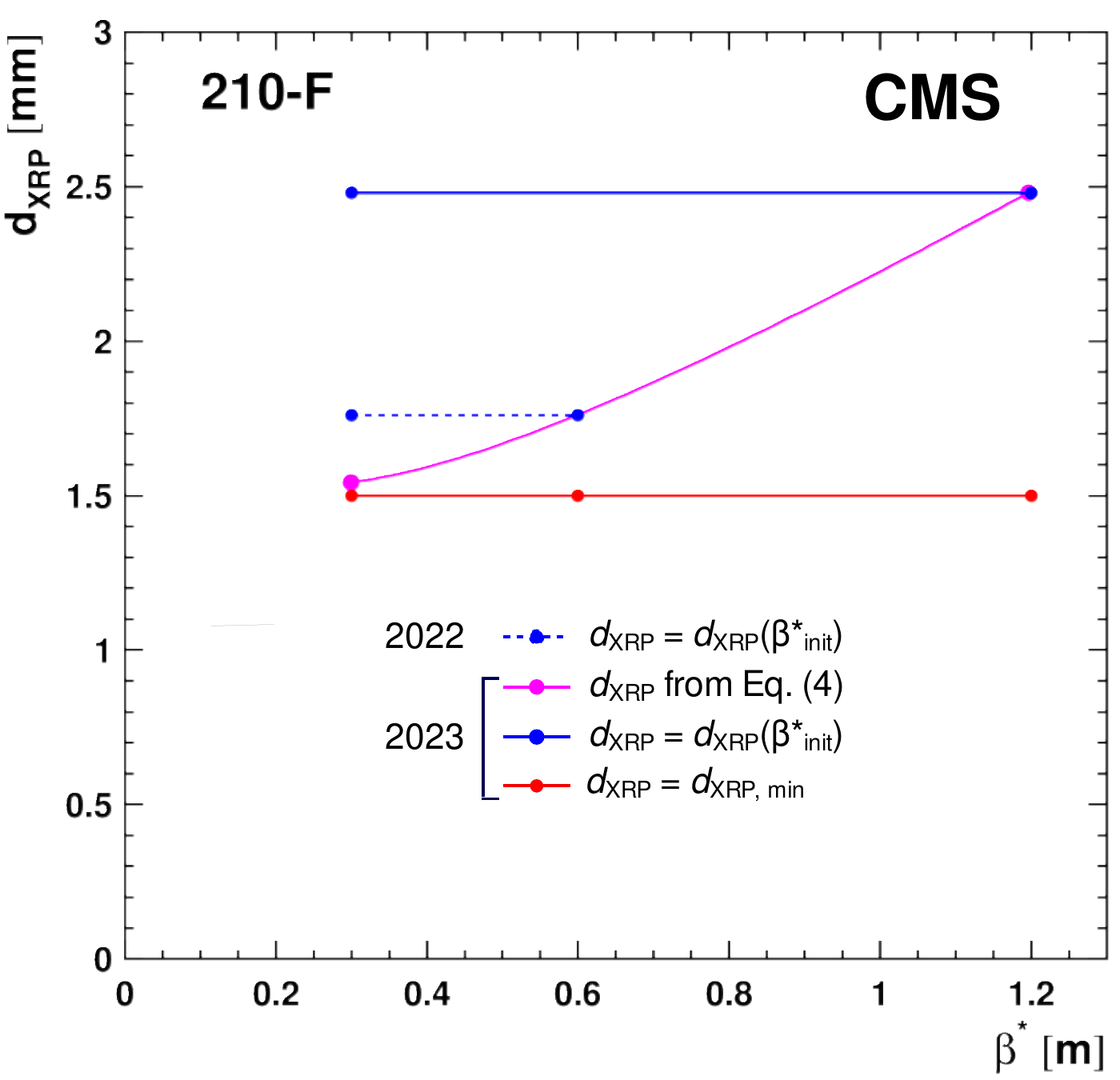}%
\hfill%
\includegraphics[width=0.48\textwidth]{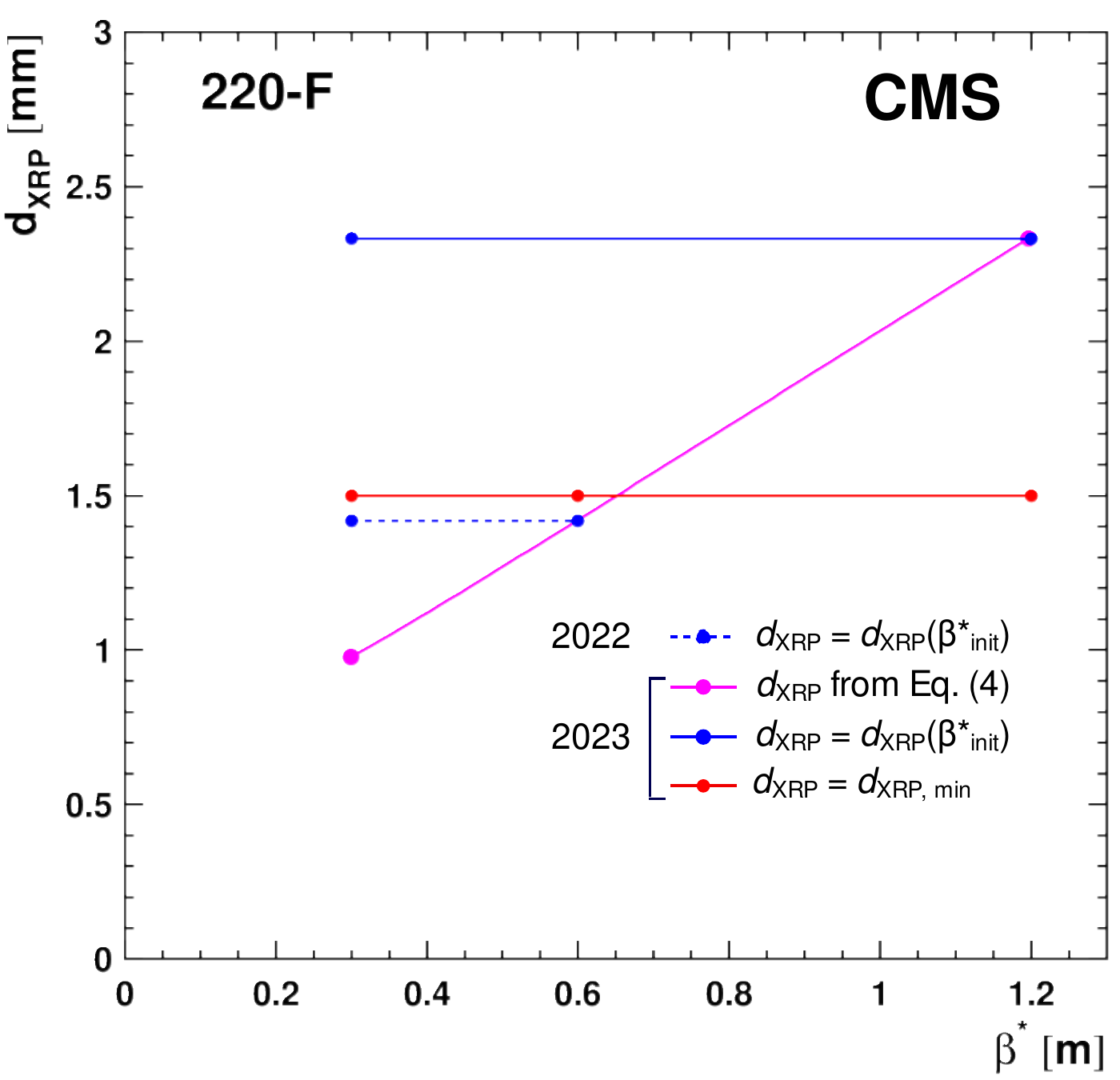}
\caption{%
    RP insertion distance \dXRP for the leveling trajectories of 2022 (dashed lines) and 2023 (continuous lines) from Fig.~\ref{fig:pps:levellingtrajectory}, evaluated for the pots 210-F (left) and 220-F (right) in the case of the TCT fixed to $\dTCT = 8.5\sigTCT(\betastar=30\cm)$.
    The magenta line shows the nominal distance according to Eq.~\eqref{eq:pps:insertionrule}.
    The blue lines show the most conservative constant RP distance, \ie, never closer to the beam than the nominal distance.
    If this blue distance is smaller than the limit of 1.5\mm (red line), which is the case for 220-F, the pot has to stay at that limit.
    The fill evolves from the right to the left.
}
\label{fig:pps:dxrpvsbeta}
\end{figure}

In the old standard collimation scheme, used until the end of 2022, the TCT did not move during stable beams, \ie, $\dTCT(\betastar)=\text{const}$, so the nominal distance \dXRP was entirely determined by the evolution of the beam widths, \sigXRP and \sigTCT, with \betastar.
With \betastar-leveling, \dXRP changes during the fill.
However, an automated RP movement synchronized with the \betastar evolution during the fill is not foreseen in the control system and would be difficult to implement, in particular because also the position limits in the interlock system would have to follow this movement.
In \Run2, this was not a problem, since the range of \betastar (25--30\cm in 2018) was so small that the variation of \dXRP was negligible.
The RPs were kept fixed at the maximum distances along their nominal trajectories.
In \Run3, the wider \betastar range complicates the situation.
The evolution of the RP distance with \betastar, for the old collimation scheme with fixed TCT positions, is shown in Fig.~\ref{fig:pps:dxrpvsbeta}, for two example RPs.

In 2022, where the \betastar leveling covered the range from 0.6 to 0.3\unit{m}, the RP with the widest nominal distance range was 210-F.
Ideally, it would have had to move by 0.22\mm.
The RP 220-F, on the other hand, could stay fixed at the limit of 1.5\mm throughout the fill.
The situation of the two other pots, 220-N and 220-C, was between these extremes, with ideal movement ranges smaller than 0.2\mm.

In 2023, the much wider \betastar range (1.2 to 0.3\unit{m}) would have led to movement amplitudes up to 1\mm.
To judge whether it is acceptable to keep the RPs fixed at the most distant points along their trajectories like in \Run2, the impact of the RP distance on the mass acceptance has to be considered.
The minimum accepted mass \Mmin of centrally produced states \PX from the process $\pp\to\Pp\PX\Pp$ with double proton detection is given by:
\begin{linenomath}\begin{equation}
    \Mmin = \sqrt{s}\,\frac{\dXRP + \delta}{\DXRP},
\end{equation}\end{linenomath}
where $\delta\approx0.5\mm$ is the insensitive margin from the outer RP window to the point of full efficiency in the detector, and $\sqrt{s}=13.6\TeV$.
The horizontal dispersion \DXRP of the LHC from IP5 to the RP depends linearly on the crossing angle, according to:
\begin{linenomath}\begin{equation}
    \DXRP(\alpha) = \DXRP(0) - D_x^\prime \alpha,
\end{equation}\end{linenomath}
with a constant $D_x^\prime > 0$, implying that smaller crossing angles yield a larger dispersion and hence a better low-mass acceptance.

\begin{figure}[!ht]
\centering
\includegraphics[width=0.48\textwidth]{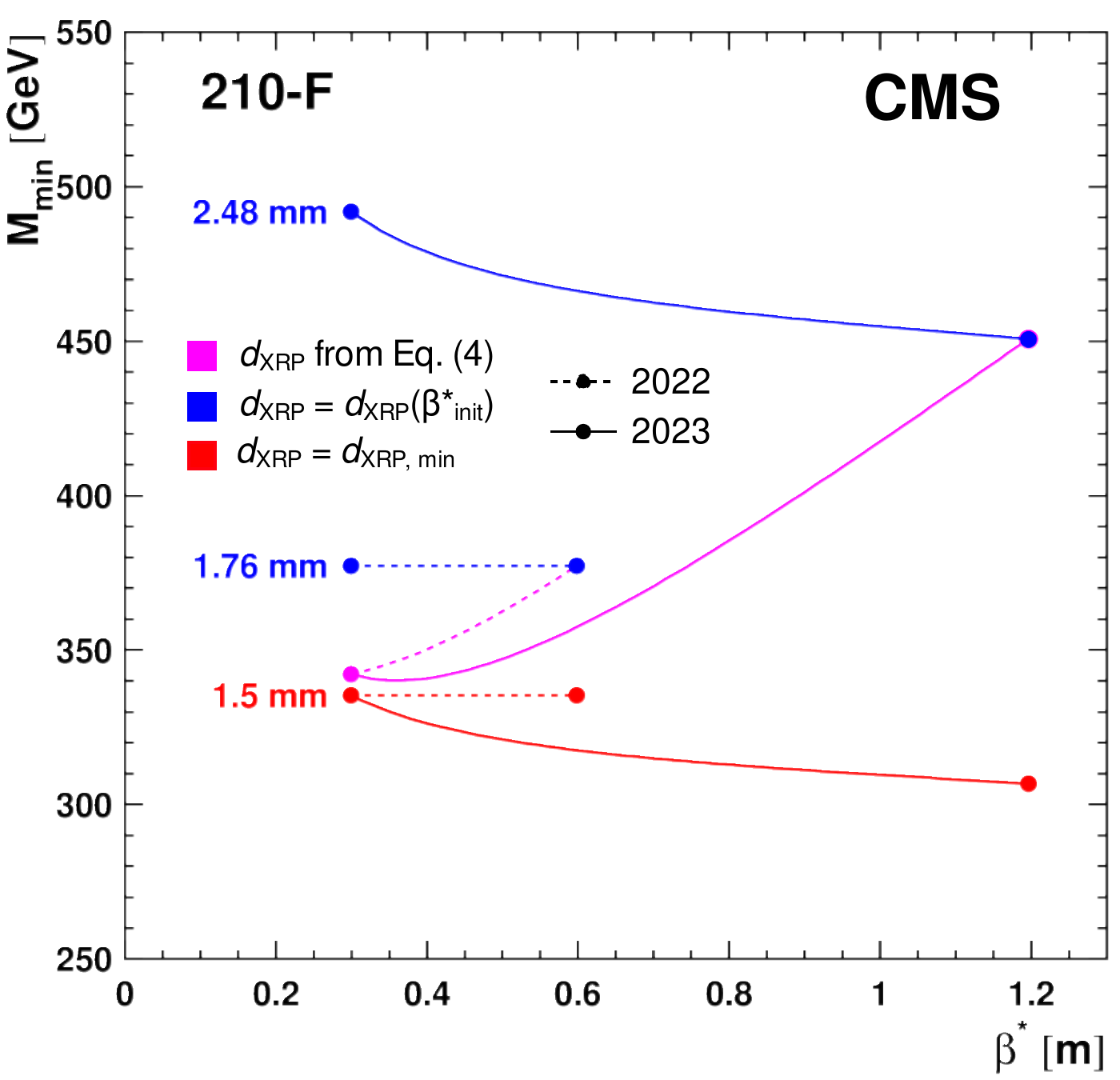}%
\hfill%
\includegraphics[width=0.48\textwidth]{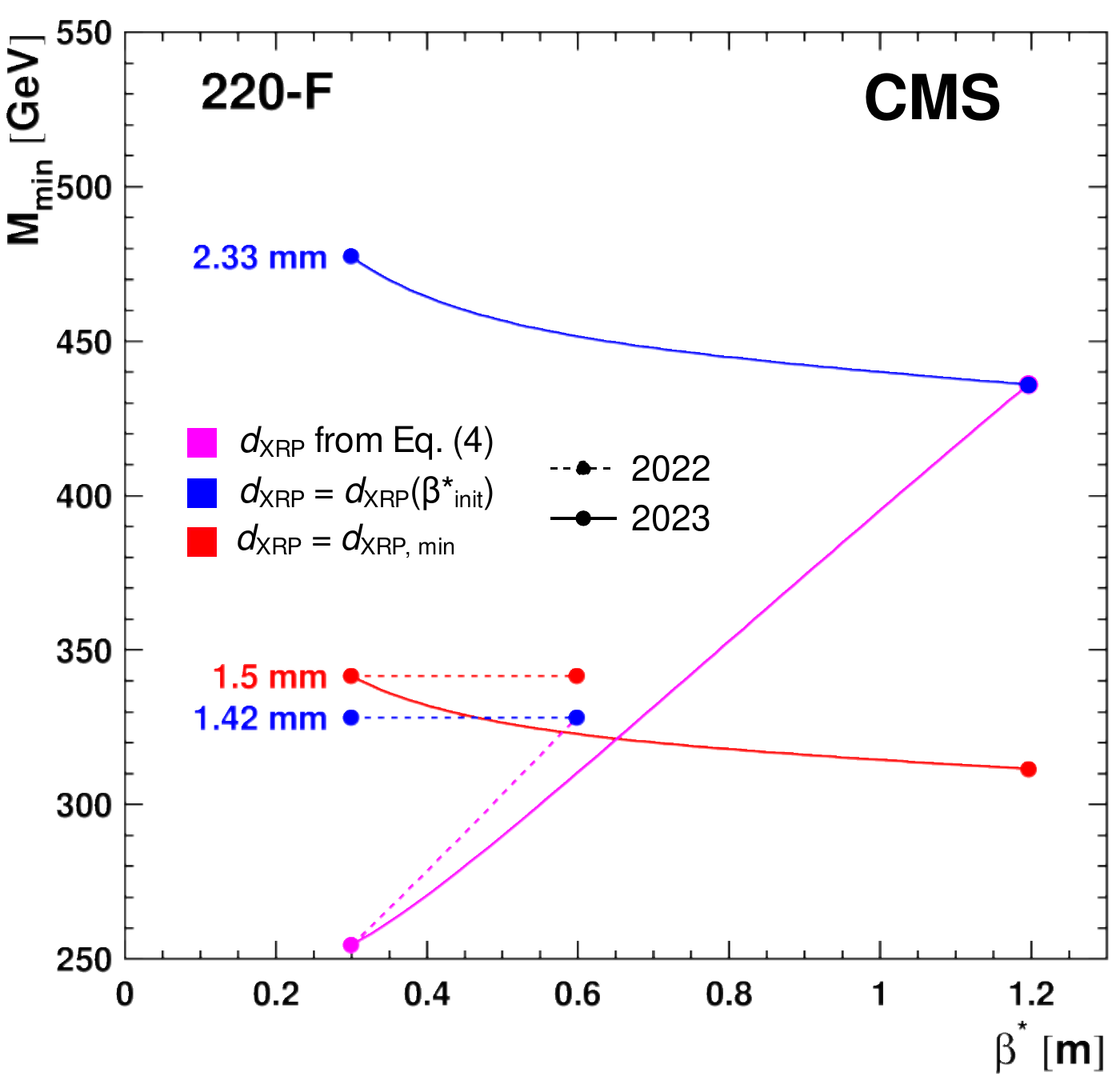}
\caption{%
    Minimum accepted central mass \Mmin in the RPs 210-F (left) and 220-F (right) for the old collimation scheme with the TCTs fixed at $\dTCT=8.5\sigTCT(\betastar=30\cm)$ and two cases for the RP positions.
    Magenta lines:\ RPs moving according to Eq.~\eqref{eq:pps:insertionrule} and Fig.~\ref{fig:pps:dxrpvsbeta}.
    Blue lines:\ RP positions fixed on the most distant point of the nominal trajectory.
    The red lines correspond to the 1.5\mm distance limit.
    The fill evolves from the right to the left.
}
\label{fig:pps:mminvsbeta}
\end{figure}

Figure~\ref{fig:pps:mminvsbeta} shows the evolution of \Mmin during \betastar leveling in 2022 and 2023 for the old collimation scheme with fixed TCTs.
The differences in \Mmin between 2022 (dashed) and 2023 (continuous) for a given \betastar are caused by the concurrent crossing-angle leveling in 2023.
The magenta lines represent the case of the RPs moving along their nominal \dXRP.
The blue lines show the effect of keeping the RPs fixed at the most distant position along their trajectories.
If a blue line crosses the limiting red line (1.5\mm distance), the RP has to be fixed at 1.5\mm, and its \Mmin follows the red line.

For 2022, the loss in \Mmin at the end of the fill, at a \betastar of 30\cm, by not following the nominal trajectory, at most 35\GeV for 210-F, was considered to be still acceptable.
Therefore, the run in 2022 proceeded along the old scenario.
Despite the slight loss from omitting the movement, the acceptance for a given $(\alpha/2, \betastar)$ point was better than in \Run2.

\begin{figure}[!htp]
\centering
\includegraphics[width=0.48\textwidth]{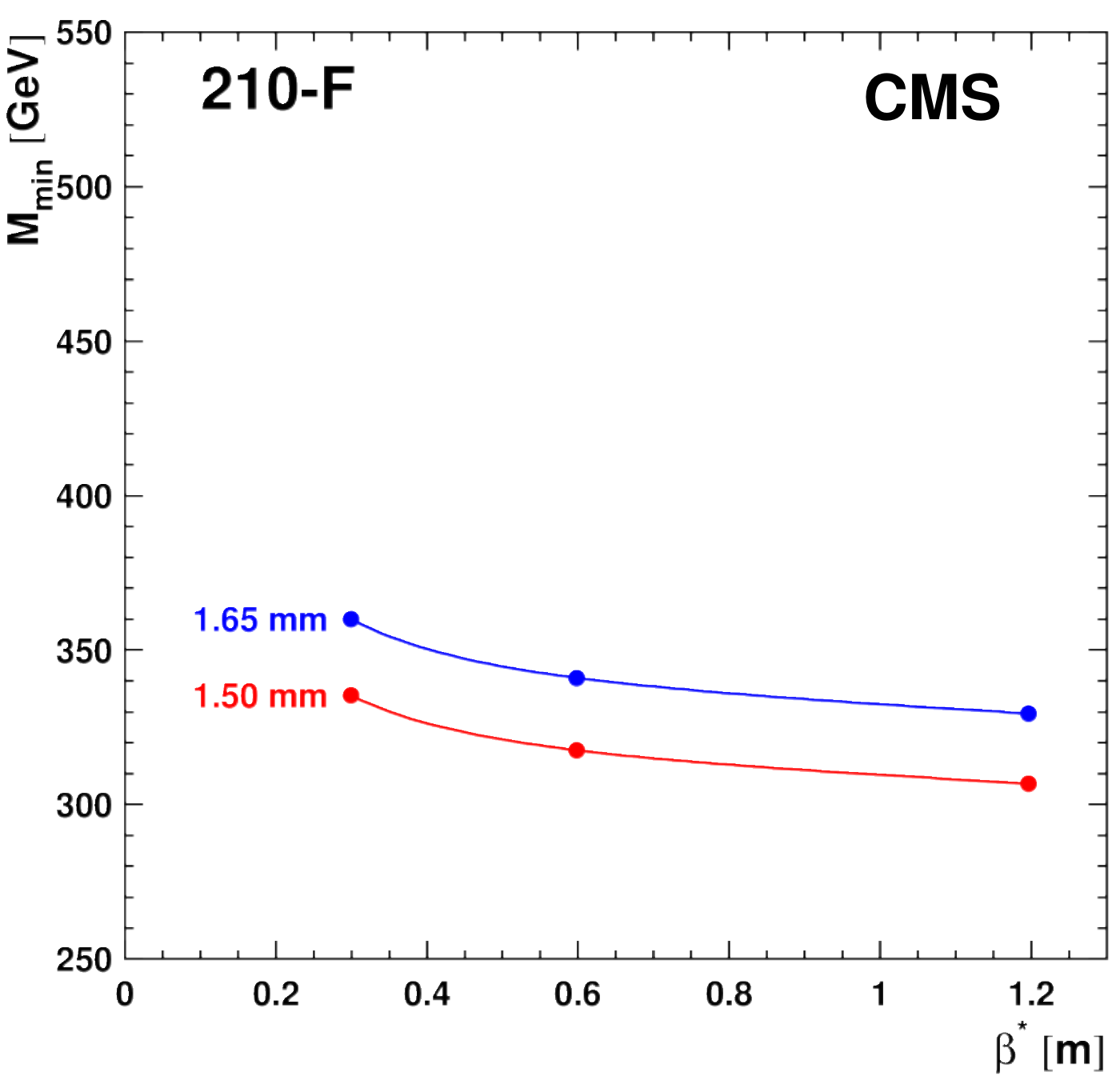}%
\hfill%
\includegraphics[width=0.48\textwidth]{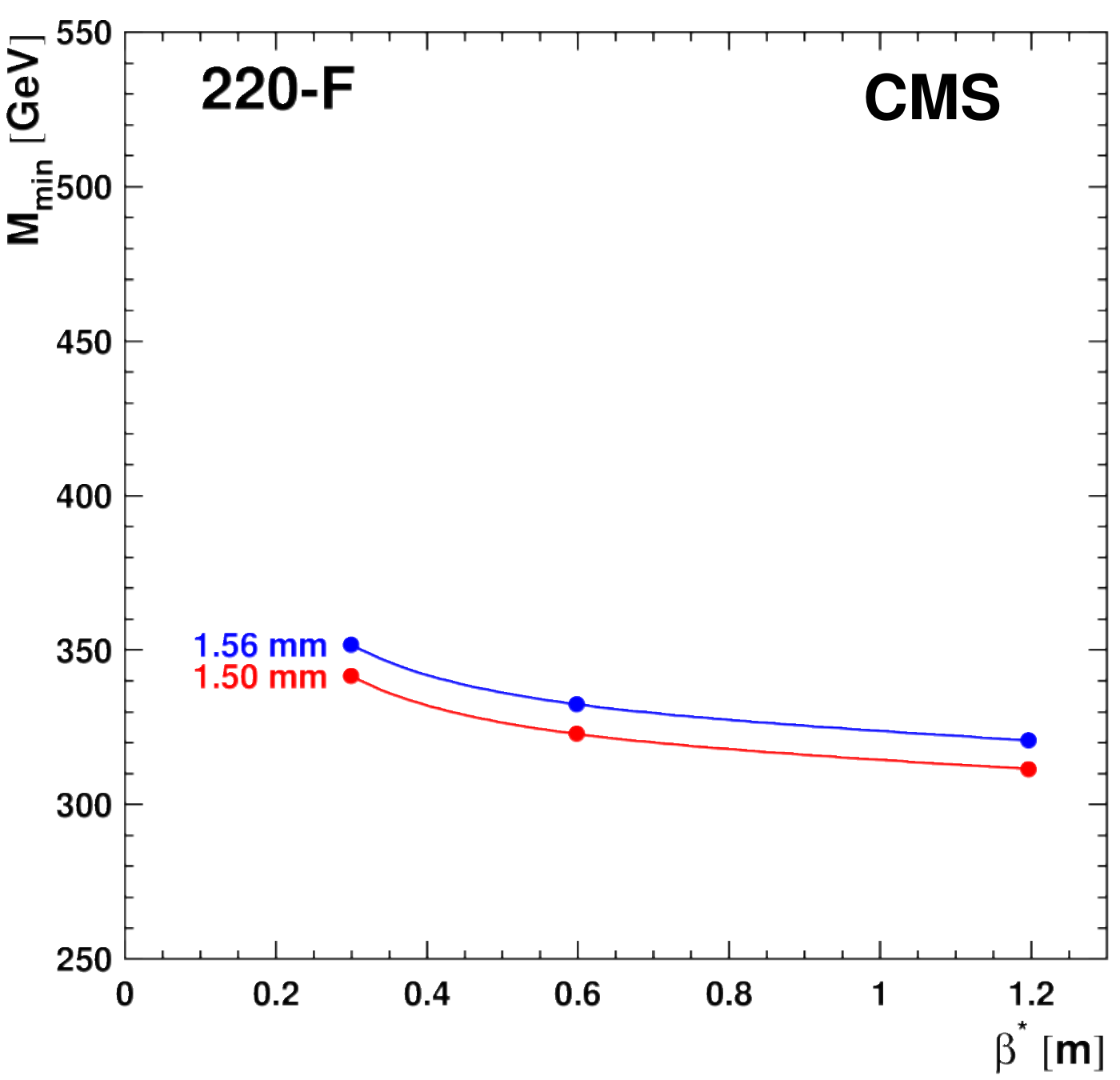}
\caption{%
    Minimum accepted central mass \Mmin in the RPs 210-F (left) and 220-F (right) for the new collimation scheme with moving TCTs and fixed RP positions.
    The red lines correspond to the 1.5\mm distance limit.
    The fill evolves from the right to the left.
}
\label{fig:pps:mminvsbetamovingtcts}
\end{figure}

For 2023, on the other hand, the \Mmin acceptance loss would have amounted to 150\unit{GeV}.
To mitigate this performance deterioration, while maintaining fixed RP positions throughout the fill, the collimation working group has elaborated an alternative scheme where the TCTs, having a more sophisticated movement control system, adapt their positions to the \betastar evolution.
In this scheme, operationally implemented in 2023, the constant RP positions are all close to the 1.5\mm limit and thus the \Mmin acceptance limits (Fig.~\ref{fig:pps:mminvsbetamovingtcts}) near the safely achievable optimum.

A quantitative mass acceptance overview for 2018, 2022, and 2023 is given in Table~\ref{tab:pps:massrange}, anticipating the upper mass limits, \Mmax, discussed in Section~\ref{sec:pps:tclsettings}.
The higher dispersion in \Run3 leads to generally lower values for both \Mmin and \Mmax than in \Run2.

\begin{table}[!htp]
\centering
\topcaption{%
    Accepted central mass range, $[\Mmin, \Mmax]$ (in {\GeVns}), for each RP at the beginning and at the end of the levelling trajectory in 2018 ($\sqrt{s}=13\TeV$), and 2022 and 2023 ($\sqrt{s}=13.6\TeV$), for the collimation schemes laid out in the text.
    Due to the coincidence requirement, the RP with the highest \Mmin (typeset in bold face) defines the spectrometer acceptance.
}
\label{tab:pps:massrange}
\renewcommand{\arraystretch}{1.1}
\begin{tabular}{ccccc}
    Year & 2018 & 2022 & \multicolumn{2}{c}{2023} \\
    $\alpha/2$ [$\mu$rad] & 160 & 160 & 135 & 160 \\
    \betastar [cm] & 30--25 & 60--30 & 120 & 30 \\ \hline
    210-F & [\textbf{449}, 2485] & [\textbf{377}, 2086] & [\textbf{329}, 1968] & [\textbf{360}, 2086] \\
    220-N & [407, 2485] & [355, 2086] & [324, 1968] & [355, 2086] \\
    220-C & [397, 2485] & [350, 2086] & [323, 1968] & [354, 2086] \\
    220-F & [347, 2485] & [342, 2086] & [321, 1968] & [352, 2086] \\
\end{tabular}
\end{table}

\subsubsection{TCL collimator insertion scheme}
\label{sec:pps:tclsettings}

The debris collimators, TCL4 and TCL5, in the outgoing beams upstream of the RPs determine the high-mass acceptance limit \Mmax of PPS.
A TCL collimator inserted to the distance \dTCL creates a mass cut-off:
\begin{linenomath}\begin{equation}
    \Mmax = \sqrt{s}\,\frac{\dTCL}{\DTCL},
\end{equation}\end{linenomath}
where \DTCL is the horizontal dispersion at the location of the TCL.
Throughout \Run2, TCL4 was inserted to a fixed distance of typically $15\sigTCL4(\betastar=30\cm)$ from the beam center.
The distance of TCL5 was chosen such that it did not introduce any tighter cut:\ $d_{\text{TCL5}}=35\sigTCL5(\betastar=30\cm)$.

\begin{figure}[!ht]
\centering
\includegraphics[width=0.48\textwidth]{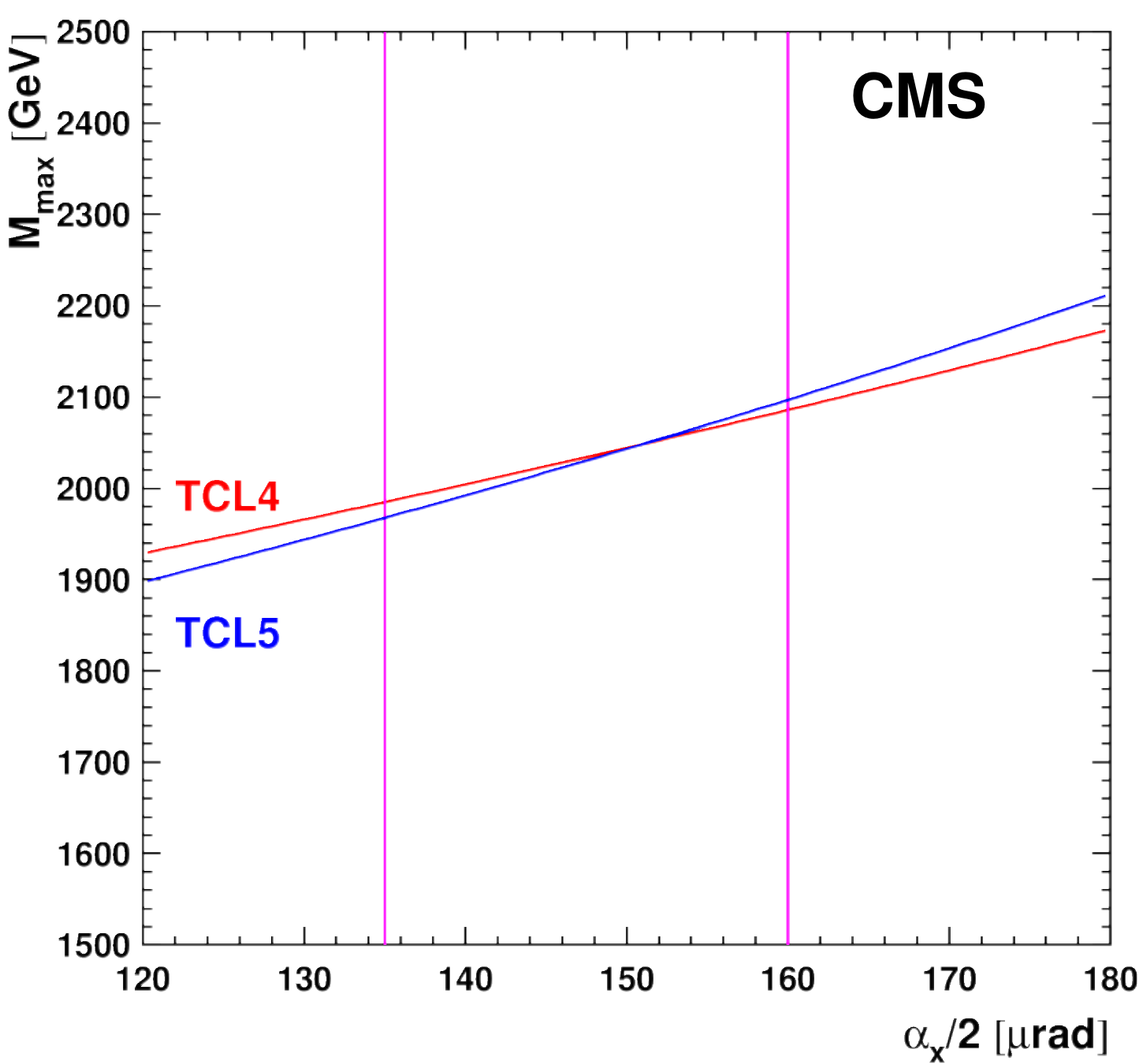}
\caption{%
    Upper mass cut-off in \Run3 caused by the debris collimators TCL4 and TCL5 for the settings explained in the text as a function of the crossing angle.
    The fill evolves from the left to the right between the magenta lines.
}
\label{fig:pps:mmaxvsalpha}
\end{figure}

In \Run3, the distance of TCL4 is unchanged with respect to \Run2.
Due to a slightly different optics, it now corresponds to $17\sigTCL4(\betastar=30\cm)$.
The TCL5 collimator is placed at $42\sigTCL5(\betastar=30\cm)$, again a position mostly in the shadow of TCL4.
Because of the crossing-angle dependence of the dispersion, \Mmax varies throughout the fill as shown in Fig.~\ref{fig:pps:mmaxvsalpha}.
The \Mmax values for 2018, 2022, and 2023 are given in Table~\ref{tab:pps:massrange}.

\clearpage
\section{Luminosity and beam conditions}
\label{sec:bril}

The main deliverables of the beam radiation instrumentation and luminosity (BRIL) system are the measurements of the online and offline luminosity, the beam-induced background, the beam losses and timing, and the radiation products in the CMS experimental cavern.
The latter are used to compare with the values from corresponding simulations.
The BRIL system delivers luminosity and beam-condition data in real time to the LHC and CMS control rooms.
In this section, the instrumentation and technologies used to carry out these measurements are described.
The software, computing, and monitoring aspects of the BRIL systems are also discussed.
The precision measurement of the integrated luminosity for the different data-taking periods of \Run2 has been described in Refs.~\cite{CMS:LUM-17-003, CMS:LUM-17-004, CMS:LUM-18-002, CMS:LUM-18-001, CMS:LUM-16-001, CMS:LUM-19-001, CMS:LUM-17-002}.
For the 2016 data, an uncertainty of 1.2\% was achieved~\cite{CMS:LUM-17-003}.
For 2017, 2018, and \Run3, improved instrumentation and refined analysis techniques are expected to lead to further reduced uncertainties.
Key ingredients of the precision are the availability of the various independently calibrated luminometers with different instrumental systematics and the in situ monitoring of their performance in terms of efficiency and linearity using short vdM-like transverse beam separation scans in physics fills.

For the online bunch-by-bunch luminosity measurement in \Run3, we rely on two fully dedicated luminosity detectors (luminometers):\ the pixel luminosity telescope (PLT) and the fast beam conditions monitor (BCM1F).
In addition, the forward hadron calorimeter (HF) is used for the online luminosity measurement.
The HF has a separate trigger-level readout dedicated to the luminosity measurement.

Both PLT and BCM1F were already operated successfully throughout \Run2.
For operation during \Run3, improved versions of these detectors were constructed and installed during LS2.
Other subsystems relevant to the luminosity measurement are the pixel detector and the muon drift tube system.
These various luminometers provide redundancy and complementarity for the precise measurement of the online and offline luminosity and the online monitoring of the backgrounds.

The three independent online luminometers, PLT, BCM1F, and HF, are based on fundamentally different technologies, and each provide precise bunch-by-bunch measurements of the instantaneous luminosity and the beam-induced backgrounds (BIB).
The BRIL online measurements are available at all times, in particular whenever there is beam in the LHC, independently of the status of other CMS subsystems and the central CMS DAQ system.
The BRIL data-acquisition system (BRILDAQ) delivers data in real time every 1.458\unit{s}, corresponding to four so-called ``lumi nibbles'' (NB).
The duration of one NB is 0.3645\unit{s}, corresponding to $2^{12}$ orbits, and 64\unit{NB} make up one luminosity section (LS) of about 23.3\unit{s}.

Data from the pixel detector and the muon systems contain important additional information that is used to ensure a precise calibration and long-term monitoring of the stability and linearity of the luminosity detectors.
Tracker data are collected via the central CMS DAQ for offline analysis, and provide pixel cluster and vertex information per event during calibration runs, and bunch-by-bunch cluster counts per LS for data-taking periods with collisions.
Muon data in \Run3 are collected via the BRILDAQ from three sources:\ (i) orbit-integrated muon track stubs from the muon barrel track finder; (ii) trigger primitives from the DT chambers that are equipped with prototype \Phase2 readout; and (iii) muon counts via the 40\MHz scouting system.
The first source is the same as that used during \Run2, while the last two are new and were introduced for \Run3 in preparation for the \Phase2 upgrade.

To monitor the beam conditions, in addition to the dedicated luminometers, we also use the beam-halo monitor (BHM) and beam-condition monitor for beam losses (BCML).

A simulation and monitoring strategy is in place for the radiation background to evaluate and, if necessary, mitigate its effects.

\subsection{Real-time bunch-by-bunch luminometers}

\subsubsection{Pixel luminosity telescope (PLT)}

\begin{figure}[!ht]
\centering
\includegraphics[width=\textwidth]{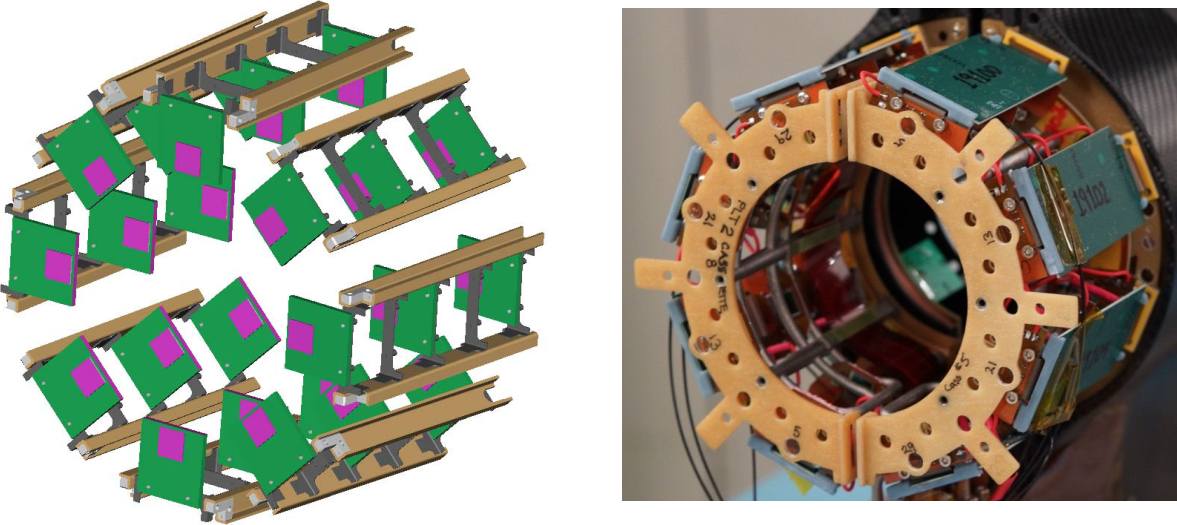}
\caption{%
    Left:\ sketch of the general PLT geometry.
    The sensors are indicated by the purple squares.
    Right:\ the actual PLT detector at one end of CMS, showing the arrangement of the eight telescopes around the beam pipe.
    Figures from Ref.~\cite{CMS:DP-2015-025}.
}
\label{fig:bril:plt}
\end{figure}

The PLT detector~\cite{Kornmayer:2016wkz, Rose:2016vok, Lujan:2017kvh, CMS:NOTE-2022-007, Karunarathna:2022tyd} is an independent system for the luminosity measurement.
It consists of 48 silicon pixel sensors arranged on 16 ``telescopes'', eight
on each side at 1.75\unit{m} from the interaction point (IP), close to the beam pipe, at a pseudorapidity $\abseta\approx4.2$.
Each telescope is 7.5\cm long and contains three planes with individual silicon sensors.
Their inner edges are as close as 4.7\cm from the beam line.
Figure~\ref{fig:bril:plt} shows a sketch of the layout of one end of the PLT (left) and a photograph of the actual detector showing the two half-frames that each side consists of (right).

The silicon sensors~\cite{Bolla:2002my, Allkofer:2007ek} are n-in-n (45 sensors) and n-in-p (3 sensors) type and consist of 80 rows and 52 columns of pixels (or 26 double columns of 160 pixels each).
The pixels are 150\mum wide and 100\mum high, and cover a total active area of 8$\times$8\mmsq with an active thickness of 285 and 150\mum, respectively.
During data taking, it is possible to set a smaller active area to decrease the contribution from accidental hits, which are triple coincidences of signals that do not actually come from a single particle track originating from the IP.
The sensors are bump-bonded to a PSI46v2 readout chip (ROC)~\cite{Kastli:2005jj, Barbero:2003rta}.
Both the n-in-n silicon sensors and the PSI46v2 ROCs are inherited from the CMS \Phase0 pixel detector, while the telescope with n-in-p sensors of 150\mum thickness uses prototype CMS \Phase2 silicon sensors.

The PLT detector is usually operated in a ``fast-or'' mode in which the data are read out at the full bunch-crossing frequency of 40\MHz.
The fast-or mode does not contain full information on hit positions.
In fast-or mode, if any pixels in a sensor register a hit over threshold, a single signal is produced.
The ROCs also have a separate data path to read out the full pixel data (hit location and pulse height) at a lower rate (up to about 3.3\kHz) upon receipt of a level-1 accept (L1A) trigger signal.
The full pixel information, read out at a lower rate, is used to measure calibration constants, corrections, and systematic uncertainties for the online and offline measurements.

The three ROCs for a single telescope are connected to a high-density interconnect (HDI) card, which contains a token bit manager (TBM) chip~\cite{Bartz:2005cds}.
The TBM chip coordinates the readout of the three individual ROCs and produces a single readout for each telescope.
Four telescopes are connected to a port card, which manages the communication and control signals for a single half-frame of the detector.
The port card is in turn connected to the opto-motherboard (OMB), which converts the electrical signals into optical signals that are then sent by fibers to the CMS service cavern, where the backend readout electronics are located.
Of the different kinds of application-specific integrated circuits (ASICs) used in the PLT, all but two (SlowHub and PLTdriver) were developed for the \Phase0 CMS pixel detector~\cite{CMS:Detector-2008}.
The SlowHub chip, on the OMB, uses blocks from the pixel TBM chip to extract the slow \ItwoC signals from the fast 40\MHz commands needed for the pixel readout, thus significantly reducing the total number of fibers needed.
The PLTdriver chip manages the fast-or signal outputs from the individual ROCs.

The backend readout electronics comprise a single frontend controller (FEC) card, which issues commands to the ROCs, TBMs, and OMB, and three frontend driver (FED) cards.
One FED is used to read out and decode the full pixel data from the ROCs, and is identical to the FEDs used by the \Phase0 pixel detector~\cite{Pernicka:2007cds}.
This data are then read out over an Slink~\cite{vanderBij:1997rc} connection and saved to a dedicated PC.
The other two are the fast-or FEDs, one for each side of the PLT.
These read out the fast-or data from the ROCs and histogram the number of triple coincidences per channel and per bunch crossing (BX).
These data are read out via a CAEN VME optical bridge to a separate dedicated PC.
A dedicated trigger is used, in which events from colliding BXs are read out without a requirement for specific activity in the event, this way equally sampling all colliding BXs in the LHC orbit.

The readout hardware counts the number of triple coincidences, events where all three planes in a telescope register a signal, to determine the instantaneous luminosity.
This fast-or readout allows the PLT to provide online per-bunch luminosity with excellent statistical precision, with the triple coincidence requirement providing a strong suppression of background from noise, residual radioactivity of the detector material (afterglow), and BIB sources from beam-gas interactions and beam halo.
The luminosity measurement is obtained using a zero-counting technique, described, \eg, in Ref.~\cite{CMS:LUM-17-003}, in which the actual rate is inferred from the measured rate of no hits, thereby correcting for the overlap of signals.

\subsubsection{Fast beam conditions monitor (BCM1F)}

The BCM1F detector is a fast particle counter installed around the beam pipe, $\pm$1.8\unit{m} away from the IP, at a radius of about 6\cm, corresponding to $\abseta\approx4.1$.
The BCM1F provides a real-time measurement of both luminosity and beam-induced backgrounds.

The experience gained from BCM1F operation in \Run2~\cite{CMS:2016tsw, Guthoff:2019cil} led to an update of the design for \Run3~\cite{Wanczyk:2022avs}.
For \Run3, the mix of silicon and poly-crystalline (pCVD) diamond sensors used in \Run2 was changed to an all-silicon sensor configuration with active cooling that has an improved stability with increasing integrated fluence and an improved linear response of the sensors as a function of the instantaneous luminosity.

\begin{figure}[!ht]
\centering
\includegraphics[width=0.569\textwidth]{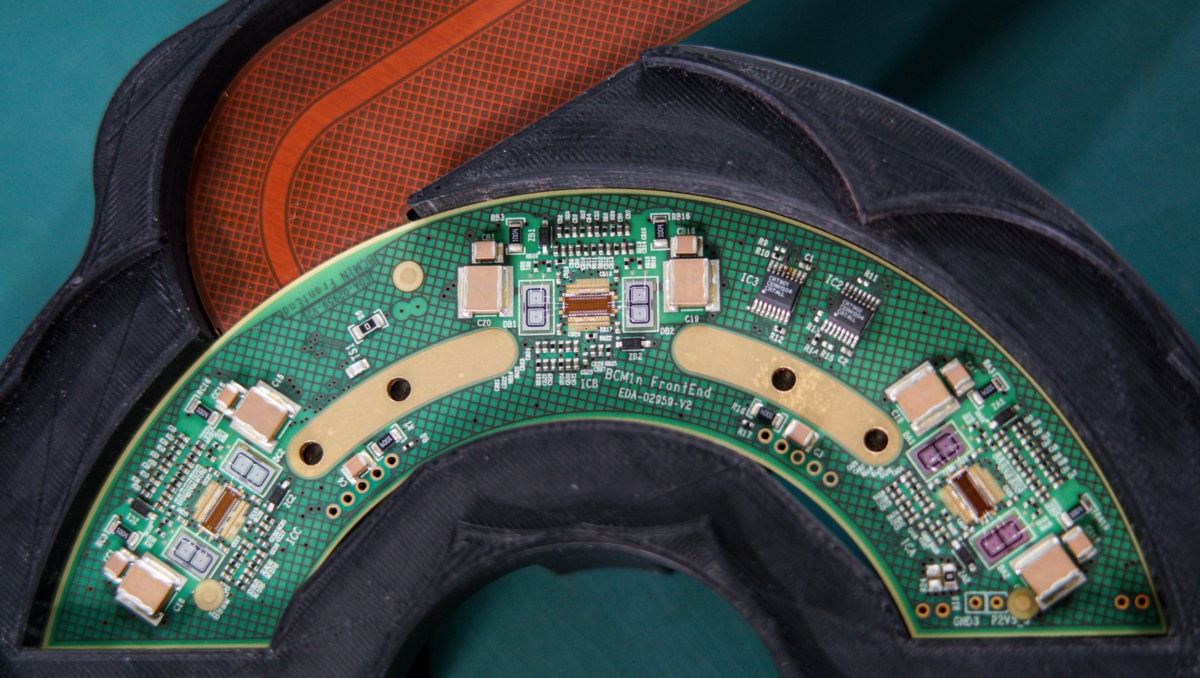}%
\hfill%
\includegraphics[width=0.39\textwidth]{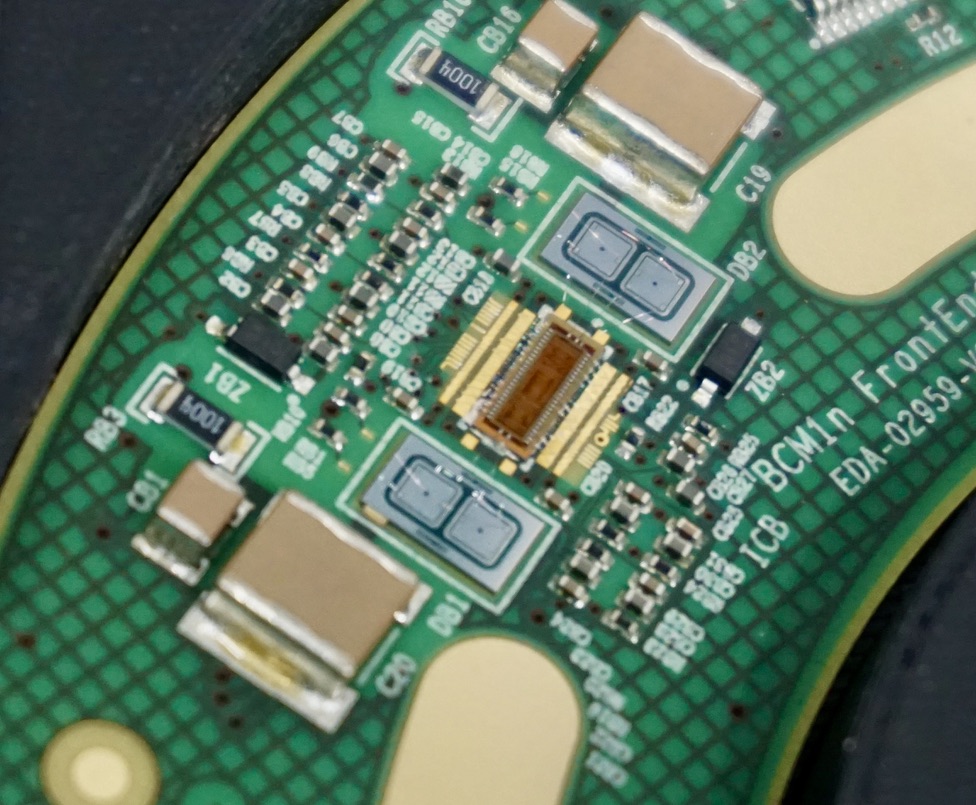}
\caption{%
    Photographs of the BCM1F detector used in \Run3, from Ref.~\cite{CMS:PHO-GEN-2022-001}.
    Left:\ one full BCM1F C-shape printed circuit board.
    Right:\ a closeup of the frontend module, which includes the processing chip with two silicon double-diode sensors on each side.
}
\label{fig:bril:bcm1f}
\end{figure}

As in \Run2, the \Run3 detector consists of four ``C-shaped'' printed circuit boards (PCB), referred to as C-shapes (Fig.~\ref{fig:bril:bcm1f}).
A total of four C-shapes form two rings, one on each side of the IP.
Each C-shape has six double-diode silicon sensors that are used to measure particle hits.
The entire system comprises 48 identical channels.
The C-shapes for \Run3 were designed for improved robustness and with additional space for active cooling contacts.

Performance studies led to the choice of a sensor per-channel with an area of 1.7$\times$1.7\mmsq, balancing the occupancy at high interaction rates with the statistical precision at low luminosity.
The sensors were designed as alternating current (AC)-coupled double-diodes and produced on the CMS \Phase2 outer tracker strip wafers~\cite{CMS:TDR-014} with a full depletion thickness of 290\mum.

Each sensor is directly connected via bond wires to one of the three fast JK16~\cite{Przyborowski:2016aco} CMS ASIC chips.
This preamplifier chip was designed to have a fast rise time of 7\ns, narrow pulses with 10\ns full width at half maximum, and a return to the baseline within 25\ns.
The connections are established via AC coupling to prevent baseline drifts due to radiation-damage-induced leakage currents that would not be compensated by the preamplifier.
The coupling capacitance is part of the sensor design, while a resistor on the PCB is used as the bias resistor.

Copper pads were added on the C-shapes for \Run3 to make the cooling contact with a new titanium 3D-printed cooling pipe, which is placed in series with the PLT cooling loop.
This improves the thermal contact of the sensors to the cooling system and prevents thermal runaway induced by radiation damage, thereby ensuring the expected lifetime of the BCM1F silicon sensors exceeds the duration of \Run3.

{\tolerance=800
Using analog opto-hybrids~\cite{Friedl:2004edms}, the AC signals from the C-shapes are propagated to the backend electronics that includes CAEN V895 discriminators, real-time histogramming units (RHU)~\cite{Penno:2013pubdb}, and a CAEN V1721 flash ADC, all of which use a VME bus.
Parallel \uTCA electronics consist of an 8-bit ADC mezzanine card with a 1.25\GBs sampling rate on a gigabit link interface board (GLIB) carrier board~\cite{Zagozdzinska:2014rqa, Zagozdzinska:2016clv}, using an FPGA-based derivative peak-finder algorithm~\cite{Ruede:2018cds}.
At the RHU, which was already used for \Run2, the data are filled into histograms with 6.25\ns granularity.
This time interval corresponds to the time of flight for relativistic particles arriving from the IP that is a quarter of the time between two collisions.
\par}

{\tolerance=800
The BCM1F detector, situated in this location and delivering a fast response with zero dead time, allows for the bunch-by-bunch separation of collision signals from backgrounds.
The luminosity is measured bunch-by-bunch using the zero-counting method mentioned above, followed by a correction for out-of-time hits and the application of a calibration constant.
The relative contribution from background can be determined using the first bunch in the train or a noncolliding bunch in the beginning of the orbit, where the background from out-of-time hits due to activation from previous collisions is small.
The background measurement is of crucial importance for the operation of the CMS detector since it is used by the silicon strip and pixel detectors as an automatic switch-on semaphore.
\par}

\subsubsection{The forward hadron calorimeter (HF)}

A comprehensive description of the HF is provided in Section~\ref{sec:hcal}.
Here, we highlight the aspects that are relevant for the luminosity measurement.

Like the PLT and BCM1F, the HF reports luminosity measurements at all times, independently of whether the main CMS DAQ system is in operation.
The outputs from the circuits used to digitize the signals from the HF photomultipliers (PMTs) are routed to FPGAs that are part of the HF readout~\cite{CMS:TDR-010}.
The FPGAs tap into the primary readout path in a noninvasive way and collect channel-occupancy (OC) and transverse-energy-sum data in histograms that have one bin for each of the 3564 BX time windows of one beam orbit.
The histograms are periodically transmitted to a central BRILDAQ processor node.
With this configuration, the HF delivers a continuous and precise measurement of the bunch-by-bunch luminosity at all times.

While all HF channels are available for use in the BRIL backend electronics, Monte Carlo simulation studies indicate that the best linearity of the luminosity measurement is obtained when using only the $\eta$ rings 31 and 32, corresponding to a pseudorapidity range $3.15<\abseta<3.5$, where the long-term radiation damage is predicted to be the smallest.

The HF luminosity system allows the extraction of a real-time instantaneous luminosity relying on two algorithms.
The first (HFOC) is based on zero counting, in which the average fraction of below threshold towers is used to infer the mean number of interactions per BX.
The second (HFET) exploits the linear relationship between the average transverse energy per tower and the luminosity.
Both the HFOC and HFET methods exhibit excellent linearity, based on the experience acquired during \Run2~\cite{CMS:LUM-17-003, CMS:LUM-17-004, CMS:LUM-18-002, CMS:LUM-19-001, CMS:LUM-18-002} and in the beginning of \Run3~\cite{CMS:DP-2022-038}.

\subsection{Additional luminometers}

\subsubsection{Tracker}

The CMS silicon tracker, described in Section~\ref{sec:tracker}, is characterized by a low occupancy in the silicon pixel detector.
This feature is exploited by the pixel cluster counting (PCC) method to measure the luminosity offline with excellent precision.
The most recent PCC luminosity measurement during \Run2~\cite{CMS:LUM-18-002} was based on data collected by zero-bias (colliding bunches only) and random triggers at recording rates of approximately 2\kHz and 400\unit{Hz}, respectively.
The zero-bias data were used for the luminosity calculation, while random-triggered events served for the determination and subtraction of the afterglow backgrounds present in the zero-bias data.
The method achieved a statistical precision of about 0.2\% per luminosity section during physics runs, with excellent stability after removing problematic sensor modules.

We applied a similar strategy for \Run3, with improvements in the data processing and recording to reduce and streamline the data.
The cluster reconstruction and counting is implemented in the HLT software, which reduces the data size by three orders of magnitude.
Potentially unstable modules are removed using an iterative method in which modules with large fluctuations relative to the average cluster count are removed.
The implementation of a new processing path is in progress during \Run3, in preparation for the \Phase2 upgrade of the luminosity instrumentation for the HL-LHC~\cite{CMS:NOTE-2019-008}.
In this path, the data from the HLT are transferred directly to the BRILDAQ.

An alternative method using information from the silicon tracker involves counting the number of primary-interaction vertices~\cite{CMS:LUM-17-003}.
This method is used during van-der-Meer (vdM) calibration fills as a tool in the measurement of beam-dependent parameters.
Vertices are also analyzed to derive the bunch density distributions of the beams, and to calibrate the distance by which the steering magnets displace the beams in the transverse direction.

\subsubsection{Muon system}

The BRIL system also makes use of level-1 trigger information from the muon barrel (MB) drift tube (DT) detectors, which deliver counts of the number of orbit-integrated muon stubs.
While the rate of muons in the barrel is low, the stability and robustness of the muon system makes the DT a valuable source for luminosity monitoring.
The muon detectors and trigger are described in Sections~\ref{sec:muon} and~\ref{sec:l1trigger:muon}, respectively.
The muon information used by BRIL is aggregated by the barrel muon track-finder algorithm (BMTF)~\cite{CMS:2017woh} of the level-1 muon trigger.
The BRILDAQ system receives the data from the BMTF via a dedicated readout system.

In addition to the orbit-integrated data, during \Run3 a new histogramming firmware module was added to the backend of the DT slice test that uses \Phase2 readout technology to test how trigger primitives (muon segments per DT chamber) can be used for bunch-by-bunch luminosity measurements.
This system serves as a demonstration of how a luminosity measurement can be made with per-bunch granularity for the CMS \Phase2 upgrade.

\begin{figure}[!ht]
\centering
\includegraphics[width=\textwidth]{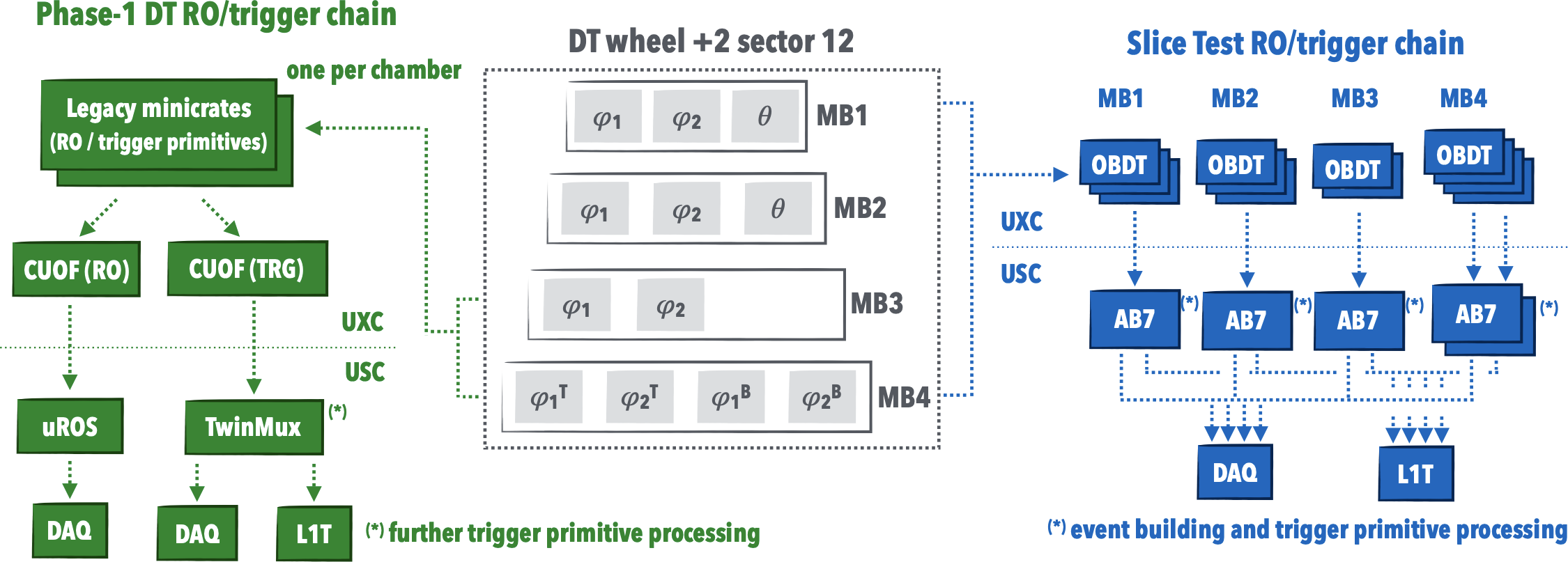}
\caption{%
    Diagram comparing the DT readout (RO) and trigger chains in the \Phase1 system used for data taking in \Run3 and in the DT slice test that acts as a demonstrator of the \Phase2 system in \Run3~\cite{CMS:TDR-021}.
    The central part of the figure indicates the number of \Phase2 on-board DT (OBDT) boards installed in each DT chamber.
    The left part of the figure shows how the \Phase1 readout and trigger-primitive generation are performed by the legacy on-board DT electronics (minicrates).
    Information, transmitted by optical fibers from the detector balconies to the counting room, is further processed independently for the readout (\uROS) and trigger (TwinMux).
    The right part of the figure illustrates how the slice test TDC data are streamed by each OBDT to AB7 boards hosted in the counting room, which are used for event building and trigger-primitive generation.
}
\label{fig:bril:dtrun3demonstrator}
\end{figure}

Figure~\ref{fig:bril:dtrun3demonstrator} illustrates the readout diagram of the DT slice test.
A detailed description is provided in Ref.~\cite{CMS:TDR-021}.
The MB1 and MB2 chambers are read out using the upgraded electronics.
Data duplication was established for the MB3 and MB4 layers, thus allowing us to evaluate the performance of the upgraded electronics with respect to the older system.
The BRIL histogramming module is placed on the so-called AB7 backend boards.
Each AB7 board produces a single histogram.
Each of the MB1, MB2, and MB3 chambers are read out by a single AB7, whereas two AB7 boards are needed to read out the MB4 chamber.

In addition, the 40\MHz scouting system in \Run3, described in detail in Section~\ref{sec:l1trigger:scouting}, provides an alternative means to use the bunch-by-bunch muon rate from the level-1 muon trigger for the luminosity measurements.
The concept of the 40\MHz scouting system is to capture part or all of the trigger data streams from different trigger layers or their subsystems using spare optical outputs.
A demonstration of the scouting system was already tested in \Run2 based on the output from the global muon trigger algorithm \uGMT described in Section~\ref{sec:l1trigger:muon}.
Due to the inclusion of the layer-2 calorimeter trigger and the BMTF data, the \Run3 scouting demonstration offers additional objects with more sophisticated customizable selection possibilities.
The histogramming for measuring the luminosity is performed per object type using the already mentioned generic histogramming firmware loaded on the scouting board.

The histogramming firmware is a generic module developed by BRIL that targets luminosity measurements using various, primarily \Phase2, CMS subsystems.
The DT muon detector and the 40\MHz scouting system are the first to use this new module, and multiple instances of the module are placed on backend FPGAs.
The module comes in two flavors:\ a synchronous type is adapted to the bunch-by-bunch luminosity measurements, when the counted objects arrive synchronously with the bunch clock, as is the case of the DT system.
An asynchronous flavor is used for the 40\MHz scouting system in which preprocessing of the trigger primitives is performed at a rate of 250\MHz, providing histogram input approximately every six clock periods.

The module is implemented for the Xilinx 7-series and newer FPGAs.
The readout of the histograms can be performed using any suitable data transfer protocol.
In particular, the IPbus and DMA readout interfaces have been implemented for the DT and 40\MHz scouting systems, respectively.

\subsubsection{\PZ boson counting}

Due to a clean signature and a relatively high cross section, the $\PZ\to\MM$ process is of particular interest for the luminosity measurement.
It provides an alternative and complementary method to transfer integrated luminosity measurements between data sets, as discussed in Refs.~\cite{Salfeld-Nebgen:2018bdp, CMS:LUM-21-001}.
In \Run3, \PZ boson counting is implemented as part of the data reproduction chain with the capability of providing precise results within about one week of the data taking.
A full initial analysis of the 2022 data has been performed~\cite{CMS:DP-2023-003} and used to confirm the early calibration of the luminometers by comparison with the predicted cross section~\cite{CMS:TOP-22-012}.
The \PZ boson rates are also used as an additional method to directly compare the integrated luminosities delivered to the ATLAS and CMS experiments~\cite{CMS:DP-2012-014, Schwick:2019mkt}.

\subsection{Beam monitoring instrumentation}

\subsubsection{Beam-halo monitor (BHM)}

To monitor the BIB at high radius, an efficient detection system was installed during LS1 and operated in 2015--2016.
It was then recommissioned for \Run3.
The BHM monitors~\cite{Orfanelli:2015qgz} are composed of a total of forty cylindrical quartz Cherenkov detectors, twenty on each end of the CMS detector, mounted on the rotating shield.
The rotating shield is situated at a distance of about 20\unit{m} from the interaction point.
The exact location was chosen to maximize the time difference between the arrival of BIB and particles from the \pp collisions.
The BHMs are mounted on the outer radius of the shield, at a distance of 1.83\unit{m} from the beam line.
The units are uniformly distributed in $\phi$, starting from $\phi=-30\de$ and continuing through $\phi=210\de$.

\begin{figure}[!ht]
\centering
\includegraphics[width=0.4\textwidth]{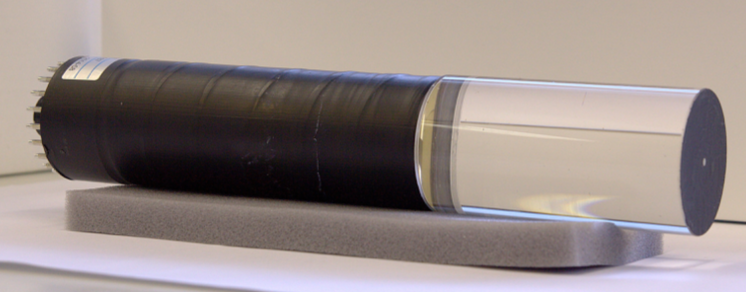}%
\hfill%
\includegraphics[width=0.56\textwidth]{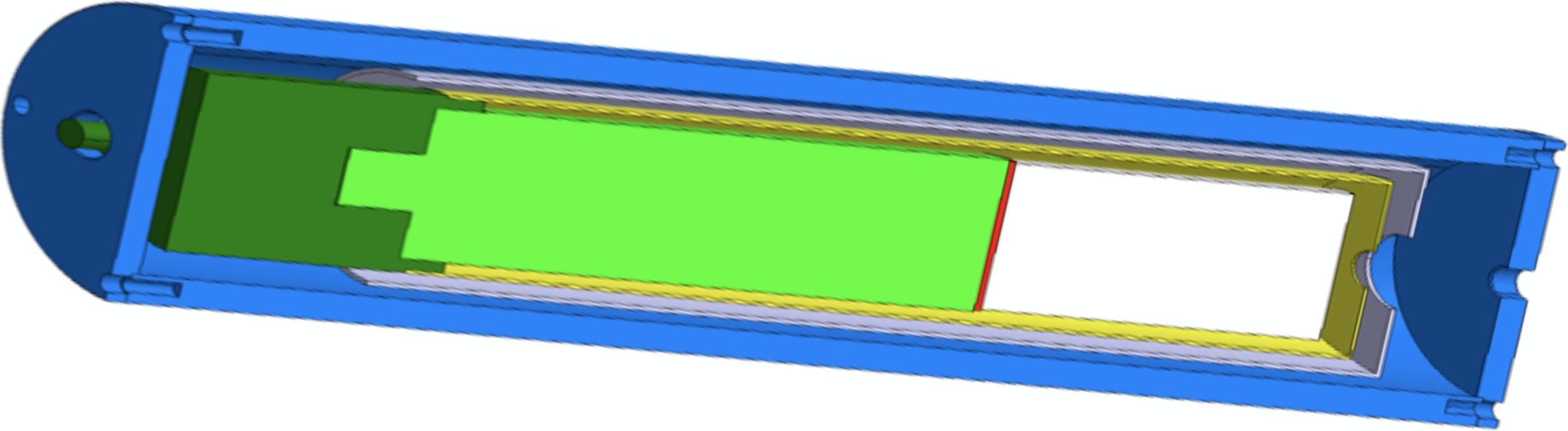}
\caption{%
    Left:\ photograph of the active element of one BHM detector unit, which is a 100\mm-long by 52\mm-diameter cylindrical quartz radiator, connected to a Hamamatsu R2059 photomultiplier.
    Right:\ shielding of a BHM detector unit, consisting of a low-carbon steel tube shown in blue, a mu-metal tube in gray, and permalloy in yellow.
    The quartz radiator, photomultiplier tube, and socket are shown in white, light green, and dark green, respectively.
    Figures from Ref.~\cite{Orfanelli:2015qgz}.
}
\label{fig:bril:bhm}
\end{figure}

The quartz radiators, shown in Fig.~\ref{fig:bril:bhm} (left), are optically coupled to a fast ultraviolet-sensitive PMT on one end and painted black on the other.
These detector units use the directional nature and timing of the BIB background and the \pp collision products.
In order to use the directional property, the PMT side of the quartz detector faces the center of the CMS detector.
When a BIB muon arrives with the incoming beam and produces Cherenkov radiation in the quartz, the light propagates forward and is collected by the PMT.
In contrast, \pp collision products arrive from the opposite direction, and the produced light is absorbed by the black paint.

The shielding of each BHM unit consists of three coaxially arranged metallic tubes, as illustrated in Fig.~\ref{fig:bril:bhm} (right).
The outer tube is a 10\mm-thick, 38\cm-long tube made of mild steel with endcaps on both sides composed of soft iron (ARMCO) with the same thickness.
The intermediate layer is a 30\cm-long, 1.5\mm-thick mu-metal tube, and the innermost tube is 27\cm long and 0.8\mm thick made of permalloy.
The shielding is designed to protect the PMT tubes from the 17\mTesla residual magnetic field and from radiation that mainly originates from low-energy electrons and positrons produced in the \pp collisions, which otherwise would form the dominant contribution to the BIB.

The readout of the BHM detector makes use of many components developed for the HCAL \Phase1 electronics upgrade~\cite{CMS:TDR-010}, with dedicated firmware and readout adapted to the beam-monitoring requirements~\cite{Tosi:2014nea}.
The PMT signal is digitized by a charge-integrating ASIC (QIE10), providing both the signal rise time and charge integrated over one bunch crossing.
It ensures dead-timeless readout of the signal amplitude and edge time information with 500\ps resolution. As a part of the readout from the backend electronics, the occupancy histograms are obtained in time bins of a configurable duration.
The histogram bins are typically 6.25\ns wide and are read out every 4 NB (about 1.458\unit{s}).
In special running periods, detailed self-triggered data containing event-level information about amplitude and time-of-arrival of each hit can be also collected.

A calibration and monitoring system was also installed for the BHMs~\cite{Tosi:2015fis} to evaluate possible changes in performance of the PMTs and quartz radiator due to aging and radiation damage.
The system uses a light signal produced by UV-emitting pulsed LEDs, which is sent to each detector unit through quartz optical fibers and optical splitters.
The monitoring system can be set to send a laser pulse periodically in an empty orbit or in periods when beams are not present.

\subsubsection{Beam-condition monitor for beam losses (BCML)}

The BCML detector~\cite{Muller:2010qza, Guthoff:2014swa, Kassel:2017hhb} is linked to the abort system of the LHC, and protects the CMS silicon tracker from beam-loss events.
The detectors simultaneously measure the integrated currents over 12 different time intervals and produce an unmaskable beam-abort trigger if one of them exceeds a threshold, which is configured to be at least three orders of magnitude less than the amount estimated to cause damage.
The BCML system is situated in four different locations within the CMS detector and includes the BCML1 detector located at $z=\pm1.8\unit{m}$ and the BCML2 detector located at $z=\pm14.4\unit{m}$ from the interaction point.
A total of 16 channels of the BCML detector are actively used in the beam-abort system, with four channels in each location, providing CMS with monitoring redundancy.

The BCML system primarily uses pCVD diamond sensors, most of which were replaced in LS2.
In addition, BCML2 has sapphire sensors installed, as described below.
Diamond detectors are typically used as robust beam monitors in locations where the radiation levels are very high.
Diamond is radiation hard and does not require active cooling.

\begin{figure}[!p]
\centering
\includegraphics[width=0.75\textwidth]{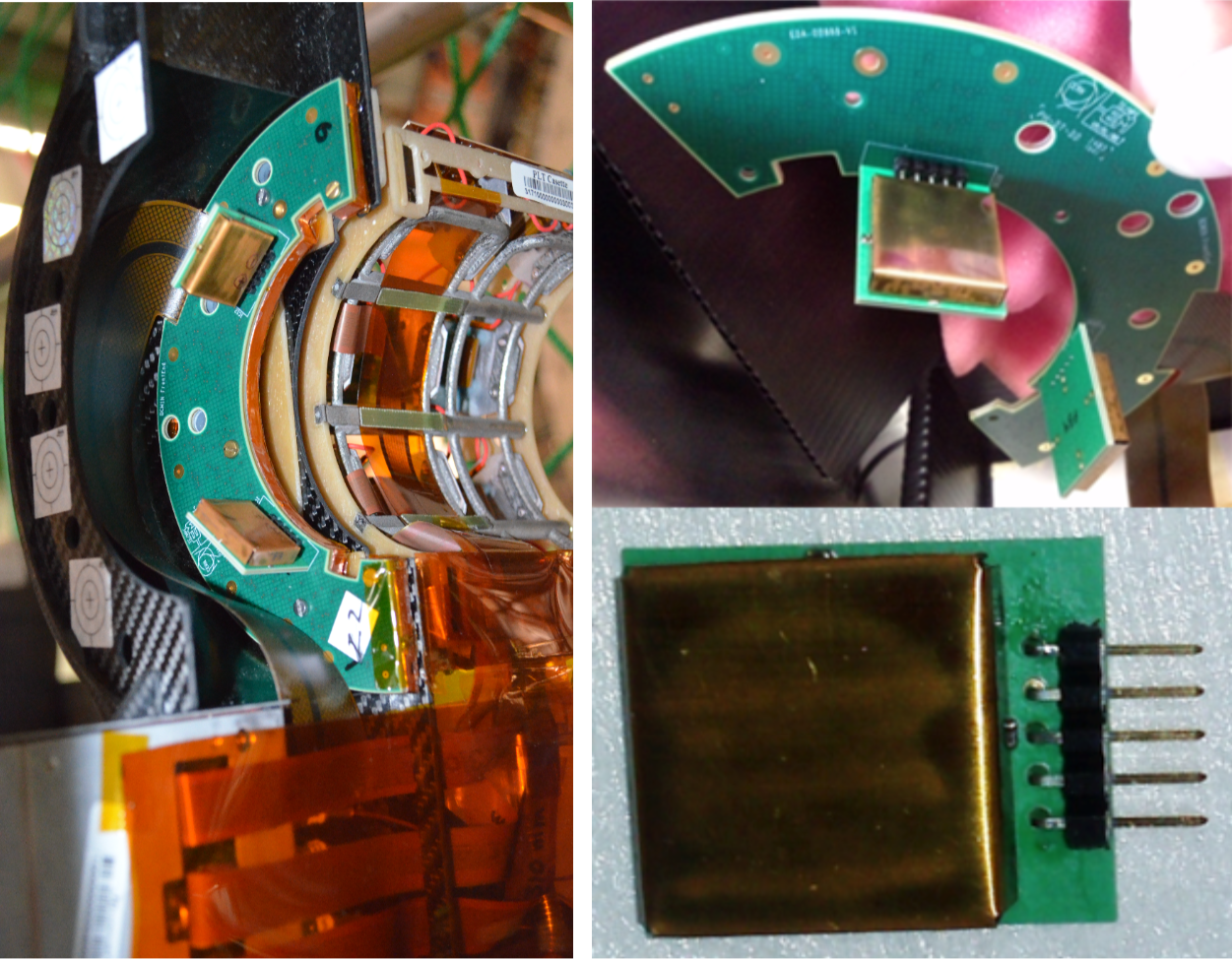}
\caption{%
    Mechanics for the BCML1 detector system.
    Left:\ a BCML1 mounted on the BCM1F C-shape PCB, attached to the PLT support structure.
    Right:\ a BCML1 mounted on a C-shape PCB (upper right), and a single sensor in a Faraday cage (lower right).
}
\label{fig:bril:bcml1}
\end{figure}

\begin{figure}[!p]
\centering
\includegraphics[width=0.4\textwidth]{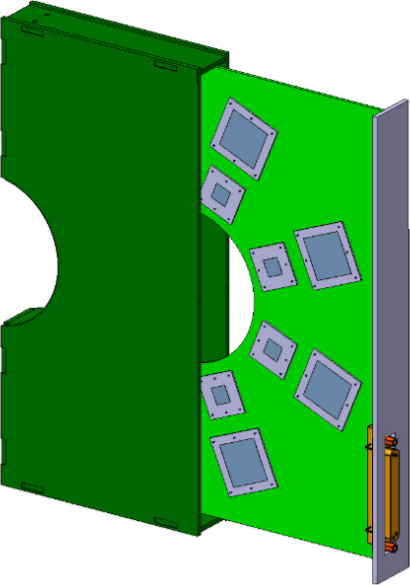}%
\hspace*{0.04\textwidth}%
\includegraphics[width=0.42\textwidth]{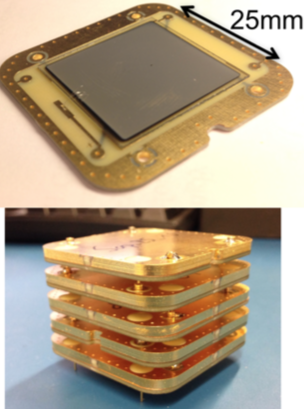}
\caption{%
    Mechanics for the BCML2 detector system.
    Left:\ the support structure of the BCML2 detectors, which also serves as mechanical protection. This structure is compatible with the \Phase2 beam pipe.
    Right:\ single-sensor base plate PCB for the large sensors used in BCML2 (upper right), and a complete sensor box assembly, with a stack of five sensors (lower right).
}
\label{fig:bril:bcml2}
\end{figure}

There are four pCVD diamond sensors installed  at each end of both the BCML1 (Fig.~\ref{fig:bril:bcml1}) and the BCML2 (Fig.~\ref{fig:bril:bcml2}) detectors, 16 sensors in total, each having an active volume of 10$\times$10$\times$0.4\mmcub.
The current created by ionization in the sensor is proportional to the ionizing particle flux through the active detector material, and it thus provides a good observable to determine the amount of radiation that a sensor receives.
The BCML sensor signals are read out using the LHC beam-loss monitor (BLM) electronics~\cite{Dehning:2002lop, Emery:2007zzb, Zamantzas:2006vk}.

If a beam loss signal, \ie, an integrated current over threshold, is detected in BCML1, the beam can be dumped within two to three LHC turns.
The threshold is defined for two different integration time intervals.
A short-duration (40\mus) threshold protects the silicon tracker from potentially damaging amounts of energy deposition in the electronics, and a longer-term ($\approx$83.3\unit{s}) integration threshold protects against high beam background conditions that can result in problematic conditions for data taking and reconstruction, as well as an increased dose rate to the inner detector region.
A machine beam interlock system~\cite{Todd:2011edms} ensures that the BCML system is fully operational before beam injection into the LHC can begin.
The BCML systems are powered independently of CMS from an uninterruptible power system, so that the LHC can run even if the local CMS power distribution is not functional.

New optical grade single-crystal sapphire sensors were installed in BCML2 during LS2 in six stacks.
These are produced industrially via the Czochralski process at low cost~\cite{Kurlov:2016book}.
Similar to diamond, sapphire, chemically aluminum oxide \AltwoOthree, behaves as a wide-band-gap semiconductor, with a band gap of 9.9\eV.
First measurements indicated a very high tolerance against radiation damage~\cite{Karacheban:2015jga, Kassel:2017hhb}, although the charge collection efficiency is low ($\approx$1--10\%)~\cite{Karacheban:2015jga}.
The sapphire \Phase2 prototype sensors used in BCML2 have an active volume of 25$\times$25$\times$0.5\mmcub and can be installed in units ranging from a single sensor to a stack of up to five sensors, as shown in Fig.~\ref{fig:bril:bcml2}.
In the stack option, the sensors are electrically mounted in parallel such that the observed current is the sum of the currents in each sensor.

To normalize the BCML2 readings and thereby monitor nonlinearities and instabilities in the response from the diamond and sapphire sensors, in addition to the solid-state detectors, two ionization chambers (one per end) of the LHC BLM type~\cite{Dehning:2002lop} were placed on top of the table originally designed to support the Castor subdetector.
This is located at about $z=1450\cm$, directly behind BCML2, and the chambers are connected to the same readout electronics as the BCML2 system.
These detectors provide a good dynamic range, demonstrate linearity with a deviation less than 1\%~\cite{CMS:DP-2021-008}, and no radiation degradation is expected.
The proximity to the BCML2 detectors, and the fact that these ionization chambers are in the same readout system as BCML2, makes them ideal for normalizing the BCML2 readings and thereby monitoring the nonlinearities and instabilities in the response from the diamond and sapphire sensors.
Since the response of these ionization chambers is also well known for a mixed radiation field, the detectors can also be used to benchmark the radiation simulations.

\subsection{Radiation instrumentation and simulation}

Radiation background can cause unwanted signals in detectors, activation of materials, damage to electronics, and detector degradation.
A radiation simulation and monitoring strategy is in place such that background radiation can be estimated, detected, analyzed, and lowered where necessary.

\subsubsection{Radiation monitoring}

The radiation monitoring strategy for \Run3 includes the use of the following systems:
\begin{itemize}
\item The HF RadMons~\cite{Gribushin:2017boy} are gas-filled proportional counters situated inside the HF regions.
There are eight in use during \Run3, four on either end of the CMS detector.
The detection principle is based on neutron capture by boron with the emission of an $\alpha$ particle in the argon-filled proportional counter.
The incident neutrons are slowed by a surrounding polyethylene moderator with a 3-inch radius.
The HF RadMons provide information on the total neutron flux for benchmarking the simulation estimates.
\item As the part of a wider radiation and environment monitoring unified supervision (REMUS) system under the responsibility of the CERN radiation protection group (HSE/RP), during \Run3 there are 12 inducted activity monitors (IAM) in the CMS cavern, read out by the radiation monitoring system for the environment and safety (RAMSES)~\cite{SeguraMillan:2005cpc, Ledeul:2019abp}.
Ten of these are in the same place as in \Run2, and two additional ones were installed in the same location as the existing monitors on the HF, to support luminosity stability studies.
The primary function of the IAMs is to measure the ambient-dose equivalent rates, $H^\ast(10)$, associated with the residual radiation field when the LHC is not operational.
However, their large dynamic range means that they can also provide reliable measurements of the prompt radiation field during collisions.
They are used to monitor any upgrades to shielding in the rotating shield region and provide benchmark data for simulations.
This system is also used by BRIL for luminosity stability and linearity studies.
\item The LHC RadMon system~\cite{Spiezia:2011jp, Martinella:2018cds}, which is maintained by the CERN EN/STI group, is designed to monitor the radiation that is related to electronics damage.
There are currently 29 LHC RadMon units situated inside the experimental cavern, and it is foreseen to add ten more during \Run3.
Each unit, upgraded to ``Type v6'', has 13 detectors, which includes two
radiation-sensitive field effect transistors (RadFETs) to measure the total ionizing dose, three Pin Diodes (in series) to measure the 1\MeV-equivalent fluence in silicon, and eight static random access memory (SRAM) modules.
The latter measure the cumulative fluence of high-energy hadrons via single-event effects and thermal neutrons using a  different voltage setting.
The BRIL group also uses the LHC RadMon data for radiation simulation benchmarking studies.
\end {itemize}
In addition to the monitoring systems listed above, several established methods are used to provide information about the radiation field.
These include activation samples that are placed in the outer cavern with subsequent measurements performed by the HSE/RP group; monitoring of CMS detector degradation; and the two LHC beam-loss monitors that are used as part of the BCML setup, described in the previous section.
Radiation monitoring is complemented by dedicated simulation studies, used to predict and understand in more detail the background radiation in CMS.

\subsubsection{Radiation simulation}

\begin{figure}[!ht]
\centering
\includegraphics[width=0.48\textwidth]{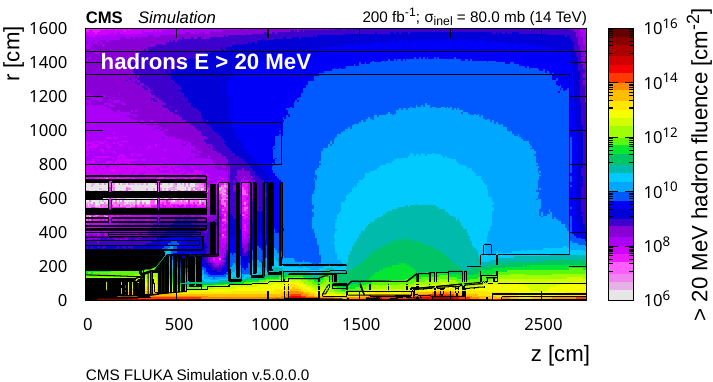}%
\hfill%
\includegraphics[width=0.48\textwidth]{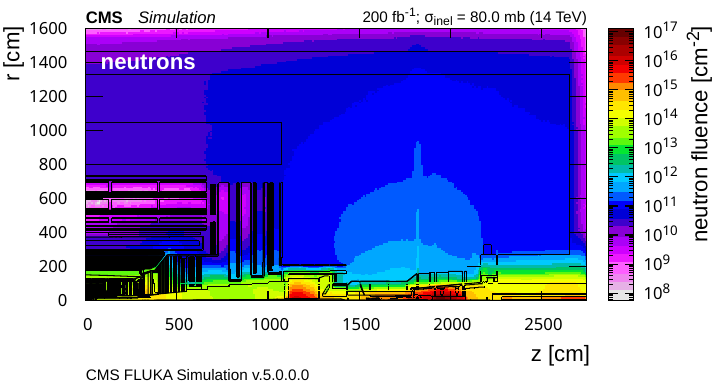} \\
\includegraphics[width=0.48\textwidth]{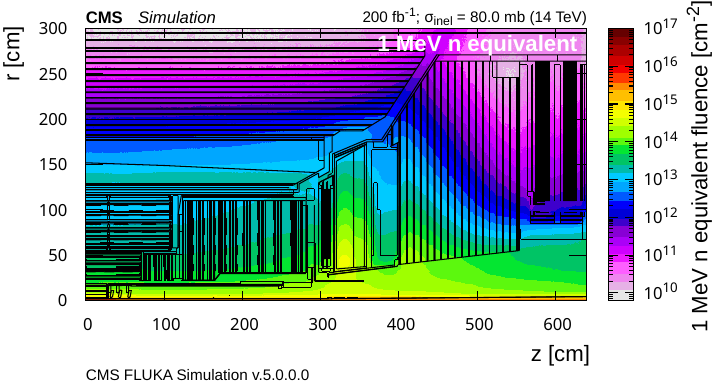}%
\hfill%
\includegraphics[width=0.48\textwidth]{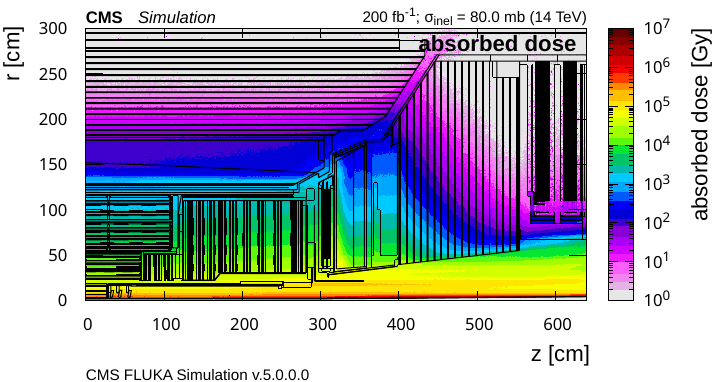} \\
\includegraphics[width=0.48\textwidth]{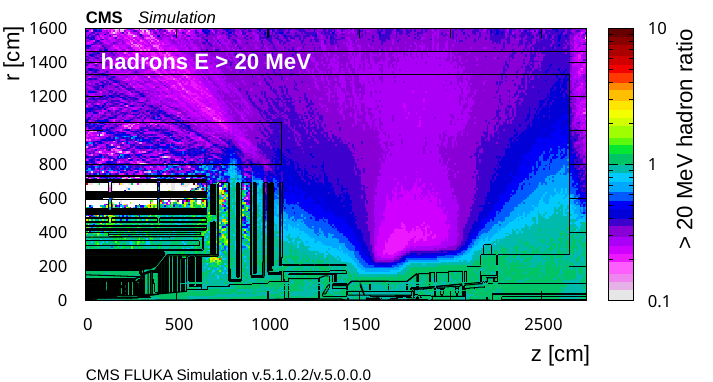}%
\hfill%
\includegraphics[width=0.48\textwidth]{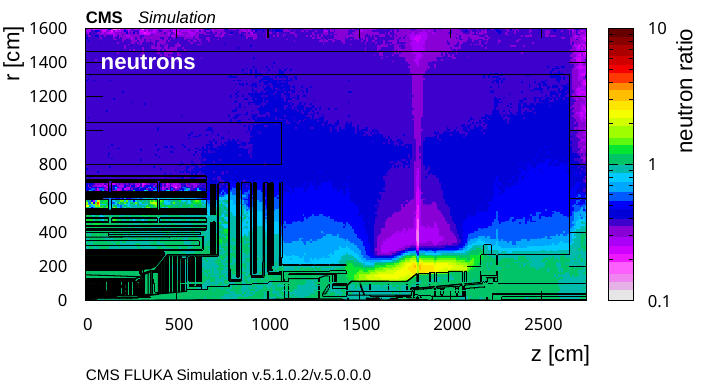}
\caption{%
    The FLUKA predictions for the expected \Run3 fluence of hadrons with energies greater than 20\MeV (upper left) and neutron fluence (upper right) normalized to an integrated luminosity of 200\fbinv at 7\TeV per beam are shown for the CMS cavern and detector.
    For the central part of CMS, 1\MeV-neutron-equivalent fluence (middle left), and absorbed-dose (middle right) are also presented.
    The lower two plots show the expected effect of the new forward shield as the ratio of hadron (left) and neutron fluences (right) in the CMS cavern comparing the \Run3 FLUKA simulation results of v5.1.0.2 with the \Run3 baseline of v5.0.0.0.
}
\label{fig:bril:fluka}
\end{figure}

Monte Carlo simulations to predict radiation levels in the experimental cavern and the CMS detector are typically performed with the CERN FLUKA radiation transport code~\cite{Bohlen:2014buj, Bohlen:2014buj}.
The maintenance of the FLUKA geometry models is the responsibility of the BRIL radiation simulation team, as well as the dissemination of the simulation results to the CMS subdetector teams via a web-based tool.
The baseline \pp collision simulation using a geometry model that reflects the current CMS \Run3 configuration is tagged v.5.0.0.0.
Relative to the \Run2 FLUKA geometry, this includes the implementation of the LS2 upgrades, the replacement of the central beam pipe from the IP to $z\approx16.7\unit{m}$, modifications of material within the rotating shield regions including reinforcements to fill existing gaps, as well as modifications to the installation of the LHC vacuum assembly for experimental area (VAX) equipment in LS3.
The radiation levels per integrated luminosity (or collision) in the central detectors for \Run3 are expected to be higher than in \Run2 by up to a factor of 1.5, depending on the location and type of particle, owing to changes in the shape of the central beam pipe.
Changes in the predicted radiation levels in the outer detectors for \Run3 are influenced by the increased showering in the central regions (resulting in a reduction downstream) and modifications inside and outside the rotating shield region where a lot of the secondary radiation is generated and leaks into the cavern.
The lighter aluminum beam pipe results in significantly lower activation levels for a given irradiation pattern.
The FLUKA predictions of the radiation environment in \Run3 are illustrated in Fig.~\ref{fig:bril:fluka} for the CMS cavern and detector.
The effects of an additional new forward shield, expected to be installed in 2024 around the existing rotating shield and included in geometry model v.5.1.0.2, are shown in Fig.~\ref{fig:bril:fluka} (lower).

\subsection{The BRIL online data acquisition and monitoring }

The BRIL data acquisition (BRILDAQ) system facilitates the real-time recording, processing, and monitoring of luminosity and beam background data at a latency of only a few seconds.
Essential BRIL data are delivered in real-time to the CMS and LHC control rooms, allowing for prompt and efficient operation of the LHC machine and CMS detector.
BRILDAQ is completely separate from the central CMS DAQ system, such that BRIL can operate continuously and independently of the CMS data-taking status even when all other CMS detector components are switched off.

Similar to the central DAQ system of CMS, BRILDAQ manages distributed and heterogeneous subsystems using a homogeneous distributed software architecture, where reliability and redundancy are important to prevent down time.
The BRIL data from different subsystems can be viewed together and correlated in real time, and new subsystems can easily be integrated.

The core software is based on the CMS online framework XDAQ~\cite{Gutleber:2003cd}.
This event-driven and message-oriented architecture enables large numbers of loosely coupled software components and services to exchange information in near real time.
Messages are exchanged via the publisher/subscriber model (eventing) where the publishers send messages that can be received by any number of subscribers.
Communication between BRILDAQ and the LHC is based on the data interchange protocol (DIP)~\cite{Copy:2018igv}, which also uses a publisher/subscriber model.

To fully exploit the redundant online luminosity and beam background monitoring systems, the histogramming of the BRIL subsystem frontends is synchronized using common timing signals distributed by the CMS TCDS system (Section~\ref{sec:daq}), defining the hit count integration interval boundaries.
Long command counters are also distributed for additional intersystem synchronization.
Using such a technique facilitates uniform accounting of the delivered luminosity and the downstream data handling by BRILDAQ.

Typically, a subsystem provides one or several source instances, which read the raw histograms from the hardware, and one single processor instance for data aggregation and calibration.
Both kinds of applications are stateless to ensure the availability of the data, regardless of the status of the LHC beams and the running state of CMS.
In \Run3, the participating subsystems are PLT, including the fast-or and the Slink data, BCM1F, with both RHU and \uTCA readouts, HF, DT, and REMUS for luminosity measurements, and other BRIL measurements from subsystems such as the beam-pickup system for timing measurements (BPTX)~\cite{CMS:NOTE-2019-008}, HF-RadMon gas-filled proportional chambers, BHM, and BCML.
High-level components include the storage manager, the vdM monitor providing the luminometer calibration constants in real time after a beam-separation scan, the luminosity monitor, and the best-luminosity selector.
The latter selects the value of the instantaneous luminosity that is delivered to the LHC.
The DIP-related components use the service provided centrally by the CMS DAQ team.
In addition, it is foreseen to process PCC data out of the HLT in near-real-time fashion, as mentioned in Section~\ref{sec:tracker}.

The run control system, responsible for controlling and tracking the configurations of the BRIL applications, is based on the CMS run control framework (RCMS)~\cite{Brigljevic:2003kv} and its configuration database.
Each BRIL process is controlled by a function manager (FM) that manages a simple finite-state machine.

The run control web frontend allows the operator to control the life cycle of BRILDAQ applications and to manage the configurations, which are versioned and can be retrieved and stored in the database via the web.
A set of other web tools completes the system by providing useful functionalities such as displaying logs of the processes and showing results of the data analysis.

The BRIL online web monitor is a single-page application built on the Angular framework~\cite{Fain:2016book} where the website interacts with the user by dynamically rewriting the current web page with new data instead of reloading the entire page.
It displays real-time and historical charts of online quantities.
Monitoring data are published by BRILDAQ applications, then stored in the ElasticSearch~\cite{Andre:2015dvu} backend database, which responds to requests from the web client. The monitoring system is nonintrusive and flexible for easy integration of new data and charts.

The raw and calibrated data are stored on local disks, then transferred to an offline storage area, and eventually moved to tape. Summary data are loaded into the BRIL database.
All types of data can be reprocessed, and multiple versions can coexist on disk and in the databases.

To provide high-quality luminosity measurements for use in physics analysis, an application toolkit is provided to the physics community, centered around the so-called ``brilcalc'' tool for luminosity calculation.
It provides delivered and recorded luminosity with different calibration sets, using a ``normtag'' that defines the best detectors and calibrations to use for each time period.
This strategy has proven to be successful and remains unchanged in \Run3.

\clearpage
\section{Data acquisition system}
\label{sec:daq}

This section describes the scope, design choices, and implementation of the experiment's central data acquisition (DAQ) system.
The present implementation for \Run3 is described in detail along with a discussion of its performance.
We also give an overview of the evolution of the system over the lifetime of the CMS experiment.

\subsection{Scope}

The CMS online event selection is performed using two trigger levels:\ the level-one (L1) trigger, described in Section~\ref{sec:l1trigger}, implemented in custom electronics, which selects approximately 100\kHz of events based on coarse information from the calorimeters and the muon detectors; and the high-level trigger (HLT), described in Section~\ref{sec:hlt}, which runs on a farm of commercial computer nodes integrated with the DAQ data flow.
The HLT processes fully assembled events, applying algorithms similar to those used in offline reconstruction, and selects a few kHz of events for storage on disk.

While this two-level approach with full event building after the first level greatly simplifies the overall system design compared to approaches with more trigger levels and/or partial event building, the resulting requirements on the data acquisition system are demanding:
The DAQ system needs to read out approximately 700 detector backend boards at a rate of $\approx$100\kHz and perform event building and distribution with a throughput of about 100\GBs---a challenging task when the DAQ system was first implemented~\cite{CMS:Detector-2008} for \Run1 with the hardware available in the late 2000s.

\begin{figure}[!ht]
\centering
\includegraphics[width=\textwidth]{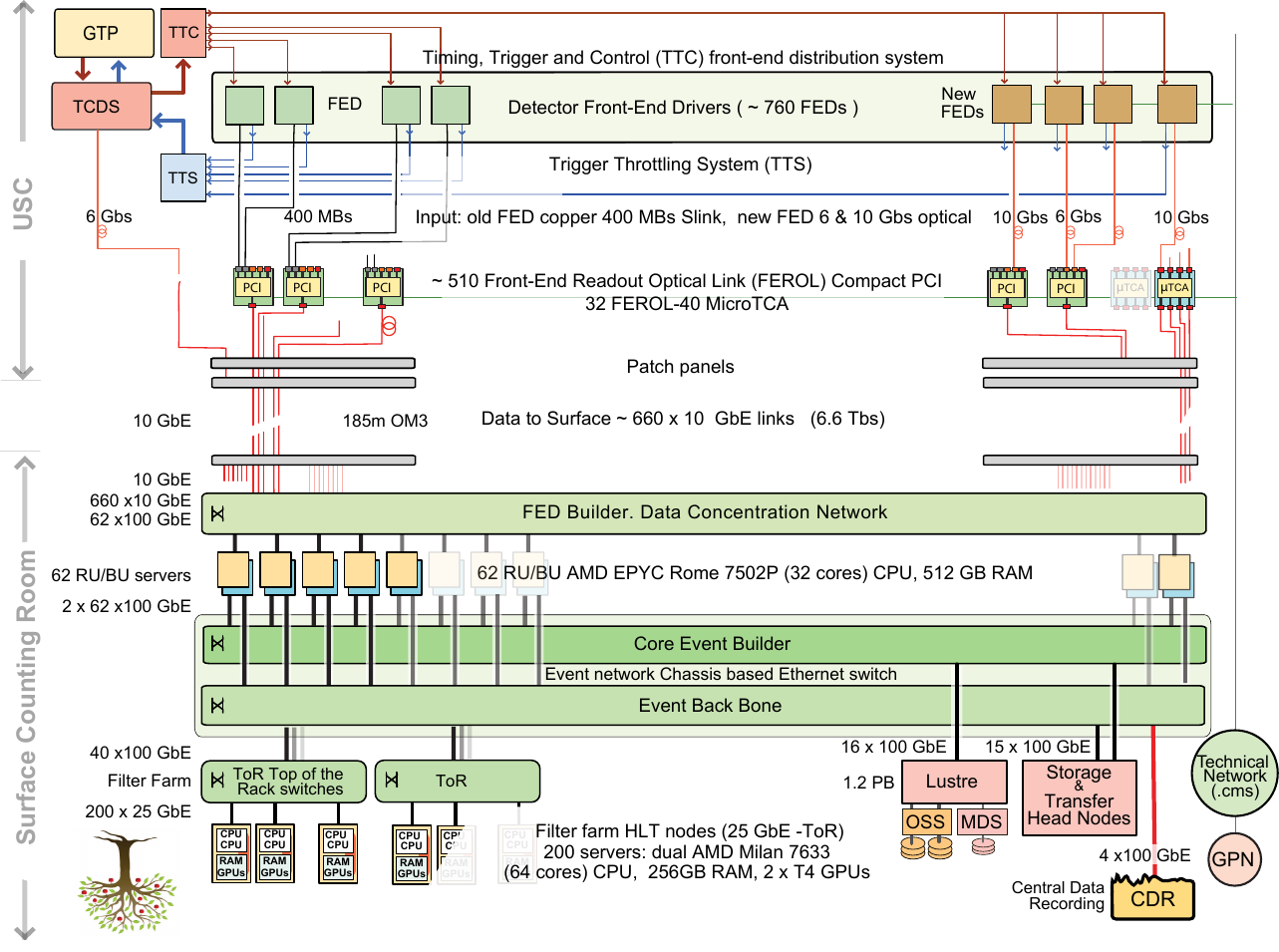}
\caption{%
    Diagram of the \Run3 DAQ system.
    The total numbers of cabled elements are given, including elements used in the MiniDAQ systems described in Section~\ref{sec:daq:minidaq} and elements installed as hot spares or for contingency.
    In typical global data-taking configurations, a subset is used, as described in the text.
}
\label{fig:daq:layout}
\end{figure}

The DAQ system is also responsible for collecting events selected by the HLT, buffering them at the experiment site, and transferring them to the \Tier0 computing center at CERN.
The general structure of the DAQ system has remained fairly constant from its original implementation for \Run1 up to its current implementation (illustrated in Fig.~\ref{fig:daq:layout}).
Across implementations, the main components of the system are:
\begin{itemize}
\item A trigger throttling system (TTS), consisting of custom electronics modules collecting fast status information from all the backend boards in order to throttle the trigger to avoid buffer overflows.
\item Since \Run2, the trigger control and distribution system (TCDS), implementing the trigger control logic, reacting to fast status signals from the TTS or collected directly from upgraded FEDs, and distributing timing and trigger signals to the entire experiment.
\item Custom electronics modules, the frontend readout (optical) link boards that receive data over a custom link from the detector backend boards (in CMS called frontend driver or FED boards), and send data over a commercial link.
\item A data concentrator network based on commercial network technology.
\item Readout unit (RU) servers that aggregate data from $\mathcal{O}(10)$ FEDs into super-frag\-ments and buffer the data.
\item An event builder network based on commercial network technology.
\item Builder unit (BU) servers that build full events from super-fragments received from all the RUs.
In \Run3, the RUs and BUs run on the same set of servers (RU/BU nodes).
\item An event backbone network, based on commercial network technology, connecting the BUs to the filter unit (FU) servers.
\item The FU servers that execute the HLT algorithms.
\item A storage and transfer system (STS), merging events selected by the HLT into larger files, buffering them locally, and transferring them to the \Tier0 at CERN via the central data-recording (CDR) network.
\end{itemize}
The DAQ system comprises software for data handling and controlling custom electronics, a hierarchical control system to manage most of the aforementioned components, a monitoring system that collects metrics, and several monitoring clients that interpret these metrics.

\subsection{Evolution}

Since its first implementation for \Run1, the CMS DAQ system has undergone multiple upgrades to keep pace with the evolving needs of the experiment.
Towards the end of \Run1, it became clear that during \Run2 the LHC would reach higher instantaneous luminosities than originally anticipated, resulting in increased pileup and larger event sizes.
In response, several subdetectors implemented upgrades and replaced VME-based backend electronics with new systems based on the \uTCA standard.
A new optical readout link with a higher bandwidth of up to 10\GBs was developed for these new backend systems.
The custom electronics of the DAQ system were upgraded to support this optical link and to support a new network technology for the data-concentrator network, from \Run2 onwards.

The bulk of the DAQ system downstream of the custom readout electronics consists of commercial computing and networking equipment, for which a regular replacement cycle must be observed.
This circumstance provided the opportunity to reimplement large parts of the DAQ system for \Run2, and again for \Run3, benefiting from advances in technology to achieve a much more compact design while doubling the event building bandwidth from \Run2 onwards.
The choice of networking technology has been adapted following trends in the supercomputing industry.
An overview of the three generations of the CMS DAQ system is provided in Table~\ref{tab:daq:parameters}.

\begin{table}[!ht]
\centering
\topcaption{%
    Key parameters of the CMS DAQ system in \Run1~\cite{CMS:Detector-2008, Bauer:2010zza}, \Run2~\cite{Bauer:2014qxa, Bawej:2014atd}, and \Run3.
}
\label{tab:daq:parameters}
\renewcommand{\arraystretch}{1.1}
\renewcommand{\thefootnote}{\alph{footnote}}
\cmsTable{\begin{tabular}{llll}
    & \Run1 & \Run2 & \Run3 \\
    \hline
    Event building rate \pp\footnotemark[1] & 100\kHz & 100\kHz & 100\kHz \\
    Event size \pp\footnotemark[1] & 1\MB & 2\MB & 2\MB \\
    Read-out links \SLINK (copper) 400\MBs\footnotemark[2] & 636\footnotemark[3] & 575\footnotemark[3]--532\footnotemark[4] & 528\footnotemark[3] \\
    Read-out links optical\footnotemark[5] 6\Gbs\footnotemark[2] & \NA & 55\footnotemarks{4}{6} & 55\footnotemarks{3}{6} \\
    Read-out links optical\footnotemark[5] 10\Gbs & \NA & 60\footnotemark[3]--167\footnotemark[4] & 176\footnotemark[3] \\
    FED builder network technology & Myrinet & Ethernet & Ethernet \\
    FED builder network speed & 2 rails of 2.5\Gbs & 10 \& 40\Gbs & 10 \& 100\Gbs \\
    Event builder \# of readout units & 640 & 108\footnotemark[4] & 57\footnotemark[7] \\
    Event builder network technology & Ethernet & Infiniband & Ethernet RoCE v2\footnotemark[8] \\
    Event builder link speed & 1--3 rails of 1\Gbs & 56\Gbs & 100\Gbs \\
    Event builder parallel slices & 8 & 1 & 1 \\
    Event builder network throughput & 1.0\Tbs & 1.6\Tbs & 1.6\Tbs \\
    Event builder \# of builder units & 1260\footnotemark[9] & 73\footnotemark[4] & 57\footnotemark[7] \\
    BU RAM disk buffer & none & 16\TB & 10\TB \\
    HLT \# of filter units & 720\footnotemarks{3}{9}--1260\footnotemarks{4}{9} & 900\footnotemark[3]--1084\footnotemark[4] & 200 \\
    HLT \# of cores & 5.8k\footnotemark[3]--13k\footnotemark[4] & 16k\footnotemark[3]--31k\footnotemark[4] & 26k\footnotemark[10] \\
    HLT computing power (MHS06) & 0.05\footnotemark[3]--0.20\footnotemark[4] & 0.34\footnotemark[3]--0.72\footnotemark[4]  & 0.65\footnotemark[10] \\
    HLT \# of NVIDIA T4 GPUs & \NA & \NA & 400 \\
    Storage system technology & 16 SAN\footnotemark[11] systems & 1 cluster file system & 1 cluster file system \\
    Storage system bandwidth write + read & 2\GBs & 9\GBs & 30\GBs \\
    Storage system capacity & 300\TB & 500\TB & 1.2\uPB \\
    Transfer system to \Tier0 speed & 2$\times$10\Gbs & 4$\times$40\Gbs & 4$\times$100\Gbs \\[1ex]
    \multicolumn{4}{@{}p{1.2\textwidth}@{}}{%
        \footnotemark[1]Design value.
        \footnotemark[2]Main data-taking configuration, excluding links from partition managers used for partitioned running.
        \footnotemark[3]At the beginning of the run.
        \footnotemark[4]At the end of the run.
        \footnotemark[5]SlinkExpress.
        \footnotemark[6]54 links from mezzanine cards with optical SlinkExpress.
        \footnotemark[7]Readout and builder unit running on the same server (``folded event builder'').
        \footnotemark[8]Remote DMA over Converged Ethernet.
        \footnotemark[9]Filter and builder units running on the same server.
        \footnotemark[10]Not including the GPU computing power.
        \footnotemark[11]Storage-area network.
    }
\end{tabular}}
\end{table}

The HLT computing capacity was scaled up according to the experiment's needs on a yearly basis over \Run1 and \Run2.
The \Run3 system includes general purpose graphics processing units (GPUs), providing cost-effective computing acceleration.
This has required a major effort in porting the HLT code.
Major changes in the software enabled better decoupling of the DAQ and HLT processes since \Run2.
Various automated features and automatic diagnostics have been added to the control system to maximize the uptime of the DAQ system.
Driven by the need for more detector partitions than the original system could support, and to host \uTCA backends, a new trigger control and distribution system (TCDS) was developed for \Run2, absorbing the functionality of the original trigger control system (TCS), and TTC system, and providing optical inputs for the TTS signals from the upgraded backends.
In the following sections, the upgrades to the DAQ system are described in more detail with an emphasis on the implementation for \Run3.

\subsection{Subdetector readout interface}

During \Run1, the CMS subdetectors were read out exclusively
through the \SLINK~\cite{vanderBij:1997rc} DAQ readout link, an LVDS-based copper link capable of transferring up to 400\MBs.
One or two such links are received by the frontend readout link (FRL)~\cite{Bauer:2007vdr}, a custom Compact-PCI card.
The FRL forwarded the data to a commercial Myrinet network interface card (NIC) via an internal PCI-64 bus at 66\MHz.
Super-fragments were built on the Myrinet NICs using custom firmware~\cite{Bauer:2007vdr}.
Assembled super-fragments were then transferred into the memory of a readout unit (RU) server via DMA.

At the start of \Run2, the Myrinet NIC on the FRL was replaced by a custom-developed PCI-X card, the frontend readout optical link (FEROL)~\cite{Bauer:2013vma}.
The FEROL acts as a 10\Gbs Ethernet NIC, sending data to the event builder.
It receives data either via the PCI-X interface from the FRL (from the bulk of the subsystems that continue to use \SLINK) or via optical SlinkExpress inputs from new or upgraded subsystems.
Up to two SlinkExpress inputs at 6\Gbs or one input at 10\Gbs are supported.
While \SLINK senders are mezzanine cards plugged onto the backend boards of the subdetectors, requiring a significant footprint, the SlinkExpress sender is a firmware IP-core that can be included in the FPGAs of the backend boards.
Only the footprint of an optical transceiver cage is needed, facilitating the move to smaller form factors such as \uTCA.
The SlinkExpress works with 8b/10b encoding at up to 6.3\Gbs or 64b/66b encoding at 10.3\Gbs, resulting in an effective bandwidth of up to 5.0 or 10.0\Gbs, respectively.
The data format (definition of headers and trailers) is identical to that of \SLINK.
The SlinkExpress is packet-based and supports re-transmission at the packet level.
Packets have a variable size of at most 4096 bytes.
Fragments up to 4096 bytes are transferred into individual packets, while larger fragments are split across multiple packets.

At the output side, the FEROL sends data via TCP/IP, employing a custom TCP/IP engine implemented in the FPGA logic.
This was achieved through a simplification of the TCP/IP protocol for unidirectional use, which reduced the number of states from 11 to 3~\cite{Bauer:2014pya}.
The TCP/IP streams (one per input) are sent via an Ethernet network to a standard Ethernet NIC in the readout unit (RU) server where they can be received with the standard Linux TCP/IP stack.
Using performance tuning as described in Ref.~\cite{Bauer:2013vma}, a sustained point-to-point throughput of 9.7\Gbs has been achieved for fragments larger than 1\kB.

For subdetectors that were potentially limited by the copper \SLINK's bandwidth, but did not upgrade their backend electronics, a new type of mezzanine card with optical transmission was developed.
It plugs onto the subdetector readout electronics in place of the original mezzanine cardand transmits data using the 6\Gbs SlinkExpress, increasing the bandwidth to 625\MBs.
A version using 10\Gbs SlinkExpress is also available.
This mezzanine card has been deployed for the ECAL subdetector with 6\Gbs SlinkExpress links.

In 2017, a new pixel detector was installed with new backend electronics requiring readout through 108 links at 10\Gbs.
Instead of producing more FEROL boards that are powered and controlled through PCI-X because of the legacy interface to the FRL, a new readout board was implemented, based on the \uTCA standard, incorporating the features of four FEROL boards.
The new FEROL-40 board~\cite{Gigi:2017bpd} receives up to four channels at 10\Gbs using the SlinkExpress and sends the data on four links of 10\Gbs Ethernet using TCP/IP.

Table~\ref{tab:daq:readout} shows the readout parameters of the CMS experiment at the end of \Run1 and \Run2, and at the start of \Run3.
Starting in \Run3, the HCAL barrel partitions are read out with higher segmentation, as described in Section~\ref{sec:hcal:backend}, resulting in a significant increase in data size.
Nine additional readout links were added for the HCAL.
The dependence of the data size on the pileup (PU) and the resulting number of vertices has been studied for each data source.
Data sizes observed at peak PU in \Run2 resulted in a total event size of 1.4\MB.
The event size in \Run3 at the planned peak luminosity of $\mathcal{L}=2\times10^{34}\percms$ ($\avPU=56$) is 1.6\MB.
An extrapolation, using a polynomial fit, to $\mathcal{L}=3\times10^{34}\percms$, $\avPU=85$, results in an estimated event size of 2.0\MB.
This is still within the design value of the DAQ system, which is therefore considered capable of handling the readout bandwidth at conditions beyond the planned peak luminosity in \Run3.

\begin{table}[!ht]
\centering
\topcaption{%
    Subdetector readout configuration.
}
\label{tab:daq:readout}
\renewcommand{\arraystretch}{1.1}
\renewcommand{\thefootnote}{\alph{footnote}}
\cmsTable{\begin{tabular}{l@{\hspace{5\tabcolsep}}cc@{\hspace{5\tabcolsep}}ccccc@{\hspace{5\tabcolsep}}cc}
    & \multicolumn{2}{c@{\hspace{5\tabcolsep}}}{\hspace*{-1.5\tabcolsep}End of \Run1} & \multicolumn{5}{c@{\hspace{5\tabcolsep}}}{\hspace*{-1.5\tabcolsep}End of \Run2 (start of \Run3)} & \multicolumn{2}{c}{\hspace*{-1.5\tabcolsep}Data size [kB]} \\
    Subdetector & \# & \# & \# & \multicolumn{3}{c}{\# FRL/FEROL} & \# & \Run2\footnotemark[2] & \Run3\footnotemark[3] \\[-2pt]
    & FED & FRL & FED & copper & 6\Gbs & 10\Gbs & F40\footnotemark[1] & $\avPU=56$ & $\avPU=56$ \\
    \hline
    Tracker pixel & 40 & 40 & 108 & \NA & \NA & \NA & 32 & 259 & \\
    Tracker strips & 440 & 250 & 440 & 250 & \NA & \NA & \NA & 731 & \\
    Preshower & 40 & 28 & 40 & 26 & \NA & \NA & \NA & 54 & \\
    ECAL & 54 & 54 & 54 & \NA & 54\footnotemark[4] & \NA & \NA & 74 & \\
    HCAL & 32 & 32 & 32 (41) & \NA & \NA & 32 (41) & \NA & 170 & $+$221 \\
    Muons CSC & 8 & 8 & 36 & 18 & \NA & \NA & \NA & 40 & \\
    Muons RPC & 3 & 3 & 3 & 3 & \NA & \NA & \NA & 0.3 & \\
    Muons DT & 10 & 10 & 9 & \NA & \NA & 9 & \NA & 22 & \\
    Trigger & 5 & 5 & 14 & \NA & \NA & 14 & \NA & 41 & \\
    CASTOR & 4 & 4 & 4 (0) & 4 (0) & \NA & \NA & \NA & \NA & \\
    TCDS & \NA & \NA & 1 & \NA & 1 & \NA & \NA & 1.0 & \\
    CTPPS & \NA & \NA & 11 & 9 & \NA & 2 & \NA & 2.2 & \\
    Muons GEM & \NA & \NA & 2 & \NA & \NA & 2 & \NA & \NA\footnotemark[5] & $+$22 \\[1ex]
    \multirow{2}{*}{Total} & \multirow{2}{*}{636} & \multirow{2}{*}{434} & 754 & 310 & \multirow{2}{*}{55} & 59 & \multirow{2}{*}{32} & \multirow{2}{*}{1.39\MB} & \multirow{2}{*}{1.63\MB} \\[-2pt]
    & & & (759) & (306) & & (68) & & & \\[1ex]
    \multicolumn{10}{@{}p{1.23\textwidth}@{}}{%
        \footnotemark[1]FEROL-40 (4$\times$10\Gbs).
        \footnotemark[2]Observed at \Run2 peak luminosity.
        \footnotemark[3]Only the increment in data size with respect to \Run2 at the same peak luminosity is shown.
        \footnotemark[4]Mezzanine card with optical SlinkExpress.
        \footnotemark[5]Not included for regular data taking.
    }
\end{tabular}}
\end{table}

For heavy ion runs, where the L1 trigger rate will be 50\kHz, the expected event size based on the latest heavy ion run in 2018, taking into account the upgrade of the HCAL readout, is 3.3\MB.
The overall throughput at the input to the DAQ system will thus be similar to that in \pp runs.
The FED sizes (FED can indicate both the subdetector electronic interface and, in this context, the data payload that is read out from that interface) is, however, distributed in a different way.
In many cases, a special optimization of zero-suppression and/or selective readout algorithms at the level of the FEDs is applied in order not to be limited by the bandwidth of individual FEDs.

\subsection{Event builder}
\label{sec:daq:eventbuilder}

The CMS experiment uses a two-stage event builder (EVB) system responsible for assembling event fragments retrieved from around 760 detector backend boards into a single event payload and delivering the built events to the HLT.
The first stage, called the FED builder, reads fragments from the underground FEROL and FEROL-40 boards and, using the switched network, aggregates them in the RU/BU nodes.
In the second stage, the core EVB, all the RU/BU nodes transfer fragments from each event into one destination node, assigned on a per-event basis.
With the \Run3 DAQ system, each node in this way handles around 2\kHz of fully-built events.
On each node, events are written into a large 200\GB RAM buffer and made available to the HLT via a dedicated data network.

Network interconnect technologies are a main driver for the EVB design and scaling.
The \Run1 FED builder system~\cite{Bauer:2010zza} was based on the Myrinet 2.5\Gbs network, comprising 640 RU nodes and split into 8 parallel slices for performance.
Prior to \Run2, Infiniband and 40\Gbs Ethernet became established mainstream standards in high-performance computing (HPC).
They were evaluated~\cite{Bauer:2014qxa, Bauer:2012na} and ultimately adopted~\cite{Bawej:2014atd, Andre:2016sne} as technologies of choice for the \Run2 DAQ upgrade.
This allowed the system to be scaled down by an order of magnitude and implemented in a single slice, while increasing the overall EVB throughput~\cite{Andre:2017ijs}.

For \Run3, the \Run2 equipment, which had reached the end of its vendor support, had to be retired.
The design choice was taken to implement the EVB with 100\Gbs Ethernet for the \Run3 system.
This system, described in the following sections, satisfies nearly identical performance requirements as the \Run2 DAQ, and, owing to hardware evolution, is reduced in scale by more than a factor of two.
Some design details of the preceding \Run2 system are also outlined.

\subsubsection{FED builder}

The FED builder transports data fragments over 10\Gbs links from the underground cavern, over a distance of around 200\unit{m}, to data concentration network Ethernet switches located on the surface.
A total of 557 links from the FEROL and FEROL-40 boards are used in a typical \Run3 data-taking configuration, as detailed in Table~\ref{tab:daq:readout}.

The data concentration network is implemented using a Juniper QFX10016~\cite{Juniper:2021web} chassis-based 100\Gbs Ethernet switch, providing a flat network, which allows full flexibility in routing the fragment traffic between readout nodes.
For inbound traffic from the FEROLs, seven QFX10000-36Q line-cards are installed, capable of supporting a total of 672 10\Gbs links, split from 24$\times$40\Gbs ports per line-card.
Three QFX10000-30C cards installed in the switch provide a capacity of up to ninety 100\Gbs ports for connections to the RU/BU nodes.
A pause frame~\cite{IEEESA:2012web} flow control mechanism is used to achieve a lossless low-latency network and ensure optimal buffer occupancy on the FEROLs.
The system is configured to support Ethernet jumbo frames of up to 9\kB to improve performance.

\subsubsection{Core event builder}

The core EVB has three distinct functional units, the RU, BU, and event manager (EVM).
They are implemented as software applications and run on a single set of RU/BU computers, interconnected using a dedicated event-building data network.
These applications are implemented within the XDAQ framework described in Section~\ref{sec:daq:software}.

The RUs unpack and merge fragments received via TCP streams from the FEROLs over the data concentration network.
Each fragment checksum is verified before being merged with others into a super-fragment, \ie, a collection of fragments from the same event aggregated by a particular RU, and buffered in memory.
The EVM, which runs on a single EVB node, orchestrates the event building process by performing a destination assignment, \ie, allocating a given event to a specific BU and sending a message to the RUs to send super-fragments to that destination.
When the BUs receive a super-fragment from each RU the event is completely built.
Completed events are written into ramdisk, a memory buffer structured as a file system, as described in Section~\ref{sec:daq:eventfilter}.
This file system is exported through the event backbone network to the FUs.
An FU group of typically three or four FUs is assigned to read the event data from each RU/BU node.

Starting in \Run2, EVB senders and receivers were implemented using the Infiniband Verbs-API, described in Section~\ref{sec:daq:software}, for communication over the network using remote direct memory access (RDMA) capabilities of network-interface cards and switches.
This technology facilitates the bypass of the operating system networking stack and delivers the payloads directly into the application-accessible memory, avoiding CPU and memory overheads associated with the handling of high bandwidth and packet rate networking in software.

For the \Run3 EVB, the cost-effective 100\Gbs Ethernet technology was chosen over the native Infiniband used in \Run2.
A second Juniper QFX10016 chassis-based switch is used as the core EVB network backbone (equipped with nine QFX10000-30C line-cards having a total of 270 100\Gbs ports).
The switch supports a lossless Ethernet network, a prerequisite for RDMA over Converged Ethernet (RoCE) v2 protocol, providing encapsulated Infiniband protocol over Ethernet hardware.
Importantly, like native Infiniband, it facilitates direct memory access between communicating nodes using offloading by NICs.
A benefit of using RoCE v2 is that the Verbs-API-based EVB applications developed in \Run2 can be reused with minor adaptations.

In addition to the core EVB network, the switch runs the event backbone network, which handles the traffic between the EVB and HLT farm.
The same chassis switch supports also the storage and transfer system (including transfers to \Tier0), traffic from the DAQ to the online data quality monitoring system, and running the online cloud on some of the legacy \Run2 HLT nodes.
The 62 RU/BU nodes are connected to the EVB network over 100\Gbs optical links, including 57 nodes as part of the nominal data-taking configuration and five as hot spares.
In addition, there are seven cold spare nodes without network interface cards.

As pointed out above, the \Run3 system also introduced a folded core EVB setup, where the RU/BU nodes serve as both RU and BU functional units (RU/BU).
This allows the bidirectional utilization of the network links, nearly halving the number of needed RU/BU nodes and the network bandwidth compared to a nonfolded setup, thus reducing the overall size and cost of the system for the same throughput requirement.
On the other hand, this design is demanding on the I/O and memory performance of the nodes, which in such a system need to receive and merge fragments, exchange super-fragments with other RU/BU nodes, build and serve full events, and pass-through the HLT output to the STS, all over multiple 100\Gbs interfaces.
Therefore, a modern server architecture was required for the task.
Dell R7515 servers, with the 32-core AMD EPYC Rome 7502P CPU running at 2.5\GHz, equipped with 512\GB of DDR4 RAM, were selected, with all memory channels populated for the maximum memory performance.
Each server is equipped with two Mellanox ConnectX6 dual-100\Gbs PCIe Gen4 NICs.
One card has both links used for the respective connections to the FED builder and event backbone networks, while the second card uses one link for connection to the core EVB network.

\subsubsection{Performance}

Detailed studies were done comparing the EVB performance of both the Intel Xeon Skylake dual-socket servers~\cite{Mommsen:2018csk} and comparable AMD EPYC Rome single-socket platforms.
The AMD platform was ultimately chosen for RU/BU nodes for \Run3 due to better performance and a simplified memory architecture.
The AMD 7502P CPU is internally assembled from multiple 4-core silicon dies (CCDs), each having a separate L3 CPU cache and an interconnect fabric providing eight memory controllers, as well as multiple PCIe Gen4 interfaces.
Due to these characteristics, the CPU internally resembles, to an extent, the nonuniform memory access (NUMA) memory architecture.
It can be configured in a 4-, 2-, and single-node NUMA mode with respect to the CPU die and memory controller topology.

The single-node NUMA mode uses interleaved access to the RAM controllers with improved maximum bandwidth, with the drawback of potentially higher memory and I/O latency.
In evaluations, this mode was found to be a well-balanced setup for folded EVB requirements, avoiding the delicate tuning of thread affinities to CPU cores, which would have been needed to optimize performance.
The setup maintains CPU affinity settings for groups of threads running similar tasks to the same or adjacent CPU cores and placing them closer to the corresponding NIC PCIe lines in the interconnect topology.

A memory-based file system based on Linux tmpfs~\cite{linux:2021tmpfs} is used as the EVB event output ramdisk.
To achieve high read/write performance, 2\MB huge-page support for tmpfs is used, available in recent Linux kernel versions.
It significantly improves the ramdisk throughput since it avoids bottlenecks when the CPU is handling a large number of 4\kB memory pages.

\begin{figure}[!ht]
\centering
\includegraphics[width=0.48\textwidth]{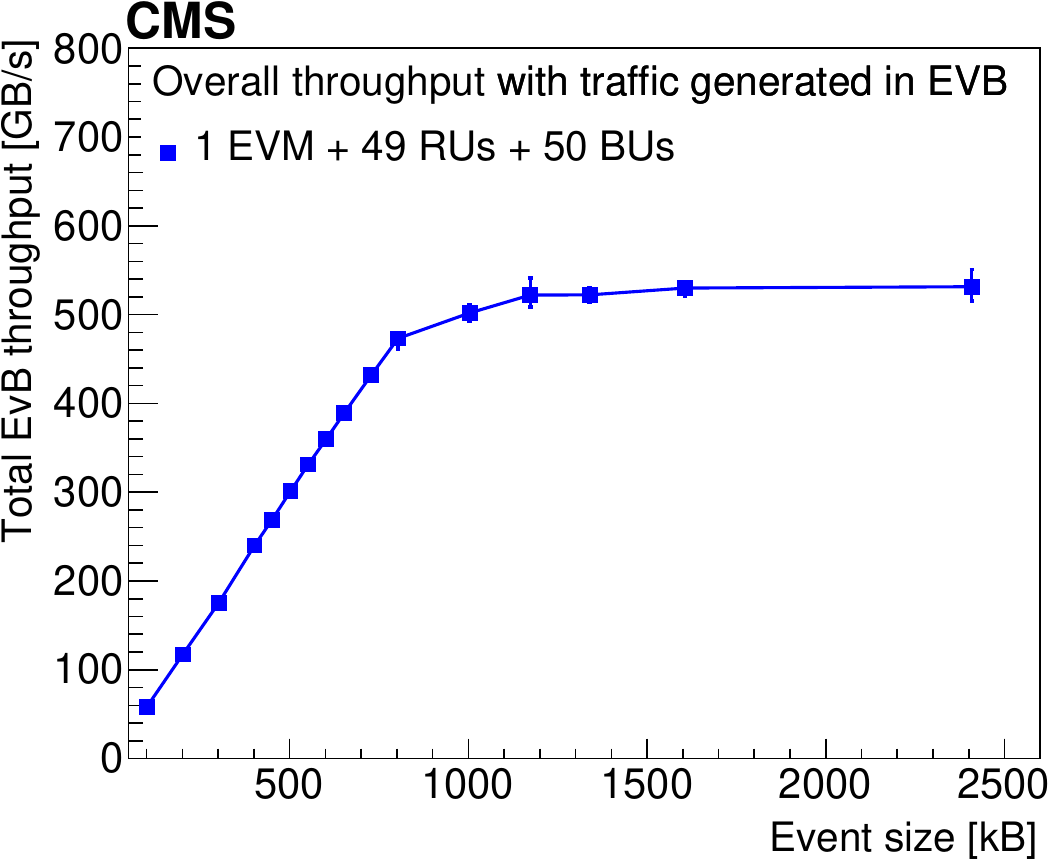}%
\hfill%
\includegraphics[width=0.48\textwidth]{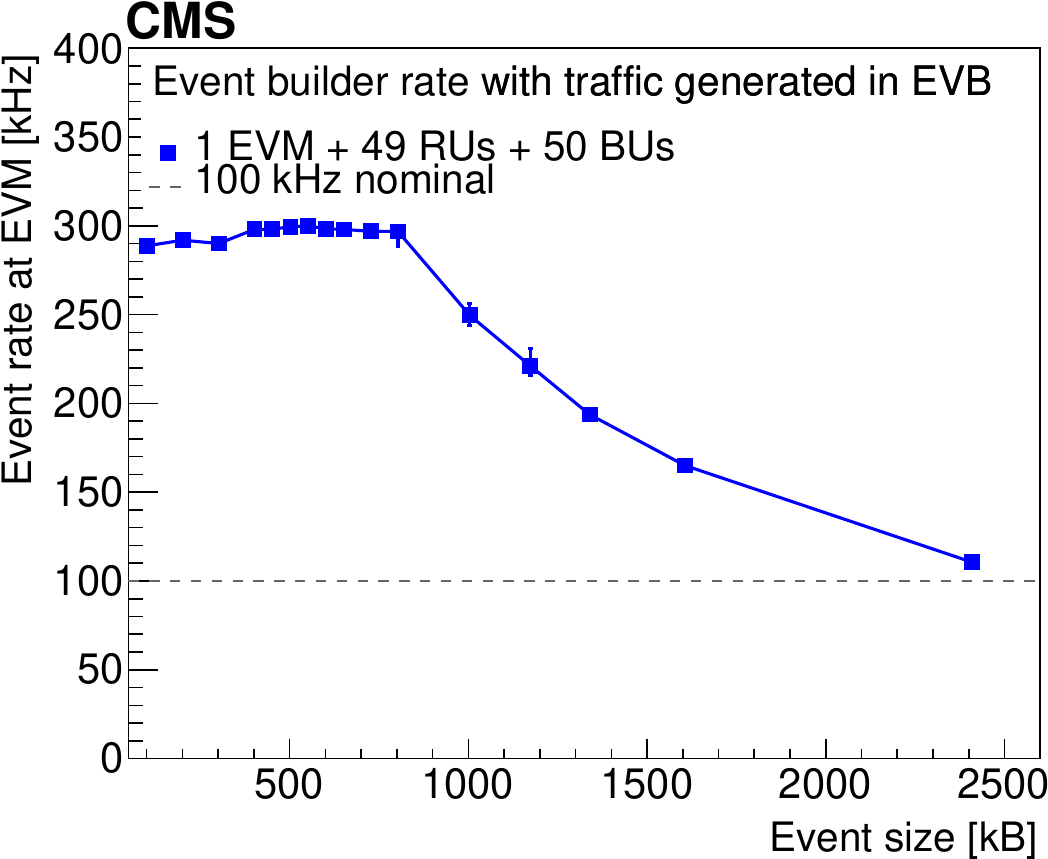}
\caption{%
    Core event building throughput (left) and event rate (right) of the full-scale RU/BU EVB setup shown for a range of built event sizes.
    Emulated input data are generated in the RU/BUs and discarded after the event building stage.
}
\label{fig:daq:evbperf}
\end{figure}

Measurement of the DAQ event building performance with emulated data generated by the RUs and discarded after event building, is shown in Fig.~\ref{fig:daq:evbperf}.
This demonstrates that the EVB system is capable of handling the nominal 100\kHz event rate from the L1 trigger with event sizes up to $\approx$2.5\MB.
A throughput of approximately 10\GBs is achieved per RU/BU node at the plateau, which is 80\% of the available network bandwidth.

\begin{figure}[!ht]
\centering
\includegraphics[width=0.48\textwidth]{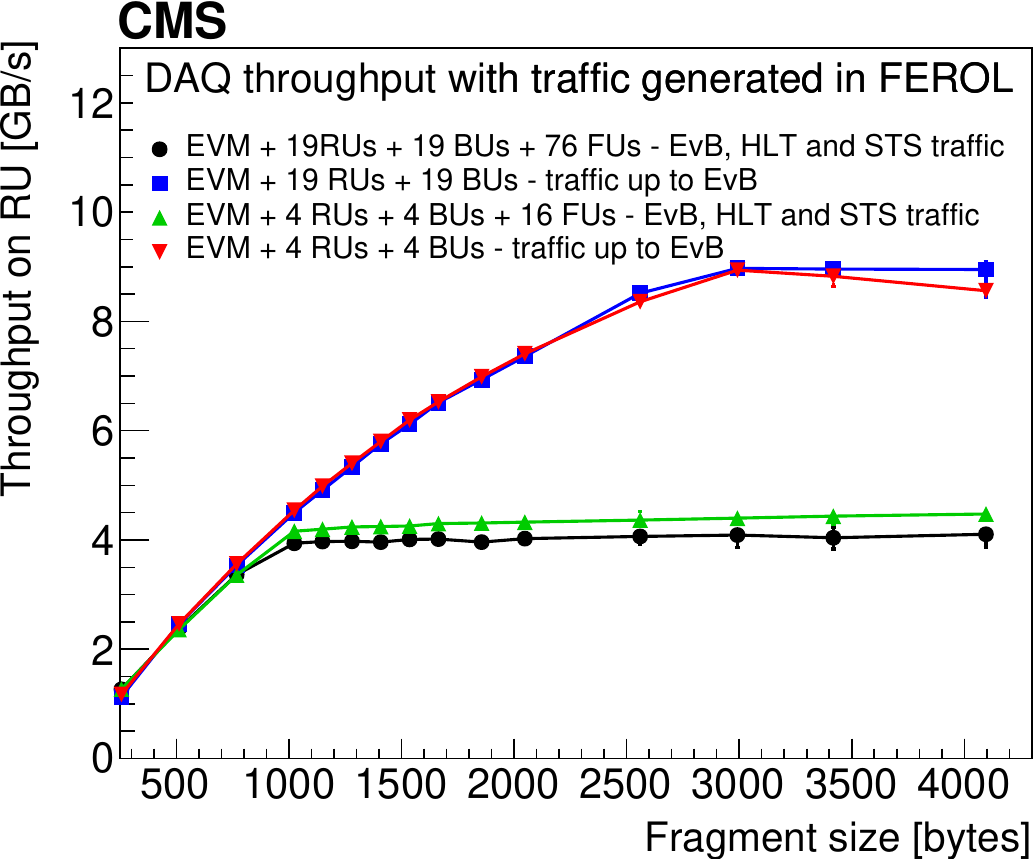}%
\hfill%
\includegraphics[width=0.48\textwidth]{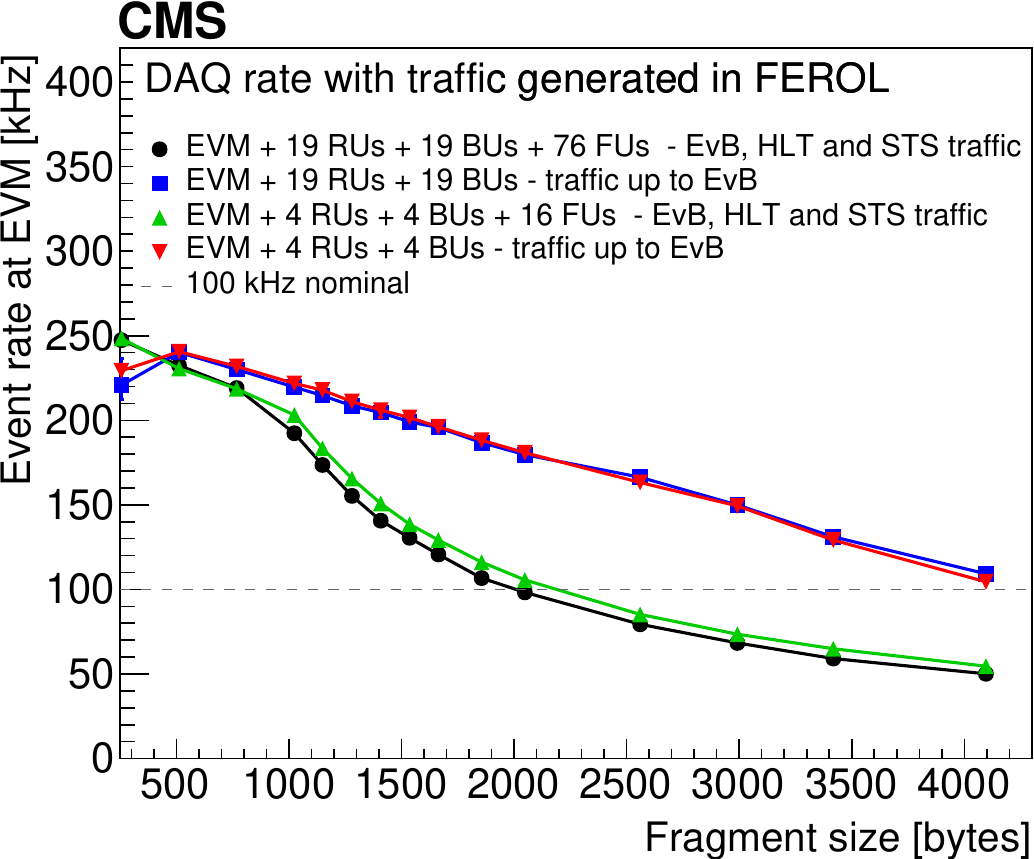}
\caption{%
    The DAQ throughput (left) and event rate (right) per RU/BU node for a range of uniform fragment sizes using a mode with the data discarded after the EVB and with the traffic flow through the HLT and STS.
    Emulated input data are generated at the FEROLs with 20 fragments concentrated per RU/BU.
}
\label{fig:daq:hltperf}
\end{figure}

The performance for input generated in the FEROLs and either discarded after the EVB or passed through the full-chain DAQ system (including the HLT and STS) is shown in Fig.~\ref{fig:daq:hltperf}.
Two setups of a smaller scale than the data-taking setup were used for this measurement, comprising either four or 19 RU/BU servers in a folded setup, with an additional machine running an EVM application.
Each RU application was concentrating 20 fragments from the FEROLs (two fragment streams per 10\Gbs FEROL link, or a total of 100\Gbs of the input network bandwidth per RU).
A total of either 16 or 76 FUs, four assigned to each RU/BU, were used.
These servers were \Run2 FU nodes equipped with 10\Gbs Ethernet NICs.
Thus, the total available bandwidth was 40\Gbs over the event backbone network to the FU group.
With each system, a plateau throughput of about 9\GBs was achieved per RU (BU), with discarding the data after the EVB, amounting to over 70\% of the available network bandwidth from the FEROLs.
About 4\GBs per server was achieved with the full-chain DAQ, which is about 80\% of the available network bandwidth to the FUs.

This result demonstrates that the system is capable of handling traffic for fragment sizes larger than 2000 bytes at rates above 100\kHz.
When extrapolated to the larger system of 50 RU/BUs (nearly the scale of the system used in \Run3), a fragment range around 2000 bytes corresponds to the total event size of over 2\MB, which is above the size expected in \Run3.
With the fixed number of FUs per RU/BU, the achieved bandwidth approximately scales from the small- to the medium-size system and gives confidence that the system comprising over 50 RU/BU nodes can robustly scale and handle the \Run3 requirements.
Furthermore, while tests were performed using a 10\Gbs network bandwidth per FU, the \Run3 FUs have more bandwidth available from the 25\Gbs Ethernet links, as mentioned in Section~\ref{sec:daq:evolution}, and, therefore, higher full-chain throughput capacity.

\subsubsection{Event builder load balancing}

The EVB protocol implements a global dynamic re-balancing mechanism of the event building throughput on the RU/BU nodes.
The assignment algorithm, running on the EVM, periodically determines the load status of each node, depending on factors such as the local output buffer
occupancy, as well as the number and type of HLT computers attached to a particular RU/BU.
From these, it determines the rate at which each node can build events to fairly distribute the load and equalize the processing latency at the EVB and HLT across the cluster.
A local throttling mechanisms on the RU/BUs can also get triggered, based on occupancy thresholds and latency information reported by the HLT infrastructure software.
This results in the BU application temporarily blocking or throttling the event building at the corresponding RU/BU.

\subsection{Event filter}
\label{sec:daq:eventfilter}

The HLT performs the second stage of event filtering, analyzing at a rate of approximately 100\kHz, the events selected by the L1 trigger, and accepting events at a rate on the order of a few kHz.
It runs on a large cluster of multicore servers (FUs) that perform the reconstruction and filtering in software, implemented within the CMS software framework (CMSSW)~\cite{Jones:2015soc}.

By using the same framework and code base for the online and offline reconstruction, most of the algorithms developed in CMSSW for the offline reconstruction can also be employed in the HLT.
This enables the reuse of the same data structures, supports a unified approach to the use of detector condition data, and allows for rapid deployment of both newly developed and established offline algorithms.

Starting from the first year of \Run2, the CMSSW and HLT implemented task-based multithreaded event processing.
This feature was instrumental in reducing the memory requirements of the HLT:
By running a smaller number of processes on the FUs, each using multiple threads, common data such as detector conditions and calibrations can be shared among threads, decreasing the overall memory footprint on the servers.
As a result, this made it possible to fully exploit all the logical cores of the CPUs, leading to increased overall computing efficiency.
At the beginning of \Run3, support for offloading to accelerators, such as GPUs, was also added.
In addition, reducing the number of running processes per server was found to be beneficial in reducing the memory footprint on the GPU.

The framework architecture is described in more detail in Section~\ref{sec:offline}.
The HLT algorithms and offloading are detailed in Section~\ref{sec:hlt}.
A dedicated software infrastructure couples the CMSSW workflow with the EVB and the storage and transfer system.
It handles the data input and output for all these processes, as well as the bookkeeping of event processing and monitoring.

\subsubsection{File-based filter farm}

The HLT interface to the DAQ, running in the same process as the HLT algorithms, is implemented in ordinary CMSSW modules and services, which facilitate the communication and reading of input data from the RU/BU nodes.
The data transfer into the HLT starts from a ramdisk file system serving as a data buffer on the RU/BU~\cite{Andre:2015xio}.
An NFS (network file system) v4 server provides access to the file system over the network.
The ramdisk is sized to buffer the data during the HLT startup time of 60--90\unit{s}, dominated by the loading of the software libraries, calibrations, and conditions, as well as to absorb fluctuations in the HLT event processing time.

The BU application writes fully built events into the ramdisk, using a custom binary file format with a header prepended to each event describing its total length and providing information such as an identification number and a checksum.
The files created by the BU application, each containing around 100 events, are distributed among CMSSW processes running on several FUs assigned to the RU/BU node.
A dedicated application running on the RU/BU nodes, called the file broker, is responsible for the unique assignment of files to the processes.
The requests for files are placed with the file broker using an HTTP-based protocol, while the standard file system access API is used to read files from the NFS.
All input-related communication and file access is implemented in the CMSSW input source module written in C++.
This module operates several dedicated threads for pre-buffering of the input data to minimize stalling the CPU cores while waiting for input.
The input source parses the assigned input files, extracts and verifies the event payload against stored fragment checksums, and hands it over to the framework, which schedules the HLT modules performing the event reconstruction, analysis, and filtering.

\subsubsection{The HLT software infrastructure}
\label{sec:daq:hltsoftware}

The HLT daemon is a system service orchestrating the HLT operation in the RU/BUs and FUs.
Implemented as a Python application, it relies on a file notification API of the Linux kernel, inotify~\cite{linux:2021inotify}, which provides low-overhead monitoring of file system events.
The start of a run is triggered by the appearance of raw event data files in the ramdisk.
Similarly, a run completes when all files have been processed by the HLT.
The service runs on all RU/BUs, where it is responsible for contacting the attached FU nodes to process a run, as well as the FUs, where it is responsible for managing the life-cycle of the locally running CMSSW processes.
The HLT daemon starts the CMSSW applications as forked child processes, monitors their status, and can restart them in case of unexpected process termination.
Each child CMSSW process is allocated a number of CPU cores or hyperthreads (in the case of using simultaneous multithreading (SMT), a technique allowing separate tasks to run on the same CPU core), also taking into account the number of parallel threads configured in the CMSSW processes, as also described in Section~\ref{sec:hlt}.
By using such an allocation strategy, servers can be used at their maximum computing capacity.

\subsubsection{The HLT menu and output data streams}
\label{sec:daq:hltmenu}

The HLT configuration, also referred to as the ``HLT menu'' and described in Section~\ref{sec:hlt}, defines a set of paths in which physics objects are reconstructed and events are filtered based on specific physics requirements.
Events that are accepted through HLT paths for similar physics processes are stored in primary data sets, defined also in the HLT menu.
The primary data sets are defined such that the total event rate of each data set is kept within the limits imposed by the offline data processing, while minimizing the overlap among different data sets.
In turn, the primary data sets are grouped into streams, corresponding to the actual files that are output by the HLT processes and transferred to the \Tier0 for offline processing.
In addition to those used to collect events for physics analyses, the HLT defines dedicated paths and streams to collect data for detector calibrations and for online data quality monitoring (DQM).
Finally, special paths and streams are used to collect statistical information about the HLT processes themselves, such as the individual trigger rates and CPU usage.

{\tolerance=800
Events selected by the HLT are serialized using ROOT~\cite{Brun:1997pa}, compressed using gzip~\cite{Deutsch:1996rfc}, LZMA~\cite{xz:2023git}, or Zstandard~\cite{Collet:2021rfc}, and written to files on the local partition.
Lossless compression used for the HLT output, applied to the entire serialized event payload, reduces the total event size to about 70\% of the input event size for proton runs (specified in Table~\ref{tab:daq:readout}), as observed in \Run2 and the first year of \Run3.
For proton runs, it is preferable to minimize the CPU usage of the compression in order to preserve resources for the HLT reconstruction.
The gzip tool, which was used in \Run2 and in 2022, typically used up to on the order of 1\% of the CPU in the HLT during \Run2 and 2022.
\par}

The application of Zstandard was evaluated after the first year of \Run3 and shown to have improved performance for a compression factor comparable to gzip, with an estimated use of less than 0.5\% of the CPU.
The HLT began using this algorithm for compression in 2023 for proton runs.
For the heavy ion run, the compressed size is estimated at about 3\MB for centrally colliding collision events and 1.2\MB for minimum-bias events, based on the 2018 heavy ion run.
The LZMA algorithm, which can typically improve the compression factor by an additional 10--15\%, but is an order of magnitude more CPU demanding than gzip or Zstandard, is being considered in order to maximize the HLT output rate within the DAQ bandwidth limits.
The approaches to size reduction of heavy-ion events are discussed in Section~\ref{sec:hlt:heavyion}.

Each output stream is written into a separate file and includes the event raw data, a selection or a summary of the HLT objects, and the trigger decision for each path specific to a stream.
The streamer file format used for output allows files belonging to the same stream, but produced by different processes and nodes, to be trivially concatenated.
This step, which is performed separately for each stream, is called merging.
In the first merging stage, streams are merged from local CMSSW processes and copied to the NFS-mounted output partition on the RU/BU.
Subsequent merging stages are handled by the storage and transfer system (STS), as described in Section~\ref{sec:daq:storagetransfer}.

The CMSSW input, output, and merging operate on sets of data and metadata files created separately for each stream at a regular time interval called the luminosity section (LS), spanning $2^{18}$ LHC orbits, which takes approximately 23.3\unit{s} as explained in Section~\ref{sec:bril}.
Metadata, written by the BU, by the CMSSW processes, and by the HLT output merging at LS intervals, specify information such as event counts, checksums, and total output size.
Bookkeeping (completion) and integrity checks are performed at the granularity of an LS by comparing input and output metadata.
In the case of a failed integrity check, alerts are raised to notify the shift crew, and the affected data are discarded and marked as missing in the bookkeeping.

The HLT daemon services handle all the local output collection and merging tasks on the FUs.
They also report the latency in event processing and merging to the BU application.
The BU uses this feedback to throttle event building in case of high delays and can activate throttling in case of high ramdisk occupancy.

\subsubsection{Monitoring}
\label{sec:daq:hltmonitoring}

{\tolerance=800
The file-based filter farm (F3) monitoring system was designed from the ground up in \Run2~\cite{Andre:2015dvu} around the Elasticsearch NoSQL database~\cite{elasticsearch:2023git}.
Structured JSON-format\-ted~\cite{Bray:2014rfc} monitoring data are injected into an Elasticsearch cluster from the HLT daemon service, STS, and other sources.
This includes bookkeeping such as input and output event counts, event file size, bandwidth, CPU usage, HLT information, and the hardware state at the granularity of each FU node.
Textual logs and error reporting from the HLT processes are also handled.
Information is stored for an extended time of up to several years, allowing real-time as well as follow-up analysis of performance and faults in the system.
Several web-based tools have been developed as frontends to the system, visualizing the information for experts and the shift crew.
Monitoring of the HLT performance is also collected using non-event streams, such as histograms, which are merged and shipped to the DQM system, and trigger statistics, which are merged and stored in a database.
\par}

\subsubsection{Evolution of the HLT farm}
\label{sec:daq:evolution}

The F3 was gradually expanded almost every year during \Run2, as the HLT processing requirements increased due to evolving LHC and detector conditions.
The FU nodes are typically replaced with a new generation of hardware after the end of their 5-year warranty periods, with the old nodes being assigned to the online cloud, as described in Section~\ref{sec:daq:overlaycloud}.
Table~\ref{tab:daq:hltevolution} summarizes the composition of the HLT farm at the end of \Run2 and at the beginning of \Run3, along with the computing power estimates based on the HEPSPEC 2006~\cite{Michelotto:2010zz} (HS06) benchmark measurements.
Prior to the start of \Run3 data taking, the entire farm was replaced with new AMD CPU nodes, each equipped with two NVIDIA T4 GPUs~\cite{Nvidia:2019web}.
In this configuration the average processing time of the HLT can be up to 500\ms per event, at the nominal L1 rate of 100\kHz.

\begin{table}[!ht]
\centering
\topcaption{%
    Summary of the HLT filter farm unit specifications, thermal design power, and performance based on HS06 in the final year of \Run2 and first year of \Run3.
}
\label{tab:daq:hltevolution}
\renewcommand{\arraystretch}{1.1}
\renewcommand{\thefootnote}{\alph{footnote}}
\cmsTable{\begin{tabular}{l@{\hspace{5\tabcolsep}}ccc@{\hspace{5\tabcolsep}}c}
    Run & & \Run2 (2018) & & \Run3 (2022) \\
    \hline
    Architecture & Intel Haswell & Intel Broadwell & Intel Skylake & AMD Milan \\
    CPU model & dual E5-2680v3 & dual E5-2680v4 & dual Gold 6130 & dual 7763 \\
    CPU cores & 2$\times$12 & 2$\times$14 & 2$\times$16 & 2$\times$64 \\
    Nominal freq. [GHz] & 2.5 & 2.4 & 2.1 & 2.45 \\
    Turbo freq. [GHz] & 3.3 & 3.3 & 3.7 & 3.5 \\
    TDP [W] & 120 & 120 & 125 & 280 \\
    Memory [GB] & 64 & 64 & 96 & 256 \\
    Nodes & 360 & 324 & 400 & 200 \\
    CPU cores (total) & 8640 & 9072 & 12800 & 25600 \\
    HS\footnotemark[1]/node & 538 & 659 & 773 & 3224 \\
    TDP [W]/kHS\footnotemark[1] & 223 & 182 & 162 & 87 \\
    kHS\footnotemark[1] & 194 & 214 & 309 & 645 \\
    GPU card & \NA & \NA & \NA & 2$\times$NVIDIA T4 \\[1ex]
    \multicolumn{5}{l}{\small\footnotemark[1]HS06 measurements only take into account the CPU performance. The precision is around 1\%.}
\end{tabular}}
\end{table}

Specific HLT reconstruction algorithms, such as those used in data scouting, described in Section~\ref{sec:hlt:scouting}, can be offloaded to these GPUs, reducing the average HLT event processing time by over 40\%, as shown in Fig.~\ref{fig:hlt:cputiming} and discussed in Section~\ref{sec:hlt}.
In comparison to an HLT farm equipped only with CPUs, this corresponds to a reduction in the farm's overall cost by approximately 15\% and its power consumption by 30\%.

On F3 nodes, CMSSW jobs are divided into two groups, and, for each group, the CPU and memory affinity is pinned to a single CPU socket NUMA domain on dual-socket AMD nodes.
Each GPU is attached to PCIe lanes on an individual CPU socket and assigned to the group of processes running on the same socket.
The NVIDIA multiprocess daemon is used to schedule GPU access between processes, providing a small performance enhancement over direct access.
A total of eight jobs with 32 threads and 24 streams each (\ie, parallel event processing pipelines in CMSSW) is used per node, since this configuration was found to fit within the memory capacity of the nodes and GPUs.
It was determined that increasing the number of threads per process in CMSSW adds a small additional overhead, and thus it was not further increased to keep the optimal processing capacity of the F3.

To support NFS data transfers in \Run3, a Juniper QFX5120 top-of-the-rack (ToR) switch is employed, which uses 8$\times$100\Gbs up-links to the event backbone network from each rack.
The network contains approximately 40 FU nodes with 25\Gbs Ethernet connections to the ToR switch.
A flat interconnect network is provided, enabling any FU to communicate with any RU/BU node.

\subsection{Storage and transfer system}
\label{sec:daq:storagetransfer}

The storage and transfer system is the last stage of the DAQ data flow.
It collects the HLT output from each RU/BU and associated FU group and writes it to a cluster file system.
It later transfers the data to \Tier0 for repacking from the streamer format into the ROOT format and to permanent storage on disk and tape servers.

The HLT daemon services on the FUs within the FU group concatenate output streamer files, analyze metadata, and copy the corresponding files into a dedicated RU/BU output partition.
This was a spinning disk RAID array in \Run2, and is a 200\GB ramdisk partition in \Run3 to support higher throughput.
The merger service running on the RU/BU nodes periodically polls the streamer and JSON metadata files in this area, written by each FU for a particular stream and luminosity section.
The JSON metadata files are used to verify the completion of per-LS output.
Once all FUs have copied a complete set of files, the service distributes tasks to several worker threads to read the output of all streamer files and append them into a single file location in the distributed file system.
For most streams and a majority of the output bandwidth, this step is done by simultaneously writing into a single file at a different offset.
This technique is used to efficiently merge data into the final file object, one per stream and LS.
This file can be transferred by a single copy operation to \Tier0, instead of requiring an additional read and write operation to perform such a concatenation.

The final stage of merging is performed on a set of dedicated STS nodes, with each handling a subset of streams.
For most streams this amounts to verification of the metadata and checking the completion of files written by the merger service from each RU/BU.
Fully merged streamer files, residing in a cluster file system, are handled and transferred to their destination by the transfer service, also running on the same nodes as the merger service.
For data destined for the \Tier0, a pool of threads starts the file copy jobs to transfer data to the EOS disk system~\cite{Peters:2015aba} using a high-speed link to the central data recording (CDR).
The transferred file size is  kept below 16 GB for optimal transfer throughput performance.
The network infrastructure used to transfer to the \Tier0 is detailed in Section~\ref{sec:daq:infrastructure}.
Streams are also delivered to several other destinations, such as the DQM and the calibration cluster, or are parsed to extract the HLT and L1 trigger rate monitoring information and inject it into the relational database.

Extensive bookkeeping is required to track all the files passing through the system.
This information is injected and visualized in the F3 monitoring system.
Metadata relevant for the transfers are also written to an SQL database for bookkeeping and to provide transfer metadata to the \Tier0.

For \Run2, Lustre~\cite{Andre:2015jty} was selected as the cluster file system, after initial evaluation in the DAQ integration system.
The production system consisted of two disk servers (OSS) and a metadata server (MDS), providing on the order of 9\GBs of total read and write throughput,
with around 500\TB of storage space provided by a redundant hard drive setup.

In \Run3, the storage system requirements are driven mainly by the heavy ion running, which, apart from centrally colliding lead ion events, aims to collect a large amount of minimum-bias events.
An estimated throughput of 17\GBs is required in a scenario that includes a trigger selection of around 1\kHz of centrally colliding heavy ion events and a minimum-bias rate of 10\kHz at the HLT output, as described in Section~\ref{sec:daq:hltmenu}.

To cover these requirements, a hardware refresh was pursued in \Run3, retaining the same file system technology while significantly improving the bandwidth capability compared to \Run2.
A new system comprising two DDN EXAScaler~\cite{DDN:2023web} SFA7990X data servers and a single SFA400NVX metadata server was acquired for the task.
Each data server consists of 124 SAS 8\TB 7.2k RPM hard disks, organized in a RAID6 array.
The metadata server comprises 23 SSDs, each of 1.8\TB capacity, in a RAID6 array.
The system is connected via the chassis-based Ethernet switch also used for the EVB and HLT.
Seven STS nodes were additionally installed and are connected to the chassis-based Ethernet switch.
Together with the RU/BU nodes, these nodes are set up as Lustre clients and the file system made accessible through mount points.
The \Run3 Lustre system storage space was scaled to provide several days of storage space for proton LHC runs or up to a day for heavy ion runs should the \Tier0 connection fail.
The usable disk capacity is 1.2\uPB, and the system is capable of simultaneously writing 24\GBs and reading 11\GBs using standard file system benchmarking tools. The TCP/IP protocol was used for communication in these tests.
The system is capable of temporarily prioritizing the write bandwidth at the expense of the read bandwidth, up to the limit of the available disk space.
This is particularly useful for runs with high peak bandwidth, such as heavy ion runs, where the read rate recovers and allows draining the accumulated files towards the end of the LHC fill and in interfill periods.

\subsection{Trigger throttling system}
\label{sec:daq:tts}

The trigger throttling system (TTS) collects fast readiness signals from all FEDs, merges them per TTC partition with a priority logic, and makes them available to the trigger control logic to avoid buffer overflows by inhibiting triggers or drive recovery actions.
While in \Run1 this trigger control logic was implemented in the trigger control system (TCS), since \Run2 it is part of the trigger control and distribution
system (TCDS), described in Section~\ref{sec:daq:tcds}.
A FED may signal the following main TTS states:\ Ready (to accept triggers), Warning/Busy (buffer fill level above high-water mark), Out-Of-Sync (synchronization loss), and Error (other error situation), in ascending order of priority.
The TCDS reacts to the highest-priority TTS state by inhibiting triggers, executing a re-synchronization sequence, or executing a reset-sequence.

For the legacy FEDs, TTS signals are sent using four LVDS pairs, which are merged using compact-PCI based fast merging modules (FMMs)~\cite{Bauer:2007vdr} that combine the TTS states of up to 20 FEDs using a priority logic.
For partitions with more than 20 FEDs, the FMMs are arranged in a tree structure.
For upgraded FEDs, TTS signals, transmitted over optical fiber, are merged with a similar logic by \uTCA-based TCDS partition interface (PI) modules, as described in Section~\ref{sec:daq:tcds}.

\subsection{Trigger control and distribution system}
\label{sec:daq:tcds}

At the start of \Run2, a new TCDS~\cite{Hegeman:2016hlt} was introduced, replacing the \Run1 trigger control system that was integrated into the L1 trigger system, and the \Run1 TTC system.
The TCDS provides support for an enlarged range of detector partitions and for detector backends sending their trigger throttling signals over optical fiber.
These were both needed to integrate the additional and upgraded subdetectors with \uTCA backends during \Run2.
The TCDS distributes timing and control (synchronization) data that are flowing to the detector backends and frontends and receives back status information related to the readiness of the detector systems to handle more triggers.
The clock reference that is distributed along with the fast control information is synchronous with the beams in the LHC, which is required to keep the data taking in step across the various detector systems.
The TCDS is implemented in the \uTCA architecture.

\begin{figure}[!t]
\centering
\includegraphics[width=\textwidth]{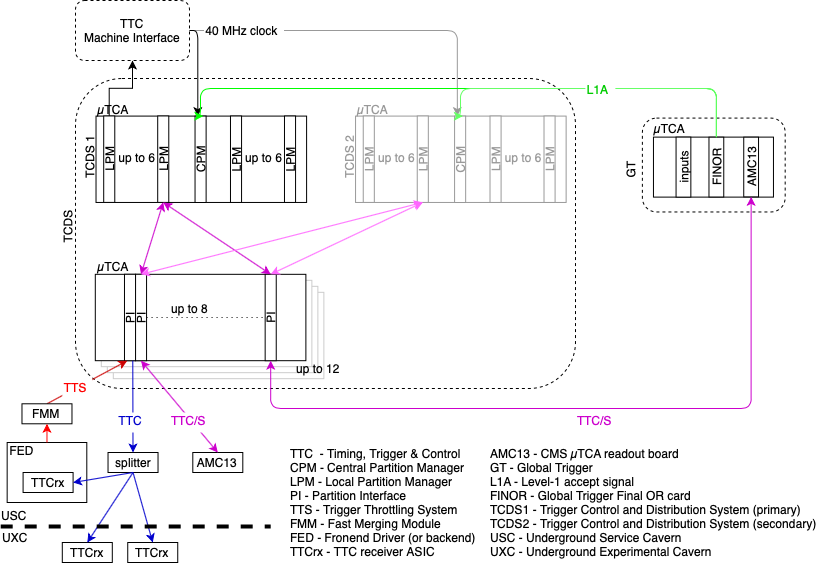}
\caption{%
    Overview of the CMS trigger control and distribution system (TCDS).
}
\label{fig:daq:tcds}
\end{figure}

As illustrated in Fig.~\ref{fig:daq:tcds}, a central crate contains a central partition manager (CPM) board and up to twelve local partition manager (LPM) boards.
The CPM receives the LHC clock from the TTC machine interface and the L1 accept signal from the global trigger, as described in Section~\ref{sec:l1trigger:mugt}).
Each of the LPM boards contains eight independent integrated CMS interface (iCI) logic blocks that are able to control a detector partition for local running.
The iCI blocks translate the generic TCDS synchronization commands to subdetector specific commands.
The iCI block also contains a partition-specific emulator of the APV25 tracker readout chip buffer levels that inhibits triggers that would lead to overflows of these buffers.
Each LPM board also contains two partition manager (PM) blocks that can orchestrate combined runs with sets of partitions in the same LPM.
The CPM contains one PM block orchestrating global runs with all partitions.
The PM blocks provide the following functionality:
\begin{itemize}
\item Trigger throttling, taking into account:
\begin{itemize}
    \item the TTS state of partitions;
    \item trigger rules, suppressing bursts of triggers by limiting the number of triggers in certain windows of bunch crossings, as required by the subdetector frontend electronics;
    \item protection against overflows in the pre-shower frontend ASIC (in a similar way to the protection against overflows in the APV25);
    \item resonant trigger protection, \ie, protection against triggers arriving at regular intervals over prolonged periods, which could give rise to resonant vibrations that might damage systems (such as wire bonds),
    \item the DAQ back pressure to the PM's readout link.
\end{itemize}
\item Bunch mask trigger veto that can be used to inhibit triggers (such as prefiring triggers) in LHC bunch crossings that are not filled.
\item Generation of random triggers.
\item Generation of (periodic) calibration triggers to trigger the readout of calibration events (interleaved with physics events).
\item Definition of sequences of synchronization commands such as run start, run stop, and detector backend recovery.
\item Automatic triggering of sequences based on the trigger throttling state, for example, for global re-synchronization or subsystem-specific recovery actions.
\item Monitoring of trigger rates, sequences, and dead times.
\end{itemize}
A secondary set of local and central partition manager hardware, connected in parallel to the primary partition managers, is available for final validation of firmware and/or software updates, or as hot spares.

Each of the iCI blocks is connected to a partition interface (PI) board, fanning out triggers and synchronization commands to a single detector partition and merging the throttling signals from the partition. These boards are located in separate \uTCA crates.
For the bulk of the backend boards, the PI board receives the TTS signal from an FMM, as described in Section~\ref{sec:daq:tts}, while triggers and synchronization commands are sent via backwards-compatible TTC fibers and the legacy TTC hardware.
For upgraded subsystems, the PI sends triggers and synchronization commands to up to 10 backend boards via bidirectional optical links.
In the reverse direction, it receives TTS signals from the backends and merges these signals with a priority logic similar to the FMM.

\subsection{Networking and computing infrastructure}
\label{sec:daq:infrastructure}

The operation of the CMS experiment is supported by the CMS service network, a high-perfor\-mance distributed network connecting all the computers directly related to the operation of the experiment.
While isolated from the CERN general purpose network (GPN), connections to the CDR and LHC technical network (LHC-TN), as well as the GPN, are allowed for use in specific cases.
The network follows a redundant design established in the early years of the experiment.
Currently, routers provide 1\Gbs Ethernet connections to the servers.
Available bandwidth is commonly used for user access, control, and monitoring.
The majority of the DAQ computers are located in racks, to which the service network is provided using ToR switches.

The DAQ data networks have been significantly redesigned and upgraded throughout the lifetime of the experiment, as described previously in this section.
To facilitate sending data taken by the experiment in \Run3 to the \Tier0, four 100\Gbs links from the event backbone network to the CDR are employed.
This facility is also used by the online cloud for accessing external data stores and services, as described in Section~\ref{sec:daq:overlaycloud}.

A network-attached storage (NAS) system is used by CMS for core storage needs.
This includes home directories and subsystem storage, exposed via standard remote file system protocols to nodes in the CMS network.
The system acquired for \Run3 provides 1.3\uPB of storage space.
The CMS online computing system runs the CERN-supported version of Linux, CERN \CentOSseven (CC7)~\cite{Centos:2020web}, as well as Red Hat Enterprise \Linuxeight (RHEL8)~\cite{RedHat:2023web}.
Both of these operating systems are also used for the DAQ operation.
Machines are initially installed (or reinstalled) using a network preboot execution environment (PXE) boot installation service.
After \Run1, the Quattor~\cite{GarciaLeiva:2004jgc} system was replaced with the Puppet~\cite{puppet:2023git} software configuration management tool, which handles the OS installation, configuration, and deployment of the online software in distributed and reproducible fashion on thousands of DAQ and subsystem computers.

\subsubsection{Virtualization}

Virtualization allows the reuse of physical computers by sharing them for multiple services that are running in virtual-machine (VM) instances, while simultaneously providing an isolated execution environment for each service.
A virtual-machine infrastructure has been set up and used to run the CMS online services based on the oVirt~\cite{oVirt:2023git} open-source virtualization management platform on the \CentOSseven OS.
The VMs running on this infrastructure are installed from OS boot images provided on the network and using the Puppet configuration management system described above.
The VMs are accessible on the service network via dedicated network names assigned by the DNS.

\subsubsection{Online cloud}
\label{sec:daq:overlaycloud}

The HLT farm consists of a large number of multicore nodes, representing significant computing power.
During interfill periods, week-long technical stops, and longer LHC and detector upgrade periods, this capacity is mostly unused for the HLT tasks.
Thus, starting in \Run2, infrastructure was developed~\cite{daSilvaGomes:2018pqz} to run an Openstack-based cloud overlay to use the computing resources for offline production jobs.
In order to not overlap with the HLT-based workflows, the FUs are able to run virtual machines (VM) that provide the necessary software environment for the worldwide LHC computing grid (WLCG) CMS jobs.
These jobs run tasks such as simulation and reconstruction in full isolation from the HLT.
In Section~\ref{sec:offline}, more details on the WLCG and workflows are provided.
An API is implemented in the HLT daemon to facilitate the automatic suspension and startup of the cloud mode or resumption of the HLT.
The cloud images can be suspended to disk and later resumed, allowing the quick save of unfinished jobs when, for example, the cloud needs to be suspended based on the LHC status.
The system can also perform the switch automatically, reacting to the LHC state or, in another mode, monitoring the HLT CPU usage and dynamically re-allocating unused fractions of the HLT to the cloud.
About 5\% of the farm remains in HLT mode at all times to provide an operable HLT for cosmic ray data taking, commissioning, and tests.

Openstack VMs use a virtual local area network (VLAN) for access to the CERN services such as EOS, where the job input data, as well as the destination of job results, are located.
The VLAN is implemented in the event backbone network and routed through CDR links to the CERN IT infrastructure.

The overlay cloud has been extensively used since \Run2 and successfully expands the CMS computing resources through opportunistic re-purposing of the hardware.
Furthermore, it consists not only of the active HLT cluster, but also computing nodes that were retired from the HLT.
They are kept in cloud operation as long as the computing infrastructure support is possible.
Overall, the online cloud, which is classified as a \Tier2 CMS site, as described in Section~\ref{sec:offline}, is one of the major contributors to CMS offline computing, comparable to the largest \Tier1 sites in the amount of CMS production workload.

\subsection{Software, control, and monitoring of the DAQ}
\label{sec:daq:software}

{\tolerance=800
Two software frameworks have been developed within CMS to implement the bulk of the experiment's online software:\ the C++ based XDAQ framework~\cite{Gutleber:2003cd}, used to implement hardware access and data transport, and the Java based run control and monitoring system (RCMS)~\cite{Brigljevic:2003kv, Bauer:2012ww}, used to implement the hierarchical control structure and main user interface.
These two frameworks, which have been adopted by the central DAQ system and by all subdetectors, are maintained and enhanced according to the evolving requirements of the experiment and continue to be used in \Run3.
\par}

The XDAQ software is a platform designed specifically for the implementation of distributed DAQ systems.
It has a layered middleware structure, providing support for communication, a web user interface, high-speed networking, hardware access, multithreading, performance tuning, monitoring, error reporting, and logging.
The XDAQ system builds upon industry standards, open protocols, and libraries, \eg, TCP, HTTP, XML, and Apache Xerces.
Notable enhancements to XDAQ include the development of new ``peer transport'' plug-ins to support new network technologies, such as RDMA using Infiniband Verbs~\cite{Bauer:2012fea, Bawej:2014fua}, and a new service-based approach to the configuration of the built-in monitoring and alarming infrastructure~\cite{Bauer:2012mz}.
New XDAQ-based applications have been added to control new types of custom hardware and existing applications, such as the event builder enhanced with features such as load balancing and fault tolerance.

{\tolerance=800
The RCMS is a framework based on web applications running inside container instances (Apache Tomcat), which provides the building blocks to compose a distributed hierarchy of nodes to control and monitor the state of XDAQ applications and other online applications used during data taking.
Control nodes are based on state machines, with system-specific control logic implemented in Java.
They are steered and monitored through web user interfaces.
The RCMS includes database schemas to hold the configuration of all software components, define hardware configurations, and manage the complex interconnects required for the two-stage event builder.
Extensive tools for configuration management are available.
In the area of the RCMS, the development of guidance systems for operators and automation have helped to make the operation of the experiment less error-prone and less reliant on the knowledge of experts.
The overall configuration of the experiment can be automatically selected based on the state of the LHC, actions needed in response to certain LHC state changes, and actions like high-voltage ramping in the detectors that are performed automatically~\cite{Bauer:2012ww}.
Recovery from regular single-event upsets and from other typical data-taking problems is fully automated~\cite{Bauer:2014eya, Andre:2017wdy}.
Configuration management tools have been enhanced according to new requirements in \Run2 and \Run3 supporting, for example, fine-grained data-flow optimization according to the network topology.
\par}

The filter farm and the storage and transfer system are controlled and monitored by online software based on Python and Elasticsearch, as described in Sections~\ref{sec:daq:hltsoftware} and~\ref{sec:daq:hltmonitoring}.

A number of monitoring clients transform the raw monitoring data from both the XDAQ monitoring system and the Elasticsearch-based monitoring system into web-based graphical and textual monitoring displays used by the shift crew.
The monitoring clients typically display instant data with a latency of a few seconds and can also be used to browse historic data to facilitate post-mortem analysis.
The Java-based DAQ expert tool~\cite{PH:2018jxj, CMS:2018paa, Badaro:2020fyf} detects all common data-taking problems by evaluating rules encapsulated in logic modules using snapshots of monitoring data.
This helps the shift-crew with the sometimes difficult task of pin-pointing the cause of data-flow problems.
With problems for which a recovery is known, the tool can drive completely automatic recovery actions.
It consists of several micro-services responsible for reasoning, notification, and control of the recovery.

A switch monitoring system with a web-based graphical representation was developed to monitor the link status and performance metrics of the DAQ data networks and assist experts in diagnosing network-related failures.

In addition to specific aggregation and presentation tools for the DAQ and HLT, a general service is provided to the subsystems and collaboration at large to aggregate and present online monitoring data stored in the different databases.
The online monitoring service (OMS) is a new \Run3 software tool replacing a set of web-based monitoring tools (WBM~\cite{Lopez-Perez:2017ixp}) used in \Run1 and \Run2 to provide unified remote access to the monitoring data.
The OMS uses a generic relational database model (Data Warehouse) and interface, and a web-based presentation framework by which information across heterogeneous data sources and formats is aggregated and presented.
The presentation is organized in a structure of folders and pages that contain portlets typically displaying information in the form of tables and graphs, showing, for example, run and fill details, trigger rates, or subdetector monitoring.

\subsection{MiniDAQ}
\label{sec:daq:minidaq}

In addition to the global DAQ system, self-service DAQ systems, called MiniDAQs, are provided for most of the CMS subdetectors.
These setups can be used at any time by the subdetector groups for calibration runs, tests, and debugging using detector partitions that are not participating in the global data taking.
These setups have proven extremely useful since they allow independent testing under almost the same conditions as in global data taking.
Trigger control for MiniDAQ systems is provided by one of the PM blocks in a TCDS LPM.
During \Run1 and \Run2, these MiniDAQ systems ran on dedicated RU, BU, and FU servers and provided limited bandwidth with respect to the global DAQ system.
In \Run3, MiniDAQ systems share the RU/BU and FU servers with the global system and provide a full bandwidth to each subdetector.
The configuration of the MiniDAQ systems is dynamically updated to follow any changes in the global system.

\clearpage
\section{Level-1 trigger}
\label{sec:l1trigger}

The level-1 (L1) trigger is implemented in custom hardware processors.
It comprises calorimeter and muon trigger systems that provide jets, \egamma, hadronic \PGt, and muon candidates, along with calculations of energy sums, to the global trigger (GT).
At the GT, the trigger decision is generated, based on the multiplicity and kinematic information of the various candidate trigger objects.
The trigger configuration is implemented in a trigger ``menu'' comprised of several hundred ``seed'' algorithms.
Upon a positive GT decision, the full detector data are read out for further filtering in the higher-level trigger (HLT).
During LS1, in 2013--2014, the L1 trigger hardware was entirely upgraded, and has subsequently been operated successfully since 2016.
A detailed report on this \Phase1 L1 trigger upgrade and performance with \Run2 data is given in Ref.~\cite{CMS:TRG-17-001}.

For \Run3, although no major trigger hardware upgrade was performed, new capabilities have become available already through new algorithmic approaches, some of which are based on machine learning (ML) techniques.
Software such as \textsc{hls4ml}~\cite{Duarte:2018ite} facilitates the use of ML techniques in FPGAs.
Developments for \Run3 within the L1 trigger mostly focus on broadening the physics reach of CMS through the addition of dedicated triggers for long-lived particle (LLP) signatures, improving object measurement and calibration, utilizing the upgraded calorimeter trigger primitives (TPs) and additional muon TPs from the new GEM muon detector, and implementing additional calculations in the global trigger to provide greater flexibility in the design of L1 trigger algorithms.

The addition of a 40\MHz scouting system, commissioned in the early stages of \Run3, that receives data from both the calorimeter and muon L1 trigger subsystems, has the potential to further broaden the physics reach of CMS.
It enables the readout of unfiltered data, reconstructed in situ at limited precision but at full bunch-crossing rate, and provides unprecedented monitoring capabilities.
The following sections describe the \Run3 developments specific to each of the L1 trigger subsystems.

\subsection{Calorimeter trigger}

\subsubsection{Calorimeter \layer1 trigger}

The calorimeter \layer1 trigger receives TPs from ECAL, HCAL, and HF, calibrates them, combines the ECAL and HCAL TPs into single trigger towers (TTs), and transmits the TTs to \layer2 for further processing.
Calorimeter TPs for triggered events are readout to DAQ, and used in the data quality monitoring (DQM) system, where they provide the input to the software emulator, such that online and emulated data can be compared in real time for monitoring purposes.
For \Run3, \layer1 receives updated HCAL TPs (Section~\ref{sec:hcal:trigger}), which improves the mitigation of out-of-time pileup, and updated ECAL TPs (Section~\ref{sec:ecal:trigger}) with improved rejection of spikes caused by particles striking the avalanche photodiodes.

The TT energies are calibrated to account for energy losses due to inactive material in front of the calorimeters.
The calibration is performed separately for each calorimeter.
Since the inactive-material map is symmetric in $\phi$, the calibration is performed as a function of $\eta$ and transverse energy, \ET, only.
Whereas for the ECAL TP calibrations, the scale factors vary by less than 20\% across the \ET range, the HCAL and HF calibration scale factors are much more \ET dependent, varying by about 50\%.
Both ECAL and HCAL scale factors vary by about 20\% across the $\eta$ range.

Due to the large volume of data produced by the TPs and limited DAQ bandwidth, only TPs for triggered events are read out by the DAQ system.
To study possible trigger bias, for approximately every 100th event that passes the full L1 trigger selection, validation data are read out, that also contain the TT information.
In \Run3, validation events contain the ECAL TP data for five bunch crossings, including the two bunch crossings before and after the triggered bunch crossing.
This information is useful for studying unexpected detector effects, and can be used to study ECAL prefiring (Section~\ref{sec:ecal:trigger}), both using the DQM and offline analysis.
This additional ECAL TP data can also be used for various optimization studies, such as monitoring the timing of the ECAL TPs through the L1 trigger path during commissioning phases.

\subsubsection{Calorimeter \layer2 trigger}

Calorimeter \layer2 receives calibrated TTs from \layer1, reconstructs jet, \egamma, and \PGt candidates, and computes energy sums.
The energies of jet, \egamma, and \PGt candidates are calibrated as a function of \pt and $\eta$, and isolation and ID criteria are applied to \egamma and \PGt candidates.
Pileup mitigation is applied to all objects to reduce the rates while maintaining high efficiencies.
The \layer2 hardware remains the same as for \Run2.
Ten main processor cards each process data from the entire calorimeter for a single bunch crossing in a time-multiplexed configuration.
A single demultiplexer processor receives data from the main processors, performs the final calculation of the energy sums, and forwards the object collections to the global trigger.
More details can be found in Ref.~\cite{CMS:TRG-17-001}.

A range of improvements to the \layer2 algorithms are being investigated for \Run3, most of which involve utilizing the updated ECAL and HCAL calorimeter TPs as discussed in Sections~\ref{sec:ecal:trigger} and~\ref{sec:hcal:trigger}, respectively.
In particular, the ability to trigger on LLP signatures by identifying displaced jets using the additional HCAL timing and depth information available for \Run3 is being pursued to help broaden the LLP physics program of CMS.

Jets are reconstructed by summing the energies of a 9$\times$9 window of TTs centered on a jet seed that must have an energy greater than 4\GeV, which corresponds to approximately the same jet size as jets reconstructed offline with $\DR=0.4$ within the barrel calorimeter.
The energy contribution due to pileup is estimated by summing the three lowest energy out of the four 3$\times$9 regions on the boundaries of the jet and subtracting this pileup estimate from the jet energy, which is then calibrated.

For \Run3, an LLP jet identification algorithm is implemented that uses the HCAL timing and depth information.
Each TT has an HCAL feature bit set in \layer1, which compresses six feature bits received from the HCAL backend, containing timing and depth information, into one feature bit.
When the jets are reconstructed from TTs at \layer2, an LLP jet ID bit is set if the jet contains more than a configurable number of TTs with the HCAL feature bit set.
Additional LLP jet algorithms have been added to the GT menu that require the LLP jet ID bit to be true.
In addition to tagging LLPs, the ability to tag boosted jets with substructure using a pattern-matching technique, and the use of ML techniques to calibrate the jet energy and perform pileup subtraction are being investigated for \Run3.

The missing transverse momentum (\ptmiss) is calculated as the vector sum of the TT energies across the full calorimeter.
During \Run2, a significant increase in the instantaneous luminosity and thus the pileup per bunch crossing revealed a nonlinear relation between pileup and L1 \ptmiss trigger rate, leading to a large increase of rate at large pileup.
To maintain the L1 \ptmiss trigger thresholds while keeping the rates manageable, a pileup mitigation procedure was implemented using a lookup table (LUT) to exclude TTs below a configurable energy threshold from the \ptmiss calculation, based on the $\eta$ of the TT, and an estimate of the pileup of the event.
The TT energy thresholds were rederived for \Run3, using Monte Carlo (MC) simulation samples containing inclusive \pp events with a pileup distribution reflecting that of \Run3.

\begin{figure}[!ht]
\centering
\includegraphics[width=0.48\textwidth]{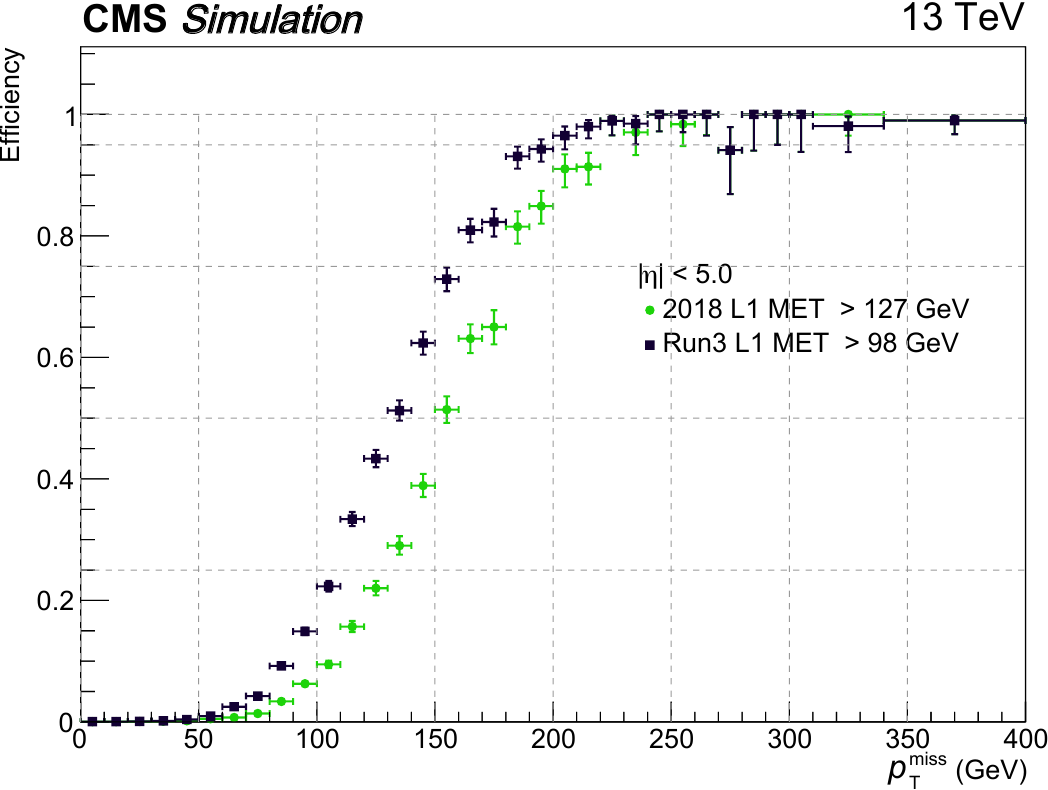}
\caption{%
    Comparison of the L1 \ptmiss trigger efficiency using pileup mitigation in 2018 (circles) and in \Run3 (squares) for thresholds that provide a rate of 4.3\kHz, for $\PZ\to\PGm\PGm$ events.
}
\label{fig:l1trigger:ptmisspu}
\end{figure}

A comparison of the L1 \ptmiss trigger efficiencies between 2018 and \Run3, for different L1 \ptmiss trigger thresholds that provide the same L1 trigger rate, is shown in Fig.~\ref{fig:l1trigger:ptmisspu}.
Compared to the 2018 LUT, the updated \Run3 LUTs show a significant improvement of the efficiency relative to the true \ptmiss in the event.
As the LHC conditions evolve, the pileup mitigation will be updated to ensure optimal performance.
In the current tuning, the energy threshold below which TTs are excluded is set such that 99.5\% of TTs are excluded for inclusive \pp events.

The \egamma and \PGt candidates are constructed by clustering TTs containing energy deposits greater than 1\GeV around a cluster seed tower with an energy deposit of at least 2\unit{GeV}.
This clustering is done dynamically using the available tower-level information.
The cluster is trimmed by removing towers, and a veto based on the trimmed cluster shape is applied, to reject pileup and reduce background rates.
A fine-grain veto is applied in the barrel calorimeter that quantifies the compactness of the electromagnetic shower within the seed tower to reject hadron-induced showers.
A veto is also applied that requires a low HCAL-to-ECAL energy ratio ($H/E$) in the seed tower, with different thresholds used in the barrel and the endcap regions.
Isolation requirements are applied to set an isolation bit and provide isolated candidates to the GT.
Merged \PGt candidates are constructed by merging nearby clusters that pass a set of proximity conditions, to capture multiprong hadronic \PGt decays.
Energy calibrations for \egamma candidates use the cluster shape and those for \PGt use $H/E$ and the presence of merged clusters, in addition to the \pt and $\eta$.
While no significant changes to either the \egamma or \PGt algorithms have been implemented for \Run3, the existing calibrations and isolation working points have been and will be rederived throughout \Run3 to reflect updated detector conditions and calorimeter TP algorithms.
Methods to improve the isolation working points of \egamma and \PGt candidates utilizing ML techniques are being implemented.

\subsection{Muon trigger}
\label{sec:l1trigger:muon}

The L1 muon trigger for \Run3 receives TPs from four partially overlapping muon subdetectors:\ DT, CSC, RPC, and GEM.
As described in detail in Section~\ref{sec:muon}, three of these subdetectors, DT, CSC, and RPC, were operated during \Run2, while the GEM detector was added as part of the \Phase1 upgrade and is used for the first time in \Run3.
The L1 muon trigger system reconstructs muon tracks and provides measurements of muon track parameters using TPs which provide position, timing, and quality information from detector hits.
In the barrel, accurate directional information is also provided.
The geometrical arrangement of the muon subdetectors, including the new GE1/1 detector in front of ME1/1, is shown in Fig.~\ref{fig:muon:quadrant}.
In this section, the changes to the muon track finders are discussed in detail.

The L1 muon trigger system in \Run3 comprises the same overall design as in \Run2.
Three muon track finders (TFs) reconstruct muon tracks in three distinct pseudorapidity regions using TPs from muon detectors.
The barrel muon track finder (BMTF) receives inputs from DT and RPC in the barrel ($\abseta<0.83$), the overlap muon track finder (OMTF) uses DT, CSC, and RPC in the overlap between barrel and endcap ($0.83<\abseta<1.2$), while the endcap muon track finder (EMTF) takes inputs from CSC, RPC, and GEM in the endcap ($1.2<\abseta<2.4$).
All three muon track finders transmit up to 36 muons each per bunch crossing to the global muon trigger (\uGMT), which resolves duplicates and transmits a maximum of eight muon tracks per bunch crossing to the GT, similar to \Run2.

In \Run3, all three muon track finders additionally provide measurements of parameters for muon tracks that are displaced from the primary interaction point.
The beamspot constraint requiring the track to originate from the interaction point is removed.
The newly available track parameters are used in the GT to provide L1 muon trigger seeds targeting displaced-muon signatures that could originate from LLPs.
Additionally, in \Run3, the EMTF receives TPs also from the GEM detectors, and this information can be used to improve both prompt and displaced muon triggering.

The new displaced-muon algorithms provide a \pt, measured without the beamspot constraint, and a transverse displacement, \dxy, from the beam line, for muon tracks obtained from propagating back to just the first muon station.
Algorithms optimized for prompt muons typically underestimate the \pt of highly displaced muons, as the displacement is mistaken as increased track curvature due to the beamspot constraint.
The new displaced-muon algorithms improve the \pt estimation for displaced muons, hence improving the efficiencies for these triggers.
The displaced TF algorithms in general do not affect the muon track building, but provide additional measurements of unconstrained quantities.
Due to this approach, the trigger efficiency is increased significantly for muon $\pt>10\GeV$ when the displacements are larger than 20\cm, while in the case of lower \pt muons, the displaced muon algorithms perform similarly to the prompt algorithms.
This is due to the fact that all the prompt TF algorithms have a minimum \pt assignment value (2--4\GeV depending on the TF), and the underestimation of \pt for displaced muons becomes less important at low \pt values.

\subsubsection{Barrel muon track finder (BMTF)}
\label{sec:l1trigger:bmtf}

In the BMTF, the kBMTF algorithm reconstructs muons in the barrel region using TPs received from the DTs and RPCs via the TwinMux concentrator cards, and has been used online since 2018~\cite{CMS:TRG-17-001}.
It is the successor of the original BMTF algorithm, used between 2016 and 2018, and based on an approximate Kalman filter algorithm in which muon tracks are reconstructed from detector hits starting from the outermost muon station and propagating inwards while updating the track parameters.

\begin{figure}[!ht]
\centering
\includegraphics[width=0.48\textwidth]{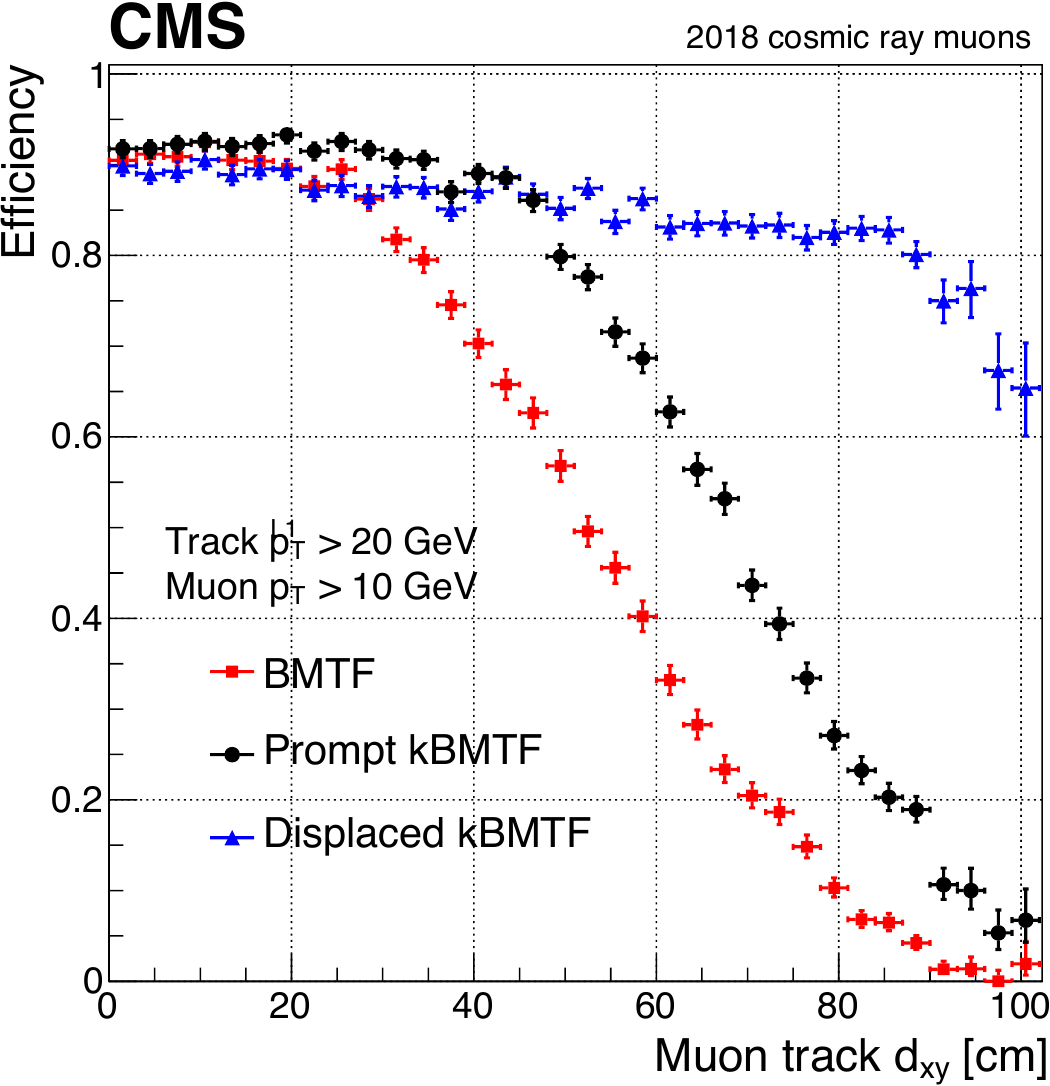}
\caption{%
    Displaced (blue) and prompt (black) kBMTF trigger efficiencies compared to the prompt BMTF (red) trigger efficiency with respect to the muon track \dxy, obtained using a sample of cosmic ray muons from 2018 data.
    The efficiencies are measured using muon candidates with $\pt>10\GeV$.
    The prompt kBMTF improves BMTF efficiencies up to about 90\% for up to 50\cm displacements, while displaced kBMTF retains efficiencies above 80\% for up to 90\cm displacements.
}
\label{fig:l1trigger:kbmtf}
\end{figure}

In \Run3, the updated version of the kBMTF algorithm mainly improves the displaced-muon triggering performance.
The new algorithm retains high efficiency for prompt muon tracks up to $\dxy\approx50\cm$, while the displaced-kBMTF algorithm increases the efficiency of muon tracks with displacements above 50\cm.
The expected performance for \Run3 was evaluated using a \Run2 cosmic ray muon data sample in which the displaced algorithm shows efficiencies around 80\% for L1 $\pt>10\GeV$, up to displacements of 100\cm (Fig.~\ref{fig:l1trigger:kbmtf}).

\subsubsection{Overlap muon track finder (OMTF)}

The OMTF builds muon tracks using the TPs from the DTs and RPCs in the barrel and CSCs and RPCs in the endcap.
The algorithm for \Run3, which is mostly identical to that of \Run2, uses a Bayes classifier algorithm based on precomputed patterns generated from simulated events to associate hits in each station with the reference hit in the pattern.
The patterns contain information about muon track propagation and the probability density function of the hit distribution in $\phi$ with respect to the reference hit.
There are 26 patterns for each muon charge, corresponding to \pt values between 2 and 140\GeV, that are then used to estimate the \pt of the muon track based on the likelihood that the track matches a pattern.
In \Run3, the OMTF includes additional patterns to improve displaced-muon triggering.
Although the general structure of the algorithm remains the same, the updated \Run3 algorithm now finds the best matching prompt and displaced patterns for a given muon track.
There are 22 displaced patterns for each muon charge corresponding to different \dxy values which are valid for high-momentum tracks ($\pt>30\GeV$).

\begin{figure}[!ht]
\centering
\includegraphics[width=0.6\textwidth]{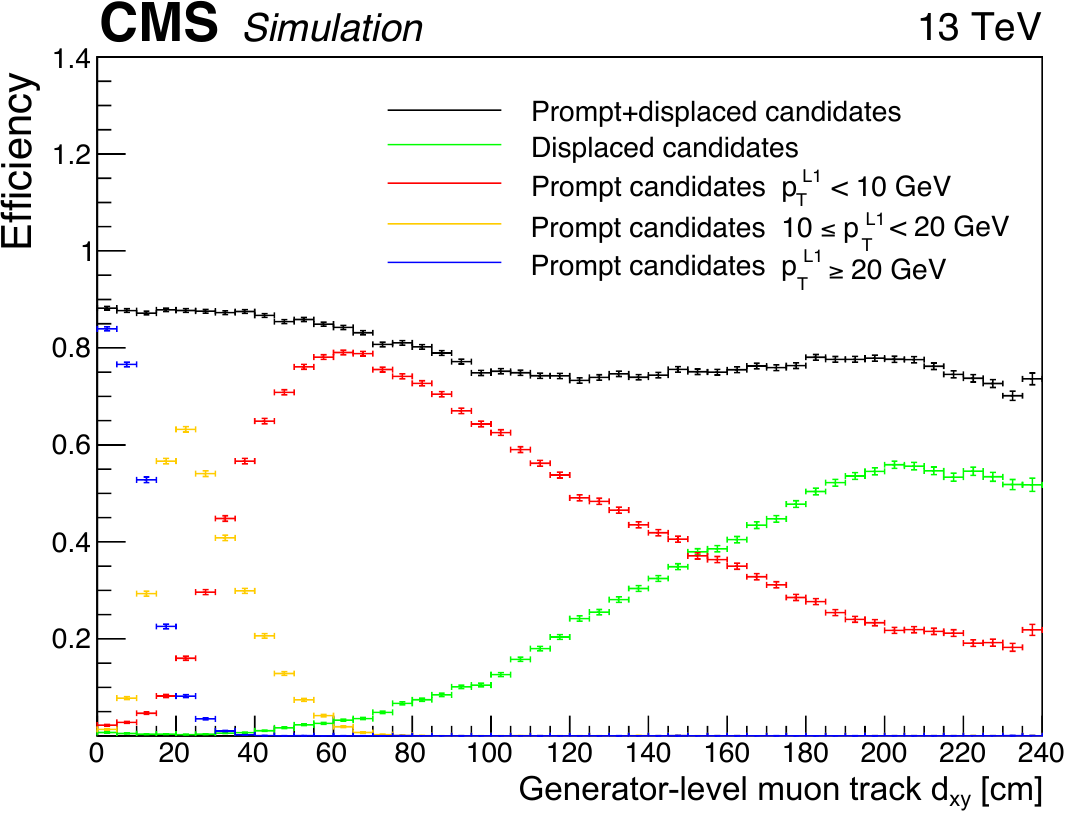}
\caption{%
    The OMTF trigger efficiencies for displaced and prompt algorithms with respect to muon track \dxy obtained using a displaced-muon gun sample.
    The efficiency curves are plotted for different values of the \pt estimate from the prompt algorithm (red, yellow, and blue), for the displaced algorithm (green), and for the combination (black).
    The prompt algorithm underestimates the \pt of displaced tracks, causing most of the tracks to have $\pt<10\GeV$.
    The displaced algorithm can recover these tracks and improve the efficiencies to be around 80\% for up to 200\cm displacements.
}
\label{fig:l1trigger:omtf}
\end{figure}

In \Run3, the OMTF uses prompt and displaced patterns in parallel for each track.
The algorithm identifies prompt-muon tracks and estimates their \pt using prompt patterns, and at the same time it estimates the \dxy of the displaced muon using the displaced patterns.
The expected performance for \Run3 was evaluated using a displaced-muon gun simulation sample, in which just muons are simulated with zero pileup, in order to compare ideal efficiencies, in which the displaced-OMTF algorithm shows efficiencies around 80\% up to displacements of 200\cm (Fig.~\ref{fig:l1trigger:omtf}).

\subsubsection{Endcap muon track finder (EMTF)}
\label{sec:l1trigger:emtf}

The EMTF builds muon tracks using TPs from CSCs, RPCs, and GEMs in both endcaps.
The \Run3 EMTF algorithm includes GEM inputs for the first time.
Already in \Run2, the algorithm used one TP per endcap muon station, with CSCs having priority over RPCs, to build muon tracks based on a set of predefined patterns.
After the patterns were found, the differences in $\phi$ and $\theta$ angles between hits in different stations were used to estimate the \pt of the muon track using a boosted decision tree (BDT).
The BDT was trained using simulated single-muon events and the output values of the BDT were stored in LUTs for fast evaluation.
Since the beginning of \Run3, the EMTF algorithm also receives the new GEM TPs in station 1, as well as the updated CSC TPs in all stations, which provide better position and bending resolution.
These TPs can be used both at the track building stage, as well as the \pt estimation stage to improve the performance of the EMTF algorithm.
Additionally, the EMTF uses a neural network (NN), implemented in FPGA logic, to estimate the \pt and \dxy of displaced-muon tracks, and forwards data to a hadronic shower trigger that uses the multiplicity of hits in the endcap CSCs to trigger on LLPs producing showers as they enter the endcap muon systems.

The first of the GEM detectors (GE1/1) is included in the EMTF algorithm in station 1 of the CMS endcap muon system.
The \Run3 EMTF algorithm can use GE1/1 hits in conjunction with ME1/1 hits to improve prompt and displaced trigger efficiencies and reduce rates caused by mismeasured muons.
Due to the placement of the GE1/1 and ME1/1 chambers in the CMS endcaps, the strong magnetic field in this region causes a larger bending of the muon track.
This bending information between GE1/1 and ME1/1 is foreseen to be used to improve the \pt assignment for both prompt and displaced-muon tracks.

\begin{figure}[!ht]
\centering
\includegraphics[width=0.48\textwidth]{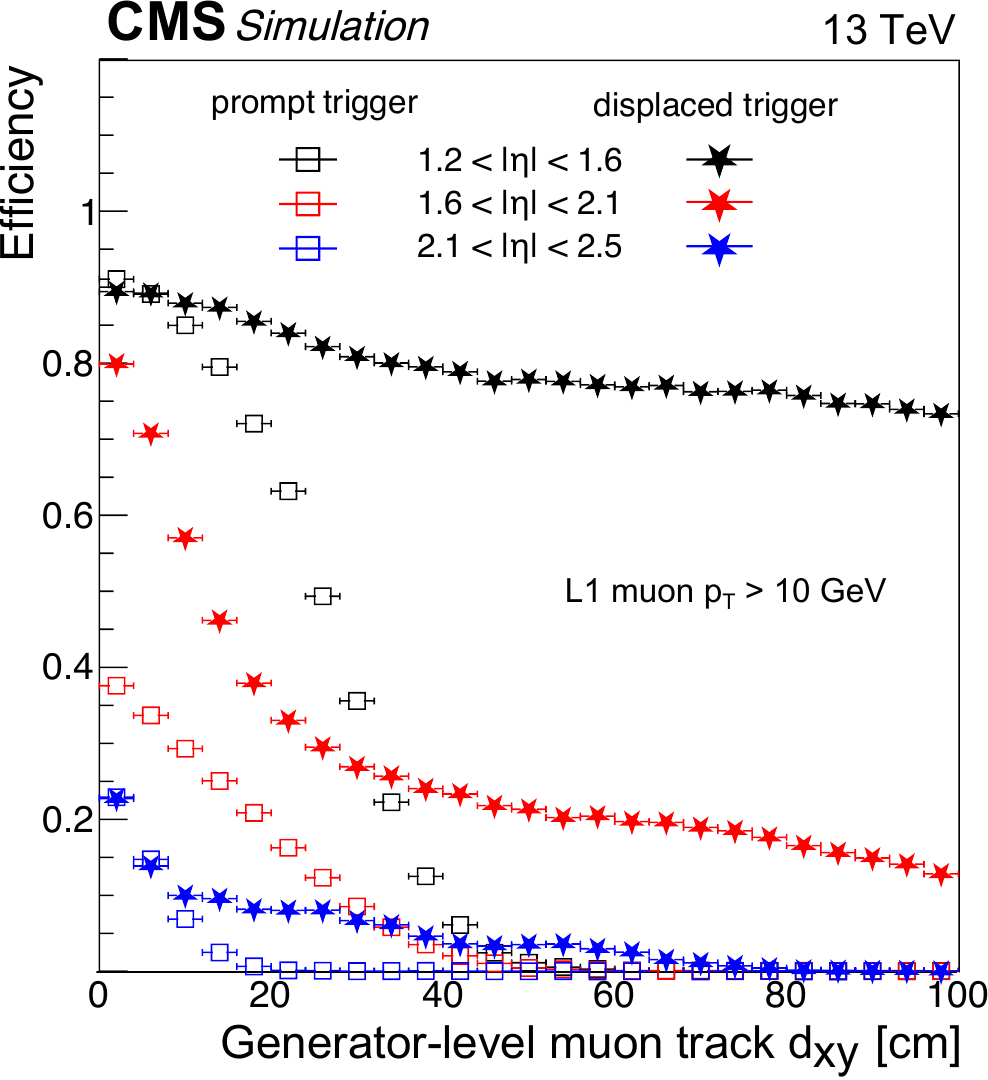}
\caption{%
    The EMTF trigger efficiencies for prompt and displaced-muon algorithms for L1 $\pt>10\GeV$ with respect to muon track \dxy obtained using a displaced-muon gun sample.
    The solid stars show displaced NN performance while hollow squares show the prompt BDT performance.
    The different colors show different $\eta$ regions:\ $1.2<\abseta<1.6$ (black), $1.6<\abseta<2.1$ (red), and $2.1<\abseta<2.5$ (blue).
}
\label{fig:l1trigger:emtf}
\end{figure}

Similar to the other track finders, the EMTF also includes a new algorithm to improve dis\-placed-muon triggering.
The EMTF for \Run3 includes a NN-based \pt and \dxy assignment algorithm, which runs in parallel to the prompt algorithm.
The NN has been directly incorporated into the EMTF firmware and estimates the \pt and \dxy of muon tracks that are built by the EMTF track building algorithm.
The EMTF performance for prompt muons originating from the primary vertex remains identical.
As shown in Fig.~\ref{fig:l1trigger:emtf}, the displaced-EMTF algorithm (NN-EMTF) shows an improved efficiency up to a \dxy of about 100\cm, while the prompt EMTF algorithm retains a high efficiency for prompt muon tracks up to about 25\cm.
The expected performance for \Run3 was evaluated using a displaced-muon gun simulation sample with zero pileup.
Zero-pileup samples were used since they are more useful in optimizing the NN and comparing ideal efficiencies.
For L1 $\pt>10\GeV$, the NN-EMTF algorithm shows efficiencies above 80\% for $1.2<\abseta<1.6$ and up to 100\cm displacements, while efficiencies for $1.6<\abseta<2.1$ and $2.1<\abseta<2.5$ at 60\cm displacement are around 20 and 5\%, respectively (Fig.~\ref{fig:l1trigger:emtf}).

Finally, the EMTF for \Run3 forwards CSC hit information to provide a standalone method for triggering on hadronic showers occurring in the CSC detectors.
An LLP decaying to hadronic particles within or slightly before the endcap muon systems can cause a shower of charged particles hitting the muon detectors, which are then recognized through a high hit multiplicity.
In \Run3, the CSC detector sends information on whether a high multiplicity is found in any given chamber by comparing the measured multiplicity to a set of predetermined thresholds, individually for each CSC station and ring combination.
The EMTF processes this information to decide whether there was a hadronic shower of a given quality in at least one of the CSC chambers in any given sector.
The expected performance of this algorithm for \Run3 was evaluated using multiple physics simulation samples containing LLPs and found to provide efficiencies around 30\%.

\subsection{Global trigger}
\label{sec:l1trigger:mugt}

\begin{figure}[!ht]
\centering
\includegraphics[width=0.5\textwidth]{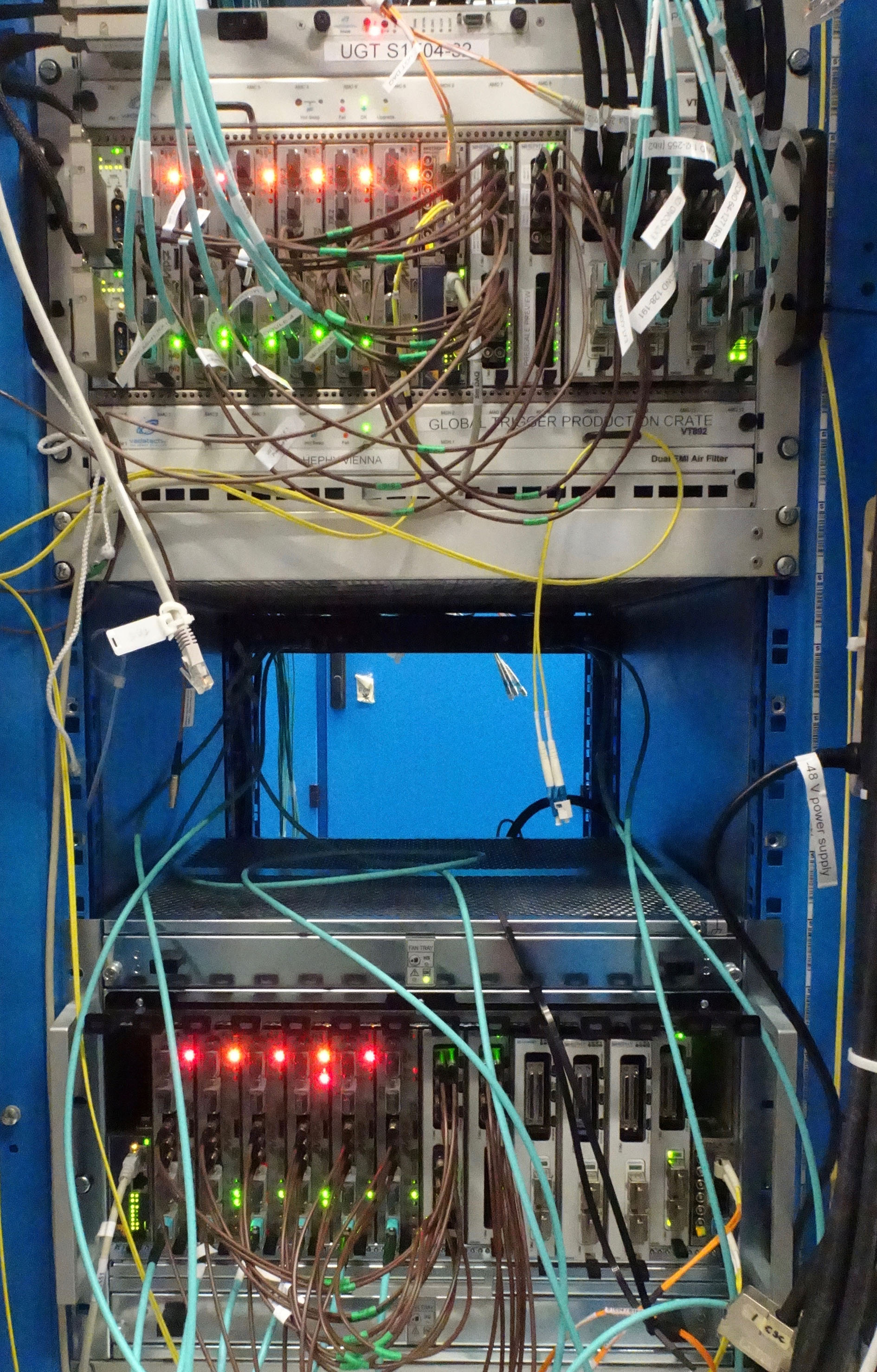}
\caption{%
    Production (upper) and the new test crate (lower) of the \uGT.
}
\label{fig:l1trigger:ugtcrates}
\end{figure}

The hardware for the present global trigger system, the \uGT, was installed as part of the \Phase1 upgrade, and was used for most of LHC \Run2 and is used for \Run3.
The flexibility of the L1 trigger system has allowed for the addition of a \uGT test crate containing the same hardware to be added to further extend the global trigger capabilities.
It is used for testing and development purposes, for example, to test new or experimental trigger menus, or to test \Phase2 algorithms using ML or autoencoding, as described in Ref.~\cite{CMS:TDR-021}.
It receives the same optical inputs as the production crate from a passive optical splitter panel.
For \Run3, this crate has been included as an optional component into the CMS data acquisition system. While the test crate does not issue triggers itself, for events triggered by the production crate, the full test crate data (inputs and results of calculations) is available in the CMS data stream for offline analysis.
The production and test crates are shown in Fig.~\ref{fig:l1trigger:ugtcrates}.

By running the test crate with the same firmware, and thus the same trigger algorithms, as the production crate, it is possible to carry out consistency and hardware checks.
More importantly, one can also run with different firmware in the test crate.
This allows trigger developers to test new trigger menus with actual data.
During normal data taking, a stream of zero-bias data is included, where the only trigger requirement is that the bunches for that crossing are filled.
The zero-bias trigger has a high prescale applied, such that only 1 in $N$ events are recorded, where $N$ is the prescale value.
This makes it possible to investigate the performance of new triggers in the test crate menu.
A test crate algorithm that is more restrictive than a specific algorithm in the production system can also be studied by using the latter algorithm as a reference, thus benefiting from a larger data sample.
The fact that the test crate can be included in data acquisition during normal running at no additional cost means that large amounts of realistic data can be collected, thus allowing for accurate tests even of very restrictive algorithms that only rarely provide a trigger.
By normalizing the offline data to the number of zero-bias triggers taken or by recording the online monitoring data, one can also check the total trigger rate of a new menu.
This is very useful, since it is not trivial to accurately estimate the total menu rate offline using the rate of individual algorithms, since multiple algorithms can fire the same event simultaneously.
This is important in order to guarantee that when running with the new menu the L1 trigger rate remains within the total bandwidth, so that deadtime is minimized.

The upgraded monitoring backend is configured to collect and store data from the test crate in order to provide integration with the central monitoring services.
Trigger menu rates are stored in the central Prometheus monitoring database, described in Section~\ref{sec:l1trigger:online}, together with those from the production crate for online monitoring and prompt offline studies.
Data are also expected to be sent to offline computing systems for long-term storage.
Alerting services are configured to warn shifters and experts with relaxed severity compared to the production system, as the system is not critical for data taking.

\subsection{Trigger menu}
\label{sec:l1trigger:menu}

Expanding the CMS physics reach significantly beyond that explored already during \Run2~\cite{CMS:TRG-17-001} requires important changes to the \Run3 trigger menu, and relies mostly on improvements to the L1 trigger algorithms.
The LHC beam conditions and filling scheme, as well as other aspects like luminosity leveling, described in more detail in Section~\ref{sec:pps:rpinsertion}, should be taken into account in order for the menu to be efficient and robust within the L1 bandwidth limit of about 100\kHz.

The updates to the calorimeter and muon trigger systems discussed earlier aim to retain the \Run2 physics coverage described in Ref.~\cite{CMS:TRG-17-001}, while providing additional access to signatures that do not originate at the primary vertex, such as displaced muons or displaced jets.
Furthermore, the GT is able to perform two new kinematic computations that can be utilized to increase the physics coverage:\ the three-body invariant mass and the di-object ratio between invariant mass and the $\DR=\sqrt{\smash[b]{(\Deta)^2+(\Dphi)^2}}$ between the objects.
Special L1 trigger menus are available in addition to the standard menus to account for various data-taking scenarios, as was done during \Run2, \eg, targeting signatures related to the physics of bottom quarks using the so-called ``\PB parking''~\cite{Bainbridge:2020pgi}, recording additional data for \PB physics and other studies, as further described in Section~\ref{sec:hlt:parking}.

\subsubsection{Trigger seeds for displaced muons}

Physics signatures involving LLPs were not well represented within the L1 trigger menu during \Run2, since specific algorithms capable of triggering on displaced objects within the detector were not yet implemented.
The changes for \Run3 provide the ability to trigger on displaced muons with higher efficiency compared to \Run2.
Updates to the muon algorithms in all three muon track finders provide unconstrained \pt and \dxy measurements for displaced muons, as discussed in Section~\ref{sec:l1trigger:muon}.

To benefit from the displaced muon algorithms, the L1 trigger menu of \Run3 extends that of \Run2 by offering seeds that use the unconstrained \pt and \dxy measurements.
These new seed features are available for muon objects with $\pt>10\GeV$, and possible additional selections are defined according to the planned physics program.
The chosen \pt threshold is based on the expected gain in the trigger efficiency described in Section~\ref{sec:l1trigger:muon}, concerning only muons with $\pt>10\GeV$ and displacements larger than approximately 20\cm.
Since the displaced muon algorithms do not perform better compared to the prompt algorithms in the case of low \pt muons, no new seed is created if the leading muon in the considered seed has a \pt below 10\unit{GeV}.

\subsubsection{Trigger seeds using new kinematic variables}

Low-mass resonance searches, for example those used to study the physics of bottom quarks, are based on targeting a final state with low \pt objects.
While the final states targeted in \PB-physics searches are often expected in the barrel region, many other interesting physics scenarios predict a wide pseudorapidity distribution for the final state objects.
Ideal seeds with a minimum object selection with low object thresholds would result in a high trigger rate that is unsustainable, exceeding the available trigger bandwidth under any LHC data taking conditions, and provide low purity.

During \Run2, the most used unprescaled seeds relying on a single object had relatively high \pt selections to keep trigger rates manageable, \eg, the single-muon trigger with $\pt>22\GeV$.
Special \PB parking triggers, used for events that are written to disk storage without full online event reconstruction at HLT, were developed to lower the L1 trigger thresholds while keeping within the HLT trigger bandwidth.
To provide reasonable rates, typical \PB parking seeds in general restrict pseudorapidity to the barrel and overlap regions, \eg, the double-muon trigger with no \pt selection but with $\abseta<1.5$ and $\DR_{\PGm\PGm}<1.4$.
However, the HLT thresholds used for the \PB parking seeds, generally above 5\GeV, would often significantly reduce the total acceptance of interesting physics signals.

The substantial L1 trigger rates related to low \pt trigger thresholds can alternatively be managed by using kinematic variables that are optimized to increase the acceptance of predicted signal events.
During \Run3, the GT is able to perform two new kinematic variable computations:\ the three-body invariant mass, and the di-object ratio between invariant mass and the \DR.
These allow the rate of seeds with low \pt thresholds to be reduced.
For example, the three-body invariant mass can be harnessed to target the decays of a \PGt into three muons, a final state with low-\pt final objects with wide pseudorapidity distributions.
Alternatively, the di-object ratio between invariant mass and the \DR\ could benefit a low-mass dimuon resonance search providing a dark photon interpretation.

\subsubsection{\Run3 trigger rates}

\begin{figure}[!ht]
\centering
\includegraphics[width=\textwidth]{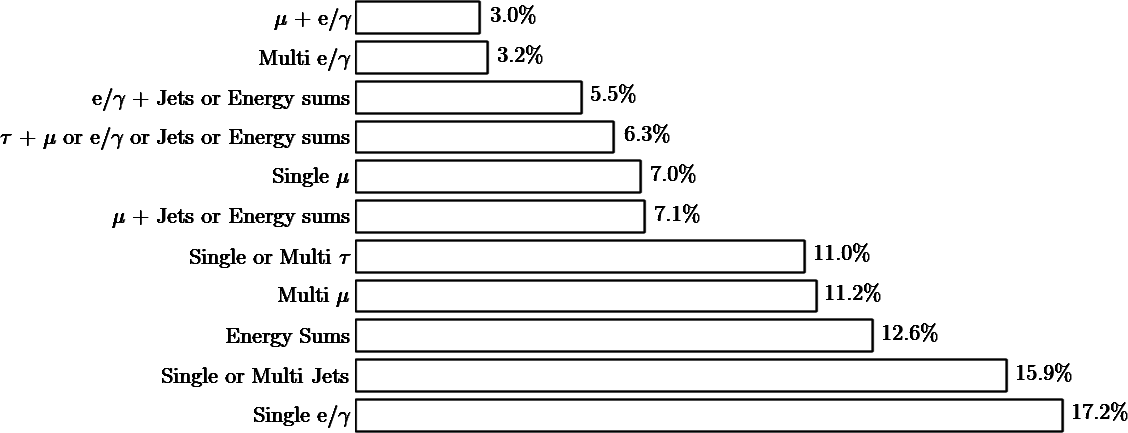}
\caption{%
    Fractions of the 100\kHz rate allocation for single- and multi-object triggers and cross triggers in the baseline \Run3 menu, calculated using \Run3 Monte Carlo simulation samples of inclusive \pp events with appropriate pileup.
}
\label{fig:l1trigger:run3rates}
\end{figure}

The baseline L1 trigger menu in \Run3 is identical to that of \Run2 detailed in Ref.~\cite{CMS:TRG-17-001}.
Since the beam and detector conditions, and upgraded TPs are different from those during \Run2, the baseline trigger menu has been reviewed using \Run3 simulation samples to understand if slight modifications to the existing seeds are necessary to respect the total rate budget of 100\kHz.
Two preliminary rate studies have been performed by reweighting the pileup distribution in \Run3 simulation samples, which has an unrealistic flat shape between 30 and 80.
Firstly, by reweighting the \Run3 MC simulation samples to the \Run2 pileup distribution, it was confirmed that the rates using \Run3 simulations with updated TPs provide the same rate as obtained from \Run2 data, for the same pileup conditions.
Secondly, the expected rates for realistic \Run3 pileup conditions were estimated by reweighting using a pileup distribution determined from the expected instantaneous luminosity during luminosity leveling.
This approach takes into account various beam conditions, such as the bunch-to-bunch variation in pileup, which can have a significant impact on the rates.
The rate allocation for single- and multi-object triggers and cross triggers is shown in Fig.~\ref{fig:l1trigger:run3rates}.
These results can be compared to a similar figure for the L1 trigger rate allocation under the \Run2 conditions, provided in Ref.~\cite{CMS:TRG-17-001}.
Additional seeds, relying on the aforementioned new features, are implemented on top of the baseline L1 trigger menu.
This menu is then tuned according to the desired physics program of CMS for \Run3.

\subsection{Online software and monitoring}
\label{sec:l1trigger:online}

During LS2, a number of improvements to the development workflow for the L1 trigger online software have been used.
Practices from the well-established methodologies DevOps~\cite{Erich:2017jsep} and Agile~\cite{Dingsoyr:2012jss} have been introduced in the \Run3 online software to simplify the development and deployment of upgrades and enable continuous improvement.
The online software is built in centrally-provided Docker images that provide a replicable development and test environment.
Every online software project employs continuous integration (CI) pipelines in GitLab to automatically build and test new commits.
In order to minimise maintenance efforts and facilitate the development of CI pipelines, an in-house Auto-DevOps platform was developed, taking inspiration from the GitLab Auto-DevOps project.
Auto-DevOps is employed to centrally provide a set of configurable and generic CI jobs that represent typical tasks in the online software development, such as building RPM package manager packages and Docker images.
Feedback from introducing standard CI pipelines in all software projects was excellent:
Given the large number of software components and teams that the online software consists of, this procedure established standard workflows that greatly increased the safety of new deployments.
New automated tests are constantly developed and integrated into the CI whenever new incidents occur, reinforcing the importance of feedback from operations to minimise disruptions of service.
Work is ongoing to develop a testing cluster managed by RedHat OpenShift running on CERN computing services where test deployments can be performed from CI pipelines.
The goal is to provide a complete environment where software checks on the entire infrastructure can be run before deploying on the CMS computing resources.

\begin{figure}[!ht]
\centering
\includegraphics[width=\textwidth]{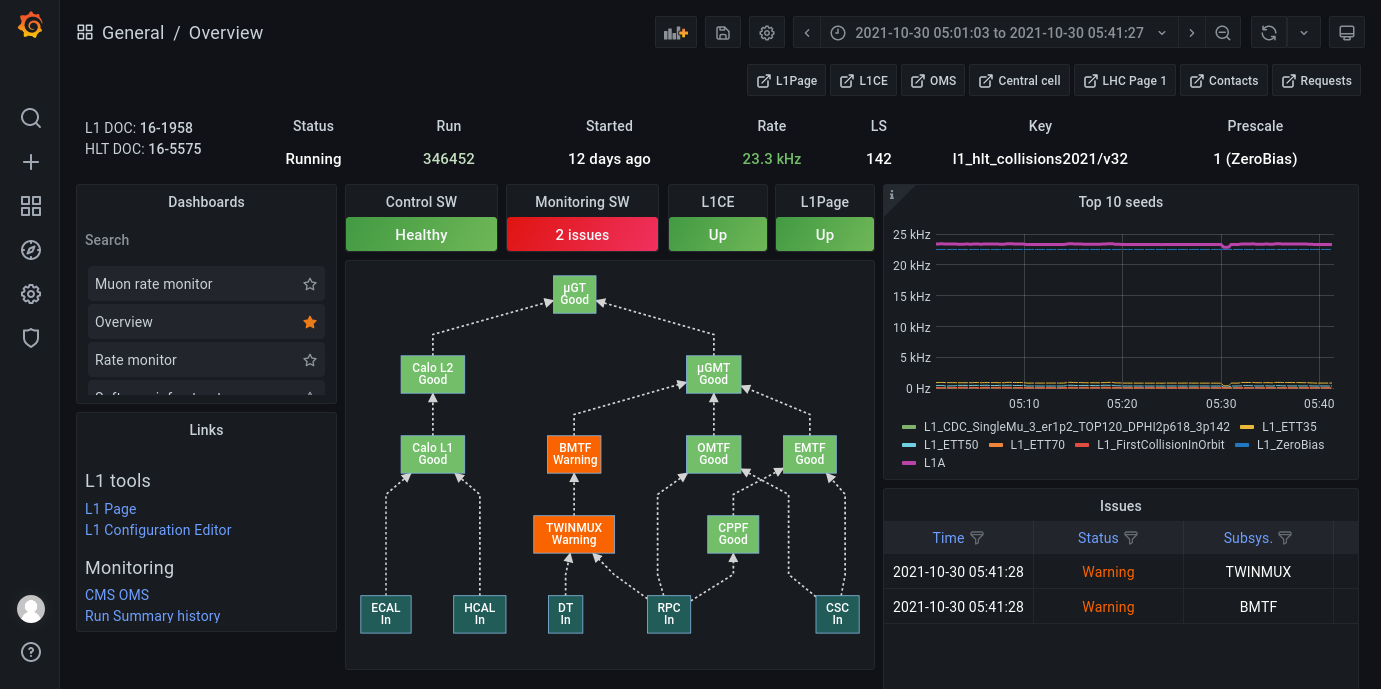}
\caption{%
    Screenshot of the Grafana L1 trigger monitoring dashboard.
}
\label{fig:l1trigger:grafana}
\end{figure}

The \Run3 monitoring backend is based on industry-standard tools and is centered around a Prometheus monitoring database instance collecting monitoring data from all trigger subsystems~\cite{prometheus:2023git}.
The database is expected to collect around 300\,000 metrics every 20\unit{s} and correlate monitoring information coming from different sources to provide a powerful monitoring and alerting system.
Alerts are handled by the Prometheus Alertmanager service~\cite{prometheus:2023git}.
The software enables binding alerts to external software based on conditions to page shifters and escalate to experts if necessary.
Actions to automatically mitigate or solve problems can also be configured to minimise downtime.
Monitoring dashboards are built using Grafana and can provide powerful inspection tools for online and prompt post-run analysis (Fig.~\ref{fig:l1trigger:grafana})~\cite{SunilKumar:2021jetir}.
Prometheus is not designed for providing long-term storage of metrics, therefore integration with offline computing systems is under consideration.

The \Run3 trigger configuration editor and main monitoring page are designed in the Vue.js JavaScript framework, which superseded the outdated and slower Polymer framework.
The \Run3 main monitoring page has also been updated to present data taken from the updated monitoring backend.

The significantly improved L1 online software and monitoring enables users to spot problems much more efficiently, and react more quickly compared to \Run2.
It also facilitates more efficient diagnosis of problems after they have occurred, since more historical information is available and is easy to access.

\subsection{The L1 scouting system}
\label{sec:l1trigger:scouting}

For \Run3, the L1 trigger project added dedicated hardware for the triggerless recording of objects reconstructed in the level-1 trigger, at a rate of 40\MHz without trigger or filtering.
The data are received by dedicated FPGA-based processing cards, either housed in powerful servers (I/O nodes) or operating standalone in custom enclosures.
Once received and pre-processed, the information is provided to a computing farm via Ethernet.
Triggerless data recording, referred to in CMS as ``40\MHz scouting'' or ``L1 scouting'', facilitates improved precision of measurements such as the luminosity, as well as unprecedented levels of trigger monitoring.
The term ``scouting'' is also used in the context of HLT, described in Section~\ref{sec:hlt:scouting}, where it refers to the selection and high-rate recording of small-size analysis data sets.

The L1 scouting system deployed for \Run3 is meant as the first large-scale demonstration of a larger system planned for the \Phase2 L1 trigger upgrade~\cite{CMS:TDR-021}.
Besides providing a testing ground for the technical implementation of the readout, it provides the opportunity to study solutions for large-scale distributed processing of high-rate data, the correlation of multi-bunch-crossing signals, as well as first studies of possible physics applications using limited resolution trigger data.
The measurement of a range of physics processes, such as LLPs with displaced muons and flavor anomalies in \PGt physics, could potentially benefit from L1 scouting.

\subsubsection{Architecture of the L1 scouting system}

The L1 scouting system receives data from the level-1 trigger via spare output links and processes them quasi-online in a dedicated computing farm.
The system operates largely independently from the standard CMS trigger and data acquisition chain.
In the first test system in \Run2, the data sent from the \uGMT to the \uGT was duplicated, and final muon objects, as well as intermediate muon candidates derived from the BMTF inputs, were transmitted over eight 10\Gbs optical links to the L1 scouting system.

For \Run3, an additional set of duplicated \uGMT outputs supply a L1 scouting processor dedicated to luminosity monitoring, as described in Section~\ref{sec:l1trigger:scoutingapplications}.
Each of the twelve BMTF processors dedicates two 10\Gbs links to L1 scouting information.
L1 scouting data from the calorimeter \layer2 system also mirrors the trigger objects provided to the \uGT.
Each \uGT processor transmits 512 bits per bunch crossing over three 10\Gbs links, indicating which of the trigger algorithms has fired in a given bunch crossing.

The architecture of the initial \Run3 L1 scouting system is a scaled up version of the \Run2 system, and consists of I/O nodes housing FPGA-based input boards that receive up to eight 10\Gbs links using the L1 trigger link protocol.
Data are transmitted unidirectionally with no back-pressure to the trigger.
The L1 scouting system therefore does not interfere with the standard trigger system in any way.
The FPGA logic performs both zero suppression and preprocessing of the data such as reformatting or recalibration.
The use of fast NN algorithms, implemented within FPGA resources to improve recalibration performance has been demonstrated (see Section~\ref{sec:l1trigger:scoutingapplications}).
Data are transferred by direct memory access (DMA) via a Gen-3 PCIe x16 bus into the memory of the I/O nodes, from where they are sent to dedicated processing units in the surface data center via 100\Gbs Ethernet over coarse wavelength division multiplexing CWDM4 single mode optical infrastructure.
Even after full zero-suppression, the long-term storage of the huge amount of raw data produced by the trigger processors, in view of a subsequent ``classic'' multitiered offline analysis and reduction, does not represent a viable approach.
Data taken in the early months of LHC \Run3 are being used to investigate various methods for a real-time analysis.

In the \Run2 demonstrator system, a KCU1500 Xilinx development kit, equipped with a KU15P FPGA, was used to capture the \uGMT inputs.
Additional I/O nodes for \Run3 are equipped with more powerful boards, such as the Micron SB-852, using a large Xilinx Ultrascale+ VU9P FPGA and also providing access to the  Micron Deep Learning Accelerator (MDLA)~\cite{MingChang:2020jsa}, which is a proprietary compiler that translates pre-trained ML networks into instructions for an FPGA-based hardware implementation.
Monitoring of the readout board is being developed using the AXI-lite interface provided by the Xilinx xDMA core.
Custom software exposes access to monitoring and control registers via a RESTful interface and monitoring and diagnostic data are exported to the Prometheus server described in Section~\ref{sec:l1trigger:online}.

\begin{figure}[!ht]
\centering
\includegraphics[width=\textwidth]{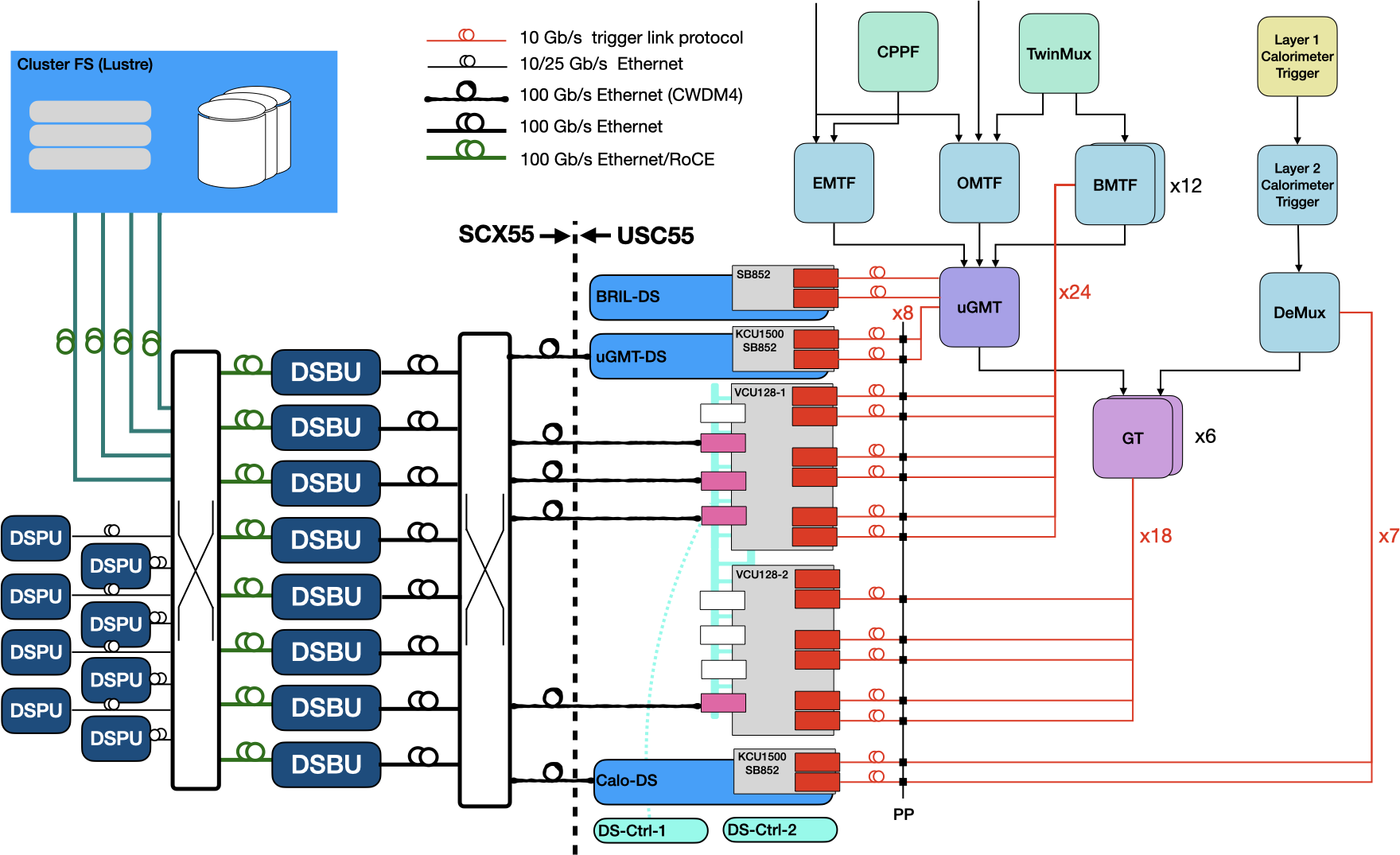}
\caption{%
    Architecture of the \Run3 L1 scouting prototype system.
    The input system, located in the experiment service cavern, consists of a combination of different types of FPGA receiver boards, hosted in the PCIe bus of I/O server nodes or extender.
    The boards receive and pre-process data from the different trigger systems.
    Two of the boards (KCU1500 and SB852) use DMA to move their data to host memory, from where they are transmitted to the surface data center over 100\Gbs Ethernet links.
    The VCU128 implements a TCP/IP core in the FPGA and directly trasmits data to the surface system.
    In the surface data center, links from the L1 scouting input system are connected to a switched network.
    Data streams are received through the said network by L1 scouting buffer servers (DSBU) and buffered in files on large RAMdisks.
    The L1 scouting processing units (DSPU) access buffered data from the DSBUs to perform data reduction and analysis.
    The processed data are finally moved to a Lustre cluster file system for long-term storage.
}
\label{fig:l1trigger:scoutingarchitecture}
\end{figure}

In addition to the I/O nodes with dedicated receiver boards, Xilinx VCU128 development kits~\cite{Xilinx:2022web} are used as standalone receivers for L1 scouting data.
These boards, which are housed in a custom enclosure with PCIe extender buses for control and monitoring, are equipped with a VU37P FPGA with 8\GB of high bandwidth memory.
The VU35P, with a very similar architecture, is planned for use on the DAQ-800 board currently being designed for the CMS \Phase2 data acquisition~\cite{CMS:TDR-022}, and is the anticipated readout board for the L1 scouting system of the \Phase2 upgrade.
The VCU128 boards in \Run3 thus allow a realistic test of the \Phase2 design, albeit with reduced input bandwidth.
Equipped with an additional FMC mezzanine to provide up to 32 10\Gbs input links, they transmit L1 scouting data directly over TCP/IP to the surface computing farm, providing more efficient utilization of the bandwidth available on the 100\Gbs Ethernet links.
The \Run3 scouting architecture is illustrated in Fig.~\ref{fig:l1trigger:scoutingarchitecture}.

\subsubsection{Applications of the L1 scouting system}
\label{sec:l1trigger:scoutingapplications}

The use of ML algorithms to improve the physics potential of L1 trigger objects captured by the triggerless recording of trigger data at 40\MHz has been investigated.
An example of this is the use of deep NNs, optimized for a throughput of around 1\MHz, to recalibrate L1 muon objects in real time, such that the accuracy of the \pt, $\eta$, and $\phi$ parameters of the muons can be improved over the standard L1 trigger reconstruction, which is optimized for efficiency at threshold, not for a full physics analysis.
Neural networks have been trained on zero-bias triggered data from LHC \Run2, to predict the offline fully reconstructed muon parameters of matched L1 muon objects.
This recalibration has been shown to improve the accuracy of the L1 muon parameters when compared to the standard L1 reconstruction, both for \uGMT and BMTF L1 muon parameters~\cite{Golubovic:2021cds, CMS:DP-2022-066}.
Additionally, neural networks that are trained to reject pairs of L1 muon candidates that do not correspond to matched fully reconstructed muons, and networks for anomaly detection have also been implemented in the Xilinx VU9P FPGA with a throughput of about 1\MHz, using the MDLA.

A copy of the \uGMT scouting system will be used to provide direct per-bunch measurements of the L1 muon multiplicity to the CMS BRIL system described in Section~\ref{sec:bril}.
FPGA firmware has been developed and implemented in both the KCU1500 and SB-852 boards to histogram the muon multiplicity per bunch over a period of four lumi nibbles, corresponding to 1.458\unit{s}.
These histograms are read out to the host PC in real time via the PCIe DMA implementation.

\begin{figure}[!htp]
\centering
\includegraphics[width=0.25\textwidth]{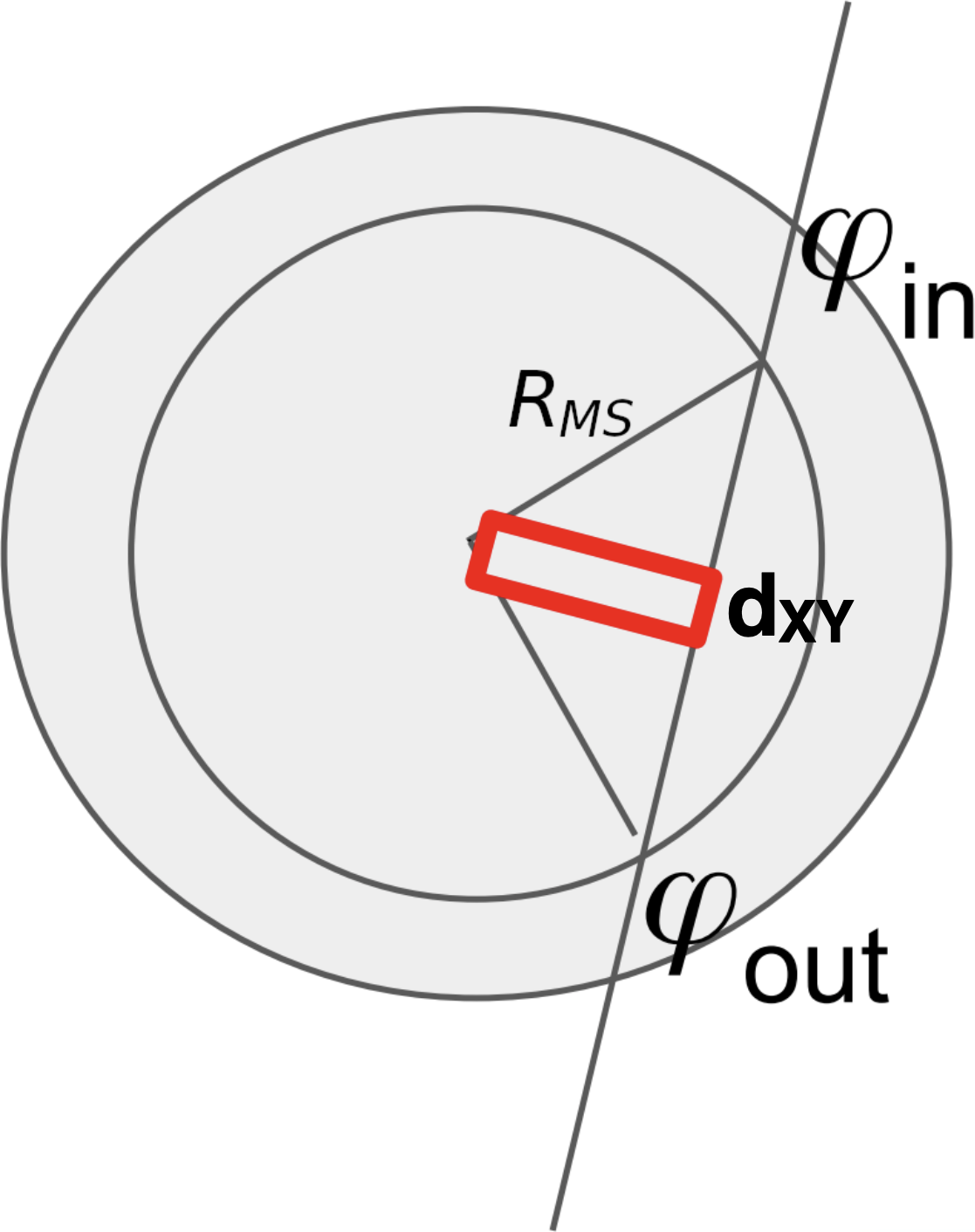}
\caption{%
    Illustration of the correlation between the impact parameter \dxy, highlighted in red, and the difference between the angles measured for the incoming and outgoing legs of a cosmic muon.
}
\label{fig:l1trigger:scoutingillustration}
\end{figure}

\begin{figure}[!htp]
\centering
\includegraphics[width=0.8\textwidth]{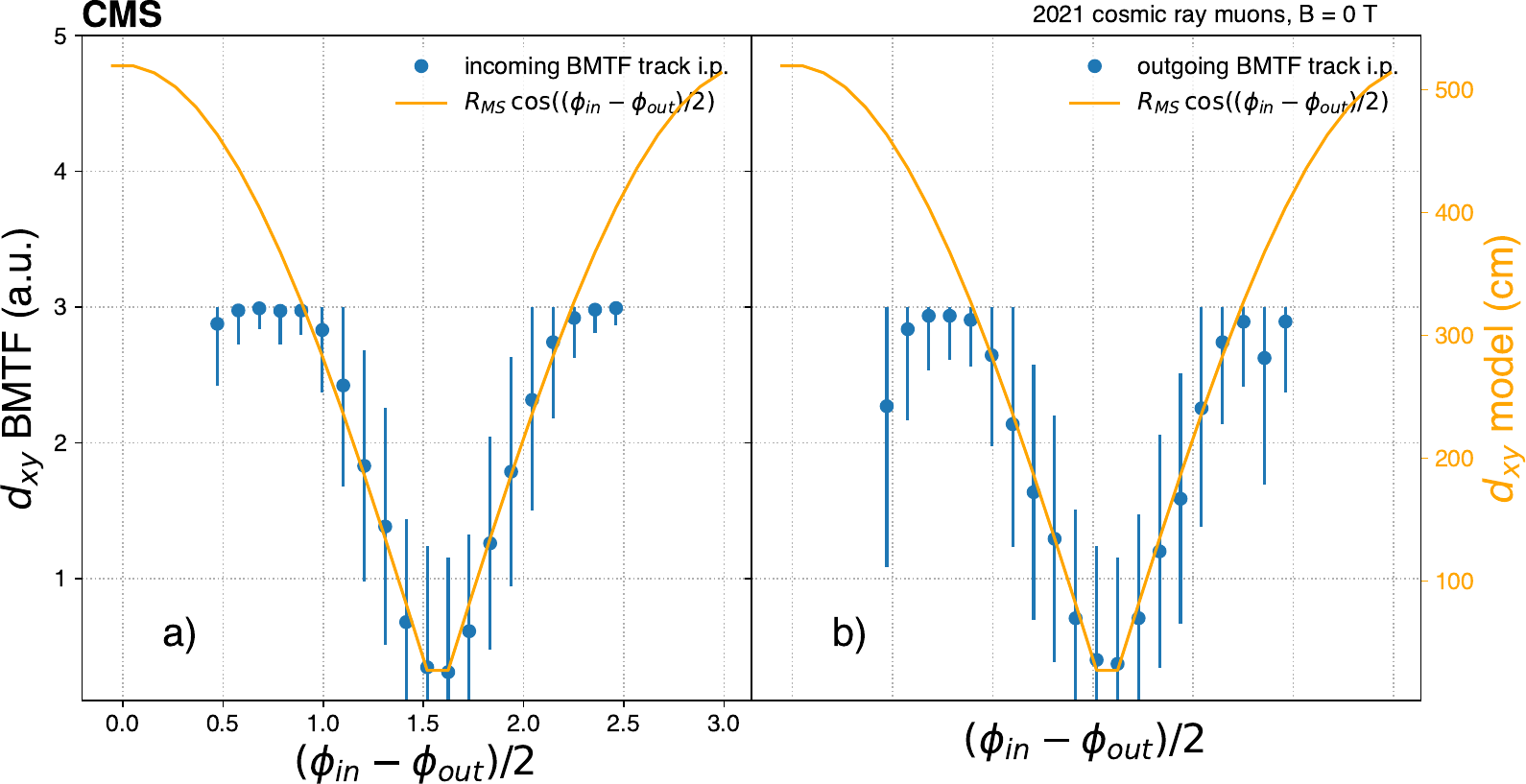}
\caption{%
    The impact parameter \dxy of the incoming (left) and outgoing (right) cosmic ray muon tracks as measured by the BMTF as a function of the difference of azimuthal coordinates of the incoming and outgoing legs.
    The BMTF firmware encodes the impact parameter in two bits, hence the range of values on the left $y$ axis.
    The orange curves model this dependence of the actual impact parameter as $R_{MS}\cos{(\phi_{in}-\phi_{out})/2}$, where $R_{MS}$ is the radius at which the BMTF measures the $\phi$ coordinate of the track.
    The right hand side $y$ axis shows the \dxy values (in cm) as predicted by this model, which exhibits remarkable consistency with the measurement (for the values within the range), if one assumes one unit of the left axis to correspond to an impact parameter of about 100\cm.
    Figure from Ref.~\cite{CMS:2023ukn}.
}
\label{fig:l1trigger:scoutingresult}
\end{figure}

Cosmic muon data taken with the L1 scouting system in special runs during LS2 have been used to analyse the performance of the new kBMTF algorithm.
The L1 track finding algorithms assume that muons always originate from the interaction region.
Therefore, a cosmic muon traversing the full detector appears as two back-to-back muons, usually separated in time by one or two bunch crossings.
The data were taken when the magnet was off.
In the absence of a magnetic field, muons traverse the detector in a straight line, leaving two muon tracks in the barrel drift tubes, typically two bunch-crossings apart ($\Delta\text{BX}=2$).
Figures~\ref{fig:l1trigger:scoutingillustration}--\ref{fig:l1trigger:scoutingresult} show the relationship between the L1 reconstructed impact parameters, \dxy, and the angle ($\phi_{\text{in}}-\phi_{\text{out}}$) of the two corresponding L1 tracks.

The \Run3 demonstrator system was used to systematically collect and store data during collision runs throughout 2022 from the \uGMT and calorimeter \layer2.
The offline analysis of these data provided invaluable insight into the reliability of the acquisition system and in particular of the hardware, firmware and software.
The results of this work have been and are being used to guide the design of further extensions of the demonstrator and the design of the \Phase2 systems.
The quality of the data was studied in view of assessing their usability for trigger diagnostics, luminosity measurement, and physics studies.
This work will continue throughout the rest of \Run3, along with the completion of the demonstrator with the commissioning of the VCU128 acquisition boards, collecting data from the uGT and BMTF systems, which represents a major step forward in validating the \Phase2 design.

\clearpage
\section{High-level trigger}
\label{sec:hlt}

\subsection{Overview}

To select events of potential physics interest, the CMS trigger system divides the processing into two levels:\ the level-1 (L1) trigger, implemented in custom hardware as described earlier in Section~\ref{sec:l1trigger}, and the high-level trigger (HLT), implemented in software, running on a farm of commercial computers.
The HLT refines the purity of the physics objects selected by the L1 trigger, with a maximum input rate increased to 110\kHz in 2023 from 100\kHz earlier.
In 2022, for standard \pp collisions at an average instantaneous luminosity of $1.5\times10^{34}\percms$, the average rate to offline storage for promptly reconstructed physics events was approximately 1.7\kHz.
In 2023, for a peak luminosity of $2.0\times10^{34}\percms$, the rate of promptly reconstructed physics events was about 2.6\kHz.
Additional data streams for calibration purposes or ``HLT data scouting'' are stored at higher rates with a smaller event content.
The HLT data scouting differs from the L1 scouting (described in Section~\ref{sec:l1trigger:scouting}) in that events must still satisfy a subset of L1 triggers before being written to disk at a high rate.

The HLT runs on a cluster of 200 nodes, each equipped with two AMD EPYC ``Milan'' 7763 CPUs, two NVIDIA T4 GPUs, and 256\GB of memory, running Red Hat Enterprise \Linuxeight.
The HLT farm is described in Section~\ref{sec:daq:eventfilter}, with the details specific to \Run3 highlighted in Section~\ref{sec:daq:evolution}.

The HLT data processing uses the concept of ``paths'' to structure its workflow.
These paths are sequences of algorithmic steps designed to reconstruct physics objects and make selections based on specific physics requirements.
Steps within a path are typically organized in ascending order of complexity, reconstruction refinement, and physics sophistication.
For example, the resource-intensive track reconstruction process is usually carried out after completing a series of initial reconstruction and selection steps involving the data from the calorimeters and muon detectors.

The reconstruction modules and selection filters of the HLT use the same software framework that is also used for offline simulation, reconstruction, and analysis (CMSSW~\cite{Jones:2015soc}).
As noted in Section~\ref{sec:daq:hltmenu}, HLT paths selecting similar physics object topologies are grouped into primary data sets for subsequent offline processing, and collections of primary data sets are organized into streams for efficient handling.

In preparation for \Run3, the HLT software was adapted to make use of heterogeneous computing architectures, and several reconstruction modules were developed to take advantage of that to meet the challenges of processing data at ever increasing luminosity and pileup.
Algorithms implemented to run on both CPUs and GPUs, are automatically directed to run on a GPU if a GPU is available; otherwise, the CPU-based version of the algorithm is executed.
The Patatrack project~\cite{Bocci:2020pmi} has created parallelized versions of pixel track and vertex reconstruction algorithms that can run on an NVIDIA GPU and were written using the NVIDIA CUDA language.
The data structures are optimized for GPU, and the entire reconstruction chain is executed on the GPU to minimize time-consuming data transformations and transfers.
A subset of ECAL and HCAL local reconstruction algorithms have also been ported to GPU, also using CUDA.
Based on these efforts a reduction in overall event processing time of about 40\% has been achieved.
More details can be found in Section~\ref{sec:hlt:rates}.
To make full use of the gain, CMS deployed and commissioned a filter farm composed of nodes comprising two GPUs in addition to two CPUs, as noted in Section~\ref{sec:daq:eventfilter}.
More information about multithreaded processing and GPU offloading is given in Section~\ref{sec:offline:framework}.

The HLT selects data for storage through the application of a trigger ``menu'', in which the collection of individual HLT paths is configured.
The trigger path definitions, physics object thresholds, and rate allocations are set to meet the physics objectives of the experiment.
In 2022, the HLT menus for \pp data taking typically contained around 600 paths.
This includes the primary HLT paths for analysis as well as paths for calibration and efficiency measurements that are typically looser than the primary paths.
These latter HLT paths are often ``prescaled'', \ie, only a fraction of the events that pass the requirements are actually accepted, to limit processing time and storage rate.
Different trigger menus are used for the recording of heavy-ion collision data.
The rates, physics breakdown, and CPU timing of the \pp menu are described further in Section~\ref{sec:hlt:rates}.

\subsection{HLT reconstruction}
\label{sec:hlt:reconstruction}

The HLT paths in the menu depend on the modules that produce the physics objects from the all-silicon inner tracker and from the crystal electromagnetic and brass-scintillator hadron calorimeters, operating inside a 3.8\unit{T} superconducting solenoid, together with data from the gas-ionization muon detectors embedded in the flux-return yoke outside the solenoid.
A foundation to many of the specific object reconstructions is the particle-flow algorithm~\cite{CMS:PRF-14-001}, which uses information from these systems to identify candidates for charged and neutral hadrons, electrons, photons, and muons.
The main features of the HLT physics object reconstruction and improvements for \Run3 are described in the following subsections.

\subsubsection{Tracking}
\label{sec:hlt:tracking}

Tracking using the hits recorded by the pixel and strip trackers is generally performed iteratively using a combinatorial Kalman filter, starting with tight requirements for the track seeds that become looser for each subsequent iteration.
Hits in the tracking detectors that have already been used in a track are removed at the beginning of the next iteration.
For \Run2, initially, the track reconstruction in the HLT consisted of three iterations.
The first two iterations required four consecutive hits in the pixel detector to seed the tracking.
These iterations target first higher \pt tracks and then lower \pt ones, and use the full volume of the pixel detector.
The third iteration relaxes the requirement on the number of hits in the pixel detector to three, and is restricted to the vicinity of jet candidates identified from calorimeter information and the tracks reconstructed in the two previous iterations.
The track reconstruction is limited to tracks that are consistent with the leading vertices reconstructed with the pixel detector (those vertices with the largest summed $\pt^2$ of pixel tracks).

In 2017, several issues with the \Phase1 pixel detector were identified that led to a nonnegligible fraction of inactive pixel modules (Section~\ref{sec:tracker:design}).
During the 2017--2018 year-end technical stop, the performance of the pixel detector was restored.
Nevertheless, an additional recovery iteration was added to the tracking, to safeguard against a recurrence of this or possible other detector failures.
In particular, track seeds consisting of just two pixel hits were allowed to be reconstructed in regions of the detector where two inactive modules overlap.

\begin{figure}[!t]
\centering
\includegraphics[width=0.48\textwidth]{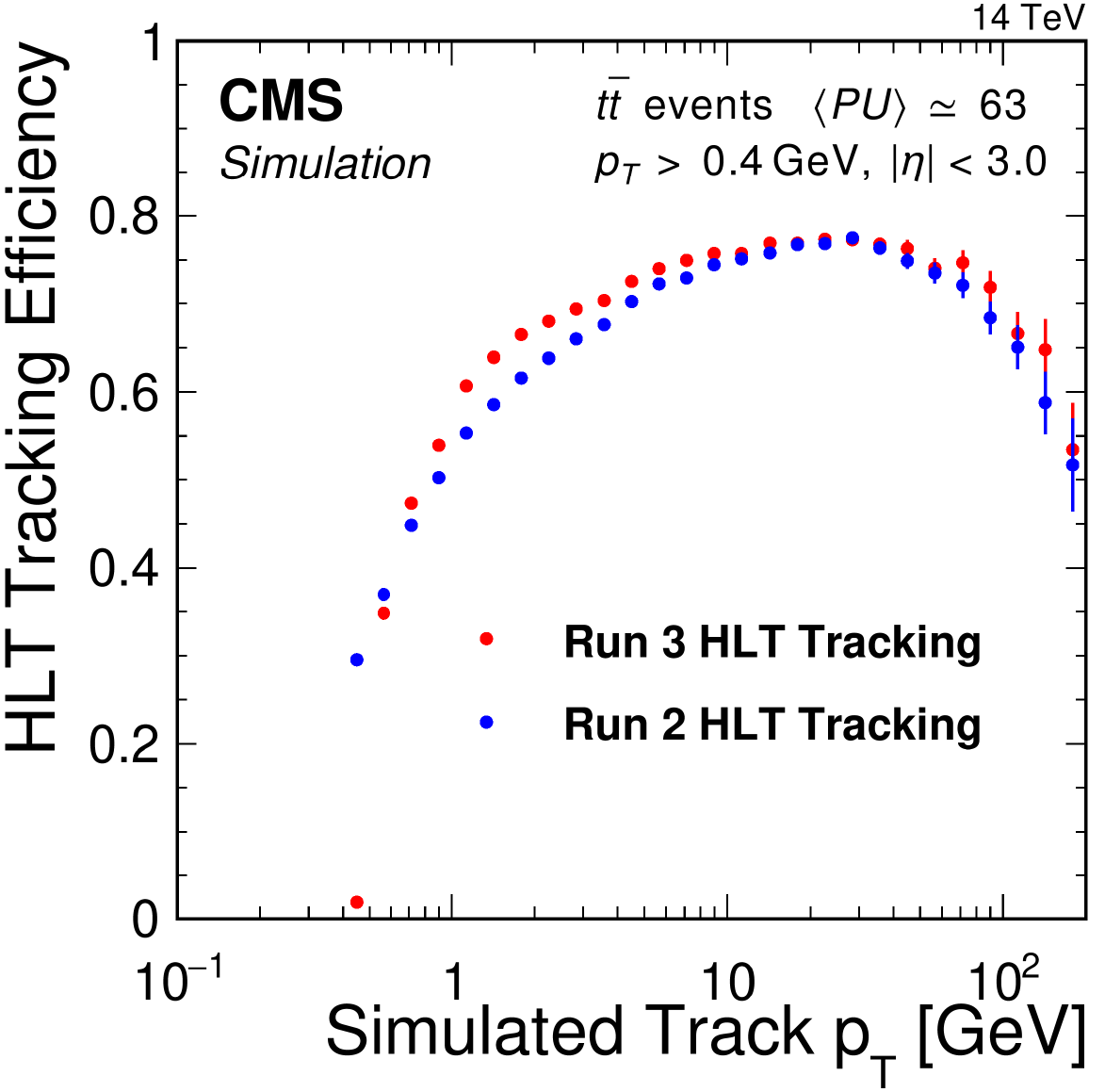}%
\hfill%
\includegraphics[width=0.48\textwidth]{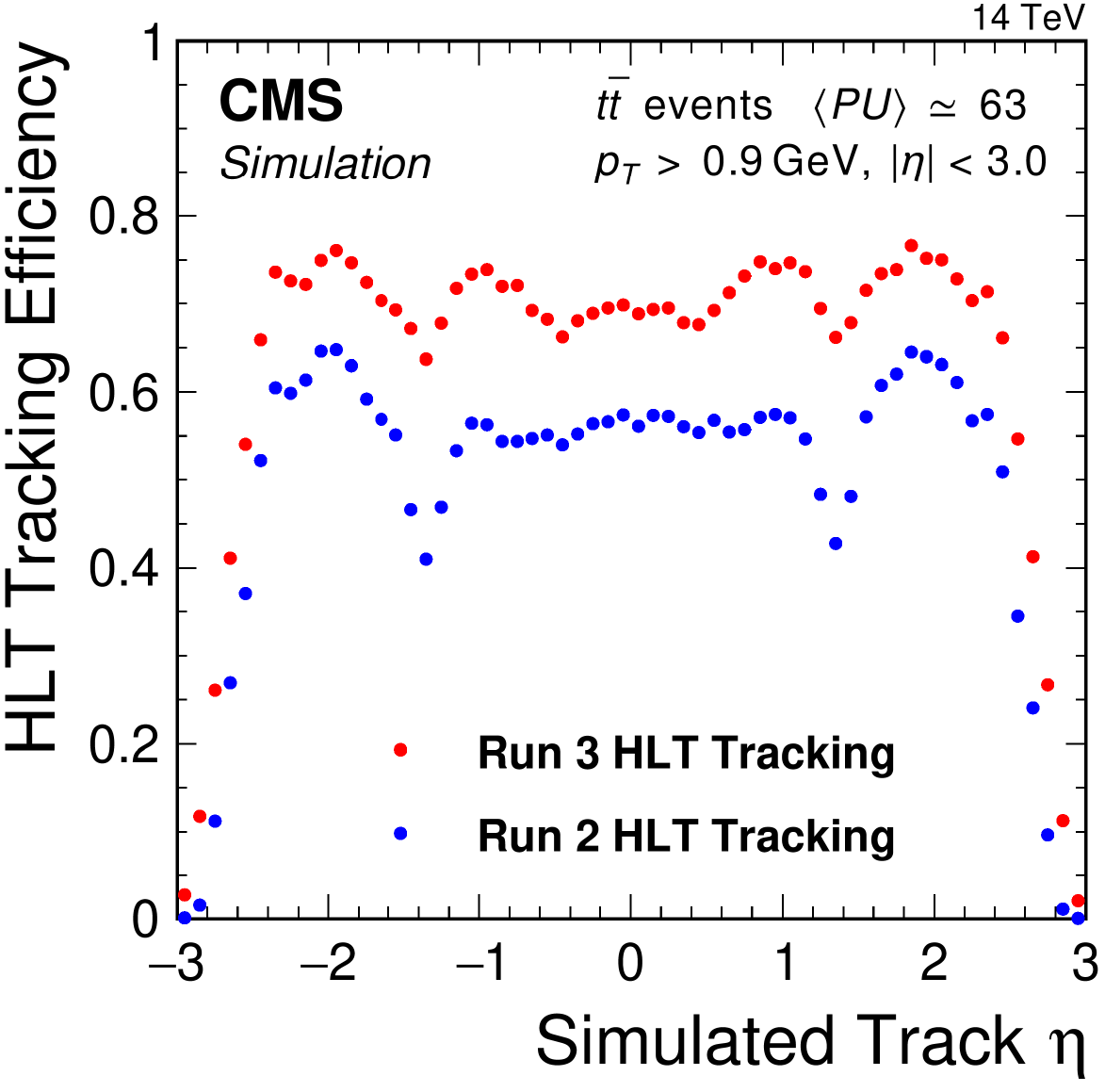} \\[1ex]
\includegraphics[width=0.48\textwidth]{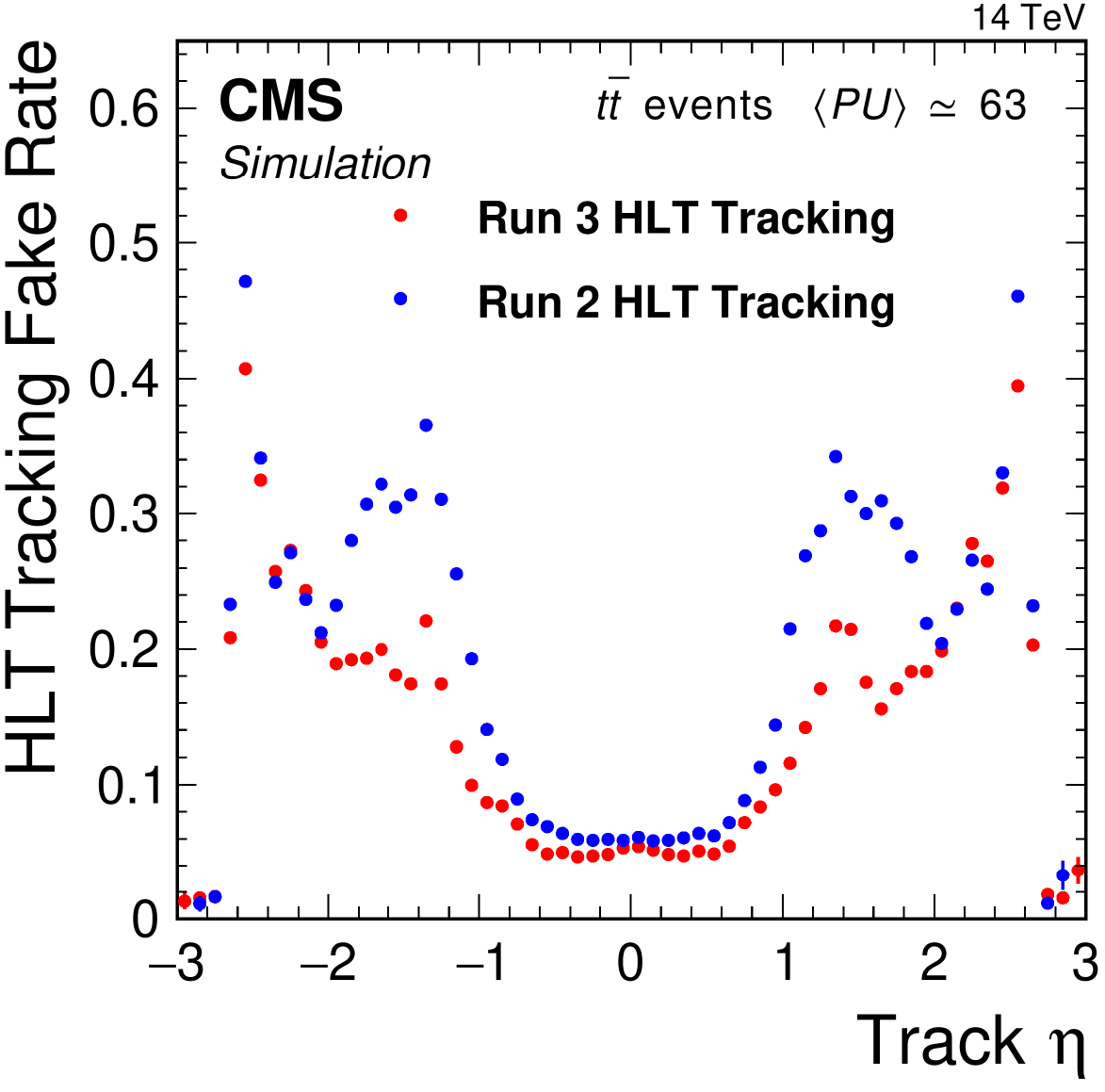}
\caption{%
    Tracking efficiency for the \Run2 HLT tracking (blue) and the \Run3 HLT single-iteration tracking (red) as a function of the simulated track \pt (upper left) and track $\eta$ (upper right).
    Only simulated tracks with $\abseta<3.0$ are considered in the efficiency measurement, with $\pt>0.4$ (0.9)\GeV required for the upper left (right) plots.
    The tracking fake rate (lower) is shown as a function of the reconstructed track $\eta$ for the \Run2 HLT tracking (blue) and the \Run3 HLT single-iteration tracking (red).
}
\label{fig:hlt:tracking}
\end{figure}

For \Run3 the seeding and tracking were significantly revised.
Tracking is now performed using only a single global iteration, and is seeded by a loose selection of the pixel tracks reconstructed by the Patatrack algorithm, which offers improved performance over the four-hit pixel tracking used for data taking in 2018~\cite{Bocci:2020pmi}.
To be used as track seeds, Patatrack pixel tracks must satisfy a loose selection requiring three or more pixel hits, $\pt>0.3\GeV$, and be compatible with a primary vertex candidate.
Despite fewer iterations and less CPU time required, the \Run3 tracking has improved performance over that used for \Run2.
The tracking efficiency and fake rate are measured in simulated \ttbar events with an average pileup of 63.
Figure~\ref{fig:hlt:tracking} shows the \Run3 tracking efficiency and fake rate as determined from Monte Carlo (MC) simulation.
The tracking efficiency is defined as the fraction of simulated particles from the signal interaction within the considered \pt and $\eta$ regions, and longitudinal (transverse) impact parameters $<$35 (70)\cm that are matched to a reconstructed track.
The fake rate is defined as the fraction of reconstructed tracks that could not be matched to a simulated particle.
The \Run3 tracking efficiency is higher than that of \Run2 for $\pt>0.7\GeV$, mostly in the central tracking region and the overall fake rate is lower, particularly around the transition region between the barrel and the endcap pixel detectors ($0.9<\abseta<2.1$).

\subsubsection{Muons}

Tracking algorithms are also deployed to identify and reconstruct muons measured in the muon detectors and in combination with the pixel and strip trackers.
While the algorithms used during \Run2 are described in more detail in Ref.~\cite{CMS:MUO-19-001}, a brief summary is given here along with the changes implemented for \Run3.

Muon track reconstruction at the HLT takes place in two steps:\ first using hits only in the muon system (L2 reconstruction), followed by a combination with hits in the inner tracking system (L3 reconstruction).
The L2 reconstruction is equivalent to the offline standalone muon reconstruction.
The L3 reconstruction is seeded by an L2 muon and follows an iterative track reconstruction similar to that described in the previous section in a region around the seed starting from the outer tracking layers and working inward  (``outside-in'') or from the inner tracking layers working out (``inside-out'').
The inside-out approach can also be seeded directly by L1 trigger muons.
The L3 reconstruction is essentially 100\% efficient with respect to that in the L1 trigger, while consuming only about 30\% of the overall HLT CPU time.
After the muon track reconstruction, identification criteria are applied, as well as isolation criteria for the isolated muon category.
The isolation is based on the sum of \pt from additional tracks associated with the primary vertex and calorimeter energy deposits that are clustered using a particle-flow algorithm in a cone of radius $\DR=0.3$ around the muon.
The estimated contribution from pileup to the energy deposits in the calorimeter is subtracted.

For \Run3, several modifications to the muon reconstruction were made to improve the HLT performance, in particular with respect to CPU timing.
The L2 muon reconstruction was extended to include hits from the GEM detectors, described in Section~\ref{sec:gem}.
At L3, the tracking was adapted to make use of the Patatrack pixel track seeds, followed by a single iteration with the full tracker.
The efficiency for low \pt muons was improved by optimizing the search regions around seeds in which muon tracks are reconstructed in the tracker, achieving an efficiency above 80\% for $\pt>2\GeV$.
Additionally, machine-learning algorithms were incorporated:\ a boosted decision tree (BDT) classifier for the search algorithm and a neural network (NN) algorithm for seeding strategy.
The inside-out algorithm uses the BDT classifier to consider only the seeds with high quality.
The NN algorithm was developed to choose the best seeding strategy for the outside-in reconstruction from the L2 muon and thus limit the total number of seeds considered.
With these changes, the L3 efficiency remains unchanged, while the CPU time was reduced by 15\%.

\subsubsection{Electrons and photons}

The HLT electron and photon identification uses the L1 calorimeter trigger candidates as a starting point to perform a ``regional'' reconstruction of the energies deposited in the ECAL crystals around them.
Subsequently, superclusters (clusters of ECAL deposits within a certain geometric area around the seed cluster) are built using the same reconstruction algorithm as used offline~\cite{CMS:EGM-13-001}.
The energy correction applied to HLT superclusters is simpler than the one used offline in that it uses ECAL information only.
Requirements are then imposed on the minimal energy, as well as other properties of the energy deposits in the ECAL and HCAL subdetectors.

For electrons, the ECAL supercluster is associated to a reconstructed track with a direction compatible with its location.
The first step is a match with pixel detector hits.
Since 2017, the pixel matching algorithm requires three pixel hits, to maximize early background rejection, while a hit doublet is accepted only if the trajectory passes through a maximum of three active modules.
If the supercluster is successfully matched with the pixel hit seeds, the electron track is reconstructed using a Gaussian sum filter (GSF) tracking algorithm~\cite{Adam:2003kg}.

Variables to enhance the identification of true electrons and photons are applied based on the shower shape in the ECAL, the energy deposition in the HCAL, and, in the case of electrons, the matching between the track and the ECAL supercluster, as well as the quality of the GSF track.
As with the other leptons, isolation criteria are generally applied, except for some specific paths, to electrons and photons based on the calorimeter energy deposits in a cone of radius $\DR=0.3$ around the electron or photon and the sum of \pt from additional tracks associated with the primary vertex.
Several HLT paths with different isolation criteria and \pt thresholds are defined to provide a range of efficiencies and rates for specific physics analyses.

\subsubsection{Tau leptons}

The reconstruction of hadronic tau-lepton decays (\tauh) at the HLT is also of crucial importance for the physics program.
During \Run2, it was performed in three steps.
The first step, the L2 reconstruction, is seeded by L1 \tauh candidates.
The energy depositions in the calorimeter towers around the candidates within a cone of radius 0.8 are clustered, and L2 \tauh candidates are reconstructed by using the anti-\kt algorithm~\cite{Cacciari:2008gp, Cacciari:2011ma} with a distance parameter of 0.2.

In the second step, known as L2.5, a charged particle isolation criterion, based on pixel detector information, is implemented.
Pixel tracks are reconstructed around L2 \tauh candidates with $\pt>20\GeV$ and $\abseta<2.5$ in a region of $\Deta{\times}\Dphi=0.5{\times}0.5$.
Tracks originating from the primary vertex with a transverse impact parameter $\dxy<0.2\cm$, at least three hits, and a trajectory in an isolation cone of $0.15<\DR<0.4$ around an L2 \tauh candidate, are considered for the isolation sum.
An L2 \tauh candidate is considered isolated if the scalar sum of the \pt of the associated pixel tracks is less than 4.5\unit{GeV}.

The final step, the L3 reconstruction, includes track reconstruction using the full tracker.
For \Run2, tracking used a reduced number of iterations to fit into CPU time budget.
Moreover, the track reconstruction was performed regionally around the L2 \tauh candidates.
Until mid 2018, the L3 reconstruction was performed using a cone-based algorithm.
It was then upgraded to the hadrons-plus-strips (HPS) algorithm~\cite{CMS:TAU-14-001} that is also used in offline reconstruction.
Both algorithms start with jets reconstructed by the anti-\kt algorithm with a distance parameter of 0.4.

In \Run3, the L2 and L2.5 sequences were replaced by a convolutional neural network, in which the pixel tracks from the Patatrack algorithm and the calorimeter candidates are used as input.
This improved the rejection rate at L2 by about a factor of 2 for similar efficiency.
Additionally, the efficiency of the L3 reconstruction was increased by introducing a neural network, DeepTau~\cite{CMS:TAU-20-001}, adapted from the offline reconstruction such that it matched HLT requirements for speed and performance.

\subsubsection{Jets and global energy sums}

Jets are reconstructed at the HLT using the anti-\kt clustering algorithm~\cite{Cacciari:2008gp, Cacciari:2011ma} with a nominal distance parameter of 0.4, or of 0.8 in the case of wide jets used in boosted topologies and multijet triggers.
Inputs to the jet algorithm are usually particle-flow candidates, or alternatively calorimeter towers.
Depending on detector and beam conditions, corrections are applied to the jet energy scale, as well as the measured particle-flow hadron and average pileup energies.

Triggers using jets and global energy sums are used in CMS across a wide spectrum of physics analyses.
Multijet trigger paths, for example, are key to select vector boson fusion event candidates which contain two very forward jets in opposite endcaps, with a large angular separation and a large dijet invariant mass.
Final states with boosted multijet signatures can also be identified using dedicated jet substructure techniques such as the soft drop approach~\cite{Larkoski:2014wba}.
Signatures with many jets in the final state can also be triggered using \HT, the transverse energy sum of all jets, or $S_{\mathrm{T}}$ that combines jets with leptons.

Paths based on missing transverse momentum \ptvecmiss, defined as the negative vector sum of the \pt of input objects, also exist.
For these paths, the accounting for noise and beam-induced backgrounds is especially important in order to keep rates and resolutions under control.
Trigger paths based on \ptmiss alone, or in combination with jets, leptons, or photons in the event, are also used, \eg, to search for weakly interacting particles.

\subsubsection{\PQb jet tagging}

The identification of \PQb jets at the HLT is essential in order to enhance the fraction of events containing heavy flavor jets from processes like vector-boson associated Higgs boson production where the Higgs boson decays into a pair of \PQb quarks.
Such processes would otherwise be unlikely to pass the standard thresholds for leptons, jets, or missing transverse momentum.

Since \PQb tagging relies on the measurement of tracks that are displaced with respect to the primary vertex, both the pixel and the silicon strip trackers are used to improve the spatial and momentum resolutions of such tracks.
In 2016, the combined secondary vertex algorithm CSVv2 was used.
Subsequently, in 2017 and 2018, the multiclassifier neural network DeepCSV was implemented~\cite{CMS:BTV-16-002}.
With DeepCSV, the \PQb jet tagging efficiency was improved by 5--15\% at constant gluon or light-quark misidentification rates.

For \Run3, two new neural network taggers, DeepJet~\cite{Bols:2020bkb} and ParticleNet~\cite{Qu:2019gqs}, were deployed in 2022, with further improved performance.
In addition to tracks, the DeepJet algorithm also uses information from neutral and charged particle-flow jet constituents.
The ParticleNet algorithm provides multiclass jet-flavor classification for categories of \PQb, \PQc, and light quarks, gluons, and hadronically decaying tau leptons.
For use in HLT, the \Run3 tagging algorithms were trained on dedicated HLT-reconstructed simulation samples.

\begin{figure}[!ht]
\centering
\includegraphics[width=0.6\textwidth]{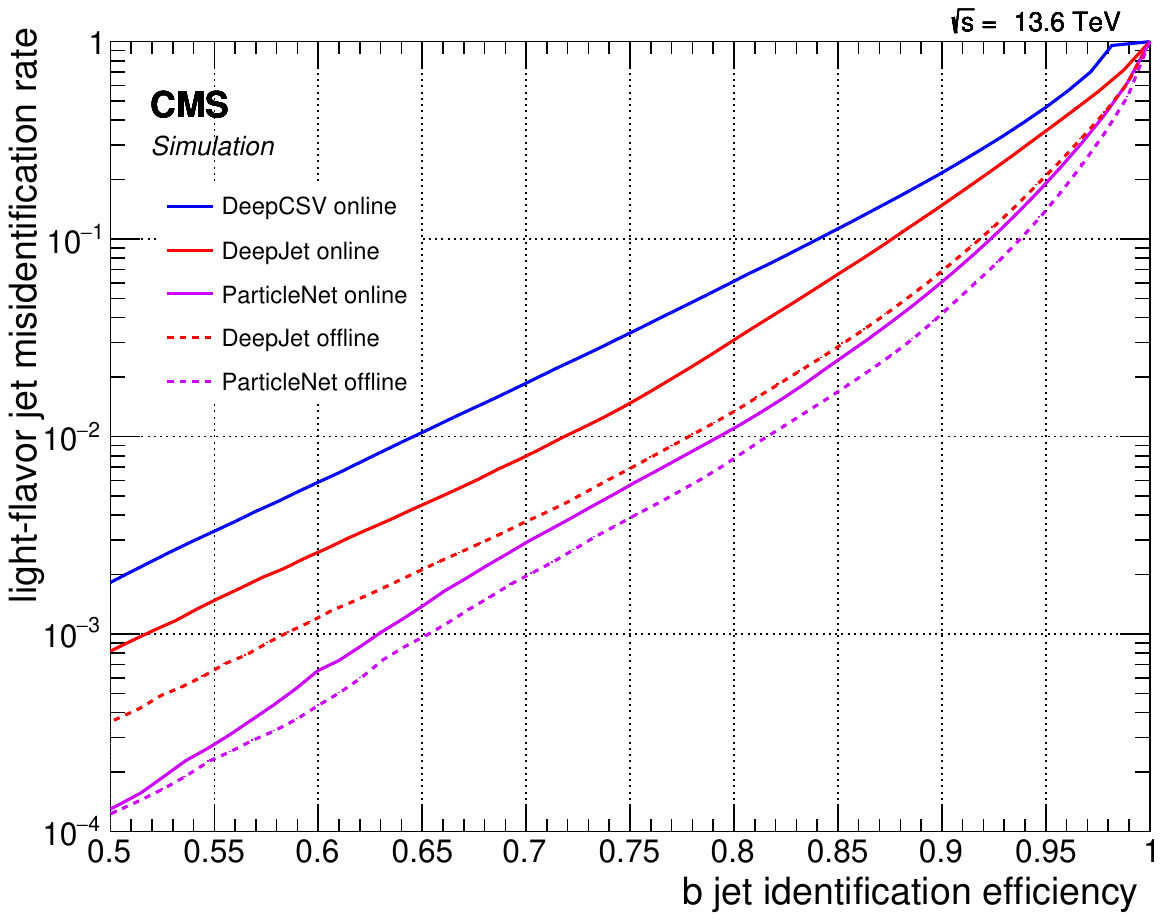}
\caption{%
    Light-flavor jet misidentification rate versus the \PQb jet efficiency for the various \PQb tagging algorithms.
    The solid curves show the performance of the DeepCSV (blue), DeepJet (red), and ParticleNet (magenta) algorithms in the HLT.
    The dashed curves show the corresponding offline performance for DeepJet (red) and ParticleNet (magenta) taggers using offline reconstruction and training.
}
\label{fig:hlt:deeptaggers}
\end{figure}

Figure~\ref{fig:hlt:deeptaggers} shows the light-flavor jet misidentification rate versus the \PQb jet efficiency for the different tagging algorithms, evaluated on simulated top-quark pair production events with an HLT jet selection of $\pt>30\GeV$ and $\abseta<2.5$.
Compared to the performance of the DeepCSV algorithm used during \Run2, the DeepJet algorithm trained using HLT quantities has a light-flavor jet misidentification rate that is lower by about a factor of 3 (up to efficiencies of about 75\%).
The ParticleNet algorithm reduces the misidentification rate by another factor of 2.5.

\subsubsection{New HLT paths for long-lived particles}

In addition to the improved reconstruction algorithms discussed in Section~\ref{sec:hlt:reconstruction}, the \Run3 HLT menu has been significantly expanded to explore new and unconventional physics signatures.
The \Run3 HLT menu includes new dedicated triggers targeting long-lived particle (LLP) signatures, such as displaced jets, displaced leptons and photons, and delayed jets using timing information from ECAL and HCAL, as well as HCAL depth information, as detailed in Section~\ref{sec:hcal:trigger}.
Furthermore, it features high-multiplicity trigger (HMT) paths that trigger on hadronic showers in the muon system, and new triggers for \PB physics, as described in Section~\ref{sec:hlt:parking}.
In 2023, further resources were deployed to enhance the sensitivity for demanding event topologies, such as the production of Higgs boson pairs decaying with at least two \PQb jets, vector boson fusion (VBF) processes, and events with LLP signatures.
Several of these new HLT paths are also seeded by new dedicated L1 trigger candidates (Section~\ref{sec:l1trigger:menu}), further improving the signal acceptance for their target searches.
The new triggers for \PB and VBF physics will be described in more detail in Section~\ref{sec:hlt:parking}.

There are now several flavors of displaced and delayed jet paths available at the HLT in \Run3.
First, there is a suite of displaced dijet paths that were already available to a certain extent in \Run2, but that have gone through major improvements over time.
These displaced dijet paths are either inclusive, in that they select events with calorimeter $\HT>650\GeV$ and two jets with less than two prompt tracks each, or they are more exclusive and require events with calorimeter $\HT>430\GeV$ and two jets with less than two prompt tracks and at least one displaced track.
For \Run3, the displaced jet selections have been improved at the HLT.
In particular, the selection on the number of prompt tracks was tightened in order to reduce the rate, while at the same time the definitions of prompt and displaced tracks were loosened in order to improve the signal efficiency for low-mass LLPs.
Furthermore, additional L1 trigger seeds have been added.
An L1 seed that requires a single muon in addition to a small amount of \HT helps to reduce the displaced dijet \HT threshold at the HLT.
In addition, the new HCAL timing and depth seeds, described in Section~\ref{sec:hcal:trigger}, bring improved efficiency for LLPs with $c\tau>0.5\unit{m}$.
All of these improvements to the displaced dijet paths provide better efficiency to trigger on low-mass LLPs, especially those with heavy-flavor decays.

In addition to delayed jet paths that use HCAL timing, there are also new HLT paths that exploit the ECAL timing.
For LLPs that produce jets with delays of about 1\ns or more, the signal efficiency is improved by an order of magnitude, with respect to the MET triggers that were available for this analysis in \Run2.
In particular, there are two different kinds of delayed jet triggers that use ECAL timing at the HLT; there are paths that are seeded by \HT, and there are paths that are seeded by L1 \PGt objects.
For both seeds, different requirements are made at the HLT, namely one or two jets, independently of whether those jets are trackless or not, and the amount of timing delay.
The paths seeded by \HT improve the sensitivity to low-mass LLPs with respect to the MET paths, and the paths that are seeded by L1 \PGt objects increase the efficiency to trigger on Higgs boson decays to long-lived scalars that decay to four \PQb jets as well as to four \PGt leptons.

Neutral LLPs with particularly long lifetimes could decay hadronically beyond the calorimeters, creating a high-multiplicity shower in the muon system.
Such showers are expected to consist of hundreds of hits, but no tracks or jets reconstructed in the inner detectors.
Essentially, the CMS muon system would act as a sampling calorimeter.
As mentioned in Section~\ref{sec:l1trigger:emtf}, new L1 seeds have been developed to collect these high-multiplicity events in the CSCs.
The high-multiplicity triggers (HMTs) at L1 are used to seed several HLT paths.
At the HLT, a clustering of hits in the muon system is performed using the Cambridge--Aachen algorithm~\cite{Dokshitzer:1997in, Wobisch:1998wt}.
The first HMT HLT path reconstructs a single CSC cluster, with stricter cluster requirements than at L1 in order to control the rate.
The second HMT HLT path reconstructs a CSC cluster as at L1, and then additionally requires a cluster in the DTs with at least 50 hits.
The last available HMT HLT path reconstructs a single DT cluster with 50 hits, makes no requirements on CSC clusters, and is seeded by MET triggers at the L1 trigger.
As compared with the MET triggers that were available in \Run2, the trigger efficiency for these unique signals is improved by factors of 3 to 20 depending on the path.

Paths for displaced muons at the HLT have also been improved in \Run3.
Displaced dimuon paths are seeded by two L1 muons with low \pt thresholds and by new displaced kBMTF double muon seeds with unconstrained \pt and \dxy, as described in Section~\ref{sec:l1trigger:bmtf}.
These seeds feed into several types of displaced dimuon paths at the HLT.
There are L2 double-muon paths that require an in-time collision based on the beam pickup timing device with a veto on prompt muons, complemented by paths that make use of a seed developed for cosmic ray muons.
These two paths require displacements of at least 1\cm.
Lastly, there are L3 double-muon paths that require displacements of at least 100\mum.
This suite of HLT paths covers a wide range of displacements and improves the signal efficiency over that of \Run2.
The efficiency, measured in a cosmic ray muon sample recorded in 2022, was measured to be 100\% for displacements $\dxy<100\cm$ for the L2 \pp seed + prompt-veto path, and 90\% for $\dxy<350\cm$ for the L2 cosmic seed + prompt veto path.
The efficiency of the L3 displaced dimuon path is measured to be 85\% in the data sample of non-prompt \PJGy events.
At the same time, the background efficiency is small:\ it is measured to be $<$1\% in Drell--Yan data events for all three displaced dimuon HLT paths.

\subsection{\Run3 HLT menu composition, rates, and timing}
\label{sec:hlt:rates}

While the complete list of HLT paths in the \Run3 HLT menu is too long to be listed here, a representative sample of some standard triggers with their HLT thresholds and rates is provided in Table~\ref{tab:hlt:menurates}.

\begin{table}[!p]
\centering
\topcaption{%
    HLT thresholds and rates of some generic triggers in the \Run3 HLT menu.
    The rates were obtained from measurements during an LHC fill in November 2022 and have been scaled to a luminosity of $2.0\times10^{34}\percms$.
}
\label{tab:hlt:menurates}
\renewcommand{\arraystretch}{1.1}
\begin{tabular}{lr}
    HLT algorithm & \multicolumn{1}{c}{Rate} \\
    \hline
    Isolated muon with $\pt>24\GeV$ & 250\uHz \\
    Isolated electron with $\ET>32\GeV$ & 182\uHz \\
    Particle-flow based $\ptmiss>110\GeV$ & 81\uHz \\
    4 PF jets with $\pt>70$, 50, 40, and 35\GeV with two \PQb tags & 57\uHz \\
    Two isolated tau leptons with $\pt>35\GeV$ & 54\uHz \\
    Muon with $\pt>50\GeV$ & 51\uHz \\
    Two electrons with $\ET>25\GeV$ & 21\uHz \\
    AK4 PF jet with $\pt>500\GeV$ & 16\uHz \\
    Two same-sign muons with $\pt>18$ and 9\GeV & 10\uHz \\
\end{tabular}
\end{table}

\begin{figure}[!p]
\centering
\includegraphics[width=0.9\textwidth]{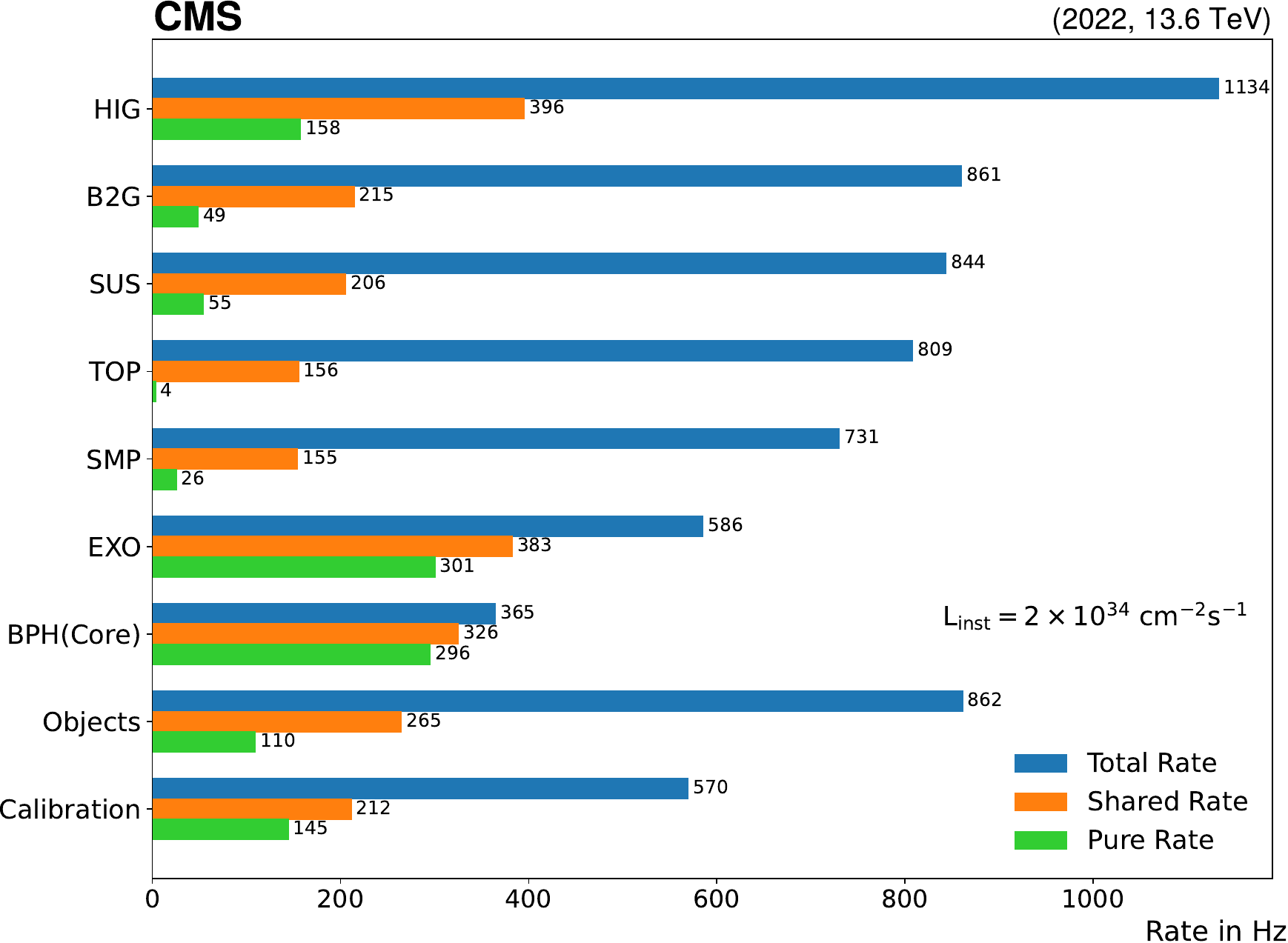}
\caption{%
    The HLT rate allocation by physics group for the \Run3 menu deployed in November 2022, scaled to a luminosity of $2.0\times10^{34}\percms$.
    The total rate (blue bar) is the inclusive rate of all triggers used by a physics group and the pure rate (green bar) is the exclusive rate of all triggers unique to that group.
    The shared rate (orange bar) is the rate calculated by dividing the rate of each trigger equally among all physics groups that use it, before summing the total group rate.
}
\label{fig:hlt:menurates}
\end{figure}

Figure~\ref{fig:hlt:menurates} shows the ``standard physics'' HLT rates consumed by each CMS physics group, estimated from a fraction of events recorded by CMS from a fill taken in November 2022.
``Standard physics'' refers to the collection of triggers included in the primary data sets whose reconstruction starts within 48 hours after the data were recorded.
This does not include data scouting at the HLT or parking triggers, as detailed in Sections~\ref{sec:hlt:scouting}--\ref{sec:hlt:parking}.
The average delivered instantaneous luminosity during this fill was $1.8\times10^{34}\percms$, but the measured rates have been scaled to correspond to a luminosity of $2.0\times10^{34}\percms$.
For the rate measurement, events are assigned to a physics group if the group uses at least one of the HLT algorithms that triggered the event.
The group categories correspond to the physics analysis working groups:\ Higgs boson physics (HIG),
searches for new physics in boosted signatures (B2G),
searches for new physics in final states with imbalanced transverse momentum (SUS),
top quark physics (TOP),
standard model physics (SMP),
\PB physics (BPH),
and searches for exotica (EXO).
The calibration category corresponds to all HLT algorithms used for subdetector alignment and calibration purposes.
The ``objects'' category corresponds to HLT algorithms used for monitoring and calibration by the so-called physics object groups.

What can be seen from Fig.~\ref{fig:hlt:menurates} is that roughly one third of the menu rate has been devoted to standard model physics processes (HIG, SMP, TOP), one third to searches for physics processes beyond the standard model (EXO, SUS, B2G), and the remaining one third to \PB physics processes, physics objects (for monitoring and calibration purposes for example), and subdetector calibration.
The trigger selection for \PB physics and LLP (included in EXO) are largely unique to those groups.

\begin{figure}[!ht]
\centering
\includegraphics[width=0.48\textwidth]{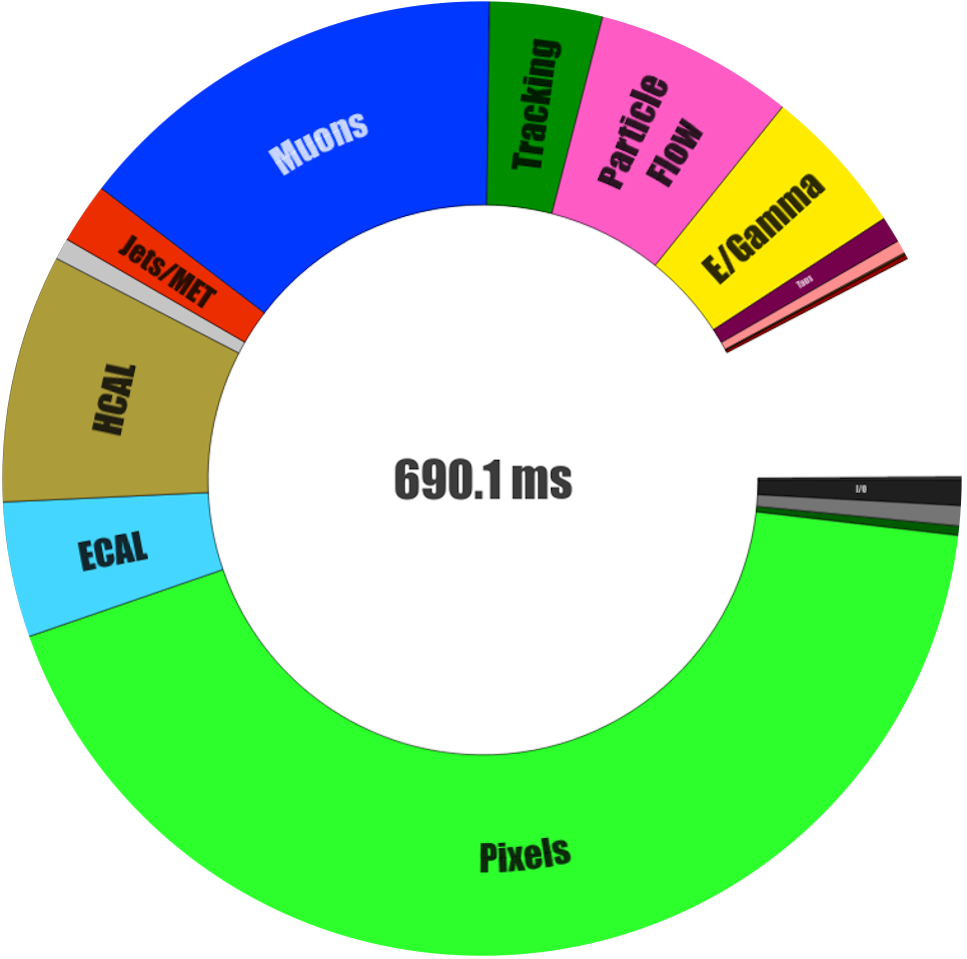}%
\hfill%
\includegraphics[width=0.48\textwidth]{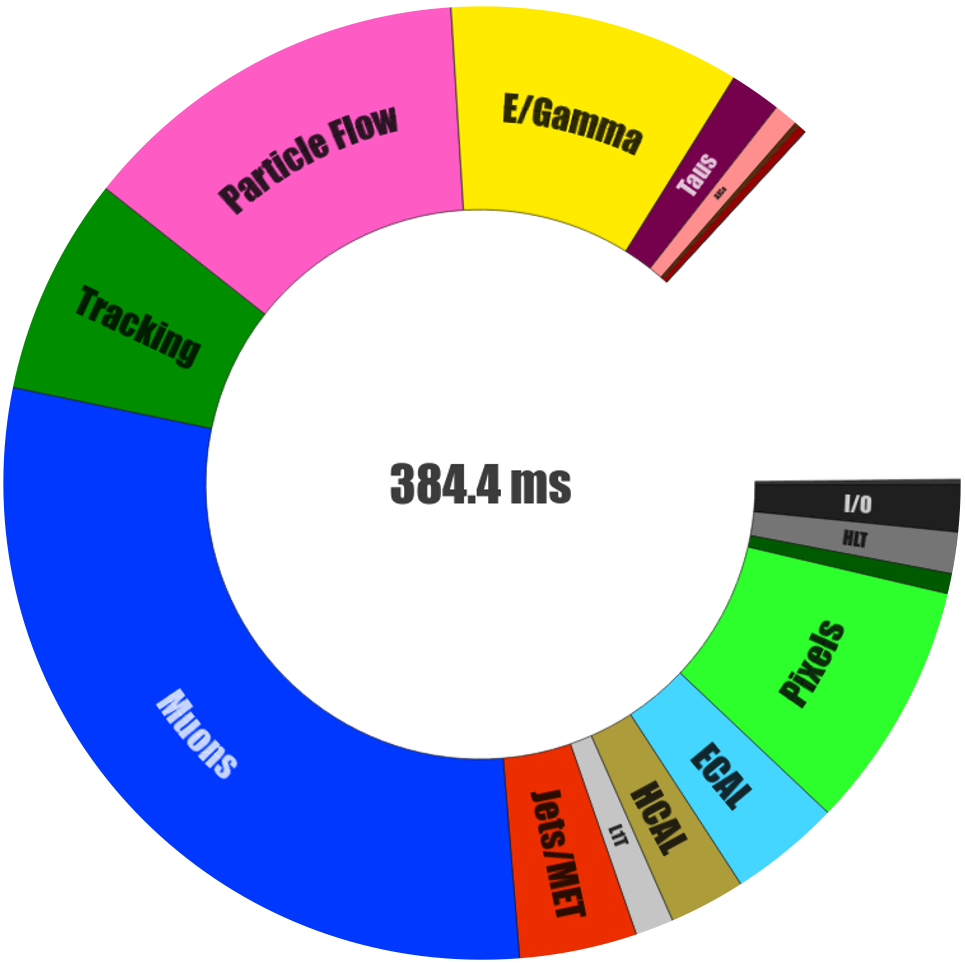}
\caption{%
    Pie chart distributions of the processing time for the HLT reconstruction running only on CPUs (left) and offloading part of the reconstruction to GPUs (right).
    The slices represent the time spent in different physics object or detector reconstruction modules.
    The empty slice indicates the time spent outside of the individual algorithms.
}
\label{fig:hlt:cputiming}
\end{figure}

The distribution of HLT time spent processing the data is shown in Fig.~\ref{fig:hlt:cputiming}.
The processing time running only on CPUs (left) is compared to that when part of the reconstruction is offloaded to GPUs (right).
These results were obtained for an HLT configuration representative of the 2022 conditions, running over a sample of 64\,000 \pp collision events with an average pileup of 56 collisions.
The measurements were performed on a machine identical to those used in the HLT farm, as described in Section~\ref{sec:daq:eventbuilder}, equipped with 2$\times$~AMD EPYC Milan 7763 CPUs and 2$\times$~NVIDIA T4 GPUs.
The node was configured identically to the HLT farm, with simultaneous multithreading (SMT) enabled, NVIDIA multiprocess server (MPS) enabled, and running eight jobs in parallel with 32 CPU threads and 24 concurrent events each.
The average processing time per event is 690\ms when running only on CPUs and 384\ms when offloading part of the reconstruction to GPUs, corresponding to a speedup of over 40\%.
The maximum processing time per event for the initial \Run3 event filter farm configuration is 500\ms, as noted in Section~\ref{sec:daq:evolution}.

\subsection{Data scouting at the HLT}
\label{sec:hlt:scouting}

A limiting factor for the data-acquisition rate is the bandwidth of the data to record on disk (a few $\GBs$), not the event rate per se.
Thus, if the size of the data per event is reduced, a higher rate of events can be recorded, \ie, using significantly lower trigger thresholds.
In the so-called ``HLT data scouting'', only the most relevant physics information is stored, as reconstructed by the HLT, and not the complete set of raw data.
Scouting was first implemented in \Run1~\cite{CMS:EXO-11-094}, and was developed further during \Run2, for selected physics objects, such as jets~\cite{CMS:EXO-14-005} and dimuons~\cite{CMS:EXO-19-018}.
For instance, the \HT threshold in the HLT scouting data was reduced from 800 to 410\GeV, and this data was used for a search for three-jet resonances~\cite{CMS:EXO-17-030}.

A further benefit of HLT data scouting is that events are reconstructed only once, using the resources of the HLT, and thus do not require further computing resources to perform the offline reconstruction step, which is described in Section~\ref{sec:offline}.
On the other hand, a complete reprocessing of HLT scouting data not possible.
The quality of the scouting data depends on the calibrations and alignments used in the HLT, but these must anyway be precise to maintain good trigger performance.
Furthermore, the CPU processing time budget per event is limited at the HLT (less than about 300\ms per CPU core during \Run2) and scouting requires extra processing, which must fit in the constraints.

During the beginning of \Run3, HLT scouting data was recorded with a rate of up to 30 (22)\kHz for the 2022 (2023) data-taking periods, respectively, and an event size of about 13\kB, compared to the full raw data event size of about 1\MB.
Events are reconstructed in the HLT scouting scheme if they are accepted by an array of L1 triggers targeting one or two electrons, muons or jets or a moderate amount of \HT with thresholds lower than for other HLT paths.
For \Run3, a special version of the particle-flow reconstruction algorithm using pixel tracks reconstructed with Patatrack (Section~\ref{sec:hlt:tracking}) was deployed.
This special version allowed offloading to GPUs at the expense of a slightly degraded parameter resolution in comparison to tracks reconstructed using the full tracker information.
The reduced reconstruction time made it possible to use particle-flow reconstruction on a larger fraction of input events.

In addition to the objects from the particle-flow reconstruction~\cite{CMS:PRF-14-001} that were already stored during \Run2 (muons, jets, and particle-flow candidates), HLT scouting in \Run3 includes the reconstruction and storage of electrons, photons, and tracks.
At the beginning of the 2023 data-taking period, the electron reconstruction was further optimized by loosening the L1 seeding requirement, resulting in an increased efficiency for low \pt electron reconstruction.
No event selection is applied after the reconstruction.
Reconstructed objects are stored if they fulfill relatively loose criteria, such as $\pt>20\GeV$ and $\abseta<3$ for jets, and $\pt>0.6\GeV$ and $\abseta<3$ for particle-flow candidates.
In addition to kinematic quantities like \pt and $\eta$, other information to facilitate offline analyses is also stored.
In the case of charged particle-flow candidates, this includes parameters of associated tracks to facilitate, \eg, the training of jet tagging using machine-learning algorithms.
Muons are stored not only with track parameters but also with information about the hits in the tracker, allowing the possibility to refit the dimuon vertices.

Since HLT scouting is able to access low-momentum objects with higher rates than conventional HLT trigger paths, it is well suited for analyses targeting low momenta and low-mass particles.
Current studies include, for example, analyses of low-mass dimuons, diphotons, and dielectrons as well as a $\PH\to\bbbar$ analysis that benefits from the decreased threshold on \HT.

\subsection{Data parking}
\label{sec:hlt:parking}

One limiting factor for the HLT output rate is the bandwidth of the prompt event reconstruction at \Tier0.
An alternative approach to increase the amount of data available for physics analysis is to increase the storage rate on disk, while delaying the reconstruction of the data until a later time, when the necessary computing resources are available.
The reconstruction can be scheduled during a year-end technical stop or a long shutdown, for instance.
This concept, known as ``data parking'', was already implemented in \Run1~\cite{CMS:TRG-12-001} and \Run2 to record additional data for \PB physics and other studies.
During an LHC fill, as the luminosity decreases, the bandwidth to trigger additional events increases, and the trigger thresholds for data-parking events are gradually relaxed to record parking data.

\begin{figure}[!ht]
\centering
\includegraphics[width=0.8\textwidth]{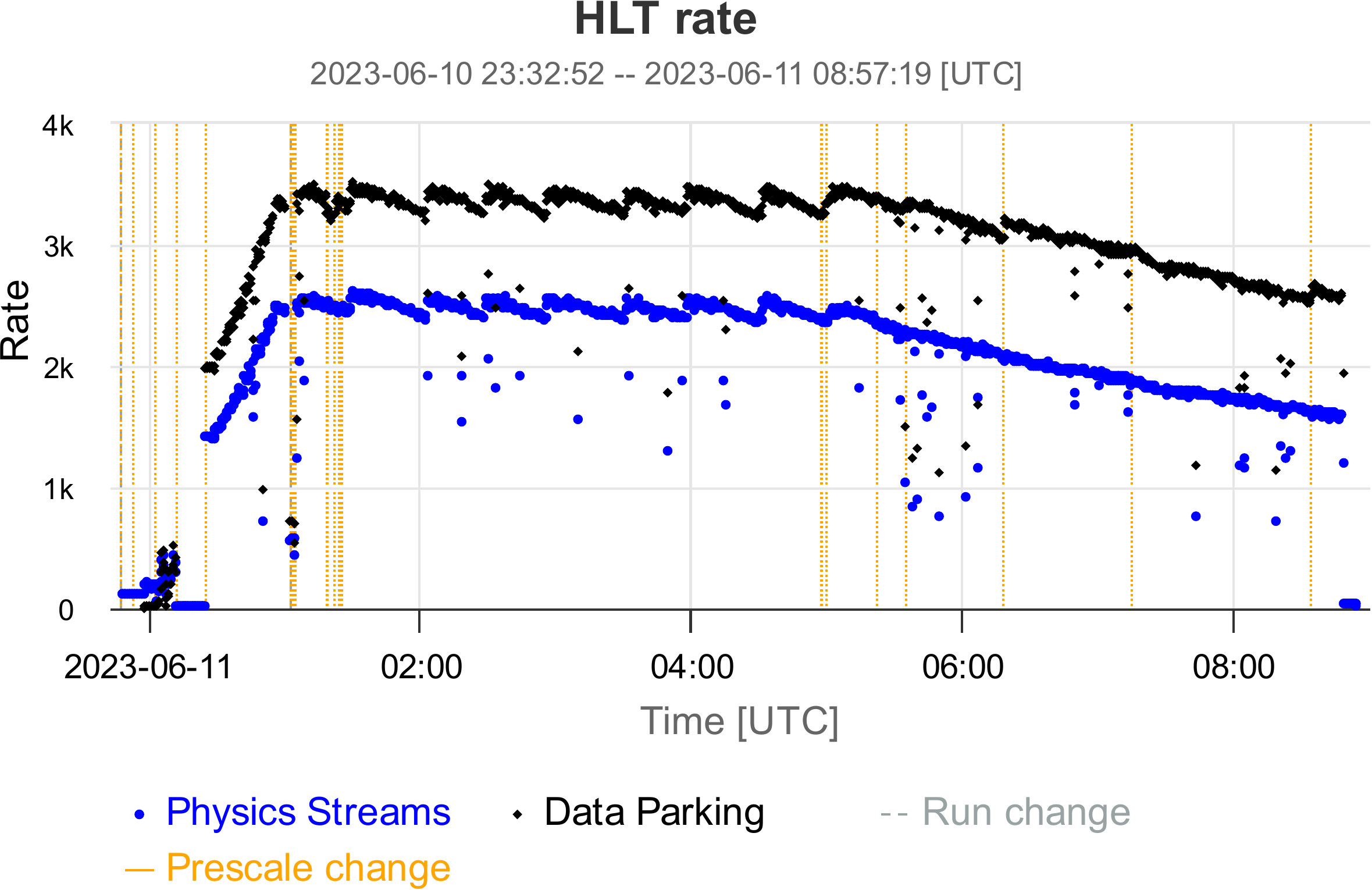}
\caption{%
    The HLT rates for promptly reconstructed data streams (blue) and parked data (black) as a function of time during an LHC fill in 2023.
}
\label{fig:hlt:fillrates}
\end{figure}

In 2018, for example, the collection of \bbbar events was enhanced by tagging and storing events containing at least one displaced muon, \eg, from a semileptonic \PB decay.
The \pt threshold at the HLT for a single isolated muon was 24\GeV for the standard physics menu.
For the parked \PB data the \pt threshold was as low as 7\GeV and the HLT output rate reached 5\kHz at the end of fills, enabling CMS to accumulate about $10^{10}$ \PQb hadrons~\cite{Bainbridge:2020pgi}.
In \Run3, data parking still targets \PB physics, but it also includes a rich set of other physics data.
As of the end of 2022, the parking streams record events with
at least one muon candidate with $\pt>12\GeV$ and a transverse impact parameter significance larger than 6;
events with two muons with $\pt>4$ and 3\GeV with an invariant mass less than 8.5\GeV;
and events with two electrons with $\abseta<1.22$, $\pt<4\GeV$, and an invariant mass less than 6\unit{GeV}.
With the addition of the parking streams, the total HLT reaches peak output rates of 6\kHz.
This can be seen in Fig.~\ref{fig:hlt:fillrates}, which shows (separately) the HLT output rates for promptly reconstructed events and for parked data for an LHC fill recorded in 2023 with a peak levelled luminosity of about $2\times10^{34}\percms$.

\begin{figure}[!ht]
\centering
\includegraphics[width=0.48\textwidth]{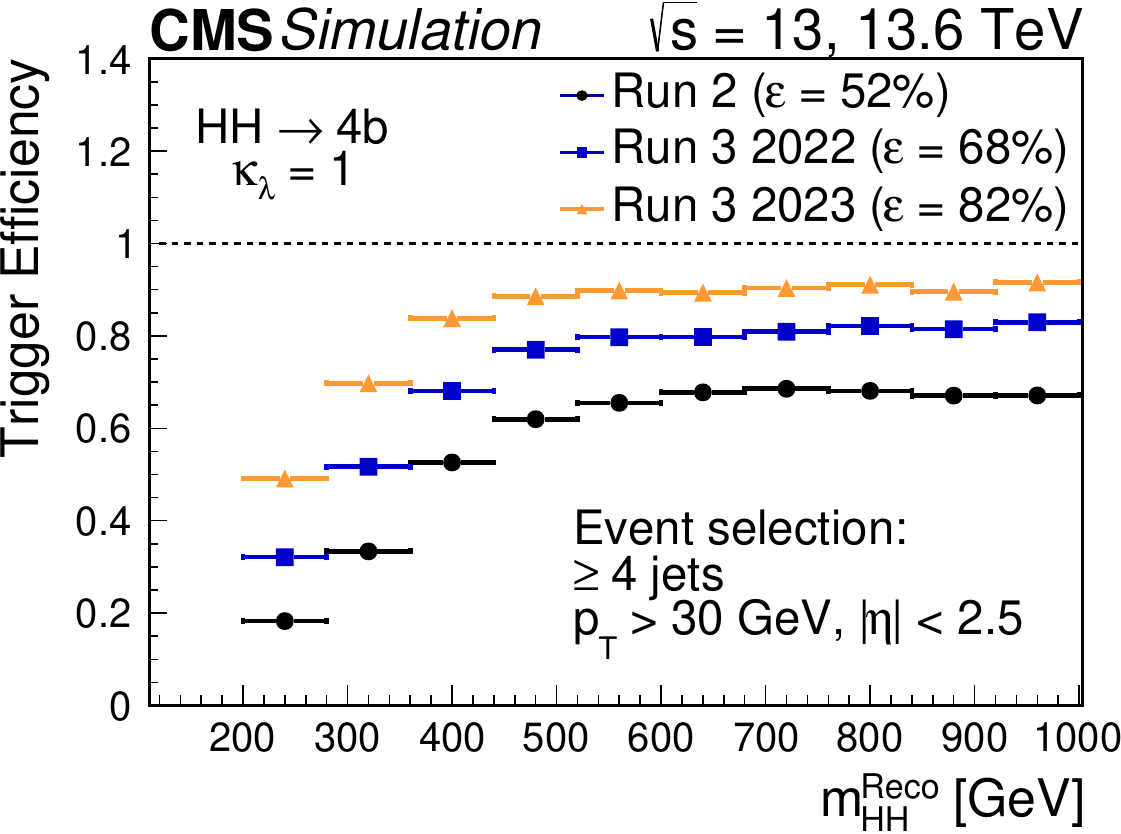}%
\hfill%
\includegraphics[width=0.48\textwidth]{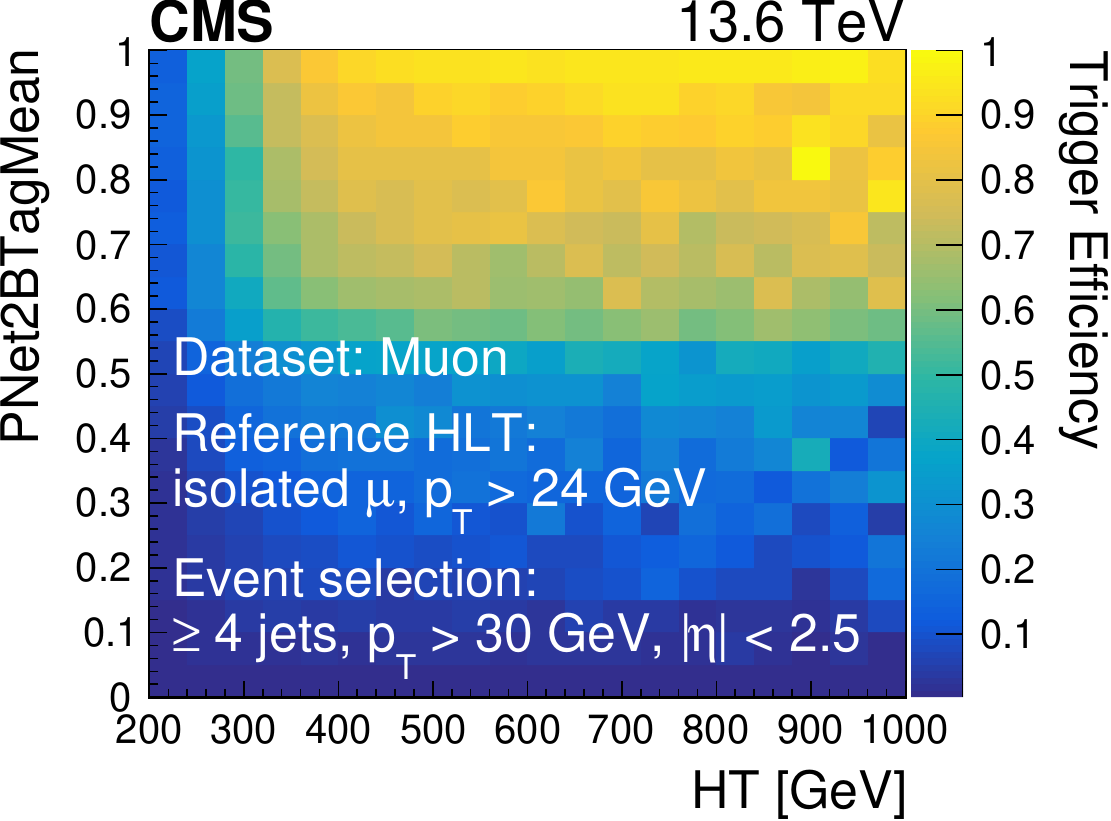}
\caption{%
    Left:\ comparison of the trigger efficiency of the $\PH\PH\to\bbbar$ trigger among the three different strategies used in \Run2 (black), 2022 (blue), and 2023 (orange) using the signal MC sample.
    Right:\ trigger efficiency of the $\PH\PH\to\bbbar$ trigger using events collected by the single muon trigger in 2023.
}
\label{fig:hlt:hh}
\end{figure}

In 2023 the parking strategy was extended to improve the signal acceptance for critical Higgs boson measurements and searches.
By dropping the single-muon parking approach, which was limited to instantaneous luminosities lower than $1.7\times10^{34}\percms$, and by improving the purity of the dielectron triggers, CMS is now able to dedicate bandwidth for events targeting final states with two \PQb-tagged jets, the VBF production mechanism, and LLP signatures.
The two \PQb-tagged jet criterion, mainly designed for the production and decay of $\PH\PH$ into four \PQb quarks, relies on the presence of four jets with $\pt>30\GeV$, two loose \PQb-tagged jets using the ParticleNet tagger, and an aggressive threshold $\HT>280\GeV$.
Figure~\ref{fig:hlt:hh} (left) shows the trigger efficiencies for prompt and parking data.
A clear improvement of more than 20\% is observed with respect to the trigger efficiency of \Run2.
The measured efficiency in a single-muon dataset recorded in 2023 confirms a plateau efficiency of more than 90\% (Fig.~\ref{fig:hlt:hh}, right).

For VBF event candidates a twofold strategy, one inclusive and one exclusive, is pursued.
The inclusive approach relies only on the invariant mass of two forward jets and applies a tight selection of $\mjj>1000\GeV$, while in the exclusive approach a looser threshold on the invariant mass is affordable because there are additional requirements on the central objects in the events.
Table~\ref{tab:hlt:vbf} summarizes the details of the two complementary approaches.

\begin{table}[!ht]
\centering
\topcaption{%
    HLT thresholds and rates of the VBF triggers, as obtained from measurements during an LHC fill in June 2023 at an average instantaneous luminosity of $2.0\times10^{34}\percms$, corresponding to a pileup of 61.
}
\label{tab:hlt:vbf}
\renewcommand{\arraystretch}{1.1}
\cmsTable{\begin{tabular}{l<{,}l<{,}l<{,}l@{}r}
    \multicolumn{4}{l}{HLT algorithm} & Rate [Hz] \\
    \hline
    2 jets with $\pt>105$/40\GeV & $\mjj>1000\GeV$ & \multicolumn{1}{l}{$\detajj>3.5$} & & 720 \\
    2 jets with $\pt>70$/40\GeV & $\mjj>600\GeV$ & $\detajj>2.5$ & 2 central jets $\pt>60\GeV$ & 430 \\
    2 jets with $\pt>105$/40\GeV & $\mjj>1000\GeV$ & $\detajj>3.5$ & 3 central jets & 120 \\
    2 jets with $\pt>90$/40\GeV & $\mjj>600\GeV$ & $\detajj>2.5$ & isolated muon $\pt>3\GeV$ & 110 \\
    2 jets with $\pt>75$/40\GeV & $\mjj>500\GeV$ & $\detajj>2.5$ & $\ptmiss>85\GeV$ & 110 \\
    2 jets with $\pt>45\GeV$ & $\mjj>500\GeV$ & $\detajj>2.5$ & tau $\pt>45\GeV$ & 40 \\
    2 jets with $\pt>90$/40\GeV & $\mjj>600\GeV$ & $\detajj>2.5$ & isol.\ \PGm $\pt>3\GeV$, 3 jets & 12 \\
    2 jets with $\pt>45\GeV$& $\mjj>500\GeV$ & $\detajj>2.5$ & electron $\pt>12\GeV$ & 5 \\
    2 jets with $\pt>75$/40\GeV& $\mjj>500\GeV$ & $\detajj>2.5$ & $\ptmiss >85\GeV$, 3 jets & 5 \\
    2 jets with $\pt>70$/40\GeV& $\mjj>600\GeV$ & $\detajj>2.5$ & 2 jets $\pt>60\GeV$ & 3 \\
\end{tabular}}
\end{table}

\subsection{Heavy ion physics}
\label{sec:hlt:heavyion}

Heavy-ion collisions impose unique challenges on the DAQ and HLT systems, as also noted in Section~\ref{sec:daq}.
A completely custom HLT menu was developed with almost no overlap in content and paths with the \pp menu.
Additionally, because of the dense environment, the physics object reconstruction algorithms used by the HLT are generally customized for heavy-ion running.
For example, while jet reconstruction remains based on the anti-\kt algorithm, the underlying event energy subtraction differs from that used to handle pileup in \pp running.

In preparation for heavy-ion data taking in \Run3, significant effort is devoted to increasing the available L1 bandwidth.
In 2018 up to 30\kHz was achieved, whereas we plan up to 50\kHz in \Run3 conditions.
The most critical components to study include the ECAL, pixel, and tracker detectors.
During a heavy ion test run in 2022, a scan of different ECAL readout settings was performed to study the readout size and potential impact on physics objects.
A size reduction is crucial to achieve the target L1 rate.
In order to fully utilize the L1 bandwidth, fractional prescale factors will be used to operate the trigger at the optimal point between rate and dead time.

The target for the PbPb collision run in 2023 is to record about $5\times10^9$  events.
The HLT will be operated at an output rate above 10\kHz.
About 10\kHz are ``minimum bias'' events, collision events with only a minimal L1 selection.
The budget also contains 1--2\kHz of triggers of selected physics objects, such as muons, electrons, photons, jets, and track multiplicity conditions in central collisions.
A significant fraction of the latter events will also be contained in the minimum bias data set.

To use the bandwidth between the HLT and the \Tier0 center in an optimal way, all triggers are collected into a single ``HIPhysics'' data set.
Only later, during the offline processing step, which is described in Section~\ref{sec:offline}, will they be split into secondary data sets.
Tools have been developed to automatically configure the splitting of large data sets into multiple outputs.

In addition to hadronic collision events, a suite of ultra-peripheral collision triggers will also be deployed.
These target, for example, processes where the two ultrarelativistic nuclei do not collide directly, but their electromagnetic fields interact with each other.
Such events are typically very clean, with only a few physics objects in the detector and no other visible activity.
The size of the event data is minimal and contributes little to the bandwidth.

With the aging of the detector, the efficiency to trigger on some of the more peripheral collisions drops.
Therefore a task force was formed to understand how the decrease in efficiency can be avoided, including, for example, combining information with other detector parts like the Zero Degree Calorimeter.
The heavy ion event size for hadronic collisions is comparable to that for \pp collisions, and a sustained throughput of 17\GBs is anticipated.
Even though the bandwidth is much higher than that in \Run2, given the event size, the target number of recorded minimum bias events will be hard to achieve.
To increase the event rate further, a new approach has been developed in which one of the most significant components of the event record, the tracker information, is removed.
Instead, for each strip cluster, only summary information is written out.
This approach is expected to reduce the event size by about 30\%.
The algorithm was commissioned during the 2022 heavy-ion test run, where both event format contents (the reduced format, ``RawPrime'', and the complete information) were written out.

\clearpage
\section{Offline software and computing}
\label{sec:offline}

\subsection{Overview}

CMS offline computing has evolved over the past 15 years to support the ever-growing needs to trigger, filter, store, transfer, calibrate, reconstruct, and analyze the recorded and simulated data of the experiment.
The offline system receives a subset of the real-time detector information from the data acquisition system, once it is filtered by the high-level trigger at the experimental site, ensures safe curation of the raw data, and produces data for physics analysis.
Major activities are also the production and distribution of Monte Carlo (MC) simulation data, as well as the processing of conditions and calibration information, and other nonevent data.
The data input and output layer of the CMS data processing software is provided by the ROOT framework~\cite{Brun:1997pa}.

Key components of the offline computing system, described in the sections that follow, include an event data model and corresponding application framework, the processing chain and data tiers, computing centers, referred to as \Tier1, \Tier2, and \Tier3, which provide storage and processing resources all connected through a distributed world-wide computing grid, and a set of computing services that provide tools to transfer, locate, and process the data.

\begin{figure}[!ht]
\centering
\includegraphics[width=0.8\textwidth]{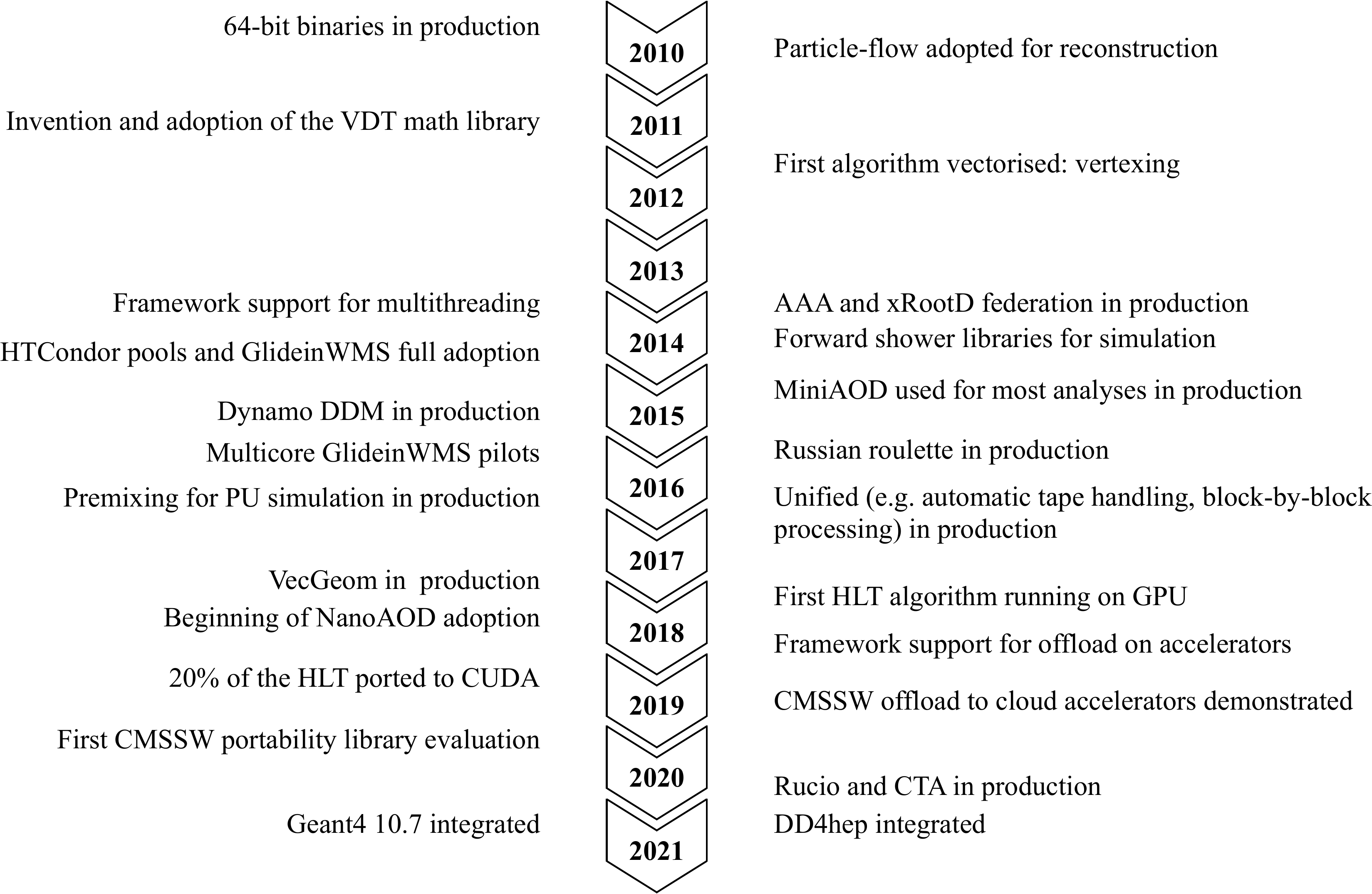}
\caption{%
    Timeline of the major data processing and computing software improvements put in production since 2010.
}
\label{fig:offline:innovations}
\end{figure}

A timeline of the major data processing and computing software improvements put into production over the past decade or so is shown in Fig.~\ref{fig:offline:innovations}.
CMS developed a highly flexible computing model, with the goal of using efficiently all of the resources available to the experiment while minimizing both hardware and personnel needs and maximizing overall throughput.
The innovations led to an increase in efficiency in the use of computing resources, and enabled the experiment to use additional new and diverse types of computing resources that were not available a decade ago.

\subsection{Detector simulation}

The CMS detector simulation is based on the \GEANTfour~\cite{GEANT4:2002zbu} toolkit.
It is augmented with computationally efficient techniques, such as shower libraries for the forward calorimeters, and specific identification criteria per particle type and detector region for neutrons~\cite{Lange:2015sba}, which guarantee high fidelity of the simulation.
The ``Full MC'' simulation chain includes execution of the standard reconstruction chain.
A subset of physics analyses, such as scans of new-physics signatures over large parameter space, use the ``Fast MC'' chain or ``Fast Simulation'' application.
This does not rely on \GEANTfour, but rather a parameterized approach~\cite{Sekmen:2016iql} with a simplified CMS geometry.
The Fast MC chain also uses a faster version of the reconstruction step.
The performance of the CMS simulation is detailed in Ref.~\cite{Pedro:2018jqu}.

\subsection{Event reconstruction}

Event reconstruction is the processing step that transforms the event information contained in the raw data, \ie, packed detector readout data, into high-level physics objects, such as electrons, muons, photons, or jets, which are used in physics data analysis.
In CMS, the event reconstruction is logically partitioned into several steps, starting from the local reconstruction in which the data are processed individually by a single detector component, and ending with global particle-flow reconstruction~\cite{CMS:PRF-14-001} and object identification.
The particle-flow algorithm aims to reconstruct and identify each individual particle in an event, with an optimized combination of information from the various elements of the CMS detector.

A great deal of attention is dedicated to the computing performance of the event reconstruction, both for \pp and heavy-ion collision data.
The run-time optimization is a priority for CMS, and is achieved through algorithmic and technical improvements.
Examples of technical improvements are the adoption of a more recent compiler version, utilization of special compilation flags and other code optimizations, without impact on physics performance.
The timing is benchmarked continuously and in great detail in order to identify and remove performance degrading patterns as early as possible in the release integration process~\cite{Caputo:2023cds}.

Run-time performance improvements must be paired with precise physics validation of the software, delivered swiftly to the code authors.
Physics validation happens at multiple levels during the software integration process, mainly through automatic comparison of data quality monitoring (DQM) histograms of relevant distributions and physics quantities.
Before new reconstruction code is merged into the central software repository, a limited number of events are processed and the quality of the reconstruction is assessed.
During twice-daily integration builds of the CMS software, a growing level of detail is added to the automatic validation, and a larger number of events are processed to obtain the DQM histograms.

Every major release, which may be intended for data processing or mass production of MC simulation samples, is preceded by weekly or biweekly pre-releases.
The validation of pre-releases involves confirmation by detector and physics object experts who assess the quality of the results obtained with the software compared to actual Monte Carlo and data processing campaigns of a few million events.
Final validation, referred to as release validation~\cite{Piparo:2012nx}, is required before new CMS software releases are put into production.

\subsection{Computer architectures and platforms}

CMS strives to use efficiently all of the computing resources at the experiment's disposal.
To this end, the data processing software supports a generic mechanism, described in Section~\ref{sec:offline:framework}, to offload work onto accelerators such as GPUs, described in Section~\ref{sec:offline:offloading}.

Our builds support three different CPU architectures, x86\_64, ARM, and IBM~\textsc{Power}, driven by the opportunities to obtain allocations at HPC centers where these architectures are available (Section~\ref{fig:offline:hpc}).
We perform integration and unit testing of the CMS code, and create regular releases of the whole CMS software suite CMSSW~\cite{Jones:2015soc} for these architectures~\cite{Rodozov:2019zen}.
ARM support began in 2012 and \textsc{Power} in 2016.
Installations of non-x86\_64 releases are performed since 2014 on the file system used as a vector for CMS software, CVMFS, an aggressively cached distributed read-only file system~\cite{Blomer:2011zz, Blomer:2020zen}.

CMS not only builds, tests, and runs code regularly on different CPU architectures, but also employs two different compilers, GCC and Clang, as well as different operating systems.
The combination of a CPU architecture, a compiler, and an operating system is commonly referred to as a ``platform''.
The ability to build, test, and run CMSSW and perform integration tests on several platforms significantly contributes to achieving top code quality, a capability which CMS plans to preserve for the entire \Run3 and beyond.
There is an increased risk of bugs and unstable algorithms silently altering the results of computations without being noticed if the code is tested and executed only on one platform.

\subsection{Application framework}
\label{sec:offline:framework}

\subsubsection{Multithreading}

For LHC \Run2, the application memory usage was foreseen to increase beyond the 2\GB-per-CPU core limit of the computing grid worker nodes, due to increased pileup compared with \Run1.
In order to reduce memory usage per CPU core, the CMS application framework was enhanced to support parallelism through a multithreading paradigm~\cite{Jones:2014uza, Jones:2015soc}.
The multithreaded framework follows a task-parallel paradigm implemented with Intel's oneTBB~\cite{oneTBB:2023git} library, which expresses concurrent units of work as tasks and passes them to oneTBB's task scheduler to run.
The multithreaded framework was introduced for production jobs in 2015 for \Tier1 computing resources, and during 2016 for the majority of other resources.
Currently, production jobs use eight threads by default, which is a good compromise between application memory usage and CPU efficiency.
In addition, 8-core slots have been agreed as a standard job size on the shared worldwide LHC computing grid (WLCG)~\cite{Bird:2005js} resources.

The initial version of the multithreaded framework processed only separate events concurrently.
The framework had to synchronize between the worker threads at specific stages of the data processing, and combined with large variability between the processing times of different collision events, this led to noticeable inefficiencies in the CPU utilization.
The threading efficiency has been gradually improved~\cite{Jones:2017iww}, and currently many levels of concurrency are exploited.
Within one event, independent modules are run simultaneously whenever possible according to their data dependencies.
Modules can use oneTBB's parallel constructs in their internal work including so-called event setup modules that process conditions data.
Events from multiple luminosity sections (Section~\ref{sec:daq}) and interval-of-validity ranges of the conditions data can be processed in parallel.
The framework supports physics modules with different levels of thread-safety guarantee, which are also associated with their threading efficiency or memory usage.

\subsubsection{Offloading to accelerators}
\label{sec:offline:offloading}

The framework has generic support for offloading computations from the CPU worker threads, which allows those worker threads to continue to work on other computations~\cite{Bocci:2020olh}.
A module that offloads computations has its event processing function split into two stages, where the first function is to offload the computations, and the second function is called when the offloaded computations have been completed.

Support for specific offloading technologies is implemented on top of the framework's generic mechanisms.
The Nvidia GPUs on the same computing node are supported with the CUDA API~\cite{Bocci:2020olh}.
Utility classes help with asynchronous execution, sharing resources between modules via the event, and minimizing data movements.
At the time of writing, pixel local reconstruction, pixel track and vertex reconstruction~\cite{Bocci:2020pmi}, ECAL unpacking and local reconstruction, and HCAL local reconstruction have GPU implementations.
Offloading them to a GPU reduces the HLT CPU usage by about 40\%~\cite{CMS:TDR-022}.

Historically, the accelerator vendors have provided their own APIs.
However, developing and maintaining separate versions of algorithms for each platform is unsustainable and, therefore, CMS investigated ways to achieve performance portability with a single code base.
A performance portability framework makes it possible to have one single code base and to build libraries for different classes of hardware, such as CPUs or different vendors of GPUs and accelerators.
The suitability of Alpaka~\cite{Zenker:2016ieee} and Kokkos~\cite{Edwards:2014pdc} libraries for the CMS data processing model and software were explored in detail~\cite{Kortelainen:2021usk, Bocci:2023iiz}, with the conclusion that Alpaka was better suited for CMS for LHC \Run3.

The ability to use accelerators on a remote computing node, or in separate processes on the same node, could allow more flexible use of accelerators.
Using remote accelerators for machine learning (ML) inference has already been demonstrated from the CMS application framework~\cite{Duarte:2019fta, Rankin:2020usv, Krupa:2020bwg} using inference servers such as Nvidia Triton~\cite{Nvidia:2023web}.
In general, ML algorithms are expected to be easily portable between various accelerators.

\subsubsection{Geometry}

Geometry information is fundamental for several CMS applications, such as simulation, reconstruction, and event visualization.
For the start of \Run3, the in-house developed geometry description tool used during \Run1 and \Run2, DDD~\cite{Case:2005cds}, was replaced with the community tool, DD4Hep~\cite{Frank:2014zya}, which is also used by the LHCb experiment and others.
The selection of DD4Hep was made because its library is well behaved in multithreaded environments and because the replacement of an in-house solution with a community-supported tool improves the sustainability of our software stack.
The migration of the geometry for \Run3 took place during LS2 and was an opportunity to review the entire description of the CMS detector, even improving it in some respects.
The migration to DD4Hep of the \Phase2 geometry was completed during 2022.

\subsection{Data formats and processing}

Data and simulation processing workflows are broken into several steps, each defined by the output data structures per event it produces, referred to as a ``data tier''.
The data tiers in use for centrally produced simulation and reconstruction workflows are listed in Table~\ref{tab:offline:datatiers}.
No changes in the data formats were made between \Run2 and \Run3.
A single executable process may produce multiple outputs corresponding to different data tiers, which reduces I/O operations when the necessary data structures are already in process memory.

\begin{table}[!ht]
\centering
\topcaption{%
    Description of the data tiers regularly produced by centrally managed workflows.
    The ROOT framework is used to write and read the data.
}
\label{tab:offline:datatiers}
\renewcommand{\arraystretch}{1.1}
\begin{tabular}{cp{0.82\textwidth}}
    Name & Description \\
    \hline
    GEN & Intermediate and outgoing stable ($c\tau\gtrsim1\cm$) particles from the collision simulation.
    May include Les Houches accord event (LHE) data from the matrix-element generator, if applicable. \\
    SIM & Detailed description of energy deposits left by stable outgoing particles in the detector material.
    Two options are available:\ a highly-accurate \GEANTfour-based application (Full MC); and a parametric fast simulation application (Fast MC), which trades accuracy for a 100-fold decrease in detector simulation time or 10-fold decrease in total CPU time per simulated event.
    The level of inaccuracy introduced by Fast~MC is typically a difference of less than 10\% in final analysis observables. \\
    DIGI & Digitized detector readout or simulation thereof.
    In simulation, the effect of additional collision events (pileup) is folded into the event description in this step.
    In \Run2, a ``premixing'' technique was introduced, where the additional events are summed in a separate processing step and then applied to the simulated primary event. \\
    RAW & Packed detector readout data. \\
    RECO & Detailed description of calibrated detector hits and low-level physics objects.\\
    AOD & Reduced description of calibrated detector hits and low-level physics objects, uncalibrated high-level physics objects. \\
    MiniAOD & Reduced low-level physics objects and calibrated high-level physics objects.
    A truncated floating-point representation is used for most object attributes.
    Introduced for \Run2 to reduce the number of analyses requiring AOD inputs. \\
    NanoAOD & Compact data format containing only high-level physics object attributes stored as (arrays of) primitive data types.
    Introduced during \Run2 to reduce the number of analyses requiring MiniAOD inputs. \\
\end{tabular}
\end{table}

The smaller-size analysis formats are key to reducing both the overall amount of data stored by CMS and analysis processing time per event.
In \Run2, approximate event sizes in each format are 400\kB for AOD (Analysis Object Data), 40\kB for MiniAOD, and 1--2\kB for NanoAOD.
The goal is for 50\% of CMS analyses to use only NanoAOD data sets as input before the end of \Run3.
Although this format is not suitable for every analysis, the increase in usability and speed should appeal to many users.

Nonevent data are used to interpret and reconstruct events~\cite{CMS:Detector-2008}.
Four types of nonevent data remain in use:\ construction data, generated during the construction of the detector; equipment management data; configuration data, comprising programmable parameters related to detector operation; and conditions data, including calibrations, alignments, and detector status information.
A procedure for deriving a selection of online calibration constants has been in place since \Run1.
This has since been consolidated and the ability to create calibration constants from data collected across multiple runs was developed along with new calibration workflows.

A typical \Run3 simulation workflow will be a so-called ``step-chain'' job composing GEN-SIM-DIGI/PU MIX-RAW-RECO-AOD steps and the reduction to MiniAOD and NanoAOD formats.
For detector data, only the RECO-AOD step is performed, followed by the reduction to Mini and NanoAOD.
For a given data-taking period, AOD is produced 1--3 times in large-scale processing campaigns, while MiniAOD~\cite{Petrucciani:2015gjw} and NanoAOD~\cite{Rizzi:2019rsi} data tiers are reproduced more frequently as high-level physics object calibrations are updated.

In the CMS workflow management system, a processing step applied to a given set of inputs (or requested number of events in the case of GEN) forms a task.
Tasks are chained together to form a complete workflow, with intermediate tasks writing their output to site-local storage, and optionally registering the output in the data management system, as discussed in Section~\ref{sec:offline:datamanagement}.
To reduce data transfer, task chains are converted to step-chains when possible, with intermediate output kept only on job-local scratch disk and all subsequent steps executed in a single job.
As discussed in Section~\ref{sec:offline:framework}, the steps are typically executed multithreaded, using up to eight cores per executable, which reduces the workflow management overhead.
The GEN step uses software tools from the HEP theory community, and although some tools may force a step to be executed single-threaded, work is ongoing to improve per-executable parallelism in this context.

\subsubsection{Premixing}

In \Run2 the increase in pileup events resulted in a more I/O and computing-intensive pileup simulation.
A number of individual minimum-bias events comparable to the pileup level had to be read from local or remote disk pools and superimposed to the hard scatter event.
For this reason, a ``premixing'' simulation method~\cite{Lange:2015sba, Hildreth:2017vpw} was introduced to drastically reduce the I/O by summarising all the parasitic pileup collisions in one single ``pileup-only'' event.
Hard-scatter events are generated and simulated without pileup; separately, a sample of MC pileup-only events is also produced, using the pileup distribution for a certain running period and considering in-time and out-of-time interactions in all subdetectors.
A selected premixed set of pileup events is then overlaid on the hard scatter events.
This approach reduces I/O operations by 90\% compared with the previous method of overlaying individual minimum-bias events.
On average, digitization and reconstruction of simulated events with pileup is twice as fast~\cite{Hildreth:2017vpw}.
The size of the premix library is proportional to the number of simulated events and contained about 200 million events in \Run2.
For \Run3, the integrated luminosity is expected to be higher, and larger premixed samples will be generated.

\subsection{Computing centers}

The organization of the distributed computing infrastructure used by CMS was initially based on a model described by MONARC~\cite{Aderholz:2000nk}, where the sites were organized in ``Tiers'', pledging CPU, disk storage, and tape resources, proportional to the commitment of each funding agency within the CMS Collaboration.
The roles of the various Tiers were established as follows:
\begin{itemize}
\item A \Tier0 center close to the experiment (CERN) to execute a first  calibration and reconstruction pass, the so-called ``prompt reconstruction'', as well as to maintain a custodial copy of all RAW data on tape; 24h/7 support is guaranteed;
\item \Tier1 regional centers (six in use by CMS), which maintain a second distributed custodial copy of the RAW data on tape, and provide CPU for re-reconstruction and MC simulation; 24h/7 support is guaranteed;
\item \Tier2 local centers (about 50 used by CMS), providing support for analysis activity and MC simulation; guaranteed support only during working hours.
\end{itemize}
The \Tier0 is the largest of the CMS sites.
It provides computing capacity, disk and tape storage and is hosted by CERN.
At the \Tier0, prompt reconstruction starts 48 hours after the data are acquired.
This delay is necessary for an initial set of detector calibrations to be bootstrapped, starting from the execution of alignment and calibration processing sequences on the newly acquired data---the so-called ``express'' reconstruction.
One such workflow is described in Ref.~\cite{Futyan:2010zz}.
Once the calibrations are derived, they are consolidated in a payload written to a database that is then read by prompt reconstruction at a later stage.
The flexibility of the CMS central job submission infrastructure makes it possible to produce substantial amounts of simulated samples at the \Tier0.

\begin{figure}[!ht]
\centering
\includegraphics[width=\textwidth]{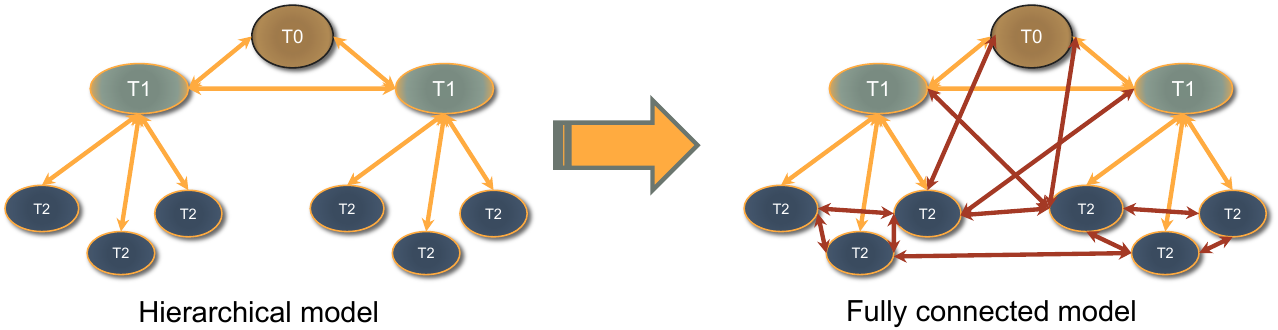}
\caption{%
    The evolution of the CMS computing model from a hierarchical (left) to fully connected structure (right).
}
\label{fig:offline:computingmodel}
\end{figure}

France, Germany, Italy, the US, the UK, and Spain provide the six \Tier1 sites of CMS.
These sites have a primary role in the computing model, offering the precious combination of computing capacity, disk storage and tape archival space, all together with high bandwidth connectivity to CERN and \Tier1s through the LHCOPN private IP network, as well as to other sites through the LHCOne network.
\Tier1s are used to store the active copy of the RAW data on tape, to perform re-reconstruction passes when data processing algorithms and calibrations improve so much to require it.
The \Tier1 centers have been adapted to support the \Tier0 in the task of prompt data processing, whenever needed.
In the current computing model a job can now run, in principle, wherever free CPU is available while accessing input data through the WAN (Fig.~\ref{fig:offline:computingmodel}).
The total amount of computing resources pledged by the largest and smallest of the \Tier1 centers of CMS differs by an order of magnitude, still these special sites are equally critical for the support of the CMS physics program.

The computing model used in \Run2 and \Run3 is a significant evolution of the MONARC hierarchical model.
The roles of the \Tier1/2s have become more similar in order to optimize the usage of the resources with high efficiency, and following collaboration-wide priorities.
The presence of high-speed wide-area network (WAN) connections, the development of advanced data federation~\cite{Bloom:2015ieee, Bauerdick:2012st} and caching~\cite{CMS:2014cof} technologies and the creation of optimized data tiers such as MiniAOD and NanoAOD, allow distribution of the tasks among the \Tier1's and \Tier2's following CMS priorities, with minimal net distinction between analysis, MC production, and data (re-)processing.
However, \Tier1 sites still tend to be assigned a larger proportion of central production jobs.

\Tier3 sites are those which do not pledge resources, but nevertheless provide CPU and storage in varying amounts.
These could be sites with batch farms for mainly local use, or those that predominantly support another experiment, but allow CMS to take CPU slots opportunistically when they are available.

An important aspect of the computing model in CMS is the continuous monitoring of the distributed data centers, both via recording of CPU and data metrics of the workloads, and with automatic tests checking periodically the health and reliability of the various services.
These tests allow the production system to temporarily disable sites with transient issues and to re-enable them when working again.

In order to easily run computing jobs on diverse resources, the environment was standardized during \Run2 via the use of virtualization and containerization in Singularity~\cite{Kurtzer:2017pone}.

\subsubsection{High performance computing (HPC)}
\label{fig:offline:hpc}

{\tolerance=800
National and supranational bodies are investing substantial resources in supercomputers.
These machines are part of our computing infrastructure and, given the advantage they represent for many scientific and industrial applications, they are here to stay.
In the exascale era, HEP might be compelled to count on these resources for the bulk of its computing capacity.
CMS has already started to take advantage of HPCs, adapting its software and computing tools.
However, the integration of individually highly optimized HPC systems with the CMS experiment computing environment is not a trivial task.
From a technical perspective, HPC centers differ in a variety of specialized hardware setups and strict usage policies can apply, particularly related to security.
As a consequence, a variety of conditions can be encountered when attempting to integrate these centers, spanning from less common operating systems, absence of local scratch disk space on computing nodes, low memory availability per core, limited or even absent outbound network connectivity, and, of course, distinct architectures.
Given the unique nature of HPC systems, CMS has invested a substantial amount of effort to develop solutions case by case.
\par}

CMS has deployed several solutions which can be classified broadly into two categories:\ a transparent site-extension model where HPC computing resources are seamlessly integrated into an existing CMS grid site, and the HEPCloud model in which an additional layer of job submission infrastructure sits between the experiment's workflow management system and the HPC computing resources.
In both cases, the storage of existing CMS grid sites is used through remote access from the HPC, complemented by on-site caches or local storage at some HPC centers.

Integration of HPCs that are transparent to central computing operations have been performed in several countries in Europe, \eg, Italy, Switzerland, Germany, and Spain, for example expanding existing \Tier1 or \Tier2 sites elastically into these machines~\cite{Boccali:2021wil, Boccali:2020lcd, Acosta-Silva:2021ouk}, or by deploying so-called overlay batch systems~\cite{Fischer:2020bqx}.
It is also through this model that a validation of the physics performance of CMS software on the PowerPC architecture could be performed~\cite{Boccali:2023jin}.

The HPC resources located in the US are integrated into the CMS systems via HEPCloud~\cite{Hufnagel:2018zzy}, a portal to an ecosystem of diverse computing resources, commercial or academic, hosted at FNAL.
Since 2020, HEPCloud has provisioned resources from seven different HPC centers located for example at the NERSC, PSC, TACC, or SDSC computing centers.

\subsubsection{Data archive at CERN}

CMS migrated to the CERN tape archive (CTA)~\cite{Davis:2019jbi} system in December 2020.
The CTA system provides the tape backend to the CERN EOS disk system~\cite{Peters:2015aba}, and together EOS and CTA replace CASTOR.
This upgrade was necessary to prepare for the higher data rates of \Run3 and beyond, as well as to provide a uniform API for disk and tape operations, with support for newer protocols and authentication methods.
It is difficult to achieve high throughput using a traditional buffer made up of spinning disks, due to contention on the drives from multiple simultaneous streams.
Therefore a new approach was taken, with a small, fast SSD buffer in front of the physical tapes.
This allows the tape drives to operate close to their nominal speed of 400\MBs.

\subsection{Computing services}

\subsubsection{Data management}
\label{sec:offline:datamanagement}

During \Run1 and \Run2, CMS used PhEDEx~\cite{Rehn:2006chep} and Dynamo~\cite{Iiyama:2020biz} as data management tools.
However, for \Run3 and beyond it was necessary to adopt a more scalable, flexible, and powerful system to increasingly automate the data management, and include the possibility to scale up transfers to around 100 petabytes per day by the late 2020s for the start of the High-Luminosity LHC (HL-LHC)~\cite{ZurbanoFernandez:2020cco}.
The new software should also support future technologies such as token authorization and non-FTS (File Transfer Service)~\cite{Karavakis:2020cnw} transfers.
The Rucio system~\cite{Barisits:2019fyl}, a data management project for scientific communities, was adopted by CMS at the end of 2020.
It can perform all of the functions of PhEDEx and Dynamo, and also make higher-level decisions about data placement.
Rucio is run centrally, without an agent at every grid site as required by PhEDEx.

The CMS Rucio infrastructure is based on Helm~\cite{helm:2023git}, Kubernetes~\cite{kubernetes:2023git}, and Docker~\cite{docker:2023git}, which are industry standards.
All of the Rucio services are built into a single Kubernetes cluster which can be brought up from scratch in under an hour.
Rucio removes data as additional space is needed at a site.
Only data that is not held in place by one or more rules is eligible to be removed.
To make this decision, Rucio uses the last access time of the data.
This information is taken from job reports, CMSSW file reads, and monitoring of the AAA (Any data, Anywhere, Any time) system, which is described in the following subsection.

CMS developed a mechanism to check the consistency of the Rucio database with the actual state of files on disk.
Remote disks are scanned using XRootD tools, comparing the results with the state of the database, and generating lists of missing and unneeded files.
These lists are passed to additional Rucio components for file removal or re-transfer.

\subsubsection{Data transfer protocols}

During \Run1 and \Run2, CMS transferred data among sites via protocols based on the grid security infrastructure (GSI)~\cite{Foster:1998acm, Butler:2000comp}, with GSIFTP the most common protocol used and GridFTP its most common implementation.
When the end of support for GSI was announced, the LHC community started looking for a replacement both for the protocol and an authentication mechanism.
The WebDAV protocol~\cite{Dusseault:2007rfc}, an extension of HTTP which supports third party copy (TPC) transfers and tokens for authentication, was selected.
The adoption of the WebDAV protocol started early in 2020 and was completed in 2022.

The ``Any Data, Anywhere, Anytime'' (AAA) data federation~\cite{Bloom:2015ieee}
was introduced in 2014--2015 during LS1.
This is a model for effective federation of distributed storage resources via an XRootD cluster at each computing center, allowing for remote access to any file within the CMS namespace.
The XRootD framework supports partial reads of files and is commonly used by analysis jobs.
The combination of AAA and caching~\cite{CMS:2014cof} technologies allows nonlocality between data and CPU, and is expected to lead to entirely storage-free computing centers in \Run3.
These are the first steps towards a ``Data Lakes'' architecture~\cite{Espinal:2020fyj}, a centralized data repository, as envisioned for the HL-LHC era.

\subsubsection{Central processing and production}

Large-scale MC sample production and data event reconstruction activities are performed in a distributed computing infrastructure, coupled to a specialized workload management system (WMS), described in Section~\ref{sec:offline:wms}.
A global batch queue manages the distribution of production and analysis jobs to the CMS distributed computing system in an optimized and flexible way.
The submission infrastructure (SI) employs GlideinWMS~\cite{Sfiligoi:2009cct} and HTCondor~\cite{Thain:2005cpe} software suites in order to build and manage a ``Global Pool''~\cite{Balcas:2015rbw} of computing resources where the majority of the CMS tasks are executed.

\begin{figure}[!ht]
\centering
\includegraphics[width=0.8\textwidth]{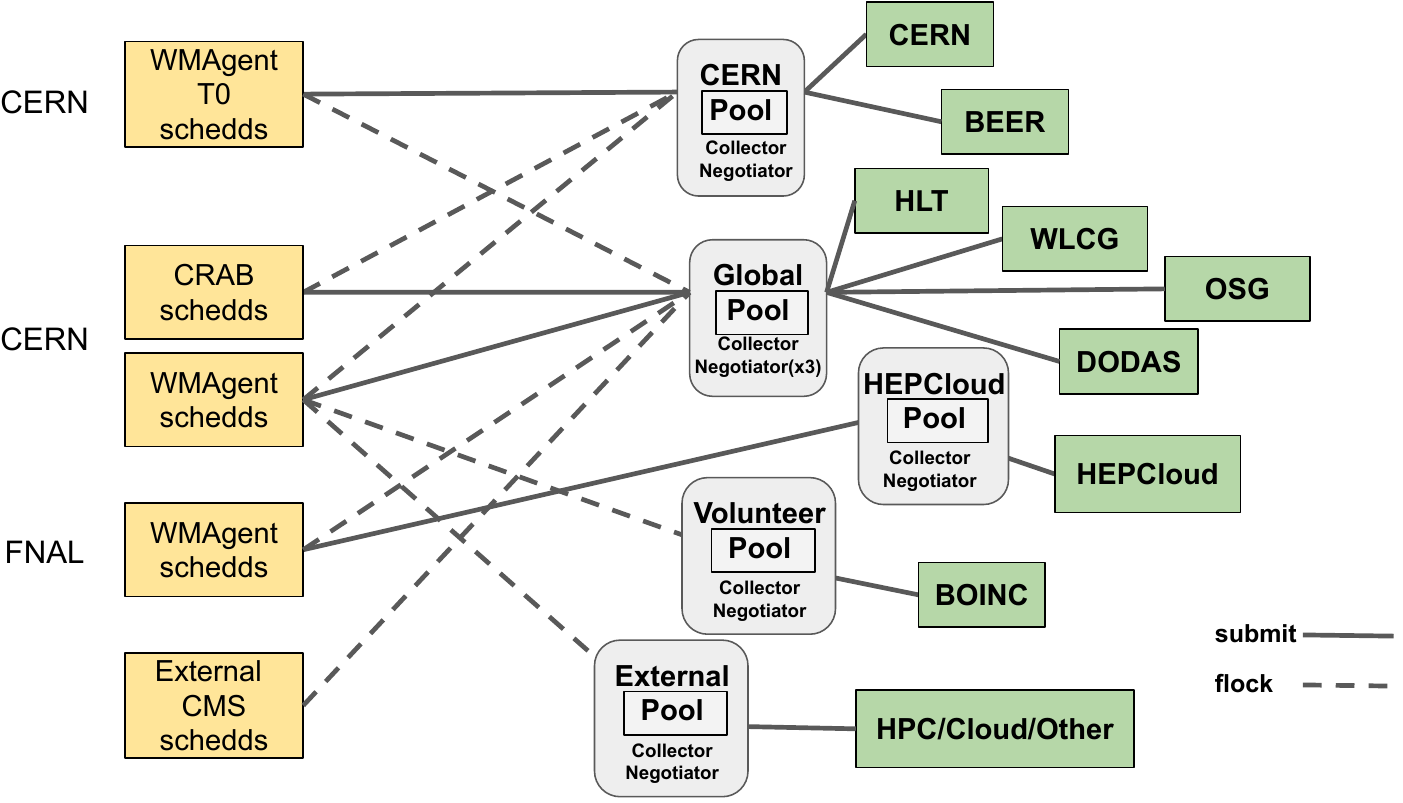}
\caption{%
    Schematic diagram of the submission infrastructure, including multiple distributed central processing and production (WMAgent) and analysis (CRAB) job submission agents (schedds), from Ref.~\cite{CMS:2020qqu}.
    Computing resources allocated from diverse origins (green boxes) are grouped into HTCondor pools (gray boxes), federated via workload flocking.
    The collector and negotiator agents (yellow boxes) keep the state of each pool and perform the workload-to-resource matchmaking.
}
\label{fig:offline:sicomplex}
\end{figure}

The SI comprises multiple interconnected HTCondor pools~\cite{CMS:2020qqu}, as shown in Fig.~\ref{fig:offline:sicomplex}, redundantly deployed at CERN and FNAL in order to ensure a high-availability service.
The main component of the SI, the Global Pool, obtains the majority of its resources via the submission of pilot jobs to WLCG~\cite{Bird:2005js} and open science grid (OSG) sites.
However, locally instantiated processing nodes via, for example, DODAS~\cite{Spiga:2019fwz} or BOINC~\cite{Anderson:2020jgc}, as well as opportunistic resources, such as the HLT filter farm~\cite{daSilvaGomes:2018pqz} when not in use for data taking, can also be employed.
As described in Section~\ref{fig:offline:hpc}, the SI computing capacity has recently expanded into HPC facilities which are integrated as part of the Global or HEPCloud pools.
CERN on-site CMS resources, along with opportunistic local (BEER~\cite{Smith:2018jph}) and cloud~\cite{Giordano:2016zen, Cordeiro:2017bha} computing slots, are organized into a third HTCondor pool, built on a dedicated set of hosts to isolate it from potential issues in the main Global Pool, given its critical role in supporting \Tier0 tasks during data-taking periods.
Specialized nodes known as ``schedds'' control workload submission.
While being primarily attached to one ``pool'', these schedds can interact with other federated pools, requesting additional resources when demands are not covered in the primary pool, as indicated in Fig.~\ref{fig:offline:sicomplex}.

The SI can leverage any grid, cloud, HPC, or opportunistic resource available to CMS, which in total have tripled in size over the last five years, currently amounting to around 400k CPU cores.
The Global Pool matches diverse CMS workloads, which include single-core as well as multicore requests~\cite{Perez-CaleroYzquierdo:2017pky} to these resources.
A successful scheduling is achieved by simultaneously ensuring that all available resources are efficiently used~\cite{Bockelman:2019leu}, a fair share of resources between users is reached, and the completion of CMS tasks follows their prioritization, minimizing job failures and manual intervention.
While the SI typically manages 100k to 150k simultaneously executing tasks, recent scalability tests~\cite{Perez-CaleroYzquierdo:2021sjh} have demonstrated the capacity of the infrastructure to sustain in excess of half a million concurrently running jobs.

\subsubsection{Workload management system}
\label{sec:offline:wms}

The CMS production WMS, based on the WMCore framework, comprises many services, including WMAgent, the most important and complex system in this architecture.
It is a stateful software, responsible for the work splitting, event and data bookkeeping, job management and submission, while interfacing with the data management tools for global data bookkeeping and grid data transfers.
The WMAgent system is tightly coupled to the schedds in the SI, enabling it to make use of the Global Pool resources, handling thousands of requests in parallel and enforcing the workflow prioritization.
Other components in the production WMS include the WMStats system, responsible for the workflow and job monitoring, and providing logging and recovery features, and the request manager, which manages and stores workflow descriptions, submitting workloads for processing on grid resources.
A suite of ``Microservices'' has been developed to perform very specific tasks such as automatic input and output data placement, monitoring, and cleanup of the data used in central production.
The introduction of Microservices has resulted in multiple features of the system being automated, thus reducing the required operational effort.

\subsubsection{Distributed analysis}

For CMS scientists, a dedicated tool known as CMS remote analysis builder (CRAB) has been available since the start of \Run1, to manage the submission on the grid of data analysis applications or small-scale productions of data samples that are not centrally managed.
CRAB offers the same functionalities as originally planned~\cite{CMS:Detector-2008}.
Users interact with a thin client running inside the CMSSW environment, which packs the user analysis environment into a sandbox and uploads it together with a configuration file to a central, database-centric service hosted in the CMSWeb services framework described in Section~\ref{sec:offline:cmsweb}.
Centralized components run either as Docker containers, or under the control of HTCondor~\cite{Thain:2005cpe} or DAGMan~\cite{Couvares:2007book}.
These components handle preparation, execution, bookkeeping, error recovery, monitoring, and output delivery for the user application.
CRAB has been continuously improved to provide increased automation and scaling, leaner operations, better error recovery, access to data archived on tape and integration with the Rucio data management system, while keeping the same basic interface for the user.
For users needing a simpler, more interactive service to access resources from the Global Pool to make histograms, for example, the CMS Connect~\cite{Balcas:2017wse} service is available as an alternative to CRAB.

\subsubsection{DBS database}

The data bookkeeping service (DBS) is a catalog holding event metadata for all simulated and experimental data processed and stored by the CMS experiment.
DBS contains all necessary information for tracking the data sets, including provenance information, processing history and parentage relationships between files and data sets.
DBS is used by all computing jobs in CMS.
The current version (version 3) was completely redesigned and reimplemented following a project review in 2009, in order to better fit the evolved CMS data processing model, to better integrate with the evolved CMS data management and workflow management (DMWM) projects, and to ensure a higher scalability.
DBS3 is implemented as a Python-based web service using a standardized architecture provided by the CMS DMWM project.
Client-server communication in combination with thin client APIs using the JSON format as a lightweight replacement of XML-RPC led to better scalability.
In addition, the database schema has been streamlined and fine-tuned in order to avoid both excessive table joins, reduce query latency, and improve server stability, which enhance data accessibility.

The data aggregation system (DAS)~\cite{Kuznetsov:2010pcs} provides access to distributed CMS metadata via a common query language (QL).
Users may access DBS, Rucio and other metadata sources by placing simple QL queries without prior knowledge of service APIs, database schemas, and implementation or location of the services.

\subsubsection{Web services and security}
\label{sec:offline:cmsweb}

In order to guarantee successful data taking, CMS uses a series of web applications, which perform various tasks related to job submission, job monitoring and bookkeeping, and the location of the data sets.
These services are provided and maintained by several teams and centrally deployed under the umbrella CMSWeb.

To achieve optimal resource utilization and high availability, the services have been containerized and are now being deployed in a Kubernetes cluster~\cite{Imran:2021vfp}.
Detailed monitoring, for all the services and the nodes they run on, has been built on top, using the central CMS computing monitoring tools.
Authorization and authentication (A\&A) is also being offered as a service in the CMSWeb cluster.
Currently, A\&A is using X509 certificates, but is rapidly migrating to authentication tokens, following the shift to the OAuth standard initiated by the WLCG.
This migration should take place by 2024, according to current plans~\cite{WLCGAuthorizationWG:2022zen}.

Security is a very important aspect of web services.
A great effort is required to stay up-to-date with the latest security recommendations.
The web services and security teams work closely with the CERN IT security team, following up  vulnerabilities.
In a joint effort, proactive penetration tests are organized several times a year to discover potential issues ahead of time and to ensure that the infrastructure remains safe.
As part of improving security, the logs of all the CMSWeb services are now pushed to the security operations center (SOC), which analyses them and identifies possible threats.

\subsubsection{Monitoring and analytics}

A scalable and reliable monitoring system is required to ensure efficient operation of the CMS distributed computing services, and to provide comprehensive measurements of the system performance.
Metrics such as wall-time consumption of computing resources, memory, CPU, storage usage, and data access patterns are monitored to study the evolution of the performance over time and allow for in-depth analyses of the main system parameters.

The main components of the CMS monitoring infrastructure are presented in Ref.~\cite{Ariza-Porras:2020sly}.
CMS relies on central CERN~\cite{Aimar:2017nfu} and dedicated CMS monitoring infrastructures.
The former, supported by the CERN IT department, is used extensively to store data from CMS computing subsystems, such as HTCondor, submission infrastructure, CMSWeb user activities, analysis and production workflows coming from the WMAgent and CRAB job submission tools, and data-transfer and storage information from Rucio.
The status of all CMS computing systems is monitored in real time using predefined views and dashboards.

The dedicated CMS monitoring infrastructure is composed of several Kubernetes clusters.
It is mostly used to monitor computing nodes and services, and to provide additional features such as a sophisticated alert and notification system, data aggregation, annotations, and automation of several workflows.
Several hundred nodes and services are monitored, and a fault-tolerant infrastructure is provided with minimal operational and maintenance effort.

\clearpage
\refstepcounter{section}
\addcontentsline{toc}{section}{Summary}
\section*{Summary}
\label{sec:summary}

Since the beginning of the LHC operation in 2009, the CMS detector has undergone a number of changes and upgrades, adapting the experiment to operating conditions at luminosities well beyond the original design.
In 2022, the LHC \Run3 began and CMS successfully recorded its first 40\fbinv of proton-proton data at a center-of-mass energy of 13.6\TeV with an operation efficiency of 92\%.
This paper describes the modifications, as installed and commissioned for LHC \Run3.

The upgraded pixel tracking detector was installed in early 2017.
In the new detector, the number of barrel layers was increased from three to four, and the number of disks in each endcap from two to three, whereas the material budget was reduced, leading to a better tracking performance up to an absolute pseudorapidity of 3.0 for pixel tracks.
The upgrade also involved a new readout chip enabling increased hit detection efficiencies at higher occupancy.

In the electromagnetic calorimeter, measures were taken to improve the monitoring and calibration of effects caused by irradiation, leading to a loss in the \PbWOfour crystal transparency and an increase of the leakage current in the avalanche photodiodes.
In the second long shutdown (LS2) the calibration algorithms were refined to identify and remove spurious spike signals and to determine time-dependent correction factors for the laser monitoring system.

The upgrade of the hadron calorimeter included new readout electronics with finer granularity, leading to an increase in the number of channels and longitudinal segmentation.
The previous generation of photosensors was replaced by silicon photomultipliers, which measure the scintillator light output with a better signal-to-noise ratio.

In the muon system, a gas electron multiplier (GEM) detector, consisting of four gas gaps separated by three GEM foils, was added in the endcaps.
The other subsystems, drift tubes (DT), cathode strip chambers (CSC), and resistive-plate chambers (RPC), underwent several upgrades.
In the DT, the muon trigger logic was replaced by a new data concentrator based on \uTCA architecture.
The top of CMS was covered with a neutron shield to reduce the background in the top external DT chambers.
An outer ring of CSCs (ME4/2) was added in LS1, and in view of the High-Luminosity LHC, the bulk of the CSC electronics upgrades that required chamber access were performed already during LS2.
An outer rings of the RPC chambers in station four (RE4/2 and RE4/3) were added as well.
The endcap muon track finder of the L1 trigger was upgraded to utilize GEM-CSC joint track segments to optimize the final track reconstruction and resolution at the trigger level.

The precision proton spectrometer was upgraded significantly.
Its tracker radiation-damaged sensors and chips were replaced.
The mechanics of the detector, as well as the front-end electronics, were completely redesigned to add a novel internal motion system designed to mitigate the effects of radiation damage.
In the timing system a second station was installed in each arm.
All detector modules were replaced by new double-diamond modules with the aim of further improving the timing resolution.

In the beam radiation instrumentation and luminosity system, new versions of the pixel luminosity telescope (PLT), the fast beam conditions monitor (BCM1F), and the beam conditions monitor for losses (BCML) were installed for \Run3.

To cope with increasing instantaneous luminosities, the CMS data acquisition (DAQ) system underwent multiple upgrades.
The backend technology was gradually moved to the more powerful \uTCA standard.
A new optical readout link with a higher bandwidth of 10\Gbs was developed.
The bulk of the DAQ system downstream from the custom readout benefited from advances in technology to achieve a much more compact design, while doubling the event building bandwidth.

The first level (L1) trigger, composed of custom hardware processors, uses information from the calorimeters and muon detectors to select events at a rate of up to 110\kHz within a fixed latency of about 4\mus.
The developments in the L1 trigger mostly focused on the addition of dedicated triggers that became possible due to enhanced capabilities of the global trigger logic and increased trigger information delivered by the calorimeters and muon systems.
Among other applications, new triggers for long-lived particle signatures were implemented.
The addition of a 40\MHz scouting system that receives data from both the calorimeter and muon systems, further broadens the physics reach of CMS.

The high-level trigger (HLT) performs the second stage of event filtering and accepts events at a sustained rate of the order of 5\kHz.
Since \Run3 began, an additional 30\kHz of HLT scouting data is recorded.
Since 2016, the HLT has been operated using multithreaded event processing software, minimizing memory requirements through reduction of the number of processes concurrently running.
For \Run3, GPUs were successfully deployed in the HLT.

Substantial improvements were achieved in the physics performance and speed of the software, as well as in the computing infrastructure.
Some of the major changes are:\ support for multithreaded processes and utilization of GPUs; direct remote data access; and usage of high-performance computing centers.
New tools such as Rucio for data management were adopted with future data rates in mind.
Considerable effort was put into the automation of the workflows and the validation of the software.
Physics analyses have been moved to smaller and smaller formats for centrally produced and experiment-wide shared data samples, the most recent of which is the NanoAOD.

The development of the CMS detector, as described in this paper, constitutes a solid basis for future data taking.

\clearpage
\refstepcounter{section}
\addcontentsline{toc}{section}{Acknowledgments}
\begin{acknowledgments}
\hyphenation{Bundes-ministerium Forschungs-gemeinschaft Forschungs-zentren Rachada-pisek} We congratulate our colleagues in the CERN accelerator departments for the excellent performance of the LHC and thank the technical and administrative staffs at CERN and at other CMS institutes for their contributions to the success of the CMS effort. In addition, we gratefully acknowledge the computing centers and personnel of the Worldwide LHC Computing Grid and other centers for delivering so effectively the computing infrastructure essential to our analyses. Finally, we acknowledge the enduring support for the construction and operation of the LHC, the CMS detector, and the supporting computing infrastructure provided by the following funding agencies: the Armenian Science Committee, project no. 22rl-037; the Austrian Federal Ministry of Education, Science and Research and the Austrian Science Fund; the Belgian Fonds de la Recherche Scientifique, and Fonds voor Wetenschappelijk Onderzoek; the Brazilian Funding Agencies (CNPq, CAPES, FAPERJ, FAPERGS, and FAPESP); the Bulgarian Ministry of Education and Science, and the Bulgarian National Science Fund; CERN; the Chinese Academy of Sciences, Ministry of Science and Technology, the National Natural Science Foundation of China, and Fundamental Research Funds for the Central Universities; the Ministerio de Ciencia Tecnolog\'ia e Innovaci\'on (MINCIENCIAS), Colombia; the Croatian Ministry of Science, Education and Sport, and the Croatian Science Foundation; the Research and Innovation Foundation, Cyprus; the Secretariat for Higher Education, Science, Technology and Innovation, Ecuador; the Ministry of Education and Research, Estonian Research Council via PRG780, PRG803 and PRG445 and European Regional Development Fund, Estonia; the Academy of Finland, Finnish Ministry of Education and Culture, and Helsinki Institute of Physics; the Institut National de Physique Nucl\'eaire et de Physique des Particules~/~CNRS, and Commissariat \`a l'\'Energie Atomique et aux \'Energies Alternatives~/~CEA, France; the Bundesministerium f\"ur Bildung und Forschung, the Deutsche Forschungsgemeinschaft (DFG), under Germany's Excellence Strategy -- EXC 2121 ``Quantum Universe" -- 390833306, and under project number 400140256 - GRK2497, and Helmholtz-Gemeinschaft Deutscher Forschungszentren, Germany; the General Secretariat for Research and Innovation and the Hellenic Foundation for Research and Innovation (HFRI), Project Number 2288, Greece; the National Research, Development and Innovation Office (NKFIH), Hungary; the Department of Atomic Energy and the Department of Science and Technology, India; the Institute for Studies in Theoretical Physics and Mathematics, Iran; the Science Foundation, Ireland; the Istituto Nazionale di Fisica Nucleare, Italy; the Ministry of Science, ICT and Future Planning, and National Research Foundation (NRF), Republic of Korea; the Ministry of Education and Science of the Republic of Latvia; the Lithuanian Academy of Sciences; the Ministry of Education, and University of Malaya (Malaysia); the Ministry of Science of Montenegro; the Mexican Funding Agencies (BUAP, CINVESTAV, CONACYT, LNS, SEP, and UASLP-FAI); the Ministry of Business, Innovation and Employment, New Zealand; the Pakistan Atomic Energy Commission; the Ministry of Education and Science and the National Science Center, Poland; the Funda\c{c}\~ao para a Ci\^encia e a Tecnologia, grants CERN/FIS-PAR/0025/2019 and CERN/FIS-INS/0032/2019, Portugal; the Ministry of Education, Science and Technological Development of Serbia; MCIN/AEI/10.13039/501100011033, ERDF ``a way of making Europe", Programa Estatal de Fomento de la Investigaci{\'o}n Cient{\'i}fica y T{\'e}cnica de Excelencia Mar\'{\i}a de Maeztu, grant MDM-2017-0765, projects PID2020-113705RB, PID2020-113304RB, PID2020-116262RB and PID2020-113341RB-I00, and Plan de Ciencia, Tecnolog{\'i}a e Innovaci{\'o}n de Asturias, Spain; the Ministry of Science, Technology and Research, Sri Lanka; the Swiss Funding Agencies (ETH Board, ETH Zurich, PSI, SNF, UniZH, Canton Zurich, and SER); the Ministry of Science and Technology, Taipei; the Ministry of Higher Education, Science, Research and Innovation, and the National Science and Technology Development Agency of Thailand; the Scientific and Technical Research Council of Turkey, and Turkish Energy, Nuclear and Mineral Research Agency; the National Academy of Sciences of Ukraine; the Science and Technology Facilities Council, UK; the US Department of Energy, and the US National Science Foundation.

Individuals have received support from the Marie-Curie program and the European Research Council and Horizon 2020 Grant, contract Nos.\ 675440, 724704, 752730, 758316, 765710, 824093, and COST Action CA16108 (European Union) the Leventis Foundation; the Alfred P.\ Sloan Foundation; the Alexander von Humboldt Foundation; the Belgian Federal Science Policy Office; the Fonds pour la Formation \`a la Recherche dans l'Industrie et dans l'Agriculture (FRIA-Belgium); the Agentschap voor Innovatie door Wetenschap en Technologie (IWT-Belgium); the F.R.S.-FNRS and FWO (Belgium) under the ``Excellence of Science -- EOS" -- be.h project n.\ 30820817; the Beijing Municipal Science \& Technology Commission, No. Z191100007219010; the Ministry of Education, Youth and Sports (MEYS) of the Czech Republic; the Shota Rustaveli National Science Foundation, grant FR-22-985 (Georgia); the Hungarian Academy of Sciences, the New National Excellence Program - \'UNKP, the NKFIH research grants K 124845, K 124850, K 128713, K 128786, K 129058, K 131991, K 133046, K 138136, K 143460, K 143477, 2020-2.2.1-ED-2021-00181, and TKP2021-NKTA-64 (Hungary); the Council of Scientific and Industrial Research, India; the Latvian Council of Science; the Ministry of Education and Science, project no. 2022/WK/14, and the National Science Center, contracts Opus 2021/41/B/ST2/01369 and 2021/43/B/ST2/01552 (Poland); the Funda\c{c}\~ao para a Ci\^encia e a Tecnologia, grant FCT CEECIND/01334/2018; the National Priorities Research Program by Qatar National Research Fund; the Programa Estatal de Fomento de la Investigaci{\'o}n Cient{\'i}fica y T{\'e}cnica de Excelencia Mar\'{\i}a de Maeztu, grant MDM-2017-0765 and projects PID2020-113705RB, PID2020-113304RB, PID2020-116262RB and PID2020-113341RB-I00, and Programa Severo Ochoa del Principado de Asturias (Spain); the Chulalongkorn Academic into Its 2nd Century Project Advancement Project, and the National Science, Research and Innovation Fund via the Program Management Unit for Human Resources \& Institutional Development, Research and Innovation, grant B05F650021 (Thailand); the Kavli Foundation; the Nvidia Corporation; the SuperMicro Corporation; the Welch Foundation, contract C-1845; and the Weston Havens Foundation (USA).\end{acknowledgments}

\clearpage
\refstepcounter{section}
\addcontentsline{toc}{section}{References}
\bibliography{auto_generated}
\cleardoublepage \appendix\section{The CMS Collaboration \label{app:collab}}\begin{sloppypar}\hyphenpenalty=5000\widowpenalty=500\clubpenalty=5000
\cmsinstitute{Yerevan Physics Institute, Yerevan, Armenia}
{\tolerance=6000
A.~Hayrapetyan, A.~Tumasyan\cmsAuthorMark{1}\cmsorcid{0009-0000-0684-6742}
\par}
\cmsinstitute{Institut f\"{u}r Hochenergiephysik, Vienna, Austria}
{\tolerance=6000
W.~Adam\cmsorcid{0000-0001-9099-4341}, J.W.~Andrejkovic, B.~Arnold, H.~Bergauer, T.~Bergauer\cmsorcid{0000-0002-5786-0293}, S.~Chatterjee\cmsorcid{0000-0003-2660-0349}, K.~Damanakis\cmsorcid{0000-0001-5389-2872}, M.~Dragicevic\cmsorcid{0000-0003-1967-6783}, A.~Escalante~Del~Valle\cmsorcid{0000-0002-9702-6359}, P.S.~Hussain\cmsorcid{0000-0002-4825-5278}, M.~Jeitler\cmsAuthorMark{2}\cmsorcid{0000-0002-5141-9560}, N.~Krammer\cmsorcid{0000-0002-0548-0985}, D.~Liko\cmsorcid{0000-0002-3380-473X}, I.~Mikulec\cmsorcid{0000-0003-0385-2746}, J.~Schieck\cmsAuthorMark{2}\cmsorcid{0000-0002-1058-8093}, R.~Sch\"{o}fbeck\cmsorcid{0000-0002-2332-8784}, D.~Schwarz\cmsorcid{0000-0002-3821-7331}, M.~Sonawane\cmsorcid{0000-0003-0510-7010}, S.~Templ\cmsorcid{0000-0003-3137-5692}, W.~Waltenberger\cmsorcid{0000-0002-6215-7228}, C.-E.~Wulz\cmsAuthorMark{2}\cmsorcid{0000-0001-9226-5812}
% ...
\par}
\vskip\cmsinstskip
\dag:~Deceased\\
$^{1}$Also at Yerevan State University, Yerevan, Armenia\\
$^{2}$Also at TU Wien, Vienna, Austria\\
% ...
\end{sloppypar}
%%% END EDITABLE REGION %%%
% skeleton_end
\end{document}